\documentclass[11pt]{article}
\usepackage{eurosym}
\usepackage{amsfonts}
\usepackage{amssymb}
\usepackage{graphicx}
\usepackage{amsmath}
\usepackage{makeidx}
\usepackage{indentfirst}
\usepackage[T1]{fontenc}
\usepackage[utf8]{inputenc}

\newcounter{resultnum}[section]
\newcounter{conclusionnum}[section]
\newcounter{conditionnum}[section]
\newcounter{conjecturenum}[section]
\newcounter{examplenum}[section]
\newcounter{exercisenum}[section]
\newcounter{lemmanum}[section]
\newcounter{notationnum}[section]
\newcounter{theoremnum}[section]
\newcounter{definitionnum}[section]
\newcounter{corollarynum}[section]
\newcounter{remarknum}[section]
\newcounter{propositionnum}[section]
\newcounter{acknowledgementnum}[section]
\newcounter{algorithmnum}[section]
\newcounter{axiomnum}[section]
\newcounter{casenum}[section]
\newcounter{claimnum}[section]
\newcounter{summarynum}[section]
\newcounter{problemnum}[section]
\begin{document}

\title{The Anholonomic Frame and Connection Deformation Method for
constructing off-diagonal solutions in (modified) Einstein gravity and
nonassociative geometric flows and Finsler-Lagrange-Hamilton theories}
\date{March 20 and published online September 22, 2025}
\author{{\textbf{Lauren\c{t}iu Bubuianu}\thanks{%
email: laurentiu.bubuianu@tvr.ro and laurfb@gmail.com}} \and {\small \textit{%
SRTV - Studioul TVR Ia\c{s}i} and \textit{University Appolonia}, 2 Muzicii
street, Ia\c{s}i, 700399, Romania} \vspace{.1 in} \and {\textbf{Julia O.
Seti }} \thanks{%
email: yuliia.o.seti@lpnu.ua and jseti18@gmail.com} \\
{\small \textit{\ Department of Applied Mathematics, Lviv Polytechnic
National University, Lviv, 79013, Ukraine}} \vspace{.1 in} \\
{\textbf{Douglas Singleton }} \thanks{%
email: dougs@mail.fresnostate.edu} \\
{\small \textit{Department of Physics,\ California State University Fresno,
Fresno, CA 93740-8031, USA}} \vspace{.1 in} \\
{\textbf{Panayiotis Stavrinos}} \thanks{%
email: pstavrin@math.uoa.gr} \\
{\small \textit{Department of Mathematics, National \& Kapodistrian
University, Panepistimiopolis 15784, Athens, Greece}} \vspace{.1 in} \\
\textbf{Sergiu I. Vacaru} \thanks{%
emails: sergiu.vacaru@fulbrightmail.org ; sergiu.vacaru@gmail.com } \\
{\small \textit{Department of Physics, California State University at
Fresno, Fresno, CA 93740, USA; }} \\
{\small \textit{\ Department of Physics,\ Kocaeli University, 41380,
Izmit,Turkey}} \vspace{.1 in} \\
{\textbf{El\c{s}en Veli Veliev }} \thanks{%
email: elsen@kocaeli.edu.tr and elsenveli@hotmail.com} \\
{\small \textit{\ Department of Physics,\ Kocaeli University, 41380,
Izmit,Turkey }} }
\maketitle

\begin{abstract}
This article is a status report on the Anholonomic Frame and Connection
Deformation Method, AFCDM, for constructing generic off-diagonal exact and
parametric solutions in general relativity, GR, relativistic geometric flows
and modified gravity theories, MGTs. Such models can be generalized to
nonassociative and noncommutative star products on phase spaces and modelled
equivalently as nonassociative Finsler-Lagrange-Hamilton geometries. Our
approach involves a nonholonomic geometric reformulation of classical models
of gravitational and matter fields described by Lagrange and Hamilton
densities on relativistic phase spaces. Using nonholonomic dyadic variables,
the Einstein equations in GR and MGTs can be formulated as systems of nonlinear
partial differential equations (PDEs), which can be decoupled and integrated
in some general off-diagonal forms. In this approach, the Lagrange and
Hamilton dynamics and related models of classical and quantum evolution, are
equivalently described in terms of generalized Finsler-like or canonical
metrics and (nonlinear) connection structures on deformed phase spaces
defined by solutions of modified Einstein equations. New classes of exact
and parametric solutions in (nonassociative) MGTs are formulated in terms of
generating and integration functions and generating effective/ matter
sources. The physical interpretation of respective classes of solutions
depends on the type of (non) linear symmetries, prescribed boundary/
asymptotic conditions, or posed Cauchy problems.

We consider possible applications of the AFCDM with explicit examples of
off-diagonal deformations of black holes, cylindrical metrics and wormholes,
black ellipsoids, and torus configurations. In general, such solutions encode
nonassociative and/or nonholonomic geometric flow variables. For another types of
generic off-diagonal (nonassociative) solutions, we study models with
nonholonomic cosmological solitonic and spheroid deformations involving
vertices and solitonic vacua for voids. We emphasize that such new classes
of generic off-diagonal solutions can not be considered, in general, in the
framework of the Bekenstein-Hawking entropy paradigm. This motivates
relativistic/ nonassociative phase space extensions of the G. Perelman
thermodynamic approach to geometric flows and MGTs defined by nonholonomic
Ricci solitons. In the Appendix, Tables 1-16, summarize the AFCDM for various
classes of quasi-stationary and cosmological solutions in MGTs with 4-d and
10-d spacetimes and (nonassociative) phase space variables on (co) tangent
bundles.

\vskip5pt

\end{abstract}

\tableofcontents




\section{Introduction, historical remarks and preliminaries}

\label{sec1} It is well known that exact solutions play a very important
role in gravity theories. For Einstein's gravity, there are important
textbooks \cite{kramer03,misner,hawking73,wald82} summarizing most important
physical solutions and the methods for constructing such solutions which
typically are defined by certain diagonalizable ansatz for metrics with
certain prescribed global and local symmetries. In this work, we review and
provide a series of new results on geometric and analytic methods for
constructing exact and parametric generic off-diagonal solutions in general
relativity, GR, and modified gravity theories, MGTs. We note that a metric
is generic off-diagonal if it can't be diagonalized by coordinate transforms
in finite spacetime/phase space region. Such extra dimension (super)
gravity, string gravity or nonassociative and noncommutative theories can be
elaborated as effective ones defined on pseudo-Riemannian or metric-affine
(with independent metric and linear connection structures) spaces of
dimensions 4-11 and on relativistic eight-dimensional, 8-d, phase spaces. A
spacetime in GR is modelled as a four-dimensional, 4-d, Lorentz spacetime
manifold enabled with a symmetric metric field signature $(+,+,+,-) $ which
must be a solution of the Einstein equations. MGTs can be formulated in
abstract and adapted frame forms, in general, on higher dimension Lorentz
manifolds and/or on (co) tangent Lorentz bundles. They may involve
nonholonomic metric-affine structures with general nonsymmetric metrics and
nonlinear connections, N-connections, characterized by nontrivial torsion
and nonmetricity fields. For reviews of such results and methods, we cite 
\cite{partner01,partner02,vacaru18,bubuianu18a}. Respective geometric
constructions can be performed in coordinate free (or in some special
coordinates which allow us to find solutions in explicit forms) for various
low and high dimensions when physical theories are defined on phase spaces
enabled with conventional velocity/ momentum coordinates.

\vskip5pt In this work, a geometric formalism with nonholonomic
distributions defining N-connections structures and associated nonholonomic
(co) frames defining conventional dyadic decompositions, i.e.
(2+2)-splitting is outlined. Here we note that in mathematical and physical
literature, there are used equivalent terms like anholonomic, i.e.
non-integrable, variables/coordinates. In the first part of this paper, the
most important results are formulated in abstract geometric form when
necessary details and dyadic frame formulas are provided for 4-d spacetimes
and (modified) gravity theories. In the second part of the paper and in
appendix, we also explain how using abstract geometric/symbolic
constructions, the approach can be extended for extra dimensions and/or on
(co) tangent Lorentz bundles. In all cases, respective classes of
off-diagonal solutions are generated for oriented shells of dyadic
decompositions of type (2+2+2+...). For our approach to nonassociative and
noncommutative phase space theories, the geometric constructions are defined
by star product R-flux deformations in string theory. To characterize the
physical properties of new classes of solutions of physically important
systems of nonlinear partial differential equations, PDEs, on such
nonholonomic spacetime manifolds and phase spaces the geometric
constructions have to be extended for theories of nonholonomic geometric
flows on a real (temperature-like) parameter and corresponding statistical
and geometric thermodynamic models. There are defined nonholonomic frame
transforms and canonical deformations of linear connections for geometric
constructions adapted to an N-connection splitting. In such nonholonomic
variables, various physically important systems of nonlinear PDEs (for
instance, modified Einstein and geometric flow equations) can be decoupled
and integrated, i.e. solved, in certain general forms defining exact or
parametric solutions determined by generic off-diagonal metrics and
generalized connections. The Levi-Civita, LC, configurations with zero
torsion can be extracted by imposing additional nonholonomic constraints.
The coefficients of nonholonomic geometric objects constructed for such
classes of generic off-diagonal solutions depend, in general, on all
spacetime coordinates. During the last 25 years, in a series of tenths of
our and co-authors' works, such a geometric technique was concluded as the
Anholonomic Frame and Connection Method, AFCDM, for constructing solutions
in geometric flow and gravity theories.

\vskip5pt We note that the AFCDM \cite%
{partner02,vacaru18,bubuianu18a,partner03,partner04} is very different from
the other well-known geometric, analytic and numeric methods on constructing
exact solutions in gravity outlined, for instance, in \cite{kramer03} and 
\cite{misner,hawking73,wald82}. Usually, the books on GR and higher
dimension gravity theories summarize the methods and physically important
results developed for some diagonal ansatz of metrics when the Einstein
equations are transformed into some systems of nonlinear ordinary
differential equations, ODEs. In our approach, we elaborated on more general
geometric and analytic methods for generating directly (not reducing to
ODEs) off-diagonal solutions of nonlinear PDEs encoding general classes of
nonholonomic deformations of gravitational and matter field equations in GR
and MGTs. The main goal of this status report is to outline the AFCDM and
related constructions for 4-d (modified) Einstein gravity and analyze a
series of new and physically important examples of generic off-diagonal
exact and parametric solutions. We also show how the approach can be
extended in abstract geometric form to higher dimensions and on (co) tangent
Lorentz bundles, for more general MGTs when generating functions and
effective sources may encode nonassociative and noncommutative data for
nonholonomic geometric flows and generalized Finsler variables. Such
constructions are outlined in Tables 1-16 in the Appendix. The data from
such tables can be generalized for other classes of effective sources (they
may encode quantum deformations; supersymmetric or spinor variables;
functional dependencies; thermodynamic variables; other type nonlinear and
algebraic or group structures etc.) we can generate off-diagonal solutions
for corresponding physically important systems of nonlinear PDEs.
Additionally to the bibliography presented in reviews \cite%
{partner01,partner02,vacaru18,bubuianu18a}, we cite a series of works
published during the last 30 years by authors from Eastern Europe. Such
results and methods are less known in Western Countries and we include and
summarize them in this status report (that why almost a half of citations
are related to our contributions and collaborations). For nonassociative
phase space theories and related canonical Finsler-Lagrange-Hamilton
variables, we shall cite only the works (by other authors), which are
closely related to applications and developments of the AFCDM.

\vskip5pt Following only standard Lagrange or Hamilton formulations (for
instance, using the Wheeler - DeWitt, or Ashtekar formulations) to construct
in explicit form certain off-diagonal solutions of nonlinear PDEs and
consider quantum deformations of such systems are not possible. In canonical
dyadic and Finsler like variables, we can formulate new geometric and
analytic methods of finding solutions of nonlinear classical and quantum
functional physically important systems of nonlinear PDEs. The AFCDM
provides a new geometric technique for constructing general forms of various
classes of solutions of nonlinear systems PDEs for (nonassociative)
geometric flows and MGTs. This approach uses not only special diagonal
ansatz for metrics transforming PDEs into ordinary differential equations,
ODEs, but also uses various off-diagonal ansatz for metrics and auxiliary
connections, allowing direct integration of physically important systems.
The geometric constructions are performed in abstract and adapted frame
forms for 4-d and higher dimension Lorentz manifolds and their (co) tangent
bundles. Such spacetimes and generalized (star product deformed) phase
spaces can be endowed with conventional nonholonomic distributions defining
dyadic splitting of type 2+2+2+... of the total phase space and spacetime
dimensions. The main idea is to define and use for such a splitting an
auxiliary canonical distinguished connection, d-connection, and respective
nonholonomic frames which allow us to decouple and integrate (modified)
Einstein equations in general forms. Off-diagonal Levi-Civita, LC,
configurations with zero torsion can be also extracted by imposing
additional nonholonomic constraints on some more general classes of
solutions.

\subsection{Diagonal ansatz reducing (modified) Einstein equations to
nonlinear ODEs}

The most important geometric and analytic methods for constructing exact
solutions (and/or with some decompositions on constant parameters) in GR are
summarized and discussed in standard monographs, for instance, \cite%
{kramer03,misner,hawking73,wald82}. In \cite{misner}, the Einstein equations
are formulated in abstract geometric form, 
\begin{equation}
En=Ric-\frac{1}{2}\mathbf{g}Rsc+\Lambda \mathbf{g}=\frac{8\pi G}{c^{4}}Tm,
\label{en1}
\end{equation}%
with a cosmological constant $\Lambda $ and an energy-momentum tensor $Tm$.
In 4-d, this consists a system of nonlinear PDEs for six independent
components of the metric tensor $\mathbf{g}=g_{\alpha \beta
}(u^{\gamma})e^{\alpha }\otimes e^{\beta }$. In equations (\ref{en1}), the
metric compatible and zero torsion Levi-Civita, LC, connection, $\nabla
\lbrack \mathbf{g}]$, is used. We follow such conventions (see details in
next sections):\ The systems of coordinates and indices are labeled as $%
u^{\gamma}=(u^{1},u^{2},u^{3},u^{4}=t)$. For higher dimensions, we can write 
$u^{5},u^{6},...$ considering metrics of different signatures etc. We shall
use also $u^{\gamma }=\allowbreak (x^{i},u^{5},u^{6}),$ for $i=1,2,3,4;$ or $%
u=(x,u^{5},u^{6}).$ Usually we state that the light velocity constant $c=1$
excepting some formulas when it will be physically important to write $c$.
Various types (not) primed, underlined etc. indices may run values of type $%
\alpha ,\beta ,...,\alpha ^{\prime },\beta ^{\prime },..=1,2,3,4$. The
Einstein convention on up-low indices is used. Frame transforms are defined
as $e^{\alpha }=e_{\ \alpha ^{\prime }}^{\alpha }(u^{\gamma ^{\prime
}})du^{\alpha ^{\prime }}$ and $e_{\beta }=e_{\beta }^{\ \beta ^{\prime
}}(u^{\gamma ^{\prime }})\partial _{\beta ^{\prime }},$ for $\partial
_{\beta ^{\prime }}=\partial /\partial u^{\beta ^{\prime }}$ and $e_{\
\alpha ^{\prime }}^{\alpha }e_{\beta }^{\ \alpha ^{\prime }}=\delta
_{\beta}^{\alpha },$ where $\delta _{\beta }^{\alpha }\,$ is the Kronecker
symbol. From 10 components of a symmetric tensor $g_{\alpha \beta
}(u^{\gamma }),$ there are 6 independent ones because 4 of them can
transformed in zero using coordinate transforms as consequence of the
Bianchi identities for (pseudo) Riemannian spaces. The coefficients of the
Ricci tensor for $\nabla $ are $Ric=R_{\alpha \beta }e^{\alpha }\otimes
e^{\beta },$ the curvature scalar $Rsc:=g^{\alpha \beta }R_{\alpha \beta },$
and $Tm=\{T_{\alpha \beta }\}$ is the symmetric energy-momentum tensor for
matter (with $G$ being the gravitational/Newton constant). Usually, we shall
use abstract index, or abstract not index, formulas as in \cite{misner}.
Nevertheless certain necessary coordinate and abstract index formulas for
general and N-adapted frames will be used when certain special types of
indices/ coordinates are important for constructing explicit classes of
solutions. The bulk of known and physically important exact solutions of (%
\ref{en1}) were constructed for diagonal ansatz of metrics, motivated by
certain assumptions on symmetries of gravitational and matter field
interactions, when corresponding systems of nonlinear PDEs are transformed
into systems of nonlinear ordinary differential equations, ODEs. Such
equations can be integrated (i.e. solved) in certain general forms depending
on respective integration constants and physical parameters. The integration
constants can be related to certain physical constants using respective
boundary/asymptotic conditions, some prescribed data for Cauchy systems etc.
Physically important solutions are selected to define some well-defined and
verifiable physical models. For instance, such constructions must be with
relativistic causality, to satisfy some positive entropy conditions and
allow to define thermodynamic variables. Typical models are elaborated, for
instance, for some positive energy conditions, with the goal to avoid
singularities at least in some observable regions, etc.

\vskip3pt As the most important example of a solution generated by a
diagonal ansatz and ODEs in GR, we can consider the Schwarzschild black
hole, BH, metric. It is for the vacuum Einstein spaces, when (\ref{en1})
transforms into $Ric=0$. Using spherical coordinates $u^{\alpha}=(r,\theta,%
\varphi ,t)$, such a diagonal static solution can be written as%
\begin{equation}
g_{\alpha \beta }=diag[g_{1}(r)=(1-\frac{r_{s}}{r})^{-1},g_{2}(r,\theta
)=r^{2},g_{3}(u^{\gamma })=r^{2}\sin ^{2}\theta ,g_{4}(u^{\gamma})=-f(r)=-(1-%
\frac{r_{s}}{r})],  \label{sch}
\end{equation}%
for the quadratic line element 
\begin{equation*}
ds_{Sch}^{2}=g_{\alpha }(u^{\gamma
})[du^{\alpha}]^{2}=g_{1}dr^{2}+g_{2}d\theta ^{2}+g_{3}d\varphi
^{2}+g_{4}dt^{2}.
\end{equation*}%
In (\ref{sch}), the Schwarzschild (horizon) radius $r_{s}=2Gm/c^{2}$ is
determined by the condition that for $r\gg r_{s}$ such a metric defined the
Newton gravitational potentials for a point mass $m.$ We emphasize that in
corresponding chosen coordinate bases the coefficients of the Schwarzschild
metric do not depend on the time coordinate $t,$ i.e. it posses a Killing
symmetry on time like vector $\partial _{4}=\partial _{t}.$

\vskip3pt Another important example of a diagonal ansatz used for
constructing homogeneous and isotropic cosmological models in GR and MGTs is
that for the Friedman-Lema\^{\i}tre-Robertson-Walker, FLRW, spaces,%
\begin{equation}
g_{\alpha \beta }=diag[g_{1}(r,t)=a^{2}(t)/(1-\epsilon
r^{2}),g_{2}(r,t)=a^{2}(t)r^{2},g_{3}(r,\theta ,t)=a^{2}(t)r^{2}\sin
^{2}\theta ,g_{4}=-1.  \label{flrw}
\end{equation}%
In this quadratic line element, the constant $\epsilon $ represents the
curvature of the space (it can be taken $0,\pm 1$) and the "scale factor" $%
a(t)$ should be, for instance, a solution of the Einstein equations (\ref%
{en1}), when the energy- momentum tensor is taken in a form 
\begin{equation}
T_{\alpha \beta }=diag[P,P,P,\rho ]  \label{fluidm}
\end{equation}%
for a fluid type matter with pressure $P$ and energy density $\rho .$

\vskip3pt In higher dimension gravity theories, the diagonal ansatz (\ref%
{sch}) and/or (\ref{flrw}) were correspondingly generalized for extra
dimension coordinates (spherical, cylindrical and other higher symmetry
ones) but keeping the property to be diagonalizable by certain coordinate
transforms. That allowed to construct a number of BH, wormhole (WH) and
cosmological solutions in (super) string and MGTs and exploited, for
instance, in modern cosmology and astrophysics.

\vskip3pt It is very difficult to construct in explicit forms exact or
parametric off-diagonal solutions of systems of coupled nonlinear PDEs of
type (\ref{en1}) and their higher dimension generalizations, or in MGTs. The
parametric solutions may be also exact for a fixed value and order of a
physical parameter (like the Planck and string constants, or other ones with
possible polarizations; in this work, we consider only classical models even
certain nonassociative contributions may be determined by star products
involving the imaginary unity). The main property of ansatz of types (\ref%
{sch}) and/or (\ref{flrw}) is that they reduce the gravitational field
equations to some systems of nonlinear ODEs, which can be integrated in
certain general or approximate forms determined by integration constants.
Their physical interpretation depends on the types of assumptions on
symmetries, boundary/assimptotic conditions and/or how a corresponding
Cauchy problem is solved (all such conditions are stated following certain
geometric/ physical considerations). The cosmological constant and the data
for an energy momentum tensor can be considered respectively as effective /
matter generating sources. Usually, this type of solutions in GR involve
certain singularities and horizons.

\vskip3pt The constructions can be generalized to spaces of higher dimension
and for various modifications of gravity and matter field theories, with
possible quantum corrections, additional terms with supersymmetric and
superstring contributions, nonassociative/noncommutative generalizations
etc. Such solutions were found, generalized, and studied intensively in GR
and MGTs during the last 100 years. There were formulated a series of
geometric and physically important theorems on BH singularities, cosmic
censorships, conditions of stability, scenarios of evolution/ inflation /
acceleration etc. In this approach, the basic ideas and principles for
constructing off-diagonal solutions of (modified) Einstein equations can be
stated in this form:%
\begin{eqnarray*}
&&%
\mbox{\bf Principles 1 (reducing PDEs to ODEs and constructing diagonal
solutions):} \\
&&\left[ 
\begin{array}{c}
\mbox{system of  nonlinear PDEs}, \\ 
\mbox{(modified) Einstein eqs. }(\ref{en1})%
\end{array}%
\right] \Rightarrow \\
&&\left[ 
\begin{array}{c}
\mbox{frame/coordinate transforms and symmetries:} \\ 
\mbox{ spherical/cylindrical, Killing, Lie algebras,} \\ 
\mbox{diagonal ansatz}\ g_{\alpha }(u^{\gamma })%
\end{array}%
\right] \Rightarrow \left[ 
\begin{array}{c}
\mbox{\bf integrable systems of} \\ 
\mbox{\bf nonlinear ODEs}%
\end{array}%
\right] \\
&&{\qquad }{\qquad }{\qquad }{\qquad }{\qquad }{\qquad }{\qquad }{\qquad }{%
\qquad }{\qquad }\mathbf{\Downarrow } \\
&&\left[ 
\begin{array}{c}
\mbox{special generating functions }f(r),\mbox{or }a(t), \\ 
\begin{array}{c}
\mbox{ non-structure vacuum, or special generating sources}:\ \Lambda
,diag[P,P,P,\rho ];%
\end{array}
\\ 
\begin{array}{c}
\mbox{ integration constants determined by boundary/asymptotic conditions,
Cauchy problems, }%
\end{array}
\\ 
\begin{array}{c}
\mbox{horizons, singularities, BH theorems, hyper-surface  thermodynamics,
cosmic censorship etc.}%
\end{array}%
\end{array}%
\right] .
\end{eqnarray*}

Such methods allow us to elaborate on nonlinear physical models for
gravitational and matter fields interactions determined by solutions of some
classes of ODEs, when from six independent components of metrics there are
chosen only a few diagonal components (maximum 4, for a 4-d spacetime) of
"diagonalizable metrics". For instance, such a metric is defined by a
function $f(r)$, or $a(t)$, when the presence of other coordinates is
motivated by choosing certain spherical/ cylindrical/ ellipsoidal/ toroid
systems of coordinates and/or considering some frame/coordinate transforms.
We emphasize that prescribing such an ansatz, we "cut" other possibilities
to find more general classes of solutions depending, for instance, on all
spacetime coordinates and when the metrics contain generic off-diagonal
terms (for instance, with 6 independent coefficients). In GR, any metric can
be diagonalized in a point, or along a geodesics, and represented as a
standard diagonal Minkowski metric, $\eta _{\alpha \beta}=diag[1,1,1,-1]$. A
general pseudo-Riemannian metric can be represented in a diagonal form with
coefficients depending on spacetime coordinated with respect to certain
nonholonomic frames. Some such generalized ansatz, for instance, with
necessary 2+2 decompositions can be convenient for constructing new classes
of diagonal and off-diagonal solutions. This is the price we should pay in
order to solve systems of nonlinear PDEs by reducing them to more simple
systems of nonlinear ODEs. Nevertheless, even in such cases, there were
elaborated a number of on physically important gravitational models with
applications in modern astrophysics and cosmology. The bulk of experimental
and observational verifications, theoretical constructions and applications
in modern gravity / particle physics / cosmology were performed using
systems of (non) linear wave/oscilator equations, respective BH and
cosmological ODEs, and their (superpositions of) solutions. Modern
approaches to accelerating cosmology, dark matter and energy physics and
related plethora of MGTs request more advanced geometric, analytic and
numeric methods for generating exact and parametric solutions in physically
important nonlinear systems of PDEs not constraining the "geometric optics
and methodology" only via ODEs.

\subsection{Physical and geometric motivations for constructing off-diagonal
solutions}

In GR and MGTs, there were found more general classes of solutions with
geometric and physical properties which are different than those stated
above. For instance, the Kerr solution for rotating BHs contains
odd-diagonal terms and ellipsoidal ergo-spheres, which are induced in
rotation frames. There were found examples of solutions for black rotoids/
toroids, wormholes etc. with various types of broken/nonlinear symmetries
describing locally anisotropic matter interactions and respective
inhomogeneous/ anisotropic cosmological models (a number of examples are
reviewed in monograph \cite{kramer03}). We cite \cite%
{misner,hawking73,wald82} for the main concepts, methods, interpretations
and discussions of most physical important solutions and models. In such
monographs, there are provided certain examples of solutions for
gravitational nonlinear waves and solitons when the coefficients of metrics
depend on 2 or 3 spacetime coordinates, with parametric dependencies, and
may involve certain off-diagonal terms. They were constructed using some
special methods for generating solutions of nonlinear PDEs, for instance,
with so-called LA symmetries and solitonic hierarchies, see details in \cite%
{av06,sv08,sv15,biv17} and references therein.

\vskip3pt Nevertheless, during many years of research on mathematics and
physics of gravitational field equations, it was not formulated a general
geometric and analytic method for constructing generic off-diagonal
solutions with dependence on all spacetime coordinates in GR and MGTs. The
solutions with maximal, or with many, degrees of freedom (for 4-d gravity
theories, being considered 6 independent components of metrics) are of
crucial importance if we try to explore and solve a series of fundamental
problems in nonlinear physics and elaborate quasi-classical models of
gravity an quantum gravity, QG. For such models, generic off-diagonal
interactions and nonholonomic constraints are important in the
non-perturbative and nonlinear regimes which should test QG and higher
dimension theories. Off-diagonal symmetric and nonsymmetric metrics and
generalized (non) linear connections are used for elaborating realistic and
modified gravity models for inhomogeneous/ anisotropic/ acceleration
cosmology; to construct dark energy and dark matter theories with
quasi-periodic structure and pattern forming, filaments, vortices, solitons
etc. In such cases, we can't work only with "simplified" diagonal ansatz
reducing gravitational and matter field equations to certain systems of
nonlinear ODEs. We have to elaborate new methods which allow to construct
generic off-diagonal solutions, with constraints and generating functions
and sources, solving in direct form respective systems of nonlinear PDEs.

\vskip3pt A series of our works were devoted to constructing new classes of
exact solutions in GR and MGTs of 4-d, 5-d spacetimes and 8-d phase spaces
with warped dimensions, and further nonassociative/ supersymmetric
generalizations for string and generalized Finsler geometry \cite%
{sv00,sv00a,vp,vt,vs01a,vs01b,sv03a,sv03b,sv05,vmon3,sv07,sv11,vvy13,bubuianu17,vacaru18,bubuianu18a,bubuianu20,bubuianu19, partner01,partner02, partner03,partner04}%
. Considering nonholonomic 2(3)+2+2+.... splitting of dimensions by a
so-called nonlinear connection, N-connection, structure, $\mathbf{N,}$ we
defined an auxiliary connection, called as the canonical distinguished, d,
connection, 
\begin{equation}
\widehat{\mathbf{D}}[\mathbf{g}]=\nabla \lbrack \mathbf{g}]+\widehat{\mathbf{%
Z}}[\mathbf{g}],  \label{cdist}
\end{equation}%
with the canonical distortion d-tensor, $\widehat{\mathbf{Z}}$, when all
three geometric objects are determined by the same metric structure $\mathbf{%
g}$. For geometric objects adapted to a N-connection, we use bold face
symbols. In next section, we provide all necessary definitions and
abstract/index formulas. With respect to so-called nonholonomic N-adapted
frames, we can decouple in some general forms the modified Einstein
equations, 
\begin{equation}
\widehat{\mathbf{E}}n[\mathbf{g,}\widehat{\mathbf{D}}]=\widehat{\mathbf{R}}%
ic-\frac{1}{2}\mathbf{g}\widehat{R}sc=\widehat{\mathbf{Y}}[\mathbf{g,}%
\widehat{\mathbf{D}}].  \label{cdeq}
\end{equation}%
The geometric objects in such a nonlinear system of PDEs with nonholonomic
variables and respective constraints, and canonical distortion (\ref{cdeq})
of the LC-connection $\nabla $, are determined by geometric data $[\mathbf{g}%
,\widehat{\mathbf{D}}].$ The canonical effective source $\widehat{\mathbf{Y}}
$ encodes possible deformations of the standard energy momentum tensor in (%
\ref{en1}), terms like $-\Lambda \mathbf{g}$ and distortions of the Einstein
tensor determined by (\ref{cdist}). For MGTs, there are included
contributions from generalized gravitational and matter fields Lagrangians,
higher dimension corrections, (non) associative/ (non) commutative /
supersymmetric contributions from string/M-theory, in
Finsler-Lagrange-Hamilton gravity etc.

\vskip5pt Having decoupled in a general form the system of nonlinear and
nonholonomic PDEs (\ref{cdeq}), it is possible to solve it in certain exact
or parametric forms. This way, we can construct generic off-diagonal
solutions $\mathbf{g}$ determined by corresponding classes of generating and
integration functions, effective generating sources $\widehat{\mathbf{Y}},$
and corresponding N-connection, $\mathbf{N,}$ splitting. It should be noted
that such solutions involve a canonical d-torsion structure, $\widehat{%
\mathbf{T}}s$, which is determined by nonzero anholonomy coefficients if $%
\mathbf{N}$ are nontrivial, and related off-diagonal terms. Such
nonholonomic torsions are different from the torsion fields, for instance,
in the Einstein-Cartan and/or string gravity. We can extract, in general,
off-diagonal solutions $\mathbf{g}$ for the LC-connection $\nabla \lbrack 
\mathbf{g}]$ if we impose additional constraints on generating and
integration functions which result in zero distortion d-tensors $\widehat{%
\mathbf{Z}}[\mathbf{g}]$ in (\ref{cdist}),%
\begin{equation}
\widehat{\mathbf{Z}}=0,\mbox{ which is equivalent to }\ \widehat{\mathbf{D}}%
_{\mid \widehat{\mathbf{Ts}}=0}=\nabla .  \label{lccond}
\end{equation}

The basic ideas and principles are stated as 
\begin{eqnarray*}
&&%
\mbox{\bf Principles 2 - AFCDM: off-diagonal solutions, generalized
connections \& LC-connections} \\
&&\left[ 
\begin{array}{c}
\mbox{system of  nonlinear PDEs}, \\ 
\mbox{distorted Einstein eqs. }(\ref{cdeq})%
\end{array}%
\right] \Rightarrow \\
&&\left[ 
\begin{array}{c}
\mbox{frame/coordinate transforms, N-adapted } \widehat{\mathbf{D}}[\mathbf{g%
}]\ (\ref{cdist}) \\ 
\mbox{ nonlinear and Killing symmetries,  effective sources } \\ 
\mbox{off-diagonal ansatz}\ g_{\alpha \beta }(u^{\gamma })%
\end{array}%
\right] \Rightarrow \left[ 
\begin{array}{c}
\mbox{\bf decoupling and integrable } \\ 
\mbox{\bf systems of nonlinear PDEs}%
\end{array}%
\right] \\
&&{\qquad }{\qquad }{\qquad }{\qquad }{\qquad }{\qquad }{\qquad }{\qquad }{%
\qquad }{\qquad }\mathbf{\Downarrow } \\
&&\left[ 
\begin{array}{c}
\mbox{ generating and integration functions depending on spacetime
coordinates } \\ 
\begin{array}{c}
\mbox{ nontrivial vacuum, generating sources, effective cosmologial
constants};%
\end{array}
\\ 
\begin{array}{c}
\mbox{ integration functions determined by boundary/asymptotic conditions,
Cauchy problems, }%
\end{array}
\\ 
\begin{array}{c}
\mbox{horizons, singularities,  G. Perelman  thermodynamics, stability and
flow evolution, etc.}%
\end{array}%
\end{array}%
\right] \\
&&{\qquad }{\qquad }{\qquad }{\qquad }{\qquad }{\qquad }{\qquad }{\qquad }{%
\qquad }{\qquad }\mathbf{\Downarrow } \mbox{\ nonholonomic sLC-conditions } (%
\ref{lccond}) \\
&& {\qquad }{\qquad }{\qquad }{\qquad }{\qquad }{\qquad }{\qquad } \left[ 
\begin{array}{c}
\mbox{system of  nonlinear PDEs}, \\ 
\mbox{Einstein eqs. } (\ref{en1})%
\end{array}%
\right].
\end{eqnarray*}

\vskip3pt The first general goal of this article is to show how Principles 2
- AFCDM can be performed in explicit form for 4-d Lorentz manifolds with
nonholonomic 2+2 splitting and canonical distortion of the LC-connection. A
corresponding new methodology of constructing generic off-diagonal solutions
analyzing their possible physical implications in (modified) gravity will be
outlined. We shall provide and discuss a series of important physical
solutions related to BH physics and modern cosmology. Then, the second
general goal is to outline in brief (using abstract geometric methods) that
the AFCDM can be extended to higher dimensions using nonholonomic 2+2+2+
splitting. In the case of phase space theories, the geometric constructions
involve additional velocity/ momentum variables which is similar to
Finsler-Lagrange-Hamiton geometry. Here, we emphasize that if we restrict
our research only with "pure" Lagrange or Hamilton phase space theories
(which are very important for quantization) we are not able to unify the
spacetime and phase space geometry and physics and describe the
constructions in terms of off-diagonal solutions of (modified) Einstein
equations. In canonical nonholonomic variables, our approach can be
generalized for nonassociative and noncommutative gravity and geometric flow
theories.

\vskip3pt We review an unified geometric abstract formalism for 4-d gravity
theories in the Part 1 and extend the methods in abstract geometric forms
for 8-d phase spaces with nonassociative geometric flows and
Finsler-Hamilton-Lagrange variables in Part II. Finally, we outline and
summarize in Tables 1 - 16 from the Appendix how the AFCDM can be applied
for generating off-diagonal solutions in higher dimension theories and for
(nonassociative) phase space models, with corresponding formulas for 8-d and
10-d quasi-stationary and locally anisotropic cosmological configurations.

\subsection{Nonassociative (co) tangent Lorentz bundles and
Fins\-ler-Lagrange-Hamilton geometry}

The 4-d nonholonomic geometric constructions and the AFCDM can be extended
to nonassociative and noncommutative theories defined on generalized
spacetime and/ or phase space models. We cite here a series of (related to
our purposes) works on nonassociative gauge/ membrane theory and double
field theory constructed in the framework of string theory \cite%
{alvarez06,luest10,blumenhagen10,condeescu13,blumenhagen13,kupriyanov19a,szabo19,blumenhagen16,aschieri17}%
. Such nonassociative structures also arise in the world volume of a
D-brane, for open strings, and for the models with flux compactification for
closed strings.

\vskip3pt Our research program \cite%
{partner01,partner02,partner03,partner04,partner05,partner06} on
nonassociative geometry, physics and quantum information theory is based on
the approach to nonassociative gravity formulated for $\star $-product (%
\textit{i.e.} star-product) deformations determined by R-flux backgrounds in
string gravity \cite{blumenhagen16,aschieri17}. In a self-consistent form,
such nonassociative and noncommutative theories are modelled on a
conventional phase space $\ ^{\shortparallel }\mathcal{M}$. To include the
general relativity (GR) as a particular case we use model star deformed
phase spaces on cotangent bundle, $\ ^{\shortparallel }\mathcal{M}=T^{\ast }%
\mathbf{V,}$ on spacetime Lorentz manifold, $\mathbf{V}$. In our works, the
phase space dimension is $\dim \mathcal{M}=8,$ with total phase space local
coordinates labelled in the form $\ ^{\shortparallel }u^{\alpha
_{s}}=(x^{i_{s}},\ ^{\shortparallel }p_{a_{s}}),$ where $\ ^{\shortparallel }
$ on the left indicates a phase space with spacetime coordinates, $%
x^{j_{s}}=(x^{j},t)$, and \textit{complex} momentum coordinates, $\
^{\shortparallel }p_{a_{s}}=\ ip_{a_{s}}=(ip_{a},iE$), with $i^{2}=-1$.
Alternatively, we can consider also real momentum coordinates with $\
^{\shortmid }p_{a_{s}}=\ p_{a_{s}}=(p_{a},E$) as introduced in \cite%
{aschieri17} which we modified for nonholonomic configurations with
corresponding labels and boldface symbols. For classical models, we can
study only real deformations of the geometric and physical objects which are
nontrivial even, in general form, the star product structure involves the
complex unity. The phase space will be denoted $\ ^{\shortmid }\mathcal{M}$
if the momentum-like coordinates are real ones and labelled in the form $\
^{\shortmid }u^{\alpha _{s}}=(x^{i_{s}},\ ^{\shortmid }p_{a_{s}})$ (in
brief, $\ ^{\shortmid }u=(x,p))$. In original form, the nonassociative,
vacuum, gravitational equations, $\ ^{\shortparallel }Ric^{\star }[\
^{\shortparallel }\nabla ^{\star }]=0,$ were postulated as phase space $%
\star $-deformations of the standard Ricci tensor $Ric$ in GR. Here we note
that well-defined nonassociative Ricci tensors allow us to formulate
corresponding models of geometric flows as in \cite%
{partner03,partner04,partner05,partner06}. For a prescribed star product
(see definitions in next sections), the nonassociative tensor $\
^{\shortparallel }Ric^{\star }$ can be constructed for a unique
nonassociative Levi-Civita (LC) connection, $\ ^{\shortparallel }\nabla
^{\star },$ which is torsionless and metric compatible with the respective $%
\star $-deformed symmetric, $\ _{\star }^{\shortparallel }\mathbf{g,}$ and
nonsymmetric, $\ _{\star }^{\shortparallel }\mathbf{q},$ metric structures.
Using real phase space variables, we can write in the symbolic form $\
^{\shortmid }Ric^{\star }(\ ^{\shortmid }u)=\ ^{\shortmid }Ric^{\star }(x,p)$
for $\left( \ _{\star }^{\shortmid }\mathbf{g}(\ ^{\shortmid }u), \ _{\star
}^{\shortmid }\mathbf{q}(\ ^{\shortmid }u)\right) $ and $\
^{\shortmid}\nabla ^{\star }(\ ^{\shortmid }u),$ when the geometric objects
depend additionally on momentum-like coordinates $\ ^{\shortmid }u^{\alpha
_{s}}=(x^{i_{s}},p_{a_{s}})$. Such dependencies on velocity or momentum like
coordinated are considered in Finsler-Lagrange-Hamilton geometry and gravity 
\cite{vacaru96b,vacaru09a,bubuianu19} (when the generating functions and
respective nonlinear and linear connections are subjected to certain
homogeneity and nonholonomic conditions on $\mathcal{M}$ or $\ ^{\shortmid }%
\mathcal{M}$ and even on $\mathbf{V}$ with a corresponding nonholonomic). In
this work, we use and abstract geometric formalism when those constructions
can be extended by respective star products on $\mathcal{M}^{\star }$ or $\
^{\shortmid }\mathcal{M}^{\star }.$

\vskip3pt Nonassociative and noncommutative modified MGTs were formulated as
a type of bimetric gravity theory \cite{gheor14,luest21}. We cite \cite%
{rosen40,rosen77}, for commutative bimetric theories, and \cite%
{einstein25,einstein45,eisenhart51,eisenhart52}, for constructions when the
second metric structure can be nonsymmetric. Further developments for
commutative and nonassociative gravity were performed in \cite%
{moffat95,partner01}; when the geometric constructions on phase spaces $%
\mathcal{M}^{\star }$ or $\ ^{\shortmid }\mathcal{M}^{\star }$ enabled with $%
\star $-product structure. Such theories may involve also nonassociative
generalizations of relativistic and supersymmetric/ (non) commutative
Finsler-Lagrange-Hamilton spaces \cite{vacaru96b,vacaru09a,bubuianu19}. In
this paper, we consider Finsler-like geometric objects and variables and
generalize in nonassociative and noncommutative form such geometric models
with but re-define the formulas for nonlinear quadratic elements in a form
so that nonassociative phase space BH solutions can be generated as real
configurations coming from star R-flux deformations. We also compute
nontrivial phase space components of the nonsymmetric parts of the metric
which can occur in nonassociative gravity.

\vskip3pt The works \cite%
{blumenhagen16,aschieri17,partner01,partner02,partner04} raised three
important questions in nonassociative gravity:

\begin{enumerate}
\item How to formulate and understand the physical properties of generic
off-diagonal solutions for 4-d (\ref{en1}) and nonholonomic (\ref{cdeq}) and
their nonassociative phase space 8-d generalizations involving $\
^{\shortmid }Ric^{\star }(x,p)$ and respective $\ ^{\shortmid }\widehat{%
\mathbf{R}}ic^{\star }(x,p)$ (in general, phase space systems of nonlinear
PDEs may encode nontrivial nonassociative sources $\ ^{\shortmid }\mathbf{J}%
^{\star }(x,p)$)?

\item How to construct in explicit form classes of exact, physically
important solutions in nonassociative gravity and determine the physical
meaning of such solutions? \ Here we note that such exact or parametric
solutions, in general, are generic off-diagonal and when certain nontrivial
effective sources $\ ^{\shortmid }\mathbf{J}^{\star }(x,p)$ contain terms
defined by nonassociative star product and R-flux data and, via nonlinear
symmetries, relate various classes of generating functions and generating
sources to certain effective cosmological constants $\ ^{\shortmid }\Lambda. 
$

\item In \cite{partner02,partner04,partner05,partner06}, we proved that the
AFCDM can generalized for nonassociative gravity but new classes of
solutions are generic off-diagonal with nonassociative geometric objects of
type $\left(\ _{\star }^{\shortmid }\mathbf{g}(x,p),\ _{\star }^{\shortmid }%
\mathbf{q}(x,p)\right) $ and $\ ^{\shortmid }\widehat{\mathbf{D}}%
^{\star}(x,p)$ with generic dependence on extra-dimension and/ or momentum
like variables. Such solutions can't be interpreted in the framework of the
Bekenstein-Hawking paradigm \cite{bek1,bek2,haw1,haw2} because, in general,
they do not involve certain hypersurface, duality, or holographic
properties. To characterize the thermodynamic and informational properties
of such nonassociative solutions we must generalize the approach for
nonassociative and quantum information flows and study generalized models of
G. Pereman thermodynamics for Ricci flows \cite{perelman1}.
\end{enumerate}

\subsection{Motivations and the main hypothesis on Finsler-Lagrange-Hamilton
phase space geometries}

\label{motivhypoth}Various types of phase space gravity theories have been
elaborated with the aim of quantizing gravity and formulating quantum field
theories, QGTs, on curved spaces using certain classical and quantum
Lagrange and/or Hamilton formulations. Here we note the
Arnowit-Deser-Misner, ADM, approach with 3+1 spacetime splitting (for an
abstract geometric formulation see \cite{misner,wheeler68}) used in
canonical quantum gravity and further developments \cite%
{dirac69,dewitt67,ashtekar91,isham92}. There are quite different geometric
and quantum theoretic formalisms involving corresponding Lagrange density, $%
L(x,v)$, and/or Hamilton density, $H(x,p)$, on respective phase spaces. Such
constructions can be related via Legendre transforms, re-defined for
Poisson/ almost symplectic structures etc. which was applied with certain
success in formulating canonical, string, loop, gauge-like and other types
of approaches to QG, geometric quantization and deformation quantization
(DQ) etc. Nevertheless, even in the semi-classical limits, such theories
involve certain variants of modified Lagrange, Hamilton, Hamilton-Jacoby,
Wheeler -- De Witt, WDW, gauge gravity and other type of nonlinear
equations. It is not possible to decouple and solve in certain general forms
such as nonlinear and/or quantum systems of (functional) PDEs. For instance,
the WDW equation is a very sophisticated functional equation in the space of
metrics, which has deep consequences of the arrow of time problem in quantum
cosmology. A. Ashtekar introduced new variables by analogy to quantum
electrodynamics for certain connections like gravitational potentials which
was exploited in loop gravity and related theories. The existing methods do
not convert, or connect, the above mentioned approaches with conventional
phase space Lagrangians and Hamiltonians to the problem of constructing
generic off-diagonal solutions of gravitational field equations in GR, or
MGTs, formulated on phase spaces in certain analogous forms to the Einstein
equations. For constructing models of QG, it is important to consider also
quantum deformations and respective nonlinear functional equations, which in
Finsler-Lagrange-Hamilton variables can be performed in explicit form as
exact and parametric solutions of certain systems of nonlinear classical or
quantum PDEs.

\vskip3pt In a series of works \cite%
{vacaru18,bubuianu18a,vmon3,bubuianu17,bubuianu20,bubuianu19,stavr14,bubuianu20}%
, we proved that GR can be extended in off-diagonal integrable forms on
phase spaces using nonholonomic dyadic variables with 2 (3)+2+...
decompositions and auxiliary connections. The auxiliary connections are
adapted to a N-connection structure as in modified Finsler geometry on (co)
tangent Lorentz bundles, and defined as distortions of the so-called Cartan/
Berwald / Chern or other connections in Finsler geometry, and can be
nonholonomically constrained to the Levi-Civita, LC, connection. The main
point is that we can consider toy models with nonholonomic 2+2 splitting
with N-connection and certain canonical connections defined by
pseudo-Riemannian metrics in GR, but with distorted LC connections to
nonholonomic data which allow construction of off-diagonal solutions in very
general forms. Having defined a general class of solutions in "distorted
GR", we can impose additional nonholonomic constraints and extract LC
configurations for the standard GR. The approach works also for theories of
higher dimension and for various models on phase spaces, including
nonassociative star product deformations to $\mathcal{M}^{\star }$ or $\
^{\shortmid }\mathcal{M}^{\star }$. For all such MGTs and generalizations to
(non) associative geometric flow evolution models, we can introduce
canonical type dyadic variables and, equivalently, Finsler-Lagrange-Hamilton
variables. If we work directly only with the Lagrange or Hamilton phase
space configurations, we are not able to derive standard geometric metric
and connection structures as in the (pseudo) Riemannian or metric-affine
geometry described by corresponding Ricci, torsion, nonmetricity etc.
tensors. Nevertheless, if we introduce the Hessians of $L(x,v)$ and or $%
H(x,p)$ as corresponding (co) vertical metrics, and corresponding Sasaki
lifts to total metrics on $\mathcal{M}$ or $\ ^{\shortmid }\mathcal{M}$, we
preserve the priorities of phase space constructions in the Lagrange and
Hamilton mechanics (with extensions to classical and quantum field theories)
and the possibility to work with metrics, adapted frames, connections,
curvatures etc. as in higher dimension gravity, which may have certain extra
dimension coordinates as generalized velocity/ momentum ones.

\vskip3pt We explain the details of our alternative geometrization of
mechanics and field theories and nonassociative star product deformation in
section \ref{sec5}. Here we emphasize that our geometric formalism involves
modified Finsler connections (using canonical "hat" d-connections) which
allow to apply the AFCDM and integrate in general form physically important
systems of nonlinear PDEs. Working only with $L$- and/or $H$-models on phase
spaces, without a definition of N-connections and related canonically
deformed Finsler-like connections, we are not able to apply the AFCDM for
finding off-diagonal solutions in GR and MGTs. Theories with conventional $L$%
- and/or $H$, and correspondingly related Poisson/ almost symplectic
formalisms are efficient for elaborating various methods of quantization
when the quantum field theory (QFT) was developed using corresponding
methods of linear functional analysis. From a rigorous mathematical
viewpoint, such approaches do not work for QG because a general theory of
nonlinear functional analysis has not been formulated yet and the existing
methods have not been applied directly to physically important systems of
PDEs like (modified) Einstein equations and geometric flow models. We note
that QG has the property of asymptotic safety (also referred to as
nonperturbative renormalizability) related to the modern Wilsonian viewpoint
on QFT involving functional renormalization group equations \cite{reuter}.
Nevertheless, even in this approach, there are not used in direct form
certain exact and parametric solutions for metric and connection variables;
the flow equations of such a QG are the results of a higher-derivative
Einstein-Hilbert truncation.

\vskip3pt In this work, we follow the idea that Lagrangians and/ or
Hamiltonians can be used on phase spaces for defining new type geometric
objects such as the N-connection structures (via corresponding semi-spray,
i.e. nonlinear geodesic equations, which are equivalent to the
Euler-Lagrange, or Hamilton equations) and certain total phase space
s-metrics and canonical s-connections. Such a Finsler-Lagrange-Hamilton
phase space gravity, with generalized Ricci s-tensors and scalar curvatures,
can be formulated as a nonholonomic extra dimension generalization of the
Einstein gravity (we can add also nonassociative star product deformations),
which can be integrated in certain off-diagonal forms. Using corresponding
classes of off-diagonal parametric solutions of (nonassociative) geometric
flow and phase-modified gravitational equations, we can apply various
methods of quantization which can be selected to be parametric
renormalizable for a respective family of metric-affine configurations. This
is not possible if we work only with some effective Lagrangians and
Hamiltonians for quantizing undefined classes of metrics and connections, or
for certain WDW functional equations, or the Ashtekar variables.

\vskip3pt The main \textbf{Hypothesis} in this work on 4-d nonholonomic
Einstein gravity and generalizations on 8-d co-tangent Lorentz bundle (phase
space) for nonassociative and noncommutative gravity and geometric flow
theories is that such \textit{models can be formulated in nonholonomic
canonical variables and, equivalently, in Finsler-Hamilton variables.} The
corresponding N-connection structures can be defined from respective
nonlinear geodesic equations which are equivalent to the Hamilton equations
on a phase space, but there are derived also Ricci s-tensors for
nonholonomic metric-affine phase spaces. In the so-called canonical form
(with "hat" variables), physically important systems of nonlinear PDEs (for
instance, modified geometric flow and Ricci soliton equations) possess
general decoupling and integrability properties. This allows us to construct
and study the physical properties of various classes of exact and parametric
solutions defined by off-diagonal symmetric and nonsymmetric metrics. The
corresponding AFCDM works for general connections (nonlinear and adapted
ones, encoding Finsler-like and nonassociative and noncommutative
distortions), when LC-configurations can be extracted for additional
nonholonomic constraints which can be solved in explicit form. We argue (and
we shall provide explicit examples in respective sections) that explicit
criteria can be formulated when such solutions define 4-d , 8-d and 10-d
nonholonomic Lorentz spacetime models and (nonassociative and/ or
noncommutative) black hole/ ellipsoid, wormhole, toroid, and cosmological
configurations described by modified dispersion relations, MDRs, and
encoding nonholonomic Finsler-Hamilton structures, in general, in
nonassociative forms.

\subsection{The objectives and structure of the paper}

This paper provides a status report on the AFCDM and applications with
geometric computation examples on 4-d and extra dimension gravity theories
(including 8-d nonassociative phase space models with geometric evolution).
Details of the proofs are provided in Part I for nonholonomic (2+2)
decompositions when the constructions for higher dimensions, in Part II, are
derived in abstract geometric form to higher dimensions including and phase
space models. The most important formulas and physically important systems
of nonlinear PDEs can be modelled equivalently in canonical and/or
Finsler-Hamilton variables; and for nonassociative theories of geometric and
information flows and gravity, see related results and methods in \cite%
{bubuianu20,bubuianu19,partner01,partner02,partner03,partner04,partner05,partner06}%
. In Appendix, we summarize the approach and provide abstract and N-adapted
formulas for higher dimensions and (co) tangent Lorentz bundles which may
encode Finsler-Lagrange-Hamilton structure and/or nonassociative data.

\vskip5pt The main objectives are stated for respective parts of the
article. For the Part I, there are three objectives:

\vskip5pt The \textbf{first objective}, Obj1, is to review in section \ref%
{sec2} the geometry of N-connections defining (dyadic) nonholonomic
(2+2)-splitting, and related adapted frames and distinguished connection,
d-connection, structures. There are defined canonical d-connection and
LC-connection determined by the same metric structure with a N-connection
splitting and corresponding curvature and torsion d-tensors, Ricci and
Einstein d-tensors. The (modified) gravitational equations are formulated in
canonical nonholonomic variables. We show how to compute in explicit forms
the N-adapted formulas for the canonical d-connections and derive the
coefficient formulas for the canonical torsion and Ricci d-tensors,
canonical distortion of the scalar curvature. There are considered necessary
parametrizations in N-adapted forms of the effective and energy-momentum
tensors.

\vskip5pt We prove the general decoupling and integration properties of the
Einstein equations for the canonical d-connection in section \ref{sec3}
(this consists of the \textbf{second objective}, Obj2, of our work. For
simplicity, the general formulas for generic off-diagonal solutions are
derived for nonholonomic spacetimes with Killing d-vector symmetry.
Corresponding N-adapted coefficients are expressed in terms of generating
and integration functions and generating sources. We show that such classes
of solutions possess nonlinear symmetries which allow to re-define the
generating functions and introduce effective cosmological constants. There
are provided the quadratic linear elements for off-diagonal solutions in
using geometric data with 1] generating functions and generating sources; 2]
re-defined generating functions and effective cosmological constants; 3]
nonlinearly related to "original" generating functions and effective
sources/ cosmological constants. We also study deformations of prime
d-metrics by so-called 4] gravitational polarization functions (which can be
also considered as generating functions) into target d-metrics defining
off-diagonal solutions of the Einstein equations. For 5] small parametric
deformations of the polarization functions, we consider parametric
deformations of the d-metrics and respective solutions. We consider a toy
2+2 model with effective momentum variables which allow a straightforward
geometric generalizations for (co) tangent phase space models with
conventional (2+2)+(2+2) splitting considered in Tables 7-16 from the
Appendix. There are analyzed the conditions, and the possibility to solve
such nonholonomic constraints, which are necessary for extracting
LC-configurations.

\vskip5pt The \textbf{third objective}, Obj3, stated for section \ref{sec4},
is to provide and study explicit examples of new classes of exact/
parametric generic off-diagonal solutions constructed by using the AFCDM.
For the first class of such quasi-stationary solutions, we consider examples
of new Kerr de Sitter solution and their nonholonomic deformations to
spheroidal configurations. Then the procedure of nonholonomic off-diagonal
deformations of cylindrical systems in GR is formulated. Such constructions
are applied for generating solutions describing locally anisotropic
wormholes. We also provide new classes of generic off-diagonal solutions
describing locally anisotropic black torus, BT, and black ellipsoid, BE,
configurations. There are analyzed solutions involving both BT and BE
nonholonomically deformed geometric objects. Then, we construct and analyze
new classes of nonholonomic cosmological solitonic and spheroid deformations
involving 2-d vertices and elaborate on models with small parametric
off-diagonal cosmological deformations with solitonic vacua for voids.

\vskip5pt For the Part II devoted to (nonassociative) canonical and
Finsler-Hamilton phase space generalizations the main objectives are:

\vskip5pt

The \textbf{forth objective}, Obj4, is to outline in section \ref{sec5} the
nonassociative geometric flow theory in canonical nonholonomic variables.
Such models are determined by star product R-flux deformations in string
theory which, in our approach, are formulated in canonical nonholonomic
variables with dyadic (shell) splitting. This allows us to prove general
decoupling and integrability properties of such theories using abstract
geometric methods. We explain why "pure" Lagrange and/or Hamilton phase
space models can't be used for generating integrable MGTs when equivalent
formulations in terms of generalized Finsler-Lagrange-Hamilton phase space
models, with N-connections and canonical s-connections, allow to construct
off-diagonal solutions of (modified) Einstein equations.

\vskip5pt The nonassociative Finsler-Lagrange-Hamilton geometric flow
theories are formulated and studied in section \ref{sec6}. This consists the 
\textbf{fifth objective}, Obj5, which aims to characterize such geometries
using G. Perelman's F- and W-functionals (with respective generalizations
and nonholonomic modifications) and derived statistical and geometric
thermodynamic models encoding nonassociative and/or nonholonomic
Finsler-Hamilton like data. The \textbf{sixth objective}, Obj6, in the same
section, is to formulate the theory of nonassociative
Finsler-Lagrange-Hamilton geometric flows. We define star product R-flux
versions of such nonassociatie generalized Finsler geometries and postulate
nonassociative versions of geometric evolution equations. Another very
important \textbf{seventh objective}, Obj7, also in section \ref{sec6}, is
to formulate nonassociative Finsler like generalizations of G. Perelman
thermodynamics for geometric flows which is important for future
developments of thermofield and quantum gravity theories encoding
nonassociative and modified dispersion data.

\vskip5pt In section \ref{sec8}, we show how the AFCDM for nonassociative
geometric flows and gravity can be applied (using respective distortions of
adapted Finsler-Lagrange-Hamilton structures and nonlinear transforms) to
decouple and solve physically important systems of nonlinear PDEs for such
models and physical theories. This consists the \textbf{eights objective},
Obj8, of this work. We distinguish two general classes of nonassociative
Finsler like quasi-stationary solutions and locally anisotropic cosmological
solutions. Respective nonlinear symmetries encoding nonassociative and
noncommutative modified Finsler-Hamilton data are analyzed. In the same
section, there is the \textbf{ninths objective}, Obj9, to construct and
analyze explicit examples of physically important nonassociative and
Finsler-Hamilton configurations. We provide exact parametric solutions
describing nonassociative Finsler-Hamilton black ellipsoids and compute
respective Bekenstein-Hawking entropy and G. Perelman thermodynamic
variables.

\vskip5pt We discuss and conclude the results and methods in section \ref%
{sec8}. Here we also note that the \textbf{tenth objective}, Obj10, is
stated for the Appendix. It aims a summary of the AFCDM (from 4-d GR till
10-d and nonassociative phase spaces of 8-d) stated as Tables 1-16. Such
tables can be used for generating exact and parametric solutions in various
types of gravity theories by prescribing physically important generating and
integration functions and effective and matter field sources. They outline
respective geometric ideas, typical off-diagonal ansatz, and the main
results on decoupling of respective modified Einstein equations and
classifications into quasi-stationary and locally anisotropic cosmological
solutions for gravity theories on 4-d Lorentz manifolds (see Tables 1-3).
Then the results are extended for 10-d Lorentz manifolds in Tables 4-6.
There are also considered geometric summaries of the AFCDM for phase space
models elaborated on (co) tangent Lorentz bundles of total dimension 8-d and
for a 4-d Lorentz base spacetime manifold. The formulas are provided for
(effective) sources which, in general, encode nonassociative
Finsler-Lagrange-Hamilton data which allow us to construct exact/parametric
solutions using canonical variables. The Tables 7-11 classify and outline
possible versions for solutions determined by generic off-diagonal metrics
and generalized connections depending both on base spacetime and velocity
type coordinates.

\vskip5pt In the final subsection of the Appendix, we summarize the results
on solutions for gravity like models elaborated on nonholonomic cotangent
Lorentz bundles encoding momentum like variables which may be with a fixed
energy like parameter or depend like a "rainbow" on an energy type
coordinate. Such solutions can be formulated both for commutative and
nonassociative models, in general, with nonassociative
Finsler-Lagrange-Hamilton geometric flow dependence. Projections of such
nonholonomic phase space solutions on the base Lorentz spacetime manifolds
may model quasi-stationary and/ or locally anisotropic cosmological
solutions. Such generic off-diagonal phase space solutions may encode
various types of nonassociative/ noncommutative / supersymmetric / nonmetric
/ nonsymmetric data / quasi-classical contributions computed (or introduced
phenomenologically) for different models of modified geometric and
information flow theories, superstring/ M-theory and other MGTs.

\subsection{Remarks on abbreviations and notations}

In this review work, we use many abbreviations and notations which are
standard for certain researchers in some directions on geometry and physics
but unkown for others working in particle physics, cosmology and other
directions. The principles and main formulas of abstract geometric and
N-adapted coefficient calculus are outlined in respective subsections with
explanations of Tables 1 - 16 from the Appendix. We have to dub some terms
and symbols in certain sections containing inter-disciplinary methods. For
instance, the abbreviations introduced new geometric methods of constructing
parametric solutions are dubbed and explained again for applications in
modern cosmology or nonassociative geometric flow thermodynamics. For
convenience, we present here a list of abbreviations mentioning them in the
order as they were introduced in a respective paragraph of a (sub) section.

\begin{description}
\item GR - \textit{general relativity} - section 1, paragraph 1

\item MGTs - \textit{modified gravity theories} - section 1, paragraph 1

\item 2-d, 4-d, or 8-d etc. - \textit{two, four, or eight etc. dimenisonal/
dimensions} - section 1, paragraph 1

\item N-connection(s) - \textit{nonlinear connection(s) }-section 1,
paragraph 1

\item 2+2+2+... -\textit{\ dyadic decompostions/ splitting of phase space or
spacetime dimensions} - section 1, paragraph 2

\item PDEs -\textit{\ partial differential equations} - section 1, paragraph
2

\item LC configuration/ connection - \textit{Levi-Civita configuration/
connection} - section 1, paragraph 2

\item AFCDM - the \textit{Anholonomic Frame and Connection Deformation Method%
} (of constructing off-diagonal solutions in geometric flow and gravity
theories) - section 1, paragraph 2

\item ODEs - \textit{ordinary differential equations }- section 1, paragraph
3

\item BH - \textit{black hole }- section 1.1, paragraph 2

\item FLRW - \textit{Friedman - Lema\^{\i}tre - Robertson - Walker}
(cosmology) - section 1.1, paragraph 3, formulas (3)

\item WH - \textit{wormhole} - section 1.1, paragraph 4

\item LA - \textit{LA symmetries} (a standard term in the theory of
solitons) - section 1.2, paragraph 2

\item QG - \textit{quantum gravity }- section 1.2, paragraph 3

\item $\star $-product/ -deformation - \textit{star product / deformation }-
section 1.3, paragraph 1

\item D-brane - \textit{Dirichlet brane} (a standard term in string/
M-theory) - section 1.3, paragraph 1

\item R-flux - a \textit{standard term} in string/ M-theory (explained in
paragraphs related to formulas (155) - (157)) - section 1.3, paragraph 2

\item ADM - \textit{Arnowit - Deser - Misner} (a method with 3+1 splitting
in GR and MGTs) - section 1.4, paragraph 1

\item WDW - \textit{Wheeler de Wit} (an important equation in GR and QG) -
section 1.4, paragraph 1

\item QFT - \textit{quantum field theory} - section 1.4, paragraph 3

\item Obj1 - \textit{objective 1 }(of this work) - section 1.5, paragraph 1

\item d-connection, d-tensor, d-object - \textit{geometric objects adapted
to a N-connection structure} - section 1.5, paragraph 3

\item BT - \textit{black torus} - section 1.5, paragraph 6

\item BE - \textit{black ellipsoid} - section 1.5, paragraph 6

\item F- and W-functionals - introduced by G. Perelman together with the
concept of W-entropy (in his geometric flow thermodynamics) - section 1.5,
paragraph 8, related to Obj5, see details in section 6.3

\item MDR - \textit{modified dispersion relations} - Part II, paragraph 1
\end{description}

To avoid ambiguities certain abbreviations and abstract symbols are
repeated, if necessary, in different parts/ sections of the paper.

\part{Nonholonomic variables in Einstein gravity \& MGTs}

\section{The geometry of GR \& MGTs with nonholonomic (2+2)-splitting}

\label{sec2} We review the main ideas and proofs for a nonholonomic
geometric formulation of the Einstein gravity (general relativity, GR) and
modified gravity theories, MGTs, on 4-d metric-affine manifolds, which in
canonical nonholonomic variables will allow us to prove in next section
certain general decoupling and integrations properties of gravitational
field equations.

\subsection{Geometric and physical objects in nonholonomic 2+2 variables}

\subsubsection{Nonlinear connections and distinguished metrics}

We shall work on a Lorentz spacetime enabled with standard geometric data $%
(V,\mathbf{g}),$ where $V$ is a 4-d pseudo-Riemannian manifold of necessary
smooth (differentiability) class, defined by a symmetric metric tensor of
signature $(+++-),$ 
\begin{equation}
\mathbf{g}=g_{\alpha ^{\prime }\beta ^{\prime }}(u)e^{\alpha
^{\prime}}\otimes e^{\beta ^{\prime }}.  \label{mst}
\end{equation}%
In this formula, we consider general co-frames $e^{\alpha ^{\prime }}$ which
are dual to frame bases $e_{\alpha ^{\prime }}.$ On a coordinate
neighborhood $U\subset V,$ we can always define local coordinates $%
u=\{u^{\alpha }=(x^{i},y^{a})\}$ involving a conventional $2+2$ splitting
into h-coordinates, $x=(x^{i}),$ and v-coordinates, $y=(y^{a}),$ for indices 
$j,k,...=1,2$ and $a,b,c,...=3,4,$ when $\alpha ,\beta ,...=1,2,3,4.$ A
local coordinate basis and a co-base are written respectively as $e_{\alpha
}=\partial _{\alpha }=\partial /\partial u^{\beta} $ and $e^{\beta
}=du^{\beta }.$ Transforms to arbitrary frames (tetrads / vierbeinds) are
defined as $e_{\alpha ^{\prime }}=e_{\ \alpha ^{\prime }}^{\alpha
}(u)e_{\alpha }$ and $e^{\alpha ^{\prime }}= e_{\alpha \ }^{\ \alpha
^{\prime }}(u)e^{\alpha }.$ Such (co) bases are orthonormal if $e_{\alpha \
}^{\ \alpha ^{\prime }}e_{\ \alpha ^{\prime }}^{\beta }=\delta _{\alpha
}^{\beta },$ where $\delta _{\alpha }^{\beta }$ is the Kronecker symbol.

In coordinate free form, a 2+2 decomposition can be introduced as a
conventional nonlinear connection structure (N-connection), when for the
tangent bundle $TV$ $:=\bigcup\nolimits_{u}T_{u}V$ it is prescribed a
non--integrable (equivalently, nonholonomic/anholonomic) conventional
horizontal and vertical splitting, in brief, h- and v--decomposition into
respective 2-d and 2-d subspaces, $hV$ and $vV.$ This is equivalent to the
condition that a Whitney sum 
\begin{equation}
\mathbf{N}:\ TV=hV\oplus vV  \label{ncon}
\end{equation}%
is globally defined for $V$ and $TV.$ For instance, in Finsler geometry \cite%
{vacaru18,bubuianu20,bubuianu19}, the N-connections are defined by splitting
of type $\ TTV=hTV\oplus vTV,$ involving the second tangent bundle $TTV,$ or
(in some equivalent forms) using nonholonomic distributions and splitting in
exact sequences. On Lorentz manifolds, a N-connection (\ref{ncon}) consist
an example of nonholonomic distribution defining a fibered 2+2 structure. We
can use the term nonholonomic Lorentz / pseudo-Riemannian manifold when a
conventional h-v-splitting of some classes of local bases is defined at
least on a neighbourhood $U\subset V, e_{\alpha ^{\prime
}}=(e_{i^{\prime}},e_{a^{\prime }})$ and $e^{\beta ^{\prime }}=(e^{i^{\prime
}},e^{a^{\prime}}).$ Hereafter, we shall omit priming/underlying/overlying
etc. of indices if that will not result in ambiguities. We also note that in
our works we use "boldface" symbols in order to emphasize that certain
spaces/geometric objects are enabled/adapted with/to a N-connection
structures. In local coordinate form, a N-connection is defined by a
nonholonomic distribution stated by a set of coefficients $N_{i}^{a}(u)$
when $\mathbf{N}= N_{i}^{a}(x,y)dx^{i}\otimes \partial /\partial y^{a}.$

Using N-connection coefficients $\mathbf{N}=\{N_{i}^{a}\}$ (\ref{ncon}), we
can define N--elongated (equivalently, N-adapted) local bases (partial
derivatives), $\mathbf{e}_{\nu },$ and co-bases (differentials), $\mathbf{e}%
^{\mu },$ when 
\begin{eqnarray}
\mathbf{e}_{\nu } &=&(\mathbf{e}_{i},e_{a})=(\mathbf{e}_{i}=\partial
/\partial x^{i}-\ N_{i}^{a}(u)\partial /\partial y^{a},\ e_{a}=\partial
_{a}=\partial /\partial y^{a}),\mbox{ and  }  \label{nader} \\
\mathbf{e}^{\mu } &=&(e^{i},\mathbf{e}^{a})=(e^{i}=dx^{i},\ \mathbf{e}%
^{a}=dy^{a}+\ N_{i}^{a}(u)dx^{i}),  \label{nadif}
\end{eqnarray}%
are linear on $N_{i}^{a}.$ For instance, a N-elongated basis (\ref{nader})
satisfies the nonholonomy relations 
\begin{equation}
\lbrack \mathbf{e}_{\alpha },\mathbf{e}_{\beta }]=\mathbf{e}_{\alpha }%
\mathbf{e}_{\beta }-\mathbf{e}_{\beta }\mathbf{e}_{\alpha }=W_{\alpha \beta
}^{\gamma }\mathbf{e}_{\gamma },  \label{nonholr}
\end{equation}%
with (antisymmetric) nontrivial anholonomy coefficients 
\begin{equation}
W_{ia}^{b}=\partial _{a}N_{i}^{b},W_{ji}^{a}=\Omega _{ij}^{a}=\mathbf{e}%
_{j}\left( N_{i}^{a}\right) -\mathbf{e}_{i}(N_{j}^{a}),  \label{anhcoef}
\end{equation}%
where $\Omega _{ij}^{a}$ define the coefficients of N-connection curvature.
If all anholonomic coefficients (\ref{anhcoef}) are zero for a $\mathbf{e}%
_{\alpha },$ such a N-adapted base is holonomic and we can write it as a
partial derivative $\partial _{\alpha },$ with zero N-connection
coefficients for corresponding coordinate transforms. In curved coordinates,
for holonomic bases, the coefficients $N_{j}^{a}$ may be non-zero even all $%
W_{\alpha \beta }^{\gamma}=0.$

The geometric objects on a nonholonomic manifold $\mathbf{V}$ enabled with a
N-connection structure $\mathbf{N}$ (and on extensions to tangent, $T\mathbf{%
V,}$ and cotangent, $T^{\ast }\mathbf{V}$, bundles; and their tensor
products, for instance, $T\mathbf{V\otimes }T^{\ast }\mathbf{V}$) are called
distinguished (in brief, d-objects, d-vectors, d-tensors etc) if they are
adapted to the N--connection structure via corresponding decompositions with
respect to frames of type (\ref{nader}) and (\ref{nadif}). For instance, we
write a d--vector as $\mathbf{X}=(hX,vX)$.

Any spacetime metric $\mathbf{g}$ (\ref{mst}) can be represented
equivalently as a d--metric $\mathbf{g}=(hg,vg),$ when 
\begin{eqnarray}
\ \mathbf{g} &=&\ g_{ij}(x,y)\ e^{i}\otimes e^{j}+\ g_{ab}(x,y)\ \mathbf{e}%
^{a}\otimes \mathbf{e}^{b},\mbox{ in N-adapted form};  \label{dm} \\
&=&\underline{g}_{\alpha \beta }(u)du^{\alpha }\otimes du^{\beta },%
\mbox{
using tensor products of dual coordinate bases}.  \label{cm}
\end{eqnarray}%
In above formulas $hg=\{\ g_{ij}\}$ and $\ vg=\{g_{ab}\}.$ Introducing
coefficients of (\ref{nadif}) into (\ref{dm}) and regrouping with respect to
the coordinate dual basis, we obtain the formulas for the coefficients in (%
\ref{cm}), 
\begin{equation}
\underline{g}_{\alpha \beta }=\left[ 
\begin{array}{cc}
g_{ij}+N_{i}^{a}N_{j}^{b}g_{ab} & N_{j}^{e}g_{ae} \\ 
N_{i}^{e}g_{be} & g_{ab}%
\end{array}%
\right]  \label{ansatz}
\end{equation}%
for any prescribed set of coefficients $N_{i}^{a}.$ A metric $\mathbf{g}=\{%
\underline{g}_{\alpha \beta }\}$ (\ref{ansatz}) is generic off--diagonal if
the anholonomy coefficients $W_{\alpha \beta }^{\gamma }$ are not identical
to zero (in 4-d, such a matrix can not be diagonalized via coordinate
transforms, but we can consider such diagonalizations for 2-d and 3-d
subspaces). Parameterizations of type (\ref{ansatz}) are used 1) in
Kaluza-Klein theories when $N_{j}^{e}=A_{j}^{e}$ are identified as certain
gauge fields after compactification on $y$-coordinates (usually, there are
considered higher dimension spacetimes);\ 2) in Finsler like theories, when $%
N_{j}^{e}$ are defined in certain forms which are used in respective
Finsler-Lagrange-Hamilton theories, see details in \cite%
{vmon3,vacaru18,bubuianu18a}; \ and in GR when N-coefficients are treated as
off-diagonal terms and used for N-adapted geometric constructions.

In this work, we prefer to work with d-metrics (\ref{dm}), d-tensors,
d-connections etc., because in certain N-adapted forms for d-objects it is
possible to prove the decoupling and integration properties of (modified)
Einstein equations. In coordinate bases, the constructions are very
cumbersome and a general decoupling is not possible if we use the
LC-connection $\nabla .$ Finally, we note that the components of the inverse
metric $\underline{g}^{\alpha \beta }$ (in general, with off-diagonal terms)
are computed for nondegenerated metric structures following formulas $%
\underline{g}^{\alpha \beta }\underline{g}_{\gamma \beta }=\delta _{\beta
}^{\alpha }.$ In similar forms, there are defined and computed the inverse
d-metrics and their h- and v-coefficients, $\mathbf{g}^{\alpha \beta
}=(g^{ij},g^{ab}).$

\subsubsection{N--adapted connections, the canonical d-connection and
fundamental geometric objects}

Linear connection structures on a nonholonomic $\mathbf{V}$ can be defined
in N-adapted or in a general form, which may be not N-adapted, for instance,
in the case of the LC-connection $\nabla $).

A\textbf{\ d--connection} $\mathbf{D}=(hD,vD)$ is a linear connection
preserving under parallelism the N--connection splitting (\ref{ncon}). It
defines a covariant N--adapted derivative $\mathbf{D}_{\mathbf{X}}\mathbf{Y}$
of a d--vector field $\mathbf{Y}$ in the direction of a d--vector $\mathbf{X}%
.$ With respect to N--adapted frames (\ref{nader}) and (\ref{nadif}), any $%
\mathbf{D}_{\mathbf{X}}\mathbf{Y}$ can be computed as in GR and/or metric
affine gravity but with the coefficients decomposed defined by h- and
v-indices, 
\begin{equation}
\mathbf{D}=\{\mathbf{\Gamma }_{\ \alpha \beta }^{\gamma }=(L_{jk}^{i},\acute{%
L}_{bk}^{a};\acute{C}_{jc}^{i},C_{bc}^{a})\},\mbox{ where }hD=(L_{jk}^{i},%
\acute{L}_{bk}^{a})\mbox{ and }vD=(\acute{C}_{jc}^{i},C_{bc}^{a}).
\label{hvdcon}
\end{equation}

By definition, any d--connection is characterized by three fundamental
geometric d-objects, which (by definition in abstract forms) are: 
\begin{eqnarray}
\mathcal{T}(\mathbf{X,Y}) &:=&\mathbf{D}_{\mathbf{X}}\mathbf{Y}-\mathbf{D}_{%
\mathbf{Y}}\mathbf{X}-[\mathbf{X,Y}],\mbox{ torsion d-tensor,  d-torsion};
\label{fundgeom} \\
\mathcal{R}(\mathbf{X,Y}) &:=&\mathbf{D}_{\mathbf{X}}\mathbf{D}_{\mathbf{Y}}-%
\mathbf{D}_{\mathbf{Y}}\mathbf{D}_{\mathbf{X}}-\mathbf{D}_{\mathbf{[X,Y]}},%
\mbox{ curvature d-tensor, d-curvature} ;  \notag \\
\mathcal{Q}(\mathbf{X}) &:=&\mathbf{D}_{\mathbf{X}}\mathbf{g,}%
\mbox{nonmetricity d-fiels, d-nonmetricity}.  \notag
\end{eqnarray}%
The N--adapted coefficients of such geometric d-objects are computed by
introducing $\mathbf{X}=\mathbf{e}_{\alpha }$ and $\mathbf{Y}=\mathbf{e}%
_{\beta },$ defined by (\ref{nader}), and considering h-v-splitting (\ref%
{hvdcon}) for $\mathbf{D}=\{\mathbf{\Gamma }_{\ \alpha \beta }^{\gamma }\}$
into above formulas, see details in \cite{vmon3,vacaru18,bubuianu18a}, 
\begin{eqnarray*}
\mathcal{T} &=&\{\mathbf{T}_{\ \alpha \beta }^{\gamma }=\left( T_{\
jk}^{i},T_{\ ja}^{i},T_{\ ji}^{a},T_{\ bi}^{a},T_{\ bc}^{a}\right) \}; \\
\mathcal{R} &\mathbf{=}&\mathbf{\{R}_{\ \beta \gamma \delta }^{\alpha }%
\mathbf{=}\left( R_{\ hjk}^{i}\mathbf{,}R_{\ bjk}^{a}\mathbf{,}R_{\ hja}^{i}%
\mathbf{,}R_{\ bja}^{c}\mathbf{,}R_{\ hba}^{i},R_{\ bea}^{c}\right) \mathbf{%
\};} \\
\ \mathcal{Q} &=&\mathbf{\{Q}_{\ \alpha \beta }^{\gamma }=\mathbf{D}^{\gamma
}\mathbf{g}_{\alpha \beta }=(Q_{\ ij}^{k},Q_{\ ij}^{c},Q_{\ ab}^{k},Q_{\
ab}^{c})\}.
\end{eqnarray*}%
We say that any geometric data $\left( \mathbf{V},\mathbf{N},\mathbf{g,D}%
\right) $ define a N-adapted metric-affine structure (equivalently,
metric-affine d-structure) determined by a d-metric and a d-connection
stated independently, but both in N-adapted form $\mathbf{V}$.

Using a d-metric $\mathbf{g}$ (\ref{dm}), we can define two important linear
connection structures (the Levi-Civita, LC, connection and the canonical
d-connection): 
\begin{equation}
(\mathbf{g,N})\rightarrow \left\{ 
\begin{array}{cc}
\mathbf{\nabla :} & \mathbf{\nabla g}=0;\ _{\nabla }\mathcal{T}=0,\ 
\mbox{\
LC--connection }; \\ 
\widehat{\mathbf{D}}: & \widehat{\mathbf{Q}}=0;\ h\widehat{\mathcal{T}}=0,v%
\widehat{\mathcal{T}}=0,\ hv\widehat{\mathcal{T}}\neq 0,%
\mbox{ the canonical
d-connection}%
\end{array}%
\right.  \label{twocon}
\end{equation}%
In our work, "hat" labels are used for geometric d-objects written in such a
canonical form. For any $\widehat{\mathbf{D}},$ we can define and compute
the canonical fundamental geometric objects (\ref{fundgeom}), $\widehat{%
\mathcal{R}}=\{\widehat{\mathbf{R}}_{\ \beta \gamma \delta }^{\alpha }\}$
etc. In a similar form, we can compute the fundamental geometric objects
defined by $\mathbf{\nabla ,}$ for instance, $\ _{\nabla }\mathcal{R}=\{\
_{\nabla }R_{\ \beta \gamma \delta }^{\alpha }\}$ (in such cases, boldface
indices are not used). Considering the canonical distortion relation for
linear connections (\ref{cdist}), we can compute respective canonical
distortions of fundamental geometric d-objects (\ref{fundgeom}). Such
formulas relate, for instance, two different curvature tensors, $\ _{\nabla }%
\mathcal{R}=\{\ _{\nabla }R_{\ \beta \gamma \delta}^{\alpha }\}$ and $\ 
\widehat{\mathcal{R}}=\{\widehat{\mathbf{R}}_{\ \beta \gamma \delta
}^{\alpha }\}$ etc.

\subsubsection{The canonical Ricci and Einstein d-tensors; the canonical
d-torsion and LC-conditions}

We can define the canonical Ricci d-tensor as the contraction on the 1st and
4th indices of the canonical curvature d-tensor,%
\begin{equation}
\widehat{\mathbf{R}}ic=\{\widehat{\mathbf{R}}_{\ \beta \gamma }:=\widehat{%
\mathbf{R}}_{\ \beta \gamma \alpha }^{\alpha }\}.  \label{criccidt}
\end{equation}%
It should be noted that this d-tensor, in general, is not symmetric, i.e. $%
\widehat{\mathbf{R}}_{\ \beta \gamma }\neq \widehat{\mathbf{R}}_{\
\gamma\beta }.$ This is typical for nonholonomic geometric objects. The
canonical scalar curvature is introduced as%
\begin{equation*}
\widehat{R}sc:=\mathbf{g}^{\alpha \beta }\widehat{\mathbf{R}}_{\ \alpha
\beta }.
\end{equation*}%
These formulas allow to define the canonical (nonholonomic) Einstein
d-tensor, 
\begin{equation}
\widehat{\mathbf{E}}n:=\widehat{\mathbf{R}}ic-\frac{1}{2}\mathbf{g}\widehat{R%
}sc=\{\widehat{\mathbf{R}}_{\ \beta \gamma }-\frac{1}{2}\mathbf{g}_{\ \beta
\gamma }\widehat{R}sc\}.  \label{ceinstdt}
\end{equation}%
Adapting for $\widehat{\mathbf{D}}$ the geometric abstract principles for
deriving gravitational filed equations provided in \cite{misner} for $%
\nabla, $ and considering a nontrivial cosmological constant $\Lambda ,$ we
can derive the canonical Einstein d-equations (\ref{cdeq}) using the
canonical d-tensor (\ref{ceinstdt}). Such equations can be proved also in
N-adapted variational form if we introduce conventional gravitational and
matter fields Lagrange densities, $\ ^{g}L(\widehat{\mathbf{R}}ic)$ (as in
GR with $\ ^{g}L(R)$) and postulating a $\ ^{m}L(\varphi ^{A},\mathbf{g}%
_{\beta \gamma}).$ In N-adapted form, the stress-energy d-tensor of matter
fields $\varphi ^{A} $ (labeled by a general index $A$) is defined and
computed 
\begin{equation}
\mathbf{T}_{\alpha \beta }=-\frac{2}{\sqrt{|\mathbf{g}_{\mu \nu }|}}\frac{%
\delta (\ ^{m}L\sqrt{|\mathbf{g}_{\mu \nu }|})}{\delta \mathbf{g}^{\alpha
\beta }}.  \label{emdt}
\end{equation}%
For such sources, we can define the trace $T:=\mathbf{g}^{\alpha \beta }%
\mathbf{T}_{\alpha \beta }$ and the effective source $\widehat{\mathbf{Y}}[%
\mathbf{g,}\widehat{\mathbf{D}}]\simeq \{\mathbf{T}_{\alpha \beta }\}.$ In
various physical theories, one consider more general $\ ^{m}L,$ for
instance, depending on some covariant/spinor derivatives, and/or $\
^{g}L(f(T,R))$ determined by a functional $f(R,T)$ in MGTs. There are
considered various nonsymmetric effective source, for instance, in massive
gravity. For simplicity, in this work we omit such considerations (see
details and review of results generalized with N-connection structures, in 
\cite{vmon3,vacaru18,bubuianu18a,vvy13} and references therein).

In this work, we consider (effective) sources $\widehat{\mathbf{Y}}[\mathbf{%
g,}\widehat{\mathbf{D}}]=\{\Upsilon _{~\delta }^{\beta }\}$ parameterized
with respect to N-adapted frames (\ref{nader}) and (\ref{nadif}) in such
forms (this can be done for very general classes of energy-momentum tensors
using by respective frame/coordinate transforms): 
\begin{equation}
\widehat{\Upsilon }_{~\delta }^{\beta }=diag[\Upsilon _{\alpha }:\Upsilon
_{~1}^{1}=\Upsilon _{~2}^{2}=~^{h}\Upsilon (x^{k});\Upsilon
_{~3}^{3}=\Upsilon _{~4}^{4}=~^{v}\Upsilon (x^{k},y^{a})].  \label{esourc}
\end{equation}%
This assumption means that we shall work with such classes of nonholonomic
transforms and constraints when the effective sources are determined by 
\textbf{two generating sources} $\ ^{h}\Upsilon (x^{k})$ and $%
~^{v}\Upsilon(x^{k},y^{a})$. It imposes certain nonholonomic constraints on $%
\mathbf{T}_{\alpha \beta }$, cosmological constant $\Lambda$ and possible
splitting of such constants into h- and v-components; as well on distortion
d-tensors $\widehat{\mathbf{Z}}[\mathbf{g}]$ (\ref{cdist})\ and other values
included in $\widehat{\mathbf{Y}}.$ To decouple and integrate in general
explicit forms some physically important systems of PDEs (for geometric
flows and gravitational and matter field equations) is possible if we
consider that $\widehat{\mathbf{Y}}[\mathbf{g,}\widehat{\mathbf{D}},\kappa]$
contains a small parameter $\kappa$, or if the gravitational and matter
field dynamics is subjected to certain convenient classes of constraints. In
such cases, the solutions can be constructed exactly and/or recurrently
using power decompositions $\kappa ^{0},\kappa ^{1},\kappa ^{2},...$ (for
instance, $\kappa $ can be a string constant, or other parameter for
constructing ellipsoid deformations etc. We say that the corresponding
classes of solutions are exact/parametric, for instance, for linear
dependencies on $\kappa ^{0}$ and $\kappa ^{1}.$

With respect to N-adapted frames (\ref{nader}) and (\ref{nadif}), we can
re-write equivalently the Einstein equations (\ref{en1}) for $\nabla $ using
the canonical d-connection $\widehat{\mathbf{D}},$%
\begin{eqnarray}
\widehat{\mathbf{R}}_{\ \ \beta }^{\alpha } &=&\widehat{\mathbf{\Upsilon }}%
_{\ \ \beta }^{\alpha },  \label{cdeq1} \\
\widehat{\mathbf{T}}_{\ \alpha \beta }^{\gamma } &=&0,  \label{lccond1}
\end{eqnarray}%
where generating sources $\widehat{\mathbf{\Upsilon }}_{\ \ \beta
}^{\alpha}=[\ ^{h}\Upsilon \delta _{\ \ j}^{i}, \ ^{v}\Upsilon \delta _{\ \
b}^{a}]$ (\ref{esourc}) and the equations (\ref{lccond1}) are equivalent to (%
\ref{lccond}), when $\widehat{\mathcal{T}}=\{\widehat{\mathbf{T}}_{\ \alpha
\beta }^{\gamma }[\mathbf{g,N,}\widehat{\mathbf{D}}]\}$ is defined in
abstract form as in (\ref{fundgeom}). Here we note that, in general, $%
\widehat{\mathbf{D}}^{\beta }\widehat{\mathbf{E}}_{\ \ \beta }^{\alpha }\neq
0$ and $\widehat{\mathbf{D}}^{\beta }\widehat{\mathbf{\Upsilon }}_{\ \
\beta}^{\alpha }\neq 0,$ which is typical for nonholonomic systems. For
instance, in nonholonomic mechanics, the conservation laws are not standard
ones. We have to introduce the so-called Lagrange multiples associated to
certain classes of nonholonomic constraints. Solving the constraint
equations, it is possible to re-define the variables, then to introduce new
effective Lagrangians and, finally, to define standard conservation laws.
This can be performed in explicit general forms only for some "toy" models.\
Using distortions of connections (\ref{cdist}), we can rewrite (\ref{cdeq1})
in terms of $\nabla ,$ when $\nabla ^{\beta }E_{\ \ \beta }^{\alpha }=\nabla
^{\beta }T_{\ \ \beta }^{\alpha }=0.$ So, there are not conceptual problems
with the definition of conservation laws for matter fields using two
different linear connections (\ref{twocon}) defined by the same metric
structure $\mathbf{g.}$ This is different, for instance, from the
Einstein-Cartan and string theory with torsion fields when the second
connection may be not defined by the same metric structures but for certain
additional or effective gauge fields etc.

We conclude that all geometric constructions and physical theories derived
for the geometric data $\left( \mathbf{g},\nabla \right) $ can be
equivalently modeled by the canonical geometric data $(\mathbf{g,N},\widehat{%
\mathbf{D}})$ if we use canonical distortion relations (\ref{cdist}), or we
consider nonholonomic constraints of type (\ref{lccond1}), equivalently (\ref%
{lccond}). The main result of the AFCDM (to be proven in next section) is
that we can decouple in general forms the canonical nonholonomic Einstein
equations (\ref{cdeq1}) with $(\mathbf{g,N},\widehat{\mathbf{D}})$ and for
certain generic off-diagonal ansatz (\ref{ansatz}), re-written in adapted
form as a d-metric (\ref{dm}). For such ansatz, we cannot decouple the
standard Einstein equations (\ref{en1}) written in terms of geometric data $%
\left(\mathbf{g},\nabla \right)$. The main geometric and analytic idea of
the AFCDM is that we should search for certain classes of general solutions
of gravitational field equations written for $\widehat{\mathbf{D}}.$ After
some classes of off-diagonal metrics are constructed in terms of generated
and integration functions, we can extract LC-configurations for $\nabla $
imposing additional nonholonomic constraints (\ref{lccond1}), equivalently (%
\ref{lccond}). To solve such nonholonomic equations we have to restrict the
classes of generating and integration functions as we shall prove in next
section.

\subsection{Coefficient formulas for the canonical d-connection and Ricci
d-tensors}

\label{sec22} In this subsection, we provide some coefficient formulas which
are important for proofs of decoupling (modified) gravitational field
equations and generating solutions following the AFCDM, see details and
proofs in \cite{vmon3,vacaru18,bubuianu18a,vvy13}.

With respect to N--adapted frames (\ref{nader}) and (\ref{nadif}) for a
h-v-splitting (\ref{hvdcon}) of a d-connection$\ \mathbf{D},$ the
fundamental d-objects (\ref{fundgeom}) are defined by such coefficient
formulas: 
\begin{eqnarray}
\mbox{ d-curvature}, \mathcal{R}&=&\mathbf{\{R}_{\ \beta \gamma \delta
}^{\alpha}= (R_{\ hjk}^{i},R_{\ bjk}^{a},P_{\ hja}^{i},P_{\ bja}^{c},S_{\
hba}^{i},S_{\ bea}^{c})\},  \notag \\
\mbox{ for } R_{\ hjk}^{i} &=&\mathbf{e}_{k}L_{\ hj}^{i}-\mathbf{e}_{j}L_{\
hk}^{i}+ L_{\ hj}^{m}L_{\ mk}^{i}-L_{\ hk}^{m}L_{\ mj}^{i}-C_{\
ha}^{i}\Omega _{\ kj}^{a},  \notag \\
R_{\ bjk}^{a} &=&\mathbf{e}_{k}\acute{L}_{\ bj}^{a}-\mathbf{e}_{j}\acute{L}%
_{\ bk}^{a}+\acute{L}_{\ bj}^{c}\acute{L}_{\ ck}^{a}-\acute{L}_{\ bk}^{c}%
\acute{L}_{\ cj}^{a}-C_{\ bc}^{a}\Omega _{\ kj}^{c},  \label{dcurv} \\
P_{\ jka}^{i} &=&e_{a}L_{\ jk}^{i}-D_{k}\acute{C}_{\ ja}^{i}+\acute{C}_{\
jb}^{i}T_{\ ka}^{b},\ P_{\ bka}^{c}=e_{a}\acute{L}_{\ bk}^{c}-D_{k}C_{\
ba}^{c}+C_{\ bd}^{c}T_{\ ka}^{c},  \notag \\
S_{\ jbc}^{i} &=&e_{c}\acute{C}_{\ jb}^{i}-e_{b}\acute{C}_{\ jc}^{i}+\acute{C%
}_{\ jb}^{h}\acute{C}_{\ hc}^{i}-\acute{C}_{\ jc}^{h}\acute{C}_{\ hb}^{i},%
\hspace{0in}\ S_{\ bcd}^{a}=e_{d}C_{\ bc}^{a}-e_{c}C_{\ bd}^{a}+C_{\
bc}^{e}C_{\ ed}^{a}-C_{\ bd}^{e}C_{\ ec}^{a};  \notag \\
\mbox{ d-torsion}, \mathcal{T}&=& \{\mathbf{T}_{\ \alpha \beta }^{\gamma
}=(T_{\ jk}^{i},T_{\ ja}^{i},T_{\ ji}^{a},T_{\ bi}^{a},T_{\ bc}^{a})\}, 
\notag \\
\mbox{ for } T_{\ jk}^{i}&=& L_{jk}^{i}-L_{kj}^{i},T_{\
jb}^{i}=C_{jb}^{i},T_{\ ji}^{a}=-\Omega _{\ ji}^{a},\
T_{aj}^{c}=L_{aj}^{c}-e_{a}(N_{j}^{c}),T_{\ bc}^{a}=C_{bc}^{a}-C_{cb}^{a};
\label{dtors} \\
\mbox{d-nonmetricity}, \mathcal{Q} &=& \mathbf{\{Q}_{\gamma \alpha \beta }=
\left( Q_{kij},Q_{kab},Q_{cij},Q_{cab}\right) \},  \notag \\
\mbox{ for }
Q_{kij}&=&D_{k}g_{ij},Q_{kab}=D_{k}g_{ab},Q_{cij}=D_{c}g_{ij},Q_{cab}=D_{c}g_{ab}.
\label{dnonm}
\end{eqnarray}

The h-v-coefficients of the Ricci d-tensor, $\mathbf{R}ic=\{\mathbf{R}_{\
\beta \gamma }:=\mathbf{R}_{\ \beta \gamma \alpha }^{\alpha }\},$ split into
four groups of coefficient, respectively defined by contacting respective
indices in (\ref{dcurv}), 
\begin{equation}
\mathbf{R}_{\alpha \beta }=\{R_{ij}:=R_{\ ijk}^{k},\ R_{ia}:=-R_{\
ika}^{k},\ R_{ai}:=R_{\ aib}^{b},\ R_{ab}:=R_{\ abc}^{c}\}.  \label{driccic}
\end{equation}%
Using the inverse d-tensor of a d-metric (\ref{dm}), we can compute the
scalar curvature $\ ^{s}\widehat{R}$ of $\ \widehat{\mathbf{D}}$ is by
definition 
\begin{equation}
Rsc:=\mathbf{g}^{\alpha \beta }\mathbf{R}_{\ \alpha
\beta}=g^{ij}R_{ij}+g^{ab}R_{ab}.  \label{sdcurv}
\end{equation}

A canonical d-connection (\ref{twocon}) is defined by N-adapted coefficients 
$\widehat{\mathbf{D}}=\{\widehat{\mathbf{\Gamma }}_{\ \alpha
\beta}^{\gamma}= (\widehat{L}_{jk}^{i},\widehat{L}_{bk}^{a},\widehat{C}%
_{jc}^{i},\widehat{C}_{bc}^{a})\},$ for 
\begin{eqnarray}
\widehat{L}_{jk}^{i} &=&\frac{1}{2}g^{ir} (\mathbf{e}_{k}g_{jr}+\mathbf{e}%
_{j}g_{kr}-\mathbf{e}_{r}g_{jk}),  \label{cdc} \\
\widehat{L}_{bk}^{a} &=&e_{b}(N_{k}^{a})+\frac{1}{2}g^{ac}(\mathbf{e}%
_{k}g_{bc}-g_{dc}\ e_{b}N_{k}^{d}-g_{db}\ e_{c}N_{k}^{d}),  \notag \\
\widehat{C}_{jc}^{i} &=&\frac{1}{2}g^{ik}e_{c}g_{jk},\ \widehat{C}_{bc}^{a}=%
\frac{1}{2}g^{ad}(e_{c}g_{bd}+e_{b}g_{cd}-e_{d}g_{bc}),  \notag
\end{eqnarray}%
are computed for a d--metric $\mathbf{g}=[g_{ij},g_{ab}]$ (\ref{dm}) using
N--elongated partial derivatives (\ref{nader}). In a similar form, we can
compute the coefficients of a LC-connection $\nabla =\{\Gamma _{\
\alpha\beta }^{\gamma }\},$ see general coefficient and/or N-adapted
formulas in \cite{misner,vmon3}. We note that symbols $\Gamma _{\ \alpha
\beta }^{\gamma }$ are not boldface because $\nabla $ is not a d-connection,
i.e. it do not preserve a h- and v-splitting under parallelism. The
N-adapted coefficients of the canonical distortion d-tensor in (\ref{cdist})
can be found following formulas $\widehat{\mathbf{Z}}= \{\widehat{\mathbf{Z}}%
_{\ \alpha \beta }^{\gamma }= \widehat{\mathbf{\Gamma }}_{\ \alpha \beta
}^{\gamma }-\Gamma _{\ \alpha \beta }^{\gamma }\}.$

Finally, we note that introducing the formulas $\widehat{\mathbf{\Gamma }}%
_{\ \alpha \beta }^{\gamma }$ \ (\ref{cdc}) in (\ref{dcurv})-(\ref{sdcurv})
(instead of the coefficients of a general d-connection $\widehat{\mathbf{%
\Gamma }}_{\ \alpha \beta }^{\gamma })$), we can compute the N-adapted
coefficients of canonical fundamental d--objects, for instance, $\widehat{%
\mathcal{R}}=\{\widehat{\mathbf{R}}_{\ \beta \gamma \delta}^{\alpha }= (%
\widehat{R}_{\ hjk}^{i},\widehat{R}_{\ bjk}^{a},...)\},\ \widehat{\mathcal{T}%
}=\{\widehat{\mathbf{T}}_{\ \alpha \beta }^{\gamma }=(\widehat{T}_{\ jk}^{i},%
\widehat{T}_{\ ja}^{i},...)\},$ for $\widehat{\mathcal{Q}}=\{ \widehat{%
\mathbf{Q}}_{\gamma \alpha \beta }=(\widehat{Q}_{kij}=0, \widehat{Q}%
_{kab}=0) =0,$ and similarly for $\widehat{\mathbf{R}}_{\alpha \beta }=\{%
\widehat{R}_{ij}:=\widehat{R}_{\ ijk}^{k},\ ...\}$ and $\widehat{R}sc:=%
\mathbf{g}^{\alpha \beta }\widehat{\mathbf{R}}_{\ \alpha \beta }=g^{ij}%
\widehat{R}_{ij}+g^{ab}\widehat{R}_{ab}.$ Such formulas will be used in next
section to prove decoupling properties of (modified) Einstein equations.

\section{Decoupling and integration of (modified) Einstein equations}

\label{sec3} In this section we show how the canonical distorted Einstein
equations (\ref{cdeq1}) can be decoupled and integrated in general forms. We
provide necessary N-adapted coefficient formulas, study respective nonlinear
and dual symmetries and discuss most important and general variants for
parameterising such solutions. The conditions of extracting
LC-configurations and off-diagonal solutions in GR, i.e. for standard
Einstein equations (\ref{en1}), are stated in explicit form.

\subsection{Decoupling property}

We prove that for very general classes of off-diagonal ansatz for d-metrics,
the system of nonlinear PDEs (\ref{cdeq1}) with generating sources (\ref%
{esourc}) can be decoupled in general N-adapted form.

\subsubsection{Off-diagonal ansatz with a Killing vector}

Let us consider a quasi-stationary d-metric of type (\ref{dm}), when 
\begin{eqnarray}
\mathbf{\hat{g}} &=&g_{i}(x^{k})dx^{i}\otimes dx^{i}+h_{3}(x^{k},y^{3})%
\mathbf{e}^{3}\otimes \mathbf{e}^{3}+h_{4}(x^{k},y^{3})\mathbf{e}^{4}\otimes 
\mathbf{e}^{4},  \notag \\
&&\mathbf{e}^{3}=dy^{3}+w_{i}(x^{k},y^{3})dx^{i},\ \mathbf{e}%
^{4}=dy^{4}+n_{i}(x^{k},y^{3})dx^{i},  \label{dmq}
\end{eqnarray}%
with Killing symmetry on the time like coordinate $\partial _{4}=\partial
_{t}$. For such ansatz, the N-connection coefficients $\widehat{N}%
_{i}^{3}=w_{i}(x^{k},y^{3})$ and $\widehat{N}_{i}^{4}=n_{i}(x^{k},y^{3})$
and N-adapted coefficients of d-metric $\widehat{\mathbf{g}}_{\alpha \beta
}=[\widehat{g}_{ij}(x^{\kappa }),\widehat{g}_{ab}(x^{\kappa },y^{3})]$ are
functions of necessary smooth class. With respect to coordinate frames, the
d-metric (\ref{dmq}) transforms into an off-diagonal ansatz (\ref{ansatz}),
when the coefficients of metrics do not depend on $y^{4}=t,$ but depend in
certain general forms on space coordinates $(x^{i},y^{3}).$ We can prove
decoupling properties for more general ansatz for d-metrics, when 
\begin{eqnarray*}
\mathbf{g} &=&g_{i}(x^{k})dx^{i}\otimes dx^{i}+\omega ^{2}(x^{k},y^{a})[%
\underline{h}_{3}(x^{k},y^{4})h_{3}(x^{k},y^{3})\mathbf{e}^{3}\otimes 
\mathbf{e}^{3}+h_{4}(x^{k},y^{3})\underline{h}_{4}(x^{k},y^{4})]\ \mathbf{e}%
^{4}\otimes \mathbf{e}^{4}, \\
&&\mathbf{e}^{3}=dy^{3}+[w_{i}(x^{k},y^{3})+\underline{n}%
_{i}(x^{k},y^{4})]dx^{i},\ \mathbf{e}^{4}=dy^{4}+[n_{i}(x^{k},y^{3})+%
\underline{w}_{i}(x^{k},y^{4})]dx^{i},
\end{eqnarray*}%
does not have explicit Killing symmetries but may involve vertical co-space
conformal transforms with factor $\omega (x^{k},y^{a}),$ see examples in 
\cite{vmon3,vvy13}. Such an ansatz results in more cumbersome formulas and
additional technical difficulties for generating exact/parametric solutions.
For simplicity, we do not provide such constructions in this work. We label
a d-metric $\mathbf{\hat{g}}$ with a "hat" in order to emphasize that it is
of type (\ref{dmq}) with Killing symmetry on $\partial _{t}.$ It is supposed
that such a parametrization can be obtained under some frame/coordinate
transforms even, in general, such $\mathbf{\hat{g}}(u)$ may depend on all
spacetime coordinates.

We can consider a different ansatz for locally anisotropic cosmological
d-metrics, 
\begin{eqnarray}
\underline{\mathbf{g}} &=&g_{i}(x^{k})dx^{i}\otimes dx^{i}+\underline{h}%
_{3}(x^{k},t)\underline{\mathbf{e}}^{3}\otimes \underline{\mathbf{e}}^{3}+%
\underline{h}_{4}(x^{k},t)\underline{\mathbf{e}}^{4}\otimes \underline{%
\mathbf{e}}^{4},  \notag \\
&&\underline{\mathbf{e}}^{3}=dy^{3}+\underline{n}_{i}(x^{k},t)dx^{i},\ 
\underline{\mathbf{e}}^{4}=dy^{4}+\underline{w}_{i}(x^{k},t)dx^{i},
\label{dmc}
\end{eqnarray}%
with Killing symmetry on the space like coordinate $\partial _{3}$.
Correspondingly, we parameterize the N-connection coefficients $\underline{N}%
_{i}^{3}=\underline{n}_{i}(x^{k},t)$ and $\underline{N}_{i}^{4}=\underline{w}%
_{i}(x^{k},t)$ and the coefficients of a d-metric $\underline{\mathbf{g}}%
_{\alpha \beta }=[g_{ij}(x^{\kappa }),\underline{g}_{ab}(x^{\kappa },t)]$,
all such values being functions of necessary smooth class. With respect to
coordinate frames, the d-metric (\ref{dmc}) transforms into a different type
off-diagonal ansatz (\ref{ansatz}), when the coefficients of metrics do not
depend on $y^{3},$ but depend in certain general forms on spacetime
coordinates $(x^{i},y^{4}=t).$

For simplicity, we shall sketch proofs of general decoupling and integration
properties for quasi-stationary d-metrics (\ref{dmq}). To generate solutions
for locally anisotropic d-metrics we can change in formal symbolic forms,
respectively, $h_{3}(x^{k},y^{3})\rightarrow \underline{h}_{4}(x^{k},t),$ $%
h_{4}(x^{k},y^{3})\rightarrow \underline{h}_{3}(x^{k},t)$ and $%
w_{i}(x^{k},y^{3})\rightarrow \underline{n}_{i}(x^{k},t),$ $%
n_{i}(x^{k},y^{3})\rightarrow \underline{w}_{i}(x^{k},t).$ Such "dual"
symmetries can be prescribed only for generic off-diagonal solutions with
respective Killing symmetries on $\partial _{4},$ or $\partial _{3}.$
Nevertheless, this allow to study main geometric and physical properties of
generic off-diagonal metrics with possible nonholonomic constraints and
deformations constructed as solutions of systems of nonlinear PDEs,
depending, in principle, on 3 from 4 spacetime/space coordinates, not
reducing the problem to finding solutions of systems of nonlinear ODEs.

\subsubsection{N-adapted coefficients of quasi-stationary canonical
d-connections}

To simplify computations we use brief notations of partial derivatives, for
instance, $\partial _{1}q(u^{\alpha })=q^{\bullet },$ $\partial
_{2}q(u^{\alpha })=q^{\prime }$, $\partial _{3}q(u^{\alpha })=q^{\ast }$ and 
$\partial _{4}q(u^{\alpha})=q^{\diamond }$.

There are such nontrivial coefficients of $\widehat{\mathbf{\Gamma }}_{\
\alpha \beta }^{\gamma }$ (\ref{cdc}) computed for quasi-stationary
d-metrics (\ref{dmq}), 
\begin{eqnarray}
\widehat{L}_{11}^{1} &=&\frac{g_{1}^{\bullet }}{2g_{1}},\ \widehat{L}%
_{12}^{1}=\frac{g_{1}^{\prime }}{2g_{1}},\widehat{L}_{22}^{1}=-\frac{%
g_{2}^{\bullet }}{2g_{1}},\ \widehat{L}_{11}^{2}=\frac{-g_{1}^{\prime }}{%
2g_{2}},\ \widehat{L}_{12}^{2}=\frac{g_{2}^{\bullet }}{2g_{2}},\ \widehat{L}%
_{22}^{2}=\frac{g_{2}^{\prime }}{2g_{2}},  \label{nontrdc} \\
\widehat{L}_{4k}^{4} &=&\frac{\mathbf{\partial }_{k}(h_{4})}{2h_{4}}-\frac{%
w_{k}h_{4}^{\ast }}{2h_{4}},\widehat{L}_{3k}^{3}=\frac{\mathbf{\partial }%
_{k}h_{3}}{2h_{3}}-\frac{w_{k}h_{3}^{\ast }}{2h_{3}},\widehat{L}_{4k}^{3}=-%
\frac{h_{4}}{2h_{3}}n_{k}^{\ast },  \notag \\
\widehat{L}_{3k}^{4} &=&\frac{1}{2}n_{k}^{\ast },\widehat{C}_{33}^{3}=\frac{%
h_{3}^{\ast }}{2h_{3}},\widehat{C}_{44}^{3}=-\frac{h_{4}^{\ast }}{h_{3}},\ 
\widehat{C}_{33}^{4}=0,~\widehat{C}_{34}^{4}=\frac{h_{4}^{\ast }}{2h_{4}},%
\widehat{C}_{44}^{4}=0.  \notag
\end{eqnarray}%
We shall need also the values 
\begin{equation}
\ \widehat{C}_{3}=\widehat{C}_{33}^{3}+\widehat{C}_{34}^{4}=\frac{%
h_{3}^{\ast }}{2h_{3}}+\frac{h_{4}^{\ast }}{2h_{4}},\widehat{C}_{4}=\widehat{%
C}_{43}^{3}+\widehat{C}_{44}^{4}=0.  \label{aux3}
\end{equation}

The formulas (\ref{nontrdc}) and (\ref{aux3}) are important for computing in
explicit form the N-adapted coefficients of the canonical d-torsion and
canonical Ricci and Einstein d-tensors.

\subsubsection{Coefficients of the N-connection curvature, canonical
d-torsion, and LC-conditions}

For the N-connection coefficients in (\ref{dmq}), the coefficients of the
N-connection curvature $\widehat{\Omega }_{ij}^{a}=\widehat{\mathbf{e}}%
_{j}\left( \widehat{N}_{i}^{a}\right) -\widehat{\mathbf{e}}_{i}(\widehat{N}%
_{j}^{a})$ used in (\ref{anhcoef}) are computed 
\begin{equation*}
\widehat{\Omega }_{ij}^{a}=\mathbf{\partial }_{j}\left( \widehat{N}%
_{i}^{a}\right) -\partial _{i}(\widehat{N}_{j}^{a})-w_{i}(\widehat{N}%
_{j}^{a})^{\ast }+w_{j}(\widehat{N}_{i}^{a})^{\ast }.
\end{equation*}%
We find such nontrivial values 
\begin{eqnarray}
\widehat{\Omega }_{12}^{3} &=&-\widehat{\Omega }_{21}^{3}=\mathbf{\partial }%
_{2}w_{1}-\partial _{1}w_{2}-w_{1}w_{2}^{\ast }+w_{2}w_{1}^{\ast
}=w_{1}^{\prime }-w_{2}^{\bullet }-w_{1}w_{2}{}^{\ast }+w_{2}w_{1}^{\ast }{};
\notag \\
\widehat{\Omega }_{12}^{4} &=&-\widehat{\Omega }_{21}^{4}=\mathbf{\partial }%
_{2}n_{1}-\partial _{1}n_{2}-w_{1}n_{2}^{\ast }+w_{2}n_{1}^{\ast
}=n_{1}^{\prime }-n_{2}^{\bullet }-w_{1}n_{2}^{\ast }{}+w_{2}n_{1}^{\ast }{}.
\label{omeg}
\end{eqnarray}

As a result, we can compute the nontrivial coefficients of the canonical
d--torsion using formulas (\ref{dtors}). We have $\widehat{T}_{\
ji}^{a}=-\Omega _{\ ji}^{a}$, with nontrivial coefficients (\ref{omeg}), and 
$\widehat{T}_{aj}^{c}=\widehat{L}_{aj}^{c}-e_{a}(\widehat{N}_{j}^{c}).$ For
other types of coefficients, we express%
\begin{eqnarray*}
\widehat{T}_{\ jk}^{i} &=&\widehat{L}_{jk}^{i}-\widehat{L}_{kj}^{i}=0,~%
\widehat{T}_{\ ja}^{i}=\widehat{C}_{jb}^{i}=0,~\widehat{T}_{\ bc}^{a}=\ 
\widehat{C}_{bc}^{a}-\ \widehat{C}_{cb}^{a}=0, \\
\widehat{T}_{3k}^{3} &=&\widehat{L}_{3k}^{3}-e_{3}(\widehat{N}_{k}^{3})=%
\frac{\mathbf{\partial }_{k}h_{3}}{2h_{3}}-w_{k}\frac{h_{3}^{\ast }}{2h_{3}}%
-w_{k}^{\ast }{},\widehat{T}_{4k}^{3}=\widehat{L}_{4k}^{3}-e_{4}(\widehat{N}%
_{k}^{3})=-\frac{h_{4}}{2h_{3}}n_{k}^{\ast },\ 
\end{eqnarray*}%
\begin{eqnarray}
\widehat{T}_{3k}^{4} &=&~\widehat{L}_{3k}^{4}-e_{3}(\widehat{N}_{k}^{4})=%
\frac{1}{2}n_{k}^{\ast }-n_{k}^{\ast }=-\frac{1}{2}n_{k}^{\ast },\widehat{T}%
_{4k}^{4}=\widehat{L}_{4k}^{4}-e_{4}(N_{k}^{4})=\frac{\mathbf{\partial }%
_{k}h_{4}}{2h_{4}}-w_{k}\frac{h_{4}^{\ast }}{2h_{4}},  \notag \\
-\widehat{T}_{12}^{3} &=&w_{1}^{\prime }-w_{2}^{\bullet }-w_{1}w_{2}^{\ast
}{}+w_{2}w_{1}^{\ast },\ -\widehat{T}_{12}^{4}=n_{1}^{\prime
}-n_{2}^{\bullet }-w_{1}n_{2}^{\ast }{}+w_{2}n_{1}^{\ast }{}.
\label{nontrtors}
\end{eqnarray}

The LC-conditions (\ref{lccond1}) for zero canonical d-torsions, are
satisfied if 
\begin{equation}
\widehat{L}_{aj}^{c}=e_{a}(\widehat{N}_{j}^{c}),\ \widehat{C}_{jb}^{i}=0,\ 
\widehat{\Omega }_{\ ji}^{a}=0,  \label{lcconstr}
\end{equation}%
when in N-adapted frames $\widehat{\mathbf{\Gamma }}_{\ \alpha \beta
}^{\gamma }=\Gamma _{\ \alpha \beta }^{\gamma }$ even, in general, $\widehat{%
\mathbf{D}}\neq \nabla $. This is possible because two different linear
connections have different transformation laws under general frame/
coordinate transforms (such values are not (d) tensor objects). In the
LC--cases, all values (\ref{nontrtors}) must vanish. We obtain nontrivial
off-diagonal solutions with $h_{4}^{\ast }\neq 0$ and $w_{k}^{\ast }\neq 0$
but $n_{k}^{\ast }=0,$ for $w_{k}=\mathbf{\partial }_{k}h_{4}/h_{4}^{\ast }.$
We can search for other types of LC-configurations with $n_{k}^{\ast }\neq 0$
and/or $h_{3}^{\ast }\neq 0$ but it is difficult to obtain explicit formulas
for such classes of solutions (they may be also off-diagonal). Finally, it
should be noted that conditions of type (\ref{lcconstr}) can be imposed
after a general class of quasi-stationary off-diagonal metrics is
constructed in a general off-diagonal form involving a nonholonomic torsion.
It should be noted that it is not possible to decouple in a general form the
Einstein equations working from the very beginning with $\nabla $ defined by
a generic off-diagonal ansatz with coefficients depending on 2-4 coordinates.

\subsubsection{N--adapted coefficients of the canonical Ricci d--tensor}

The h-coefficients of a canonical variant of Ricci d-tensor (\ref{driccic})
are computed for contractions of indices in (\ref{dcurv}), when $\widehat{R}%
_{ij}=\widehat{R}_{\ ijk}^{k}$, for 
\begin{eqnarray}
\widehat{R}_{\ hjk}^{i} &=&\mathbf{e}_{k}\widehat{L}_{.hj}^{i}-\mathbf{e}_{j}%
\widehat{L}_{hk}^{i}+\widehat{L}_{hj}^{m}\widehat{L}_{mk}^{i}-\widehat{L}%
_{hk}^{m}\widehat{L}_{mj}^{i}-\widehat{C}_{ha}^{i}\widehat{\Omega }_{jk}^{a}
\notag \\
&=&\mathbf{\partial }_{k}\widehat{L}_{.hj}^{i}-\partial _{j}\widehat{L}%
_{hk}^{i}+\widehat{L}_{hj}^{m}\widehat{L}_{mk}^{i}-\widehat{L}_{hk}^{m}%
\widehat{L}_{mj}^{i}.  \label{auxcurvh}
\end{eqnarray}%
We note that these formulas are for and quasi-stationary ansatz (\ref{dmq})
and respectively computed values (\ref{nontrdc}) when $\widehat{C}_{\
ha}^{i}=0$ and 
\begin{equation*}
e_{k}\widehat{L}_{hj}^{i}=\partial _{k}\widehat{L}_{hj}^{i}+N_{k}^{a}%
\partial _{a}\widehat{L}_{hj}^{i}=\partial _{k}\widehat{L}_{hj}^{i}+w_{k}(%
\widehat{L}_{hj}^{i})^{\ast }+n_{k}(\widehat{L}_{hj}^{i})^{\diamond
}=\partial _{k}\widehat{L}_{hj}^{i}
\end{equation*}%
because $\widehat{L}_{hj}^{i}$ depend only in h-coordinates. Taking
derivatives of (\ref{nontrdc}), we obtain 
\begin{eqnarray*}
\partial _{1}\widehat{L}_{\ 11}^{1} &=&(\frac{g_{1}^{\bullet }}{2g_{1}}%
)^{\bullet }=\frac{g_{1}^{\bullet \bullet }}{2g_{1}}-\frac{\left(
g_{1}^{\bullet }\right) ^{2}}{2\left( g_{1}\right) ^{2}},\ \partial _{1}%
\widehat{L}_{\ 12}^{1}=(\frac{g_{1}^{\prime }}{2g_{1}})^{\bullet }=\frac{%
g_{1}^{\prime \bullet }}{2g_{1}}-\frac{g_{1}^{\bullet }g_{1}^{\prime }}{%
2\left( g_{1}\right) ^{2}},\  \\
\partial _{1}\widehat{L}_{\ 22}^{1} &=&(-\frac{g_{2}^{\bullet }}{2g_{1}}%
)^{\bullet }=-\frac{g_{2}^{\bullet \bullet }}{2g_{1}}+\frac{g_{1}^{\bullet
}g_{2}^{\bullet }}{2\left( g_{1}\right) ^{2}},\ \partial _{1}\widehat{L}_{\
11}^{2}=(-\frac{g_{1}^{\prime }}{2g_{2}})^{\bullet }=-\frac{g_{1}^{\prime
\bullet }}{2g_{2}}+\frac{g_{1}^{\bullet }g_{2}^{\prime }}{2\left(
g_{2}\right) ^{2}}, \\
\partial _{1}\widehat{L}_{\ 12}^{2} &=&(\frac{g_{2}^{\bullet }}{2g_{2}}%
)^{\bullet }=\frac{g_{2}^{\bullet \bullet }}{2g_{2}}-\frac{\left(
g_{2}^{\bullet }\right) ^{2}}{2\left( g_{2}\right) ^{2}},\ \partial _{1}%
\widehat{L}_{\ 22}^{2}=(\frac{g_{2}^{\prime }}{2g_{2}})^{\bullet }=\frac{%
g_{2}^{\prime \bullet }}{2g_{2}}-\frac{g_{2}^{\bullet }g_{2}^{\prime }}{%
2\left( g_{2}\right) ^{2}},
\end{eqnarray*}%
\begin{eqnarray*}
\partial _{2}\widehat{L}_{\ 11}^{1} &=&(\frac{g_{1}^{\bullet }}{2g_{1}}%
)^{\prime }=\frac{g_{1}^{\bullet \prime }}{2g_{1}}-\frac{g_{1}^{\bullet
}g_{1}^{\prime }}{2\left( g_{1}\right) ^{2}},~\partial _{2}\widehat{L}_{\
12}^{1}=(\frac{g_{1}^{\prime }}{2g_{1}})^{\prime }=\frac{g_{1}^{\prime
\prime }}{2g_{1}}-\frac{\left( g_{1}^{\prime }\right) ^{2}}{2\left(
g_{1}\right) ^{2}}, \\
\partial _{2}\widehat{L}_{\ 22}^{1} &=&(-\frac{g_{2}^{\bullet }}{2g_{1}}%
)^{\prime }=-\frac{g_{2}^{\bullet ^{\prime }}}{2g_{1}}+\frac{g_{2}^{\bullet
}g_{1}^{^{\prime }}}{2\left( g_{1}\right) ^{2}},\ \partial _{2}\widehat{L}%
_{\ 11}^{2}=(-\frac{g_{1}^{\prime }}{2g_{2}})^{\prime }=-\frac{g_{1}^{\prime
\prime }}{2g_{2}}+\frac{g_{1}^{\bullet }g_{1}^{\prime }}{2\left(
g_{2}\right) ^{2}}, \\
\partial _{2}\widehat{L}_{\ 12}^{2} &=&(\frac{g_{2}^{\bullet }}{2g_{2}}%
)^{\prime }=\frac{g_{2}^{\bullet \prime }}{2g_{2}}-\frac{g_{2}^{\bullet
}g_{2}^{\prime }}{2\left( g_{2}\right) ^{2}},\partial _{2}\widehat{L}_{\
22}^{2}=(\frac{g_{2}^{\prime }}{2g_{2}})^{\prime }=\frac{g_{2}^{\prime
\prime }}{2g_{2}}-\frac{\left( g_{2}^{\prime }\right) ^{2}}{2\left(
g_{2}\right) ^{2}}.
\end{eqnarray*}%
Introducing these values in (\ref{auxcurvh}), we obtain two types of
nontrivial components: 
\begin{eqnarray*}
\widehat{R}_{\ 212}^{1} &=&\frac{g_{2}^{\bullet \bullet }}{2g_{1}}-\frac{%
g_{1}^{\bullet }g_{2}^{\bullet }}{4\left( g_{1}\right) ^{2}}-\frac{\left(
g_{2}^{\bullet }\right) ^{2}}{4g_{1}g_{2}}+\frac{g_{1}^{\prime \prime }}{%
2g_{1}}-\frac{g_{1}^{\prime }g_{2}^{\prime }}{4g_{1}g_{2}}-\frac{\left(
g_{1}^{\prime }\right) ^{2}}{4\left( g_{1}\right) ^{2}}, \\
\widehat{R}_{\ 112}^{2} &=&-\frac{g_{2}^{\bullet \bullet }}{2g_{2}}+\frac{%
g_{1}^{\bullet }g_{2}^{\bullet }}{4g_{1}g_{2}}+\frac{\left( g_{2}^{\bullet
}\right) ^{2}}{4(g_{2})^{2}}-\frac{g_{1}^{\prime \prime }}{2g_{2}}+\frac{%
g_{1}^{\prime }g_{2}^{\prime }}{4(g_{2})^{2}}+\frac{\left( g_{1}^{\prime
}\right) ^{2}}{4g_{1}g_{2}}.
\end{eqnarray*}%
Considering that $\widehat{R}_{11}=-\widehat{R}_{\ 112}^{2}$ and $\widehat{R}%
_{22}=\widehat{R}_{\ 212}^{1},$ for $g^{i}=1/g_{i}$ and $\widehat{R}%
_{j}^{j}=g^{j}\widehat{R}_{jj}$ (in these formulas, there is no summarizing
on repeating indices), we compute 
\begin{equation}
\widehat{R}_{1}^{1}=\widehat{R}_{2}^{2}=-\frac{1}{2g_{1}g_{2}}%
[g_{2}^{\bullet \bullet }-\frac{g_{1}^{\bullet }g_{2}^{\bullet }}{2g_{1}}-%
\frac{\left( g_{2}^{\bullet }\right) ^{2}}{2g_{2}}+g_{1}^{\prime \prime }-%
\frac{g_{1}^{\prime }g_{2}^{\prime }}{2g_{2}}-\frac{(g_{1}^{\prime })^{2}}{%
2g_{1}}].  \label{hcdric}
\end{equation}

At the next step, we compute N-adapted coefficients with mixed h- and
v-indices of the canonical Ricci d-tensor. We consider the third formula in (%
\ref{dcurv}), 
\begin{equation*}
\widehat{R}_{\ bka}^{c}=\frac{\partial \widehat{L}_{bk}^{c}}{\partial y^{a}}-%
\widehat{C}_{~ba|k}^{c}+\widehat{C}_{~bd}^{c}\widehat{T}_{~ka}^{d}=\frac{%
\partial \widehat{L}_{bk}^{c}}{\partial y^{a}}-(\frac{\partial \widehat{C}%
_{ba}^{c}}{\partial x^{k}}+\widehat{L}_{dk}^{c\,}\widehat{C}_{ba}^{d}-%
\widehat{L}_{bk}^{d}\widehat{C}_{da}^{c}-\widehat{L}_{ak}^{d}\widehat{C}%
_{bd}^{c})+\widehat{C}_{bd}^{c}\widehat{T}_{ka}^{d}.
\end{equation*}%
Contracting respectively the indices, we obtain 
\begin{equation*}
\widehat{R}_{bk}=\widehat{R}_{\ bka}^{a}=\frac{\partial L_{bk}^{a}}{\partial
y^{a}}-\widehat{C}_{ba|k}^{a}+\widehat{C}_{bd}^{a}\widehat{T}_{ka}^{d},
\end{equation*}%
where for $\widehat{C}_{b}:=\widehat{C}_{ba}^{c}$ (\ref{aux3}),%
\begin{equation*}
\widehat{C}_{b|k}=\mathbf{e}_{k}\widehat{C}_{b}-\widehat{L}_{\ bk}^{d\,}%
\widehat{C}_{d}=\partial _{k}\widehat{C}_{b}-N_{k}^{e}\partial _{e}\widehat{C%
}_{b}-\widehat{L}_{\ bk}^{d\,}\widehat{C}_{d}=\partial _{k}\widehat{C}%
_{b}-w_{k}\widehat{C}_{b}^{\ast }-n_{k}\widehat{C}_{b}^{\diamond }-\widehat{L%
}_{\ bk}^{d\,}\widehat{C}_{d}.
\end{equation*}%
We consider a conventional splitting $\widehat{R}_{bk}=\ _{[1]}R_{bk}+\
_{[2]}R_{bk}+\ _{[3]}R_{bk},$ where%
\begin{eqnarray*}
\ _{[1]}R_{bk} &=&(\widehat{L}_{bk}^{3})^{\ast }+(\widehat{L}%
_{bk}^{4})^{\diamond },\ _{[2]}R_{bk}=-\partial _{k}\widehat{C}_{b}+w_{k}%
\widehat{C}_{b}^{\ast }+n_{k}\widehat{C}_{b}^{\diamond }+\widehat{L}_{\
bk}^{d\,}\widehat{C}_{d}, \\
\ _{[3]}R_{bk} &=&\widehat{C}_{bd}^{a}\widehat{T}_{ka}^{d}=\widehat{C}%
_{b3}^{3}\widehat{T}_{k3}^{3}+\widehat{C}_{b4}^{3}\widehat{T}_{k3}^{4}+%
\widehat{C}_{b3}^{4}\widehat{T}_{k4}^{3}+\widehat{C}_{b4}^{4}\widehat{T}%
_{k4}^{4}.
\end{eqnarray*}%
Using formulas (\ref{nontrdc}), (\ref{nontrtors}) and (\ref{aux3}), we
compute 
\begin{eqnarray*}
\ _{[1]}R_{3k} &=&\left( \widehat{L}_{3k}^{3}\right) ^{\ast }+\left( 
\widehat{L}_{3k}^{4}\right) ^{\diamond }=\left( \frac{\mathbf{\partial }%
_{k}h_{3}}{2h_{3}}-w_{k}\frac{h_{3}^{\ast }}{2h_{3}}\right) ^{\ast
}=-w_{k}^{\ast }\frac{h_{3}^{\ast }}{2h_{3}}-w_{k}\left( \frac{h_{3}^{\ast }%
}{2h_{3}}\right) ^{\ast }+\frac{1}{2}\left( \frac{\mathbf{\partial }_{k}h_{3}%
}{h_{3}}\right) ^{\ast }, \\
\ _{[2]}R_{3k} &=&-\partial _{k}\widehat{C}_{3}+w_{k}\widehat{C}_{3}^{\ast
}+n_{k}\widehat{C}_{3}^{\diamond }+\widehat{L}_{\ 3k}^{3\,}\widehat{C}_{3}+%
\widehat{L}_{\ 3k}^{4\,}\widehat{C}_{4}= \\
&=&w_{k}[\frac{h_{3}^{\ast \ast }}{2h_{3}}-\frac{3}{4}\frac{(h_{3}^{\ast
})^{2}}{(h_{3})^{2}}+\frac{h_{4}^{\ast \ast }}{2h_{4}}-\frac{1}{2}\frac{%
(h_{4}^{\ast })^{2}}{(h_{4})^{2}}-\frac{1}{4}\frac{h_{3}^{\ast }}{h_{3}}%
\frac{h_{4}^{\ast }}{h_{4}}]+\frac{\mathbf{\partial }_{k}h_{3}}{2h_{3}}(%
\frac{h_{3}^{\ast }}{2h_{3}}+\frac{h_{4}^{\ast }}{2h_{4}})-\frac{1}{2}%
\partial _{k}(\frac{h_{3}^{\ast }}{h_{3}}+\frac{h_{4}^{\ast }}{h_{4}}), \\
\ _{[3]}R_{3k} &=&\widehat{C}_{33}^{3}\widehat{T}_{k3}^{3}+\widehat{C}%
_{34}^{3}\widehat{T}_{k3}^{4}+\widehat{C}_{33}^{4}\widehat{T}_{k4}^{3}+%
\widehat{C}_{34}^{4}\widehat{T}_{k4}^{4} \\
&=&w_{k}\left( \frac{(h_{3}^{\ast })^{2}}{4(h_{3})^{2}}+\frac{(h_{4}^{\ast
})^{2}}{4(h_{4})^{2}}\right) +w_{k}^{\ast }\frac{h_{3}^{\ast }}{2h_{3}}-%
\frac{h_{3}^{\ast }}{2h_{3}}\frac{\mathbf{\partial }_{k}h_{3}}{2h_{3}}-\frac{%
h_{4}^{\ast }}{2h_{4}}\frac{\mathbf{\partial }_{k}h_{4}}{2h_{4}}.
\end{eqnarray*}%
Putting together above formulas (originally such computations were provided
in \cite{sv11}), we find that%
\begin{eqnarray}
\ \widehat{R}_{3k} &=&w_{k}[\frac{h_{4}^{\ast \ast }}{2h_{4}}-\frac{1}{4}%
\frac{(h_{4}^{\ast })^{2}}{(h_{4})^{2}}-\frac{1}{4}\frac{h_{3}^{\ast }}{h_{3}%
}\frac{h_{4}^{\ast }}{h_{4}}]+\frac{h_{4}^{\ast }}{2h_{4}}\frac{\mathbf{%
\partial }_{k}h_{3}}{2h_{3}}-\frac{1}{2}\frac{\partial _{k}h_{4}^{\ast }}{%
h_{4}}+\frac{1}{4}\frac{h_{4}^{\ast }\partial _{k}h_{4}}{(h_{4})^{2}}  \notag
\\
&=&\frac{w_{k}}{2h_{4}}[h_{4}^{\ast \ast }-\frac{(h_{4}^{\ast })^{2}}{2h_{4}}%
-\frac{h_{3}^{\ast }h_{4}^{\ast }}{2h_{3}}]+\frac{h_{4}^{\ast }}{4h_{4}}(%
\frac{\mathbf{\partial }_{k}h_{3}}{h_{3}}+\frac{\partial _{k}h_{4}^{\ast }}{%
h_{4}})-\frac{1}{2}\frac{\partial _{k}h_{4}^{\ast }}{h_{4}}.
\label{vhcdric3}
\end{eqnarray}

The N-adapted coefficients $\ \widehat{R}_{4k}=\ _{[1]}R_{4k}+\
_{[2]}R_{4k}+\ _{[3]}R_{4k},$ are defined by%
\begin{eqnarray*}
\ _{[1]}R_{4k} &=&(\widehat{L}_{4k}^{3})^{\ast }+(\widehat{L}%
_{4k}^{4})^{\diamond },\ _{[2]}R_{4k}=-\partial _{k}\widehat{C}_{4}+w_{k}%
\widehat{C}_{4}^{\ast }+n_{k}\widehat{C}_{4}^{\diamond }+\widehat{L}_{\
4k}^{3\,}\widehat{C}_{3}+\widehat{L}_{\ 4k}^{4\,}\widehat{C}_{4}, \\
_{\lbrack 3]}R_{4k} &=&\widehat{C}_{4d}^{a}\widehat{T}_{ka}^{d}=\widehat{C}%
_{43}^{3}\widehat{T}_{k3}^{3}+\widehat{C}_{44}^{3}\widehat{T}_{k3}^{4}+%
\widehat{C}_{43}^{4}\widehat{T}_{k4}^{3}+\widehat{C}_{44}^{4}\widehat{T}%
_{k4}^{4}.
\end{eqnarray*}%
Introducing in these formulas $\widehat{L}_{4k}^{3}$ and $\widehat{L}%
_{4k}^{4}$ from (\ref{nontrdc}), we compute%
\begin{equation*}
\ _{[1]}R_{4k}=(\widehat{L}_{4k}^{3})^{\ast }+(\widehat{L}%
_{4k}^{4})^{\diamond }=(-\frac{h_{4}}{2h_{3}}n_{k}^{\ast })^{\ast
}=-n_{k}^{\ast \ast }\frac{h_{4}}{2h_{3}}-n_{k}^{\ast }\frac{h_{4}^{\ast
}h_{3}-h_{4}h_{3}^{\ast }}{2(h_{3})^{2}}.
\end{equation*}%
The second term follows from $\widehat{C}_{3}$ and $\widehat{C}_{4},$ see (%
\ref{aux3}), and for $\ \widehat{L}_{4k}^{3}$ and $\widehat{L}_{4k}^{4},$ (%
\ref{nontrdc}), 
\begin{equation*}
\ _{[2]}R_{4k}=-\partial _{k}\widehat{C}_{4}+w_{k}\widehat{C}_{4}^{\ast
}+n_{k}\widehat{C}_{4}^{\diamond }+\widehat{L}_{\ 4k}^{3\,}\widehat{C}_{3}+%
\widehat{L}_{\ 4k}^{4\,}\widehat{C}_{4}=-n_{k}^{\ast }{}\frac{h_{4}}{2h_{3}}(%
\frac{h_{3}^{\ast }}{2h_{3}}+\frac{h_{4}^{\ast }}{2h_{4}}).
\end{equation*}%
Then we use $\widehat{C}_{43}^{3},\widehat{C}_{44}^{3},\widehat{C}_{43}^{4},%
\widehat{C}_{44}^{4}$ from (\ref{nontrdc}) and $\widehat{T}_{k3}^{3},%
\widehat{T}_{k3}^{4},\widehat{T}_{k4}^{3},\widehat{T}_{k4}^{4}$ from (\ref%
{nontrtors}) to compute the third term,%
\begin{equation*}
_{\lbrack 3]}R_{4k}=\widehat{C}_{43}^{3}\widehat{T}_{k3}^{3}+\widehat{C}%
_{44}^{3}\widehat{T}_{k3}^{4}+\widehat{C}_{43}^{4}\widehat{T}_{k4}^{3}+%
\widehat{C}_{44}^{4}\widehat{T}_{k4}^{4}=0.
\end{equation*}%
Summarizing above three terms,%
\begin{equation}
\widehat{R}_{4k}=-n_{k}^{\ast \ast }{}\frac{h_{4}}{2h_{3}}+n_{k}^{\ast
}{}\left( -\frac{h_{4}^{\ast }}{2h_{3}}+\frac{h_{4}^{\ast }h_{3}^{\ast }}{%
2(h_{3})^{\ast }}-\frac{h_{4}^{\ast }h_{3}^{\ast }}{4(h_{3})^{\ast }}-\frac{%
h_{4}^{\ast }}{4h_{3}}\right) .  \label{vhcdric4}
\end{equation}

For the N-adapted coefficients 
\begin{equation*}
\widehat{R}_{\ jka}^{i}=\frac{\partial \widehat{L}_{jk}^{i}}{\partial y^{k}}%
-(\frac{\partial \widehat{C}_{ja}^{i}}{\partial x^{k}}+\widehat{L}_{lk}^{i}%
\widehat{C}_{ja}^{l}-\widehat{L}_{jk}^{l}\widehat{C}_{la}^{i}-\widehat{L}%
_{ak}^{c}\widehat{C}_{jc}^{i})+\widehat{C}_{jb}^{i}\widehat{T}_{ka}^{b}
\end{equation*}%
from (\ref{dcurv}), we obtain zero values because $\widehat{C}_{jb}^{i}=0$
and $\widehat{L}_{jk}^{i}$ do not depend on $y^{k}.$ So, we have $\widehat{R}%
_{ja}=\widehat{R}_{\ jia}^{i}=0$.

Contracting the indices in $\widehat{R}_{\ bcd}^{a}$ from (\ref{dcurv}), we
compute the Ricci v-coefficients, 
\begin{equation*}
\widehat{R}_{bc}=\frac{\partial \widehat{C}_{bc}^{d}}{\partial y^{d}}-\frac{%
\partial \widehat{C}_{bd}^{d}}{\partial y^{c}}+\widehat{C}_{bc}^{e}\widehat{C%
}_{e}-\widehat{C}_{bd}^{e}\widehat{C}_{ec}^{d}.
\end{equation*}%
Summarizing on indices, we express 
\begin{equation*}
\widehat{R}_{bc}=(\widehat{C}_{bc}^{3})^{\ast }+(\widehat{C}%
_{bc}^{4})^{\diamond }-\partial _{c}\widehat{C}_{b}+\widehat{C}_{bc}^{3}%
\widehat{C}_{3}+\widehat{C}_{bc}^{4}\widehat{C}_{4}-\widehat{C}_{b3}^{3}%
\widehat{C}_{3c}^{3}-\widehat{C}_{b4}^{3}\widehat{C}_{3c}^{4}-\widehat{C}%
_{b3}^{4}\widehat{C}_{4c}^{3}-\widehat{C}_{b4}^{4}\widehat{C}_{4c}^{4}.
\end{equation*}%
There are nontrivial values,%
\begin{eqnarray*}
\widehat{R}_{33} &=&\left( \widehat{C}_{33}^{3}\right) ^{\ast }+\left( 
\widehat{C}_{33}^{4}\right) ^{\diamond }-\widehat{C}_{3}^{\ast }+\widehat{C}%
_{33}^{3}\widehat{C}_{3}+\widehat{C}_{33}^{4}\widehat{C}_{4}-\widehat{C}%
_{33}^{3}\widehat{C}_{33}^{3}-2\widehat{C}_{34}^{3}\widehat{C}_{33}^{4}-%
\widehat{C}_{34}^{4}\widehat{C}_{43}^{4} \\
&=&-\frac{1}{2}\frac{h_{4}^{\ast \ast }}{h_{4}}+\frac{1}{4}\frac{%
(h_{4}^{\ast })^{2}}{(h_{4})^{2}}+\frac{1}{4}\frac{h_{3}^{\ast }}{h_{3}}%
\frac{h_{4}^{\ast }}{h_{4}}, \\
\widehat{R}_{44} &=&\left( \widehat{C}_{44}^{3}\right) ^{\ast }+\left( 
\widehat{C}_{44}^{4}\right) ^{\diamond }-\partial _{4}\widehat{C}_{4}+%
\widehat{C}_{44}^{3}\widehat{C}_{3}+\widehat{C}_{44}^{4}\widehat{C}_{4}-%
\widehat{C}_{43}^{3}\widehat{C}_{34}^{3}-2\widehat{C}_{44}^{3}\widehat{C}%
_{34}^{4}-\widehat{C}_{44}^{4}\widehat{C}_{44}^{4} \\
&=&-\frac{1}{2}\frac{h_{4}^{\ast \ast }}{h_{3}}+\frac{1}{4}\frac{h_{3}^{\ast
}h_{4}^{\ast }}{(h_{3})^{2}}+\frac{1}{4}\frac{h_{4}^{\ast }}{h_{3}}\frac{%
h_{4}^{\ast }}{h_{4}}.
\end{eqnarray*}%
These formulas can be rewritten in the form%
\begin{equation}
\widehat{R}_{~3}^{3}=\frac{1}{h_{3}}\widehat{R}_{33}=\frac{1}{2h_{3}h_{4}}%
(-h_{4}^{\ast \ast }+\frac{(h_{4}^{\ast })^{2}}{2h_{4}}+\frac{h_{3}^{\ast
}h_{4}^{\ast }}{2h_{3}}),\widehat{R}_{~4}^{4}=\frac{1}{h_{4}}\widehat{R}%
_{44}=\frac{1}{2h_{3}h_{4}}(-h_{4}^{\ast \ast }+\frac{(h_{4}^{\ast })^{2}}{%
2h_{4}}+\frac{h_{3}^{\ast }h_{4}^{\ast }}{2h_{3}}).  \label{vcdric}
\end{equation}

A quasi-stationary d-metric ansatz (\ref{dmq}) is characterized by
nontrivial N-adapted coefficients of the canonical d-connection $\widehat{R}%
_{1}^{1}=\widehat{R}_{2}^{2}$ (\ref{hcdric}), $\ \widehat{R}_{3k}$ (\ref%
{vhcdric3}), $\widehat{R}_{4k}$ (\ref{vhcdric4}) and $\widehat{R}_{~3}^{3}=%
\widehat{R}_{~4}^{4}$ (\ref{vcdric}). Here we note that for such ansatz
other classes of coefficients are trivial in N-adapted frames, i.e.$\ 
\widehat{R}_{ka}\equiv 0$ for any $k=1,2$ and $a=3,4.$ Such values may be
not zero in other systems of reference/corrdinates.

The canonical Ricci d-scalar is computed using above N-adapted nontrivial
coefficients of the canonical Ricci d-tensor for formulas (\ref{sdcurv}), 
\begin{equation*}
\widehat{R}sc:=\widehat{\mathbf{g}}^{\alpha \beta }\widehat{\mathbf{R}}_{\
\alpha \beta }=\widehat{g}^{ij}\widehat{R}_{ij}+\widehat{g}^{ab}\widehat{R}%
_{ab}=\widehat{R}_{~i}^{i}+\widehat{R}_{~a}^{a}=2(\widehat{R}_{2}^{2}+%
\widehat{R}_{~4}^{4}),
\end{equation*}%
for nontrivial (\ref{hcdric}) and (\ref{vcdric}). As a result we can compute
the nontrivial components of the canonical Einstein d-tensor (\ref{ceinstdt}%
), 
\begin{equation*}
\widehat{\mathbf{E}}n:=\{\widehat{\mathbf{R}}_{\ \gamma }^{\beta }-\frac{1}{2%
}\delta ^{\beta}_{\gamma }\ \widehat{R}sc\}=\{-\widehat{R}_{~4}^{4},-%
\widehat{R}_{~4}^{4};\ \widehat{R}_{ak};\widehat{R}_{ka}\equiv 0;-\widehat{R}%
_{2}^{2},-\widehat{R}_{2}^{2}\}.
\end{equation*}%
So, in N-adapted form, the canonical Ricci and Einstein d-tensors for
quasi-stationary d-metrics posses similar but inverted symmetries for the h-
and v-components.

\subsubsection{Explicit decoupling of the modified Einstein equations for
canonical d-connections}

Let us define such N-adapted canonical parameterizations of the effective
sources (\ref{esourc}) for quasi-stationary configurations 
\begin{equation}
\widehat{\mathbf{\Upsilon }}_{\ \ \beta }^{\alpha }=[\ ^{h}\Upsilon \delta
_{\ \ j}^{i},\ ^{v}\Upsilon \delta _{\ \ b}^{a}]=[\ ^{h}\Upsilon =-\ _{1}%
\widehat{\Upsilon }(x^{k}),\ ^{v}\Upsilon =-\ _{2}\widehat{\Upsilon }%
(x^{k},y^{3})].  \label{esourcqscan}
\end{equation}
As a result, the canonical distorted Einstein equations (\ref{cdeq1}) for
the ansatz (\ref{dmq}) (using formulas (\ref{hcdric}), $\ $(\ref{vhcdric3}),
(\ref{vhcdric4}) and (\ref{vcdric}) and can be written in the form 
\begin{eqnarray}
\widehat{R}_{1}^{1} &=&\widehat{R}_{2}^{2}=\frac{1}{2g_{1}g_{2}}[\frac{%
g_{1}^{\bullet }g_{2}^{\bullet }}{2g_{1}}+\frac{(g_{2}^{\bullet })^{2}}{%
2g_{2}}-g_{2}^{\bullet \bullet }+\frac{g_{1}^{\prime }g_{2}^{\prime }}{2g_{2}%
}+\frac{\left( g_{1}^{\prime }\right) ^{2}}{2g_{1}}-g_{1}^{\prime \prime
}]=-\ _{1}\widehat{\Upsilon },  \notag \\
\widehat{R}_{3}^{3} &=&\widehat{R}_{4}^{4}=\frac{1}{2h_{3}h_{4}}[\frac{%
\left( h_{4}^{\ast }\right) ^{2}}{2h_{4}}+\frac{h_{3}^{\ast }h_{4}^{\ast }}{%
2h_{3}}-h_{4}^{\ast \ast }]=-\ _{2}\widehat{\Upsilon },  \label{riccist2} \\
\widehat{R}_{3k} &=&\frac{\ w_{k}}{2h_{4}}[h_{4}^{\ast \ast }-\frac{\left(
h_{4}^{\ast }\right) ^{2}}{2h_{4}}-\frac{(h_{3}^{\ast })(h_{4}^{\ast })}{%
2h_{3}}]+\frac{h_{4}^{\ast }}{4h_{4}}(\frac{\partial _{k}h_{3}}{h_{3}}+\frac{%
\partial _{k}h_{4}}{h_{4}})-\frac{\partial _{k}(h_{3}^{\ast })}{2h_{3}}=0; 
\notag \\
\widehat{R}_{4k} &=&\frac{h_{4}}{2h_{3}}n_{k}^{\ast \ast }+\left( \frac{3}{2}%
h_{4}^{\ast }-\frac{h_{4}}{h_{3}}h_{3}^{\ast }\right) \frac{\ n_{k}^{\ast }}{%
2h_{3}}=0.  \notag
\end{eqnarray}%
Expressing $g_{i}=e^{\psi (x^{k})};$ introducing coefficients 
\begin{equation}
\alpha _{i}=h_{4}^{\ast }\partial _{i}(\varpi ),\beta =h_{4}^{\ast }(\varpi
)^{\ast} \mbox{  and } \gamma = (\ln \frac{|h_{4}|^{3/2}}{|h_{3}|})^{\ast },
\label{coeff}
\end{equation}
for $\varpi =\ln |h_{4}^{\ast }/\sqrt{|h_{3}h_{4}}|;$ and considering $\
\Psi =\exp (\varpi )$ as a \textbf{generating function}, we simplify the
system of nonlinear PDEs (\ref{riccist2}) in the form: 
\begin{eqnarray}
\psi ^{\bullet \bullet }+\psi ^{\prime \prime }&=&2\ _{1}\widehat{\Upsilon },
\label{eq1} \\
(\varpi )^{\ast }h_{4}^{\ast }&=&2h_{3}h_{4}\ _{2}\widehat{\Upsilon },
\label{e2a} \\
\beta w_{j}-\alpha _{j}&=&0,  \label{e2b} \\
\ n_{k}^{\ast \ast }+\gamma n_{k}^{\ast }&=&0.  \label{e2c}
\end{eqnarray}%
Any solution of this system of nonlinear equations is a solution of (\ref%
{cdeq1}) parameterized a respective quasi-stationary d-metric ansatz (\ref%
{dmq}) for a canonically parameterized effective source (\ref{esourcqscan}),
where $\ _{1}\widehat{\Upsilon }(x^{k})$ and $\ _{2}\widehat{\Upsilon }%
(x^{k},y^{3})$ are \textbf{generating sources}.

Let us explain the general decoupling property of above systems of equations
for quasi-stationary configurations. The equation (\ref{eq1}) is a standard
2-d Poisson equation with source $2\ _{1}\widehat{\Upsilon }.$ We note that
instead of a nonholonomic 2+2 splitting, we can consider a 5-d spacetime of
signature $(++++-)$ with nonholonomic 3+2 splitting. In such a case,
following tedious computations which similar to above 4-d nonholonomic
geometric constructions, we obtain (\ref{eq1}) as a standard 2-d Poisson
equation. We omit such considerations in this work. If we prescribe any data 
$(h_{3},_{2}\widehat{\Upsilon }),$ we can search $h_{4}$ as a solution of a
second order (on derivative $\partial _{3}$) nonlinear PDE (\ref{e2a}). We
can consider an inverse problem with prescribed data $(h_{4},_{2}\widehat{%
\Upsilon })$ when $h_{3}$ as a solution of a first order nonlinear PDE.
Introducing a generating function\ $\ \Psi $, such equation can be
integrated in explicit form (we shall prove in next subsection). Having
defined in some general forms $h_{3}(x^{k},y^{3})$ and $h_{4}(x^{k},y^{3}),$
we can compute respective coefficients $\alpha _{i}$ and $\beta $ for (\ref%
{e2b}), which are linear equations for $w_{j}(x^{k},y^{3}).$ This mean that
such equations and respective unknown functions decoupled from the rest of
the system of nonlinear PDEs. Respectively, we have (\ref{e2c}) as a
decoupled system of PDEs which allows to find $n_{k}(x^{k},y^{3})$ (after
two general integrations on $y^{3}$) for any $\gamma (x^{k},y^{3})$
determined by $h_{3}(x^{k},y^{3})$ and $h_{4}(x^{k},y^{3})$ as in above
formulas. This way we proved that the (modified) Einstein equations written
in certain canonical d-connection variables can be decoupled in general
forms for quasi-stationary generic off-diagonal metric ansatz determined by
d-metric ansatz (\ref{dmq}) and respective generating sources $(\ _{1}%
\widehat{\Upsilon },\ _{2}\widehat{\Upsilon })$ (\ref{esourcqscan}).

\subsection{General solutions for quasi-stationary configurations}

\subsubsection{Integrating decoupled nonholonomic gravitational field
equations}

Let us show how the system of nonlinear PDEs (\ref{eq1})-(\ref{e2c}) can be
integrated step by step in certain general forms:

The coefficients $g_{i}=e^{\psi (x^{k})}$ for the h-components of d-metric (%
\ref{dmq}) \ are defined by solutions of the corresponding 2-d Poisson
equation (\ref{eq1}) for any given source $\ _{1}\widehat{\Upsilon }(x^{k}).$

Introducing explict values of coefficients (\ref{coeff}) in (\ref{e2a})-(\ref%
{e2c}), we obtain such a nonlinear system: 
\begin{eqnarray}
\Psi ^{\ast }h_{4}^{\ast } &=&2h_{3}h_{4}\ _{2}\widehat{\Upsilon }\Psi ,
\label{auxa1} \\
\sqrt{|h_{3}h_{4}|}\Psi &=&h_{4}^{\ast },  \label{auxa2} \\
\ \Psi ^{\ast }w_{i}-\partial _{i}\Psi &=&\ 0,  \label{aux1ab} \\
\ n_{i}^{\ast \ast }+\left( \ln \frac{|h_{4}|^{3/2}}{|h_{3}|}\right) ^{\ast
}n_{i}^{\ast } &=&0.\   \label{aux1ac}
\end{eqnarray}%
Prescribing a generating function, $\Psi ,$ and a generating source, $\ 
\widehat{\Upsilon },$ we can integrate recurrently these equations if $%
h_{4}^{\ast }\neq 0$ and $\ _{2}\widehat{\Upsilon }\neq 0.$ If such
conditions are not satisfied, there are necessary more special analytic
methods considered, for instance, in section 5.5 of \cite{bubuianu20} and
references therein. We introduce 
\begin{equation}
\rho ^{2}:=-h_{3}h_{4}  \label{rho}
\end{equation}%
and re-write (\ref{auxa1}) and (\ref{auxa2}), respectively, in the form%
\begin{equation}
\Psi ^{\ast }h_{4}^{\ast }=-2\rho ^{2}\ _{2}\widehat{\Upsilon }\ \Psi 
\mbox{
and }h_{4}^{\ast }=\rho \ \Psi .  \label{auxa3a}
\end{equation}%
Substituting the value of $h_{4}^{\ast }$ from the second equation into the
first equation, we express 
\begin{equation}
\rho =-\Psi ^{\ast }/2\ _{2}\widehat{\Upsilon }.  \label{rho1}
\end{equation}%
Then we introduce this $\rho $ into the second equation in (\ref{auxa3a})
and integrate on $y^{3},$ 
\begin{equation}
\ h_{4}=h_{4}^{[0]}(x^{k})-\int dy^{3}[\Psi ^{2}]^{\ast }/4(\ _{2}\widehat{%
\Upsilon }).  \label{g4}
\end{equation}%
Using this coefficient and formulas in (\ref{rho}) and (\ref{rho1}), we
compute%
\begin{equation}
h_{3}=-\frac{1}{4h_{4}}\left( \frac{\Psi ^{\ast }}{\ _{2}\widehat{\Upsilon }}%
\right) ^{2}=-\left( \frac{\Psi ^{\ast }}{2\ _{2}\widehat{\Upsilon }}\right)
^{2}\left( h_{4}^{[0]}(x^{k})-\int dy^{3}\frac{[\Psi ^{2}]^{\ast }}{4\ _{2}%
\widehat{\Upsilon }}\right) ^{-1}.  \label{g3}
\end{equation}

At the next step, we define the N-connection coefficients. Using $h_{3}$ (%
\ref{g3}) and $h_{4}$ (\ref{g4}), we can integrate two times on $y^{3}$ and
find general solutions of the equation (\ref{aux1ac}): 
\begin{eqnarray}
n_{k}(x^{k},y^{3}) &=&\ _{1}n_{k}+\ _{2}n_{k}\int dy^{3}\ \frac{h_{3}}{|\
h_{4}|^{3/2}}=\ _{1}n_{k}+\ _{2}n_{k}\int dy^{3}\left( \frac{\Psi ^{\ast }}{%
2\ _{2}\widehat{\Upsilon }}\right) ^{2}|\ h_{4}|^{-5/2}  \notag \\
&=&\ _{1}n_{k}+\ _{2}n_{k}\int dy^{3}\left( \frac{\Psi ^{\ast }}{2\ _{2}%
\widehat{\Upsilon }}\right) ^{2}\left\vert h_{4}^{[0]}(x^{k})-\int
dy^{3}[\Psi ^{2}]^{\ast }/4(\ _{2}\widehat{\Upsilon })\right\vert ^{-5/2}.
\label{gn}
\end{eqnarray}%
In these formulas, there are considered two integration functions $\
_{1}n_{k}=\ _{1}n_{k}(x^{i})$ and (we may re-define by introducing certain
coefficients) $\ _{2}n_{k}=\ _{2}n_{k}(x^{i}).$ Finally, solving the
algebraic system (\ref{aux1ab}) for $w_{i},$ we find%
\begin{equation}
w_{i}=\partial _{i}\ \Psi /(\Psi )^{\ast }.  \label{gw}
\end{equation}

Putting together above values for the coefficients of the d-metric and
N-connection, we define general solutions of the Einstein equations in
canonical nonholonomic variables.

\subsubsection{Quadratic elements for quasi-stationary off-diagonal solutions%
}

Using N-adapted h-coefficients $g_{i}=e^{\psi (x^{k})},$ determined by
solutions of 2-d Poisson equations (\ref{eq1}); v-coefficients $h_{3}$ (\ref%
{g3}) and $h_{4}$ (\ref{g4}); and N-connection coefficients $w_{i}$ (\ref{gw}%
) and $n_{k}$ (\ref{gn}), for a quasi-stationary d-metric (\ref{dmq}), we
construct a quasi-stationary nonlinear quadratic element,%
\begin{eqnarray}
ds^{2} &=&e^{\psi (x^{k})}[(dx^{1})^{2}+(dx^{2})^{2}]+\frac{[\Psi ^{\ast
}]^{2}}{4(\ _{2}\widehat{\Upsilon })^{2}\{g_{4}^{[0]}-\int dy^{3}[\Psi
^{2}]^{\ast }/4(\ _{2}\widehat{\Upsilon })\}}(dy^{3}+\frac{\partial _{i}\Psi 
}{\Psi ^{\ast }}dx^{i})^{2}+  \label{qeltors} \\
&&\{g_{4}^{[0]}-\int dy^{3}\frac{[\Psi ^{2}]^{\ast }}{4(\ _{2}\widehat{%
\Upsilon })}\}\{dt+[\ _{1}n_{k}+\ _{2}n_{k}\int dy^{3}\frac{[(\Psi
)^{2}]^{\ast }}{4(\ _{2}\widehat{\Upsilon })^{2}|g_{4}^{[0]}-\int
dy^{3}[\Psi ^{2}]^{\ast }/4(\ _{2}\widehat{\Upsilon })|^{5/2}}]dx^{k}\}. 
\notag
\end{eqnarray}%
With respect to coordinate dual frames, such a quadratic element can be
represented equivalently in the form (\ref{cm}), $\mathbf{\hat{g}}=$ $%
\underline{\widehat{g}}_{\alpha \beta }(u)du^{\alpha }\otimes du^{\beta },$
with 
\begin{eqnarray}
\widehat{\underline{g}}_{\alpha \beta } &=&\left[ 
\begin{array}{cccc}
g_{1}+(N_{1}^{3})^{2}h_{3}+(N_{1}^{4})^{2}h_{4} & 
N_{1}^{3}N_{2}^{3}h_{3}+N_{1}^{4}N_{2}^{4}h_{4} & N_{1}^{3}h_{3} & 
N_{1}^{4}h_{4} \\ 
N_{2}^{3}N_{1}^{3}h_{3}+N_{2}^{4}N_{1}^{4}h_{4} & 
g_{2}+(N_{2}^{3})^{2}h_{3}+(N_{2}^{4})^{2}h_{4} & N_{2}^{3}h_{3} & 
N_{2}^{4}h_{4} \\ 
N_{1}^{3}h_{3} & N_{2}^{3}h_{3} & h_{3} & 0 \\ 
N_{1}^{4}h_{4} & N_{2}^{4}h_{4} & 0 & h_{4}%
\end{array}%
\right]  \notag \\
&=&\left[ 
\begin{array}{cccc}
e^{\psi }+(w_{1})^{2}h_{3}+(n_{1})^{2}h_{4} & w_{1}w_{2}h_{3}+n_{1}n_{2}h_{4}
& w_{1}h_{3} & n_{1}h_{4} \\ 
w_{1}w_{2}h_{3}+n_{1}n_{2}h_{4} & e^{\psi }+(w_{2})^{2}h_{3}+(n_{2})^{2}h_{4}
& w_{2}h_{3} & n_{2}h_{4} \\ 
w_{1}h_{3} & w_{2}h_{3} & h_{3} & 0 \\ 
n_{1}h_{4} & n_{2}h_{4} & 0 & h_{4}%
\end{array}%
\right] ,  \label{qeltorsoffd}
\end{eqnarray}%
when the coefficients are respective functions as stated by formulas $%
g_{i}=e^{\psi (x^{k})}$ and (\ref{g3})-(\ref{gn}), and/or (\ref{qeltors}).
Such a metric/solution is generic off-diagonal if there are some non-zero
anholonomy coefficients (\ref{anhcoef}). In general, it is also
characterized by a nontrivial nonholonomically induced canonical d-torsion (%
\ref{nontrtors}) if additional LC-conditions (\ref{lcconstr}) are not
satisfied.

Above generic off-diagonal quasi-stationary solutions are general in the
sense that they are determined by a generating function $\Psi (x^{k},y^{3}),$
two generating effective sources $\ _{1}\widehat{\Upsilon }(x^{k})$ and $\
_{2}\widehat{\Upsilon }(x^{k},y^{3})$, and integration functions $\
_{1}n_{k}(x^{k}),\ _{2}n_{k}(x^{k})$ $\ $and $h_{4}^{[0]}(x^{k}).$ Such
values have to be found from certain boundary/asymptotic conditions and
other physical considerations (for instance, additional linear and nonlinear
symmetry conditions, causality problems, generating quasi-periodic
structure, avoiding singularities etc.). We shall analyze explicit
physically important examples in section \ref{sec4}. Such quasi-stationary
solutions are different from the former ones considered in \cite%
{misner,hawking73,wald82,kramer03}. In the case of diagonal metric ansatz,
for instance, for generating BHs (\ref{sch}) and LC-configurations, we
obtain certain systems of nonlinear second order ODE when the solutions are
determined by two integration constants. One of the constants is stated to
be zero in order to get at asymptotic the Minkowski spacetime and the second
integration constant is identified with the BH mass. For quasi-stationary
solutions (\ref{qeltors}), equivalently (\ref{qeltorsoffd}), we have (in
general) six independent coefficients for the off-diagonal metric (four of
them transforms into N-connection coefficients in N-adapted frame) which
describes more "rich" gravitational configurations.

\subsubsection{Nonlinear symmetries of quasi-stationary off-diagonal
solutions}

The solutions (\ref{qeltors}) posses important nonlinear shell symmetries
which allow to transform generating functions and effective sources into
other types of generating functions and effective cosmological constants. We
change the generating data, $(\Psi ,\ \ _{2}\widehat{\Upsilon }%
)\leftrightarrow (\Phi ,\ _{2}\Lambda =const\neq 0),$ following formulas%
\begin{eqnarray}
\frac{\lbrack \Psi ^{2}]^{\ast }}{\ _{2}\widehat{\Upsilon }} &=&\frac{[\Phi
^{2}]^{\ast }}{\ _{2}\Lambda },\mbox{ which can be
integrated as  }  \label{ntransf1} \\
\Phi ^{2} &=&\ _{2}\Lambda \int dy^{3}(\ _{2}\widehat{\Upsilon })^{-1}[\Psi
^{2}]^{\ast }\mbox{ and/or }\Psi ^{2}=(\ _{2}\Lambda )^{-1}\int dy^{3}(\ _{2}%
\widehat{\Upsilon })[\Phi ^{2}]^{\ast }.  \label{ntransf2}
\end{eqnarray}%
Using (\ref{ntransf1}), we can simplify the formula (\ref{g4}) and write $%
h_{4}=h_{4}^{[0]}-\frac{\ \Phi ^{2}}{4\ _{2}\Lambda }.$ To express formulas (%
\ref{g3}) and (\ref{gn}) in terms of new generating data, we have to write $%
(\Psi )^{\ast }/\ _{2}\widehat{\Upsilon }$ in terms of such $(\Phi ,\
_{2}\Lambda ).$ We re-write (\ref{ntransf1}) and the second equation in (\ref%
{ntransf2}) as%
\begin{equation*}
\frac{\Psi (\ \Psi )^{\ast }}{\ _{2}\widehat{\Upsilon }}=\frac{(\Phi
^{2})^{\ast }}{2(\ _{2}\Lambda )}\mbox{ and }\ \Psi =|\ _{2}\Lambda |^{-1/2}%
\sqrt{|\int dy^{3}\ _{2}\widehat{\Upsilon }\ (\Phi ^{2})^{\ast }|}.
\end{equation*}%
Introducing $\Psi $ from the second equation in the first equation, we
re-define $\Psi ^{\ast }$ in terms of generating data $(\ _{2}\widehat{%
\Upsilon },\Phi ,\ _{2}\Lambda ),$ when 
\begin{equation}
\frac{\Psi ^{\ast }}{\ _{2}\widehat{\Upsilon }}=\frac{[\Phi ^{2}]^{\ast }}{2%
\sqrt{|\ _{2}\Lambda \int dy^{3}(\ _{2}\widehat{\Upsilon })[\Phi ^{2}]^{\ast
}|}}.  \label{ntransf3}
\end{equation}
So, the solutions of nonlinear equations (\ref{e2a}) determined by (\ref{g3}%
) and (\ref{g4}) can be written in two equivalent functional forms, 
\begin{eqnarray*}
h_{3}[\Psi ] &=&-\frac{[\Psi ^{\ast }]^{2}}{4(\ _{2}\widehat{\Upsilon }%
)^{2}h_{4}[\Psi ]}=h_{3}[\Phi ]=-\frac{1}{h_{4}[\Phi ]}\frac{\Phi ^{2}[\Phi
^{\ast }]^{2}}{|\ _{2}\Lambda \int dy^{3}\ _{2}\widehat{\Upsilon }[\Phi
^{2}]^{\ast }|},\mbox{ where } \\
h_{4}[\Psi ] &=&h_{4}^{[0]}-\int dy^{3}\frac{[\Psi ^{2}]^{\ast }}{4\ _{2}%
\widehat{\Upsilon }}=h_{4}[\Phi ]=g_{4}^{[0]}-\frac{\Phi ^{2}}{4\
_{2}\Lambda }.
\end{eqnarray*}

In similar forms, the N--connection coefficients can be re-defined for data $%
(\Phi ,\ _{2}\Lambda )$ using respectively the nonlinear transforms ,%
\begin{eqnarray*}
w_{i}(x^{k_{1}},y^{3}) &=&\frac{\partial _{i}\Psi }{\Psi ^{\ast }}=\frac{%
\partial _{i}[\Psi ^{2}]}{[\Psi ^{2}]^{\ast }}=\frac{\partial _{i}\ \int
dy^{3}\ _{2}\widehat{\Upsilon }\ [\Phi ^{2}]^{\ast }}{_{2}\widehat{\Upsilon }%
\ [\Phi ^{2}]^{\ast }};\mbox{
and } \\
n_{k}(x^{k_{1}},y^{3}) &=&\ _{1}n_{k}+\ _{2}n_{k}\int dy^{3}\ \frac{%
h_{3}[\Phi ]}{|\ h_{4}[\Phi ]|^{3/2}} \\
&=&\ _{1}n_{k}+\ _{2}n_{k}\int dy^{3}\left( \frac{\Psi ^{\ast }}{2\ _{2}%
\widehat{\Upsilon }}\right) ^{2}\left\vert h_{4}^{[0]}(x^{k})-\int dy^{3}%
\frac{[\Psi ^{2}]^{\ast }}{4\ _{2}\widehat{\Upsilon }}\right\vert ^{-5/2} \\
&=&\ _{1}n_{k}+\ _{2}n_{k}\int dy^{3}\frac{\Phi ^{2}[\Phi ^{\ast }]^{2}}{|\
_{2}\Lambda \int dy^{3}\ _{2}\widehat{\Upsilon }[\Phi ^{2}]^{\ast }|}%
\left\vert h_{4}^{[0]}-\frac{\Phi ^{2}}{4\ _{2}\Lambda }\right\vert ^{-5/2}.
\end{eqnarray*}

We conclude that any quasi-stationary solution (\ref{qeltors}) possess
important nonlinear symmetries of type (\ref{ntransf1}) and (\ref{ntransf2}%
). As a result, the nonlinear quadratic element for quasi-stationary
solutions (\ref{qeltors}) can be written in the form%
\begin{eqnarray}
ds^{2} &=&g_{\alpha _{s}\beta _{s}}(x^{k},y^{3},\Phi ,_{2}\Lambda
)du^{\alpha }du^{\beta }=e^{\psi (x^{k})}[(dx^{1})^{2}+(dx^{2})^{2}]
\label{offdiagcosmcsh} \\
&&-\frac{\Phi ^{2}[\Phi ^{\ast }]^{2}}{|\ _{2}\Lambda \int dy^{3}\ _{2}%
\widehat{\Upsilon }[\Phi ^{2}]^{\ast }|[h_{4}^{[0]}-\Phi ^{2}/4\ _{2}\Lambda
]}\{dy^{3}+\frac{\partial _{i}\ \int dy^{3}\ _{2}\widehat{\Upsilon }\ [\Phi
^{2}]^{\ast }}{\ _{2}\widehat{\Upsilon }\ [(\ _{2}^{\shortmid }\Phi
)^{2}]^{\ast }}dx^{i}\}^{2}-  \notag \\
&&\{h_{4}^{[0]}-\frac{\Phi ^{2}}{4\ _{2}\Lambda }\}\{dt+[\ _{1}n_{k}+\
_{2}n_{k}\int dy^{3}\frac{\Phi ^{2}[\Phi ^{\ast }]^{2}}{|\ _{2}\Lambda \int
dy^{3}\ _{2}\widehat{\Upsilon }[\Phi ^{2}]^{\ast }|[h_{4}^{[0]}-\Phi ^{2}/4\
_{2}\Lambda ]^{5/2}}]\},  \notag
\end{eqnarray}%
for indices: $i,j,k,...=1,2;a,b,c,...=3,4;$ generating functions $\psi
(x^{k})$ and$\ \Phi (x^{k_{1}}y^{3});$ generating sources $_{1}\widehat{%
\Upsilon }(x^{k})$ and$\ _{2}\widehat{\Upsilon }(x^{k_{1}},y^{3});$
effective cosmological constants $\ _{1}\Lambda $ and $\ _{2}\Lambda ;$ and
integration functions$\ _{1}n_{k}(x^{j}),\ _{2}n_{k}(x^{j})$ and $%
g_{4}^{[0]}(x^{k}).$

\subsubsection{Using d-metric coefficients as generating functions}

Taking the partial derivative on $y^{3}$ of formula (\ref{g4}), we obtain $%
h_{4}^{\ast }=-[\Psi ^{2}]^{\ast }/4\ _{2}\widehat{\Upsilon }.$ If we
prescribe $h_{4}$ and $\ _{2}\widehat{\Upsilon },$ we can compute up to
certain integration functions a $\ \Psi $ using $[\Psi ^{2}]^{\ast }=\int
dy^{3}\ _{2}\widehat{\Upsilon }h_{4}^{\ast }.$ This allows us to consider a
generating data $(h_{4},\ _{2}\widehat{\Upsilon })$ and re-write the
quasi-stationary d-metric (\ref{qeltors}) in equivalent form, 
\begin{eqnarray}
d\widehat{s}^{2} &=&\widehat{g}_{\alpha \beta }(x^{k},y^{3};h_{4},\ _{2}%
\widehat{\Upsilon })du^{\alpha }du^{\beta }  \label{offdsolgenfgcosmc} \\
&=&e^{\psi (x^{k})}[(dx^{1})^{2}+(dx^{2})^{2}]-\frac{(h_{4}^{\ast })^{2}}{%
|\int dy^{3}[\ _{2}\widehat{\Upsilon }h_{4}]^{\ast }|\ h_{4}}\{dy^{3}+\frac{%
\partial _{i}[\int dy^{3}(\ _{2}\widehat{\Upsilon })\ h_{4}^{\ast }]}{\ _{2}%
\widehat{\Upsilon }\ h_{4}^{\ast }}dx^{i}\}^{2}  \notag \\
&&+h_{4}\{dt+[\ _{1}n_{k}+\ _{2}n_{k}\int dy^{3}\frac{(h_{4}^{\ast })^{2}}{%
|\int dy^{3}[\ _{2}\widehat{\Upsilon }h_{4}]^{\ast }|\ (h_{4})^{5/2}}%
]dx^{k}\}.  \notag
\end{eqnarray}

The nonlinear symmetries (\ref{ntransf1}) and (\ref{ntransf2}) allow to
perform similar computations and express $\Phi ^{2}=-4\ _{2}\Lambda h_{4}.$
We can eliminate $\Phi $ from the nonlinear quadratic element in (\ref%
{offdiagcosmcsh}) and obtain a solution of type (\ref{offdsolgenfgcosmc})
determined by the generating data $(h_{4};\ _{2}\Lambda ,\ _{2}\widehat{%
\Upsilon }).$

\subsubsection{Gravitational polarizations}

Nonholonomic frame and connection deformations and nonlinear symmetries
allow to \ perform another types of geometric constructions:

\begin{itemize}
\item We can consider deformations of a \textbf{prime} d-metric (it can be
an arbitrary one) 
\begin{equation}
\mathbf{\mathring{g}=}[\mathring{g}_{\alpha },\ \mathring{N}_{i}^{a}]
\label{offdiagpm}
\end{equation}%
into a \textbf{target} d-metric $\mathbf{g,}$ for instance, being a
quasi-stationary solutions of type (\ref{dmq}), when \ 
\begin{equation}
\mathbf{\mathring{g}}\rightarrow \mathbf{g}=[g_{\alpha }=\eta _{\alpha }%
\mathring{g}_{\alpha },N_{i}^{a}=\eta _{i}^{a}\ \mathring{N}_{i}^{a}],
\label{offdiagdefr}
\end{equation}%
for $\eta _{\alpha }(x^{k},y^{3})$ and $\eta _{i}^{a}(x^{k},y^{3})$ called
gravitational polarization ($\eta $-polarization) functions.

\item For a target metric $\mathbf{g}$ defined as a solution of type (\ref%
{qeltors}), equivalently (\ref{offdiagcosmcsh}), we can consider
nonholonomic deformations with respective generating functions/ sources and
effective cosmological constants when 
\begin{eqnarray*}
(\Psi ,\ _{2}\widehat{\Upsilon }) &\leftrightarrow &(\mathbf{g},\ _{2}%
\widehat{\Upsilon })\leftrightarrow (\eta _{\alpha }\ \mathring{g}_{\alpha
}\sim (\zeta _{\alpha }(1+\kappa \chi _{\alpha })\mathring{g}_{\alpha },\
_{2}\widehat{\Upsilon })\leftrightarrow \\
(\Phi ,\ _{2}\Lambda ) &\leftrightarrow &(\mathbf{g},\ _{2}\Lambda
)\leftrightarrow (\eta _{\alpha }\ \mathring{g}_{\alpha }\sim (\zeta
_{\alpha }(1+\kappa \chi _{\alpha })\mathring{g}_{\alpha },\ _{2}\Lambda ),
\end{eqnarray*}%
where $\ _{2}\Lambda $ is an effective cosmological constant in the
v-subspace, $\kappa $ is a small parameter $0\leq \kappa <1,$ with some $%
\zeta _{\alpha }(x^{k},y^{3})$ and $\chi _{\alpha }(x^{k},y^{3}).$
\end{itemize}

Using above $\eta $- and/or $\chi $-polarizations, the nonlinear symmetries (%
\ref{ntransf2}) are written in the form: 
\begin{eqnarray}
\partial _{3}[\Psi ^{2}] &=&-\int dy^{3}\ _{2}\widehat{\Upsilon }\partial
_{3}h_{4}\simeq -\int dy^{3}\ _{2}\widehat{\Upsilon }\partial _{3}(\eta
_{4}\ \mathring{g}_{4})\simeq -\int dy^{3}\ _{2}\widehat{\Upsilon }\partial
_{3}[\zeta _{4}(1+\kappa \ \chi _{4})\ \mathring{g}_{4}],  \notag \\
\Phi ^{2} &=&-4\ _{2}\Lambda h_{4}\simeq -4\ _{2}\Lambda \eta _{4}\mathring{g%
}_{4}\simeq -4\ _{2}\Lambda \ \zeta _{4}(1+\kappa \chi _{4})\ \mathring{g}%
_{4}.  \label{nonlinsymrex}
\end{eqnarray}

Off-diagonal $\eta $-transforms of type (\ref{offdiagdefr}) \ can be
parameterized for $\eta $-polarizations, 
\begin{equation}
\psi \simeq \psi (\kappa ;x^{k}),\eta _{4}\ \simeq \eta _{4}(x^{k},y^{3}),
\label{etapolgen}
\end{equation}%
when the quasi-stationary nonlinear quadratic element (\ref%
{offdsolgenfgcosmc}) can be written in the form 
\begin{eqnarray}
d\widehat{s}^{2} &=&\widehat{g}_{\alpha \beta }(x^{k},y^{3};\mathring{g}%
_{\alpha };\psi ,\eta _{4};\ _{2}\Lambda ,\ _{2}\widehat{\Upsilon }%
)du^{\alpha }du^{\beta }=e^{\psi }[(dx^{1})^{2}+(dx^{2})^{2}]
\label{offdiagpolfr} \\
&&-\frac{[\partial _{3}(\eta _{4}\ \mathring{g}_{4})]^{2}}{|\int dy^{3}\ _{2}%
\widehat{\Upsilon }\partial _{3}(\eta _{4}\ \mathring{g}_{4})|\ \eta _{4}%
\mathring{g}_{4}}\{dy^{3}+\frac{\partial _{i}[\int dy^{3}\ _{2}\widehat{%
\Upsilon }\ \partial _{3}(\eta _{4}\mathring{g}_{4})]}{\ _{2}\widehat{%
\Upsilon }\partial _{3}(\eta _{4}\mathring{g}_{4})}dx^{i}\}^{2}  \notag \\
&&+\eta _{4}\mathring{g}_{4}\{dt+[\ _{1}n_{k}+\ _{2}n_{k}\int dy^{3}\frac{%
[\partial _{3}(\eta _{4}\mathring{g}_{4})]^{2}}{|\int dy^{3}\ _{2}\widehat{%
\Upsilon }\partial _{3}(\eta _{4}\mathring{g}_{4})|\ (\eta _{4}\mathring{g}%
_{4})^{5/2}}]dx^{k}\}^{2}.  \notag
\end{eqnarray}%
For $\Phi ^{2}=-4\ _{2}\Lambda h_{4},$ we can transform (\ref{offdiagcosmcsh}%
) in a variant of (\ref{offdiagpolfr}) with $\eta $-polarizations determined
by the generating data $(h_{4};\ _{2}\Lambda ,\ _{2}\widehat{\Upsilon }).$

\subsubsection{Generating solutions with small parametric off-diagonal
decompositions}

Considering $\kappa $-linear functions for $\eta $-polarizations in (\ref%
{offdiagpolfr}), we can define small nonholonomic deformations of a prime
d-metric $\mathbf{\mathring{g}}$ into so-called $\kappa $-parametric
solutions with $\zeta $- and $\chi $-coefficients, 
\begin{eqnarray}
\psi &\simeq &\psi (x^{k})\simeq \psi _{0}(x^{k})(1+\kappa \ _{\psi }\chi
(x^{k})),\mbox{ for }\   \label{epsilongenfdecomp} \\
\ \eta _{2} &\simeq &\eta _{2}(x^{k_{1}})\simeq \zeta _{2}(x^{k})(1+\kappa
\chi _{2}(x^{k})),\mbox{ we can consider }\ \eta _{2}=\ \eta _{1};  \notag \\
\eta _{4} &\simeq &\eta _{4}(x^{k},y^{3})\simeq \zeta
_{4}(x^{k},y^{3})(1+\kappa \chi _{4}(x^{k},y^{3})),  \notag
\end{eqnarray}%
where $\psi $ and $\eta _{2}=\ \eta _{1}$ are such way chosen to be related
to the solutions of the 2-d Poisson equation $\partial _{11}^{2}\psi
+\partial _{22}^{2}\psi =2\ _{1}\widehat{\Upsilon }(x^{k}),$ see (\ref{eq1}%
). For other type signatures of d-metrics, it can be a 2-d wave equation
with respective source.

Using (\ref{epsilongenfdecomp}), we can compute $\kappa $-parametric
deformations to quasi-stationary d-metrics with $\chi $-generating
functions: 
\begin{equation*}
d\ \widehat{s}^{2}=\widehat{g}_{\alpha \beta }(x^{k},y^{3};\psi ,g_{4};_{2}%
\widehat{\Upsilon })du^{\alpha }du^{\beta }=e^{\psi _{0}}(1+\kappa \ ^{\psi
}\chi )[(dx^{1})^{2}+(dx^{2})^{2}]
\end{equation*}%
\begin{eqnarray*}
&&-\{\frac{4[\partial _{3}(|\zeta _{4}\mathring{g}_{4}|^{1/2})]^{2}}{%
\mathring{g}_{3}|\int dy^{3}\{\ \ _{2}\widehat{\Upsilon }\partial _{3}(\zeta
_{4}\mathring{g}_{4})\}|}-\kappa \lbrack \frac{\partial _{3}(\chi _{4}|\zeta
_{4}\mathring{g}_{4}|^{1/2})}{4\partial _{3}(|\zeta _{4}\mathring{g}%
_{4}|^{1/2})}-\frac{\int dy^{3}\{\ _{2}\widehat{\Upsilon }\partial
_{3}[(\zeta _{4}\mathring{g}_{4})\chi _{4}]\}}{\int dy^{3}\{\ _{2}\widehat{%
\Upsilon }\partial _{3}(\zeta _{4}\mathring{g}_{4})\}}]\}\mathring{g}_{3} \\
&&\{dy^{3}+[\frac{\partial _{i}\ \int dy^{3}\ _{2}\widehat{\Upsilon }\
\partial _{3}\zeta _{4}}{(\mathring{N}_{i}^{3})\ _{2}\widehat{\Upsilon }%
\partial _{3}\zeta _{4}}+\kappa (\frac{\partial _{i}[\int dy^{3}\ _{2}%
\widehat{\Upsilon }\ \partial _{3}(\zeta _{4}\chi _{4})]}{\partial _{i}\
[\int dy^{3}\ _{2}\widehat{\Upsilon }\partial _{3}\zeta _{4}]}-\frac{%
\partial _{3}(\zeta _{4}\chi _{4})}{\partial _{3}\zeta _{4}})]\mathring{N}%
_{i}^{3}dx^{i}\}^{2}
\end{eqnarray*}%
\begin{eqnarray}
&&+\zeta _{4}(1+\kappa \ \chi _{4})\ \mathring{g}_{4}\{dt+[(\mathring{N}%
_{k}^{4})^{-1}[\ _{1}n_{k}+16\ _{2}n_{k}[\int dy^{3}\frac{\left( \partial
_{3}[(\zeta _{4}\mathring{g}_{4})^{-1/4}]\right) ^{2}}{|\int dy^{3}\partial
_{3}[\ _{2}\widehat{\Upsilon }(\zeta _{4}\mathring{g}_{4})]|}]
\label{offdncelepsilon} \\
&&+\kappa \frac{16\ _{2}n_{k}\int dy^{3}\frac{\left( \partial _{3}[(\zeta
_{4}\mathring{g}_{4})^{-1/4}]\right) ^{2}}{|\int dy^{3}\partial _{3}[\ _{2}%
\widehat{\Upsilon }(\zeta _{4}\mathring{g}_{4})]|}(\frac{\partial
_{3}[(\zeta _{4}\mathring{g}_{4})^{-1/4}\chi _{4})]}{2\partial _{3}[(\zeta
_{4}\mathring{g}_{4})^{-1/4}]}+\frac{\int dy^{3}\partial _{3}[\ _{2}\widehat{%
\Upsilon }(\zeta _{4}\chi _{4}\mathring{g}_{4})]}{\int dy^{3}\partial _{3}[\
_{2}\widehat{\Upsilon }(\zeta _{4}\mathring{g}_{4})]})}{\ _{1}n_{k}+16\
_{2}n_{k}[\int dy^{3}\frac{\left( \partial _{3}[(\zeta _{4}\mathring{g}%
_{4})^{-1/4}]\right) ^{2}}{|\int dy^{3}\partial _{3}[\ _{2}\widehat{\Upsilon 
}(\zeta _{4}\mathring{g}_{4})]|}]}]\mathring{N}_{k}^{4}dx^{k}\}^{2}.  \notag
\end{eqnarray}%
Such off-diagonal parametric solutions allow to define, for instance,
ellipsoidal deformations of BH metrics into BE ones.

\subsection{Space and time duality of generic off-diagonal solutions}

\label{ssstdual}We can repeat all computations presented for
quasi-stationary metrics (\ref{dmq}) with nontrivial partial derivatives $%
\partial _{3}$ presented above in this section for locally anisotropic
cosmological solutions (\ref{dmc}) with nontrivial partial derivatives $%
\partial _{4}=\partial _{t}$ . In abstract symbolic form, we can formulate a 
\textbf{principle of space and time duality }of such different generic
off-diagonal configurations: 
\begin{eqnarray*}
y^{3} &\longleftrightarrow &y^{4}=t,h_{3}(x^{k},y^{3})\longleftrightarrow 
\underline{h}_{4}(x^{k},t),h_{4}(x^{k},y^{3})\longleftrightarrow \underline{h%
}_{3}(x^{k},t), \\
N_{i}^{3} &=&w_{i}(x^{k},y^{3})\longleftrightarrow N_{i}^{4}=\underline{n}%
_{i}(x^{k},t),N_{i}^{4}=n_{i}(x^{k},y^{3})\longleftrightarrow N_{i}^{3}=%
\underline{w}_{i}(x^{k},t).
\end{eqnarray*}%
Such duality conditions are considered also for prime d-metrics and
respective generating functions/ sources and gravitational polarization
functions, when 
\begin{equation*}
\Upsilon _{~3}^{3}=\Upsilon _{~4}^{4}=~^{v}\Upsilon (x^{k},y^{3})=\ _{2}%
\widehat{\Upsilon }\longleftrightarrow \underline{\Upsilon }_{~4}^{4}=%
\underline{\Upsilon }_{~3}^{3}=~^{v}\underline{\Upsilon }(x^{k},t)=\ _{2}%
\widehat{\underline{\Upsilon }},\mbox{ see }(\ref{esourcqscan});
\end{equation*}%
\begin{equation*}
\begin{array}{ccc}
\begin{array}{c}
(\Psi ,\ _{2}\widehat{\Upsilon })\leftrightarrow (\mathbf{g},\ _{2}\widehat{%
\Upsilon })\leftrightarrow \\ 
(\eta _{\alpha }\ \mathring{g}_{\alpha }\sim (\zeta _{\alpha }(1+\kappa \chi
_{\alpha })\mathring{g}_{\alpha },\ _{2}\widehat{\Upsilon })\leftrightarrow%
\end{array}
& \Longleftrightarrow & 
\begin{array}{c}
(\underline{\Psi },\ _{2}\widehat{\underline{\Upsilon }})\leftrightarrow (%
\underline{\mathbf{g}},\ _{2}\widehat{\underline{\Upsilon }})\leftrightarrow
\\ 
(\underline{\eta }_{\alpha }\ \underline{\mathring{g}}_{\alpha }\sim (%
\underline{\zeta }_{\alpha }(1+\kappa \underline{\chi }_{\alpha })\underline{%
\mathring{g}}_{\alpha },\ _{2}\widehat{\underline{\Upsilon }})\leftrightarrow%
\end{array}
\\ 
\begin{array}{c}
(\Phi ,\ _{2}\Lambda )\leftrightarrow (\mathbf{g},\ _{2}\Lambda
)\leftrightarrow \\ 
(\eta _{\alpha }\ \mathring{g}_{\alpha }\sim (\zeta _{\alpha }(1+\kappa \chi
_{\alpha })\mathring{g}_{\alpha },\ _{2}\Lambda ),%
\end{array}
& \Longleftrightarrow & 
\begin{array}{c}
(\underline{\Phi },\ _{2}\underline{\Lambda })\leftrightarrow (\underline{%
\mathbf{g}},\ _{2}\underline{\Lambda })\leftrightarrow \\ 
(\underline{\eta }_{\alpha }\ \underline{\mathring{g}}_{\alpha }\sim (%
\underline{\zeta }_{\alpha }(1+\kappa \underline{\chi }_{\alpha })\underline{%
\mathring{g}}_{\alpha },\ _{2}\underline{\Lambda }),%
\end{array}%
\end{array}%
\end{equation*}%
and 
\begin{equation}
\begin{array}{ccc}
\Psi ^{\ast }h_{4}^{\ast }=2h_{3}h_{4}\ _{2}\widehat{\Upsilon }\Psi , & 
\longleftrightarrow & \sqrt{|\underline{h}_{3}\underline{h}_{4}|}\underline{%
\Psi }=\underline{h}_{3}^{\diamond }, \\ 
\sqrt{|h_{3}h_{4}|}\Psi =h_{4}^{\ast }, & \longleftrightarrow & \underline{%
\Psi }^{\diamond }\underline{h}_{3}^{\diamond }=2\underline{h}_{3}\underline{%
h}_{4}\ _{2}\widehat{\underline{\Upsilon }}\underline{\Psi }, \\ 
\Psi ^{\ast }w_{i}-\partial _{i}\Psi =\ 0, & \longleftrightarrow & 
\underline{n}_{i}^{\diamond \diamond }+\left( \ln \frac{|\underline{h}%
_{3}|^{3/2}}{|\underline{h}_{4}|}\right) ^{\diamond }\underline{n}%
_{i}^{\diamond }=0, \\ 
\ n_{i}^{\ast \ast }+\left( \ln \frac{|h_{4}|^{3/2}}{|h_{3}|}\right) ^{\ast
}n_{i}^{\ast }=0 & \longleftrightarrow & \underline{\Psi }^{\diamond }%
\underline{w}_{i}-\partial _{i}\underline{\Psi }=\ 0,%
\end{array}%
\mbox{ see }(\ref{auxa1})-(\ref{aux1ac}).  \label{dualcosm}
\end{equation}

For locally anisotropic cosmological configurations, the nonlinear
symmetries (\ref{ntransf1}) and (\ref{ntransf2}) are written in respective
dual forms, 
\begin{eqnarray*}
\frac{\lbrack \underline{\Psi }^{2}]^{\diamond }}{\ _{2}\widehat{\underline{%
\Upsilon }}} &=&\frac{[\underline{\Phi }^{2}]^{\diamond }}{\ _{2}\underline{%
\Lambda }},\mbox{ which can be
integrated as  } \\
\underline{\Phi }^{2} &=&\ _{2}\underline{\Lambda }\int dt(\ _{2}\widehat{%
\underline{\Upsilon }})^{-1}[\underline{\Psi }^{2}]^{\diamond }%
\mbox{ and/or
}\underline{\Psi }^{2}=(\ _{2}\underline{\Lambda })^{-1}\int dt(\ _{2}%
\widehat{\underline{\Upsilon }})[\underline{\Phi }^{2}]^{\diamond }.
\end{eqnarray*}%
As a result, there are similar duality properties of solutions determined by
quasi-stationary d-metrics (\ref{qeltors}), (\ref{offdiagcosmcsh}), (\ref%
{offdsolgenfgcosmc}), (\ref{offdiagpolfr}) and (\ref{offdncelepsilon}) and
their locally anisotropic cosmological analogs. For instance, the d-metric (%
\ref{qeltors}) transforms into 
\begin{eqnarray}
d\underline{s}^{2} &=&e^{\psi (x^{k})}[(dx^{1})^{2}+(dx^{2})^{2}]
\label{qeltorsc} \\
&&+ \{g_{3}^{[0]}-\int dt\frac{[\underline{\Psi }^{2}]^{\diamond }}{4(\ _{2}%
\widehat{\underline{\Upsilon }})}\}\{dy^{3}+[\ _{1}n_{k}+\ _{2}n_{k}\int dt%
\frac{[(\underline{\Psi })^{2}]^{\diamond }}{4(\ _{2}\widehat{\underline{%
\Upsilon }})^{2}|g_{3}^{[0]}-\int dt[\underline{\Psi }^{2}]^{\diamond }/4(\
_{2}\widehat{\underline{\Upsilon }})|^{5/2}}]dx^{k}\}  \notag \\
&&+\frac{[\underline{\Psi }^{\diamond }]^{2}}{4(\ _{2}\widehat{\underline{%
\Upsilon }})^{2}\{g_{3}^{[0]}-\int dt[\underline{\Psi }^{2}]^{\diamond }/4(\
_{2}\widehat{\underline{\Upsilon }})\}}(dt+\frac{\partial _{i}\underline{%
\Psi }}{\underline{\Psi }^{\diamond }}dx^{i})^{2}.  \notag
\end{eqnarray}%
Other d-metrics can be also derived in abstract dual form changing
corresponding indices 3 into 4, 4 into 3, underlying the respective
generating functions / effective sources / gravitational polarizations for
dependencies on $(x^i,t)$ and changing v-partial derivatives, $\ast
\rightarrow \diamond $, i.e. $\partial _3 \rightarrow \partial _4$.

\subsection{A toy 2+2 model with effective momentum variables}

\label{sstoy}The geometric method for decoupling and integrating
gravitational field equations can be re-defined for 4-d nonholonomic "phase"
space $\ ^{\shortmid }V$ gravitational equations with conventional 2+2
splitting. Such constructions are considered in so-called
Hamilton-Finsler-Cartan geometry \cite%
{vacaru18,bubuianu18a,bubuianu20,bubuianu19,partner03,partner04} when
nonholonomic co-fibered structures on $\ ^{\shortmid }V,$ $\dim \
^{\shortmid }V=4,$ and such a pseudo-Riemanian manifold is enabled with a
metric $\ ^{\shortmid }\mathbf{g}$ with signature $(+-+-)$ and the geometric
objects are adapted to certain "dual" fibration structures. A similar
geometric formulation is possible for cotangent bundles $T^{\ast }V^{(2)}=h$ 
$T^{\ast }V^{(2)}\oplus v$ $T^{\ast }V^{(2)},$ $\dim $ $V^{(2)}=2,$ and the
metrics on $V^{(2)}$ are of signature (+-). In relativistic forms, such
constructions and respective classes of off-diagonal solutions are performed
on $T^{\ast }V,$ with $\dim T^{\ast }V=8.$ In this paper, we outline only
the main ideas and methods for 4-d total phase spaces when the nonholonomic
geometry with conventional 2+2 splitting is similar and dual as for $TV^{(2)}
$ and $T^{\ast }V^{(2)}.$ Let us explain the notations for this subsection,
when the left abstract label "$\ ^{\shortmid }$" states that additionally to
(\ref{ncon}) it is considered a dual nonholonomic distribution 
\begin{equation}
\ ^{\shortmid }\mathbf{N}:\ T^{\ast }V=hV\oplus cV,  \label{cncon}
\end{equation}%
where $cV$ is stated as a conventional 2-d subspace which is dual to $vV$. \
Locally, we can consider on $cV$ certain systems of coordinates $%
p_{b}(x^{i},y^{a}).$ For the models of Lagrange-Hamilton mechanics on $\ TV$
and, respectively, $\ T^{\ast }V,$ we can consider $y^{a}=v^{a}$ and $p_{b}$
as certain velocity and momentum like variables related via Legendre
transforms etc. But for the purposes of this work, it is enough to see $\
^{\shortmid }p=p=(p_{3},p_{4}=E)$ as a part of $\ ^{\shortmid
}u=(x,p)=\{x^{i},p_{a}\}$ defined as local coordinates on $\ ^{\shortmid }V.$

In local dual coordinate form, a N-connection (\ref{cncon}) can be written
as $\ ^{\shortmid }\mathbf{N}=\ ^{\shortmid }N_{ia}(x,p)dx^{i}\otimes
\partial /\partial p_{a},$ when the N--elongated (equivalently, N-adapted)
local bases (partial derivatives), $\ ^{\shortmid }\mathbf{e}_{\nu },$ and
co-bases (differentials), $\ ^{\shortmid }\mathbf{e}^{\mu },$ (compare,
respectively, to (\ref{nader}) and (\ref{nadif})) are defined 
\begin{eqnarray}
\ ^{\shortmid }\mathbf{e}_{\nu } &=&(\ ^{\shortmid }\mathbf{e}_{i},\
^{\shortmid }e^{a})=(\ ^{\shortmid }\mathbf{e}_{i}=\partial /\partial
x^{i}-\ \ ^{\shortmid }N_{ib}(\ ^{\shortmid }u)\partial /\partial p_{b},\ \
^{\shortmid }e^{a}=\ ^{\shortmid }\partial ^{a}=\partial /\partial p_{a}),%
\mbox{ and  }  \label{naderd} \\
\ ^{\shortmid }\mathbf{e}^{\mu } &=&(e^{i},\ ^{\shortmid }\mathbf{e}%
^{a})=(e^{i}=dx^{i},\ \ ^{\shortmid }\mathbf{e}_{a}=dp_{a}+\ ^{\shortmid
}N_{ia}(\ ^{\shortmid }u)dx^{i}),  \label{nadifd}
\end{eqnarray}

Any phase space metric$\ ^{\shortmid }\mathbf{g}$ on $\ ^{\shortmid }V$ can
be represented equivalently as a d--metric $\ ^{\shortmid }\mathbf{g}=(h\
^{\shortmid }g,c\ ^{\shortmid }g),$ when 
\begin{eqnarray}
\ \ ^{\shortmid }\mathbf{g} &=&\ \ ^{\shortmid }g_{ij}(x,p)\ e^{i}\otimes
e^{j}+\ ^{\shortmid }g^{ab}(x,p)\ \ ^{\shortmid }\mathbf{e}_{a}\otimes \
^{\shortmid }\mathbf{e}_{b},  \label{dmd} \\
&=&\ ^{\shortmid }\underline{g}_{\alpha \beta }(\ ^{\shortmid }u)d\
^{\shortmid }u^{\alpha }\otimes d\ ^{\shortmid }u^{\beta },  \notag
\end{eqnarray}%
where $h\ ^{\shortmid }g=\{\ ^{\shortmid }g_{ij}\}$ and $\ c\ ^{\shortmid
}g=\{\ ^{\shortmid }g^{ab}\}.$

In abstract geometric form, we can define on $\ ^{\shortmid }V$ a \textbf{\
d--connection} structure $\ ^{\shortmid }\mathbf{D}=(h\ ^{\shortmid }D,c\
^{\shortmid }D)$ is a linear connection preserving under parallelism the
N--connection splitting (\ref{cncon}), 
\begin{equation}
\ ^{\shortmid }\mathbf{D}=\{\ ^{\shortmid }\mathbf{\Gamma }_{\ \alpha \beta
}^{\gamma }=(\ ^{\shortmid }L_{jk}^{i},\ ^{\shortmid }\acute{L}_{a\ k}^{\
b};\ ^{\shortmid }\acute{C}_{j}^{i\ c},\ ^{\shortmid }C_{a}^{\ bc})\},%
\mbox{
where }h\ ^{\shortmid }D=(\ ^{\shortmid }L_{jk}^{i},\ ^{\shortmid }\acute{L}%
_{a\ k}^{\ b})\mbox{ and }v\ ^{\shortmid }D=(\ ^{\shortmid }\acute{C}%
_{j}^{i\ c},\ ^{\shortmid }C_{a}^{\ bc}).  \label{dcond}
\end{equation}%
So, the c-indices in such N-adapted formulas are inverse to v-indices in
N-adapted formulas for $V.$ Using d-operator $\ ^{\shortmid }\mathbf{D,}$ we
can define respective fundamental geometric d-objects as we considered on $%
V, $ but with abstract symbolic definitions on $\ ^{\shortmid }V:$ 
\begin{eqnarray*}
\ ^{\shortmid }\mathcal{T}(\ ^{\shortmid }\mathbf{X,\ ^{\shortmid }Y})&:=& \
^{\shortmid } \mathbf{D}_{\ ^{\shortmid }\mathbf{X}}\ ^{\shortmid }\mathbf{Y}%
-\ ^{\shortmid }\mathbf{D}_{\ ^{\shortmid }\mathbf{Y}}\ ^{\shortmid }\mathbf{%
X}-[\ ^{\shortmid }\mathbf{X,\ ^{\shortmid }Y}],%
\mbox{ torsion d-tensor,
d-torsion}; \\
\ ^{\shortmid }\mathcal{R}(\ ^{\shortmid }\mathbf{X,\ ^{\shortmid }Y})&:= &\
^{\shortmid }\mathbf{D}_{\ ^{\shortmid }\mathbf{X}}\ ^{\shortmid }\mathbf{D}%
_{\ ^{\shortmid }\mathbf{Y}}-\ ^{\shortmid }\mathbf{D}_{\ ^{\shortmid }%
\mathbf{Y}}\ ^{\shortmid }\mathbf{D}_{\ ^{\shortmid }\mathbf{X}}-\
^{\shortmid }\mathbf{D}_{\mathbf{[\ ^{\shortmid }X,\ ^{\shortmid }Y]}},%
\mbox{ curvature d-tensor, d-curvature}; \\
\ ^{\shortmid }\mathcal{Q}(\ ^{\shortmid }\mathbf{X})&:= &\ ^{\shortmid }%
\mathbf{D}_{\ ^{\shortmid }\mathbf{X}}\ ^{\shortmid }\mathbf{g,}%
\mbox{
nonmetricity d-fields, d-nonmetricity},
\end{eqnarray*}%
where d-vectors $\ ^{\shortmid }\mathbf{X}$ and$\mathbf{\ ^{\shortmid }Y,}$
and their duals as 1-forms, can be decomposed respectively to N-linear
frames (\ref{naderd}) and (\ref{nadifd}).

Using geometric objects and formulas (\ref{cncon})-(\ref{dcond}), we can
re-define all geometric constructions and formulas for nonholonomic
manifolds $\mathbf{V}$ and tangent bundles $T\mathbf{V}$ on $\ ^{\shortmid }%
\mathbf{V}$ and $T\ ^{\shortmid }\mathbf{V.}$ In general abstract and
N-adapted form, corresponding geometric, gravity and geometric flow models
for (non) associative/commutative phase spaces are studies in \cite%
{vacaru18,bubuianu18a,bubuianu20,bubuianu19,partner03,partner04}, where
various applications in MGTs and geometric and quantum information flow
theories are elaborated.

The nonholonomic Einstein equations on 4-d phase spaces can be defined and
proven using sympolic re-definitions of variables and geometric d-objects in
(\ref{cdeq1}) and (\ref{lccond1}), 
\begin{eqnarray}
\ ^{\shortmid }\widehat{\mathbf{R}}_{\ \ \beta }^{\alpha } &=&\ ^{\shortmid }%
\widehat{\mathbf{\Upsilon }}_{\ \ \beta }^{\alpha },  \label{cdeq1c} \\
\ ^{\shortmid }\widehat{\mathbf{T}}_{\ \alpha \beta }^{\gamma } &=&0,
\label{lccond1c}
\end{eqnarray}%
with effective generating sources $\ ^{\shortmid }\widehat{\mathbf{\Upsilon }%
}_{\ \ \beta }^{\alpha }=[\ _{1}^{\shortmid }\Upsilon \delta _{\ \ j}^{i},\
_{2}^{\shortmid }\Upsilon \delta _{\ \ b}^{a}].$

The equations (\ref{cdeq1c}) and (\ref{lccond1c}) can be solved by generic
iff-diagonal ansatz with a Killing vector. For instance, the phase space
analog of a quasi-stationary d-metric of type (\ref{dm}) is parameterized 
\begin{eqnarray}
\ ^{\shortmid }\mathbf{\hat{g}} &=&g_{i}(x^{k})dx^{i}\otimes dx^{i}+\
^{\shortmid }h^{3}(x^{k},p_{3})\ ^{\shortmid }\mathbf{e}_{3}\otimes \
^{\shortmid }\mathbf{e}_{3}+\ ^{\shortmid }h^{4}(x^{k},p_{3})\ ^{\shortmid }%
\mathbf{e}_{4}\otimes \ ^{\shortmid }\mathbf{e}_{4},  \notag \\
&&\ ^{\shortmid }\mathbf{e}_{3}=dp_{3}+\ ^{\shortmid
}w_{i}(x^{k},p_{3})dx^{i},\ \ ^{\shortmid }\mathbf{e}_{4}=dE+\ ^{\shortmid
}n_{i}(x^{k},p_{3})dx^{i},  \label{dmqc}
\end{eqnarray}%
with Killing symmetry on the time like coordinate $\ ^{\shortmid }\partial
^{4}=\partial ^{E}$. For such ansatz, the N-connection coefficients $\
^{\shortmid }\widehat{N}_{i}^{3}=\ ^{\shortmid }w_{i}(x^{k},p_{3})$ and $\
^{\shortmid }\widehat{N}_{i}^{4}=\ ^{\shortmid }n_{i}(x^{k},p_{3})$ and
N-adapted coefficients of d-metric $\ ^{\shortmid }\widehat{\mathbf{g}}%
_{\alpha \beta }=[\widehat{g}_{ij}(x^{\kappa }),\ ^{\shortmid }\widehat{g}%
^{ab}(x^{\kappa },p_{3})]$ are functions of necessary smooth class.

The phase space analog of locally anisotropic cosmological d-metrics \ (\ref%
{dmc}) \ is stated by formulas, 
\begin{eqnarray}
\ ^{\shortmid }\underline{\mathbf{g}} &=&g_{i}(x^{k})dx^{i}\otimes dx^{i}+\
^{\shortmid }\underline{h}^{3}(x^{k},E)\ ^{\shortmid }\underline{\mathbf{e}}%
_{3}\otimes \ ^{\shortmid }\underline{\mathbf{e}}_{3}+\ ^{\shortmid }%
\underline{h}^{4}(x^{k},E)\ ^{\shortmid }\underline{\mathbf{e}}_{4}\otimes \
^{\shortmid }\underline{\mathbf{e}}_{4},  \notag \\
&&\ ^{\shortmid }\underline{\mathbf{e}}_{3}=dp_{3}+\ ^{\shortmid }\underline{%
n}_{i}(x^{k},E)dx^{i},\ \ ^{\shortmid }\underline{\mathbf{e}}_{4}=dE+\
^{\shortmid }\underline{w}_{i}(x^{k},E)dx^{i},  \label{dmcc}
\end{eqnarray}%
with Killing symmetry on the time like coordinate $\ ^{\shortmid }\partial
^{3}$. For such a d-metric, the N-connection coefficients $\ ^{\shortmid }%
\underline{N}_{i3}=\ ^{\shortmid }\underline{n}_{i}(x^{k},E)$ and $\
^{\shortmid }\underline{N}_{i4}=\ ^{\shortmid }\underline{w}_{i}(x^{k},E)$
and the N-adapted coefficients of d-metrics are of type $\ ^{\shortmid }%
\underline{\mathbf{g}}_{\alpha \beta }=[g_{ij}(x^{\kappa }),\ ^{\shortmid }%
\underline{g}^{ab}(x^{\kappa },E)].$

In this subsection, we consider toy 2+2 phase space model with momentum like
variables in order to show how the AFCDM can be extended on such spaces when
in abstract geometric form we can generate exact and parametric solutions.
For instance, the quasi-stationary solution (\ref{qeltors}) for a
nonholonomic phase space can be represented in a form (\ref{dmqc}), 
\begin{eqnarray}
d\ ^{\shortmid }s^{2} &=&e^{\psi (x^{k})}[(dx^{1})^{2}+(dx^{2})^{2}]+\frac{%
[\ ^{\shortmid }\partial ^{3}\ ^{\shortmid }\Psi ]^{2}}{4(\ _{2}^{\shortmid }%
\widehat{\Upsilon })^{2}\{\ ^{\shortmid }g_{[0]}^{4}-\int dp_{3}\
^{\shortmid }\partial ^{3}[\Psi ^{2}]/4(\ _{2}\widehat{\Upsilon })\}}(dy^{3}+%
\frac{\partial _{i}\ ^{\shortmid }\Psi }{\ ^{\shortmid }\partial ^{3}\
^{\shortmid }\Psi }dx^{i})^{2}+  \label{qeltorsd} \\
&&\{\ ^{\shortmid }g_{[0]}^{4}-\int dp_{3}\frac{\ ^{\shortmid }\partial
^{3}[\ ^{\shortmid }\Psi ^{2}]}{4(\ _{2}^{\shortmid }\widehat{\Upsilon })}%
\}\{dE+[\ _{1}n_{k}+\ _{2}n_{k}\int dp_{3}\frac{\ ^{\shortmid }\partial
^{3}[(\ ^{\shortmid }\Psi )^{2}]}{4(\ _{2}^{\shortmid }\widehat{\Upsilon }%
)^{2}|\ ^{\shortmid }g_{[0]}^{4}-\int dp_{3}\ ^{\shortmid }\partial ^{3}[\
^{\shortmid }\Psi ^{2}]/4(\ _{2}^{\shortmid }\widehat{\Upsilon })|^{5/2}}%
]dx^{k}\}^{2}.  \notag
\end{eqnarray}%
Such a d-metric is a phase space quasi-stationary one if we perform an
pseudo-Euclidean rotation from 4-metrics of signature (+-+-) to (+++-).

For toy 2+2 phase space models, we can generate so-called "rainbow"
d-metrics, see reviews in which are of type (\ref{dmcc}) and can be also of
cosmological type if we fix, for instance, $x^{2}=t.$ In abstract symbolic
form, we can formulate a \textbf{principle of phase space, time, momentum
and energy duality }of such different generic off-diagonal configurations: 
\begin{eqnarray*}
p_{3} &\longleftrightarrow &p_{4}=E,\ ^{\shortmid
}h^{3}(x^{k},p_{3})\longleftrightarrow \ ^{\shortmid }\underline{h}%
^{4}(x^{k},E),\ ^{\shortmid }h^{4}(x^{k},p_{3})\longleftrightarrow \
^{\shortmid }\underline{h}^{3}(x^{k},E), \\
\ ^{\shortmid }N_{i3} &=&\ ^{\shortmid
}w_{i}(x^{k},p_{3})\longleftrightarrow \ ^{\shortmid }N_{i4}=\ ^{\shortmid }%
\underline{n}_{i}(x^{k},E),\ ^{\shortmid }N_{i4}=\ ^{\shortmid
}n_{i}(x^{k},p_{3})\longleftrightarrow \ ^{\shortmid }N_{i3}=\ ^{\shortmid }%
\underline{w}_{i}(x^{k},E).
\end{eqnarray*}%
We can also formulate duality conditions for phase prime d-metrics and
respective generating functions/ sources and gravitational polarization
functions, when 
\begin{equation*}
\ ^{\shortmid }\Upsilon _{~3}^{3}=\ ^{\shortmid }\Upsilon
_{~4}^{4}=~_{\shortmid }^{c}\Upsilon (x^{k},p_{3})=\ _{2}^{\shortmid }%
\widehat{\Upsilon }\longleftrightarrow \ ^{\shortmid }\underline{\Upsilon }%
_{~4}^{4}=\ ^{\shortmid }\underline{\Upsilon }_{~3}^{3}=~_{\shortmid }^{c}%
\underline{\Upsilon }(x^{k},E)=\ _{2}^{\shortmid }\widehat{\underline{%
\Upsilon }},\mbox{ see }(\ref{esourcqscan});
\end{equation*}%
\begin{equation*}
\begin{array}{ccc}
\begin{array}{c}
(\ ^{\shortmid }\Psi ,\ _{2}^{\shortmid }\widehat{\Upsilon })\leftrightarrow
(\ ^{\shortmid }\mathbf{g},\ _{2}^{\shortmid }\widehat{\Upsilon }%
)\leftrightarrow \\ 
(\ ^{\shortmid }\eta _{\alpha }\ \ ^{\shortmid }\mathring{g}_{\alpha }\sim
(\ ^{\shortmid }\zeta _{\alpha }(1+\ ^{\shortmid }\kappa \ ^{\shortmid }\chi
_{\alpha })\mathring{g}_{\alpha },\ _{2}^{\shortmid }\widehat{\Upsilon }%
)\leftrightarrow%
\end{array}
& \Longleftrightarrow & 
\begin{array}{c}
(\ ^{\shortmid }\underline{\Psi },\ _{2}^{\shortmid }\widehat{\underline{%
\Upsilon }})\leftrightarrow (\ ^{\shortmid }\underline{\mathbf{g}},\
_{2}^{\shortmid }\widehat{\underline{\Upsilon }})\leftrightarrow \\ 
(\ ^{\shortmid }\underline{\eta }_{\alpha }\ \ ^{\shortmid }\underline{%
\mathring{g}}_{\alpha }\sim (\ ^{\shortmid }\underline{\zeta }_{\alpha }(1+\
^{\shortmid }\kappa \underline{\chi }_{\alpha })\ ^{\shortmid }\underline{%
\mathring{g}}_{\alpha },\ _{2}^{\shortmid }\widehat{\underline{\Upsilon }}%
)\leftrightarrow%
\end{array}
\\ 
\begin{array}{c}
(\ ^{\shortmid }\Phi ,\ _{2}^{\shortmid }\Lambda )\leftrightarrow (\mathbf{g}%
,\ _{2}^{\shortmid }\Lambda )\leftrightarrow \\ 
(\ ^{\shortmid }\eta _{\alpha }\ \ ^{\shortmid }\mathring{g}_{\alpha }\sim
(\ ^{\shortmid }\zeta _{\alpha }(1+\ ^{\shortmid }\kappa \ ^{\shortmid }\chi
_{\alpha })\ ^{\shortmid }\mathring{g}_{\alpha },\ _{2}^{\shortmid }\Lambda
),%
\end{array}
& \Longleftrightarrow & 
\begin{array}{c}
(\underline{\Phi },\ _{2}^{\shortmid }\underline{\Lambda })\leftrightarrow (%
\underline{\mathbf{g}},\ _{2}^{\shortmid }\underline{\Lambda }%
)\leftrightarrow \\ 
(\ ^{\shortmid }\underline{\eta }_{\alpha }\ \ ^{\shortmid }\underline{%
\mathring{g}}_{\alpha }\sim (\ ^{\shortmid }\underline{\zeta }_{\alpha }(1+\
^{\shortmid }\kappa \ ^{\shortmid }\underline{\chi }_{\alpha })\ ^{\shortmid
}\underline{\mathring{g}}_{\alpha },\ _{2}^{\shortmid }\underline{\Lambda }).%
\end{array}%
\end{array}%
\end{equation*}

For locally anisotropic E-depending phase configurations, the nonlinear
symmetries (\ref{ntransf1}) and (\ref{ntransf2}) are written in respective
dual forms, 
\begin{eqnarray}
\frac{\ ^{\shortmid }\partial ^{4}[\ ^{\shortmid }\underline{\Psi }^{2}]}{\
_{2}^{\shortmid }\widehat{\underline{\Upsilon }}} &=&\frac{\ ^{\shortmid
}\partial ^{4}[\ ^{\shortmid }\underline{\Phi }^{2}]}{\ _{2}^{\shortmid }%
\underline{\Lambda }},\mbox{ which can be
integrated as  }  \label{nonlinsymcosm} \\
\ ^{\shortmid }\underline{\Phi }^{2} &=&\ _{2}\underline{\Lambda }\int dE(\
_{2}^{\shortmid }\widehat{\underline{\Upsilon }})^{-1}\ ^{\shortmid
}\partial ^{4}[\ ^{\shortmid }\underline{\Psi }^{2}]\mbox{ and/or
}\ ^{\shortmid }\underline{\Psi }^{2}=(\ _{2}^{\shortmid }\underline{\Lambda 
})^{-1}\int dE(\ _{2}^{\shortmid }\widehat{\underline{\Upsilon }})\
^{\shortmid }\partial ^{4}[\ ^{\shortmid }\underline{\Phi }^{2}].  \notag
\end{eqnarray}%
In a similar form, we can derive respective phase space formulas with labels
"$\ ^{\shortmid }$" for solutions determined by quasi-stationary d-metrics (%
\ref{qeltors}), (\ref{offdiagcosmcsh}), (\ref{offdsolgenfgcosmc}), (\ref%
{offdiagpolfr}) and (\ref{offdncelepsilon}) and their locally anisotropic
cosmological analogs. For instance, the phase space analog of the locally
anisotropic d-metric (\ref{qeltorsc}) transforms into 
\begin{eqnarray}
d\ ^{\shortmid }\underline{s}^{2} &=&e^{\psi
(x^{k})}[(dx^{1})^{2}+(dx^{2})^{2}]  \label{4ds} \\
&&+\{\ ^{\shortmid }g_{[0]}^{3}-\int dE\frac{\ ^{\shortmid }\partial ^{4}[\
^{\shortmid }\underline{\Psi }^{2}]}{4(\ _{2}^{\shortmid }\widehat{%
\underline{\Upsilon }})}\}\{dp_{3}+[\ _{1}n_{k}+\ _{2}n_{k}\int dE\frac{\
^{\shortmid }\partial ^{4}[(\ ^{\shortmid }\underline{\Psi })^{2}]}{4(\
_{2}^{\shortmid }\widehat{\underline{\Upsilon }})^{2}|g_{[0]}^{3}-\int dE\
^{\shortmid }\partial ^{4}[\ ^{\shortmid }\underline{\Psi }^{2}]/4(\
_{2}^{\shortmid }\widehat{\underline{\Upsilon }})|^{5/2}}]dx^{k}\}  \notag \\
&&+\frac{[\ ^{\shortmid }\partial ^{4}\ ^{\shortmid }\underline{\Psi }]^{2}}{%
4(\ _{2}^{\shortmid }\widehat{\underline{\Upsilon }})^{2}\{\ ^{\shortmid
}g_{[0]}^{3}-\int dE\ ^{\shortmid }\partial ^{4}[\ ^{\shortmid }\underline{%
\Psi }^{2}]/4(\ _{2}^{\shortmid }\widehat{\underline{\Upsilon }})\}}(dE+%
\frac{\partial _{i}\ ^{\shortmid }\underline{\Psi }}{\ ^{\shortmid }\partial
^{4}\ ^{\shortmid }\underline{\Psi }}dx^{i})^{2}.  \notag
\end{eqnarray}%
Other classes of phase space d-metrics with different types of generating
functions, generating sources and effective cosmological constants can be
also derived in abstract dual form changing corresponding indices 3 into 4,
4 into 3, underlying the respective generating functions / effective sources
/ gravitational polarizations for dependencies on $(x^{i},E)$ and changing
v-partial derivatives $\ ^{\shortmid }\partial ^{3}\rightarrow \ ^{\shortmid
}\partial ^{4}$.

\subsection{Extracting LC-configurations}

\label{sslcconf}The generic off--diagonal solutions constructed in previous
subsections are constructed for a canonical d--connection $\widehat{\mathbf{D%
}}$ or corresponding phase space variants $\ ^{\shortmid }\widehat{\mathbf{D}%
}.$ In general, such solutions are characterized by nonholonomically induced
d--torsion coefficients $\ \widehat{\mathbf{T}}_{\ \alpha \beta }^{\gamma }\ 
$(\ref{dtors}) completely defined by the N--connection and d--metric
structures. We can extract zero torsion LC-configurations if we impose
additionally the conditions (\ref{lcconstr}). By straightforward
computations for quasi-stationary configurations, we can verify that all
d-torsion coefficients vanish if the coefficients of N--adapted frames and $%
v $--components of d--metrics are subjected to respective conditions, 
\begin{eqnarray}
\ w_{i}^{\ast } &=&\mathbf{e}_{i}\ln \sqrt{|\ h_{3}|},\mathbf{e}_{i}\ln 
\sqrt{|\ h_{4}|}=0,\partial _{i}w_{j}=\partial _{j}w_{i}\mbox{ and }%
n_{i}^{\ast }=0;  \notag \\
n_{k}(x^{i}) &=&0\mbox{ and }\partial _{i}n_{j}(x^{k})=\partial
_{j}n_{i}(x^{k}).  \label{zerot1}
\end{eqnarray}%
The solutions for necessary type of $w$- and $n$-functions depend on the
class of vacuum or non--vacuum metrics which we attempt to construct. Such
classes of generating functions and generating sources and N-connection
coefficients can be constructed following such steps of finding solutions (%
\ref{zerot1}):

Prescribing a generating function $\Psi =\check{\Psi}(x^{i_{1}},y^{3}),$ for
which $[\partial _{i}(\ _{2}\check{\Psi})]^{\ast }=\partial _{i}(\ _{2}%
\check{\Psi})^{\ast },$we solve the equations for $w_{j}$ from (\ref{zerot1}%
) in explicit form if $\ _{2}\widehat{\Upsilon }=const,$ or if $\ $such an
effective source can be expressed as a functional $\ \ _{2}\widehat{\Upsilon 
}(x^{i},y^{3})=\ _{2}\widehat{\Upsilon }[\ _{2}\check{\Psi}].$ Then, the
third conditions $\partial _{i}w_{j}=\partial _{j}w_{i},$ are solved by any
generating function $\ \check{A}=\check{A}(x^{k},y^{3})$ for which $\ $%
\begin{equation*}
w_{i}=\check{w}_{i}=\partial _{i}\ \check{\Psi}/(\check{\Psi})^{\ast
}=\partial _{i}\check{A}.
\end{equation*}%
The equations for $n$-functions in (\ref{zerot1}) are solved for any $%
n_{i}=\partial _{i}[\ ^{2}n(x^{k})].$

Putting together above formulas for respective classes of generating
functions in , we construct a nonlinear quadratic element for
quasi-stationary solutions with zero canonical d-torsions, (\ref{qeltors}), 
\begin{eqnarray}
d\check{s}^{2} &=&\check{g}_{\alpha \beta }(x^{k},y^{3})du^{\alpha
}du^{\beta }  \label{qellc} \\
&=&e^{\psi (x^{k})}[(dx^{1})^{2}+(dx^{2})^{2}]+\frac{[\check{\Psi}^{\ast
}]^{2}}{4(\ _{2}\widehat{\Upsilon }[\check{\Psi}])^{2}\{h_{4}^{[0]}-\int
dy^{3}[\check{\Psi}]^{\ast }/4\ _{2}\widehat{\Upsilon }[\check{\Psi}]\}}%
\{dy^{3}+[\partial _{i}(\check{A})]dx^{i}\}^{2}  \notag \\
&&+\{h_{4}^{[0]}-\int dy^{3}\frac{[\check{\Psi}^{2}]^{\ast }}{4(\ _{2}%
\widehat{\Upsilon }[\check{\Psi}])}\}\{dt+\partial _{i}[\
^{2}n(x^{k})]dx^{i}\}^{2}.  \notag
\end{eqnarray}%
In a similar form, we can extract LC-configurations for all classes of
quasi-stationary, locally anisotropic cosmologic and toy phase space
solutions considered in this section. This is always possible for
generically off-diagonal metrics with nontrivial canonical d-torsion if we
impose respective conditions for generating data of type $(\check{\Psi}%
(x^{i_{1}},y^{3}),\ _{2}\widehat{\Upsilon }[\ _{2}\check{\Psi}],\check{A}).$

\section{Examples of off-diagonal quasi-stationary or cosmological solutions}

\label{sec4}We show how choosing respective classes of generating and
integration functions we can construct certain physically important examples
of quasi-stationary generic off-diagonal solutions. They describe BHs,
nonholonomic cylindrical systems, WHs, BT and BE configurations in MGTs
modelled on 4-d Lorentz manifolds by effective sources determined by
nonholonomic distortions of d-connections and off-diagonal deformations of
metrics. In some dual on-the-time coordinates forms, the AFCDM allows us to
construct off-diagonal cosmological solutions describing nonholonomic
cosmological solitonic evolution scenarios and spheroid deformations
involving 2-d vertices. Constraining the generating and integration
functions to some subclasses of coefficients resulting in zero nonholonomic
torsions, such generic off-diagonal solutions can be generated in the
framework of GR theory.

\subsection{New Kerr de Sitter solutions and nonholonomic deformations to
spheroidal configurations}

Nonholonomic off-diagonal deformations of the Kerr and Schwarzschild - (a)
de Sitter, K(a)dS, and other types BH metrics determined by (effective)
(non) associative/ commutative sources in string theory, MGTs and geometric
information flow were studied in a series of works \cite%
{biv17,sv00a,sv03a,sv03b,sv05,sv07,bubuianu17,vi17,vacaru18,bubuianu19,partner03,partner04}%
. For spherical rotating configurations of KdS, such metrics can be
described by various families of rotating diagonal metrics involving, or not
warping effects of curvature, see details in \cite{ovalle21}. In this
subsection, we analyse how rotating BHs can be nonholonomically deformed to
parametric quasi-stationary d-metrics of type (\ref{offdncelepsilon}). There
are also computed in explicit form spheroidal rotoid deformations.

\subsubsection{Prime d-metrics as off-diagonal new KdS solutions}

We consider a prime quadratic line element of type (\ref{offdiagpm}) for
spherical coordinates parameterized in the form $x^{1}=r,x^{2}=\varphi
,y^{3}=\theta ,y^{4}=t,$ when 
\begin{equation}
d\breve{s}^{2}=\breve{g}_{\alpha }(r,\varphi ,\theta )(\mathbf{\breve{e}}%
^{\alpha })^{2},  \label{offdiagpm1}
\end{equation}%
with such nontrivial coefficients of the d-metric and N-connection: 
\begin{eqnarray*}
\breve{g}_{1} &=&\frac{\breve{\rho}^{2}}{\triangle _{\Lambda }},\breve{g}%
_{2}=\frac{\sin ^{2}\theta }{\breve{\rho}^{2}}[\Sigma _{\Lambda }-\frac{%
(r^{2}+a^{2}-\triangle _{\Lambda })^{2}}{a^{2}\sin ^{2}\theta -\triangle
_{\Lambda }}],\breve{g}_{3}=\breve{\rho}^{2},\breve{g}_{4}=\frac{a^{2}\sin
^{2}\theta -\triangle _{\Lambda }}{\breve{\rho}^{2}},\mbox{ and } \\
\breve{N}_{2}^{4} &=&\breve{n}_{2}=-a\sin \theta \frac{r^{2}+a^{2}-\triangle
_{\Lambda }}{a^{2}\sin ^{2}\theta -\triangle _{\Lambda }}.
\end{eqnarray*}%
If the functions and parameters are chosen in the form 
\begin{eqnarray*}
\Sigma _{\Lambda } &=&(r^{2}+a^{2})^{2}-\triangle _{\Lambda }a^{2}\sin
^{2}\theta ,\triangle _{\Lambda }=r^{2}-2Mr+a^{2}-\frac{\Lambda _{0}}{3}%
r^{4}, \\
\breve{\rho}^{2} &=&r^{2}+a^{2}\cos ^{2}\theta ,\mbox{ for constants  }%
a=J/M=const,
\end{eqnarray*}%
where $J$ is the angular momentum and $M$ is the total mass of the system,
and the cosmological constant $\Lambda _{0}>0,$ the d-metric (\ref%
{offdiagpm1}) define a new KdS solution reported in \cite{ovalle21} (see
also relevant details and references in that work). We note that in this
paper we use a different system of notations stated for another signature of
the metrics. It is different from the standard KdS metrics, called also $%
\Lambda $-vacuum solutions, because the scalar curvature 
\begin{equation*}
R(r,\theta )=4\widetilde{\Lambda }(r,\theta )=4\Lambda _{0}\frac{r^{2}}{%
\breve{\rho}^{2}}\neq 4\Lambda _{0}.
\end{equation*}%
Such a new KdS solution posses a warped effect when the curvature is warped
every were excepting the equatorial plane. The effective polarization of the
cosmological constant also shows a rotational effect on the vacuum energy.
Mentioned effect disappear for $r\gg a.$ A d-metric (\ref{offdiagpm1}) can
be considered as a rotating version of the Schwarzschild de Sitter metric
and represents a new solution describing the exterior of a BH with
cosmological constant. It contains certain bonds for $M(a,\Lambda _{0})$ for
existence of a BH solution. In explicit form, the respective upper, $M_{\max
}:=M_{+}$ and lower, $M_{\min }:=M,$ bounds, when 
\begin{equation}
18\Lambda _{0}M_{\pm }^{2}=1+12\Lambda _{0}a^{2}\pm (1-4\Lambda
_{0}a^{2})^{3/2}.  \label{bonds}
\end{equation}%
Such a d-metric defines a LC-configuration for the standard Einstein
equations (\ref{en1}) with fluid type energy momentum tensor (\ref{fluidm}),
when 
\begin{equation}
\breve{T}_{\alpha \beta }(r,\theta )=diag[p_{r},p_{\varphi }=p_{\theta
},p_{\theta }=\rho -2\Lambda _{0}r^{2}/\mathring{\rho}^{2},\rho =-p_{r}=%
\widetilde{\Lambda }^{2}/\Lambda _{0}].  \label{efmt1}
\end{equation}%
So, we have two possibilities to interpret that such primary d-metrics: the
first one is to consider that they are defined as solutions of some vacuum
locally anisotropic polarizations on $(r,\theta )$ of the cosmological
constant, $\Lambda _{0}\rightarrow \widetilde{\Lambda }(r,\theta ),$ or to
consider that they consist a result of some locally anisotropic tensors of
type $\breve{T}_{\alpha \beta }(r,\theta )$, or more general (effective)
sources.

\subsubsection{Nonholonomic quasi-stationary gravitational polarizations of
KdS configurations}

The goal of this subsection is to study more general off-diagonal
deformations of the standard Kerr solution when there are involved vacuum
polarizations both of effective cosmological constants and d-metric
coefficients depending on all space coordinates $(r,\varphi ,\theta )$ not
only on $(r,\theta )$ as we considered for above prime d-metrics. Such new
classes of quasi-stationary nonholonomic spacetimes posses nonlinear
symmetries of type (\ref{ntransf1}) and (\ref{ntransf2}), defined by
respective classes of nonholonomic quasi-stationary deformations and
constraints. This class of target solutions of type (\ref{dmq}), when $%
\mathbf{\hat{g}}(r,\varphi ,\theta )$ is defined equivalently by respective
generating sources of type (\ref{esourcqscan})%
\begin{equation*}
\mathbf{\breve{\Upsilon}}_{\ \ \beta }^{\alpha }(r,\varphi ,\theta )=[\
^{h}\Upsilon \delta _{\ \ j}^{i},\ ^{v}\Upsilon \delta _{\ \ b}^{a}]=[\
^{h}\Upsilon =-\ _{1}\mathbf{\breve{\Upsilon}}(r,\varphi ),\ ^{v}\Upsilon
=-\ _{2}\mathbf{\breve{\Upsilon}}(r,\varphi ,\theta )].
\end{equation*}%
Let us first consider solutions with $\eta $-polarization functions for
d-metrics in the form (\ref{offdiagpolfr}) when 
\begin{eqnarray}
d\widehat{s}^{2} &=&\widehat{g}_{\alpha \beta }(x^{k},y^{3};\breve{g}%
_{\alpha };\psi ,\eta _{4};\ _{2}\Lambda =\widetilde{\Lambda },\ _{2}\mathbf{%
\breve{\Upsilon}})du^{\alpha }du^{\beta }=e^{\psi (r,\varphi
)}[(dx^{1}(r,\varphi ))^{2}+(dx^{2}(r,\varphi ))^{2}]  \label{nkernew} \\
&&-\frac{[\partial _{\theta }(\eta _{4}\ \breve{g}_{4})]^{2}}{|\int d\theta
\ _{2}\mathbf{\breve{\Upsilon}}\partial _{\theta }(\eta _{4}\ \breve{g}%
_{4})|\ \eta _{4}\breve{g}_{4}}\{dy^{3}+\frac{\partial _{i}[\int d\theta \
_{2}\mathbf{\breve{\Upsilon}}\ \partial _{3}(\eta _{4}\breve{g}_{4})]}{\ _{2}%
\mathbf{\breve{\Upsilon}}\partial _{\theta }(\eta _{4}\breve{g}_{4})}%
dx^{i}\}^{2}  \notag \\
&&+\eta _{4}\breve{g}_{4}\{dt+[\ _{1}n_{k}(r,\varphi )+\ _{2}n_{k}(r,\varphi
)\int d\theta \frac{\lbrack \partial _{\theta }(\eta _{4}\breve{g}_{4})]^{2}%
}{|\int d\theta \ _{2}\mathbf{\breve{\Upsilon}}\partial _{3}(\eta _{4}\breve{%
g}_{4})|\ (\eta _{4}\breve{g}_{4})^{5/2}}]dx^{k}\}^{2}  \notag
\end{eqnarray}%
is determined by a generating function $\eta _{4}=\eta _{4}(r,\varphi
,\theta )$ and respective integration functions like $\ _{1}n_{k}(r,\varphi
) $ and $\ _{2}n_{k}(r,\varphi ).$ The locally anisotropic vacuum effects in
such a d-metric with anisotropic vertical coordinate $\theta $ is very
complex and it is difficult to state conditions when it defines, for
instance, BH configurations. Nevertheless, a quasi-stationary d-metric (\ref%
{nkernew}) can be characterized by nonlinear symmetries of type (\ref%
{nonlinsymrex}), 
\begin{eqnarray}
\partial _{\theta }[\Psi ^{2}] &=&-\int d\theta \ _{2}\mathbf{\breve{\Upsilon%
}}\partial _{\theta }h_{4}\simeq -\int d\theta \ _{2}\mathbf{\breve{\Upsilon}%
}\partial _{\theta }(\eta _{4}\ \breve{g}_{4})\simeq -\int d\theta \ _{2}%
\mathbf{\breve{\Upsilon}}\partial _{\theta }[\zeta _{4}(1+\kappa \ \chi
_{4})\ \breve{g}_{4}],  \label{nlims2} \\
\Psi &=&|\ \ \widetilde{\Lambda }|^{-1/2}\sqrt{|\int d\theta \ _{2}\mathbf{%
\breve{\Upsilon}}\ (\Phi ^{2})^{\ast }|},\Phi ^{2}=-4\ \widetilde{\Lambda }%
h_{4}\simeq -4\ _{2}\Lambda \eta _{4}\breve{g}_{4}\simeq -4\ \widetilde{%
\Lambda }\ \zeta _{4}(1+\kappa \chi _{4})\ \breve{g}_{4}.  \notag
\end{eqnarray}%
We note that in a series of our former works \cite%
{sv03a,sv03b,sv05,sv07,bubuianu17,biv17,bubuianu19,partner03} we constructed
and studied physical properties of d-metrics when K(a)dS and other type BH
solutions are nonholonomically deformed for $y^{3}=\varphi ,$ when
respective effective sources $\ _{2}\widehat{\Upsilon }$ are generated by
certain extra dimension (super) string contributions, nonassociative and/or
noncommutative terms, generalized Finsler and modified dispersion
deformations, or other type MGTs. There were stated explicit conditions when
off-diagonal $\varphi $-deformations may result in black ellipsoid, BE,
configurations which can be quasi-stationary and for solutions which are
different from the Kerr-Newmann-(a)dS configurations. It was proven that
such locally anisotropic configurations can be stable, or stabilized by
imposing corresponding nonholonomic constraints. In a similar form, we can
construct gravitational $\eta $-polarizations when $y^{3}=\theta ,$ which
results in different classes of solutions constructed for other types of
effective sources and, related via nonlinear symmetries, polarized or fixed
values of some prescribed cosmological constants.

\subsubsection{Off-diagonal quasi-stationary small parametric deformations
of new KdS d-metrics}

We can search for more clear physical interpretation of nonholonomic
deformations of the class of prime metrics (\ref{nkernew}) and any class of
similar BH ones if we consider small parametric decompositions with $\kappa $%
-linear terms as we considered in (\ref{epsilongenfdecomp}). To avoid
possible singular off-diagonal frame/coordinate deformations we consider a
new system of coordinates when there are nontrivial terms $\breve{N}_{i}^{a}$
both for $a=3,$ with some $\breve{N}_{i}^{3}=\breve{w}_{i}(r,\varphi ,\theta
),$ which can be zero in certain rotation frames, and, for $a=4,$ $\breve{N}%
_{i}^{4}=\breve{n}_{i}(r,\varphi ,\theta )$ which may be with a nontrivial $%
\breve{n}_{2}=-a\sin \theta (r^{2}+a^{2}-\triangle _{\Lambda })/(a^{2}\sin
^{2}\theta -\triangle _{\Lambda })$ as we considered above. Applying the
AFCDM, we construct a d-metric of type (\ref{offdncelepsilon}) determined by 
$\chi $-generating functions: 
\begin{equation*}
d\ \widehat{s}^{2}=\widehat{g}_{\alpha \beta }(r,\varphi ,\theta ;\psi
,g_{4};\ _{2}\mathbf{\breve{\Upsilon}})du^{\alpha }du^{\beta }=e^{\psi
_{0}}(1+\kappa \ ^{\psi }\chi )[(dx^{1}(r,\varphi ))^{2}+(dx^{2}(r,\varphi
))^{2}]
\end{equation*}%
\begin{eqnarray*}
&&-\{\frac{4[\partial _{\theta }(|\zeta _{4}\breve{g}_{4}|^{1/2})]^{2}}{%
\breve{g}_{3}|\int d\theta \lbrack \ \ _{2}\mathbf{\breve{\Upsilon}}\partial
_{3}(\zeta _{4}\breve{g}_{4})]|}-\kappa \lbrack \frac{\partial _{\theta
}(\chi _{4}|\zeta _{4}\breve{g}_{4}|^{1/2})}{4\partial _{\theta }(|\zeta _{4}%
\breve{g}_{4}|^{1/2})}-\frac{\int d\theta \{\ _{2}\mathbf{\breve{\Upsilon}}%
\partial _{\theta }[(\zeta _{4}\breve{g}_{4})\chi _{4}]\}}{\int d\theta
\lbrack \ _{2}\mathbf{\breve{\Upsilon}}\partial _{\theta }(\zeta _{4}\breve{g%
}_{4})]}]\}\breve{g}_{3} \\
&&\{d\theta +[\frac{\partial _{i}\ \int d\theta \ _{2}\mathbf{\breve{\Upsilon%
}}\ \partial _{\theta }\zeta _{4}}{(\breve{N}_{i}^{3})\ _{2}\mathbf{\breve{%
\Upsilon}}\partial _{\theta }\zeta _{4}}+\kappa (\frac{\partial _{i}[\int
d\theta \ _{2}\mathbf{\breve{\Upsilon}}\ \partial _{\theta }(\zeta _{4}\chi
_{4})]}{\partial _{i}\ [\int d\theta \ _{2}\mathbf{\breve{\Upsilon}}\partial
_{\theta }\zeta _{4}]}-\frac{\partial _{\theta }(\zeta _{4}\chi _{4})}{%
\partial _{\theta }\zeta _{4}})]\breve{N}_{i}^{3}dx^{i}\}^{2}
\end{eqnarray*}%
\begin{eqnarray}
&&+\zeta _{4}(1+\kappa \ \chi _{4})\ \breve{g}_{4}\{dt+[(\breve{N}%
_{k}^{4})^{-1}[\ _{1}n_{k}+16\ _{2}n_{k}[\int d\theta \frac{\left( \partial
_{\theta }[(\zeta _{4}\breve{g}_{4})^{-1/4}]\right) ^{2}}{|\int d\theta
\partial _{\theta }[\ _{2}\mathbf{\breve{\Upsilon}}(\zeta _{4}\breve{g}%
_{4})]|}]  \label{offdnceleps1} \\
&&+\kappa \frac{16\ _{2}n_{k}\int d\theta \frac{\left( \partial _{\theta
}[(\zeta _{4}\breve{g}_{4})^{-1/4}]\right) ^{2}}{|\int d\theta \partial
_{\theta }[\ _{2}\mathbf{\breve{\Upsilon}}(\zeta _{4}\breve{g}_{4})]|}(\frac{%
\partial _{\theta }[(\zeta _{4}\breve{g}_{4})^{-1/4}\chi _{4})]}{2\partial
_{\theta }[(\zeta _{4}\breve{g}_{4})^{-1/4}]}+\frac{\int d\theta \partial
_{\theta }[\ _{2}\mathbf{\breve{\Upsilon}}(\zeta _{4}\chi _{4}\breve{g}_{4})]%
}{\int d\theta \partial _{\theta }[\ _{2}\mathbf{\breve{\Upsilon}}(\zeta _{4}%
\breve{g}_{4})]})}{\ _{1}n_{k}+16\ _{2}n_{k}[\int d\theta \frac{\left(
\partial _{\theta }[(\zeta _{4}\breve{g}_{4})^{-1/4}]\right) ^{2}}{|\int
d\theta \partial _{\theta }[\ _{2}\mathbf{\breve{\Upsilon}}(\zeta _{4}\breve{%
g}_{4})]|}]}]\breve{N}_{k}^{4}dx^{k}\}^{2}.  \notag
\end{eqnarray}%
We note that polarization functions $\zeta _{4}(r,\varphi ,\theta )$ and $%
\chi _{4}(r,\varphi ,\theta )$ in this d-metric can be prescribed to be a
necessary smooth class form, when $\chi _{4}$ is a generating function. The
h-coordinates and generating h-function can be chosen in such a way that
there are not introduced additional singularities. The d-metric (\ref%
{offdnceleps1}) describes small $\kappa $-parametric deformations of the of
new KdS d-metric (\ref{nkernew}) when the coefficients get additional
anisotropy on $\varphi $-coordinate.

Solutions of type (\ref{offdnceleps1}) can be generated with additional
ellipsoidal deformations on $\theta $ if we chose 
\begin{equation}
\chi _{4}(r,\varphi ,\theta )=\underline{\chi }(r,\varphi )\sin (\omega
_{0}\theta +\theta _{0}),  \label{rotoid}
\end{equation}%
where $\underline{\chi }(r,\varphi )$ is a smooth function and $\omega _{0}$
and $\theta _{0}$ are some constants. Really, for such generating
polarizacion functions and $\zeta _{4}(r,\varphi ,\theta )\neq 0,$ we obtain
that 
\begin{equation*}
(1+\kappa \ \chi _{4})\ \breve{g}_{4}\simeq a^{2}\sin ^{2}\theta -\triangle
_{\Lambda }+\kappa \ \chi _{4}=0
\end{equation*}%
For small $a$ and $\frac{\Lambda _{0}}{3},$ we can approximate 
\begin{equation*}
r=2M/(1+\kappa \ \chi _{4}),
\end{equation*}%
which is the parametric equation defining a rotoid configuration with $%
\kappa $ being the eccentricity parameter and generating function (\ref%
{rotoid}).

In general, we can consider polarization functions when KdS BH are embedded
into a nontrivial nonholonomic quasi-stationary background. The nonholonomic
conditions can be imposed such way that the BH configuration is preserved as
conventional h- and-distributions. For small ellipsoidal deformations of
type (\ref{rotoid}), we model black ellipsoid, BE, objects. They can be
stable \cite{sv03a,sv03b,sv05} for certain classes of nonholonomic
constraints. Imposing respective classes of generating and integration
functions of type (\ref{zerot1}), we extract LC-configurations, when the
scalar curvature is of type $R(r,\varphi ,\theta )\simeq \Lambda (r,\varphi
,\theta ),$ which is determined by nonlinear symmetries (\ref{nlims2}). This
modifies on $\kappa $ the boundary conditions (\ref{bonds}) for the
effective mass $M$ and cosmological constant $\Lambda _{0}$, when such
values are with local anisotropic polarization because of vacuum
gravitational background. The phenomenon of warped curvature described in 
\cite{ovalle21} can be preserved for some subclasses of nonholonomic
deformations but the gravitational vacuum became more complex and effective
matter tensor (\ref{efmt1}) became of type $Y_{\alpha \beta }(r,\theta
,\varphi ).$

\subsection{Nonholonomic deformations of cylindrical systems in GR}

The AFCDM can be applied for constructing exact/parametric solutions of
(modified) Einstein equations describing off-diagonal deformations of
cylindrical configurations. In this subsection, we study such an example
when the solutons involve a nontrivial cosmological constant $\Lambda >0$
and/or certain generating sources for (effective) matter.

\subsubsection{Prime d-metrics for cylindrical systems}

As a prime metric ansatz, we consider the cylindrical metric used for
generating Linet-Tian (LT) families of solutions \cite%
{linet85,tian86,dasilva00}, for review of results see \cite{bron20}, 
\begin{eqnarray}
d\mathring{s}^{2} &=&dr^{2}+[Q_{1}(r)]^{2(8\sigma ^{2}-4\sigma
-1)/3\varsigma }[Q_{2}(r)]^{2/3}dz^{2}+  \label{cylpr} \\
&&\frac{\lbrack Q_{2}(r)]^{2/3}}{(c_{0})^{2}[Q_{1}(r)]^{4(8\sigma
^{2}-4\sigma -1)/3\varsigma }}d\varphi ^{2}-\frac{[Q_{2}(r)]^{2/3}}{%
(c_{0})^{2}[Q_{1}(r)]^{2(8\sigma ^{2}-4\sigma -1)/3\varsigma }}dt^{2}. 
\notag
\end{eqnarray}%
In this diagonal metric, there are considered cylindrical coordinates $%
u^{\alpha }=(r,z,\varphi ,t),$ when $\varsigma =4\sigma ^{2}-2\sigma
+1=const,$ for $0<\sigma <1/2;$ $c_{0}$ is an integration constant, which
can be fixed to be positive, and the functions $Q_{1}(r)=\frac{2}{\sqrt{%
3\Lambda }}\tan (\frac{\sqrt{3\Lambda }}{2}r)$ and $Q_{2}(r)=\frac{1}{\sqrt{%
3\Lambda }}\sin (\sqrt{3\Lambda }r)$ are defined in such form that this
metric define a solution of (\ref{en1}) with $Tm=0.$

To avoid off-diagonal deformations with coordinate and frame coefficient
singularities, we can consider frame transforms to a parametrization with
trivial N-connection coefficients $\mathring{N}_{i}^{a}=$ $\mathring{N}%
_{i}^{a}(u^{\alpha }(r,z,\varphi ,t))$ and $\mathring{g}_{\beta
}(u^{j}(r,z,\varphi ),u^{3}(r,z,\varphi )).$ For instance, introducing new
coordinates $u^{1}=x^{1}=r,u^{2}=z,$ and $u^{3}=y^{3}=\varphi +\
^{3}B(r,z),u^{4}=y^{4}=t+\ ^{4}B(r,z),$ when 
\begin{eqnarray*}
\mathbf{\mathring{e}}^{3} &=&d\varphi =du^{3}+\mathring{N}%
_{i}^{3}(r,z)dx^{i}=du^{3}+\mathring{N}_{1}^{3}(r,z)dr+\mathring{N}%
_{2}^{3}(r,z)dz, \\
\mathbf{\mathring{e}}^{4} &=&dt=du^{4}+\mathring{N}%
_{i}^{4}(r,z)dx^{i}=du^{4}+\mathring{N}_{1}^{4}(r,z)dr+\mathring{N}%
_{2}^{4}(r,z)dz,
\end{eqnarray*}%
for $\mathring{N}_{i}^{3}=-\partial \ ^{3}B/\partial x^{i}$ and $\mathring{N}%
_{i}^{4}=-\partial \ ^{4}B/\partial x^{i}.$ We transform (\ref{cylpr}) into%
\begin{eqnarray}
d\mathring{s}^{2} &=&\breve{g}_{\alpha }(r)[\mathbf{\mathring{e}}^{\alpha
}(r,z)]^{2}  \label{cylpr1} \\
&=&dr^{2}+[Q_{1}(r)]^{2(8\sigma ^{2}-4\sigma -1)/3\varsigma
}[Q_{2}(r)]^{2/3}dz^{2}+  \notag \\
&&\frac{[Q_{2}(r)]^{2/3}}{(c_{0})^{2}[Q_{1}(r)]^{4(8\sigma ^{2}-4\sigma
-1)/3\varsigma }}(dy^{3}+\mathring{N}_{i}^{3}dx^{i})^{2}-\frac{%
[Q_{2}(r)]^{2/3}}{(c_{0})^{2}[Q_{1}(r)]^{2(8\sigma ^{2}-4\sigma
-1)/3\varsigma }}(dy^{4}+\mathring{N}_{i}^{4}dx^{i})^{2}.  \notag
\end{eqnarray}

Such a prime d-metric can be used for generating new classes of
quasi-stationary solutions for corresponding types of $\eta $-polarization
and/or $\ \chi $-polarization functions of type (\ref{offdiagpolfr}) and/or (%
\ref{offdncelepsilon}). The explicit formulas for the target d-metrics
depend on the type of coordinate transforms there are considered. We may
keep a system of coordinates when $u^{3}(r,z,\varphi )\simeq \varphi $ and
generate such quasi-stationary solutions with nontrivial derivatives on $%
\partial _{3}=\partial _{\varphi }$ when the coefficients do not depend on $%
y^{4}\simeq t.$ To compute such target d-metrics, we can consider $\ _{2}%
\widehat{\Upsilon }(x^{i},y^{3})=\Lambda ,$ or to study any nontrivial
(effective) matter field contributions encoded in a general $\ _{2}\widehat{%
\Upsilon }(r,z,\varphi )=\ ^{cy}\Upsilon .$ Such a generating source is
stated in cylindric coordinates but may involve other type symmetries and
contributions for various types of classical and quantum deformations,
string symmetries etc.

\subsubsection{Nonholonomic quasi-stationary gravitational polarizations of
cylindrical configurations}

Using $\eta $-polarization functions, we derived such target
quasi-stationary metrics encoding primary d-metrics' data (\ref{cylpr1}), 
\begin{eqnarray}
d\widehat{s}^{2} &=&\widehat{g}_{\alpha \beta }(r,z,\varphi ;\psi ,\eta
_{4};\ _{2}\Lambda =\Lambda ,\ \ ^{cy}\Upsilon ,\ \mathring{g}_{4}=\
^{cy}g(r))du^{\alpha }du^{\beta }  \notag \\
&=&e^{\psi (r,z)}[(dx^{1}(r,z))^{2}+(dx^{2}(r,z))^{2}]  \label{cyltarg1} \\
&&-\frac{[\partial _{\varphi }(\eta _{4}\ \ ^{cy}g_{4})]^{2}}{|\int d\varphi
\ \ ^{cy}\Upsilon \partial _{\varphi }(\eta _{4}\ \breve{g}_{4})|\ \eta
_{4}\ ^{cy}g}\{dy^{3}+\frac{\partial _{i}[\int d\varphi \ \ ^{cy}\Upsilon \
\partial _{\varphi }(\eta _{4}\ ^{cy}g)]}{\ \ ^{cy}\Upsilon \partial
_{\varphi }(\eta _{4}\ ^{cy}g)}dx^{i}\}^{2}+\eta _{4}\breve{g}_{4}  \notag \\
&&\{dt+[\ _{1}n_{k}(r,z)+\ _{2}n_{k}(r,z)\int d\varphi \frac{\lbrack
\partial _{\varphi }(\eta _{4}\ ^{cy}g)]^{2}}{|\int d\varphi \ \
^{cy}\Upsilon \partial _{\varphi }(\eta _{4}\ ^{cy}g)|\ (\eta _{4}\
^{cy}g)^{5/2}}]dx^{k}\}  \notag
\end{eqnarray}%
In these formulas $\ \mathring{g}_{4}=\
^{cy}g(r)=-[Q_{2}(r)]^{2/3}/(c_{0})^{2}[Q_{1}(r)]^{2(8\sigma ^{2}-4\sigma
-1)/3\varsigma },$ when the families of solutions are determined by
respective generating function $\eta _{4}=\eta _{4}(r,z,\varphi )$ and
integration functions $\ _{1}n_{k}(r,z)$ and $\ _{2}n_{k}(r,z).$ The
function $\psi (r,z)$ is a solution of 2-d Poisson equation $\partial
_{11}^{2}\psi +\partial _{22}^{2}\psi =2\ \ _{1}\widehat{\Upsilon }(r,z),$
when 
\begin{equation*}
e^{\psi (r,z)}[(dx^{1}(r,z))^{2}+(dx^{2}(r,z))^{2}]=\eta
_{1}(r,z)dr^{2}+\eta _{2}(r,z)[Q_{1}(r)]^{2(8\sigma ^{2}-4\sigma
-1)/3\varsigma }[Q_{2}(r)]^{2/3}dz^{2}.
\end{equation*}

The locally anisotropic vacuum effects described by a d-metric (\ref%
{cyltarg1}) with anisotropic vertical coordinate $\varphi $ are very complex
and it is difficult to state in general form explicit conditions when it
defines, for instance, BH configurations, or result some generalized
wormholes (to be studied in next subsection) etc. Any such quasi-stationary
off-diagonal solution can be characterized by corresponding nonlinear
symmetries of type (\ref{nonlinsymrex}), 
\begin{eqnarray*}
\partial _{\varphi }[\Psi ^{2}(r,z,\varphi )] &=&-\int d\varphi \
^{cy}\Upsilon \partial _{\varphi }g_{4}\simeq -\int d\varphi \ \
^{cy}\Upsilon (r,z,\varphi )\partial _{\varphi }[\eta _{4}(r,z,\varphi )\
^{cy}g(r)] \\
&\simeq &-\int d\varphi \ \ ^{cy}\Upsilon (r,z,\varphi )\partial _{\varphi
}[\zeta _{4}(r,z,\varphi )(1+\kappa \ \chi _{4}(r,z,\varphi ))\ \ ^{cy}g(r)],
\\
\Psi (r,z,\varphi ) &=&|\ \ \Lambda |^{-1/2}\sqrt{|\int d\varphi \ \
^{cy}\Upsilon (r,z,\varphi )\ \partial _{\varphi }(\Phi ^{2})|},(\Phi
(r,z,\varphi ))^{2}=-4\ \Lambda g_{4}(r,z,\varphi ) \\
&\simeq &-4\ _{2}\Lambda \eta _{4}(r,z,\varphi )\ ^{cy}g(r)\simeq -4\Lambda
\ \zeta _{4}(r,z,\varphi )(1+\kappa \chi _{4}(r,z,\varphi ))\ \ ^{cy}g(r).
\end{eqnarray*}%
We can consider that above class of nonholonomic deformations transform a
cylindrical d-metric into a "spagetti" quasi-stationary configuration (with
different sections, curved and waved, possible interruptions etc.) embedded
into locally anisotropic gravitational vacuum media. The geometry of such
objects is determined by prescribed generating functions and sources and
integration functions.

\subsubsection{Small parametric off-diagonal quasi-stationary deformations
of cylindrical d-metrics}

We can provide a more explicit physical interpretation for small parametric
quasi-stationary deformation of cylindrical systems. In terms of $\chi $%
-polarization functions, respective d-metrics can be written in the form 
\begin{equation*}
d\ \widehat{s}^{2}=\widehat{g}_{\alpha \beta }(r,z,\varphi ;\psi ,\
^{cy}\Upsilon )du^{\alpha }du^{\beta }=e^{\psi _{0}(r,z)}[1+\kappa \ ^{\psi
(r,z)}\chi (r,z)][(dx^{1}(r,z))^{2}+(dx^{2}(r,z))^{2}]
\end{equation*}%
\begin{eqnarray*}
&&-\{\frac{4[\partial _{\varphi }(|\zeta _{4}\ \ ^{cy}g|^{1/2})]^{2}}{\breve{%
g}_{3}|\int d\varphi \{\ ^{cy}\Upsilon \partial _{\varphi }(\zeta _{4}\ \
^{cy}g)\}|}-\kappa \lbrack \frac{\partial _{\varphi }(\chi _{4}|\zeta _{4}\
^{cy}g|^{1/2})}{4\partial _{\varphi }(|\zeta _{4}\ ^{cy}g|^{1/2})}-\frac{%
\int d\varphi \{\ ^{cy}\Upsilon \partial _{\varphi }[(\zeta _{4}\
^{cy}g)\chi _{4}]\}}{\int d\varphi \{\ ^{cy}\Upsilon \partial _{\varphi
}(\zeta _{4}\ ^{cy}g)\}}]\}\ ^{cy}g_{3} \\
&&\{d\varphi +[\frac{\partial _{i}\ \int d\varphi \ ^{cy}\Upsilon \ \partial
_{\varphi }\zeta _{4}}{(\mathring{N}_{i}^{3})\ ^{cy}\Upsilon \partial
_{\varphi }\zeta _{4}}+\kappa (\frac{\partial _{i}[\int d\varphi \
^{cy}\Upsilon \ \partial _{\varphi }(\zeta _{4}\chi _{4})]}{\partial _{i}\
[\int d\varphi \ ^{cy}\Upsilon \partial _{\varphi }\zeta _{4}]}-\frac{%
\partial _{\varphi }(\zeta _{4}\chi _{4})}{\partial _{\varphi }\zeta _{4}})]%
\mathring{N}_{i}^{3}dx^{i}\}^{2}
\end{eqnarray*}%
\begin{eqnarray}
&&+\zeta _{4}(1+\kappa \ \chi _{4})\ \ ^{cy}g\{dt+[(\mathring{N}%
_{k}^{4})^{-1}[\ _{1}n_{k}+16\ _{2}n_{k}[\int d\varphi \frac{\left( \partial
_{\varphi }[(\zeta _{4}\ ^{cy}g)^{-1/4}]\right) ^{2}}{|\int d\varphi
\partial _{\varphi }[\ \ ^{cy}\Upsilon (\zeta _{4}\ ^{cy}g)]|}]  \notag \\
&&+\kappa \frac{16\ _{2}n_{k}\int d\varphi \frac{\left( \partial _{\varphi
}[(\zeta _{4}\ ^{cy}g)^{-1/4}]\right) ^{2}}{|\int d\varphi \partial
_{\varphi }[\ \ ^{cy}\Upsilon (\zeta _{4}\ ^{cy}g)]|}(\frac{\partial
_{\varphi }[(\zeta _{4}\ ^{cy}g)^{-1/4}\chi _{4})]}{2\partial _{\varphi
}[(\zeta _{4}\ ^{cy}g)^{-1/4}]}+\frac{\int d\varphi \partial _{\varphi }[\ \
^{cy}\Upsilon (\zeta _{4}\chi _{4}\ ^{cy}g)]}{\int d\varphi \partial
_{\varphi }[\ \ ^{cy}\Upsilon (\zeta _{4}\ ^{cy}g)]})}{\ _{1}n_{k}+16\
_{2}n_{k}[\int d\varphi \frac{\left( \partial _{\varphi }[(\zeta _{4}\
^{cy}g)^{-1/4}]\right) ^{2}}{|\int d\varphi \partial _{\varphi }[\ \
^{cy}\Upsilon (\zeta _{4}\ ^{cy}g)]|}]}]\mathring{N}_{k}^{4}dx^{k}\}^{2}.
\label{cyltarg2}
\end{eqnarray}%
We note that the polarization functions $\zeta _{4}(r,z,\varphi )$ and $\chi
_{4}(r,z,\varphi )$ in this d-metric can be prescribed to be a necessary
smooth class form, when $\chi _{4}$ is a generating function. The trivial
prime N-connection coefficients $\mathring{N}_{i}^{a}$ are taken from (\ref%
{cylpr1}), when $\
^{cy}g_{3}=[Q_{2}(r)]^{2/3}/(c_{0})^{2}[Q_{1}(r)]^{4(8\sigma ^{2}-4\sigma
-1)/3\varsigma }.$

Quasi-stationary off-diagonal deformations of the metric (\ref{cylpr})
defines $\Lambda $-vacuum cylindrical models imbedded self-consistently in
non-trivial off-diagonal gravitational vacuum. If such solutions are
prescribed with certain deformed hypersurface/ horizon configurations, they
are different from radial cylindrical ones. We can generate, for instance,
cylindrical configurations with certain elliptic deformations constructed as
d-metrics of type (\ref{cyltarg2}) if we chose a generating function of type%
\begin{equation*}
\chi _{4}(r,z,\varphi )=\underline{\chi }(r,z)\sin (\omega _{0}\varphi
+\varphi _{0})
\end{equation*}%
as in (\ref{rotoid}) but on a different angular coordinate.
LC-configurations can be extracted by imposing additional nonholonomic
constraints of type (\ref{zerot1}). Respective nonlinear symmetries (\ref%
{nonlinsymrex}) are defined in cylindrical coordinates and allow to
introduce effective nontrivial sources $\ ^{cy}\Upsilon (r,z,\varphi ).$ In
equivalent form, such solutions are characterized by polarizations of the
cosmological constant with curvature scalar $R(r,z,\varphi )\simeq \Lambda
(r,z,\varphi )$ which reflects a cylindrical type polarization of
gravitational vacuum, in general, with local anisotropy and more degrees of
freedom.

\subsection{Locally anisotropic wormholes}

Nonholonomic deformations of wormhole solutions to locally anisotropic were
studied in \cite{v13,v14}. Let us revise those constructions and generate
new classes of off-diagonal quasi-stationary solutions derived for primary
wormhole metrics.

\subsubsection{Prime metrics as Morris-Thorne and generalized
Ellis-Bronnikov wormholes}

The generic Morris-Thorne wormhole solution \cite{morris88} is defined by a
quadratic line element 
\begin{equation*}
d\mathring{s}^{2}=(1-\frac{b(r)}{r})^{-1}dr^{2}+r^{2}d\theta ^{2}+r^{2}\sin
^{2}\theta d\varphi ^{2}-e^{2\Phi (r)}dt^{2},
\end{equation*}%
where $e^{2\Phi (r)}$ is a red-shift function and $b(r)$ is the shape
function defined in spherically polar coordinates $u^{\alpha }=(r,\theta
,\varphi ,t).$ The usual Ellis-Bronnikov, EB, wormholes are defined for $%
\Phi (r)=0$ and $b(r)=\ _{0}b^{2}/r$ characterizing a zero tidal wormhole
with $\ _{0}b$ the throat radius. We cite \cite{kar94,roy20,souza22} for
details and a recent review of results and approaches. A generalized EB is
characterized additionally by even integers $2k$ (with $k=1,2,...$) when $%
r(l)=(l^{2k}+\ _{0}b^{2k})^{1/2k}$ is a proper radial distance coordinate
(tortoise) and the cylindrical angular coordinate $\phi \in \lbrack 0,2\pi )$
is called parallel. In such coordinates, $-\infty <l<\infty $ which is
diferent from the cylindrical radial coordinate $\rho ,$ when $0\leq \rho
<\infty .$ This allows us to define a prime metric 
\begin{equation*}
d\mathring{s}^{2}=dl^{2}+r^{2}(l)d\theta ^{2}+r^{2}(l)\sin ^{2}\theta
d\varphi ^{2}-dt^{2},
\end{equation*}%
when 
\begin{equation*}
dl^{2}=(1-\frac{b(r)}{r})^{-1}dr^{2}\mbox{ and }b(r)=r-r^{3(1-k)}(r^{2k}-\
_{0}b^{2k})^{(2-1/k))}.
\end{equation*}

We can avoid off-diagonal deformations with coordinate and frame coefficient
singularities, we can consider frame transforms to a parametrization with
trivial N-connection coefficients $\check{N}_{i}^{a}=$ $\check{N}%
_{i}^{a}(u^{\alpha }(l,\theta ,\varphi ,t))$ and $\check{g}_{\beta
}(u^{j}(l,\theta ,\varphi ),u^{3}(l,\theta ,\varphi )).$ For instance,
introducing new coordinates $u^{1}=x^{1}=l,u^{2}=\theta ,$ and $%
u^{3}=y^{3}=\varphi +\ ^{3}B(l,\theta ),u^{4}=y^{4}=t+\ ^{4}B(l,\theta ),$
when 
\begin{eqnarray*}
\mathbf{\check{e}}^{3} &=&d\varphi =du^{3}+\check{N}_{i}^{3}(l,\theta
)dx^{i}=du^{3}+\check{N}_{1}^{3}(l,\theta )dl+\check{N}_{2}^{3}(l,\theta
)d\theta , \\
\mathbf{\check{e}}^{4} &=&dt=du^{4}+\check{N}_{i}^{4}(l,\theta
)dx^{i}=du^{4}+\check{N}_{1}^{4}(l,\theta )dl+\check{N}_{2}^{4}(l,\theta
)d\theta ,
\end{eqnarray*}%
for $\mathring{N}_{i}^{3}=-\partial \ ^{3}B/\partial x^{i}$ and $\mathring{N}%
_{i}^{4}=-\partial \ ^{4}B/\partial x^{i}.$ Using such nonlinear
coordinates, the quadratic elements for above wormhole solutions can be
parameterized as a prime d-metric, 
\begin{equation}
d\mathring{s}^{2}=\check{g}_{\alpha }(l,\theta ,\varphi )[\mathbf{\check{e}}%
^{\alpha }(l,\theta ,\varphi )]^{2},  \label{pmwh}
\end{equation}%
where $\check{g}_{1}=1,\check{g}_{2}=r^{2}(l),\check{g}_{3}=r^{2}(l)\sin
^{2}\theta $ and $\check{g}_{4}=-1.$

\subsubsection{Nonholonomic quasi-stationary gravitational polarizations of
wormholes}

Off-diagonal quasi-stationary deformations of wormhole (\ref{pmwh}) are
generated by introducing nontrivial sources $\ _{1}\widehat{\Upsilon }%
(l,\theta )$ and $\ _{2}\widehat{\Upsilon }(l,\theta ,\varphi )=\ \
^{wh}\Upsilon $ related to nonlinear symmetries of type (\ref{nonlinsymrex})
to a nonzero (effective) cosmological constant $\Lambda .$ Using $\eta $%
-polarization functions, we derived such target quasi-stationary metrics 
\begin{eqnarray}
d\widehat{s}^{2} &=&\widehat{g}_{\alpha \beta }(l,\theta ,\varphi ;\psi
,\eta _{4};\ _{2}\Lambda =\Lambda ,\ \ ^{wh}\Upsilon ,\ \check{g}_{\alpha
})du^{\alpha }du^{\beta }  \notag \\
&=&e^{\psi l,\theta )}[(dx^{1}(l,\theta ))^{2}+(dx^{2}(l,\theta ))^{2}]
\label{whpolf} \\
&&-\frac{[\partial _{\varphi }(\eta _{4}\ \check{g}_{4})]^{2}}{|\int
d\varphi \ \ ^{wh}\Upsilon \partial _{\varphi }(\eta _{4}\ \breve{g}_{4})|\
\eta _{4}\ \check{g}_{4}}\{dy^{3}+\frac{\partial _{i}[\int d\varphi \ \
^{wh}\Upsilon \ \partial _{\varphi }(\eta _{4}\ \check{g}_{4})]}{\ \
^{wh}\Upsilon \partial _{\varphi }(\eta _{4}\ \check{g})}dx^{i}\}^{2}+\eta
_{4}\breve{g}_{4}  \notag \\
&&\{dt+[\ _{1}n_{k}(r,z)+\ _{2}n_{k}(r,z)\int d\varphi \frac{\lbrack
\partial _{\varphi }(\eta _{4}\ \breve{g}_{4})]^{2}}{|\int d\varphi \ \
^{wh}\Upsilon \partial _{\varphi }(\eta _{4}\ \breve{g}_{4})|\ (\eta _{4}\ 
\breve{g}_{4})^{5/2}}]dx^{k}\}.  \notag
\end{eqnarray}%
This class of solutions are determined by respective generating function $%
\eta _{4}=\eta _{4}(l,\theta ,\varphi )$ and integration functions $\
_{1}n_{k}(l,\theta )$ and $\ _{2}n_{k}(l,\theta ).$ The function $\psi
(l,\theta )$ is a solution of 2-d Poisson equation $\partial _{11}^{2}\psi
+\partial _{22}^{2}\psi =2\ \ _{1}\widehat{\Upsilon }(l,\theta ).$

The target d-metrics (\ref{whpolf}) do not describe wormhole like locally
anisotropic object for general classes of generating and integrating data.

\subsubsection{Small parametric off-diagonal quasi-stationary deformations
of wormhole d-metrics}

We can define locally anisotropic wormholes if we consider for small
parametric quasi-stationary deformation of prime metrics of type (\ref{pmwh}%
). In terms of $\chi $-polarization functions, the quadratic linear elements
are computed 
\begin{equation*}
d\ \widehat{s}^{2}=\widehat{g}_{\alpha \beta }(l,\theta ,\varphi ;\psi ,\eta
_{4};\ _{2}\Lambda =\Lambda ,\ ^{wh}\Upsilon ,\ \check{g}_{\alpha
})du^{\alpha }du^{\beta }=e^{\psi _{0}(l,\theta )}[1+\kappa \ ^{\psi
(l,\theta )}\chi (l,\theta )][(dx^{1}(l,\theta ))^{2}+(dx^{2}(l,\theta
))^{2}]
\end{equation*}%
\begin{eqnarray*}
&&-\{\frac{4[\partial _{\varphi }(|\zeta _{4}\ \breve{g}_{4}|^{1/2})]^{2}}{%
\breve{g}_{3}|\int d\varphi \{\ \ ^{wh}\Upsilon \partial _{\varphi }(\zeta
_{4}\ \breve{g}_{4})\}|}-\kappa \lbrack \frac{\partial _{\varphi }(\chi
_{4}|\zeta _{4}\breve{g}_{4}|^{1/2})}{4\partial _{\varphi }(|\zeta _{4}\ 
\breve{g}_{4}|^{1/2})}-\frac{\int d\varphi \{\ ^{wh}\Upsilon \partial
_{\varphi }[(\zeta _{4}\ \breve{g}_{4})\chi _{4}]\}}{\int d\varphi \{\
^{wh}\Upsilon \partial _{\varphi }(\zeta _{4}\ \breve{g}_{4})\}}]\}\ \breve{g%
}_{3} \\
&&\{d\varphi +[\frac{\partial _{i}\ \int d\varphi \ ^{wh}\Upsilon \ \partial
_{\varphi }\zeta _{4}}{(\check{N}_{i}^{3})\ \ ^{wh}\Upsilon \partial
_{\varphi }\zeta _{4}}+\kappa (\frac{\partial _{i}[\int d\varphi \ \
^{wh}\Upsilon \ \partial _{\varphi }(\zeta _{4}\chi _{4})]}{\partial _{i}\
[\int d\varphi \ \ ^{wh}\Upsilon \partial _{\varphi }\zeta _{4}]}-\frac{%
\partial _{\varphi }(\zeta _{4}\chi _{4})}{\partial _{\varphi }\zeta _{4}})]%
\check{N}_{i}^{3}dx^{i}\}^{2}
\end{eqnarray*}%
\begin{eqnarray}
&&+\zeta _{4}(1+\kappa \ \chi _{4})\ \breve{g}_{4}\{dt+[(\check{N}%
_{k}^{4})^{-1}[\ _{1}n_{k}+16\ _{2}n_{k}[\int d\varphi \frac{\left( \partial
_{\varphi }[(\zeta _{4}\ \breve{g}_{4})^{-1/4}]\right) ^{2}}{|\int d\varphi
\partial _{\varphi }[\ ^{wh}\Upsilon (\zeta _{4}\ \breve{g}_{4})]|}]  \notag
\\
&&+\kappa \frac{16\ _{2}n_{k}\int d\varphi \frac{\left( \partial _{\varphi
}[(\zeta _{4}\ \breve{g}_{4})^{-1/4}]\right) ^{2}}{|\int d\varphi \partial
_{\varphi }[\ ^{wh}\Upsilon (\zeta _{4}\ \breve{g}_{4})]|}(\frac{\partial
_{\varphi }[(\zeta _{4}\ \breve{g}_{4})^{-1/4}\chi _{4})]}{2\partial
_{\varphi }[(\zeta _{4}\ ^{cy}g)^{-1/4}]}+\frac{\int d\varphi \partial
_{\varphi }[\ ^{wh}\Upsilon (\zeta _{4}\chi _{4}\ \breve{g}_{4})]}{\int
d\varphi \partial _{\varphi }[\ ^{wh}\Upsilon (\zeta _{4}\ \breve{g}_{4})]})%
}{\ _{1}n_{k}+16\ _{2}n_{k}[\int d\varphi \frac{\left( \partial _{\varphi
}[(\zeta _{4}\ \breve{g}_{4})^{-1/4}]\right) ^{2}}{|\int d\varphi \partial
_{\varphi }[\ ^{wh}\Upsilon (\zeta _{4}\ \breve{g}_{4})]|}]}]\check{N}%
_{k}^{4}dx^{k}\}^{2}.  \label{whpolf1}
\end{eqnarray}

We can can model elliptic deformations as a particular case of d-metrics of
type (\ref{whpolf1}) if we chose a generating function of type%
\begin{equation*}
\chi _{4}(l,\theta ,\varphi )=\underline{\chi }(l,\theta )\sin (\omega
_{0}\varphi +\varphi _{0})
\end{equation*}%
as for cylindric configurations with $\varphi $-anisotropic deformations.
Here we note that a different family of solutions of type (\ref{whpolf})
and/or (\ref{whpolf1}) can be constructed if we change the order of angular
coordinates in the primary and target d-metrics, $\theta \leftrightarrow
\varphi .$ We omit details on such applications of the AFCDM.

\subsection{Nonholonomic toroid configurations and black torus, BT}

Nonholonomic deformations of toroidal BHs were studied in \cite{v01t,vs01b}.
They provided examples of generic off-diagonal deformations of black torus,
BT, and black ring generalizations in GR and MGTs \cite%
{lemos01,peca98,emparan02,emparan08}, see \cite{astorino17} for a recent
review of results. Such different classes of new solutions can be generated
for various types of nonholonomic distributions and nonlinear transforms. In
this subsection, we study an example when the AFCDM is applied for
generating quasi-stationary locally anisotropic solutions using prime BT
metrics analyzed in \cite{astorino17}. For simplicity, we shall consider
only small parametric deformations when the physical interpretation of new
classes of solutions is very similar to the holonomic/ diagonalizable metric
ansatz.

\subsubsection{Prime metrics for AdS BH with toroidal horizon}

Let us consider a quadratic line element (see details in section 3.1 of \cite%
{astorino17}) 
\begin{eqnarray}
d\tilde{s}^{2} &=&f^{-1}(\tilde{r})d\tilde{r}^{2}+\tilde{r}^{2}(\tilde{k}%
_{1}^{2}dx^{2}+\tilde{k}_{2}^{2}dy^{2})-f(\tilde{r})d\tilde{t}^{2}
\label{prmtor1} \\
&=&\tilde{g}_{\alpha }(\tilde{x}^{1})(d\tilde{u}^{\alpha })^{2},\mbox{ for }%
f(\tilde{r})=-\epsilon ^{2}b^{2}-\tilde{\mu}/\tilde{r}-\Lambda \tilde{r}%
^{2}/3.  \notag
\end{eqnarray}%
The coordinates in this metric $\tilde{g}=\{\tilde{g}_{\alpha }\}$ are are
related via rescaling parameter $\epsilon $ to standard toroidal
"normalized" coordinates, when $r$ is a radial coordinate, with $\theta
=2\pi k_{1}x$ and $\varphi =2\pi k_{2}y$ (when $x,y\in \lbrack 0,1]$) and
rescaling 
\begin{equation*}
k_{1}=\epsilon \tilde{k}_{1},k_{2}=\epsilon \tilde{k}_{2},\mu \rightarrow 
\frac{\mu }{(2\pi )^{3}}=\tilde{\mu}/\epsilon ^{3};r\rightarrow \frac{r}{%
2\pi }=\tilde{r}/\epsilon ,t\rightarrow 2\pi t=\epsilon \tilde{r}.
\end{equation*}%
In above formulas, the parameter $b$ is a coupling constant for the energy
momentum tensor for the nonlinear SU(2) sigma model, which is parameterized 
\begin{equation}
T_{\mu \nu }=\frac{b^{2}\epsilon ^{2}}{8\pi G\tilde{r}^{2}}[f(\tilde{r}%
)\delta _{\mu }^{4}\delta _{\nu }^{4}-f^{-1}(\tilde{r})\delta _{\mu
}^{1}\delta _{\nu }^{1}],  \label{emtsm}
\end{equation}%
and $\mu $ being an integration constant which can be fixed as a mass
parameter. The value $\epsilon =0$ allows to recover in a formal way certain
toroidal vacuum solution, for instance, from \cite{lemos01,peca98}. The
toroidal metric (\ref{prmtor1}) is an exact static solution of the Einstein
equations (\ref{en1}) for the LC-connection and energy-momentum tensor (\ref%
{emtsm}). It define an AdS BH with a toroidal horizon in 4-d Einstein
gravity and nonlinear $\sigma $-model.

We consider frame transforms to an off-diagonal parametrization of (\ref%
{prmtor1}) to a form with trivial N-connection coefficients $\tilde{N}%
_{i}^{a}=$ $\tilde{N}_{i}^{a}(u^{\alpha }(\tilde{r},x,y,t))$ and $\tilde{g}%
_{\alpha \beta }(u^{j}(\tilde{r},x,y),u^{3}(\tilde{r},x,y))$ which are
defined an any form which do not involve singular frame transforms and
off-diagonal deformations. Let us introduce new coordinates $u^{1}=x^{1}=%
\tilde{r},u^{2}=x,$ and $u^{3}=y^{3}=y+\ ^{3}B(\tilde{r},x),u^{4}=y^{4}=t+\
^{4}B(\tilde{r},x),$ when 
\begin{eqnarray*}
\mathbf{\tilde{e}}^{3} &=&dy=du^{3}+\tilde{N}_{i}^{3}(\tilde{r}%
,x)dx^{i}=du^{3}+\tilde{N}_{1}^{3}(\tilde{r},x)dr+\tilde{N}_{2}^{3}(\tilde{r}%
,x)dz, \\
\mathbf{\tilde{e}}^{4} &=&dt=du^{4}+\tilde{N}_{i}^{4}(\tilde{r}%
,x)dx^{i}=du^{4}+\tilde{N}_{1}^{4}(\tilde{r},x)dr+\tilde{N}_{2}^{4}(\tilde{r}%
,x)dz,
\end{eqnarray*}%
for $\tilde{N}_{i}^{3}=-\partial \ ^{3}B/\partial x^{i}$ and $\tilde{N}%
_{i}^{4}=-\partial \ ^{4}B/\partial x^{i}.$ In such nonlinear coordinates,
the diagonal metric (\ref{prmtor1}) transforms into a toroidal d-metric 
\begin{equation}
d\tilde{s}^{2}=\tilde{g}_{\alpha }(\tilde{r},x,y)[\mathbf{\tilde{e}}^{\alpha
}(\tilde{r},x,y)]^{2},  \label{prmtor2}
\end{equation}%
where $\tilde{g}_{1}=f^{-1}(x^{1}),\tilde{g}_{2}=(x^{1})^{2}\tilde{k}%
_{1}^{2},\tilde{g}_{3}=(x^{2})^{2}\tilde{k}_{2}^{2}$ and $\tilde{g}%
_{4}=f(x^{1}).$

\subsubsection{Small parametric off-diagonal quasi-stationary deformations
of toroidal d-metrics}

We can study locally anisotropic toroidal configurations if we construct
small parametric quasi-stationary deformations of prime metrics of type (\ref%
{prmtor2}) defined by an effective source $\ ^{tor}\mathbf{Y}[\mathbf{g,}%
\widehat{\mathbf{D}}]\simeq \{-\Lambda \mathbf{g}_{\alpha \beta }+\mathbf{T}%
_{\alpha \beta }\},$ for (\ref{emtsm}), see also (\ref{emdt}).\ The left
label "tor" will be used for toroidal configurations. Respective generating
sources (\ref{esourcqscan}) are parameterized where $\ _{1}^{tor}\Upsilon (%
\tilde{r},x)$ and $\ _{2}^{tar}\Upsilon (\tilde{r},x,y).$

Any quasi-stationary off-diagonal deformation (\ref{prmtor2}) to a class of
solutions of type of (\ref{qeltors}), (\ref{offdiagcosmcsh}), (\ref%
{offdsolgenfgcosmc}), (\ref{offdiagpolfr}) or (\ref{offdncelepsilon}) can be
characterized by corresponding nonlinear symmetries of type (\ref%
{nonlinsymrex}), 
\begin{eqnarray}
\partial _{y}[\Psi ^{2}(\tilde{r},x,y)] &=&-\int dy\ \ ^{tor}\Upsilon
\partial _{y}g_{4}\simeq -\int dy\ \ ^{tor}\Upsilon (\tilde{r},x,y)\partial
_{y}[\eta _{4}(\tilde{r},x,y)\ \tilde{g}_{4}(\tilde{r})]  \label{nsymtor} \\
&\simeq &-\int dy\ \ \ ^{tor}\Upsilon (\tilde{r},x,y)\partial _{y}[\zeta
_{4}(\tilde{r},x,y)(1+\kappa \ \chi _{4}(\tilde{r},x,y))\ \ \tilde{g}_{4}(%
\tilde{r})],  \notag \\
\Psi (\tilde{r},x,y) &=&|\ \ \Lambda +\ ^{tor}\Lambda |^{-1/2}\sqrt{|\int
dy\ \ \ ^{tor}\Upsilon (\tilde{r},x,y)\ \partial _{y}(\Phi ^{2})|},(\Phi (%
\tilde{r},x,y))^{2}=-4\ \Lambda \tilde{g}_{4}(\tilde{r},x,y)  \notag \\
&\simeq &-4\ (\ \Lambda +\ ^{tor}\Lambda )\eta _{4}(\tilde{r},x,y)\ \ \tilde{%
g}_{4}(\tilde{r})\simeq -4(\ \Lambda +\ ^{tor}\Lambda )\ \zeta _{4}(\tilde{r}%
,x,y)(1+\kappa \chi _{4}(\tilde{r},x,y))\ \tilde{g}_{4}(\tilde{r}).  \notag
\end{eqnarray}%
In these formulas, we use $\ ^{tor}\Lambda $ as an effective cosmological
constant to which the energy-momentum tensor (\ref{emtsm}) encoding
nonlinear sigma interactions can be related via nonlinear symmetries. In
general, such a $\ ^{tor}\Lambda $ is different from a prescribed
cosmological constant associated to other types of gravitational and matter
interactions. It is possible to elaborate on models with nonlinear
functionals $\widetilde{\Lambda }(\Lambda ,\ ^{tor}\Lambda ),$ with in this
subsection is approximated to a as $\widetilde{\Lambda }=\Lambda +\
^{tor}\Lambda .$

For parametric deformations in terms of $\chi $-polarization functions, the
quadratic linear elements for nonholonomic toroidal solutions are computed 
\begin{eqnarray*}
d\ \widehat{s}^{2} &=&\widehat{g}_{\alpha \beta }(\tilde{r},x,y;\psi ,\eta
_{4};\ _{2}\Lambda =\Lambda +\ ^{tor}\Lambda ,\ ^{tor}\Upsilon ,\ \tilde{g}%
_{\alpha })du^{\alpha }du^{\beta } \\
&=&e^{\psi _{0}(\tilde{r},x)}[1+\kappa \ ^{\psi (\tilde{r},x)}\chi (\tilde{r}%
,x)][(dx^{1}(\tilde{r},x))^{2}+(dx^{2}(\tilde{r},x))^{2}]-
\end{eqnarray*}%
\begin{eqnarray*}
&&\{\frac{4[\partial _{y}(|\zeta _{4}\ \tilde{g}_{4}|^{1/2})]^{2}}{\tilde{g}%
_{3}|\int dy\{\ ^{tor}\Upsilon \partial _{y}(\zeta _{4}\ \tilde{g}_{4})\}|}%
-\kappa \lbrack \frac{\partial _{y}(\chi _{4}|\zeta _{4}\tilde{g}_{4}|^{1/2})%
}{4\partial _{y}(|\zeta _{4}\tilde{g}_{4}|^{1/2})}-\frac{\int dy\{\
^{tor}\Upsilon \partial _{y}[(\zeta _{4}\ \tilde{g}_{4})\chi _{4}]\}}{\int
dy\{\ ^{tor}\Upsilon \partial _{y}(\zeta _{4}\ \tilde{g}_{4})\}}]\}\ \tilde{g%
}_{3} \\
&&\{dy+[\frac{\partial _{i}\ \int dy\ ^{tor}\Upsilon \ \partial _{y}\zeta
_{4}}{(\tilde{N}_{i}^{3})\ \ ^{tor}\Upsilon \partial _{y}\zeta _{4}}+\kappa (%
\frac{\partial _{i}[\int dy\ \ ^{tor}\Upsilon \ \partial _{y}(\zeta _{4}\chi
_{4})]}{\partial _{i}\ [\int dy\ \ ^{tor}\Upsilon \partial _{y}\zeta _{4}]}-%
\frac{\partial _{y}(\zeta _{4}\chi _{4})}{\partial _{y}\zeta _{4}})]\tilde{N}%
_{i}^{3}dx^{i}\}^{2}+
\end{eqnarray*}%
\begin{eqnarray}
&&\zeta _{4}(1+\kappa \ \chi _{4})\ \tilde{g}_{4}\{dt+[(\tilde{N}%
_{k}^{4})^{-1}[\ _{1}n_{k}+16\ _{2}n_{k}[\int dy\frac{\left( \partial
_{y}[(\zeta _{4}\ \tilde{g}_{4})^{-1/4}]\right) ^{2}}{|\int dy\partial
_{y}[\ ^{tor}\Upsilon (\zeta _{4}\ \tilde{g}_{4})]|}]+  \notag \\
&&\kappa \frac{16\ _{2}n_{k}\int dy\frac{\left( \partial _{y}[(\zeta _{4}\ 
\tilde{g}_{4})^{-1/4}]\right) ^{2}}{|\int dy\partial _{y}[\ ^{tor}\Upsilon
(\zeta _{4}\ \tilde{g}_{4})]|}(\frac{\partial _{y}[(\zeta _{4}\ \tilde{g}%
_{4})^{-1/4}\chi _{4})]}{2\partial _{y}[(\zeta _{4}\ \tilde{g}_{4})^{-1/4}]}+%
\frac{\int dy\partial _{y}[\ ^{tor}\Upsilon (\zeta _{4}\chi _{4}\ \tilde{g}%
_{4})]}{\int dy\partial _{y}[\ ^{tor}\Upsilon (\zeta _{4}\ \tilde{g}_{4})]})%
}{\ _{1}n_{k}+16\ _{2}n_{k}[\int dy\frac{\left( \partial _{y}[(\zeta _{4}\ 
\tilde{g}_{4})^{-1/4}]\right) ^{2}}{|\int dy\partial _{y}[\ ^{tor}\Upsilon
(\zeta _{4}\ \tilde{g}_{4})]|}]}]\tilde{N}_{k}^{4}dx^{k}\}^{2}.
\label{tortpol2}
\end{eqnarray}%
We can use this formula in order to model elliptic deformations if we chose
a generating function of type%
\begin{equation*}
\chi _{4}(\tilde{r},x,y)=\underline{\chi }(\tilde{r},x)\sin (\omega
_{0}y+y_{0}).
\end{equation*}%
This defines a family of toroid configurations with ellipsoidal deformations
on $y$ coordinate. Above formulas can be re-defined for a different family
of quasi-stationary solutions with off-diagonal deformations on $x$%
-coordinate if we change the order of coordinates $x\leftrightarrow y.$ In
such cases, the formulas of type (\ref{nsymtor}) and (\ref{tortpol2})
involve derivatives and integrals on $x,$ with a different order of
spacetime coordinates, $u^{\alpha }(\tilde{r},y,x,t),$ when $y$ is
considered as a h-coordinate and $x$ as a v-coordinates.

Applying the AFCDM to (\ref{prmtor2}) we can construct more general classes
of off-diagonal quasi-stationary deformations of the prime toroid
configurations, which are determined by certain $\eta $-deformations and
d-metrics of type (\ref{offdiagpolfr}). For instance, we can consider a
generating function $\eta _{4}(\tilde{r},x,y)$ and take instead of nonlinear
sigma energy-momentum tensor (\ref{emtsm}) more general types of effective
matter sources $\ _{1}\Upsilon (\tilde{r},x)$ and $\ _{2}\Upsilon (\tilde{r}%
,x,y)$ parameterized in toroid coordinates. The physical interpretation of
such more general classes of generic off-diagonal solutions depend of the
type of generating functions and generating sources we consider.
Nevertheless, they are always characterized by nonlinear symmetries of type (%
\ref{nsymtor}).

Finally, we note that we can consider that various classes of nonholonomic
deformations transform a toroid prime d-metric into "spagetti"
quasi-stationary configurations (with different sections, curved and waved,
possible interruptions, singularities etc.) embedded into locally
anisotropic gravitational vacuum media. The geometry of such d-objects is
determined by respective prime metrics and prescribed generating functions
and sources and integration functions and assumptions on nonlinear
symmetries of off-diagonal gravitational and matter field interactions. The
physical meaning of such models should be determined/ analyzed for
corresponding types of geometric data and boundary/asymptotic conditions we
state for a corresponding family of solutions.

\subsection{Nonholonomic BT and BE configurations}

Using the AFCDM, generic off-diagonal quasi-stationary solutions describing
systems of black torus, BT, and black ellipsoid, BE, configurations we
constructed in 2001 \cite{v01q}. Similar classes of solutions constructed by
different methods and describing so-called "Black Saturn" were constructed
beginning 2006 \cite{elvang07,evslin08,yazadjiev08}. The goal of this
subsection is to show how the toroidal d-metric (\ref{tortpol2}) can be
generalized in such a form that for well defined conditions describes
families of BT-BE configurations derived from a primary d-metric stating a
Schwarzschild - (anti) de Sitter, A(d)S, metric imbedded into interior of a
BT.

\subsubsection{Prime metrics for systems of AdS BH with toroidal horizon \&
Schwarzschild - (a)dS BH}

Let us consider a primary metric 
\begin{eqnarray}
d\grave{s}^{2} &=&\grave{f}^{-1}(\tilde{r})d\tilde{r}^{2}+\tilde{r}^{2}(%
\tilde{k}_{1}^{2}dx^{2}+\tilde{k}_{2}^{2}dy^{2})-\grave{f}(\tilde{r})d\tilde{%
t}^{2}  \label{prmtbe1} \\
&=&\grave{g}_{\alpha }(\tilde{x}^{1})(d\tilde{u}^{\alpha })^{2},\mbox{ for }%
\grave{f}(\tilde{r})=1-\tilde{\mu}_{s}/\tilde{r}-\epsilon ^{2}b^{2}-\tilde{%
\mu}/\tilde{r}-\Lambda \tilde{r}^{2}/3,  \notag
\end{eqnarray}%
where the local coordinates a labeled as in the toroidal metric (\ref%
{prmtor1}). In this formula, $\grave{f}(\tilde{r})$ is different from $f(%
\tilde{r})$ because it contains an additional term, $1-\tilde{\mu}_{s}/%
\tilde{r},$ when $\tilde{\mu}_{s}<$ $\tilde{\mu}$ is chosen in such a form
that $1-\tilde{\mu}_{s}/\tilde{r}=0$ describes a conventional horizon in the
interior of a torus configurations when both the spherical and toroidal
objects have the same planar and axial symmetry. In principle, a metric (\ref%
{prmtbe1}) may be not a solution of Einstein equations in GR but we shall
search for quasi-stationary off-diagonal deformations, $\grave{g}_{\alpha }(%
\tilde{x}^{1})\rightarrow g_{\alpha \beta }(u^{\gamma }(\tilde{u}^{\delta
})),$ which are exact/parametric solutions of modified gravitational
equations (\ref{eq1})-(\ref{e2c}).

We re-write (\ref{prmtbe1}) in curved coordinates in a form with trivial
N-connection coefficients \newline
$\grave{N}_{i}^{a}=\grave{N}_{i}^{a}(u^{\alpha }(\tilde{r},x,y,t))$ and $%
\tilde{g}_{\alpha \beta }(u^{j}(\tilde{r},x,y),u^{3}(\tilde{r},x,y))$ which
are defined an any form which do not involve singular frame transforms and
off-diagonal deformations. Such new coordinates are defined $u^{1}=x^{1}=%
\tilde{r},u^{2}=x,$ and $u^{3}=y^{3}=y+\ ^{3}B(\tilde{r},x),u^{4}=y^{4}=t+\
^{4}B(\tilde{r},x),$ when 
\begin{eqnarray*}
\mathbf{\grave{e}}^{3} &=&dy=du^{3}+\grave{N}_{i}^{3}(\tilde{r}%
,x)dx^{i}=du^{3}+\grave{N}_{1}^{3}(\tilde{r},x)dr+\grave{N}_{2}^{3}(\tilde{r}%
,x)dz, \\
\mathbf{\grave{e}}^{4} &=&dt=du^{4}+\grave{N}_{i}^{4}(\tilde{r}%
,x)dx^{i}=du^{4}+\grave{N}_{1}^{4}(\tilde{r},x)dr+\grave{N}_{2}^{4}(\tilde{r}%
,x)dz,
\end{eqnarray*}%
for $\grave{N}_{i}^{3}=-\partial \ ^{3}B/\partial x^{i}$ and $\grave{N}%
_{i}^{4}=-\partial \ ^{4}B/\partial x^{i}.$ In such nonlinear coordinates,
we obtain an off-diagonal toroid-spheroid type metric, equivalently a
respective d-metric, 
\begin{equation}
d\grave{s}^{2}=\grave{g}_{\alpha }(\tilde{r},x,y)[\mathbf{\grave{e}}^{\alpha
}(\tilde{r},x,y)]^{2},  \label{prmtbe2}
\end{equation}%
where $\grave{g}_{1}=\grave{f}^{-1}(x^{1}),\grave{g}_{2}=(x^{1})^{2}\tilde{k}%
_{1}^{2},\grave{g}_{3}=(x^{2})^{2}\tilde{k}_{2}^{2}$ and $\grave{g}_{4}=%
\grave{f}(x^{1}).$

\subsubsection{Small parametric off-diagonal quasi-stationary deformations
of toroidal-rotoid d-metrics}

We show how to generate locally anisotropic toroid - rotoid configurations
if we construct small parametric quasi-stationary deformations of a prime
d-metric (\ref{prmtor2}) defined by an effective source $\ ^{tor}\mathbf{Y}[%
\mathbf{g,}\widehat{\mathbf{D}}]\simeq \{-\Lambda \mathbf{g}_{\alpha \beta }+%
\mathbf{T}_{\alpha \beta }\}$.\ The left label "tor" will be used for
toroidal configurations of (effective) matter, when, for simplicity, the
constant $\Lambda $ is associated to Schwarzschild - (a) dS BH. Respective
generating sources (\ref{esourcqscan}) are parameterized as in previous
subsection, $\ _{1}^{tor}\Upsilon (\tilde{r},x)$ and $\ _{2}^{tar}\Upsilon (%
\tilde{r},x,y).$

For quasi-stationary off-diagonal deformations of (\ref{prmtbe2}), we search
for solutions described by nonlinear symmetries of type (\ref{nonlinsymrex}%
), 
\begin{eqnarray}
\partial _{y}[\Psi ^{2}(\tilde{r},x,y)] &=&-\int dy\ \ ^{tor}\Upsilon
\partial _{y}g_{4}\simeq -\int dy\ \ ^{tor}\Upsilon (\tilde{r},x,y)\partial
_{y}[\eta _{4}(\tilde{r},x,y)\ \grave{g}_{4}(\tilde{r})]  \notag \\
&\simeq &-\int dy\ \ \ ^{tor}\Upsilon (\tilde{r},x,y)\partial _{y}[\zeta
_{4}(\tilde{r},x,y)(1+\kappa \ \chi _{4}(\tilde{r},x,y))\ \grave{g}_{4}(%
\tilde{r})],  \label{nsymtorot} \\
\Psi (\tilde{r},x,y) &=&|\ \ \Lambda +\ ^{tor}\Lambda |^{-1/2}\sqrt{|\int
dy\ \ \ ^{tor}\Upsilon (\tilde{r},x,y)\ \partial _{y}(\Phi ^{2})|},(\Phi (%
\tilde{r},x,y))^{2}=-4\ \Lambda \grave{g}_{4}(\tilde{r},x,y)  \notag \\
&\simeq &-4\ (\ \Lambda +\ ^{tor}\Lambda )\eta _{4}(\tilde{r},x,y)\ \ \tilde{%
g}_{4}(\tilde{r})\simeq -4(\ \Lambda +\ ^{tor}\Lambda )\ \zeta _{4}(\tilde{r}%
,x,y)(1+\kappa \chi _{4}(\tilde{r},x,y))\ \grave{g}_{4}(\tilde{r}).  \notag
\end{eqnarray}%
These formulas are similar to (\ref{nsymtor}) but with different prime
metrics which means that $\ \grave{g}_{4}(\tilde{r})$ is different from $%
\tilde{g}_{4}(\tilde{r}).$

New classes of quasi-stationary off-diagonal toroid- rotoid solutions can be
generated by $\chi $-polarization functions when the quadratic linear
elements are computed 
\begin{eqnarray*}
d\ \widehat{s}^{2} &=&\widehat{g}_{\alpha \beta }(\tilde{r},x,y;\psi ,\eta
_{4};\ _{2}\Lambda =\Lambda +\ ^{tor}\Lambda ,\ ^{tor}\Upsilon ,\ \grave{g}%
_{\alpha })du^{\alpha }du^{\beta } \\
&=&e^{\psi _{0}(\tilde{r},x)}[1+\kappa \ ^{\psi (\tilde{r},x)}\chi (\tilde{r}%
,x)][(dx^{1}(\tilde{r},x))^{2}+(dx^{2}(\tilde{r},x))^{2}]-
\end{eqnarray*}%
\begin{eqnarray*}
&&\{\frac{4[\partial _{y}(|\zeta _{4}\ \grave{g}_{4}|^{1/2})]^{2}}{\grave{g}%
_{3}|\int dy\{\ ^{tor}\Upsilon \partial _{y}(\zeta _{4}\ \grave{g}_{4})\}|}%
-\kappa \lbrack \frac{\partial _{y}(\chi _{4}|\zeta _{4}\grave{g}_{4}|^{1/2})%
}{4\partial _{y}(|\zeta _{4}\grave{g}_{4}|^{1/2})}-\frac{\int dy\{\
^{tor}\Upsilon \partial _{y}[(\zeta _{4}\ \grave{g}_{4})\chi _{4}]\}}{\int
dy\{\ ^{tor}\Upsilon \partial _{y}(\zeta _{4}\ \tilde{g}_{4})\}}]\}\grave{g}%
_{3} \\
&&\{dy+[\frac{\partial _{i}\ \int dy\ ^{tor}\Upsilon \ \partial _{y}\zeta
_{4}}{(\grave{N}_{i}^{3})\ \ ^{tor}\Upsilon \partial _{y}\zeta _{4}}+\kappa (%
\frac{\partial _{i}[\int dy\ \ ^{tor}\Upsilon \ \partial _{y}(\zeta _{4}\chi
_{4})]}{\partial _{i}\ [\int dy\ \ ^{tor}\Upsilon \partial _{y}\zeta _{4}]}-%
\frac{\partial _{y}(\zeta _{4}\chi _{4})}{\partial _{y}\zeta _{4}})]\grave{N}%
_{i}^{3}dx^{i}\}^{2}+
\end{eqnarray*}%
\begin{eqnarray}
&&\zeta _{4}(1+\kappa \ \chi _{4})\ \grave{g}_{4}\{dt+[(\grave{N}%
_{k}^{4})^{-1}[\ _{1}n_{k}+16\ _{2}n_{k}[\int dy\frac{\left( \partial
_{y}[(\zeta _{4}\ \grave{g}_{4})^{-1/4}]\right) ^{2}}{|\int dy\partial
_{y}[\ ^{tor}\Upsilon (\zeta _{4}\ \grave{g}_{4})]|}]+  \notag \\
&&\kappa \frac{16\ _{2}n_{k}\int dy\frac{\left( \partial _{y}[(\zeta _{4}\ 
\grave{g}_{4})^{-1/4}]\right) ^{2}}{|\int dy\partial _{y}[\ ^{tor}\Upsilon
(\zeta _{4}\ \grave{g}_{4})]|}(\frac{\partial _{y}[(\zeta _{4}\ \grave{g}%
_{4})^{-1/4}\chi _{4})]}{2\partial _{y}[(\zeta _{4}\ \grave{g}_{4})^{-1/4}]}+%
\frac{\int dy\partial _{y}[\ ^{tor}\Upsilon (\zeta _{4}\chi _{4}\ \grave{g}%
_{4})]}{\int dy\partial _{y}[\ ^{tor}\Upsilon (\zeta _{4}\ \grave{g}_{4})]})%
}{\ _{1}n_{k}+16\ _{2}n_{k}[\int dy\frac{\left( \partial _{y}[(\zeta _{4}\ 
\grave{g}_{4})^{-1/4}]\right) ^{2}}{|\int dy\partial _{y}[\ ^{tor}\Upsilon
(\zeta _{4}\ \grave{g}_{4})]|}]}]\grave{N}_{k}^{4}dx^{k}\}^{2}.
\label{torotpol2}
\end{eqnarray}%
These formulas are similar to (\ref{tortpol2}) but involve an additional
spheroid configuration centered inside a toroid one. In general, they are
with local anisotropic and deformed from the "perfect" torus-spherical
structure. We can use (\ref{torotpol2}) in order to model different types of
elliptic deformations. For instance, if we chose a generating function of
type%
\begin{equation*}
\chi _{4}(\tilde{r},x,y)=\underline{\chi }(\tilde{r},x)\sin (\omega
_{0}y+y_{0})
\end{equation*}%
when it performs rotoid deformations for the torus part. Another one, with
other type effective constants and $\underline{\chi }$ can be used rotoid
deformations of the Schwarzschild - (a) dS BH. For more sophisticate
constructions, we can generate prolate/oblate deformations etc. It depends
on the type of generating and integration functions we prescribe. We can
construct exact/parametric solutions with rotoid deformations on $y$
coordinate, and another type of deformations on $x$ coordinate.

\subsection{Nonholonomic cosmological solitonic and spheroid deformations
involving 2-d vertices}

In this subsection, we provide some explicit examples of locally anisotropic
cosmological solutions (\ref{qeltorsc}) and their equivalents with
gravitational $\eta $- and $\chi $-polarizations depending on a time like
coordinate. Such solutions can be generic off-diagonal and characterized by
respective nonlinear symmetries.

\subsubsection{Prime cosmological models with spheroidal symmetry and voids}

The Minkowski spacetime can be written in \textbf{prolate} spheroidal
coordinates $u^{\alpha }=(r,\theta ,\phi ,t),$ when the usual Cartezian
coordinates $u^{\alpha }=(x,y,z,t)$ are defined 
\begin{equation*}
x=r\sin \theta \cos \phi ,y=r\sin \theta \sin \phi ,z=\sqrt{%
r^{2}+r_{\lozenge }^{2}}\cos \theta ,
\end{equation*}%
where the constant parameter $r_{\lozenge }$ has the meaning of the distance
of the foci from the origin of the coordinate system. For any fixed $r=\
_{0}r,$ such coordinates define a prolate spheroid (rotoid/ ellipsoid) with
the foci along the $z$-axis, when 
\begin{equation*}
\frac{x^{2}+y^{2}}{(\ _{0}r)^{2}}+\frac{z^{2}}{(\ _{0}r)^{2}+r_{\lozenge
}^{2}}=1,
\end{equation*}%
where $\ _{0}r$ correspons to the length of its \ minor radius and the size
of its major radious is $\sqrt{(\ _{0}r)^{2}+r_{\lozenge }^{2}}.$ The flat
Minkowski spacetime metric can be written in such prolate coordinates:%
\begin{equation*}
ds^{2}=(r^{2}+r_{\lozenge }^{2}\sin ^{2}\theta )(\frac{dr^{2}}{%
r^{2}+r_{\lozenge }^{2}}+d\theta ^{2})+r^{2}\sin ^{2}\theta d\phi -dt^{2}.
\end{equation*}

In a similar form, we can introduce \textbf{oblate} coordinates, when 
\begin{equation*}
x=\sqrt{r^{2}+r_{\lozenge }^{2}}\sin \theta \cos \phi ,y=\sqrt{%
r^{2}+r_{\lozenge }^{2}}\sin \theta \sin \phi ,z=r\cos \theta ,
\end{equation*}%
which for fixed $r=\ _{0}r,$ there is defined an oblate spheroid with a $z$
symmetric axis 
\begin{equation*}
\frac{x^{2}+y^{2}}{(\ _{0}r)^{2}+r_{\lozenge }^{2}}+\frac{z^{2}}{(\
_{0}r)^{2}}=1.
\end{equation*}%
In this hypersurface formula, the value $\sqrt{r^{2}+r_{\lozenge }^{2}}$
corresponds to the major radius and $_{0}r$ is the minor one. In oblate
coordinates, the flat Minkowski spacetime metric is written 
\begin{equation*}
ds^{2}=(r^{2}+r_{\lozenge }^{2}\cos ^{2}\theta )(\frac{dr^{2}}{%
r^{2}+r_{\lozenge }^{2}}+d\theta ^{2})+r^{2}\sin ^{2}\theta d\phi -dt^{2}.
\end{equation*}

In \cite{boero16}, it was proposed that the cosmology of voids in 4-d
gravity theories can be described by such quadratic line elements (we
underline certain symbols and follow our system of notations in order to
emphasize that we study locally anisotropic cosmological configurations):%
\begin{eqnarray}
d\underline{s}^{2} &=&\frac{a^{2}(t)}{[1+\frac{\varsigma }{4}%
(r^{2}+r_{\lozenge }^{2}\cos ^{2}\theta )]^{2}}[(r^{2}+r_{\lozenge }^{2}\sin
^{2}\theta )(\frac{dr^{2}}{r^{2}-\frac{M(r)}{r}(r^{2}+r_{\lozenge }^{2}\sin
^{2}\theta )+r_{\lozenge }^{2}}+d\theta ^{2})  \notag \\
&&+r^{2}\sin ^{2}\theta d\phi ]-B(r)dt^{2},%
\mbox{ with prolate spheroidal
symmetry};  \label{cosmvoidm1} \\
d\underline{s}^{2} &=&\frac{a^{2}(t)}{[1+\frac{\varsigma }{4}%
(r^{2}+r_{\lozenge }^{2}\sin ^{2}\theta )]^{2}}[(r^{2}+r_{\lozenge }^{2}\sin
^{2}\theta )(\frac{dr^{2}}{r^{2}-\frac{M(r)}{r}(r^{2}+r_{\lozenge }^{2}\cos
^{2}\theta )+r_{\lozenge }^{2}}+d\theta ^{2})  \notag \\
&&+(r^{2}+r_{\lozenge }^{2})\sin ^{2}\theta d\phi ]-B(r)dt^{2},%
\mbox{
with oblate spheroidal symmetry}.  \notag
\end{eqnarray}%
For $B(r)=1$ and $M(r)=0,$ these formulas define FLRW cosmological quadratic
line elements (in respective prolate/oblate coordinates), where $\varsigma
=1,0,-1$ refer respectively to a positive curved, flat, hyperbolic spacial
geometry. We discussed some details with respect to formulas (\ref{flrw})
and (\ref{fluidm}).

The mass profile function $M(r)$ from (\ref{cosmvoidm1}) can be specified as
in \cite{amendola98} (in a simple choice, one states $B(r)=1$),%
\begin{equation*}
M(r)=\left\{ 
\begin{array}{cc}
\frac{4\pi }{3}\rho _{int}r^{3}, & \mbox{ for }r<\ _{v}r; \\ 
M(\ _{v}r)+\frac{4\pi }{3}\rho _{bor}(r^{3}-\ _{v}r^{3}), & \mbox{ for }\
_{v}r\leq r<\ _{v}r+\ _{w}r; \\ 
0 & \mbox{ for }\ _{v}r+\ _{w}r\leq r.%
\end{array}%
\right.
\end{equation*}%
In these formulas, $\ _{v}r$ is associated with the radius of the void, and
the parameter $\ _{w}r$ is related to the size of wall. For spherical
symmetry, such a profile is modelled in a form that the border compensates
for the amount missing in the void (i.e. it models a compensated void). The
respective internal density of the matter, $\rho _{int},$ and border density
of matter, $\rho _{bor},$ are related to the mean density outside the void, $%
\rho _{0},$ using formulas%
\begin{equation}
\rho _{int}=-\rho _{0}\xi \mbox{ and }\rho _{bor}=\rho _{0}\xi /[(1+\
_{w}r/\ _{v}r)^{3}-1],  \label{emtvoid}
\end{equation}%
for a constant parameter $\xi <1.$ A cosmological metric (\ref{cosmvoidm1})
is a solution of the Einstein equations in GR if $a(t)$ is a solution of the
Friedman equations%
\begin{equation*}
\frac{3}{a^{2}(t)}[\frac{da}{dt}+\varsigma ]=8\pi \rho _{0}.
\end{equation*}%
To describe the characteristic phenomenology observed in astrophysical
systems with dark matter \cite{galo12}, the function $B(r)$ can be
parameterized in the form%
\begin{equation*}
B(r)=B_{0}[B_{1}+\ln (\frac{r}{r_{\lozenge }})]^{2},
\end{equation*}%
for some constants $B_{0}$ and $B_{1}.$ The value of $B_{1}$ can be fixed in
a form that the component $T_{r}^{r}=T_{1}^{1}$ of the energy momentum
tensor remains of the same order as $\rho _{0}$ (they fix $B_{1}=10^{7}$).
Other phenomenological parameters are typically stated $\ _{w}r=0.3\ _{v}r,$ 
$\xi =0.1,$ $r_{\lozenge }=0.1\ _{v}r$ when a radius $\ _{v}r$ corresponds
to a physical size of 22Mpc.

To apply the AFCDM we have to re-write (\ref{cosmvoidm1}) in curved
coordinates in a form with trivial N-connection coefficients $\underline{%
\mathring{N}}_{i}^{a}=$ $\underline{\mathring{N}}_{i}^{a}(u^{\alpha
}(r,\theta ,\phi ,t))$ and \underline{$\mathring{g}$}$_{\alpha \beta
}(u^{j}(r,\theta ,\phi ,t),u^{4}(r,\theta ,\phi ,t))$ which are defined an
any form which do not involve singular frame transforms and off-diagonal
deformations. Such new coordinates are defined $u^{1}=x^{1}=r,u^{2}=\theta ,$
and $u^{3}=y^{3}=y^{3}(r,\theta ,\phi )$ and $u^{4}=y^{4}=t+\ ^{4}B(r,\theta
),$ when 
\begin{eqnarray*}
\underline{\mathbf{\mathring{e}}}^{3} &=&du^{3}+\underline{\mathring{N}}%
_{i}^{3}(r,\theta )dx^{i}=du^{3}+\underline{\mathring{N}}_{1}^{3}(r,\theta
)dr+\underline{\mathring{N}}_{2}^{3}(r,\theta )d\theta , \\
\underline{\mathbf{\mathring{e}}}^{4} &=&du^{4}+\underline{\mathring{N}}%
_{i}^{4}(r,\theta )dx^{i}=du^{4}+\underline{\mathring{N}}_{1}^{4}(r,\theta
)dr+\underline{\mathring{N}}_{2}^{4}(r,\theta )dz,
\end{eqnarray*}%
for $\underline{\mathring{N}}_{i}^{3}=-\partial \ y^{3}/\partial x^{i}$ and $%
\underline{\mathring{N}}_{i}^{4}=-\partial \ ^{4}B/\partial x^{i}.$ In such
nonlinear coordinates, we obtain an off-diagonal spheroid type cosmological
metric parameterized as a d-metric, 
\begin{eqnarray}
d\underline{\mathring{s}}^{2} &=&\underline{\mathring{g}}_{\alpha }(r,\theta
,t)[\underline{\mathbf{\mathring{e}}}^{\alpha }(r,\theta ,t)]^{2},%
\mbox{
where for }\left\{ 
\begin{array}{c}
\mbox{ prolate }: \\ 
\mbox{ oblate }:%
\end{array}%
\right.  \label{cosmvoidm2} \\
\underline{\mathring{g}}_{1}(r,\theta ,t) &=&\left\{ 
\begin{array}{c}
\frac{a^{2}(t)(r^{2}+r_{\lozenge }^{2}\sin ^{2}\theta )}{[1+\frac{\varsigma 
}{4}(r^{2}+r_{\lozenge }^{2}\cos ^{2}\theta )]^{2}[r^{2}-\frac{M(r)}{r}%
(r^{2}+r_{\lozenge }^{2}\sin ^{2}\theta )+r_{\lozenge }^{2}]} \\ 
\frac{a^{2}(t)(r^{2}+r_{\lozenge }^{2}\sin ^{2}\theta )}{[1+\frac{\varsigma 
}{4}(r^{2}+r_{\lozenge }^{2}\sin ^{2}\theta )]^{2}[r^{2}-\frac{M(r)}{r}%
(r^{2}+r_{\lozenge }^{2}\cos ^{2}\theta )+r_{\lozenge }^{2}]}%
\end{array}%
\right. ,  \notag \\
\underline{\mathring{g}}_{2}(r,\theta ,t) &=&\left\{ 
\begin{array}{c}
\frac{a^{2}(t)}{[1+\frac{\varsigma }{4}(r^{2}+r_{\lozenge }^{2}\cos
^{2}\theta )]^{2}} \\ 
\frac{a^{2}(t)}{[1+\frac{\varsigma }{4}(r^{2}+r_{\lozenge }^{2}\sin
^{2}\theta )]^{2}}%
\end{array}%
\right. ,\ \underline{\mathring{g}}_{3}(r,\theta ,t)=\left\{ 
\begin{array}{c}
\frac{a^{2}(t)r^{2}\sin ^{2}\theta }{[1+\frac{\varsigma }{4}%
(r^{2}+r_{\lozenge }^{2}\cos ^{2}\theta )]^{2}} \\ 
\frac{a^{2}(t)(r^{2}+r_{\lozenge }^{2})\sin ^{2}\theta }{[1+\frac{\varsigma 
}{4}(r^{2}+r_{\lozenge }^{2}\sin ^{2}\theta )]^{2}}%
\end{array}%
\right. ,\underline{\ \mathring{g}}_{4}(r)=-B(r).  \notag
\end{eqnarray}%
Such a prime cosmological metric can be nonholonomically deformed using
gravitational $\eta $-polarization functions in order to generate other
classes of exact and parametric solutions of nonholonomic Einstein equations
(\ref{cdeq1}) constructed as locally anisotropic cosmological d-metrics (\ref%
{dmc}).

\subsubsection{Off-diagonal cosmological solitonic evolution encoding 2-d
vertices}

We consider nonholonomic deformations of data 
\begin{equation*}
(\underline{\mathring{g}}_{\alpha },\underline{\mathring{N}}%
_{i}^{a})\rightarrow (\underline{g}_{\alpha }=\underline{\eta }_{\alpha }%
\underline{\mathring{g}}_{\alpha },\underline{N}_{i}^{a}=\underline{\eta }%
_{i}^{a}\underline{\mathring{N}}_{i}^{a}),
\end{equation*}%
where $\underline{\eta }_{i}(r,\theta ,t)=a^{-2}(t)\eta _{i}(r,\theta ),%
\underline{\eta }_{3}(r,\theta ,t)=a^{-2}(t)\underline{\eta }(r,\theta ,t)$
and $\underline{\eta }_{4}(r,\theta ,t)$ will be prescribed/computed in such
forms that 
\begin{eqnarray}
\underline{\mathbf{g}} &=&(g_{i},g_{b},\underline{N}_{i}^{3}=\underline{n}%
_{i},\underline{N}_{i}^{4}=\underline{w}_{i})  \label{lacosm1} \\
&=&g_{i}(r,\theta )dx^{i}\otimes dx^{i}+\underline{h}_{3}(r,\theta ,t)%
\underline{\mathbf{e}}^{3}\otimes \underline{\mathbf{e}}^{3}+\underline{h}%
_{4}(r,\theta ,t)\underline{\mathbf{e}}^{4}\otimes \underline{\mathbf{e}}%
^{4},  \notag \\
&&\underline{\mathbf{e}}^{3}=d\phi +\underline{n}_{i}(r,\theta ,t)dx^{i},\ 
\underline{\mathbf{e}}^{4}=dt+\underline{w}_{i}(r,\theta ,t)dx^{i},  \notag
\end{eqnarray}%
with Killing symmetry on the angular coordinate $\varphi ,$ when $\partial
_{\varphi }$ transforms into zero the N-adaped coefficients of such a
d-metric.

In terms of $\eta $-polarization functions, a (\ref{cdeq1}) can be written
in a $t$-dual form to (\ref{offdiagpolfr}) as we explain in section \ref%
{ssstdual}, when 
\begin{eqnarray}
d\widehat{s}^{2} &=&\widehat{g}_{\alpha \beta }(r,\theta ,t;\underline{%
\mathring{g}}_{\alpha };\psi ,\eta _{3};\ _{2}\underline{\Lambda },\ _{2}%
\underline{\Upsilon })du^{\alpha }du^{\beta }=e^{\psi
}[(dx^{1})^{2}+(dx^{2})^{2}]  \label{lacosm2} \\
&&+(\underline{\eta }\underline{\mathring{g}}_{3})\{d\phi +[\ _{1}n_{k}+\
_{2}n_{k}\int dt\frac{[\partial _{t}(\underline{\eta }\underline{\mathring{g}%
}_{3})]^{2}}{|\int dt\ _{2}\underline{\Upsilon }\partial _{t}(\underline{%
\eta }\underline{\mathring{g}}_{3})|\ (\underline{\eta }\underline{\mathring{%
g}}_{3})^{5/2}}]dx^{k}\}^{2}  \notag \\
&&-\frac{[\partial _{t}(\underline{\eta }\ \underline{\mathring{g}}_{3})]^{2}%
}{|\int dt\ _{2}\underline{\Upsilon }\partial _{t}(\underline{\eta }%
\underline{\mathring{g}}_{3})|\ \eta \mathring{g}_{3}}\{dt+\frac{\partial
_{i}[\int dt\ _{2}\underline{\Upsilon }\ \partial _{t}(\underline{\eta }%
\underline{\mathring{g}}_{3})]}{\ _{2}\underline{\Upsilon }\partial _{t}(%
\underline{\eta }\underline{\mathring{g}}_{3})}dx^{i}\}^{2}.  \notag
\end{eqnarray}%
For $\underline{\Phi }^{2}=-4\ _{2}\underline{\Lambda }\underline{g}_{4},$
we can transform (\ref{lacosm2}) in a variant of (\ref{qeltorsc}) with $\eta 
$-polarizations determined by the generating data $(\underline{g}_{4};\ _{2}%
\underline{\Lambda },\ _{2}\underline{\Upsilon }).$ The effective
cosmological constant $\ _{2}\underline{\Lambda }$ is chosen as effective
ones which correspond via nonlinear symmetries (\ref{nonlinsymcosm}) to a
energy-momentum tensor (\ref{emtvoid}) in a fluid type form (\ref{fluidm})
(when respective data $(\ _{1}\underline{\Upsilon },\ _{2}\underline{%
\Upsilon })$ are related to a $T_{\alpha \beta }$ via respective
frame/coordinate transforms). We can model certain locally anisotropic
cosmological scenarios which can evolve from a primary void configuration (%
\ref{cosmvoidm2}) being determined by generating polarization function, 
\begin{equation*}
\psi \simeq \psi (x^{k})\mbox{ and }\underline{\eta }\ \simeq \underline{%
\eta }(x^{k},t).
\end{equation*}

In explicit form, we consider such a variant:

The h-part of the d-metric (\ref{lacosm2}) may be prescribed to satisfy
instead of 2-d Poisson equation the generalized Taubes equation for vortices
on a curved background 2-d surface, 
\begin{equation}
_{h}\nabla ^{2}\psi =\Omega _{0}(C_{0}-C_{1}e^{2\psi }),  \label{taubeq}
\end{equation}%
where the position-dependent confromal factor $\Omega _{0}$ and effective
source $(C_{0}-C_{1}e^{2\psi })$ are prescribed as respective generating
h-function and generating h-source $\ _{1}\underline{\Upsilon }(x^{k}).$ By
rescaling, both constants $C_{0}$ and $C_{1}$ take standard values $-1,0,$
or 1, but there are only five combinations of these values allow vortex
solutons $\psi \lbrack vortex]$ without singularities \cite{manton17}.

The v-part of (\ref{lacosm2}) can be constructed if, for instance,%
\begin{equation}
\underline{\eta }\ \simeq \left\{ 
\begin{array}{ccc}
_{r}^{sol}\underline{\eta }(r,t) & 
\mbox{ as a solution of the modified KdV
equation }\frac{\partial \underline{\eta }}{\partial t}-6\underline{\eta }%
^{2}\frac{\partial \underline{\eta }}{\partial r}+\frac{\partial ^{3}%
\underline{\eta }}{\partial r^{3}}=0, & \mbox{ for radial solitons}; \\ 
_{\theta }^{sol}\underline{\eta }(\theta ,t) & 
\mbox{ as a solution of the
modified KdV equation }\frac{\partial \underline{\eta }}{\partial \theta }-6%
\underline{\eta }^{2}\frac{\partial \underline{\eta }}{\partial \theta }+%
\frac{\partial ^{3}\underline{\eta }}{\partial \,^{3}}=0, & 
\mbox{ for
angular solitons}.%
\end{array}%
\right.  \label{solitonw}
\end{equation}%
We cite \cite{doikou20} and reference therein on such types of solitonic
wave equations.

Generic off-diagonal and locally anistoropic metrics of type (\ref{lacosm2})
describe cosmological evolution scenarios with nontrivial nonholonomic
structure with conventional h- and v-splitting. Under geometric evolution
with gravitational polarizations and for respective generating sources, a
primary metric with prolate/oblate rotoid void transforms into a vertex
h-configuration (\ref{taubeq}) and, the v-part, into solitonic wave
evolution of type (\ref{solitonw}), which results also in solitonic
configurations for the N-connection coefficients. Such solitonic waves on $t$%
-variable can be with a radial space variable, $r,$ or with an angular
variable, $\theta .$ In a series of our and co-authors works, there were
constructed more general classes of generic off-diagonal cosmological and
quasi-stationary solutions with 3-d solitonic waves and solitoninc
hierarchies in GR and MGTs \cite{sv00a,vp,vs01a,vs01b,av06,sv08,sv15} and
with quasi-periodic and pattern forming structures \cite{biv17,bubuianu17},
see a review of results in appendix B to \cite{vacaru18}.

\subsubsection{Small parametric off-diagonal cosmological deformations with
solitonic vacua for voids}

Using $t$-symmetries defined in section \ref{ssstdual}, we can construct
locally anisotropic cosmological solutions with off-diagonal small $\kappa $%
--parametric deformations of (\ref{cosmvoidm2}). In terms of $\chi $%
-polarization functions, respective d-metrics can be written in the form 
\begin{equation*}
d\ \widehat{s}^{2}=\widehat{g}_{\alpha \beta }(r,\theta ,t;\psi ,\ \ _{2}%
\underline{\Lambda },\ _{2}\underline{\Upsilon })du^{\alpha }du^{\beta
}=e^{\psi _{0}(r,\theta )}[1+\kappa \ ^{\psi }\chi (r,\theta
)][(dx^{1}(r,\theta ))^{2}+(dx^{2}(r,\theta ))^{2}]
\end{equation*}%
\begin{eqnarray}
&&+\zeta _{3}(1+\kappa \ \underline{\chi })\underline{\mathring{g}}%
_{3}\{d\phi +[(\underline{\mathring{N}}_{k}^{3})^{-1}[\ _{1}n_{k}+16\
_{2}n_{k}[\int dt\frac{\left( \partial _{t}[(\underline{\zeta }_{3}%
\underline{\mathring{g}}_{3})^{-1/4}]\right) ^{2}}{|\int d\varphi \partial
_{\varphi }[\ \ _{2}\underline{\Upsilon }(\zeta _{4}\ ^{cy}g)]|}]  \notag \\
&&+\kappa \frac{16\ _{2}n_{k}\int dt\frac{\left( \partial _{t}[(\underline{%
\zeta }_{3}\ \underline{\mathring{g}}_{3})^{-1/4}]\right) ^{2}}{|\int
dt\partial _{t}[\ \ _{2}\underline{\Upsilon }(\underline{\zeta }_{3}\ 
\underline{\mathring{g}}_{3})]|}(\frac{\partial _{t}[(\underline{\zeta }%
_{3}\ \underline{\mathring{g}}_{3})^{-1/4}\underline{\chi })]}{2\partial
_{t}[(\underline{\zeta }_{3}\ \underline{\mathring{g}}_{3})^{-1/4}]}+\frac{%
\int dt\partial _{t}[\ _{2}\underline{\Upsilon }(\underline{\zeta }_{3}%
\underline{\chi }\ \underline{\mathring{g}}_{3})]}{\int dt\partial _{t}[\ \
_{2}\underline{\Upsilon }(\underline{\zeta }_{3}\ \underline{\mathring{g}}%
_{3})]})}{\ _{1}n_{k}+16\ _{2}n_{k}[\int dt\frac{\left( \partial _{t}[(%
\underline{\zeta }_{3}\ \underline{\mathring{g}}_{3})^{-1/4}]\right) ^{2}}{%
|\int dt\partial _{t}[\ \ _{2}\underline{\Upsilon }(\underline{\zeta }_{3}\ 
\underline{\mathring{g}}_{3})]|}]}]\underline{\mathring{N}}%
_{k}^{3}dx^{k}\}^{2}.  \label{paramsoliton}
\end{eqnarray}%
\begin{eqnarray*}
&&-\{\frac{4[\partial _{t}(|\underline{\zeta }_{3}\ \underline{\mathring{g}}%
_{3}|^{1/2})]^{2}}{\ \underline{\mathring{g}}_{4}|\int dt\{\ _{2}\underline{%
\Upsilon }\partial _{t}(\underline{\zeta }_{3}\ \underline{\mathring{g}}%
_{3})\}|}-\kappa \lbrack \frac{\partial _{t}(\underline{\chi }|\underline{%
\zeta }_{3}\ \underline{\mathring{g}}_{3}|^{1/2})}{4\partial _{t}(|%
\underline{\zeta }_{3}\ \underline{\mathring{g}}_{3}|^{1/2})}-\frac{\int
dt\{\ _{2}\underline{\Upsilon }\partial _{t}[(\underline{\zeta }_{3}\ 
\underline{\mathring{g}}_{3})\underline{\chi }]\}}{\int dt\{\ _{2}\underline{%
\Upsilon }\partial _{t}(\underline{\zeta }_{3}\ \underline{\mathring{g}}%
_{3})\}}]\}\ \ \underline{\mathring{g}}_{4} \\
&&\{dt+[\frac{\partial _{i}\ \int dt\ _{2}\underline{\Upsilon }\ \partial
_{t}\underline{\zeta }_{3}}{(\underline{\mathring{N}}_{i}^{3})\ _{2}%
\underline{\Upsilon }\partial _{t}\underline{\zeta }_{3}}+\kappa (\frac{%
\partial _{i}[\int dt\ _{2}\underline{\Upsilon }\ \partial _{t}(\underline{%
\zeta }_{3}\ \underline{\mathring{g}}_{3})]}{\partial _{i}\ [\int dt\ _{2}%
\underline{\Upsilon }\partial _{t}\zeta _{4}]}-\frac{\partial _{t}(%
\underline{\zeta }_{3}\ \underline{\mathring{g}}_{3})}{\partial _{t}%
\underline{\zeta }_{3}})]\underline{\mathring{N}}_{i}^{4}dx^{i}\}^{2}
\end{eqnarray*}%
In above formulas, $\psi _{0}(r,\theta )$ and $\ ^{\psi }\chi (r,\theta )$
are chosen in such a form that they define solutions of 2-d Poisson
equations, or certain $\kappa $-parametric solutions of (\ref{taubeq}) with
some small parametric generated vortices. The generating function $%
\underline{\chi }=\underline{\chi }_{3}(r,\theta ,t)$ can be taken as a
soluton of solitonic wave equation (\ref{solitonw}), $\underline{\eta }\
\longleftrightarrow $ $\underline{\chi },$ when $\underline{\zeta }%
_{3}(r,\theta ,t)$ is also prescribed in a form for $\kappa ^{0}.$ Such
d-metrics define a v-solitonic gravitational structure of voids with $\kappa 
$--parametric and $t$-evolution. For certain explicit configurations, such
parametric gravitational void vacuum posses a nontrivial solitonic energy.
The solutions can be characterized by nonlinear symmetries relating the
effective generating source $\ _{2}\underline{\Upsilon }$ to a respective
cosmological constant $\ _{2}\underline{\Lambda }$

The vertex - solitonic wave locally anisotropic cosmological d-metrics with
respective prime prolate/oblate symmetry encode a nonholonomic vacuum
structure with nontrivial canonical d-torsion. Imposing additional
constraints, we can extract LC-configuration cosmological models if we
follow the procedure described in subsection \ref{sslcconf}.

\part{Nonassociative phase space and Finsler-Lagrange-Hamilton MGTs}

In this part, we show how the AFCDM can be generalized for 8-d phase spaces
modelled on (co) tangent Lorentz bundles and study explicit examples of
quasi-stationary solutions. Such phase spaces present natural geometric
arenas for nonassociative gravity theories determined by star products
deformations and elaborating relativistic physically important models of
Finsler-Lagrange-Hamilton geometries and theories of nonholonomic geometric
flows of nonholonomic geometric objects depending on spacetime and velocity/
momentum coordinates and on temperature like $\tau $-parameter \cite%
{bubuianu20,bubuianu19,vacaru18,bubuianu18a,stavr14,bubuianu20}. There are
four important motivations to study such theories:

\begin{enumerate}
\item Modified dispersion relations, MDRs, result in nonholonomic
generalized Finsler structures on phase spaces \cite%
{ac97,ac98,barc28,basil13,kost11,kost12,ibvv20}. Star product R-flux
deformations can be also characterized by MDRs encoding nonassociative and
noncommutative data.

\item Non-geometric star product R-flux deformations in string theory \cite%
{partner04,partner05,partner06} can be geometrized in nonassociative and
nonholonomic forms on 8-d phase spaces involving complex or real momentum
variables \cite{blumenhagen16,aschieri17,partner01,partner02}. To prove
general decoupling and integration properties of physically important
systems of nonlinear PDEs in such theories, we have to consider nonholonomic
dyadic decompositions and certain classes of generalized metrics and linear
connections adapted to N-connection structures.

\item In (nonassociative) MGTs, we can define N-connection structure
determined by semi-spray equations, i.e. nonlinear geodesic equations, which
are equivalent to the Euler-Lagrange and/or Hamilton equations. This
provides new ideas and methods for formulating generally integrable
classical and quantum gravity theories when nonperturbative quantization
methods are related to generic off-diagonal exact and parametric solutions
in phase space gravity theories. Such geometric and quantum information
formalisms can't be developed in the framework of the well-known approaches 
\cite{dirac69,dewitt67,ashtekar91,isham92}.

\item New classes of generic off-diagonal solutions in such (nonassociative)
phase spaces are characterized by G. Perelman statistical and geometric
thermodynamic models which are generalized for nonassociative
Finsler-Lagrange-Hamilton geometric flow and nonholonomic Ricci soliton
theories. In Part II, we show how such theories can be described
equivalently in canonical dyadic variables (to derive important decoupling
and integrating properties) and in generalized Finsler-Hamilton variables
which can be used in our future works for elaborating quantum models
encoding nonassociative geometric data and elaborating on new methods of
quantization of gravity and matter field theories.
\end{enumerate}

\section{N-connections and Finsler-Lagrange-Hamilton phase space geometry}

\label{sec5} A series of important works on nonassociative geometry and
physics \cite%
{luest10,blumenhagen10,condeescu13,blumenhagen13,kupriyanov19a,szabo19,blumenhagen16,aschieri17}
are based on the concept of nonassociative star product with R-flux
considered in string theory. Such twisted algebraic and geometric structures
result in nonassociative modifications of GR to nonholonomic geometries
involving extra-dimension coordinates considered as momentum-like variables.
This is similar and, for some well-defined conditions, equivalent to certain
versions of nonassociative and noncommutative Finsler-Lagrange-Hamilton,
FLH, geometries studied in details our former works \cite%
{vacaru96b,vacaru18,bubuianu18a,bubuianu20,bubuianu19,vmon3}. So, in Part II
of this review, nonassociative FLH gravitational theories are defined as
minimal nonholonomic modifications of GR because of the nonassociative star
product used in \cite{blumenhagen16,aschieri17}.

In a series of our partner works \cite%
{partner01,partner02,partner03,partner04,partner05,partner06,bgrg24,bapny24,vapny24,vcqg25}%
, we elaborated on new nonholonomic geometric methods of constructing exact
and parametric solutions in nonassociative gravity. To decouple and solve in
some general forms certain nonassociative generalizations of the Einstein
equations using only the formalism elaborated in \cite%
{szabo19,blumenhagen16,aschieri17} was not possible. So, we had to perform a
new research program on constructing off-diagonal solutions in MGTs by
applying and developing our former results on (noncommutative/
supersymmetric/ string) generalized Finsler geometry \cite%
{vacaru09a,gheor14,vacaru18,bubuianu18a,biv17,sv00a,vmon3} as a
generalization of the AFCDM in GR (reviewed in Part I).

In this section, we summarize the necessary definitions and methods from the
nonholonomic geometry of associative and commutative phase space geometry
and relativistic models of Finsler-Lagrange-Hamilton geometry. Such
geometric and physical models are elaborated on an 8-d phase space modelled
as a cotangent Lorentz bundle $\ ^{\shortmid }\mathcal{M} =T^{\ast }V$ on a
spacetime manifold $V$ of signature $(+++-)$; such a phase space is dual to $%
\mathcal{M}=TV.$ In this approach, the GR theory on $V$ is generalized on
total phase spaces with conventional extra dimension velocity/ momentum type
coordinates.

\subsection{Nonlinear connections and canonical nonholonomic (2+2)+(2+2)
splitting}

A nonlinear connection, N-connection, structure defining a 4+4 splitting is
by definition a Whitney sum 
\begin{equation}
\ ^{\shortmid }\mathbf{N}:\ \ T\mathbf{T}^{\ast }\mathbf{V}=\ hT^{\ast
}V\oplus \ cT^{\ast }V,\mbox{ which is dual to }\ \mathbf{N}:\ T\mathbf{TV}%
=hTV\oplus vTV.  \label{ncon44}
\end{equation}%
These nonholonomic distributions provide a phase space extension of formulas
(\ref{ncon}), and related formulas and derived geometric and physical
equations, when $\ ^{\shortmid }\mathbf{N}=\{\ ^{\shortmid }N_{\ ia}(\
^{\shortmid }u)\},$ for$\ ^{\shortmid }u=(\ x,\ p)=\{\ ^{\shortmid
}u^{\alpha }=(x^{i},p_{a})\};$ and, respectively, $\mathbf{N}=\{N_{\
i}^{a}(u)\},$ for $u=(x,y=v)=\{u^{\alpha}=(x^{i},y^{a}=v^{a})\}.$ The 8-d
indices split into 4+4 ones when, for instance, $\alpha ,\beta
,...=1,2,...8; $ $i,j,...=1,2,3,4$ and $a,b,...=5,6,7,8.$

To generalize and apply the AFCDM we have to consider conventional
(2+2)+(2+2) splitting on respective phase spaces stated as a nonholonomic
(equivalently, anholonomic/non-integrable) dyadic, 2-d, decomposition into
four oriented shells $s=1,2,3,4.$ In brief, we shall say that this is a
s-decomposition and use respective $s$-labels in abstract form, or for
indices and coordinates when it will be necessary. The nonholonomic
s-splitting is defined by respective N-connection (equivalently,
s-connection), structure: 
\begin{eqnarray}
\ _{s}^{\shortmid }\mathbf{N}:\ \ _{s}T\mathbf{T}^{\ast }\mathbf{V} &=&\
^{1}hT^{\ast }V\oplus \ ^{2}vT^{\ast }V\oplus \ ^{3}cT^{\ast }V\oplus \
^{4}cT^{\ast }V,\mbox{ which is dual to }  \notag \\
\ _{s}\mathbf{N}:\ \ _{s}T\mathbf{TV} &=&\ ^{1}hTV\oplus \ ^{2}vTV\oplus \
^{3}vTV\oplus \ ^{4}vTV,\mbox{  for }s=1,2,3,4.  \label{ncon2222}
\end{eqnarray}%
In these formulas, we write use $\ ^{1}h$ for a conventional 2-d shell
(dyadic) splitting on (co) tangent bundle, with $x^{i_{1}}$ local
coordinates and $\ ^{2}v$ for a 2-d vertical like splitting with $y^{a_{2}}$
coordinates on the shell $s=2.$ On the (co) fiber shell $s=3,$ the splitting
is conventional (co) vertical, when we write $\ ^{3}v$ (or $\ ^{3}c$) and
use local coordinates $v^{a_{3}}$ $($or $p_{a_{3}}).$ Similarly, on the 4th
shell $s=4,$ the respective symbols are $\ ^{4}v$ and $v^{a_{4}}$ (or $\
^{4}c $ and $p_{a_{4}}).$ Hereafter, we shall write typically the formulas
of s-geometric objects on $\ ^{\shortmid }\mathcal{M}=T^{\ast }V,$ when the
formulas for similar ones on$\ \mathcal{M}=TV$ can be formulated to encode
velocity type coordinates with necessary shell indices.

Using a set of N-connection coefficients, we can construct N-elongated bases
(N-/ s-adapted bases) as linear N-operators: 
\begin{eqnarray}
\ ^{\shortmid }\mathbf{e}_{\alpha _{s}}[\ ^{\shortmid }N_{\ i_{s}a_{s}}]
&=&(\ ^{\shortmid }\mathbf{e}_{i_{s}}=\ \frac{\partial }{\partial x^{i_{s}}}%
-\ ^{\shortmid }N_{\ i_{s}a_{s}}\frac{\partial }{\partial p_{a_{s}}},\ \
^{\shortmid }e^{b_{s}}=\frac{\partial }{\partial p_{b_{s}}})\mbox{ on }\
_{s}T\mathbf{T}_{\shortmid }^{\ast }\mathbf{V,}  \label{nadapbdsc} \\
\ ^{\shortmid }\mathbf{e}_{\alpha }[\ ^{\shortmid }N_{\ ia}] &=&(\
^{\shortmid }\mathbf{e}_{i}=\ \frac{\partial }{\partial x^{i_{s}}}-\
^{\shortmid }N_{\ ia}\frac{\partial }{\partial p_{a}},\ \ ^{\shortmid }e^{b}=%
\frac{\partial }{\partial p_{b}})\mbox{ on }\ T\mathbf{T}_{\shortmid }^{\ast
}\mathbf{V,}  \notag
\end{eqnarray}%
and, dual s-adapted bases, s-cobases,%
\begin{eqnarray}
\ ^{\shortmid }\mathbf{e}^{\alpha _{s}}[\ ^{\shortmid }N_{\ i_{s}a_{s}}]
&=&(\ ^{\shortmid }\mathbf{e}^{i_{s}}=dx^{i_{s}},\ ^{\shortmid }\mathbf{e}%
_{a_{s}}=d\ p_{a_{s}}+\ ^{\shortmid }N_{\ i_{s}a_{s}}dx^{i_{s}})\mbox{ on }\
\ _{s}T^{\ast }\mathbf{T}_{\shortmid }^{\ast }\mathbf{V,}  \label{nadapbdss}
\\
\ ^{\shortmid }\mathbf{e}^{\alpha }[\ ^{\shortmid }N_{\ ia}] &=&(\
^{\shortmid }\mathbf{e}^{i}=dx^{i},\ ^{\shortmid }\mathbf{e}_{a}=d\ p_{a}+\
^{\shortmid }N_{\ ia}dx^{i})\mbox{ on }\ \ T^{\ast }\mathbf{T}_{\shortmid
}^{\ast }\mathbf{V.}  \notag
\end{eqnarray}%
Such s-frames are not integrable, i.e. nonholonomic (equivalently,
anholonomic) because, in general, they satisfy certain anholonomy
conditions, 
\begin{equation}
\ ^{\shortmid }\mathbf{e}_{\beta _{s}}\ ^{\shortmid }\mathbf{e}_{\gamma
_{s}}-\ ^{\shortmid }\mathbf{e}_{\gamma _{s}}\ ^{\shortmid }\mathbf{e}%
_{\beta _{s}}=\ ^{\shortmid }w_{\beta _{s}\gamma _{s}}^{\tau _{s}}\
^{\shortmid }\mathbf{e}_{\tau _{s}},  \label{anholcond8}
\end{equation}%
see details in \cite{vacaru18,bubuianu18a,partner01,partner02}.

The geometric s-objects and respective formulas (\ref{ncon})-(\ref{nadapbdss}%
) can be generalized for additional running on a geometric flow evolution
parameter $\tau ,$ which is used in geometric flow theories, see details and
references in \cite{partner04,partner05,partner06}. In our works, $\tau$ can
be considered as a temperature like parameter (as in G. Perelman's geometric
flow thermodynamics \cite{perelman1}). In such cases, we write, for
instance, $\ ^{\shortmid }\mathbf{N(}\tau \mathbf{)\simeq }\ ^{\shortmid }%
\mathbf{N(}\tau ,\ ^{\shortmid }u)=\{\ ^{\shortmid }N_{\ ia}(\tau )\mathbf{%
\simeq }\ ^{\shortmid }N_{\ ia}(\tau ,x^{j},p_{b})\}$ and, respectively, $\
^{\shortmid }\mathbf{e}_{\alpha _{s}}(\tau ),\ ^{\shortmid }\mathbf{e}%
^{\alpha _{s}}(\tau ),$ etc., which will be used in next sections. For $\tau$%
-running of geometric/ physical objects, we shall write only the $\tau $%
-dependence if that will not result in ambiguities. Here, we note that in a
similar form we can introduce and write formulas for geometric objects on $\
_{s}T\mathbf{TV,}$ i.e. when the total space coordinates are of
spacetime-velocity type. In such case, we omit the labels "$\ ^{\shortmid }"$
and write, for instance, $\ \mathbf{e}_{\alpha _{s}}(\tau )$ and$\ \mathbf{e}%
^{\alpha _{s}}(\tau ).$ In general, the local coordinates are not just dual
like fiber and co-fiber ones but may include certain Legendre transforms and
symplectomorphisms \cite{bubuianu19}. We work on nonassociative phase spaces
as in \cite{blumenhagen16,aschieri17} and \cite%
{partner01,partner02,partner03,partner04,partner05,partner06} using labels "$%
\ ^{\shortmid }"$ in order to follow an unified system of notations which
will allow in next section works to elaborate on nonassociative models of
Finsler-Lagrange spaces, which are important in quantum information theory.

A metric field in a phase space $\ ^{\shortmid }\mathcal{M}$ is a second
rank symmetric tensor $\ ^{\shortmid }g=\{\ ^{\shortmid }g_{\alpha \beta
}\}\in TT^{\ast }V\otimes TT^{\ast }V$ of local signature $%
(+,+,+,-;+,+,+,-). $ It can be written in equivalent form as a s-metric $\
_{s}^{\shortmid }\mathbf{g}=\{\ ^{\shortmid }\mathbf{g}_{\alpha _{s}\beta
_{s}}\}$ for $\ _{s}^{\shortmid }\mathcal{M}$ which is a 8-d phase space
generalization of (\ref{dm}). For $\tau $-families of phase space metrics
(d-metrics for 4+4 splitting) and s-metrics, we shall use notations of type $%
\ ^{\shortmid }g(\tau )=\{\ ^{\shortmid }g_{\alpha \beta }(\tau )\}$ and,
respectively, $\ _{s}^{\shortmid }\mathbf{g}(\tau )=\{\ ^{\shortmid }\mathbf{%
g}_{\alpha _{s}\beta _{s}}(\tau )\}.$

Another important geometric concept is that of s-connection with a
(2+2)+(2+2) splitting (the term distinguished connection, d-connection, is
considered for a (4+4)-splitting). Such linear connections preserve
respective shell or h-c structures under parallel transports a corresponding
s-/ N-connection splitting (\ref{ncon2222}), or (\ref{ncon44}):%
\begin{equation}
\ _{s}^{\shortmid }\mathbf{D} = (h_{1}\ ^{\shortmid }\mathbf{D},\ v_{2}\
^{\shortmid }\mathbf{D},\ c_{3}\ ^{\shortmid }\mathbf{D},\ c_{4}\
^{\shortmid }\mathbf{D})=\{\ ^{\shortmid }\Gamma _{\ \ \beta _{s}\gamma
_{s}}^{\alpha _{s}}\}, \mbox{ or } \ ^{\shortmid }\mathbf{D} = (h\
^{\shortmid }\mathbf{D},\ \ c\ ^{\shortmid }\mathbf{D})=\{\ ^{\shortmid
}\Gamma _{\ \ \beta \gamma }^{\alpha }\}.  \label{sdc}
\end{equation}%
Hereafter we shall provide only s-adapted or N-adapted formula not dubbing
them using typical s-labels if that will not result in ambiguities.

Using standard definitions from differential geometry, we can introduce in
abstract form and compute the coefficient formulas for any s-connection $\
_{s}^{\shortmid }\mathbf{D}$ and for such fundamental geometric s-objects:%
\begin{eqnarray}
\ _{s}^{\shortmid }\mathcal{T} &=&\{\ ^{\shortmid }\mathbf{T}_{\ \beta
_{s}\gamma _{s}}^{\alpha _{s}}\},\mbox{ the s-torsion };\ \ _{s}^{\shortmid }%
\mathcal{R} = \{\ ^{\shortmid }\mathbf{R}_{\ \beta _{s}\gamma _{s}\delta
_{s}}^{\alpha _{s}}\},\mbox{ the Riemannian s-curvature };  \label{mafgeomob}
\\
\ _{s}^{\shortmid }\mathcal{R}ic &=&\{\ ^{\shortmid }\mathbf{R}_{\ \beta
_{s}\gamma _{s}}:=\ ^{\shortmid }\mathbf{R}_{\ \beta _{s}\gamma _{s}\alpha
_{s}}^{\alpha _{s}}\neq \ ^{\shortmid }\mathbf{R}_{\ \gamma _{s}\beta
_{s}}\},\mbox{ the Ricci s-tensor};  \notag \\
\ _{s}^{\shortmid }\mathcal{R}sc &= &\{\ ^{\shortmid }\mathbf{g}^{\beta
_{s}\gamma _{s}}\ ^{\shortmid }\mathbf{R}_{\ \beta _{s}\gamma _{s}}\},%
\mbox{the Riemannian scalar }.  \notag
\end{eqnarray}%
Geometric data $(\ _{s}^{\shortmid }\mathbf{g,}\ _{s}^{\shortmid }\mathbf{D)}
$ enable a $\ _{s}\mathcal{M}$ with a dyadic metric-affine s-structure which
is a s-adapted phase space version of metric-affine geometry \cite%
{misner,kramer03,vacaru18,bubuianu18a}. In general, such nonholonomic phase
spaces are characterized by a respective nonmetricity s-tensor, $\
_{s}^{\shortmid }\mathcal{Q}=\{\ ^{\shortmid }\mathbf{Q}_{\gamma _{s}\alpha
_{s}\beta _{s}\ }=\ ^{\shortmid }\mathbf{D}_{\gamma _{s}}\ ^{\shortmid }%
\mathbf{g}_{\alpha _{s}\beta _{s}}\}.$ In Appendix, we provide additional
abstract and s-adapted formulas (\ref{cncon8})-(\ref{dcond8}) explaining how
such values can be computed in explicit form.

Using a s-metric $\ ^{\shortmid }g= \ _{s}^{\shortmid }\mathbf{g}$, we can
define and compute in abstract and component forms 8-d generalizations of
the formulas (\ref{twocon}) for two important linear connection structures
(the Levi-Civita, LC, connection and the canonical s-connection): 
\begin{equation}
(\ _{s}^{\shortmid }\mathbf{g,\ _{s}^{\shortmid }N})\rightarrow \left\{ 
\begin{array}{cc}
\ ^{\shortmid }\mathbf{\nabla :} & \ ^{\shortmid }\mathbf{\nabla }\ \
_{s}^{\shortmid }\mathbf{g}=0;\ _{\nabla }^{\shortmid }\mathcal{T}=0,\ %
\mbox{\  LC--connection }; \\ 
\ _{s}^{\shortmid }\widehat{\mathbf{D}}: & 
\begin{array}{c}
\ _{s}^{\shortmid }\widehat{\mathbf{Q}}=0;\ h_{1}\ ^{\shortmid }\widehat{%
\mathcal{T}}=0,v_{2}\ ^{\shortmid }\widehat{\mathcal{T}}=0,c_{3}\
^{\shortmid }\widehat{\mathcal{T}}=0,c_{4}\ ^{\shortmid }\widehat{\mathcal{T}%
}=0, \\ 
h_{1}v_{2}\ ^{\shortmid }\widehat{\mathcal{T}}\neq 0,h_{1}c_{s}\ ^{\shortmid
}\widehat{\mathcal{T}}\neq 0,v_{2}c_{s}\ ^{\shortmid }\widehat{\mathcal{T}}%
\neq 0,c_{3}c_{4}\ ^{\shortmid }\widehat{\mathcal{T}}\neq 0,%
\end{array}%
\begin{array}{c}
\mbox{ canonical } \\ 
\mbox{ s-connection  }.%
\end{array}%
\end{array}%
\right.  \label{twocon8}
\end{equation}%
So, for higher dimensions, we can also use "hat" labels for geometric
s-objects written in canonical form, for instance, $\ _{s}^{\shortmid }%
\widehat{\mathbf{D}}, \ _{s}^{\shortmid }\widehat{\mathcal{R}}=\{\
^{\shortmid }\widehat{\mathbf{R}}_{\ \beta _{s}\gamma _{s}\delta
_{s}}^{\alpha _{s}}\}$ etc. In similar forms we can define and computed the
canonical distortion relations for linear connections (of type $\
_{s}^{\shortmid }\widehat{\mathbf{D}}=\ ^{\shortmid }\nabla +\
_{s}^{\shortmid }\widehat{\mathbf{Z}},$ with a distortions s-tensor $\
_{s}^{\shortmid }\widehat{\mathbf{Z}}$ defined by N-coefficients) which
allow to compute canonical distortions of fundamental geometric objects (\ref%
{mafgeomob}). For instance, we can consider distortions of curvature tensors
and s-tensors, for instance, $\ _{\nabla }^{\shortmid }\mathcal{R}=\{\
_{\nabla }^{\shortmid }R_{\ \beta _{s}\gamma _{s}\delta _{s}}^{\alpha
_{s}}\} $ and $\ _{s}^{\shortmid }\widehat{\mathcal{R}}=\{\ ^{\shortmid }%
\widehat{\mathbf{R}}_{\ \beta _{s}\gamma _{s}\delta _{s}}^{\alpha _{s}}\}; \
_{\nabla }^{\shortmid }\mathcal{R}ic$ and $\ _{s}^{\shortmid }\widehat{%
\mathcal{R}}ic$ etc. For $\tau $-families such formulas can written, for
instance, $\ ^{\shortmid } \nabla (\tau \mathbf{),}\ _{s}^{\shortmid }%
\widehat{\mathbf{D}}(\tau ), \ _{s}^{\shortmid }\widehat{\mathcal{R}}(\tau
)=\{\ ^{\shortmid }\widehat{\mathbf{R}}_{\ \beta _{s}\gamma _{s}\delta
_{s}}^{\alpha _{s}}(\tau ) \}, \ _{\nabla }^{\shortmid }\mathcal{R}ic(\tau )$%
, etc.

The modified Einstein equations for $\ _{s}^{\shortmid }\widehat{\mathbf{D}}$
(\ref{twocon8}) can be derived in abstract geometric form as in GR \cite%
{misner} but on phase space $\ _{s}\mathcal{M}$ and following respective
conventions on s-adapted indices,%
\begin{equation}
\ ^{\shortmid }\widehat{\mathbf{R}}ic_{\alpha _{s}\beta _{s}}=\ ^{\shortmid
}\Upsilon _{\alpha _{s}\beta _{s}}.  \label{seinsta}
\end{equation}%
In such formulas, the s-tensor for effective and/or matter field sources $\
^{\shortmid }\Upsilon _{\alpha _{s}\beta _{s}}$ can be postulated (or
derived following a conventional s-variational calculus extending the
constructions in GR or certain MGTs) in the forms 
\begin{equation}
\ ^{\shortmid }\Upsilon _{_{\beta _{s}\gamma _{s}}}=\left\{ 
\begin{array}{c}
\ _{s}^{\shortmid }\Lambda _{0}\ ^{\shortmid }\mathbf{g}_{\alpha _{s}\beta
_{s}}=\frac{1}{2}\ ^{\shortmid }\mathbf{g}_{\alpha _{s}\beta _{s}}\
_{s}^{\shortmid }\widehat{\mathbf{R}}sc+\ _{s}^{\shortmid }\lambda \ \
^{\shortmid }\mathbf{g}_{\alpha _{s}\beta _{s}},%
\mbox{vacuum with shell
cosmological constants}\ _{s}^{\shortmid }\Lambda _{0}\mbox{ or  }\
_{s}^{\shortmid }\lambda ; \\ 
\ _{s}^{\shortmid }\Lambda (\tau ,\ ^{\shortmid }u)\ ^{\shortmid }\mathbf{g}%
_{\alpha _{s}\beta _{s}},%
\mbox{ for  polarized constants from geometric
flow/ string / quantum theories}; \\ 
\ ^{\shortmid }\mathbf{Y}_{_{\beta _{s}\gamma _{s}}},%
\mbox{ from
variational/ geometric  principles of interactions on }\ _{s}\mathcal{M}; \\ 
\ ^{\shortmid }\mathbf{K}_{_{\beta _{s}\gamma _{s}}}\left\lceil \hbar
,\kappa \right\rceil ,%
\mbox{ for effective parametric star R-flux
corrections, in this work and \cite{partner02,partner03,partner04} }.%
\end{array}%
\right.  \label{sourca}
\end{equation}%
The phase space gravitational field equations (\ref{seinsta}) can be written
in terms of the LC-connection $\ ^{\shortmid }\nabla _{\alpha }$\ if we
consider distortion relations. Imposing additional zero s-torsion
conditions, 
\begin{equation}
\ _{s}^{\shortmid }\widehat{\mathbf{Z}}=0,\mbox{ which is equivalent to }\
_{s}^{\shortmid }\widehat{\mathbf{D}}_{\mid \ \ _{s}^{\shortmid }\widehat{%
\mathbf{T}}=0}=\ ^{\shortmid }\nabla ,  \label{lccond38}
\end{equation}%
we can extract LC-configurations from canonical nonholonomic classes of
solutions. Here we note that various conservation laws can be formulated by
extending in s-adapted form the formulas from GR using $\ ^{\shortmid
}\nabla $ on $\mathcal{M},$ for instance, $\ ^{\shortmid }\nabla (\ _{\nabla
}^{\shortmid }\mathcal{R}ic_{\alpha _{s}\beta _{s}}-\frac{1}{2}\ ^{\shortmid
}\mathbf{g}_{\alpha _{s}\beta _{s}}\ \ _{\nabla }^{\shortmid }Rsc)=0,$ but
such laws are written in more cumbersome forms if we distort the geometrical
objects and this equations in terms of $\ _{s}^{\shortmid }\widehat{\mathbf{D%
}}.$ This is a typical property of nonholonomic systems in geometric
mechanics and gravity theories. Here we note that notations for nonholonomic
constraints of type $\ _{s}^{\shortmid }\widehat{\mathbf{D}}_{\mid \ \
_{s}^{\shortmid }\widehat{\mathbf{T}}=0}$ (\ref{lccond}).

\subsection{Modified dispersion relations and phase space
Finsler-Lagrange-Hamilton geometry}

For semi-classical commutative MGTs and nonassociative / noncommutative
models and in QG, modified dispersion relations, MDRs, can be parameterized
locally in the form 
\begin{equation}
c^{2}\overrightarrow{\mathbf{p}}^{2}-E^{2}+c^{4}m^{2}=\varpi (E,%
\overrightarrow{\mathbf{p}},m;\ell _{P},\kappa ,...).  \label{mdrg}
\end{equation}%
An indicator $\varpi (...)$ encodes in a functional form possible
contributions of MGTs which, in general, can be with local Lorentz symmetry
violation etc. Such MDRs can be extended to dependencies on 4-d spacetime
coordinates $x^{i}=(x^{1},x^{2},x^{3},x^{4}=ct)$ and extended to higher
dimensions and for various phase space models. In explicit form, certain
classes of $\varpi (...)$ are prescribed following theoretical/
phenomenological/ arguments, or determined experimentally. We can compute
such values in the framework of certain classical/ quantum theories of
gravity and matter field interactions. If $\varpi =0,$ the equation (\ref%
{mdrg}) transforms into a standard quadratic dispersion relation for a
relativistic point particle with mass $m,$ energy $E,$ and momentum $p_{%
\acute{\imath}}$ (for $\acute{\imath}=1,2,3$), when such a particle
propagates in a 4-d, flat Minkowski spacetime. A $\varpi $ (\ref{mdrg}) may
involve dependencies on a conventional energy-momentum $p_{a}=(p_{\acute{%
\imath}},p_{4}=E),\overrightarrow{\mathbf{p}}=\{p_{\acute{\imath}}\},$ (for $%
a=1,2,3,4$), at the Planck scale $\ell _{p}:=\sqrt{\hbar G/c^{3}}\sim
10^{-33}cm$ and $\kappa :=\ell _{s}^{3}/6\hbar $ being a string constant,
were $\ell _{s}$ is a length parameter. In this work, the light velocity is
fixed $c=1$ for a respective system of physical units. Different types of $%
\varpi $ are considered in various approaches to QG and (non) commutative
MGTs, supergravity and (super) string models etc. Here we note that MGTs
with MDRs are studied also as candidates for explaining acceleration
cosmology and dark energy, DE, and dark matter, DM, physics, see \cite%
{ac97,ac98,barc28,basil13,kost11,kost12,basilakos13,stavr99,stavr04,stavr05,svcqg13}
and references therein.

We follow the Assumption 2.1 from \cite{vacaru18,bubuianu18a} that the
standard gravity and particle physics theories based on the special
relativity and Einstein gravity principles and axioms can be generalized
from a 4-d Lorentz spacetime manifold $V$ on phase spaces $TV$ or $T^{\ast
}V $ for total phase space metrics with signature $(+++-;+++-)$, 
\begin{eqnarray}
ds^{2} &=&g_{\alpha \beta }(x^{k})du^{\alpha }du^{\beta
}=g_{ij}(x^{k})dx^{i}dx^{j}+\eta _{ab}dy^{a}dy^{b},\mbox{ for }y^{a}\sim
dx^{a}/d\zeta ;\mbox{ and/ or }  \label{lqe} \\
d\ ^{\shortmid }s^{2} &=&\ ^{\shortmid }g_{\alpha \beta }(x^{k})d\
^{\shortmid }u^{\alpha }d\ ^{\shortmid }u^{\beta
}=g_{ij}(x^{k})dx^{i}dx^{j}+\eta ^{ab}dp_{a}dp_{b},\mbox{ for }p_{a}\sim
dx_{a}/d\zeta ,  \label{lqed}
\end{eqnarray}%
when certain curves $x^{a}(\zeta )$ on $V$ are parameterized by a positive
parameter $\zeta .$ A pseudo--Riemannian spacetime metric $g=\{g_{ij}(x)\}$
can be a solution of the Einstein equations for the Levi-Civita connection $%
\nabla $ as we considered in Part I. In diagonal form, the vertical metric $%
\eta _{ab}$ and its dual $\eta ^{ab}$ are standard Minkowski metrics, $\eta
_{ab}=diag[1,1,1,-1].$ The geometric and physical phase space models are
elaborated for general frame/ coordinate transforms on the base spacetime
and in total spaces when the metric structures can be parameterized
equivalently by the same h-components of $g_{\alpha \beta }(x^{k})$ and $\
^{\shortmid }g_{\alpha \beta }(x^{k})=g_{\alpha \beta }(x^{k})$,
respectively, in quadratic elements (\ref{lqe}) and (\ref{lqed}).

We suppose that the M-theory and string gravity related MGTs, and
quasi-classical limits of QG, can be characterized by MDRs (\ref{mdrg}) can
be modelled with (small) values of and indicator $\varpi $ are described by
basic Lorentzian and non-Riemannian total phase space geometries determined
by nonlinear quadratic line elements for Lagrange-Hamilton spaces: 
\begin{eqnarray}
ds_{L}^{2} &=&L(x,y),\mbox{ for models on  }TV;  \label{nqe} \\
d\ ^{\shortmid }s_{H}^{2} &=&H(x,p),\mbox{ for models on  }T^{\ast }V.
\label{nqed}
\end{eqnarray}
For localized $\varpi =0,$ the nonlinear quadratic line elements (\ref{nqe})
and (\ref{nqed}) transform correspondingly into linear quadratic elements (%
\ref{lqe}) and (\ref{lqed}). For any MDR (\ref{mdrg}), we can model a
Hamilton space $H^{3,1}$ with an Hamilton function $\ H(p):=E=\pm (c^{2}%
\overrightarrow{\mathbf{p}} ^{2}+c^{4}m^{2}-\varpi (E,\overrightarrow{%
\mathbf{p}},m;\ell _{P}))^{1/2}.$ Changing the system of frames/ coordinates
on total space, we obtain generating functions $H(x,p)$ depending also on
spacetime coordinates. We can use for phase space geometric modeling certain
general generating functions $H(x,p)$ on $T^{\ast }V$ (for simplicity, we
can work with regular configurations for nonzero Hessians of $H$). Here, we
note that there are Legendre transforms $L\rightarrow H,$ with $%
H(x,p):=p_{a}y^{a}-L(x,y)$ and $y^{a}$ determining solutions of the
equations $p_{a}=\partial L(x,y)/\partial y^{a}.$ In a similar manner, the
inverse Legendre transforms can be introduced, $H\rightarrow L,$ for $\
L(x,y):=p_{a}y^{a}-H(x,p)$ and $p_{a}$ determining solutions of the
equations $y^{a}=\partial H(x,p)/\partial p_{a}.$ For regular
configurations, we can work equivalently both with Largange and/or Hamilton
spaces. In this section, we provide the formulas for Hamilton type models on
phase spaces which admit straightforward generalizations to nonassociative
geometry determined by star products and R-flux deformations.

A relativistic 4-d model of Lagrange space $L^{3,1}=(TV,L(x,y))$ on a 8-d
phase space with velocity type conventional coordinates $y\approx v$ is
defined by a fundamental function (equivalently, generating function) $TV\ni
(x,y)\rightarrow L(x,y)\in \mathbb{R},$ which is a real valued function,
differentiable on $\widetilde{TV}:=TV/\{0\},$ for $\{0\}$ being the null
section of $TV,$ and continuous on the null section of $\pi :TV\rightarrow
V. $ Such a model is regular if the Hessian (v-metric) 
\begin{equation}
\widetilde{g}_{ab}(x,y):=\frac{1}{2}\frac{\partial ^{2}L}{\partial
y^{a}\partial y^{b}}  \label{hessls}
\end{equation}%
is non-degenerate, i.e. $\det |\widetilde{g}_{ab}|\neq 0,$ and of constant
signature.

In a similar form, a 4-d relativistic model of Hamilton space $%
H^{3,1}=(T^{\ast }V,H(x,p))$ can be constructed for a fundamental function
(equivalently, generating Hamilton function) on a Lorentz manifold $V;$ when 
$T^{\ast }V\ni (x,p)\rightarrow H(x,p)\in \mathbb{R}$ is defines by a real
valued function being differentiable on $\widetilde{T^{\ast }V}:=T^{\ast
}V/\{0^{\ast }\},$ for $\{0^{\ast }\}$ being the null section of $T^{\ast
}V, $ and continuous on the null section of $\pi ^{\ast }:\ T^{\ast
}V\rightarrow V.$ We say that such a model is regular if the Hessian
(cv-metric) 
\begin{equation}
\ \ ^{\shortmid }\widetilde{g}^{ab}(x,p):=\frac{1}{2}\frac{\partial ^{2}H}{%
\partial p_{a}\partial p_{b}}  \label{hesshs}
\end{equation}%
is non-degenerate, i.e. $\det |\ ^{\shortmid }\widetilde{g}^{ab}|\neq 0,$
and of constant signature.

The v-metric $\widetilde{g}_{ab}$ and $\ $c-metric$\ ^{\shortmid }\widetilde{%
g}^{ab}$ are labeled by tilde "\symbol{126}" in order to emphasize that such
conventional v--metrics are defined canonically by respective Lagrange and
Hamilton generating functions. In general, such functions encode various
types of MDRs and contributions for MGTs on phase spaces. General frame/
coordinate transforms on $TV$ and/or $T^{\ast }V$ allow us to express any
"tilde" Hessian in a general quadratic form, respectively as a vertical
metric (v-metric), $g_{ab}(x,y),$ and/or co-vertical metric (cv-metric), $\
^{\shortmid }g^{ab}(x,p).$ We can works also with inverse transforms by
prescribing any v-metric (cv-metric). In general, a $g_{ab}$ is different
from the inverse of $\ ^{\shortmid }g^{ab},$ i.e. from $\ ^{\shortmid}g_{ab} 
$. Lagrange and/or Hamilton models on corresponding $\mathcal{M}$ and/or $\
^{\shortmid }\mathcal{M}$ can be always constructed by prescribing certain
generating functions $L(x,y)$ and/or $H(x,p).$ We shall omit tildes on
geometrical/ physical objects if certain formulas hold in general (not only
canonical) forms and that will not result in ambiguities.

Let us consider an important geometric example: A relativistic 4-d model of
Finsler space is defined as a particular case of Lagrange space when a
regular $L=F^{2}$ is defined by a fundamental (generating) Finsler function
subjected to such three conditions: 1) a generating $F$ is a real positive
valued function which is differential on $\widetilde{TV}$ and continuous on
the null section of the projection $\pi :TV\rightarrow V;$ 2) it satisfies
also the homogeneity condition $F(x,\lambda y)=|\lambda | F(x,y),$ for a
nonzero real value $\lambda ;$ and 3) for such a fundamental function, the
Hessian (\ref{hessls}) is defined by $F^{2}$ in such a form that in any
point $(x_{(0)},y_{(0)})$ the v-metric is of signature $(+++-).$ In a
similar form, we can define relativistic 4-d Cartan spaces $%
C^{3,1}=(V,C(x,p)),$ when $H=C^{2}(x,p)$ is 1-homogeneous on co-fiber
coordinates $p_{a}.$ This is a 4-d Finsler space but with momentum like
variables. Here we note that general MDRs encoding data for certain general
MGTs do not involve certain homogeneity conditions. For the purposes of this
work, we shall not use standard examples of Lagrange-Finsler spaces but
elaborate on generalized Finsler like models with generating functions $%
H(x,p)$ which transform (for general nonholonomic frame transforms and
distortions of connections) into certain metric-affine theories on $\
^{\shortmid }\mathcal{M}.$

On a Hamiltonian phase space $\widetilde{H}$, we can define canonical
symplectic structure $\theta :=dp_{i}\wedge dx^{i}$ and a unique vector
filed $\ \widetilde{X}_{H}:=\frac{\partial \widetilde{H}}{\partial p_{i}}%
\frac{\partial }{\partial x^{i}}-\frac{\partial \widetilde{H}}{\partial x^{i}%
} \frac{\partial }{\partial p_{i}}$ determined by the equation $i_{%
\widetilde{X}_{H}}\theta =-d\widetilde{H}.$ In these formulas, $\wedge $ is
the antisymmetric product and $i_{\widetilde{X}_{H}}$ denotes the interior
produce defined by $\widetilde{X}_{H}.$ This allows to formulate explicit
Hamilton calculus for any functions $\ ^{1}f(x,p)$ and $\ ^{2}f(x,p)$ and
respective canonical Poisson structure $\{\ ^{1}f,\ ^{2}f\}:=\theta (%
\widetilde{X}_{^{1}f},\widetilde{X}_{^{2}f}).$ Let us consider how such a
structure is related to respective Hamilton-Jacobi configurations. Any
regular curve $c(\zeta ),$ when $c:\zeta \in \lbrack 0,1]\rightarrow
x^{i}(\zeta )\subset U\subset V,$ for a real parameter $\zeta ,$ can be
lifted to $\pi ^{-1}(U)\subset \widetilde{TV}$ defining a curve in the total
space, when $\widetilde{c}(\zeta ):\zeta \in \lbrack 0,1]\rightarrow \left(
x^{i}(\zeta ),y^{i}(\zeta )=dx^{i}/d\zeta \right) $ with a non-vanishing
v-vector field $dx^{i}/d\zeta .$ For any effective Hamilton phase space
model, one holds the canonical Hamilton-Jacobi equations, 
\begin{equation*}
\frac{dx^{i}}{d\zeta }=\{\widetilde{H},x^{i}\}\mbox{ and }\frac{dp_{a}}{%
d\zeta }=\{\widetilde{H},p_{a}\}.
\end{equation*}

Equivalent Lagrange and Hamilton models of relativistic phase spaces can be
formulated as $L$-dual effective phase spaces $\widetilde{H}^{3,1}$ and $%
\widetilde{L}^{3,1}$ described by fundamental generating functions $%
\widetilde{H}$ and $\widetilde{L}$ which satisfy respectively: the
Hamilton-Jacobi equations written equivalently as 
\begin{equation*}
\frac{dx^{i}}{d\zeta }=\frac{\partial \widetilde{H}}{\partial p_{i}}%
\mbox{
and }\frac{dp_{i}}{d\zeta }=-\frac{\partial \widetilde{H}}{\partial x^{i}},
\end{equation*}%
or as the Euler-Lagrange equations, 
\begin{equation*}
\frac{d}{d\zeta }\frac{\partial \widetilde{L}}{\partial y^{i}}-\frac{%
\partial \widetilde{L}}{\partial x^{i}}=0.
\end{equation*}%
The last system of equations, in their turn, are equivalent to the nonlinear
geodesic (semi-spray) equations 
\begin{equation}
\frac{d^{2}x^{i}}{d\zeta ^{2}}+2\widetilde{G}^{i}(x,p)=0,\mbox{ for }%
\widetilde{G}^{i}=\frac{1}{2}\widetilde{g}^{ij}(\frac{\partial ^{2}%
\widetilde{L}}{\partial y^{i}}y^{k}-\frac{\partial \widetilde{L}}{\partial
x^{i}}),  \label{ngeqf}
\end{equation}
with $\widetilde{g}^{ij}$ being inverse to $\widetilde{g}_{ij}$ (\ref{hessls}%
). These equations state that point like probing particles move not along
usual geodesics as on Lorentz manifolds but follow some nonlinear geodesic
equations.

Using (\ref{ngeqf}), we can define a canonical N--connection in $L$--dual
form following formulas 
\begin{equation}
\ \ \ \ ^{\shortmid }\widetilde{\mathbf{N}}=\left\{ \ ^{\shortmid }%
\widetilde{N}_{ij}:=\frac{1}{2}\left[ \{\ \ ^{\shortmid }\widetilde{g}_{ij},%
\widetilde{H}\}-\frac{\partial ^{2}\widetilde{H}}{\partial p_{k}\partial
x^{i}}\ ^{\shortmid }\widetilde{g}_{jk}-\frac{\partial ^{2}\widetilde{H}}{%
\partial p_{k}\partial x^{j}}\ ^{\shortmid }\widetilde{g}_{ik}\right]
\right\} \mbox{ and }\widetilde{\mathbf{N}}=\left\{ \widetilde{N}_{i}^{a}:=%
\frac{\partial \widetilde{G}}{\partial y^{i}}\right\} .  \label{cartnc}
\end{equation}%
For general frame transforms on $\ ^{\shortmid }\mathcal{M}, \ ^{\shortmid}%
\widetilde{\mathbf{N}}\rightarrow \ ^{\shortmid }\mathbf{N}$ (\ref{ncon44})
and for dyadic constructions, $\ ^{\shortmid }\widetilde{\mathbf{N}}%
\rightarrow \ _{s}^{\shortmid }\mathbf{N}$ (\ref{ncon2222}). Any "tilde"
N-connection allows to define respective systems of N--adapted (co) frames
of type (\ref{nadapbdss}), when 
\begin{eqnarray}
\ ^{\shortmid }\widetilde{\mathbf{e}}_{\alpha } &=&(\ ^{\shortmid }%
\widetilde{\mathbf{e}}_{i}=\frac{\partial }{\partial x^{i}}-\ ^{\shortmid }%
\widetilde{N}_{ia}(x,p)\frac{\partial }{\partial p_{a}},\ ^{\shortmid }e^{b}=%
\frac{\partial }{\partial p_{b}}),\mbox{ on }T^{\ast }V;  \label{ccnadap} \\
\ \ ^{\shortmid }\widetilde{\mathbf{e}}^{\alpha } &=&(\ ^{\shortmid
}e^{i}=dx^{i},\ ^{\shortmid }\mathbf{e}_{a}=dp_{a}+\ ^{\shortmid }\widetilde{%
N}_{ia}(x,p)dx^{i})\mbox{ on }(T^{\ast }V)^{\ast }.  \notag
\end{eqnarray}

There are canonical d-metric structures (d-metrics) $\widetilde{\mathbf{g}}$
and $\ ^{\shortmid }\widetilde{\mathbf{g}}$ completely determined by
respective data $(\widetilde{L},\ \ \widetilde{\mathbf{N}};\widetilde{%
\mathbf{e}}_{\alpha },\widetilde{\mathbf{e}}^{\alpha };\widetilde{g}_{jk},%
\widetilde{g}^{jk})$ and/or $(\widetilde{H},\ ^{\shortmid }\widetilde{%
\mathbf{N}};\ ^{\shortmid }\widetilde{\mathbf{e}}_{\alpha },\ ^{\shortmid }%
\widetilde{\mathbf{e}}^{\alpha };\ \ ^{\shortmid }\widetilde{g}^{ab},\ \
^{\shortmid }\widetilde{g}_{ab}),$ 
\begin{eqnarray}
\widetilde{\mathbf{g}} &=&\widetilde{\mathbf{g}}_{\alpha \beta }(x,y)%
\widetilde{\mathbf{e}}^{\alpha }\mathbf{\otimes }\widetilde{\mathbf{e}}%
^{\beta }=\widetilde{g}_{ij}(x,y)e^{i}\otimes e^{j}+\widetilde{g}_{ab}(x,y)%
\widetilde{\mathbf{e}}^{a}\otimes \widetilde{\mathbf{e}}^{a}\mbox{
and/or }  \label{cdms8} \\
\ ^{\shortmid }\widetilde{\mathbf{g}} &=&\ ^{\shortmid }\widetilde{\mathbf{g}%
}_{\alpha \beta }(x,p)\ ^{\shortmid }\widetilde{\mathbf{e}}^{\alpha }\mathbf{%
\otimes \ ^{\shortmid }}\widetilde{\mathbf{e}}^{\beta }=\ \ ^{\shortmid }%
\widetilde{g}_{ij}(x,p)e^{i}\otimes e^{j}+\ ^{\shortmid }\widetilde{g}%
^{ab}(x,p)\ ^{\shortmid }\widetilde{\mathbf{e}}_{a}\otimes \ ^{\shortmid }%
\widetilde{\mathbf{e}}_{b}.  \label{cdmds8}
\end{eqnarray}

Using frame transforms, the d-metric structures [with tildes] (\ref{cdms8})
and (\ref{cdmds8}) can be written, respectively, in general d-metric forms
without tildes. General vierbein transforms can be parameterized
respectively as $e_{\alpha}= e_{\ \alpha }^{\underline{\alpha }}(u)\partial
/\partial u^{\underline{\alpha }}$ and $e^{\beta }=e_{\ \underline{\beta }%
}^{\beta }(u)du^{\underline{\beta }}$, where the local coordinate indices
are underlined in order to distinguish them from arbitrary abstract ones. In
such formulas, the matrix $e_{\ \underline{\beta }}^{\beta }$ is inverse to $%
e_{\ \alpha }^{\underline{\alpha }}$ for orthonormalized bases. For Hamilton
like configurations, one writes $\ ^{\shortmid }e_{\alpha }=\ ^{\shortmid
}e_{\ \alpha }^{\underline{\alpha }}(\ ^{\shortmid }u)\partial /\partial \
^{\shortmid }u^{\underline{\alpha }}$ and $\ ^{\shortmid }e^{\beta }=\
^{\shortmid }e_{\ \underline{\beta }}^{\beta }(\ ^{\shortmid }u)d\
^{\shortmid }u^{\underline{\beta }}.$ It should be noted that there are not
used boldface symbols for such transforms because we can consider arbitrary
decompositions. In particular, we can consider diadic 2+2+2+2 splitting
which is important for decoupling of nonlinear systems of physically
important PDEs. With respect to local coordinate frames, any d--metric
structures on $\mathbf{TV}$ and/or $\mathbf{T}^{\ast }\mathbf{V}$ (in
particular, we can consider Lagrange and/or Hamilton models) can be written
in the form%
\begin{eqnarray*}
\mathbf{g} &=&\mathbf{g}_{\alpha \beta }(x,y)\mathbf{e}^{\alpha }\mathbf{%
\otimes e}^{\beta }=g_{\underline{\alpha }\underline{\beta }}(x,y)du^{%
\underline{\alpha }}\mathbf{\otimes }du^{\underline{\beta }}\mbox{
and/or } \\
\ ^{\shortmid }\mathbf{g} &=&\ ^{\shortmid }\mathbf{g}_{\alpha \beta }(x,p)\
^{\shortmid }\mathbf{e}^{\alpha }\mathbf{\otimes \ ^{\shortmid }e}^{\beta
}=\ ^{\shortmid }g_{\underline{\alpha }\underline{\beta }}(x,p)d\
^{\shortmid }u^{\underline{\alpha }}\mathbf{\otimes }d\ ^{\shortmid }u^{%
\underline{\beta }},
\end{eqnarray*}%
when for respective frame transforms, $\mathbf{g}_{\alpha \beta }=e_{\
\alpha }^{\underline{\alpha }}e_{\ \beta }^{\underline{\beta }}g_{\underline{%
\alpha }\underline{\beta }}$ and $\ ^{\shortmid }\mathbf{g}_{\alpha \beta
}=\ ^{\shortmid }e_{\ \alpha }^{\underline{\alpha }}\ ^{\shortmid }e_{\
\beta }^{\underline{\beta }}\ ^{\shortmid }g_{\underline{\alpha }\underline{%
\beta }},$ there are obtained corresponding off-diagonal forms: 
\begin{eqnarray}
g_{\underline{\alpha }\underline{\beta }} &=&\left[ 
\begin{array}{cc}
g_{ij}(x)+g_{ab}(x,y)N_{i}^{a}(x,y)N_{j}^{b}(x,y) & g_{ae}(x,y)N_{j}^{e}(x,y)
\\ 
g_{be}(x,y)N_{i}^{e}(x,y) & g_{ab}(x,y)%
\end{array}%
\right] \mbox{
and/or }  \notag \\
\ ^{\shortmid }g_{\underline{\alpha }\underline{\beta }} &=&\left[ 
\begin{array}{cc}
\ ^{\shortmid }g_{ij}(x)+\ ^{\shortmid }g^{ab}(x,p)\ ^{\shortmid
}N_{ia}(x,p)\ ^{\shortmid }N_{jb}(x,p) & \ ^{\shortmid }g^{ae}\ ^{\shortmid
}N_{je}(x,p) \\ 
\ ^{\shortmid }g^{be}\ ^{\shortmid }N_{ie}(x,p) & \ ^{\shortmid
}g^{ab}(x,p)\ 
\end{array}%
\right] .  \label{offd}
\end{eqnarray}%
Such formulas get respective labels $s$ for the abstract geometric objects
or indices if we work in s-adapted variables. Parameterizations of type (\ref%
{offd}) are considered, for instance, in the Kaluza--Klein theory and
various string theories with extra dimension coordinates. Such metrics are
generic off-diagonal if the corresponding N-adapted structure is not
integrable, see (\ref{anholcond8}).

The formulas (\ref{offd}) can be written in "tilde" nonholonomic variables
which allows us to define, for instance, $\ ^{\shortmid }\tilde{g}_{%
\underline{\alpha }\underline{\beta }}$ as off-diagonal coefficients of (\ref%
{cdmds8}) with respective Hessian (\ref{hesshs}) and semi-spray N-connection
(\ref{cartnc}) on $\ \ ^{\shortmid }\widetilde{\mathcal{M}}$. We can
consider arbitrary frame transforms, $\ ^{\shortmid }\widetilde{\mathbf{g}}%
_{\alpha \beta }=\ ^{\shortmid }e_{\ \alpha }^{\underline{\alpha }}\
^{\shortmid }e_{\ \beta }^{\underline{\beta }}\ ^{\shortmid }\widetilde{g}_{%
\underline{\alpha }\underline{\beta }}$ and/or s-adapted ones, $\
^{\shortmid }\widetilde{\mathbf{g}}_{\alpha _{s}\beta _{s}}=\ ^{\shortmid
}e_{\ \alpha _{s}}^{\underline{\alpha }}\ ^{\shortmid }e_{\ \beta _{s}}^{%
\underline{\beta }}\ ^{\shortmid }\widetilde{g}_{\underline{\alpha }%
\underline{\beta }},$ and elaborate a Finsler-Hamilton model of phase space
with geometric data $( \ ^{\shortmid }\widetilde{\mathcal{M}}:\ ^{\shortmid }%
\widetilde{\mathbf{g}},\ \ ^{\shortmid }\widetilde{\mathbf{N}}) ,$ which
corresponding versions of generalized Einstein-Finsler-Hamilton equations
can be integrated in general forms for corresponding nonholonomic dyadic
configurations $( \ _{s}^{\shortmid }\widetilde{\mathcal{M}}:\
_{s}^{\shortmid }\widetilde{\mathbf{g}}, \ _{s}^{\shortmid }\widetilde{%
\mathbf{N}}) .$ In the next sections, we shall define corresponding
canonical s-connections and related Cartan-Finsler-Hamilton connections in
next sections.

\subsection{Almost K\"{a}hler Lagrange--Hamilton structures on phase spaces}

Spacetime models encoding MDRs and formulated as almost K\"{a}hler
geometries for relativistic phase space Lagrange-Hamilton configurations
were reviewed and studied in \cite{vmon3,vacaru18,bubuianu18a}. In those
works, further developments for almost symplectic (algebroid, commutative
and noncommutative) models of deformation and geometric quantization and
various geometric flow theories are reviewed. Originally, the main ideas on
almost K\"{a}hler realisation of Finsler and Lagrange geometry were proposed
in \cite{matsumoto66,matsumoto86,kern74}. We cite \cite{mdleon85} for a
review of standard approaches to modern geometric mechanics. In our
approach, fundamental (generating) Lagrange-Hamilton functions and MDRs
determine canonical models of almost K\"{a}hler geometry. Such nonholonomic
variables can be introduced in classical and quantum MGTs on (co) tangent
bundles.

Let us explain how MDRs (\ref{mdrg}) and related canonical N--connections $%
\widetilde{\mathbf{N}}$ and $\ ^{\shortmid }\widetilde{\mathbf{N}}$ define
respectively canonical almost complex structures $\widetilde{\mathbf{J}},$
on $\mathbf{TV},$ and $\ ^{\shortmid }\widetilde{\mathbf{J}},$ on $\mathbf{T}%
^{\ast }\mathbf{V}.$ We introduce the linear operator $\widetilde{\mathbf{J}}
$ acting on $\widetilde{\mathbf{e}}_{\alpha }=(\widetilde{\mathbf{e}}%
_{i},e_{b})$ in the form: $\widetilde{\mathbf{J}}(\mathbf{e}_{i})=-%
\widetilde{\mathbf{e}}_{n+i}$ and $\widetilde{\mathbf{J}}(e_{n+i})=%
\widetilde{\mathbf{e}}_{i}$. This defines globally an almost complex
structure ($\widetilde{\mathbf{J}}\mathbf{\circ \widetilde{\mathbf{J}}=-I}$
for $\mathbf{I}$ being the unity matrix) on $\mathbf{TV}$ completely
determined by a generating function $\widetilde{L}(x,y).$ Similarly, on $%
\mathbf{T}^{\ast }\mathbf{V,}$ we can consider a linear operator $\
^{\shortmid }\widetilde{\mathbf{J}}$ acting on$\ ^{\shortmid }\mathbf{e}%
_{\alpha }=(\ ^{\shortmid }\mathbf{e}_{i},\ ^{\shortmid }e^{b})$ (\ref%
{ccnadap}) when $\ ^{\shortmid }\widetilde{\mathbf{J}}(\ ^{\shortmid }%
\mathbf{e}_{i})=-\ ^{\shortmid }e^{n+i}$ and $\ ^{\shortmid }\widetilde{%
\mathbf{J}}(\ ^{\shortmid }e^{n+i})=\ ^{\shortmid }\mathbf{e}_{i}$. Such a $%
\ ^{\shortmid }\widetilde{\mathbf{J}}$ defines globally an almost complex
structure ($\ ^{\shortmid }\widetilde{\mathbf{J}}\ \circ \ ^{\shortmid }%
\widetilde{\mathbf{J}}= - \ ^{\shortmid }I $ for $\ ^{\shortmid }\mathbf{I}$
being the unity matrix) on $\mathbf{T} ^{\ast }\mathbf{V}$ completely
determined by a $\widetilde{H}(x,p).$ We note that $\widetilde{\mathbf{J}}$
and $\ ^{\shortmid }\widetilde{\mathbf{J}}$ are standard almost complex
structures only for the Euclidean signatures, respectively, on $\mathbf{TV}$
and $\mathbf{T}^{\ast }\mathbf{V}$. For the pseudo-Euclidean signature, we
define such operators in abstract geometric forms which needs additional
assumptions and results in a different type of relativistic classical and
quantum models. Considering arbitrary frame/coordinate transforms, we write $%
\mathbf{J}$ and $\ ^{\shortmid }\mathbf{J.}$

The canonical Neijenhuis tensor fields determined by MDRs and respective
Lagrange and Hamilton phase space structures and canonical almost complex
structures $\widetilde{\mathbf{J}}$ on $\mathbf{TV}$ and/or $\ ^{\shortmid }%
\widetilde{\mathbf{J}}$ on $\mathbf{T}^{\ast }\mathbf{V},$ are considered as
curvatures of respective N--connections: 
\begin{eqnarray}
\widetilde{\mathbf{\Omega }}(\widetilde{\mathbf{X}},\widetilde{\mathbf{Y}})
&:=&\mathbf{\ -[\widetilde{\mathbf{X}}\mathbf{,}\widetilde{\mathbf{Y}}]+[%
\widetilde{\mathbf{J}}\widetilde{\mathbf{X}},\widetilde{\mathbf{J}}%
\widetilde{\mathbf{Y}}]-\widetilde{\mathbf{J}}[\widetilde{\mathbf{J}}%
\widetilde{\mathbf{X}},\widetilde{\mathbf{Y}}]-\widetilde{\mathbf{J}}[%
\widetilde{\mathbf{X}},\widetilde{\mathbf{J}}\widetilde{\mathbf{Y}}]}%
\mbox{
and/or }  \notag \\
\ ^{\shortmid }\widetilde{\mathbf{\Omega }}(\ ^{\shortmid }\widetilde{%
\mathbf{X}},\ ^{\shortmid }\widetilde{\mathbf{Y}}) &:=&\ \mathbf{-[\
^{\shortmid }\widetilde{\mathbf{X}}\mathbf{,}\ ^{\shortmid }\widetilde{%
\mathbf{Y}}]+[\ ^{\shortmid }\widetilde{\mathbf{J}}\ ^{\shortmid }\widetilde{%
\mathbf{X}},\ ^{\shortmid }\widetilde{\mathbf{J}}\ ^{\shortmid }\widetilde{%
\mathbf{Y}}]-\ ^{\shortmid }\widetilde{\mathbf{J}}[\ ^{\shortmid }\widetilde{%
\mathbf{J}}\ ^{\shortmid }\widetilde{\mathbf{X}},\ ^{\shortmid }\widetilde{%
\mathbf{Y}}]-\ ^{\shortmid }\widetilde{\mathbf{J}}[\ ^{\shortmid }\widetilde{%
\mathbf{X}},\ ^{\shortmid }\widetilde{\mathbf{J}}\ ^{\shortmid }\widetilde{%
\mathbf{Y}}],}  \label{neijt}
\end{eqnarray}%
for any d--vectors $\mathbf{X,}$ $\mathbf{Y}$ and $\ ^{\shortmid }\mathbf{%
X,\ ^{\shortmid }Y.}$ In general frame/coordinates, the curvatures (\ref%
{neijt}) can be written in general form without tilde values and/or in index
form: 
\begin{equation*}
\Omega _{ij}^{a}=\frac{\partial N_{i}^{a}}{\partial x^{j}}-\frac{\partial
N_{j}^{a}}{\partial x^{i}}+N_{i}^{b}\frac{\partial N_{j}^{a}}{\partial y^{b}}%
-N_{j}^{b}\frac{\partial N_{i}^{a}}{\partial y^{b}},\mbox{\ or\ }\ \ \mathbf{%
\ ^{\shortmid }}\Omega _{ija}=\frac{\partial \mathbf{\ ^{\shortmid }}N_{ia}}{%
\partial x^{j}}-\frac{\partial \mathbf{\ ^{\shortmid }}N_{ja}}{\partial x^{i}%
}+\ \mathbf{^{\shortmid }}N_{ib}\frac{\partial \mathbf{\ ^{\shortmid }}N_{ja}%
}{\partial p_{b}}-\mathbf{\ ^{\shortmid }}N_{jb}\frac{\partial \mathbf{\
^{\shortmid }}N_{ia}}{\partial p_{b}}.
\end{equation*}%
We obtain almost complex structures $\mathbf{J}$ and $\ ^{\shortmid }\mathbf{%
J}$ transform into standard complex structures for Euclidean signatures if $%
\mathbf{\Omega }=0$ and/or $\ ^{\shortmid }\mathbf{\Omega }=0.$

The priority of the Lagrange-Hamilton models is that they allow certain
equivalent descriptions as almost symplectic geometries which can be used,
for instance, for performing deformation quantization. Almost symplectic
structures on $\mathbf{TV}$ and $\mathbf{T}^{\ast }\mathbf{V}$ are defined
by respective nondegenerate N-adapted 2--forms 
\begin{equation*}
\ \theta =\frac{1}{2}\ \theta _{\alpha \beta }(u)\ \mathbf{e}^{\alpha
}\wedge \mathbf{e}^{\beta }\mbox{ and }\ \ ^{\shortmid }\theta =\frac{1}{2}\
\ ^{\shortmid }\theta _{\alpha \beta }(\ ^{\shortmid }u)\ \ ^{\shortmid }%
\mathbf{e}^{\alpha }\wedge \ ^{\shortmid }\mathbf{e}^{\beta }.
\end{equation*}%
For instance, in h-c components, 
\begin{equation}
\mathbf{\ }\ ^{\shortmid }\theta =\frac{1}{2}\mathbf{\ }\ ^{\shortmid
}\theta _{ij}(\ ^{\shortmid }u)e^{i}\wedge e^{j}+\frac{1}{2}\mathbf{\ }\
^{\shortmid }\theta ^{ab}(\ ^{\shortmid }u)\mathbf{\ ^{\shortmid }e}%
_{a}\wedge \mathbf{\ }\ ^{\shortmid }\mathbf{e}_{b}.  \label{aux03}
\end{equation}

We emphasize that a N--connection $\ ^{\shortmid }\mathbf{N}$ defines a
unique decomposition of a d--vector $\ ^{\shortmid }\mathbf{X=\ }X^{h}+\
^{\shortmid }X^{cv}$ on $T^{\ast }\mathbf{V},$ for $\mathbf{\ }X^{h}=h\
^{\shortmid }\mathbf{X}$ and $\mathbf{\ }^{\shortmid }X^{cv}=cv\ \
^{\shortmid }\mathbf{X.}$ Respective projectors $h$ and $cv$ are related to
a dual distribution $\ ^{\shortmid }\mathbf{N}$ on $\mathbf{V,}$ when the
properties $h+cv=\mathbf{I},h^{2}=h,(cv)^{2}=cv,h\circ cv=cv\circ h=0$ are
satisfied. The almost product operator $\ ^{\shortmid }\mathbf{P:=}%
I-2cv=2h-I $ acting on $\ ^{\shortmid }\mathbf{e}_{\alpha }=(\ ^{\shortmid }%
\mathbf{e}_{i},\ ^{\shortmid }e^{b})$ is defined by formulas 
\begin{equation*}
\ \mathbf{\ }^{\shortmid }\mathbf{P}(\ ^{\shortmid }\mathbf{e}_{i})=\
^{\shortmid }\mathbf{e}_{i}\mbox{\ and  }\ ^{\shortmid }\mathbf{P}(\
^{\shortmid }e^{b})=-\ \ ^{\shortmid }e^{b}.
\end{equation*}%
In a similar form, a N--connection $\ \mathbf{N}$ induces an almost product
structure $\mathbf{P}$ on $T\mathbf{V}.$

Another important d-geometric operators are the almost tangent (co) ones
constructed to satisfy the conditions 
\begin{eqnarray*}
\mathbb{J(}\mathbf{e}_{i}\mathbb{)} &=&e_{4+i}\mbox{\ and
\ }\ \mathbb{J}\left( e_{a}\right) =0,\mbox{ \ or \ }\mathbb{J=}\frac{%
\partial }{\partial y^{i}}\otimes dx^{i}; \\
\mathbf{\ }^{\shortmid }\mathbb{J}(\mathbf{\ }^{\shortmid }\mathbf{e}_{i}%
\mathbb{)} &=&\mathbf{\ }^{\shortmid }g_{ib}\mathbf{\ }^{\shortmid }e^{b}%
\mbox{\ and
\ }\ \mathbf{\ }^{\shortmid }\mathbb{J}\left( \mathbf{\ }^{\shortmid
}e^{b}\right) =0,\mbox{ \ or \ }\mathbf{\ }^{\shortmid }\mathbb{J=}\mathbf{\ 
}^{\shortmid }g_{ia}\frac{\partial }{\partial p_{a}}\otimes dx^{i}.
\end{eqnarray*}%
Above introduced d-operators $\ ^{\shortmid }\mathbf{P,}\ \mathbf{\ }%
^{\shortmid }\mathbf{J}$ and $\ ^{\shortmid }\mathbb{J}$ are respectively $%
\mathcal{L}$--dual to $\ \mathbf{P,}\ \mathbf{J}$ and $\ \mathbb{J}$ if and
only if $\ \mathbf{\ }^{\shortmid }\mathbf{N}$ and $\ \mathbf{N}$ are $%
\mathcal{L}$--dual and there are constructed respective (co) frame
transforms to canonical values $[\mathbf{\ }^{\shortmid }\widetilde{\mathbf{P%
}}\mathbf{,}\ \mathbf{\ }^{\shortmid }\widetilde{\mathbf{J}}\mathbf{,\ }%
^{\shortmid }\widetilde{\mathbb{J}}]$ and $[\widetilde{\mathbf{P}}\mathbf{,}%
\widetilde{\mathbf{J}}\mathbf{,}\widetilde{\mathbb{J}}].$ We can verify by
straightforward computations that there are satisfied for pairs of $\mathcal{%
L}$--dual N--connections $\left( \mathbf{N,\ \mathbf{\ }^{\shortmid }N}%
\right) $ the properties: 
\begin{equation*}
\mathbf{J}=-\delta _{i}^{a}e_{a}\otimes e^{i}+\delta _{a}^{i}\mathbf{e}%
_{i}\otimes \mathbf{e}^{a},\ \ \mathbf{\ }^{\shortmid }\mathbf{J}=-\
^{\shortmid }g_{ia}\ ^{\shortmid }e^{a}\otimes \ ^{\shortmid }e^{i}+\
^{\shortmid }g^{ia}\ ^{\shortmid }\mathbf{e}_{i}\otimes \ ^{\shortmid }%
\mathbf{e}_{a}
\end{equation*}%
correspond to a $\mathcal{L}$--dual pair of almost complex structures $%
\left( \mathbf{J},\ ^{\shortmid }\mathbf{J}\right) ;$ 
\begin{equation*}
\mathbf{P}=\mathbf{e}_{i}\otimes e^{i}-e_{a}\otimes \mathbf{e}^{a},\ \
^{\shortmid }\mathbf{P}=\ ^{\shortmid }\mathbf{e}_{i}\otimes \ ^{\shortmid
}e^{i}-\ ^{\shortmid }e^{a}\otimes \ ^{\shortmid }\mathbf{e}_{a}
\end{equation*}%
correspond to a $\mathcal{L}$--dual pair of almost product structures $%
\left( \mathbf{P},\ \ ^{\shortmid }\mathbf{P}\right) $, and respective
almost symplectic structures 
\begin{equation}
\theta =g_{aj}(x,y)\mathbf{e}^{a}\wedge e^{i}\mbox{ and }\ ^{\shortmid
}\theta =\delta _{i}^{a}\ ^{\shortmid }\mathbf{e}_{a}\wedge \ ^{\shortmid
}e^{i}  \label{sympf}
\end{equation}%
Such operators can be re-written in canonical form by considering canonical
N-adapted bases with tilde, for instance, we can write (\ref{sympf}) (using
frame transforms) as $\widetilde{\theta }=\widetilde{g}_{aj}(x,y)\widetilde{%
\mathbf{e}}^{a}\wedge e^{i}$ and $\ ^{\shortmid }\widetilde{\theta }=\delta
_{i}^{a}\ ^{\shortmid }\widetilde{\mathbf{e}}_{a}\wedge \ ^{\shortmid
}e^{i}. $ For instance, a N-connection $\mathbf{\ }^{\shortmid }\widetilde{%
\mathbf{N}}$ and/or $\mathbf{\ }_{s}^{\shortmid }\widetilde{\mathbf{N}}\ $%
can be used to N-elongate corresponding frames and define tilde data $\left( 
\mathbf{\ }^{\shortmid }\widetilde{\mathbf{J}},\mathbf{\ }^{\shortmid }%
\widetilde{\mathbb{J}},\ ^{\shortmid }\widetilde{\mathbf{P}},^{\shortmid }%
\widetilde{\theta }\right) $ and re-define them for nonholonomic dyadic
splitting of type $\left( \mathbf{\ }_{s}^{\shortmid }\widetilde{\mathbf{J}},%
\mathbf{\ }_{s}^{\shortmid }\widetilde{\mathbb{J}},\ _{s}^{\shortmid }%
\widetilde{\mathbf{P}},_{s}^{\shortmid }\widetilde{\theta }\right) .$

For modeling of (co) tangent bundle N-connection and almost symplectic
geometries on (co) tangent bundles of total dimension 8, we can formulate
the phase space nonholonomic geometry as an almost Hermitian model of a
tangent Lorentz bundle $T\mathbf{V}$\ equipped with a N--connection
structure $\mathbf{N}$\ is defined by a triple $\mathbf{H}^{8}=(T\mathbf{V}%
,\theta ,\mathbf{J}),$ where $\theta \mathbf{(X,Y)}:=\mathbf{g}\left( 
\mathbf{JX,Y}\right) .$ On a cotangent Lorentz bundle $T^{\ast }\mathbf{V}$
with a (or $\ ^{\shortmid }\mathbf{N}),$ we can define a triple $\
^{\shortmid }\mathbf{H}^{8}=(T^{\ast }\mathbf{V},\ ^{\shortmid }\theta ,\
^{\shortmid }\mathbf{J}),$ where $\ ^{\shortmid }\theta (\ ^{\shortmid }%
\mathbf{X},\ ^{\shortmid }\mathbf{Y}):=\ ^{\shortmid }\mathbf{g}\left( \
^{\shortmid }\mathbf{J}\ ^{\shortmid }\mathbf{X},\ ^{\shortmid }\mathbf{Y}%
\right) ).$ A space $\mathbf{H}^{8}$ (or $\ ^{\shortmid }\mathbf{H}^{8})$ is
almost K\"{a}hler and denoted $\mathbf{K}^{8}$ if $d\ \theta =0$ (or $\
^{\shortmid }\mathbf{K}^{8}$ if $d\ ^{\shortmid }\theta =0).$ In tilde
variables with 1--forms, respectively, defined by a regular Lagrangian $L$
and Hamiltonian $H$ (related by a Legendre transform), we have $\widetilde{%
\omega }=\frac{\partial L}{\partial y^{i}}e^{i}$ and $\ ^{\shortmid }%
\widetilde{\omega }=p_{i}dx^{i},$ for which $\widetilde{\theta }=d\widetilde{%
\omega }$ and $\ ^{\shortmid }\widetilde{\theta }=d\ ^{\shortmid }\widetilde{%
\omega }$. As a result, we get that $d\widetilde{\theta }=0$ and $d\
^{\shortmid }\widetilde{\theta }=0.$ If such conditions are satisfied, for
instance, for $\mathbf{\ }^{\shortmid }\widetilde{\mathbf{N}},$ we can
consider arbitrary or nonholonomic dyadic structures with $\mathbf{\ }%
_{s}^{\shortmid }\widetilde{\mathbf{N}}$ and $d\mathbf{\ }_{s}\widetilde{%
\theta }=0$ and $d\ \mathbf{\ }_{s}^{\shortmid }\widetilde{\theta }=0.$ But
such properties do not hold true for arbitrary $\mathbf{\ }^{\shortmid }%
\mathbf{N}$ and $\ ^{\shortmid }\theta ,$ when, in general, $d\ ^{\shortmid
}\theta \neq 0.$ We have to introduce a special distribution $\mathbf{\ }%
^{\shortmid }\widetilde{\mathbf{N}}$ determined by a $\widetilde{H}.$ Such
geometric and physical objects may have certain geometric or physical
motivation but can be also prescribed to define a subclass of N-elongated
frames $\ ^{\shortmid }\widetilde{\mathbf{e}}_{\alpha }=(\ ^{\shortmid }%
\widetilde{\mathbf{e}}_{i},\ ^{\shortmid }e^{b}).$

\subsection{Canonical and Lagrange-Hamilton connections and curvatures}

We are not able to motivate and elaborate on self-consistent phase space
generalizations of the Einstein gravity if we work only with Finsler like
metrics determined by nonlinear quadratic forms $L(x,y)$ (\ref{nqe}) and/or $%
H(x,p)$ (\ref{nqed}) (or with arbitrary nonholonomic fibered 4+4 structures)
and (co) vector/ tangent bundles. Viable Lagrange-Hamilton theories encoding
(\ref{mdrg})\ can be formulated on $\mathbf{TV}$ and $\mathbf{T}^{\ast }%
\mathbf{V}$ for additional assumptions on choosing certain types of
N-connection and d-connection structures. This is different from the
geometry of a (pseudo) Riemannian spacetime $(V,\{g_{ij}(x)\}),$ which is
completely determined by its metric structure $\{g_{is}\}$; and when the
Levi-Civita (LC) connection, $\nabla $, is uniquely defined by $\{g_{is}\}.$
In non-relativistic form, there were developed certain approaches related to
Finsler geometry and semi-spray configurations \cite{shen01,shen01a}, where
the priority was given to the Chern-Rund connection for Finsler spaces. In
another class of Finsler gravity models, the priority was given to the
Berwald connection in Finsler geometry \cite{schreck15,schreck15a,schreck16}%
. Such a d-connection is not compatible with the metric structure on the
total bundle. This creates a number of ambiguities related to elaborating
metric noncompatible Finsler gravity theories (including definition of
spinors, definition of compatible motion equations and conservation laws),
see explicit results, critics and discussions in Refs. \cite%
{vmon3,vacaru18,bubuianu18a,partner01,partner02}.

\subsubsection{N-adapted distortions of s-connections, and s-curvatures}

On phase spaces, we can elaborate on geometric models with affine (linear)
connections and respective covariant derivatives in certain forms which are,
or not, adapted to a chosen N--connection structure. Such constructions can
be defined for a general nonholonomic 4+4 splitting.

A distinguished connection (d--connection) can be defined as a linear
connection $\mathbf{D}$ on $\mathbf{TV}$ (or $\ ^{\shortmid }\mathbf{D}$ on $%
\mathbf{T}^{\ast }\mathbf{V})$ which is compatible with the almost product
structure $\mathbf{DP}=0$ (or $\ ^{\shortmid }\mathbf{D\ ^{\shortmid }P}=0)$%
. In equivalent form, such a d--connection can be defined to preserve under
parallelism a respective N--connection splitting (\ref{ncon44}), which can
be prescribed to be a more special N-connection $\ ^{\shortmid }\mathbf{%
\tilde{N}}$ and then related to a nonholonomic dyadic decomposition (\ref%
{ncon2222}).

For instance, the coefficients of d--connection $\ ^{\shortmid }\mathbf{D}$
can be defined with respect to N--adapted frames (\ref{nadapbdss}) using
equations%
\begin{equation*}
\ \ ^{\shortmid }\mathbf{D}_{\ ^{\shortmid }\mathbf{e}_{k}}\ ^{\shortmid }%
\mathbf{e}_{j}:=\ ^{\shortmid }L_{\ jk}^{i}\ ^{\shortmid }\mathbf{e}_{i},\
^{\shortmid }\mathbf{D}_{\mathbf{e}_{k}}\ ^{\shortmid }e^{b}:=-\ ^{\shortmid
}\acute{L}_{a\ k}^{\ b}\ ^{\shortmid }e^{a},\ ^{\shortmid }\mathbf{D}_{\
^{\shortmid }e^{c}}\ ^{\shortmid }\mathbf{e}_{j}:=\ ^{\shortmid }\acute{C}%
_{\ j}^{i\ c}\ ^{\shortmid }\mathbf{e}_{i},\ ^{\shortmid }\mathbf{D}_{\
^{\shortmid }e^{c}}\ ^{\shortmid }e^{b}:=-\ ^{\shortmid }C_{a}^{\ bc}\
^{\shortmid }e^{a}.
\end{equation*}%
Using respective labeling of h- and v-indices, such equations can be
considered for $\mathbf{D.}$ In brief, the N-adapted coefficients of
d-connections on a cotangent Lorentz bundles can be respectively
parameterized 
\begin{equation}
\mathbf{\Gamma }_{\ \beta \gamma }^{\alpha }=\{L_{\ jk}^{i},\acute{L}_{\
bk}^{a},\acute{C}_{\ jc}^{i},C_{\ bc}^{a}\}\mbox{ and }\ ^{\shortmid }%
\mathbf{\Gamma }_{\ \beta \gamma }^{\alpha }=\{\ ^{\shortmid }L_{\ jk}^{i},\
^{\shortmid }\acute{L}_{a\ k}^{\ b},\ ^{\shortmid }\acute{C}_{\ j}^{i\ c},\
^{\shortmid }C_{a}^{\ bc}\},  \label{componscon}
\end{equation}%
which allows to define h-- and c-splitting of covariant derivatives $\
^{\shortmid }\mathbf{D}=\left( \ _{h}^{\shortmid }\mathbf{D,\ }%
_{v}^{\shortmid }\mathbf{D}\right) ,$ where $\ _{h}\mathbf{D}= \{L_{\
jk}^{i}, \acute{L}_{\ bk}^{a}\},\ $ and $\ _{c}^{\shortmid }\mathbf{D}=\{\
^{\shortmid }\acute{C}_{\ j}^{i\ c},\ ^{\shortmid }C_{a}^{\ bc}\}.$ For
dyadic decompositions, the symbols of geometric objects and/or indices of
such objects are labelled additionally with a shell label, for instance, $\
_{cs}^{\shortmid }\mathbf{D}=\{\ ^{\shortmid }\acute{C}_{\ j_{s}}^{i_{s}\
c_{s}},\ ^{\shortmid }C_{a_{s}}^{\ b_{s}c_{s}}\},$ when, for instance, $%
j_{2}=1,2,3,4$ and $a_{3}=5,6.$ In such case, we use the terms s-connection
instead of d-connection (respectively, s-tensor instead of d-tensor). In
result, we formulate 8-d s-adapted phase space variants of (4-d) formulas (%
\ref{hvdcon}), (\ref{fundgeom}) and, respectively, (\ref{dcurv}), (\ref%
{dtors}) and (\ref{dnonm}). All higher dimension formulas on $\mathcal{M}$
and/or $\ ^{\shortmid }\mathcal{M}$ can be proven in abstract geometric and
s-adapted forms. We omit such details in this work.

\subsubsection{Physically important Filsler-Lagrange-Hamilton and canonical
d-connections}

For elaborating classical and quantum MGTs, and alternative geometrization
of mechanics and nonholonomic geometric flow theories \cite{vmon3}, we can
consider more specials classes of d--connections which can be defined
completely by a d-metric/ almost symplectic structure determined by a
respective Lagrange-Finsler and/or Hamilton-Cartan fundamental form.

The almost K\"{a}hler-Lagrange and/or almost K\"{a}hler-Hamilton phase
spaces (determined, or not, by respective MDRs (\ref{mdrg}) and a possible $%
\mathcal{L}$--duality) are characterized respectively by such geometric and
physically important linear connections and canonical/ almost symplectic
connections: 
\begin{eqnarray}
\lbrack \mathbf{g,N]} &\mathbf{\simeq }&\mathbf{[}\widetilde{\mathbf{g}},%
\widetilde{\mathbf{N}}]\mathbf{\simeq \lbrack }\widetilde{\theta }:=%
\widetilde{\mathbf{g}}(\widetilde{\mathbf{J}}\cdot ,\cdot ),\widetilde{%
\mathbf{P}}\mathbf{,}\widetilde{\mathbf{J}}\mathbf{,}\widetilde{\mathbb{J}}]
\label{canondcl} \\
&\Longrightarrow &\left\{ 
\begin{array}{ccccc}
\nabla : &  & \nabla \mathbf{g}=0;\ \mathbf{T[\nabla ]}=0, &  & %
\mbox{Lagrange LC--connection}; \\ 
\widehat{\mathbf{D}}: &  & \widehat{\mathbf{D}}\ \mathbf{g}=0;\ h\widehat{%
\mathbf{T}}=0,\ v\widehat{\mathbf{T}}=0. &  & 
\mbox{canonical Lagrange
d-connection}; \\ 
\widetilde{\mathbf{D}}: &  & \widetilde{\mathbf{D}}\widetilde{\theta }=0,%
\widetilde{\mathbf{D}}\widetilde{\theta }=0 &  & 
\mbox{almost symplectic
Lagrange d-connection.};%
\end{array}%
\right.  \notag
\end{eqnarray}%
and/or 
\begin{eqnarray}
\lbrack \ ^{\shortmid }\mathbf{g,\ ^{\shortmid }N]} &\mathbf{\simeq }&%
\mathbf{[}\ ^{\shortmid }\widetilde{\mathbf{g}},\ ^{\shortmid }\widetilde{%
\mathbf{N}}]\mathbf{\simeq \lbrack }\ ^{\shortmid }\widetilde{\theta }:=\
^{\shortmid }\widetilde{\mathbf{g}}(\ ^{\shortmid }\widetilde{\mathbf{J}}%
\cdot ,\cdot ),\ ^{\shortmid }\widetilde{\mathbf{P}},\ ^{\shortmid }%
\widetilde{\mathbf{J}},\ ^{\shortmid }\widetilde{\mathbb{J}}]
\label{canondch} \\
&\Longrightarrow &\left\{ 
\begin{array}{ccccc}
\ ^{\shortmid }\nabla : &  & \ ^{\shortmid }\nabla \ ^{\shortmid }\mathbf{g}%
=0;\ \ ^{\shortmid }\mathbf{T[\ ^{\shortmid }\nabla ]}=0, &  & %
\mbox{Hamilton LC-connection}; \\ 
\ ^{\shortmid }\widehat{\mathbf{D}}: &  & \ ^{\shortmid }\widehat{\mathbf{D}}%
\ \mathbf{g}=0;\ h\ ^{\shortmid }\widehat{\mathbf{T}}=0,\ cv\ ^{\shortmid }%
\widehat{\mathbf{T}}=0. &  & \mbox{canonical Hamilton d-connection}; \\ 
\ ^{\shortmid }\widetilde{\mathbf{D}}: &  & \ ^{\shortmid }\widetilde{%
\mathbf{D}}\ ^{\shortmid }\widetilde{\theta }=0,\ ^{\shortmid }\widetilde{%
\mathbf{D}}\ ^{\shortmid }\widetilde{\theta }=0 &  & 
\mbox{almost symplectic
Hamilton d-connection.}%
\end{array}%
\right.  \notag
\end{eqnarray}

The formulas (\ref{canondcl}) and (\ref{canondch}) consist 8-d
Lagrange-Hamilton analogs of the 4-d canonical d-connection and
LC-connection structure (\ref{twocon}). We can use different "tilde, hat,
shell, duality" and other type labels for corresponding s-connections $\
^{\shortmid }\widetilde{\mathbf{D}},\ ^{\shortmid }\widehat{\mathbf{D}},$ or 
$\ ^{\shortmid }\nabla $. Respective fundamental s-tensor objects are
labeled with respective "tilde, hat,...", for instance $\ _{\nabla }\mathcal{%
R}=\{\ _{\nabla }R_{\ \beta \gamma \delta }^{\alpha }\},  \ ^{\shortmid }%
\widehat{\mathcal{R}}=\{\ ^{\shortmid }\widehat{\mathbf{R}}_{\ \beta \gamma
\delta }^{\alpha }\}, \ ^{\shortmid }\widetilde{\mathcal{R}}=\{\ ^{\shortmid
}\widetilde{\mathbf{R}}_{\ \beta \gamma \delta }^{\alpha }\} $ etc. \ Such
s-tensors can be related via distortion relations and nonholonomic frame
transforms. To derive exact and parametric solutions in certain geometric
flow and MGTs is important to transform, for instance, a Ricci tensor $\
_{\nabla }R_{\ \beta \gamma },$ or d-tensor $\ ^{\shortmid }\widetilde{%
\mathbf{R}}_{\ \beta \gamma },$ into a  $\ ^{\shortmid }\widehat{\mathbf{R}}%
_{\ \beta \gamma }.$ The priority of the canonical d-connection $\
^{\shortmid }\widehat{\mathbf{D}}$ is that it can be written in s-form as a $%
\ _{s}^{\shortmid }\widehat{\mathbf{D}}$ when $\ ^{\shortmid }\widehat{%
\mathbf{R}}_{\ \beta \gamma }\rightarrow \ ^{\shortmid }\widehat{\mathbf{R}}%
_{\ \beta _{s}\gamma _{s}}$. In dyadic form, corresponding physically
important equations can be decoupled for certain general off-diagonal
metrics (quasi-stationary, or locally anisotropic ones). Details on such
transforms are provided in (nonassociative) Finsler-Lagrange-Hamilton forms
in Refs. \cite{vmon3,vacaru18,bubuianu18a,partner01,partner02} for
respective generalizations of the Finsler geometry (in relativistic
Lagrange-Hamilton forms and dropping the condition of homogeneity used in
Finsler geometry).

It is well-known Chern's definition \cite{chern48,bao00} that Finsler
geometry is an example of geometry when the assumption on quadratic linear
elements is dropped. But this is not enough for constructing physically
viable Finser generalizations of the Einstein gravity theory. We need
certain additional assumptions for elaborating self-consistent geometric
constructions determined by nonlinear quadratic line elements. Here we note
that the first self-consistent model of Finsler geometry (with local
geometric constructions with generalized metric, N-connection and
d-connection structures, and associated N-frames) was elaborated by E.
Cartan \cite{cartan35}, see and citations therein. In those works, there
were defined thee coordinate transforms of nonlinear and linear connections.
The original constructions with nonlinear quadratic elements were elaborated
in the famous habilitation thesis of B. Riemann \cite{riem1854}, but E.
Cartan introduced the term of Finsler geometry using the original work \cite%
{finsler18} and completing the Finsler geometry with the concepts of
N-connection and Cartan d-connection. Conventionally, that model of
Finsler-Cartan geometry, which is metric compatible, can be described on
tangent bundles (or on manifolds with fibred structure) by a triple of
fundamental geometric structures $(F:\widetilde{\mathbf{g}},\widetilde{%
\mathbf{N}},\widetilde{\mathbf{D}}).$ Such constructions can be performed in
relativistic forms on a phase space $\widetilde{\mathcal{M}}$ and re-defined
in equivalent dual form on $\ ^{\shortmid }\widetilde{\mathcal{M}}$ if we
use momentum like coordinates. J. Kern \cite{kern74} defined the Lagrange
geometry as a model Finsler geometry without homogeneity conditions, which
also can be re-defined (using more sophisticate geometric constructions) on $%
\ ^{\shortmid }\mathcal{M}$ as respective Hamilton and Cartan (phase) spaces.

\subsubsection{Distortion s-tensors and curvature and Ricci s-tensors}

There are unique distortion relations for any type of prescribed canonical
d-connection, or the Cartan d-connection, and LC-connection: 
\begin{eqnarray}
\widehat{\mathbf{D}} &=&\nabla +\widehat{\mathbf{Z}},\widetilde{\mathbf{D}}%
=\nabla +\widetilde{\mathbf{Z}},\mbox{ and }\widehat{\mathbf{D}}=\widetilde{%
\mathbf{D}}+\mathbf{Z,}\mbox{  determined by }(\mathbf{g,N)};  \notag \\
&&\mbox{ and }\ _{s}\widehat{\mathbf{D}}=\widetilde{\mathbf{D}}+\ _{s}%
\mathbf{Z,}\mbox{ 
determined by }(\ _{s}\mathbf{g,\ _{s}N)};  \label{candistr} \\
\ ^{\shortmid }\widehat{\mathbf{D}} &=&\ ^{\shortmid }\nabla +\ ^{\shortmid }%
\widehat{\mathbf{Z}},\ ^{\shortmid }\widetilde{\mathbf{D}}=\ ^{\shortmid
}\nabla +\ ^{\shortmid }\widetilde{\mathbf{Z}},\mbox{ and }\ ^{\shortmid }%
\widehat{\mathbf{D}}=\ ^{\shortmid }\widetilde{\mathbf{D}}+\ ^{\shortmid }%
\mathbf{Z,}\mbox{ determined by }(\ ^{\shortmid }\mathbf{g,\ ^{\shortmid }N)}%
;  \notag \\
&&\mbox{ and }\ _{s}^{\shortmid }\widehat{\mathbf{D}}=\ ^{\shortmid }%
\widetilde{\mathbf{D}}+\ _{s}^{\shortmid }\mathbf{Z,}\mbox{  determined by }%
(\ _{s}^{\shortmid }\mathbf{g,\ _{s}^{\shortmid }N)},  \notag
\end{eqnarray}%
for distortion s- and d-tensors $\widehat{\mathbf{Z}},\widetilde{\mathbf{Z}}%
, $ and $\mathbf{Z;}$ and $\ ^{\shortmid }\widehat{\mathbf{Z}},\ ^{\shortmid
}\widetilde{\mathbf{Z}},$ and $\ ^{\shortmid }\mathbf{Z}$ etc. For such
formulas, we can associate some MDRs (\ref{mdrg}) (this is important for
elaborating physical models, but not obligatory for geometric constructions)
are characterized by respective canonical and/or almost symplectic
distortion d-tensors $\widehat{\mathbf{Z}}[\widetilde{\mathbf{g}},\widetilde{%
\mathbf{N}}],\widetilde{\mathbf{Z}}[\widetilde{\mathbf{g}},\widetilde{%
\mathbf{N}}],$ and $\mathbf{Z}[\widetilde{\mathbf{g}},\widetilde{\mathbf{N}}%
],$ for (almost symplectic) Lagrange models, and $\ ^{\shortmid }\widehat{%
\mathbf{Z}}[\ ^{\shortmid }\widetilde{\mathbf{g}},\ ^{\shortmid }\widetilde{%
\mathbf{N}}],\ ^{\shortmid }\widetilde{\mathbf{Z}}[\ ^{\shortmid }\widetilde{%
\mathbf{g}},\ ^{\shortmid }\widetilde{\mathbf{N}}],$ and $\ ^{\shortmid }%
\mathbf{Z}[\ ^{\shortmid }\widetilde{\mathbf{g}},\ ^{\shortmid }\widetilde{%
\mathbf{N}}],$ for (almost symplectic) Hamilton models.

Using distortions relations (\ref{candistr}), the phase space geometry can
be described in different equivalent forms (up to respective nonholonomic
deformations of the linear connection and s-connection structures and
nonholonomic frame transforms) by such data{\small 
\begin{equation}
\begin{array}{ccccc}
\mbox{MDRs} & \nearrow & (\mathbf{g,N,}\widehat{\mathbf{D}})\leftrightarrows
(L:\widetilde{\mathbf{g}}\mathbf{,}\widetilde{\mathbf{N}},\widetilde{\mathbf{%
D}}) & \leftrightarrow (\widetilde{\theta },\widetilde{\mathbf{P}},%
\widetilde{\mathbf{J}},\widetilde{\mathbb{J}},\widetilde{\mathbf{D}}) & 
\leftrightarrow \lbrack (\mathbf{g[}N],\nabla )],\mbox{ on }\mathbf{TV} \\ 
\mbox{indicator }\varpi &  & \updownarrow \mbox{ possible }\mathcal{L}%
\mbox{-duality }\& & \mbox{symplectomorphisms } & \updownarrow 
\mbox{ not
N-adapted } \\ 
\mbox{ see (\ref{mdrg})} & \searrow & (\ ^{\shortmid }\mathbf{g,\
^{\shortmid }N,}\ ^{\shortmid }\widehat{\mathbf{D}})\leftrightarrows (H:\
^{\shortmid }\widetilde{\mathbf{g}},\ ^{\shortmid }\widetilde{\mathbf{N}},\
^{\shortmid }\widetilde{\mathbf{D}}) & \leftrightarrow (\ ^{\shortmid }%
\widetilde{\theta },\ ^{\shortmid }\widetilde{\mathbf{P}},\ ^{\shortmid }%
\widetilde{\mathbf{J}},\ ^{\shortmid }\widetilde{\mathbb{J}},\ ^{\shortmid }%
\widetilde{\mathbf{D}}) & \leftrightarrow \lbrack (\ ^{\shortmid }\mathbf{g}%
[\ ^{\shortmid }N],\ ^{\shortmid }\nabla )],\mbox{on}\mathbf{T}^{\ast }%
\mathbf{V}.%
\end{array}
\label{phspgd}
\end{equation}%
}

We can prove in abstract and N-adapted forms that there are canonical
distortion relations for respective Lagrange-Finsler nonholonomic variables:

\begin{eqnarray*}
\widehat{\mathcal{R}}[\mathbf{g},\widehat{\mathbf{D}} &=&\nabla +\widehat{%
\mathbf{Z}}]=\mathcal{R}[\mathbf{g},\nabla ]+\widehat{\mathcal{Z}}[\mathbf{g}%
,\widehat{\mathbf{Z}}], \\
\ ^{\shortmid }\widehat{\mathcal{R}}[\ ^{\shortmid }\mathbf{g},\ ^{\shortmid
}\widehat{\mathbf{D}} &=&\ ^{\shortmid }\nabla +\ ^{\shortmid }\widehat{%
\mathbf{Z}}]=\ ^{\shortmid }\mathcal{R}[\ ^{\shortmid }\mathbf{g},\
^{\shortmid }\nabla ]+\ ^{\shortmid }\widehat{\mathcal{Z}}[\ ^{\shortmid }%
\mathbf{g},\ ^{\shortmid }\widehat{\mathbf{Z}}],
\end{eqnarray*}%
\begin{eqnarray*}
\widehat{R}ic[\mathbf{g},\widehat{\mathbf{D}} &=&\nabla +\widehat{\mathbf{Z}}%
]=Ric[\mathbf{g},\nabla ]+\widehat{Z}ic[\mathbf{g},\widehat{\mathbf{Z}}], \\
\ ^{\shortmid }\widehat{R}ic[\ ^{\shortmid }\mathbf{g},\ ^{\shortmid }%
\widehat{\mathbf{D}} &=&\ ^{\shortmid }\nabla +\ ^{\shortmid }\widehat{%
\mathbf{Z}}]=\ ^{\shortmid }Ric[\ ^{\shortmid }\mathbf{g},\ ^{\shortmid
}\nabla ]+\ ^{\shortmid }\widehat{Z}ic[\ ^{\shortmid }\mathbf{g},\
^{\shortmid }\widehat{\mathbf{Z}}],
\end{eqnarray*}%
\begin{eqnarray*}
\ _{s}^{\shortmid }\widehat{R}[\mathbf{g},\widehat{\mathbf{D}} &=&\nabla +%
\widehat{\mathbf{Z}}]=\mathcal{R}[\mathbf{g},\nabla ]+\ _{s}\widehat{Z}[%
\mathbf{g},\widehat{\mathbf{Z}}], \\
\ _{s}^{\shortmid }\widehat{R}[\ ^{\shortmid }\mathbf{g},\ ^{\shortmid }%
\widehat{\mathbf{D}} &=&\ ^{\shortmid }\nabla +\ ^{\shortmid }\widehat{%
\mathbf{Z}}]=\ _{s}^{\shortmid }R[\ ^{\shortmid }\mathbf{g},\ ^{\shortmid
}\nabla ]+\ _{s}^{\shortmid }\widehat{Z}[\ ^{\shortmid }\mathbf{g},\
^{\shortmid }\widehat{\mathbf{Z}}],
\end{eqnarray*}%
Such distortion formulas can be considered for the almost symplectic
Lagrange, or Finsler, d-connections, 
\begin{eqnarray*}
\widetilde{\mathcal{R}}[\widetilde{\mathbf{g}} &\simeq &\widetilde{\theta },%
\widetilde{\mathbf{D}}=\nabla +\widetilde{\mathbf{Z}}]=\mathcal{R}[%
\widetilde{\mathbf{g}}\simeq \widetilde{\theta },\nabla ]+\widetilde{%
\mathcal{Z}}[\widetilde{\mathbf{g}}\simeq \widetilde{\theta },\widetilde{%
\mathbf{Z}}], \\
\ ^{\shortmid }\widetilde{\mathcal{R}}[\ ^{\shortmid }\widetilde{\mathbf{g}}
&\simeq &\ ^{\shortmid }\widetilde{\theta },\ ^{\shortmid }\widetilde{%
\mathbf{D}}=\ ^{\shortmid }\nabla +\ ^{\shortmid }\widetilde{\mathbf{Z}}]=\
^{\shortmid }\mathcal{R}[\ ^{\shortmid }\widetilde{\mathbf{g}}\simeq \
^{\shortmid }\widetilde{\theta },\ ^{\shortmid }\nabla ]+\ ^{\shortmid }%
\widetilde{\mathcal{Z}}[\ ^{\shortmid }\mathbf{g}\simeq \ ^{\shortmid }%
\widetilde{\theta },\ ^{\shortmid }\widetilde{\mathbf{Z}}],
\end{eqnarray*}%
and any geometric d-objects with "tilde" symbols.

Finally, we note that similar distortions can be defined and computed, for
instance, for the Chern d-connection, Berwald d-connection (which are not
metric compatible) and any d-connection structure considered in Finsler
geometry \cite{chern48,bao00} and can be introduced via corresponding
nonholonomic deformations of 4-d Lorentz manifold \cite{vmon3}. The physical
importance of such d- and s-connections is not clear (for instance, to it is
an unsolved problem how to define in a unique and self-consistent form the
Dirac equations on nonmetric curved spaces \cite{vacaru18,bubuianu18a}) and
how to solve respective geometric flow and modified Chern-/ Berwald-
Einstein equations is a very difficult technical problem. Re-defining the
construction in nonholonomic canonical variables with "hat" distortions, we
can prove the decoupling property of modified/ generalized Einstein
equations on $\mathbf{TV}$ and $\mathbf{T}^{\ast }\mathbf{V.}$

\section{Nonassociative star product deformed Finsler-Hamilton phase space
geometric flows}

\label{sec6} The goal of this section is to extend definition of
nonassociative star product introduced in \cite{mylonas12,mylonas13} for
nonassociative phase spaces enabled with canonical nonholonomic and/or
Finsler-Lagrange-Hamilton variables. We follow the approach with s-adapted
frames \cite{partner01,partner02} modifying for nontrivial N-connection
structures the constructions from section 2 of \cite{blumenhagen16} and
section 2 of \cite{aschieri17}. Such constructions provide also
nonassociative generalizations of the models of noncommutative gauge gravity
and generalized Finsler geometry and their deformation quantization (using
N-adapted Moyal--Weyl star products) were considered in \cite%
{vacaru01,vacaru03,vacaru09a,vacaru16}.

\subsection{Nonassociative star products and nonsymmetric metrics}

\subsubsection{Definition of star products with Finsler-Hamilton and dyadic
N-adapted frames}

Using the canonical frame structure $^{\shortmid }\widetilde{\mathbf{e}}%
_{\alpha }=(\ ^{\shortmid }\widetilde{\mathbf{e}}_{i},\ ^{\shortmid }e^{b})$
(\ref{ccnadap}), we can define a nonassociative star product \ $\widetilde{%
\star }$ on the phase space $\mathcal{M}$ modelled as Hamilton space 
\begin{eqnarray}
f\ \widetilde{\star }q &:=&\cdot \lbrack \exp (-\frac{1}{2}i\hbar (\
^{\shortmid }\widetilde{\mathbf{e}}_{i}\otimes \ ^{\shortmid }e^{i}-\
^{\shortmid }e^{i}\otimes \ ^{\shortmid }\widetilde{\mathbf{e}}_{i})+\frac{i%
\mathit{\ell }^{4}}{12\hbar }\tilde{R}^{ija}(p_{a}\ ^{\shortmid }\widetilde{%
\mathbf{e}}_{i}\otimes \ ^{\shortmid }\widetilde{\mathbf{e}}_{j}-\
^{\shortmid }\widetilde{\mathbf{e}}_{j}\otimes p_{a}\ ^{\shortmid }%
\widetilde{\mathbf{e}}_{i}))]f\otimes q  \notag \\
&=&f\cdot q-\frac{i}{2}\hbar \lbrack (\ ^{\shortmid }\widetilde{\mathbf{e}}%
_{i}f)(\ ^{\shortmid }e^{i}q)-(\ ^{\shortmid }e^{i}f)(\ ^{\shortmid }%
\widetilde{\mathbf{e}}_{i}q)]+\frac{i\mathit{\ell }^{4}}{6\hbar }%
R^{ija}p_{a}(\ ^{\shortmid }\widetilde{\mathbf{e}}_{i}f)(\ ^{\shortmid }%
\widetilde{\mathbf{e}}_{j}q)+\ldots ,  \label{starphamilt} \\
&&\mbox{ and/ or, for  }\ ^{\shortmid }\widetilde{\mathbf{e}}_{\alpha }
\rightarrow \ ^{\shortmid }\mathbf{e}_{\alpha }=e_{\ \alpha }^{\beta }(\
^{\shortmid }u)^{\shortmid }\widetilde{\mathbf{e}}_{\beta },  \notag \\
f\star q &:=&\cdot \lbrack \exp (-\frac{1}{2}i\hbar (\ ^{\shortmid }\mathbf{e%
}_{i}\otimes \ ^{\shortmid }e^{i}-\ ^{\shortmid }e^{i}\otimes \ ^{\shortmid }%
\mathbf{e}_{i})+\frac{i\mathit{\ell }^{4}}{12\hbar }R^{ija}(p_{a}\
^{\shortmid }\mathbf{e}_{i}\otimes \ ^{\shortmid }\mathbf{e}_{j}-\
^{\shortmid }\mathbf{e}_{j}\otimes p_{a}\ ^{\shortmid }\mathbf{e}%
_{i}))]f\otimes q  \notag \\
&=&f\cdot q-\frac{i}{2}\hbar \lbrack (\ ^{\shortmid }\mathbf{e}_{i}f)(\
^{\shortmid }e^{i}q)-(\ ^{\shortmid }e^{i}f)(\ ^{\shortmid }\mathbf{e}%
_{i}q)]+\frac{i\mathit{\ell }^{4}}{6\hbar }R^{ija}p_{a}(\ ^{\shortmid }%
\widetilde{\mathbf{e}}_{i}f)(\ ^{\shortmid }\widetilde{\mathbf{e}}%
_{j}q)+\ldots ,  \label{starp44}
\end{eqnarray}%
where $\ f(x,p)$ and $\ q(x,p)$ are functions on phase space coordinates;
the constant $\mathit{\ell }$ characterizes the R-flux contributions
determined by an antisymmetric $\tilde{R}^{ija},$ or $R^{ija},$ background
in string theory; where $\otimes$ is the tensor product. For small
parametric decompositions on $\hbar $ and $\kappa =\mathit{\ell }%
_{s}^{3}/6\hbar ,$ the tensor products turn into usual multiplications as in
the second line (\ref{starphamilt}).

A phase Hamilton space $\widetilde{\mathcal{M}}$ enabled with a star product
(\ref{starp44}) transforms into a nonassociative Hamilton one $\widetilde{%
\mathcal{M}}^{\star }$ (a star label can be used any form "up/low and or
left/right", for instance, $^{\star }\widetilde{\mathcal{M}}$). We can
re-define \ $\widetilde{\star }$ in s-adapted form considering frame
transforms $\ ^{\shortmid }\widetilde{\mathbf{e}}_{\alpha }\rightarrow \
^{\shortmid }\mathbf{e}_{\alpha _{s}}=e_{\ \alpha _{s}}^{\beta }(\
_{s}^{\shortmid }u)$ $^{\shortmid }\widetilde{\mathbf{e}}_{\beta }$ for an
s-adapted basis $\ ^{\shortmid } \mathbf{e}_{\alpha _{s}}$ (\ref{nadapbdsc})
used instead of $\ ^{\shortmid }\mathbf{e}_{\alpha },$ with $\widetilde{%
\star }\rightarrow $\ $\star _{s},$ which allows to work with a s-adapted
star product $\star _{s}$ on $\ _{s}^{\star } \mathcal{M}$ as in \cite%
{partner01,partner02}. For coordinate frames, $\ ^{\shortmid }\mathbf{e}%
_{\alpha }=\ ^{\shortmid }\mathbf{\partial }_{\alpha },$ such star products
transform into that considered in \cite{blumenhagen16,aschieri17}. The
priority of $\star _{s}$is that such a dyadic star product structure allows
to decouple and solve in general forms physically important systems of
nonlinear PDEs. Nonassociative deformations $\widetilde{\star }$ can be used
for elaborating nonassociative generalizations of almost Kaehler-Hamilton
geometry (\ref{phspgd}). To elaborate on models of geometric flows on a
(temperature like) parameter $\tau $ we can consider families of s-frames $\
^{\shortmid }\widetilde{\mathbf{e}}_{i_{s}}(\tau )$ and/or $\ ^{\shortmid }%
\widetilde{\mathbf{e}}_{i}(\tau )$, we obtain define respective flow
families s-adapted star products (with respective $\widetilde{\star }(\tau
),\star (\tau ), \star _{s} (\tau )($ even the functions $\ f$ and $q$ may
not depend on evolution parameter. Similar $\tau $-dependencies of
geometric/ physical s-objects and structures have to be defined for
evolution on nonassociative and associative geometric models.

In s-adapted form, the star product is written 
\begin{eqnarray}
f\star _{s}q &:=&\cdot \lbrack \exp (-\frac{1}{2}i\hbar (\ ^{\shortmid }%
\mathbf{e}_{i_{s}}\otimes \ ^{\shortmid }e^{i_{s}}-\ ^{\shortmid
}e^{i_{s}}\otimes \ ^{\shortmid }\mathbf{e}_{i_{s}})+\frac{i\mathit{\ell }%
_{s}^{4}}{12\hbar }R^{i_{s}j_{s}a_{s}}(p_{a_{s}}\ ^{\shortmid }\mathbf{e}%
_{i_{s}}\otimes \ ^{\shortmid }\mathbf{e}_{j_{a}}-\ ^{\shortmid }\mathbf{e}%
_{j_{s}}\otimes p_{a_{s}}\ ^{\shortmid }\mathbf{e}_{i_{s}}))]f\otimes q 
\notag \\
&=&f\cdot q-\frac{i}{2}\hbar \lbrack (\ ^{\shortmid }\mathbf{e}_{i_{s}}f)(\
^{\shortmid }e^{i_{s}}q)-(\ ^{\shortmid }e^{i_{s}}f)(\ ^{\shortmid }\mathbf{e%
}_{i_{s}}q)]+\frac{i\mathit{\ell }_{s}^{4}}{6\hbar }%
R^{i_{s}j_{s}a_{s}}p_{a_{s}}(\ ^{\shortmid }\mathbf{e}_{i_{s}}f)(\
^{\shortmid }\mathbf{e}_{j_{s}}q)+\ldots .  \label{starpn}
\end{eqnarray}%
Such R-flux deformations are computed in s-adapted from and allow to us
develop nonassociative versions of the AFCDM \cite%
{partner02,partner04,partner05,partner06}. Using corresponding adapted frame
transforms, we can define equivalent nonassociative star product operations
(structures) of type (\ref{starphamilt}), (\ref{starp44}) and (\ref{starpn}%
), when $\ \widetilde{\star }\approx \star \approx \star _{s}.$ We shall use
tilde and/or $s$-labels to emphasize that we work with canonical
Lagrange-Hamilton structures and/or shell configurations (which can be
defined as double geometric structures with corresponding purposes).

\subsubsection{Nonassociative star product symmetric and nonsymmetric
d-metrics}

For $\widetilde{\mathcal{M}}\rightarrow \widetilde{\mathcal{M}}^{\star },$ a 
$\widetilde{\star }$-structure transforms a symmetric metric $\ ^{\shortmid }%
\widetilde{\mathbf{g}}$ (\ref{cdmds8}) into a nonsymmetric one with
respective symmetric, $\ _{\star }^{\shortparallel }\mathbf{\tilde{g},}$ and
nonsymmetric, $\ _{\star }^{\shortparallel }\mathfrak{\tilde{g},}$
components. We can use labels $\ ^{\shortparallel }$ instead of $\
^{\shortmid }\,\ $ because such nonassociative Finsler-Hamilton metrics may
contain complex terms. Nevertheless, the nonholonomic s-structure can be
always prescribed in such a form which allow to work with real terms and
with quasi-Hopf s-structure determined by a nonassociative algebra $\mathcal{%
A}_{s}^{\star }$ (generalizing the constructions from \cite{aschieri17,drinf}%
). Such nonassociative d-objects can be represented in the forms 
\begin{eqnarray*}
\ _{\star }^{\shortmid }\mathbf{\tilde{g}} &=&\ _{\star }^{\shortmid }%
\mathbf{\tilde{g}}_{\alpha \beta }\tilde{\star}(\ ^{\shortmid }\mathbf{%
\tilde{e}}^{\alpha }\otimes \ ^{\shortmid }\mathbf{\tilde{e}}^{\beta }),%
\mbox{ where }\ _{\star }^{\shortmid }\mathbf{\tilde{g}}(\ ^{\shortmid }%
\mathbf{\tilde{e}}_{\alpha },\ ^{\shortmid }\mathbf{\tilde{e}}_{\beta })=\
_{\star }^{\shortmid }\mathbf{\tilde{g}}_{\alpha \beta }=\ _{\star
}^{\shortmid }\mathbf{\tilde{g}}_{\beta \alpha }\in \mathcal{A}_{s}^{\star }
\\
\ \ _{\star }^{\shortmid }\mathfrak{\tilde{g}}_{\alpha \beta } &=&\ \
_{\star }^{\shortmid }\mathbf{\tilde{g}}_{\alpha \beta }-\kappa \mathcal{R}%
_{\quad \alpha }^{\tau \xi }\ \ ^{\shortmid }\mathbf{\tilde{e}}_{\xi }\ \
_{\star }^{\shortmid }\mathbf{\tilde{g}}_{\beta \tau }=\ _{\star
}^{\shortmid }\mathfrak{\tilde{g}}_{\alpha \beta }^{[0]}+\ \ _{\star
}^{\shortmid }\mathfrak{\tilde{g}}_{\alpha \beta }^{[1]}(\kappa )=\ \
_{\star }^{\shortmid }\widetilde{\mathfrak{\check{g}}}_{\alpha \beta }+\ \
_{\star }^{\shortmid }\widetilde{\mathfrak{a}}_{\alpha \beta }.
\end{eqnarray*}%
In these formulas, we consider that $\ _{\star }^{\shortmid }\widetilde{%
\mathfrak{\check{g}}}_{\alpha \beta }$ (we use a inverse hat and tilde
labels to emphasize that we encoded certain Finsler-Hamilton structures and
symmetrization after R-flux deformation) is the symmetric part and $\
_{\star }^{\shortmid }\widetilde{\mathfrak{a}}_{\alpha \beta }$ is the
anti-symmetric part computed respectively: 
\begin{eqnarray}
\ _{\star }^{\shortmid }\widetilde{\mathfrak{\check{g}}}_{\alpha \beta }:= &&%
\frac{1}{2}(\ _{\star }^{\shortmid }\widetilde{\mathfrak{g}}_{\alpha \beta
}+\ _{\star }^{\shortmid }\widetilde{\mathfrak{g}}_{\beta \alpha })=\
_{\star }^{\shortmid }\widetilde{\mathbf{g}}_{\alpha \beta }-\frac{\kappa }{2%
}\left( \mathcal{R}_{\quad \beta }^{\tau \xi }\ \ ^{\shortmid }\widetilde{%
\mathbf{e}}_{\xi }\ _{\star }^{\shortmid }\widetilde{\mathbf{g}}_{\tau
\alpha }+\mathcal{R}_{\quad \alpha }^{\tau \xi }\ \ ^{\shortmid }\widetilde{%
\mathbf{e}}_{\xi }\ _{\star }^{\shortmid }\widetilde{\mathbf{g}}_{\beta \tau
}\right)  \label{aux40b} \\
&=&\ _{\star }^{\shortmid }\widetilde{\mathfrak{\check{g}}}_{\alpha \beta
}^{[0]}+\ _{\star }^{\shortmid }\widetilde{\mathfrak{\check{g}}}_{\alpha
\beta }^{[1]}(\kappa ),  \notag \\
&&\mbox{ for }\ _{\star }^{\shortmid }\widetilde{\mathfrak{\check{g}}}%
_{\alpha \beta }^{[0]}=\ \ _{\star }^{\shortmid }\widetilde{\mathbf{g}}%
_{\alpha \beta }\mbox{ and }\ \ _{\star }^{\shortmid }\widetilde{\mathfrak{%
\check{g}}}_{\alpha \beta }^{[1]}(\kappa )=-\frac{\kappa }{2}\left( \mathcal{%
R}_{\quad \beta }^{\tau \xi }\ \ ^{\shortmid }\widetilde{\mathbf{e}}_{\xi }\
\ _{\star }^{\shortmid }\widetilde{\mathbf{g}}_{\tau \alpha }+\mathcal{R}%
_{\quad \alpha }^{\tau \xi }\ ^{\shortmid }\widetilde{\mathbf{e}}_{\xi }\ \
_{\star }^{\shortmid }\widetilde{\mathbf{g}}_{\beta \tau }\right) ;  \notag
\\
\ _{\star }^{\shortmid }\widetilde{\mathfrak{a}}_{\alpha \beta }:= &&\frac{1%
}{2}(\ _{\star }^{\shortmid }\widetilde{\mathfrak{g}}_{\alpha \beta }-\
_{\star }^{\shortmid }\widetilde{\mathfrak{g}}_{\beta \alpha })=\frac{\kappa 
}{2}\left( \mathcal{R}_{\quad \beta }^{\tau \xi }\ ^{\shortmid }\widetilde{%
\mathbf{e}}_{\xi }\ \ _{\star }^{\shortmid }\widetilde{\mathbf{g}}_{\tau
\alpha }-\mathcal{R}_{\quad \alpha }^{\tau \xi }\ ^{\shortmid }\widetilde{%
\mathbf{e}}_{\xi }\ _{\star }^{\shortmid }\widetilde{\mathbf{g}}_{\beta \tau
}\right)  \notag \\
&=&\ _{\star }^{\shortmid }\widetilde{\mathfrak{a}}_{\alpha \beta
}^{[1]}(\kappa )=\frac{1}{2}(\ _{\star }^{\shortmid }\widetilde{\mathfrak{g}}%
_{\alpha \beta }^{[1]}(\kappa )-\ _{\star }^{\shortmid }\widetilde{\mathfrak{%
g}}_{\beta \alpha }^{[1]}(\kappa )).  \label{aux40aa}
\end{eqnarray}

We emphasize that choosing a primary nonholonomic distribution when $\
_{\star }^{\shortmid }\widetilde{\mathfrak{\check{g}}}_{\alpha \beta
}^{[0]}= \ ^{\shortmid }\widetilde{\mathbf{g}}_{\alpha \beta }$ (\ref{cdmds8}%
), we can work on effective nonassociative spaces with symmetric s- and
d-metrics, when the nonsymmetric coefficients are induced (and can be
recurrently computed on higher orders of $\kappa $) by data $(\mathcal{R}%
_{\quad \beta }^{\tau \xi },_{\xi },\ ^{\shortmid }\widetilde{\mathbf{e}}%
_{\xi },\ \ ^{\shortmid }\widetilde{\mathbf{g}}_{\tau \alpha })$ (\ref%
{aux40aa}). Here we note also that nonsymmetric inverse d-metrics can be
parameterized in the form $\ _{\star }^{\shortmid }\widetilde{\mathfrak{g}}%
^{\alpha \beta }=\ _{\star }^{\shortmid }\widetilde{\mathfrak{\check{g}}}%
^{\alpha \beta }+ \ _{\star }^{\shortmid }\widetilde{\mathfrak{a}}^{\alpha
\beta }$, but in nonassociative geometry the procedure of computing inverse
matrices and metrics is more sophisticate than in the commutative and
noncommutative cases, see details in \cite%
{blumenhagen16,aschieri17,partner01,partner02}. For nonassociative
constructions, $\ _{\star }^{\shortmid }\widetilde{\mathfrak{\check{g}}}%
^{\alpha \beta }$ is not the inverse to $\ _{\star }^{\shortmid }\mathfrak{%
\check{g}}_{\alpha \beta }$ and $\ _{\star }^{\shortmid }\mathfrak{a}%
^{\alpha \beta }$ is not inverse to $\ _{\star }^{\shortmid }\mathfrak{a}%
_{\alpha \beta }.$ To model nonassociative geometric flow evolution of
symmetric and nonsymmetric components of star product deformed
Finsler-Hamilton d-metrics, we have to consider respective families of
d-objects and their d-adapted components which can be written, for instance, 
$\ _{\star }^{\shortmid }\widetilde{\mathbf{g}}(\tau ), \ _{\star
}^{\shortmid }\widetilde{\mathbf{g}}_{\beta \alpha }( \tau ),  \ _{\star
}^{\shortmid }\widetilde{\mathfrak{g}}_{\alpha \beta } (\tau )=  \ _{\star
}^{\shortmid }\widetilde{\mathfrak{\check{g}}}_{\alpha \beta } (\tau )+ \
_{\star}^{\shortmid }\widetilde{\mathfrak{a}}_{\alpha _{s}\beta _{s}}(\tau )$
etc.

\subsection{Star product deformations of geometric s-objects on
Finsler-Hamilton phase spaces}

In \cite{partner01,partner02,partner04,partner06}, we used the Convention 2
for nonassociative star product deformations of geometric objects on $\
^{\shortmid }\mathcal{M}$ into respective s-objects on $\ _{s}^{\shortmid }%
\mathcal{M}^{\star }.$ In this subsection, the Convention 2 is generalized
in such forms when nonassociative s-objects and be generated by star product
deformations of d-objects for Finsler-Hamilton phase spaces (for
Finsler-Lagrange configurations the constructions are dual). All such
nonassociative and nonholonomic geometric constructions can be performed in
abstract geometric form when coefficient formulas are derived with
respective d- and/or s-adapted frames.

\subsubsection{Canonical and Finsler-Hamilton d-connections and
LC-configurations}

Using corresponding nonassociative star product (\ref{starphamilt}), (\ref%
{starp44}) and (\ref{starpn}) and d-metric (\ref{cdmds8}) , any linear /
d-connection / s-connection structure from (\ref{candistr}) can be deformed
into nonassociative d-/ s-geometric objects: 
\begin{eqnarray}
\ ^{\shortmid }\widetilde{\mathbf{D}} &=&\ ^{\shortmid }\nabla +\
^{\shortmid }\widetilde{\mathbf{Z}}\rightarrow \ ^{\shortmid }\widetilde{%
\mathbf{D}}^{\star }=\ ^{\shortmid }\nabla ^{\star }+\ ^{\shortmid }%
\widetilde{\mathbf{Z}}^{\star },\ ^{\shortmid }\widehat{\mathbf{D}}=\
^{\shortmid }\widetilde{\mathbf{D}}+\ ^{\shortmid }\mathbf{Z\rightarrow }\
^{\shortmid }\widehat{\mathbf{D}}^{\star }=\ ^{\shortmid }\widetilde{\mathbf{%
D}}^{\star }+\ ^{\shortmid }\mathbf{Z}^{\star }\ ,  \notag \\
&&\mbox{ and }\ _{s}^{\shortmid }\widehat{\mathbf{D}}=\ ^{\shortmid }%
\widetilde{\mathbf{D}}+\ _{s}^{\shortmid }\mathbf{Z\rightarrow \
_{s}^{\shortmid }\widehat{\mathbf{D}}^{\star }=\ ^{\shortmid }\widetilde{%
\mathbf{D}}^{\star }+\ _{s}^{\shortmid }\mathbf{Z}^{\star }.}
\label{stardistransf}
\end{eqnarray}%
We write conventionally $\ _{s}^{\shortmid }\widetilde{\mathbf{D}}=\
_{s}^{\shortmid }\widehat{\mathbf{\tilde{D}}}$ (to avoid using double hat
and tilde labels if that will not result in ambiguities). Here we note that,
in principle, we can re-define the geometric constructions in any convenient
"tilde" or "hat" variables, considering that $\ ^{\shortmid }\widetilde{%
\mathbf{D}}$ can be transformed into s-adapted configurations \ for some
nonholonomic s-adapted frame transforms when $(\ _{s}^{\shortmid }\mathbf{%
g,\ _{s}^{\shortmid }N)\approx }(\ ^{\shortmid }\mathbf{g,\ ^{\shortmid
}N)\approx }(\ ^{\shortmid }\mathbf{\tilde{g},\ ^{\shortmid }\tilde{N}).}$
Such constructions are important if we want to generate nonassociative
Lagrange-Hamilton phase configurations subjected to the conditions that
certain models may elaborated for some classes of exact/ parametric
solutions.

With respect to N-/ s-adapted bases, $\ ^{\shortmid }\widetilde{\mathbf{e}}%
_{\alpha },$ $\ ^{\shortmid }\mathbf{e}_{\alpha }$ or $\ ^{\shortmid }%
\mathbf{e}_{\alpha _{s}},$ we can compute corresponding coefficient formulas
for (\ref{stardistransf}). For instance, we can write in coefficient forms%
\begin{eqnarray*}
\ ^{\shortmid }\widetilde{\mathbf{D}} &=&\{\ ^{\shortmid }\widetilde{\mathbf{%
\Gamma }}_{\ \ \beta \gamma }^{\alpha }(\ ^{\shortmid }u)\}\rightarrow \
^{\shortmid }\widetilde{\mathbf{D}}^{\star }=\{\ ^{\shortmid }\widetilde{%
\mathbf{\Gamma }}_{\star \ \ \beta \gamma }^{\alpha }(\ ^{\shortmid }u)\},%
\mbox{ after }\ ^{\shortmid }\widetilde{\mathbf{e}}_{\alpha }\rightarrow \
^{\shortmid }\mathbf{e}_{\alpha _{s}}=e_{\ \alpha _{s}}^{\beta }(\
_{s}^{\shortmid }u)^{\shortmid }\widetilde{\mathbf{e}}_{\beta }, \\
\ _{s}^{\shortmid }\widetilde{\mathbf{D}} &=&\{\ ^{\shortmid }\widetilde{%
\mathbf{\Gamma }}_{\ \ \beta _{s}\gamma _{s}}^{\alpha _{s}}(\
_{s}^{\shortmid }u)\}\rightarrow \ _{s}^{\shortmid }\widetilde{\mathbf{D}}%
^{\star }=\{\ ^{\shortmid }\widetilde{\mathbf{\Gamma }}_{\star \ \ \beta
_{s}\gamma _{s}}^{\alpha _{s}}(\ _{s}^{\shortmid }u)\},
\end{eqnarray*}%
see details in \cite{partner01,partner02} redefined for Finsler-Hamilton
configurations in \cite{vmon3,vacaru18,bubuianu18a}.

The N- and s-adapted metric affine data on phase spaces are transformed
under star product transforms as 
\begin{equation*}
(\ ^{\shortmid }\widetilde{\mathbf{N}}\mathbf{,}\ ^{\shortmid }\widetilde{%
\mathbf{g}}\mathbf{,}\ ^{\shortmid }\widetilde{\mathbf{D}})\approx (\
_{s}^{\shortmid }\widetilde{\mathbf{N}}\mathbf{,}\ _{s}^{\shortmid }%
\widetilde{\mathbf{g}}\mathbf{,}\ _{s}^{\shortmid }\widetilde{\mathbf{D}}%
)\rightarrow (\ _{s}^{\shortmid }\widetilde{\mathbf{N}}\mathbf{,}\ \ _{\star
}^{\shortmid }\widetilde{\mathfrak{g}}\mathbf{=}(\ _{\star }^{\shortmid }%
\widetilde{\mathfrak{\check{g}}},\ \ _{\star }^{\shortmid }\widetilde{%
\mathfrak{a}})\mathbf{,}\ ^{\shortmid }\widetilde{\mathbf{D}}^{\star
})\approx (\ _{s}^{\shortmid }\widetilde{\mathbf{N}}\mathbf{,}\ \
_{s}^{\shortmid }\widetilde{\mathfrak{g}}^{\star }\mathbf{=}(\
_{s}^{\shortmid }\widetilde{\mathfrak{\check{g}}}^{\star },\ \
_{s}^{\shortmid }\widetilde{\mathfrak{a}}^{\star })\mathbf{,}\
_{s}^{\shortmid }\widetilde{\mathbf{D}}^{\star }).
\end{equation*}%
Such relations define respective nonholonomic and nonassociative
Finsler-Hamilton spaces depending on the type of generating functions and
star product deformations. For any $\ _{s}^{\shortmid }\widetilde{\mathbf{D}}
$ and $\ _{s}^{\shortmid }\widetilde{\mathbf{D}}^{\star },$ we can define
and compute in abstract/ coefficient forms the corresponding star
deformations of torsion s-tensors, $\ _{s}^{\shortmid }\widetilde{\mathcal{T}%
}\rightarrow \ _{s}^{\shortmid }\widehat{\mathcal{T}}^{\star },$ and Riemann
curvature s-tensors, $\ _{s}^{\shortmid }\widetilde{\mathcal{R}}\rightarrow
\ _{s}^{\shortmid }\widehat{\mathcal{R}}^{\star }.$ Originally, such
computations in coordinate frames were performed in \cite%
{blumenhagen16,aschieri17} there were considered star product deformations
without N-connection structure and for (pseudo) Riemannian data.
Corresponding nonassociative geometric objects were defined for $(\
^{\shortmid }g,\ ^{\shortmid }\nabla )\rightarrow (\ _{\star }^{\shortmid }%
\mathfrak{g}= (\ \ _{\star }^{\shortmid }\mathfrak{\check{g}},\ \
_{\star}^{\shortmid }\mathfrak{a}),\ ^{\shortmid }\nabla ^{\star }),$ where $%
\ ^{\shortmid }\nabla $ and $\ ^{\shortmid }\nabla ^{\star }$ are respective
Levi-Civita, LC, connections. Unfortunately, it is not possible to decouple
nonholonomic and nonassociative (modified) Einstein equations using $\
^{\shortmid }\nabla ,\ ^{\shortmid }\nabla ^{\star },$ and (for
Finsler-Hamilton gravity theories) \ for $\ ^{\shortmid }\widetilde{\mathbf{D%
}},$ or $\ ^{\shortmid }\widetilde{\mathbf{D}}^{\star }.$

In \cite{vacaru09a}, there were studied noncommutative black hole solutions
for the Cartan-Finsler d-connection but those classes of solutions can't be
generalized in a direct form for nonassociative models because R-flux terms
induce additional coupling. The main result of \cite{partner02} consisted in
a proof that we can decouple and integrate such important systems of
nonlinear PDEs if we use the canonical s-connections, $\ _{s}^{\shortmid }%
\widehat{\mathbf{D}}$ and $\ _{s}^{\shortmid }\widehat{\mathbf{D}}^{\star }.$
The main idea of the Part II is that we consider certain classes of
nonholonomic distributions and frame transforms when $\ ^{\shortmid }%
\widetilde{\mathbf{D}}^{\star }\rightarrow \ _{s}^{\shortmid }\widehat{%
\mathbf{D}}^{\star },$ which allows us to construct more general classes of
generic off-diagonal solutions. Then we can impose additional nonholonomic
constraints on distortion s-tensors $\ _{s}^{\shortmid }\widetilde{\mathbf{Z}%
},$ when $\ _{s}^{\shortmid }\widehat{\mathbf{D}}_{\mid \ _{s}^{\shortmid }%
\widetilde{\mathbf{Z}}=0}=$ $\ ^{\shortmid }\widetilde{\mathbf{D}}^{\star },$
and extract Finsler-Hamilton configurations. Typically, general Finsler
geometries are characterized by certain Finsler d-connection structures
which involve nontrivial d-torsion and nonmetricity d-tensor fields.
LC-configurations are not considered for such models even they can be
extracted by constraining respective distortion d-tensors. Star product
deformations preserve such properties.

For any nonassociative geometric data which include corresponding frame
transforms (we write $\mathbf{\approx }$) when $(\ ^{\shortmid }\widetilde{%
\mathbf{N}}\mathbf{\approx }\ _{s}^{\shortmid }\mathbf{N,}\ ^{\shortmid }%
\widetilde{\mathbf{g}}^{\star }\mathbf{\approx }\ \ _{s}^{\shortmid }\mathbf{%
g}^{\star },\ _{s}^{\shortmid }\widetilde{\mathbf{D}}^{\star }\mathbf{=}\
_{s}^{\shortmid }\widehat{\mathbf{D}}^{\star }+\ _{s}^{\shortmid }\widetilde{%
\mathbf{Z}}^{\star }),$ we can define such canonical s-connection,
Cartan-Finsler-Hamilton d-connection, and LC-connection structures, 
\begin{equation}
\ \ _{s}^{\shortmid }\widehat{\mathbf{D}}^{\star } = (h_{1}\ ^{\shortmid }%
\widehat{\mathbf{D}}^{\star },\ v_{2}\ ^{\shortmid }\widehat{\mathbf{D}}%
^{\star },\ c_{3}\ ^{\shortmid }\widehat{\mathbf{D}}^{\star },\ c_{4}\
^{\shortmid }\widehat{\mathbf{D}}^{\star })=\ ^{\shortmid }\nabla ^{\star
}+\ _{s}^{\shortmid }\widehat{\mathbf{Z}}^{\star }=\ _{s}^{\shortmid }%
\widetilde{\mathbf{D}}^{\star }-\ _{s}^{\shortmid }\widetilde{\mathbf{Z}}%
^{\star }, \ ^{\shortmid }\widetilde{\mathbf{D}}^{\star } = \ ^{\shortmid }%
\widetilde{\nabla }^{\star }+\ ^{\shortmid }\widetilde{\mathbf{Z}}^{\star },
\label{candistrnas}
\end{equation}%
where the canonical distortion s-tensors $\ _{s}^{\shortmid }\widehat{%
\mathbf{Z}}^{\star }[\ _{s}^{\shortmid }\widehat{\mathcal{T}}^{\star }[\
_{s}^{\shortmid }\mathbf{N,}\ _{s}^{\shortmid }\mathbf{g}^{\star }]]$ is an
algebraic functional of the canonical s-torsion $\ _{s}^{\shortmid }\widehat{%
\mathcal{T}}^{\star },$ and this involves additional distortion relations
with $\ ^{\shortmid }\widetilde{\mathbf{Z}}^{\star }$ and $\ ^{\shortmid }%
\widetilde{\mathcal{T}}^{\star }.$ The corresponding linear connections are
defined by the conditions 
\begin{equation}
\begin{array}{c}
(\ ^{\shortmid }\widetilde{\mathbf{g}}^{\star },\ ^{\shortmid }\widetilde{%
\mathbf{N}}) \\ 
\updownarrow \\ 
(\ _{s}^{\shortmid }\mathbf{g}^{\star }\mathbf{,}\ _{s}^{\shortmid }\mathbf{N%
})%
\end{array}%
\rightarrow \left\{ 
\begin{array}{cc}
\ _{\star }^{\shortmid }\mathbf{\nabla :} & 
\begin{array}{c}
\fbox{\ $\ \ ^{\shortmid }\mathbf{\nabla }^{\star }{}^{\ \shortmid }%
\widetilde{\mathbf{g}}^{\star }=0$;\ $_{\nabla }^{\shortmid }\mathcal{T}%
^{\star }=0$}\mbox{\ star
LC-connection}; \\ 
\end{array}
\\ 
\ ^{\shortmid }\widetilde{\mathbf{D}}^{\star }: & \fbox{$%
\begin{array}{c}
\ ^{\shortmid }\widetilde{\mathbf{D}}^{\star }\ ^{\ \shortmid }\widetilde{%
\mathbf{g}}^{\star }=0;\ h\ ^{\shortmid }\widetilde{\mathcal{T}}^{\star
}=0,c\ ^{\shortmid }\widetilde{\mathcal{T}}^{\star }=0,hc\ ^{\shortmid }%
\widetilde{\mathcal{T}}^{\star }\neq 0 \\ 
\mbox{ Cartan-Finsler-Hamilton d-connection};%
\end{array}%
$} \\ 
\ _{s}^{\shortmid }\widehat{\mathbf{D}}^{\star }: & \fbox{$%
\begin{array}{c}
\ _{s}^{\shortmid }\widehat{\mathbf{D}}^{\star }\ _{s}^{\ \shortmid }\mathbf{%
g}^{\star }=0;\ h_{1}\ ^{\shortmid }\widehat{\mathcal{T}}^{\star }=0,v_{2}\
^{\shortmid }\widehat{\mathcal{T}}^{\star }=0,c_{3}\ ^{\shortmid }\widehat{%
\mathcal{T}}^{\star }=0,c_{4}\ ^{\shortmid }\widehat{\mathcal{T}}^{\star }=0,
\\ 
h_{1}v_{2}\ ^{\shortmid }\widehat{\mathcal{T}}^{\star }\neq 0,h_{1}c_{s}\
^{\shortmid }\widehat{\mathcal{T}}^{\star }\neq 0,v_{2}c_{s}\ ^{\shortmid }%
\widehat{\mathcal{T}}^{\star }\neq 0,c_{3}c_{4}\ ^{\shortmid }\widehat{%
\mathcal{T}}^{\star }\neq 0,%
\end{array}%
$}\mbox{ can. s-connect.}%
\end{array}%
\right.  \label{threeconstar}
\end{equation}%
We note that in the definition of linear connections $\ \
_{\star}^{\shortmid }\mathbf{\nabla ,}\ ^{\shortmid }\widetilde{\mathbf{D}}%
^{\star } $ and $\ _{s}^{\shortmid }\widehat{\mathbf{D}}^{\star }$ we use
the s-tensor$\mathbf{\ ^{\shortmid }\widetilde{\mathbf{g}}^{\star }\mathbf{%
\approx }\ \ _{s}^{\shortmid }\mathbf{g}^{\star }.}$ The coefficient
s-adapted formulas are provided in \cite{partner01,partner02}. For defining
linear connections (\ref{threeconstar}), we can use also a nonsymmetric
metric $\ _{\star }^{\shortmid }\mathfrak{g}_{\alpha _{s}\beta _{s}}$ which
can be nonholonomically transformed and constrained to a $\mathbf{\
^{\shortmid }\widetilde{\mathbf{g}}^{\star }}$ but such a choice results in
a more strong coupling of tensor s-objects which does not allow decoupling
of physically important systems of nonlinear PDEs.

\subsubsection{ Convention 2 for Finsler-Hamilton structures}

To define and compute geometric and physical objects on a nonassociative
phase space $\ _{s}^{\star }\mathcal{M}$ we formulated the Convention 2 (see
details in \cite{partner01,partner02,partner04,partner06}). Here we
reformulate that convention in a form including "tilde" variables.

\textbf{Convention 2FH (for Finsler-Hamiton variables)} :\ The commutative
and nonassociative geometric data derived for corresponding star products \ (%
\ref{starphamilt}), (\ref{starp44}) and (\ref{starpn}), when $\ \widetilde{%
\star }\approx \star \approx \star _{s}$ can be expressed in such
abstract/symbolic s-adapted forms: 
\begin{equation}
\begin{array}{ccc}
\begin{array}{c}
(\ \widetilde{\star },\ \ \widetilde{\mathcal{A}}^{\star },\ ^{\shortmid }%
\widetilde{\mathbf{g}}^{\star }\mathbf{,\mathfrak{\ ^{\shortmid }\widetilde{%
\mathbf{g}}^{\star },}\ ^{\shortmid }}\widetilde{\mathbf{N}},\mathbf{\
^{\shortmid }}\widetilde{\mathbf{e}}_{\alpha }\mathbf{,\ ^{\shortmid }}%
\widetilde{\mathbf{D}}^{\star }) \\ 
\updownarrow \\ 
(\star ,\ \ \mathcal{A}^{\star },\ ^{\shortmid }\mathbf{g}^{\star }\mathbf{%
,\ ^{\shortmid }\mathfrak{g}}^{\star }\mathbf{\mathfrak{,}\ ^{\shortmid }N},%
\mathbf{\ ^{\shortmid }e}_{\alpha }\mathbf{,\ ^{\shortmid }}\widehat{\mathbf{%
D}}^{\star }))%
\end{array}
& \Leftrightarrow & 
\begin{array}{c}
(\ \widetilde{\star }_{s},\ \ \widetilde{\mathcal{A}}^{\star },\
_{s}^{\shortmid }\widetilde{\mathbf{g}}^{\star }\mathbf{,}\ _{s}^{\shortmid }%
\mathbf{\mathfrak{\widetilde{\mathbf{g}}^{\star },}}\ _{s}^{\shortmid }%
\widetilde{\mathbf{N}},\mathbf{\ ^{\shortmid }}\widetilde{\mathbf{e}}%
_{\alpha _{s}}\mathbf{,}\ _{s}^{\shortmid }\widetilde{\mathbf{D}}^{\star })
\\ 
\updownarrow \\ 
(\star _{s},\ \ \mathcal{A}_{s}^{\star },\ _{s}^{\shortmid }\mathbf{g}%
^{\star }\mathbf{,}\ _{s}^{\shortmid }\mathbf{\mathfrak{g}}^{\star }\mathbf{%
\mathfrak{,}}\ _{s}^{\shortmid }\mathbf{N},\mathbf{\ ^{\shortmid }e}_{\alpha
_{s}}\mathbf{,}\ _{s}^{\shortmid }\widehat{\mathbf{D}}^{\star })%
\end{array}
\\ 
& \Uparrow &  \\ 
\begin{array}{c}
(\ \widetilde{\mathcal{A}},\ ^{\shortmid }\widetilde{\mathbf{g}}\mathbf{,%
\mathfrak{\ ^{\shortmid }\widetilde{\mathbf{g}},}\ ^{\shortmid }}\widetilde{%
\mathbf{N}},\mathbf{\ ^{\shortmid }}\widetilde{\mathbf{e}}_{\alpha }\mathbf{%
,\ ^{\shortmid }}\widetilde{\mathbf{D}}) \\ 
\updownarrow \\ 
(\mathcal{A},\ ^{\shortmid }\mathbf{g,\ ^{\shortmid }\mathfrak{g,}\
^{\shortmid }N},\mathbf{\ ^{\shortmid }e}_{\alpha }\mathbf{,\ ^{\shortmid }}%
\widehat{\mathbf{D}})%
\end{array}
& \Leftrightarrow & 
\begin{array}{c}
(\ \widetilde{\mathcal{A}}^{\star },\ _{s}^{\shortmid }\widetilde{\mathbf{g}}%
\mathbf{,}\ _{s}^{\shortmid }\mathbf{\mathfrak{\widetilde{\mathbf{g}},}}\
_{s}^{\shortmid }\widetilde{\mathbf{N}},\mathbf{\ ^{\shortmid }}\widetilde{%
\mathbf{e}}_{\alpha _{s}}\mathbf{,}\ _{s}^{\shortmid }\widetilde{\mathbf{D}})
\\ 
\updownarrow \\ 
(\ \mathcal{A}_{s}^{\star },\ _{s}^{\shortmid }\mathbf{g,}\ _{s}^{\shortmid }%
\mathbf{\mathfrak{g,}}\ _{s}^{\shortmid }\mathbf{N},\mathbf{\ ^{\shortmid }e}%
_{\alpha _{s}}\mathbf{,}\ _{s}^{\shortmid }\widehat{\mathbf{D}})%
\end{array}%
\end{array}
\label{conv2FH}
\end{equation}%
for certain canonical distortions (\ref{candistrnas}).

Following the Convention 2FH, we can define and compute star product
deformations of fundamental geometric s-objects,%
\begin{eqnarray}
\ _{s}^{\shortmid }\mathcal{T} &\rightarrow &\ _{s}^{\shortmid }\widehat{%
\mathcal{T}}^{\star }=\{\ ^{\shortmid }\widehat{\mathbf{T}}_{\ \star \beta
_{s}\gamma _{s}}^{\alpha _{s}}\}\mbox{ and }\ _{s}^{\shortmid }\widetilde{%
\mathcal{T}}\rightarrow \ _{s}^{\shortmid }\widetilde{\mathcal{T}}^{\star
}=\{\ ^{\shortmid }\widetilde{\mathbf{T}}_{\ \star \beta _{s}\gamma
_{s}}^{\alpha _{s}}\},\mbox{ nonassociative  canonical
s-torsion };\   \label{mafgeomobn} \\
\ _{s}^{\shortmid }\mathcal{R} &\rightarrow &\ _{s}^{\shortmid }\widehat{%
\mathcal{R}}^{\star }=\{\ ^{\shortmid }\widehat{\mathbf{R}}_{\ \beta
_{s}\gamma _{s}\delta _{s}}^{\star \alpha _{s}}\}\mbox{ and },%
\mbox{nonassociative
canonical Riemannian s-curvature };  \notag \\
\ _{s}^{\shortmid }\mathcal{R}ic &\rightarrow &\ _{s}^{\shortmid }\widehat{%
\mathcal{R}}ic^{\star }=\{\ ^{\shortmid }\widehat{\mathbf{R}}_{\ \beta
_{s}\gamma _{s}}^{\star }:=\ ^{\shortmid }\widehat{\mathbf{R}}_{\ \beta
_{s}\gamma _{s}\alpha _{s}}^{\star \alpha _{s}}\neq \ ^{\shortmid }\widehat{%
\mathbf{R}}_{\ \gamma _{s}\beta _{s}}^{\star }\}\mbox{ and }  \notag \\
\ _{s}^{\shortmid }\widetilde{\mathcal{R}}ic &\rightarrow &\ _{s}^{\shortmid
}\widetilde{\mathcal{R}}ic^{\star }=\{\ ^{\shortmid }\widetilde{\mathbf{R}}%
_{\ \beta _{s}\gamma _{s}}^{\star }:=\ ^{\shortmid }\widetilde{\mathbf{R}}%
_{\ \beta _{s}\gamma _{s}\alpha _{s}}^{\star \alpha _{s}}\neq \ ^{\shortmid }%
\widetilde{\mathbf{R}}_{\ \gamma _{s}\beta _{s}}^{\star }\},%
\mbox{ nonassociative
canonical Ricci s-tensor};  \notag
\end{eqnarray}%
\begin{eqnarray}
\ _{s}^{\shortmid }\mathcal{R}sc &\rightarrow &\ _{s}^{\shortmid }\widehat{%
\mathcal{R}}sc^{\star }=\{\ ^{\shortmid }\mathbf{g}^{\beta _{s}\gamma _{s}}\
^{\shortmid }\widehat{\mathbf{R}}_{\ \beta _{s}\gamma _{s}}^{\star }\}%
\mbox{
and }  \notag \\
\ _{s}^{\shortmid }\widetilde{\mathcal{R}}sc &\rightarrow &\ _{s}^{\shortmid
}\widetilde{\mathcal{R}}sc^{\star }=\{\ ^{\shortmid }\mathbf{g}^{\beta
_{s}\gamma _{s}}\ ^{\shortmid }\widetilde{\mathbf{R}}_{\ \beta _{s}\gamma
_{s}}^{\star }\}\mbox{  nonassociative  canonical  Riemannian scalar }; 
\notag \\
\ _{s}^{\shortmid }\mathcal{Q} &\rightarrow &\ _{s}^{\shortmid }\mathcal{Q}%
^{\star }=\{\ ^{\shortmid }\widehat{\mathbf{Q}}_{\gamma _{s}\alpha _{s}\beta
_{s}\ }^{\star }=\ ^{\shortmid }\widehat{\mathbf{D}}_{\gamma _{s}}^{\star }\
^{\shortmid }\mathbf{g}_{\alpha _{s}\beta _{s}}^{\star }\}\mbox{ and } 
\notag \\
\ _{s}^{\shortmid }\widetilde{\mathcal{Q}} &\mathcal{=}&0\rightarrow \
_{s}^{\shortmid }\widetilde{\mathcal{Q}}^{\star }=\{\ ^{\shortmid }%
\widetilde{\mathbf{Q}}_{\gamma _{s}\alpha _{s}\beta _{s}\ }^{\star }=\
^{\shortmid }\widetilde{\mathbf{D}}_{\gamma _{s}}^{\star }\ ^{\shortmid }%
\widetilde{\mathbf{g}}_{\alpha _{s}\beta _{s}}^{\star }\}=0%
\mbox{ zero
nonassociative  canonical nonmetricity s-tensor }.  \notag
\end{eqnarray}%
For instance, the nonassociative Riemann s-tensor for Finsler-Hamilton phase
geometry $\ _{s}^{\shortmid }\widetilde{\mathcal{\Re }}_{\quad }^{\star
}=\{\ ^{\shortmid }\widetilde{\mathcal{\Re }}_{\quad \alpha _{s}\beta
_{s}\gamma _{s}}^{\star \mu _{s}}\}$ from (\ref{mafgeomobn}) can be defined
and computed for the data $(\ _{s}^{\shortmid }\mathfrak{g}^{\star }%
\mathfrak{=\{\ \ ^{\shortmid }\mathbf{g}_{\alpha _{s}\beta _{s}}^{\star }\}}%
,\ _{s}^{\shortmid }\widetilde{\mathbf{D}}^{\star }=\{\ ^{\shortmid }%
\widetilde{\mathbf{\Gamma }}_{\star \alpha _{s}\beta _{s}}^{\gamma _{s}}\})$
and written in a form with $\kappa $-linear decomposition, 
\begin{eqnarray}
\mathbf{\mathbf{\mathbf{\mathbf{\ ^{\shortmid }}}}}\widetilde{\mathbf{R}}%
_{\quad \alpha _{s}\beta _{s}\gamma _{s}}^{\star \mu _{s}} &=&\mathbf{%
\mathbf{\mathbf{\mathbf{\ _{1}^{\shortmid }}}}}\widetilde{\mathbf{R}}_{\quad
\alpha _{s}\beta _{s}\gamma _{s}}^{\star \mu _{s}}+\mathbf{\mathbf{\mathbf{%
\mathbf{\ _{2}^{\shortmid }}}}}\widetilde{\mathbf{R}}_{\quad \alpha
_{s}\beta _{s}\gamma _{s}}^{\star \mu _{s}},\mbox{ where }
\label{nadriemhopffhg} \\
\mathbf{\mathbf{\mathbf{\mathbf{\ _{1}^{\ \shortmid }}}}}\widetilde{\mathbf{R%
}}_{\quad \alpha _{s}\beta _{s}\gamma _{s}}^{\star \mu _{s}} &=&\
^{\shortmid }\mathbf{e}_{\gamma _{s}}\ ^{\shortmid }\widetilde{\Gamma }%
_{\star \alpha _{s}\beta _{s}}^{\mu _{s}}-\ ^{\shortmid }\mathbf{e}_{\beta
_{s}}\ ^{\shortmid }\widetilde{\Gamma }_{\star \alpha _{s}\gamma _{s}}^{\mu
}+\ ^{\shortmid }\widetilde{\Gamma }_{\star \nu _{s}\tau _{s}}^{\mu
_{s}}\star _{s}(\delta _{\ \gamma _{s}}^{\tau _{s}}\ ^{\shortmid }\widetilde{%
\Gamma }_{\star \alpha _{s}\beta _{s}}^{\nu _{s}}-\delta _{\ \beta
_{s}}^{\tau _{s}}\ ^{\shortmid }\widetilde{\Gamma }_{\star \alpha _{s}\gamma
_{s}}^{\nu _{s}})+\ ^{\shortmid }w_{\beta _{s}\gamma _{s}}^{\tau _{s}}\star
_{s}\ ^{\shortmid }\widetilde{\Gamma }_{\star \alpha _{s}\tau _{s}}^{\mu
_{s}},  \notag \\
\ _{2}^{\shortmid }\widetilde{\mathbf{R}}_{\quad \alpha _{s}\beta _{s}\gamma
_{s}}^{\star \mu _{s}} &=&i\kappa \ ^{\shortmid }\widetilde{\Gamma }_{\star
\nu _{s}\tau _{s}}^{\mu _{s}}\star _{s}(\mathcal{R}_{\quad \gamma
_{s}}^{\tau _{s}\xi _{s}}\ ^{\shortmid }\mathbf{e}_{\xi _{s}}\ ^{\shortmid }%
\widetilde{\Gamma }_{\star \alpha _{s}\beta _{s}}^{\nu _{s}}-\mathcal{R}%
_{\quad \beta _{s}}^{\tau _{s}\xi _{s}}\ ^{\shortmid }\mathbf{e}_{\xi _{s}}\
^{\shortmid }\widetilde{\Gamma }_{\star \alpha _{s}\gamma _{s}}^{\nu _{s}}).
\notag
\end{eqnarray}%
Such formulas are provided in abstract from for LC-configurations in \cite%
{blumenhagen16,aschieri17} and generalized for nonholonomic canonical
s-connections in \cite{partner01,partner02,partner04,partner05,partner06}.
The abstract and s-adapted formulas from those papers can be redefined for
"tilde" s-objects and when there is dependence on a geometric/ information
flow $\tau $-parameter.

\subsubsection{Parametric decomposition of fundamental d-objects on
Finsler-Hamilton phase spaces}

Hereafter, for simplicity, we shall omit s-labels and s-indices for the
geometric objects with tilde considering that s-adapted constructions can be
always performed using corresponding s-frames and canonical distortions. In
this subsection, we explain how $\kappa $-parametric decompositions of
fundamental geometric d-objects in Finsler-Hamilton phase space geometry can
be derived from respective decompositions of d-metrics and canonical
d-connections.

We can consider a parametric decomposition of the star Cartan-Hamilton
d-connection $\mathbf{\ ^{\shortmid }}\widetilde{\mathbf{D}}^{\star }$ (\ref%
{canondch}) 
\begin{equation*}
\ ^{\shortmid }\widetilde{\mathbf{\Gamma }}_{\star \alpha \beta }^{\gamma
}=\ _{[0]}^{\shortmid }\widetilde{\mathbf{\Gamma }}_{\star \alpha \beta
}^{\gamma }+i\kappa \ _{[1]}^{\shortmid }\widetilde{\mathbf{\Gamma }}_{\star
\alpha \beta }^{\gamma }=\ _{[00]}^{\shortmid }\widetilde{\mathbf{\Gamma }}%
_{\star \alpha \beta }^{\gamma }+\ _{[01]}^{\shortmid }\widetilde{\mathbf{%
\Gamma }}_{\star \alpha \beta }^{\gamma }(\hbar )+\ _{[10]}^{\shortmid }%
\widetilde{\mathbf{\Gamma }}_{\star \alpha \beta }^{\gamma }(\kappa )+\
_{[11]}^{\shortmid }\widetilde{\mathbf{\Gamma }}_{\star \alpha \beta
}^{\gamma }(\hbar \kappa )+O(\hbar ^{2},\kappa ^{2}...).
\end{equation*}%
Introducing such parametric d-coefficients in (\ref{nadriemhopffhg}), we can
compute respective parametric decompositions of the nonassociative tilde
curvature tensor,%
\begin{equation}
\ ^{\shortmid }\widetilde{\mathbf{R}}_{\star \alpha \beta \gamma }^{\mu }=%
\mathbf{\mathbf{\mathbf{\mathbf{\ }}}}\ _{[00]}^{\shortmid }\widetilde{%
\mathbf{R}}_{\star \alpha \beta \gamma }^{\mu }+\mathbf{\mathbf{\mathbf{%
\mathbf{\ }}}}\ _{[01]}^{\shortmid }\widetilde{\mathbf{R}}_{\star \alpha
\beta \gamma }^{\mu }(\hbar )+\mathbf{\mathbf{\mathbf{\mathbf{\ }}}}\
_{[10]}^{\shortmid }\widetilde{\mathbf{R}}_{\star \alpha \beta \gamma }^{\mu
}(\kappa )+\ _{[11]}^{\shortmid }\widetilde{\mathbf{R}}_{\star \alpha \beta
\gamma }^{\mu }(\hbar \kappa )+O(\hbar ^{2},\kappa ^{2},...).
\label{dcurvfh}
\end{equation}

Contracting the first and forth indices in (\ref{dcurvfh}), we define the
nonassociative canonical Ricci s-tensor, 
\begin{eqnarray*}
\mathbf{\mathbf{\mathbf{\mathbf{\ ^{\shortmid }}}}}\widetilde{\mathcal{\Re }}%
ic^{\star } &=&\mathbf{\mathbf{\mathbf{\mathbf{\ ^{\shortmid }}}}}\widetilde{%
\mathbf{\mathbf{\mathbf{\mathbf{R}}}}}ic_{\alpha \beta }^{\star }\widetilde{%
\star }(\ \mathbf{^{\shortmid }}\widetilde{\mathbf{e}}^{\alpha _{s}}\otimes
\ \mathbf{^{\shortmid }}\widetilde{\mathbf{e}}^{\beta _{s}}),\mbox{ where }
\\
&&\mathbf{\mathbf{\mathbf{\mathbf{\ ^{\shortmid }}}}}\widetilde{\mathbf{%
\mathbf{\mathbf{\mathbf{R}}}}}ic_{\alpha \beta }^{\star }:=\ ^{\shortmid }%
\widetilde{\mathcal{\Re }}ic^{\star }(\ ^{\shortmid }\widetilde{\mathbf{e}}%
_{\alpha },\ ^{\shortmid }\widetilde{\mathbf{e}}_{\beta })=\mathbf{\langle }%
\ \mathbf{\mathbf{\mathbf{\mathbf{\ ^{\shortmid }}}}}\widetilde{\mathbf{%
\mathbf{\mathbf{\mathbf{R}}}}}ic_{\mu \nu }^{\star }\widetilde{\star }(\ 
\mathbf{^{\shortmid }}\widetilde{\mathbf{e}}^{\mu }\otimes \ ^{\shortmid }%
\widetilde{\mathbf{e}}^{\nu }),\mathbf{\mathbf{\ }\ ^{\shortmid }}\widetilde{%
\mathbf{\mathbf{e}}}_{\alpha }\mathbf{\otimes \ ^{\shortmid }}\widetilde{%
\mathbf{\mathbf{e}}}_{\beta }\mathbf{\rangle }_{\widetilde{\star }},
\end{eqnarray*}%
when the coefficients can be computed in parametric form: 
\begin{eqnarray}
&\ ^{\shortmid }\widetilde{\mathbf{R}}ic_{\alpha \beta }^{\star }:=&\mathbf{%
\mathbf{\mathbf{\mathbf{\ ^{\shortmid }}}}}\widetilde{\mathcal{\Re }}_{\star
\alpha \beta \mu }^{\mu }=\ _{[00]}^{\shortmid }\widetilde{\mathbf{\mathbf{%
\mathbf{\mathbf{R}}}}}ic_{\alpha \beta }^{\star }+\mathbf{\mathbf{\mathbf{%
\mathbf{\ }}}}_{[01]}^{\shortmid }\widetilde{\mathbf{\mathbf{\mathbf{\mathbf{%
R}}}}}ic_{\alpha \beta }^{\star }(\hbar )+\mathbf{\mathbf{\mathbf{\mathbf{\ }%
}}}_{[10]}^{\shortmid }\widetilde{\mathbf{\mathbf{\mathbf{\mathbf{R}}}}}%
ic_{\alpha \beta }^{\star }(\kappa )+\mathbf{\mathbf{\mathbf{\mathbf{\ }}}}%
_{[11]}^{\shortmid }\widetilde{\mathbf{\mathbf{\mathbf{\mathbf{R}}}}}%
ic_{\alpha \beta }^{\star }(\hbar \kappa )+O(\hbar ^{2},\kappa ^{2},...), 
\notag \\
&\mbox{where}&\ _{[00]}^{\shortmid }\widetilde{\mathbf{R}}ic_{\alpha \beta
}^{\star }=\ _{[00]}^{\shortmid }\mathbf{\mathbf{\mathbf{\mathbf{\
^{\shortmid }}}}}\widetilde{\mathcal{\Re }}_{\star \alpha \beta \mu }^{\mu }%
\mathbf{\mathbf{\mathbf{\mathbf{\ ,}}}}\ _{[01]}^{\shortmid }\widetilde{%
\mathbf{\mathbf{\mathbf{\mathbf{\mathbf{\mathbf{\mathbf{\mathbf{R}}}}}}}}}%
ic_{\alpha \beta }^{\star }=\ _{[01]}^{\shortmid }\mathbf{\mathbf{\mathbf{%
\mathbf{\ ^{\shortmid }}}}}\widetilde{\mathcal{\Re }}_{\star \alpha \beta
\mu }^{\mu },  \label{driccifhstar1} \\
&&\ _{[10]}^{\shortmid }\widetilde{\mathbf{\mathbf{\mathbf{\mathbf{\mathbf{%
\mathbf{\mathbf{\mathbf{R}}}}}}}}}ic_{\alpha \beta }^{\star }=\
_{[10]}^{\shortmid }\widetilde{\mathbf{\mathbf{\mathbf{\mathbf{\mathcal{\Re }%
}}}}}_{\quad \alpha \beta \mu }^{\star \mu },\ _{[11]}^{\shortmid }%
\widetilde{\mathbf{\mathbf{\mathbf{\mathbf{R}}}}}ic_{\alpha \beta }^{\star
}=\ _{[11]}^{\shortmid }\widetilde{\mathbf{\mathbf{\mathbf{\mathbf{\mathcal{%
\Re }}}}}}_{\quad \alpha \beta \mu }^{\star \mu }.  \notag
\end{eqnarray}%
Such Ricci d-tensors for are not symmetric for general nonassociative cases
even for terms proportional to $\hbar ^{0}$ and/or $\kappa ^{0}$. This is a
\ typical property of nonholonomic configurations and deformations even in
GR.

\subsubsection{Nonassociative Finsler-Hamilton generalization of the
Einstein equations}

Considering the inverse d-metric $\ _{\star }^{\shortmid }\widetilde{%
\mathfrak{g}}^{\mu \nu }$ (it is tedious procedure similar to that first
introduced in \cite{aschieri17}, see also s-adapted formulas in \cite%
{partner01,partner02}) we can constact the indices with $\ ^{\shortmid }%
\widetilde{\mathbf{R}}ic_{\alpha \beta }^{\star }$ (\ref{driccifhstar1}) to
define and compute the nonassociative Finsler-Hamilton Ricci scalar
curvature:%
\begin{eqnarray}
&\ ^{\shortmid }\widetilde{\mathbf{R}}sc^{\star }:=&\ _{\star }^{\shortmid }%
\widetilde{\mathfrak{g}}^{\mu \nu }\mathbf{\mathbf{\mathbf{\mathbf{\
^{\shortmid }}}}}\widetilde{\mathbf{\mathbf{\mathbf{\mathbf{R}}}}}ic_{\mu
\nu }^{\star }=\left( \ _{\star }^{\shortmid }\mathfrak{\check{g}}^{\mu \nu
}+\ _{\star }^{\shortmid }\widetilde{\mathfrak{a}}^{\mu \nu }\right) \left( 
\mathbf{\mathbf{\mathbf{\mathbf{\ ^{\shortmid }}}}}\widetilde{\mathbf{%
\mathbf{\mathbf{\mathbf{R}}}}}ic_{(\mu \nu )}^{\star }+\mathbf{\mathbf{%
\mathbf{\mathbf{\ ^{\shortmid }}}}}\widetilde{\mathbf{\mathbf{\mathbf{%
\mathbf{R}}}}}ic_{[\mu \nu ]}^{\star }\right) =\ ^{\shortmid }\widetilde{%
\mathbf{\mathbf{\mathbf{\mathbf{R}}}}}ss^{\star }+\ ^{\shortmid }\widetilde{%
\mathbf{\mathbf{\mathbf{\mathbf{R}}}}}sa^{\star },  \notag \\
&\mbox{where}&\ ^{\shortmid }\widetilde{\mathbf{\mathbf{\mathbf{\mathbf{R}}}}%
}ss^{\star }=:\ _{\star }^{\shortmid }\mathfrak{\check{g}}^{\mu \nu }\mathbf{%
\mathbf{\mathbf{\mathbf{\ ^{\shortmid }}}}}\widetilde{\mathbf{\mathbf{%
\mathbf{\mathbf{R}}}}}ic_{(\mu \nu )}^{\star }\mbox{ and }\ ^{\shortmid }%
\widetilde{\mathbf{\mathbf{\mathbf{\mathbf{R}}}}}sa^{\star }:=\ _{\star
}^{\shortmid }\widetilde{\mathfrak{a}}^{\mu \nu }\mathbf{\mathbf{\mathbf{%
\mathbf{\ ^{\shortmid }}}}}\widetilde{\mathbf{\mathbf{\mathbf{\mathbf{R}}}}}%
ic_{[\mu \nu ]}^{\star },  \label{ricciscsymnonsym}
\end{eqnarray}%
where respective symmetric $\left( ...\right) $ and anti-symmetric $\left[
...\right] $ operators are defined using the multiple $1/2.$ For instance,
we define and compute $\ ^{\shortmid }\widetilde{\mathbf{R}}ic_{\mu \nu
}^{\star }=\mathbf{\mathbf{\mathbf{\mathbf{\ ^{\shortmid }}}}}\widetilde{%
\mathbf{\mathbf{\mathbf{\mathbf{R}}}}}ic_{(\mu \nu )}^{\star }+\mathbf{%
\mathbf{\mathbf{\mathbf{\ ^{\shortmid }}}}}\widetilde{\mathbf{\mathbf{%
\mathbf{\mathbf{R}}}}}ic_{[\mu \nu ]}^{\star }.$

We can follow abstract geometric principles formulated in \cite{misner} but
generalizing the constructions for nonholonomic phase spaces. A.\ Einstein
postulated his gravitational field equations in a similar geometric form on
pseudo-Riemannian spaces. This allows us to postulate for the
Finsler-Cartan-Hamilton phase space data $( \ ^{\star }\widetilde{\mathcal{M}%
},\ _{\star }^{\shortmid }\widetilde{\mathfrak{g}},\ ^{\shortmid }\widetilde{%
\mathbf{D}}^{\star })$ the nonassociatve and noncommutative modified vacuum
Einstein equations, 
\begin{equation}
\ ^{\shortmid }\widetilde{\mathbf{R}}ic_{\alpha \beta }^{\star }-\frac{1}{2}%
\ _{\star }^{\shortmid }\widetilde{\mathfrak{g}}_{\alpha \beta }\
^{\shortmid }\widetilde{\mathbf{R}}sc^{\star }=\ ^{\shortmid }\lambda \
_{\star }^{\shortmid }\widetilde{\mathfrak{g}}_{\alpha \beta }.
\label{nonassocfheinsteq}
\end{equation}%
Such systems of nonlinear of PDEs can't be decoupled and integrated in
certain general off-diagonal forms with 4+4 splitting for nonassociative
star product deformations. In \cite{vacaru09a}, we constructed
nonassociative Finsler BH solutions for the Cartan d-connection. R-flux
modifications introduce additional coupling into modified Einstein equations
which makes more cumbersome the procedure of finding exact and parametric
solutions. Here we note that (\ref{nonassocfheinsteq}) allows a formulation
in (nonassociative and noncommutative) almost Kaehler variables which can be
used for deformation quantization of such phase space theories, see \cite%
{vacaru16} and references therein.

We can generate solutions of (\ref{nonassocfheinsteq}) if we introduce
additional dyadic decompositions ($\ ^{\shortmid }\widetilde{\mathbf{e}}%
_{\alpha }\rightarrow \ ^{\shortmid }\mathbf{e}_{\alpha _{s}}=e_{\ \alpha
_{s}}^{\beta }(\ _{s}^{\shortmid }u)^{\shortmid }\widetilde{\mathbf{e}}%
_{\beta }$ with $\ _{\star }^{\shortmid }\widetilde{\mathfrak{g}}_{\alpha
\beta }\rightarrow \ _{\star }^{\shortmid }\widetilde{\mathfrak{g}}_{\alpha
_{s}\beta _{s}})$ and canonical s-distortions, $\ ^{\shortmid }\widetilde{%
\mathbf{D}}^{\star }\rightarrow $ $\ _{s}^{\shortmid }\widehat{\mathbf{D}}%
^{\star }.$ For such nonholonomic transforms, distortions of d- and
s-connections (\ref{stardistransf}) result in distortion relations for
respective nonassociative Ricci tensors/ d-tensors/ s-tensor, which in
abstract index form can be expressed as 
\begin{equation*}
\ ^{\shortmid }\widetilde{\mathbf{R}}ic_{\alpha _{s}\beta _{s}}^{\star }=\
^{\shortmid }\widehat{\widetilde{\mathbf{R}}}ic_{\alpha _{s}\beta
_{s}}^{\star }[\ ^{\shortmid }\widetilde{\mathbf{D}}^{\star },\mathbf{\
_{s}^{\shortmid }\mathbf{Z}^{\star }},\ _{\star }^{\shortmid }\widetilde{%
\mathfrak{g}}_{\alpha \beta }]+\ ^{\shortmid }\widetilde{\mathbf{Z}}%
ic_{\alpha _{s}\beta _{s}}^{\star }[\ ^{\shortmid }\widetilde{\mathbf{D}}%
^{\star },\mathbf{\ _{s}^{\shortmid }\mathbf{Z}^{\star }},\ _{\star
}^{\shortmid }\widetilde{\mathfrak{g}}_{\alpha \beta }],
\end{equation*}%
where $[...]$ are used to emphasize that such value are determined as
functionals of certain geometric d- and/or s-objects. Both the hat and tilde
labels are kept in order to emphasize that we shall use a canonical Ricci
s-tensor generated from a Finsler-Hamilton structure. As a result, we can
re-write (\ref{nonassocfheinsteq}) in an equivalent form:%
\begin{eqnarray}
\ ^{\shortmid }\widehat{\widetilde{\mathbf{R}}}ic_{\alpha _{s}\beta
_{s}}^{\star } &=&\ ^{\shortmid }\widetilde{\Upsilon }_{\alpha _{s}\beta
_{s}}^{\star },\mbox{ where }  \label{fhcanoneinst} \\
\ ^{\shortmid }\widetilde{\Upsilon }_{\alpha _{s}\beta _{s}}^{\star } &=&\
^{\shortmid }\lambda \ _{\star }^{\shortmid }\widetilde{\mathfrak{g}}%
_{\alpha _{s}\beta _{s}}+\frac{1}{2}\ _{\star }^{\shortmid }\widetilde{%
\mathfrak{g}}_{\alpha \beta }\ ^{\shortmid }\widetilde{\mathbf{R}}sc^{\star
}-\ ^{\shortmid }\widetilde{\mathbf{Z}}ic_{\alpha _{s}\beta _{s}}^{\star } 
\notag
\end{eqnarray}%
where the effective sources $\ ^{\shortmid }\widetilde{\Upsilon }_{\alpha
_{s}\beta _{s}}^{\star }$ involves a star - scalar functional $\ ^{\shortmid
}\widetilde{\mathbf{R}}sc^{\star }[\ ^{\shortmid }\widetilde{\mathbf{D}}%
^{\star },\mathbf{\ _{s}^{\shortmid }\mathbf{Z}^{\star }},\ _{\star
}^{\shortmid }\widetilde{\mathfrak{g}}_{\alpha \beta }].$ We do not provide
explicit formulas for $\ ^{\shortmid }\widetilde{\Upsilon }_{\alpha
_{s}\beta _{s}}^{\star }$ because we prove in section \ref{ssnonlsym} that
such s-tensors are related via certain nonlinear symmetries to certain
effective cosmological constants, $\ _{s}\widetilde{\Lambda },$ when $\
^{\shortmid }\widetilde{\Upsilon }_{\alpha _{s}\beta _{s}}^{\star
}\rightarrow \ _{s}\widetilde{\Lambda }\ ^{\shortmid }\widetilde{\mathbf{g}}%
_{\alpha _{s}\beta _{s}}$.

The system of nonlinear PDEs (\ref{fhcanoneinst}) can be generalized for
nonassociative geometric flows and decoupled and integrated in general
parametric form using the AFCDM as it was proven in nonassociative form in 
\cite{partner01,partner02,partner04,partner05,partner06}, where we
considered a different type of effective and matter field sources. The main
assumptions to get explicit off-diagonals decoupling are that the
nonholonomic s-adapted frame structure is chosen in such a form that 
\begin{eqnarray}
\ _{\star }^{\shortmid }\mathfrak{\check{g}}_{\alpha _{s}\beta _{s}}^{[0]}
&=&\ _{\star }^{\shortmid }\widetilde{\mathbf{g}}_{\alpha _{s}\beta _{s}}=\
^{\shortmid }\widetilde{\mathbf{g}}_{\alpha _{s}\beta _{s}};\ _{\star
}^{\shortmid }\mathfrak{\check{g}}_{\beta _{s}\gamma _{s}}^{[1]}(\kappa
)=0,\ \ _{\star }^{\shortmid }\widetilde{\mathfrak{a}}_{\alpha _{s}\beta
_{s}}^{[0]}=0,\ _{\star }^{\shortmid }\widetilde{\mathfrak{a}}_{\alpha
_{s}\beta _{s}}^{[1]}=i\kappa \mathcal{R}_{\quad \lbrack \alpha _{s}}^{\tau
_{s}\xi _{s}}\ \mathbf{^{\shortmid }}\widetilde{\mathbf{e}}_{|\xi _{s}}\
^{\shortmid }\widetilde{\mathbf{g}}_{\tau _{s}|\beta _{s}]},\mbox{ and } 
\notag \\
\ ^{\shortmid }\widetilde{\Upsilon }_{\star \ \beta _{s}}^{\alpha _{s}}
&=&[~_{1}^{\shortmid }\widetilde{\Upsilon }(\hbar ,\kappa ,x^{k_{1}})\delta
_{i_{1}}^{j_{1}},~_{2}^{\shortmid }\widetilde{\Upsilon }(\hbar ,\kappa
,x^{k_{1}},x^{3})\delta _{b_{2}}^{a_{2}},~_{3}^{\shortmid }\widetilde{%
\Upsilon }(\hbar ,\kappa ,x^{k_{2}},\ ^{\shortmid }p_{6})\delta
_{a_{3}}^{b_{3}},~_{4}^{\shortmid }\widetilde{\Upsilon }(\hbar ,\kappa
,x^{k_{3}},\ ^{\shortmid }p_{8})\delta _{a_{4}}^{b_{4}}].
\label{fhamiltsourcpar}
\end{eqnarray}%
The effective s-sources $~_{s}^{\shortmid }\widetilde{\Upsilon }$ can be
prescribed as generating sources for some classes of off-diagonal solutions 
or considered in recurrent form for effective parametric sources with
coefficients proportional to $\hbar ,\kappa $ and $\hbar \kappa ,$ when  $\
^{\shortmid }\widetilde{\Upsilon }_{_{\beta _{s}\gamma _{s}}}=\
_{[0]}^{\shortmid }\widetilde{\Upsilon }_{_{\beta _{s}\gamma _{s}}}+\
_{[1]}^{\shortmid }\widetilde{\Upsilon }_{_{\beta _{s}\gamma
_{s}}}\left\lceil \hbar ,\kappa \right\rceil $ as in \cite{aschieri17}.
General frame transforms on $\ _{s}^{\shortmid }\widetilde{\mathcal{M}}%
^{\star }$ transforms (\ref{fhamiltsourcpar}) into off-diagonal sources
encoding nonassociative Finsler-Hamilton data which also are contained in
nontrivial N-connection coefficients and respective coefficients of
s-metrics.

The system of nonassociative modified Einstein equations on phase spaces
does not have a variational proof for general twist product (this problem
exists in nonassociative and noncommutative theories when we do not fix a
unique differential and integral calculus). In parametric form, we can fix a
N-adapted or s-adapted variational calculus on $\ _{s}^{\shortmid }%
\widetilde{\mathcal{M}}^{\star }$ and then perform a star product
deformation procedure (for $\hbar ,\kappa $ and $\hbar \kappa $ R-flux
deformations). The general decoupling and integration properties of
parametric (\ref{fhcanoneinst}) with effective sources of type (\ref%
{fhamiltsourcpar}) can be proven for 8-d phase spaces with nonholonomic
canonical 2+2+2+2 decompositons as we considered in section \ref{sec3} for
nonholonomic 2+2 splitting on Lorentz spacetime manifolds. In abstract
geometric form, such nonassociative and Finsler-Hamilton parametric
constructions and generation of higher dimension classes of solution can be
performed by geometric analogy and extending with momentum shell variable
the dependence of generating and integration function and sources.

\subsection{Nonassociative Finsler-Lagrange-Hamilton geometric flows}

The theory of nonassociative geometric flows on phase spaces $\ ^{\shortmid }%
\mathcal{M}^{\star }$ and $\ _{s}^{\shortmid }\mathcal{M}^{\star }$ was
elaborated in off-diagonally integrable form using canonical s-variables in 
\cite{partner04,partner05,partner06}. The goal of this subsection is to
study models of nonassociative Finsler-Cartan-Hamilton flows in tilde
variables for $\ _{s}^{\shortmid }\widetilde{\mathcal{M}}^{\star }.$ Such a
formulation is important for connecting the AFCDM to theories of deformation
quantization and other models of quantum phase space. In tilde variables,
the nonassociative Finsler-Hamilton flow theory allows an equivalent
formulation in almost sympectic variables (\ref{sympf}) when exists
deformation quantization procedure outlined in \cite{vacaru16}. We show how
the abstract geometric and N-adapted formalism can be applied to define
cotangent Lorentz bundle generalizations of the R. Hamilton \cite{hamilton82}
and D. Friedan \cite{friedan80}) geometric flow equations and G. Perelman 
\cite{perelman1} thermodynamics for Ricci flows. Comprehensive mathematical
reviews of results on geometric flows of Riemannian and Kahler metrics and
related issues on Thurston-Poincar\'{e} conjecture (i.e. theorem, after
Perelman's proof) are presented in \cite{kleiner06,morgan06,cao06}. For
applications in modern mathematical particle physics, cosmology and quantum
information flows, we cite \cite%
{v07,v08,sv19,bvv21,ibv22,kehagias19,biasio21,lueben21,biasio22}.

\subsubsection{Finsler variables for Perelman's functionals and
nonassociative geometric flows}

Let us consider a nonassociative star product R-flux deformed phase space $\
^{\shortmid }\widetilde{\mathcal{M}}^{\star }$ enabled with d-objects $[\
^{\shortmid }\widetilde{\mathfrak{g}}^{\star },\ ^{\shortmid }\widetilde{%
\mathbf{D}}^{\star }]$ for a star product $\star $ structure (\ref%
{starphamilt}) N-adapted to a nonholonomic (4+4) decompositon which can be
also associated to a necessary nonholonomic shell (2+2)+(2+2) decomposition.
We follow the \textbf{Convention 2FH} (\ref{conv2FH}) with a $\kappa $%
-linear parametric decomposition of nonholonomic structures and geometric
d-objects when $\ _{\star }^{\shortmid }\mathfrak{\check{g}}_{\alpha \beta
}^{[0]}=\ _{\star }^{\shortmid }\widetilde{\mathbf{g}}_{\alpha \beta }= \
^{\shortmid }\widetilde{\mathbf{g}}_{\alpha \beta }$ as in (\ref%
{fhamiltsourcpar}).

Nonassociative Finsler-Cartan-Hamilton will be modelled for as flows on
temperature like parameter $\tau $ (when $0\leq \tau \leq \tau _{0}$) of
d-objects on $\ ^{\shortmid }\widetilde{\mathcal{M}}^{\star }$ when in the $%
[0]$-approximation, i.e. zero power on $\kappa ,$ are defined flows of
volume elements 
\begin{equation}
d\ \ ^{\shortmid }\widetilde{\mathcal{V}}ol(\tau )=\sqrt{|\ \ ^{\shortmid }%
\widetilde{\mathbf{g}}_{\alpha \beta }\ (\tau )|}\ \widetilde{\delta }^{8}\
^{\shortmid }u^{\gamma _{s}}(\tau ).  \label{volformfh}
\end{equation}%
Such a value is computed using N-elongated s-differentials $\widetilde{%
\delta }^{8}\ ^{\shortmid }u^{\gamma _{s}}(\tau )$ which are linear on $\
^{\shortmid }\widetilde{N}_{\ i_{s}a_{s}}\ (\tau )$ as in $\ ^{\shortmid }%
\widetilde{\mathbf{e}}_{i}(\tau ).$ The nonassociative geometric flow
constructions from \cite{partner04,partner05,partner06} can be reformulated
for the geometric data $[\ ^{\shortmid }\widetilde{\mathbf{g}}^{\star }(\tau
),\ ^{\shortmid }\widetilde{\mathbf{D}}^{\star }(\tau )],$ when the Perelman
type functionals are postulated: 
\begin{eqnarray}
\ ^{\shortmid }\widetilde{\mathcal{F}}^{\star }(\tau ) &=&\int_{\
^{\shortmid }\widetilde{\Xi }}(\ ^{\shortmid }\widetilde{\mathbf{R}}%
sc^{\star }+|\ ^{\shortmid }\widetilde{\mathbf{D}}^{\star }\ ^{\shortmid }%
\widetilde{f}|^{2})\widetilde{\star }e^{-\ \ ^{\shortmid }\widetilde{f}}\ d\
^{\shortmid }\widetilde{\mathcal{V}}ol(\tau ),\mbox{ and }
\label{naffunctfh} \\
\ ^{\shortmid }\widetilde{\mathcal{W}}^{\star }(\tau ) &=&\int_{\
^{\shortmid }\widetilde{\Xi }}\left( 4\pi \tau \right) ^{-4}\ [\tau (\
^{\shortmid }\widetilde{\mathbf{R}}sc^{\star }+\sum\nolimits_{s}|\
^{\shortmid }\widetilde{\mathbf{D}}^{\star }\widetilde{\star }\ ^{\shortmid }%
\widetilde{f}|)^{2}+\ ^{\shortmid }\widetilde{f}-8]\widetilde{\star }e^{-\ \
^{\shortmid }\widetilde{f}}\ d\ ^{\shortmid }\widetilde{\mathcal{V}}ol(\tau
).  \label{nawfunctfh}
\end{eqnarray}%
The 8-d hypersurfrace integrals for such F- and W-functionals are determined
by a volume element (\ref{volformfh}) and the h-c-normalizing functions $\
^{\shortmid }\widetilde{f}(\tau ,\ ^{\shortmid }u)$ can be stated to satisfy
the condition 
\begin{equation}
\int_{\ ^{\shortmid }\widetilde{\Xi }}\ ^{\shortmid }\widetilde{\nu }\ \ d\
^{\shortmid }\widetilde{\mathcal{V}}ol(\tau ):=\int_{t_{1}}^{t_{2}}\int_{%
\widetilde{\Xi }_{t}}\ \int_{\ ^{\shortmid }\widetilde{\Xi }_{E}}\
^{\shortmid }\widetilde{\nu }\ \ d\ ^{\shortmid }\widetilde{\mathcal{V}}%
ol(\tau )=1.  \label{normcond}
\end{equation}%
In these formulas, where the integration measures $\ \ ^{\shortmid }%
\widetilde{\nu }=\left( 4\pi \tau \right) ^{-4}e^{-\ ^{\shortmid }\widetilde{%
f}}$ are parameterized for the h- and c-components, with shell further
parameterizations if necessary. For general topological considerations, such
conditions may be not considered. We can consider also star-deformations of
the volume form when 
\begin{eqnarray*}
e^{-\ ^{\shortmid }\widetilde{f}}d\ ^{\shortmid }\widetilde{\mathcal{V}}%
ol(\tau ) &\rightarrow &e^{-\ _{s}^{\shortmid }\widehat{f}}d\ ^{\shortmid }%
\mathcal{V}ol(\tau )\rightarrow e^{-\ \ _{s}^{\shortmid }\widehat{f}}d\
^{\shortmid }\mathcal{V}ol^{\star }(\tau )=\newline
e^{-\ \ _{s}^{\shortmid }\widehat{f}}\sqrt{|\ _{\star }^{\shortmid }\mathbf{g%
}_{\alpha _{s}\beta _{s}}\ (\tau )|}\delta \ ^{\shortmid }u^{\gamma
_{s}}(\tau ) \\
&\rightarrow &e^{-\ \ _{s}^{\shortmid }\widehat{f}}d\ ^{\shortmid }\mathcal{V%
}ol^{\star }(\tau )=\newline
e^{-\ \ ^{\shortmid }\widetilde{f}}\sqrt{|\ _{\star }^{\shortmid }\widetilde{%
\mathbf{g}}_{\alpha _{s}\beta _{s}}\ (\tau )|}\widetilde{\delta }\
^{\shortmid }u^{\gamma _{s}}(\tau ).
\end{eqnarray*}%
Other types of adapted integration measures and nonholonomic s-shells, for
instance, involving $\ _{s}^{\shortmid }\mathfrak{g}^{\star }\approx \
^{\shortmid }\widetilde{\mathfrak{g}}^{\star }.$ Such transforms can be
encoded into \ respective normalizing functions and adapted to a respective
separation of nonsymmetric components of s-metrics for $\kappa $--linear
parameterizations. .

The nonassociative geometric flow evolution equation of the Finsler-Hamilton
data $[\ ^{\shortmid }\widetilde{\mathbf{g}}^{\star }(\tau ),\ ^{\shortmid }%
\widetilde{\mathbf{D}}^{\star }(\tau ),\ ^{\shortmid }\widetilde{f}(\tau )]$
are postulated in the form%
\begin{eqnarray}
\partial _{\tau }\ ^{\shortmid }\widetilde{\mathfrak{g}}_{\alpha \beta
}^{\star }(\tau ) &=&-2\ ^{\shortmid }\widetilde{\mathbf{R}}_{\ \alpha \beta
}^{\star }(\tau ),  \label{nonassocfhfl} \\
\partial _{\tau }\ ^{\shortmid }\widetilde{f}(\tau ) &=&\ ^{\shortmid }%
\widetilde{\mathbf{R}}sc^{\star }(\tau )-\ ^{\star }\widetilde{%
\bigtriangleup }(\tau )\widetilde{\star }\ \ ^{\shortmid }\widetilde{f}(\tau
)+(\ ^{\shortmid }\widetilde{\mathbf{D}}^{\star }(\tau )\widetilde{\star }\
\ ^{\shortmid }\widetilde{f}(\tau ))^{2}.  \notag
\end{eqnarray}%
In (\ref{nonassocfhfl}), $\ \ ^{\star }\widetilde{\bigtriangleup }(\tau )=[\
^{\shortmid }\widetilde{\mathbf{D}}^{\star }(\tau )]^{2}$ are families of
the Laplace d-operators and the nonsymmetric components of $\
_{\star}^{\shortmid }\widetilde{\mathfrak{g}}_{\alpha _{s}\beta _{s}}(\tau )$
are computed using $\kappa $-linear parameterizations (\ref{aux40b})--(\ref%
{aux40aa}). In commutative versions, this system of such nonlinear PDEs can
be derived in variational forms from the F- and W-potentials, respectively, (%
\ref{naffunctfh}) and (\ref{nawfunctfh}) generalizing the proofs provided in 
\cite{perelman1}, see details in monographs \cite{kleiner06,morgan06,cao06}
and, for various nonassociative, nonholonomic non-Riemannian
generalizations, \cite{partner04,partner05,partner06}. Applying abstract
geometric methods, we can derive (\ref{nonassocfhfl}) as a generalization of
the relativistic canonical evolution equations following the \textbf{%
Convention 2FH} (\ref{conv2FH}). Here we note that for $\kappa $-linear
parametric decompositions such nonlinear geometric evolution equations can
be derived in a variational form from $\kappa $-linear parameterizations of (%
\ref{naffunctfh}) and (\ref{nawfunctfh}). Nonassociative geometric flow
equations can be also motivated as star product R-flux deformations of a
two-dimensional sigma model with beta functions and dilaton field as it was
stated by equations (79) and (80) in \cite{kehagias19}. Nevertheless,
variational proofs are not possible for general twist products as we
discussed in details in \cite{partner04,partner05}. when abstract geometric
and N-adapted methods became very important.

Nonassociative Ricci solitons for the Finsler-Hamilton d-connection $\
^{\shortmid }\widetilde{\mathbf{D}}^{\star }$ are defined as self-similar
configurations of gradient geometric flows (\ref{nonassocfhfl}) for a fixed
parameter $\tau _{0}.$ On $\ $ $\ ^{\shortmid }\widetilde{\mathcal{M}}%
^{\star },$ the Ricci soliton d-equations are of type 
\begin{equation}
\ ^{\shortmid }\widetilde{\mathbf{R}}_{\ \alpha \beta }^{\star }+\
^{\shortmid }\widetilde{\mathbf{D}}_{\alpha }^{\star }\ ^{\shortmid }%
\widetilde{\mathbf{D}}_{\beta }^{\star }\ _{s}^{\shortmid }\widetilde{\varpi 
}(\ ^{\shortmid }u)=\ _{s}^{\shortmid }\lambda \ \ \ _{\star }^{\shortmid }%
\widetilde{\mathfrak{g}}_{\alpha \beta }.  \label{naricsolfh}
\end{equation}%
where $\ _{s}^{\shortmid }\widetilde{\varpi }$ is a smooth potential
function on every shell $s=1,2$ and $\lambda =const.$ The the nonassociatve
Finsler-Hamilton modified Einstein equations (\ref{nonassocfheinsteq})
consist an example of nonassociative Ricci soliton ones (\ref{naricsolfh}).

We emphasize that the nonlinear systems of PDEs (\ref{nonassocfhfl}) and (%
\ref{naricsolfh}) can be decoupled and integrated in general form, applying
a generalized AFCDM, after introdcing a double nonholonomic splitting, with
N-connections and nonholonomic dyadic structures, writing, for instance, 
\begin{eqnarray}
\partial _{\tau }\ ^{\shortmid }\widetilde{\mathfrak{g}}_{\alpha _{s}\beta
_{s}}^{\star }(\tau ) &=&-2\ ^{\shortmid }\widetilde{\mathbf{R}}_{\ \alpha
_{s}\beta _{s}}^{\star }(\tau ),  \label{nonassocfhflcan} \\
\partial _{\tau }\ \ ^{\shortmid }\widehat{f}(\tau ) &=&\ _{s}^{\shortmid }%
\widetilde{\mathbf{R}}sc^{\star }(\tau )-\ _{s}^{\star }\widetilde{%
\bigtriangleup }(\tau )\widetilde{\star }\ \ \ ^{\shortmid }\widehat{f}(\tau
)+(\ _{s}^{\shortmid }\widetilde{\mathbf{D}}^{\star }(\tau )\widetilde{\star 
}\ \ ^{\shortmid }\widehat{f}(\tau ))^{2}.  \notag
\end{eqnarray}%
with re-definition of normalizing functions for s-shells, $\ ^{\shortmid }%
\widetilde{f}(\tau )\rightarrow \ ^{\shortmid }\widehat{f}(\tau )$.

\subsubsection{Thermodynamic models for nonassociative Finsler-Hamilton flow
thermodynamics}

Let \ us consider such geometric data: a family of d-metrics $\ ^{\shortmid }%
\widetilde{\mathbf{g}}_{\alpha \beta }\ (\tau )$ \ used for nonassociative
star product deformations; a closed hypersurface $^{\shortmid }\widetilde{%
\Xi }$ in the nonassociative phase space $\ ^{\shortmid }\widetilde{M}%
^{\star }\subset \ ^{\shortmid }\widetilde{\mathcal{M}}^{\star };$ and the
volume form $d\ \ ^{\shortmid }\widetilde{\mathcal{V}}ol(\tau )$ (\ref%
{volformfh}). We can introduce the partition function for nonassociative
Finsler-Hamilton phase spaces of dimension $n=8$, 
\begin{equation}
\ ^{\shortmid }\widetilde{\mathcal{Z}}^{\star }(\tau )=\exp
[\int_{^{\shortmid }\widetilde{\Xi }}[-\ ^{\shortmid }\widetilde{f}+4]\
\left( 4\pi \tau \right) ^{-4}e^{-\ ^{\shortmid }\widetilde{f}}\ d\
^{\shortmid }\widetilde{\mathcal{V}}ol(\tau ),  \label{spffh}
\end{equation}%
for which using standard statisical and geometric mechanics computations 
\cite{perelman1,partner04,partner05,partner06} we can define and compute
such thermodynamic variables: 
\begin{eqnarray}
\mbox{ average energy },\ ^{\shortmid }\widetilde{\mathcal{E}}^{\star }(\tau
)\ &=&-\tau ^{2}\int_{^{\shortmid }\widetilde{\Xi }}\ \left( 4\pi \tau
\right) ^{-4}\left( \ ^{\shortmid }\widetilde{\mathbf{R}}sc^{\star }+|\
^{\shortmid }\widetilde{\mathbf{D}}^{\star }\ ^{\shortmid }\widetilde{f}%
|^{2}-\frac{4}{\tau }\right) \widetilde{\star }e^{-\ ^{\shortmid }\widetilde{%
f}}\ d\ ^{\shortmid }\widetilde{\mathcal{V}}ol(\tau );  \label{nagthermodfh}
\\
\mbox{ entropy },\ ^{\shortmid }\widetilde{\mathcal{S}}^{\star }(\tau )\
&=&-\int_{^{\shortmid }\widetilde{\Xi }}\left( 4\pi \tau \right) ^{-4}\left(
\tau (\ ^{\shortmid }\widetilde{\mathbf{R}}sc^{\star }+|\ ^{\shortmid }%
\widetilde{\mathbf{D}}^{\star }\ ^{\shortmid }\widetilde{f}|^{2})+\
^{\shortmid }\tilde{f}-8\right) \widetilde{\star }e^{-\ ^{\shortmid }%
\widetilde{f}}\ d\ ^{\shortmid }\widetilde{\mathcal{V}}ol(\tau );  \notag \\
\mbox{ fluctuation },\ _{s}^{\shortmid }\widehat{\sigma }^{\star }(\tau )
&=&2\ \tau ^{4}\int_{^{\shortmid }\widetilde{\Xi }}\left( 4\pi \tau \right)
^{-4}|\ \ ^{\shortmid }\widetilde{\mathbf{R}}_{\alpha \beta }^{\star }+\
^{\shortmid }\widetilde{\mathbf{D}}_{\alpha }^{\star }\ \ ^{\shortmid }%
\widetilde{\mathbf{D}}_{\beta }^{\star }\ \ ^{\shortmid }\tilde{f}-\frac{1}{%
2\tau }\ ^{\shortmid }\widetilde{\mathbf{g}}_{\alpha \beta }^{\star }|^{2}%
\widetilde{\star }e^{-\ ^{\shortmid }\widetilde{f}}\ d\ ^{\shortmid }%
\widetilde{\mathcal{V}}ol(\tau ).  \notag
\end{eqnarray}%
These formulas can be derived generalizing in s-adapted form the commutative
variational procedure with further twisted product deformations \cite%
{perelman1,kleiner06,morgan06,cao06} using $\ ^{\shortmid }\widetilde{%
\mathcal{W}}^{\star }(\tau )=-\ ^{\shortmid }\widetilde{\mathcal{S}}%
^{\star}(\tau )$ (\ref{nawfunctfh}) and $\ \ ^{\shortmid }\widetilde{%
\mathcal{Z}}^{\star }(\tau )$ (\ref{spffh}).

For applications in modern physics, we can consider $\ \kappa $-linear
parametric decompositions of nonassociative geometric thermodynamic
variables (\ref{nagthermodfh}). Corresponding formulas can be derived in
variational or abstract geometric form using the F- and W-functionals (\ref%
{naffunctfh}) and (\ref{nawfunctfh}), 
\begin{eqnarray}
\ ^{\shortmid }\widetilde{\mathcal{F}}_{\kappa }^{\star }(\tau )
&=&\int_{^{\shortmid }\widetilde{\Xi }}(\ ^{\shortmid }\widetilde{\mathbf{R}}%
sc+\ ^{\shortmid }\widetilde{\mathbf{K}}sc+|\ \ ^{\shortmid }\widetilde{%
\mathbf{D}}\ \ ^{\shortmid }\tilde{f}|^{2})e^{-\ \ \ ^{\shortmid }\tilde{f}%
}\ d\ ^{\shortmid }\widetilde{\mathcal{V}}ol(\tau ),\mbox{
and }  \label{naffunctpfh} \\
\ \ ^{\shortmid }\widetilde{\mathcal{W}}_{\kappa }^{\star }(\tau )
&=&\int_{^{\shortmid }\widetilde{\Xi }}\left( 4\pi \tau \right) ^{-4}\ [\tau
(\ \ ^{\shortmid }\widetilde{\mathbf{R}}sc+\ ^{\shortmid }\widetilde{\mathbf{%
K}}sc+|\ ^{\shortmid }\widetilde{\mathbf{D}}\ \ ^{\shortmid }\tilde{f}|^{2}\
)^{2}+\ \ ^{\shortmid }\tilde{f}-8]e^{-\ \ \ ^{\shortmid }\tilde{f}}\ d\
^{\shortmid }\widetilde{\mathcal{V}}ol(\tau ).  \notag
\end{eqnarray}%
In (\ref{naffunctpfh}), the tilde Ricci scalar splits into two components, $%
^{\shortmid }\widetilde{\mathbf{R}}sc^{\star }=\ ^{\shortmid }\widetilde{%
\mathbf{R}}sc+\ ^{\shortmid }\widetilde{\mathbf{K}}sc,$ where\newline
$\ ^{\shortmid }\widetilde{\mathbf{K}}sc:=\ _{\star }^{\shortmid }\widetilde{%
\mathfrak{g}}^{\mu \nu }\ ^{\shortmid }\widetilde{\mathbf{K}}_{\ \beta
\gamma }\left\lceil \hbar ,\kappa \right\rceil $ contains the coefficients
proportional to $\hbar $ and $\kappa .$ The normalizing function $\ \
^{\shortmid }\tilde{f}$ is re-defined to include $\left\lceil \hbar ,\kappa
\right\rceil $-terms from $\ ^{\shortmid }\widetilde{\mathbf{D}}^{\star
}\rightarrow \ ^{\shortmid }\widetilde{\mathbf{D}}$ and other terms
including $\kappa $-parametric decompositions.

Using (\ref{naffunctpfh}), we derive such phase geometric flow equations
with $\kappa $-terms encoding star product R-flux deformations, 
\begin{eqnarray}
\partial _{\tau }\ ^{\shortmid }\widetilde{\mathbf{g}}_{\alpha \beta }(\tau
) &=&-2(\ ^{\shortmid }\widetilde{\mathbf{R}}_{\ \alpha \beta }(\tau )+\
^{\shortmid }\widetilde{\mathbf{K}}_{\ \alpha \beta }(\tau ,\left\lceil
\hbar ,\kappa \right\rceil )),  \label{nonassocgeomflfhp} \\
\partial _{\tau }\ ^{\shortmid }\widetilde{f}(\tau ) &=&\ ^{\shortmid }%
\widetilde{\mathbf{R}}sc(\tau )+\ ^{\shortmid }\widetilde{\mathbf{K}}sc(\tau
)-\widetilde{\bigtriangleup }(\tau )\ \ ^{\shortmid }\widetilde{f}(\tau )+(\
^{\shortmid }\widetilde{\mathbf{D}}(\tau )\ \ ^{\shortmid }\widetilde{f}%
(\tau ))^{2}(\tau ),  \notag
\end{eqnarray}%
where the Laplace operator $\widetilde{\bigtriangleup }$ is constructed from
the canonical s-connection $\ ^{\shortmid }\widetilde{\mathbf{D}}.$ For
self-similar configurations with $\tau =\tau _{0},$ the parametric
Finsler-Hamilton equations (\ref{nonassocgeomflfhp}) transform into a system
of nonlinear PDEs for $\kappa $--parametric generalized Finsler--Ricci
solitons.

Let us express the parametric Finsler-Hamilton flow equations as a $\tau $%
-family which similar to modified Einstein equations. We introduce $\tau $%
-depending sources $\ ^{\shortmid }\widetilde{\Im }_{\alpha \beta }^{\star
}(\tau )=-(\ ^{\shortmid }\widetilde{\mathbf{K}}_{\alpha _{s}\beta
_{s}}(\tau )+\frac{1}{2}\partial _{\tau }\ ^{\shortmid }\widetilde{\mathbf{g}%
}_{\alpha \beta }(\tau))$ and write (\ref{nonassocgeomflfhp}) in the form 
\begin{equation}
\ ^{\shortmid }\widetilde{\mathbf{R}}ic_{\alpha \beta }(\tau )=\ ^{\shortmid
}\widetilde{\Im }_{\alpha \beta }^{\star }(\tau ),
\label{nonassocgeomflefhf}
\end{equation}%
which are similar to the modified Einstein equations (\ref{nonassocfheinsteq}%
). Using frame transforms $\ ^{\shortmid }\widetilde{\Im }_{\alpha ^{\prime
}\beta ^{\prime }}^{\star }=e_{\ \alpha ^{\prime }}^{\alpha _{s}}e_{\ \beta
^{\prime }}^{\beta _{s}}\ ^{\shortmid }\widetilde{\Im }_{\alpha _{s}\beta
_{s}}^{\star },$ the effective sources can be parameterized in the form 
\begin{equation}
\ ^{\shortmid }\widetilde{\Im }_{\star \ \beta _{s}}^{\alpha _{s}}~(\tau ,\
^{\shortmid }u^{\gamma _{s}})=[~_{1}^{\shortmid }\widetilde{\Im }^{\star
}(\kappa ,\tau ,x^{k_{1}})\delta _{i_{1}}^{j_{1}},~_{2}^{\shortmid }%
\widetilde{\Im }^{\star }(\kappa ,\tau ,x^{k_{1}},y^{c_{2}})\delta
_{b_{2}}^{a_{2}},~_{3}^{\shortmid }\widetilde{\Im }^{\star }(\kappa ,\tau
,x^{k_{2}},p_{c_{3}})\delta _{a_{3}}^{b_{3}},~_{4}^{\shortmid }\widetilde{%
\Im }^{\star }(\kappa ,\tau ,~^{\shortmid }x^{k_{3}},p_{c_{4}})\delta
_{a_{4}}^{b_{4}}],  \label{cannonsymparamc2b}
\end{equation}%
i.e. $\ ^{\shortmid }\widetilde{\Im }_{_{\beta _{s}\gamma _{s}}}^{\star
}(\tau )=diag\{\ _{s}^{\shortmid }\widetilde{\Im }^{\star }(\tau )\}.$
Prescribed values $\ _{s}^{\shortmid }\widetilde{\Im }^{\star }~(\tau ,\
^{\shortmid }u^{\gamma _{s}})$ imposes a s-shell nonholonomic constraint for 
$\tau $-derivatives of the metrics s-coefficients $\partial _{\tau }\
^{\shortmid }\widetilde{\mathbf{g}}_{\alpha _{s}\beta _{s}}(\tau ).$ For a
fixed $\tau _{0}$ , the system of nonlinear PDEs transform into a
nonholonomic Ricci solition equation (\ref{naricsolfh}).

For small parametric deformations and a fixed $\tau _{0}$, such constraints
can be solved in explicit general forms, or allow recurrent parametric
computations of the coefficients of s-metrics and s-connection for a
corresponding class of solutions. Using effective sources (\ref%
{cannonsymparamc2b}), the $\kappa $-linear parametric geometric flow
equations (\ref{nonassocgeomflefhf}) can be written equivalently as a $\tau $%
-family of R-flux deformed Einstein equations written in canonical
s-variables, 
\begin{equation}
\ ^{\shortmid }\widetilde{\mathbf{R}}_{\ \ \gamma _{s}}^{\beta _{s}}(\tau )={%
\delta }_{\ \ \gamma _{s}}^{\beta _{s}}\ _{s}^{\shortmid }\widetilde{\Im }%
^{\star }(\tau ).  \label{nonassocrffh}
\end{equation}%
The data $\ _{s}^{\shortmid }\widetilde{\Im }(\tau )^{\star }$ are
considered as generating sources for certain re-defined effective sources
including distortions , $\ _{s}^{\shortmid }\widetilde{\mathbf{D}} ^{\star
}=\ _{s}^{\shortmid }\widehat{\mathbf{D}}^{\star }+\ _{s}^{\shortmid}%
\widetilde{\mathbf{Z}}^{\star }$ when the system (\ref{nonassocrffh}) can be
decoupled and integrated in certain off-diagonal forms using the AFCDM. For
4-d configurations, the proofs were provided in details in the Part I when
the constructions can be extended in abstract geometric form on
nonassociative 8-d phase spaces.

\section{General off-diagonal solutions for nonassociative Finsler-Hamilton
flows and examples}

\label{sec7}

The AFCDM allows us to construct general classes of off-diagonal solutions
with dependencies on all spacetime and phase space coordinates in various
geometric flow and MGTs (which nonassociative, generalized Finsler and other
types). Most general the formulas became very cumbersome and we omit such
considerations in this work. If a Killing symmetry exists, the procedure of
generating exact/parametric solutions became more simple when the extensions
of formulas to 8-d phase spaces and 10-d spacetimes consist abstract
geometric and s-adapted generalizations of the 4-d proofs and solutions
presented in the Part I. The main geometric constructions and formulas are
summarized in Tables 1-16 the Appendix.

In this section, we provide the necessary off-diagonal ansatz which allow to
generate parametric solutions of nonassociative geometric flow equations (%
\ref{nonassocrffh}). We analyze the a class of nonlinear symmetries of such $%
\tau $-families of nonassociative Finsler-Hamilton phase spaces. Then we
study three physically important examples showing how the solutions from
when the off-diagonal solutions from \cite{partner04,partner05,partner06}
can be modified to encode nonassociative Finsler like geometric data. Such
off-diagonal metrics describe respective nonassociative Finsler-Hamilon
black ellipsoids, BEs; nonassociative generalized WHs; or locally
anisotropic cosmological solutions for $\tau $-evolution and modified
Finsler configurations.

\subsection{Off-diagonal ansatz and nonlinear symmetries of nonassociative
generalized Finsler flows}

Technically, it is not possible to decouple in general form the
nonassociative Finsler-Hamilton flows (\ref{nonassocgeomflefhf}) with
nonholonomic 4+4 splitting. Introducing an additional 2+2+2+2 splitting,
when $\ ^{\shortmid }\widetilde{\mathbf{g}}_{\alpha ^{\prime }\beta
^{\prime}}= e_{\ \alpha ^{\prime }}^{\alpha _{s}}e_{\ \beta ^{\prime
}}^{\beta _{s}}\ ^{\shortmid }\widetilde{\mathbf{g}}_{\alpha _{s}\beta _{s}},
$ and the distortions $\ _{s}^{\shortmid }\widetilde{\mathbf{Z}}^{\star }$
from $\ _{s}^{\shortmid }\widetilde{\mathbf{D}}^{\star }= \ _{s}^{\shortmid }%
\widehat{\mathbf{D}}^{\star }+\ _{s}^{\shortmid }\widetilde{\mathbf{Z}}
^{\star }$ are encoded into $\ ^{\shortmid }\widetilde{\Im }_{\alpha \beta
}^{\star }$ being parameterized in a form $diag\{\ _{s}^{\shortmid }%
\widetilde{\Im }^{\star }(\tau )\}$ (\ref{cannonsymparamc2b}), we transform
the system of nonlinear PDEs in a form (\ref{nonassocrffh}). This allows us
to integrate such equations for respective ansatz with one Killing symmetry.

\subsubsection{Off-diagonal ansatz with one Killing symmetry}

Quasi-stationary solutions can be constructed using such a s-adapted ansatz 
\begin{eqnarray}
d\ ^{\shortmid }\widetilde{s}^{2}(\tau ) &=&\ \widetilde{g}_{i_{1}}(\tau
,x^{k_{1}})(dx^{i_{1}})^{2}+\widetilde{g}_{a_{2}}(\tau ,x^{i_{1}},y^{3})(%
\widetilde{\mathbf{e}}^{a_{2}}(\tau ))^{2}+ \ ^{\shortmid }\widetilde{g}%
^{a_{3}}(\tau ,x^{i_{2}},p_{6})(\ ^{\shortmid }\widetilde{\mathbf{e}}%
_{a_{3}}(\tau ))^{2}+  \notag \\
&& \ ^{\shortmid }\widetilde{g}^{a_{4}}(\tau ,\ ^{\shortmid
}x^{i_{3}},p_{7})(\ ^{\shortmid }\widetilde{\mathbf{e}}_{a_{4}}(\tau ))^{2},%
\mbox{where }  \notag \\
\widetilde{\mathbf{e}}^{a_{2}}(\tau ) &=&dy^{a_{2}}+\widetilde{N}%
_{k_{1}}^{a_{2}}(\tau ,x^{i_{1}},y^{3})dx^{k_{1}},\ ^{\shortmid }\widetilde{%
\mathbf{e}}_{a_{3}}(\tau )=dp_{a_{3}}+\ ^{\shortmid }\widetilde{N}%
_{a_{3}k_{2}}(\tau ,x^{i_{2}},p_{6})dx^{k_{2}},  \notag \\
\ ^{\shortmid }\widetilde{\mathbf{e}}_{a_{4}}(\tau ) &=&dp_{a_{4}}+\
^{\shortmid }\widetilde{N}_{a_{4}k_{3}}(\tau ,\ ^{\shortmid
}x^{i_{3}},p_{7})d\ ^{\shortmid }x^{k_{3}}.\   \label{ans1rf}
\end{eqnarray}%
Such a s-metric $\ ^{\shortmid }\widetilde{\mathbf{g}}_{\alpha _{s}\beta
_{s}}$ redefined in a coordinate base can't be diagonalized in a finite
phase space region by coordinate frame transforms and posses symmetry on a
time like Killing vector, i.e. on $\partial _{4}=\partial _{t},$ at least on
the shells $s=1$ and 2. We can prescrib a nonholonomic frame structure, when
the coefficients for respective s-metric and d-metrics, $\ ^{\shortmid }%
\widetilde{\mathbf{g}}_{\alpha \beta }$\ $\approx $ $\ ^{\shortmid }%
\widetilde{\mathbf{g}}_{\alpha _{s}\beta _{s}},$ and N-connections do not
depend on $y^{4}=t.$ The s-metric (\ref{ans1rf}) contains certain shell
Killing symmetries: on $\partial _{4}$ for the shell $s=2;$ on $\partial
_{5} $ for the shell $s=3;$ and on $\partial _{8}$ for the shell $s=4.$

Fixing $\tau _{0}$ and chosing self-similar configurations, we can generate
parametric solutions for nonassociatve Finsler-Hamilton modified Einstein
equations (\ref{nonassocfheinsteq}) or nonassociative Ricci soliton
equations (\ref{naricsolfh}). We can generate various classes of type (\ref%
{ans1rf}) quasi-stationary, or their time dual locally anisotropic
cosmological solutions, so-called rainbow metrics with Killing symmetry on $%
\partial _{7}$ and explicit dependence on an energy type coordinate $%
p_{8}=E. $

A general ansatz for generating \ off-diagonal solutions of (\ref%
{nonassocrffh}) with a phase space Killing symmetry on a$\ ^{\shortmid
}\partial ^{a_{4}},$ for $a_{4}=7,$ or 8; and a spacetime Killing symmetry
on $\partial _{a_{2}}$, for $a_{2}=3,$ or 4, (we distinguish two classes
depending on a Killing symmetry on $\partial _{4}=\partial _{t},$ or $%
\partial _{3},$ at least on the shells $s=1$ and 2) can be written in the
form 
\begin{eqnarray}
d\widetilde{s}^{2}(\tau ) &=&\widetilde{g}_{i_{1}}(\tau )(dx^{i_{1}})^{2}+%
\widetilde{g}_{a_{2}}(\tau )(\widetilde{\mathbf{e}}^{a_{2}}(\tau ))^{2}+\
^{\shortmid }\widetilde{g}^{a_{3}}(\tau )(\ ^{\shortmid }\widetilde{\mathbf{e%
}}_{a_{3}}(\tau ))^{2}+\ ^{\shortmid }\widetilde{g}^{a_{4}}(\tau )(\
^{\shortmid }\widetilde{\mathbf{e}}_{a_{4}}(\tau ))^{2},\mbox{where }
\label{ssolutions} \\
\widetilde{\mathbf{e}}^{a_{2}}(\tau ) &=&dy^{a_{2}}+\widetilde{N}%
_{k_{1}}^{a_{2}}(\tau )dx^{k_{1}},\ ^{\shortmid }\widetilde{\mathbf{e}}%
_{a_{3}}(\tau )=dp_{a_{3}}+\ ^{\shortmid }\widetilde{N}_{a_{3}k_{2}}(\tau
)dx^{k_{2}},\ ^{\shortmid }\widetilde{\mathbf{e}}_{a_{4}}(\tau
)=dp_{a_{4}}+\ ^{\shortmid }\widetilde{N}_{a_{4}k_{3}}(\tau )d\ ^{\shortmid
}x^{k_{3}},\   \notag
\end{eqnarray}%
where s-metric and N-connection coefficients (related to tilde
Finsler-Hamilton variables) are parameterized in the form 
\begin{equation*}
\frame{{\scriptsize $%
\begin{array}{ccccc}
\begin{array}{c}
{\ }\widetilde{{g}}_{i_{1}}{\ (\tau ,x}^{k_{1}}) \\ 
{\ =e}^{\widetilde{\psi }(\hbar ,\kappa ;\tau ,x^{k_{1}})}%
\end{array}
& 
\begin{array}{cc}
\begin{array}{c}
\begin{array}{c}
\widetilde{{g}}_{a_{2}}{(\tau ,x}^{i_{1}},y^{3}) \\ 
\widetilde{{N}}_{k_{1}}^{a_{2}}{(\tau ,x}^{i_{1}},y^{3})%
\end{array}
\\ 
\end{array}
& 
\begin{array}{c}
\mbox{quasi-} \\ 
\mbox{stationary}%
\end{array}
\\ 
\begin{array}{c}
\begin{array}{c}
\widetilde{{g}}_{a_{2}}(\tau ,x^{i_{1}},t) \\ 
\widetilde{{N}}_{k_{1}}^{a_{2}}(\tau ,x^{i_{1}},t)%
\end{array}
\\ 
\end{array}
& 
\begin{array}{c}
\mbox{locally anisotropic} \\ 
\mbox{cosmology}%
\end{array}%
\end{array}
& 
\begin{array}{c}
\begin{array}{c}
\ ^{\shortmid }{\ }\widetilde{{g}}^{a_{3}}(\tau ,x^{i_{2}},p_{5}) \\ 
\ ^{\shortmid }\widetilde{N}_{a_{3}k_{2}}(\tau ,x^{i_{2}},p_{5})%
\end{array}
\\ 
\\ 
\begin{array}{c}
\ ^{\shortmid }\widetilde{{g}}^{a_{3}}(\tau ,x^{i_{2}},p_{6}) \\ 
\ ^{\shortmid }\widetilde{{N}}_{a_{3}k_{2}}(\tau ,x^{i_{2}},p_{6})%
\end{array}
\\ 
\end{array}
& 
\begin{array}{c}
\ ^{\shortmid }\widetilde{{g}}^{a_{4}}(\tau ,\ ^{\shortmid }x^{i_{3}},p_{7})
\\ 
\ ^{\shortmid }\widetilde{{N}}_{a_{4}k_{3}}(\tau ,\ ^{\shortmid }{x}%
^{i_{3}},p_{7})%
\end{array}
& 
\begin{array}{c}
\mbox{fixed} \\ 
p_{8}=E_{0}%
\end{array}
\\ 
\begin{array}{c}
\tau \mbox{-flows of 2-d} \\ 
\mbox{Poisson eqs} \\ 
{\partial }_{11}^{2}\widetilde{{\psi }}{+\partial }_{22}^{2}\widetilde{{\psi 
}}{=} \\ 
{2\ }_{1}\widetilde{\Im }^{\star }{(\hbar ,\kappa ;\tau ,x}^{k_{1}})%
\end{array}
& 
\begin{array}{cc}
\begin{array}{c}
\begin{array}{c}
\widetilde{{g}}_{a_{2}}(\tau ,x^{i_{1}},y^{3}) \\ 
\widetilde{{N}}_{k_{1}}^{a_{2}}(\tau ,x^{i_{1}},y^{3})%
\end{array}
\\ 
\end{array}
& 
\begin{array}{c}
\mbox{quasi-} \\ 
\mbox{stationary}%
\end{array}
\\ 
\begin{array}{c}
\begin{array}{c}
\widetilde{{g}}_{a_{2}}(\tau ,x^{i_{1}},t) \\ 
\widetilde{{N}}_{k_{1}}^{a_{2}}(\tau ,x^{i_{1}},t)%
\end{array}
\\ 
\end{array}
& 
\begin{array}{c}
\mbox{locally anisotropic} \\ 
\mbox{cosmology}%
\end{array}%
\end{array}
& 
\begin{array}{c}
\begin{array}{c}
\\ 
\ ^{\shortmid }\widetilde{{g}}^{a_{3}}(\tau ,x^{i_{2}},p_{5}) \\ 
\ ^{\shortmid }\widetilde{N}_{a_{3}k_{2}}(\tau ,x^{i_{2}},p_{5})%
\end{array}
\\ 
\\ 
\begin{array}{c}
\ ^{\shortmid }\widetilde{{g}}^{a_{3}}(\tau ,x^{i_{2}},p_{6}) \\ 
\ ^{\shortmid }\widetilde{{N}}_{a_{3}k_{2}}(\tau ,x^{i_{2}},p_{6})%
\end{array}
\\ 
\end{array}
& 
\begin{array}{c}
\ ^{\shortmid }\widetilde{{g}}^{a_{4}}(\tau ,\ ^{\shortmid }x^{i_{3}},E) \\ 
\ ^{\shortmid }\widetilde{{N}}_{a_{4}k_{3}}(\tau ,\ ^{\shortmid }x^{i_{3}},E)%
\end{array}
& 
\begin{array}{c}
\mbox{rainbow} \\ 
\mbox{s-metrics} \\ 
\mbox{variable } \\ 
p_{8}=E%
\end{array}%
\end{array}%
$}}
\end{equation*}

The AFCDM method for ansat of type (\ref{ssolutions}) on commutative phase
spaces is summarized in Tabels 13-16 from Appendix. Those formulas are
extended on nonassociative Finsler-Hamilton phase spaces by introducing
respective nonassociative sources, 
\begin{equation*}
\ _{s}^{\shortmid }\widetilde{\Im }(\tau )\approx \ _{s}^{\shortmid }\mathbf{%
\Upsilon }(\tau )\rightarrow \ _{s}^{\shortmid }\widetilde{\Im }^{\star
}(\tau ).
\end{equation*}%
For various general classes and explicit examples of nonassociative 4d-8d
phase space solutions, s-adapted frame proofs are provided in \cite%
{partner02,partner03,partner04,partner05,partner06}. In abstract geometric
form, we can consider s-adapted generalizations of all 4-d formulas proven
in sections \ref{sec3} and \ref{sec4}. Because in this Part II we work on $\
^{\shortmid }\widetilde{\mathcal{M}}^{\star },$ the nonholonomic structures
must be adapted to certain $\tau $-running nonassociative geometric data $(\
^{\shortmid }\widetilde{\mathbf{N}}\approx \ _{s}^{\shortmid }\mathbf{N,}\
^{\shortmid }\widetilde{\mathbf{g}}^{\star }\mathbf{\approx }\ \
_{s}^{\shortmid }\mathbf{g}^{\star },\ _{s}^{\shortmid }\widetilde{\mathbf{D}%
}^{\star }= \ _{s}^{\shortmid }\widehat{\mathbf{D}}^{\star }+\
_{s}^{\shortmid }\widetilde{\mathbf{Z}}^{\star })$ even for constructing
solutions we introduce canonical s-variables (with hat labels and re-defined
effective sources).

\subsubsection{Nonassociative quasi-stationary Finsler-Hamilton evolution}

One of the main purposes of Part II of this paper is to elaborate on
geometric methods on finding physically important solutions (black holes,
BH; wormholes, WH; cosmological solutions) in nonassociative gravity and
string theory. Various nonassociative and noncommutative can be encoded into
generic off-diagonal terms of metrics and modified nonlinear and linear
connection structures. Such methods of finding solutions were developed
Finsler-like MGTs and in our works on off-diagonal solutions in 4-d GR and
MGTs (massive, quadratic, nonmetric etc, see \cite%
{av06,sv08,sv15,sv00a,vp,vt,sv03a,sv03b,sv05,sv07,sv11,vvy13,bubuianu17} and
Part I of this paper.

We extend the 4-d quasi-stationary quadratic elements (\ref{qeltors}) in
abstract geometric form to quasi-stationary 8-d phase space configurations
under nonassociative $\tau $-evolution if $\ \ _{s}\widehat{\Upsilon }%
(x^{i_{s}},y^{a_{s}})$ (for $s=1,2)\ $\ are respectively shell by shell
substituted by $\ _{s}^{\shortmid }\widetilde{\Im }^{\star }(\tau
,x^{i_{1}},y^{3},p_{a_{3}},p_{a_{4}}),$ (for $s=1,2,3,4),$ in (\ref%
{cannonsymparamc2b}) as $\kappa $-parametric geometric flow off-diagonal
solutions (\ref{ssolutions}) if 
\begin{eqnarray}
d\widetilde{s}^{2}(\tau ) &=&\ ^{\shortmid }\widetilde{\mathbf{g}}_{\alpha
\beta }^{\star }(\tau )\widetilde{\mathbf{e}}^{\alpha }(\tau )\widetilde{%
\mathbf{e}}^{\beta }(\tau )=\ ^{\shortmid }\widetilde{\mathbf{g}}_{\alpha
_{s}\beta _{s}}^{\star }(\tau )\widetilde{\mathbf{e}}^{\alpha _{s}}(\tau )%
\widetilde{\mathbf{e}}^{\beta _{s}}(\tau )=\widetilde{\mathbf{g}}_{1}^{\star
}(\tau )(dx^{1})^{2}+\widetilde{\mathbf{g}}_{2}^{\star }(\tau )(dx^{2})^{2}+%
\widetilde{\mathbf{g}}_{3}^{\star }(\tau )(\widetilde{\mathbf{e}}^{3}(\tau
))^{2}  \notag \\
&&\ +\widetilde{\mathbf{g}}_{4}^{\star }(\tau )(\widetilde{\mathbf{e}}%
^{4}(\tau ))^{2}+\ ^{\shortmid }\widetilde{\mathbf{g}}_{\star }^{5}(\tau )(%
\widetilde{\mathbf{e}}_{5}(\tau ))^{2}+\ ^{\shortmid }\widetilde{\mathbf{g}}%
_{\star }^{6}(\tau )(\widetilde{\mathbf{e}}_{6}(\tau ))^{2}+\ ^{\shortmid }%
\widetilde{\mathbf{g}}_{\star }^{7}(\tau )(\widetilde{\mathbf{e}}_{5}(\tau
))^{2}+\ ^{\shortmid }\widetilde{\mathbf{g}}_{\star }^{8}(\tau )(\widetilde{%
\mathbf{e}}_{6}(\tau ))^{2}  \notag \\
&=&e^{\psi (\hbar ,\kappa ;\tau ,x^{k_{1}})}[(dx^{1})^{2}+(dx^{2})^{2}]+
\label{sol1nhrfsmfh} \\
&&\frac{[\partial _{3}(\ _{2}\widetilde{\Psi }(\tau ))]^{2}}{%
4(~_{2}^{\shortmid }\widetilde{\Im }^{\star }(\tau ))^{2}\{g_{4}^{[0]}(\tau
)-\int dy^{3}\frac{\partial _{3}[(\ _{2}\widetilde{\Psi }(\tau ))^{2}]}{4(\
~_{2}^{\shortmid }\widetilde{\Im }^{\star }(\tau ))}\}}(\widetilde{\mathbf{e}%
}^{3}(\tau ))^{2}+(g_{4}^{[0]}(\tau )-\int dy^{3}\frac{\partial _{3}[(\ _{2}%
\widetilde{\Psi }(\tau ))^{2}]}{4(~_{2}^{\shortmid }\widetilde{\Im }^{\star
}(\tau ))})(\widetilde{\mathbf{e}}^{4}(\tau ))^{2}  \notag
\end{eqnarray}%
\begin{equation*}
+\frac{[\ ^{\shortmid }\partial ^{5}(\ _{3}^{\shortmid }\widetilde{\Psi }%
(\tau ))]^{2}}{4(~_{3}^{\shortmid }\widetilde{\Im }^{\star })^{2}\{\
^{\shortmid }g_{[0]}^{6}(\tau )-\int dp_{5}\frac{\ ^{\shortmid }\partial
^{6}[(\ _{3}^{\shortmid }\widetilde{\Psi }(\tau ))^{2}]}{4(~_{3}^{\shortmid }%
\widetilde{\Im }^{\star }(\tau ))}\}}(\ ^{\shortmid }\widetilde{\mathbf{e}}%
_{5}(\tau ))^{2}+(\ ^{\shortmid }g_{[0]}^{6}(\tau )-\int dp_{5}\frac{\
^{\shortmid }\partial ^{5}[(\ _{3}^{\shortmid }\widetilde{\Psi }(\tau ))^{2}]%
}{4(\ _{3}^{\shortmid }\widetilde{\Im }^{\star })})(\ ^{\shortmid }%
\widetilde{\mathbf{e}}_{6}(\tau ))^{2}
\end{equation*}%
\begin{equation*}
+\frac{[\ ^{\shortmid }\partial ^{7}(\ _{4}^{\shortmid }\widetilde{\Psi }%
(\tau ))]^{2}}{4(~_{4}^{\shortmid }\widetilde{\Im }^{\star }(\tau ))^{2}\{\
^{\shortmid }g_{[0]}^{8}(\tau )-\int dp_{7}\frac{\ ^{\shortmid }\partial
^{7}[(\ _{4}^{\shortmid }\widetilde{\Psi }(\tau ))^{2}]}{4(~_{4}^{\shortmid
}\Im (\tau ))}\}}(\ ^{\shortmid }\widetilde{\mathbf{e}}_{7}(\tau ))^{2}+(\
^{\shortmid }g_{[0]}^{8}(\tau )-\int dp_{7}\frac{\ ^{\shortmid }\partial
^{7}[(\ _{4}^{\shortmid }\widetilde{\Psi }(\tau ))^{2}]}{4(\ _{4}^{\shortmid
}\widetilde{\Im }^{\star }(\tau ))})(\ ^{\shortmid }\widetilde{\mathbf{e}}%
_{8}(\tau ))^{2},
\end{equation*}%
where $\widetilde{\mathbf{g}}_{1}^{\star }(\tau )=\widetilde{\mathbf{g}}%
_{2}^{\star }(\tau )=e^{\widetilde{\psi }(\hbar ,\kappa ;\tau ,x^{k_{1}})},%
\widetilde{\mathbf{g}}_{3}^{\star }(\tau )=\frac{[\partial _{3}(\ _{2}%
\widetilde{\Psi }(\tau ))]^{2}}{4(~_{2}^{\shortmid }\widetilde{\Im }^{\star
}(\tau ))^{2}\{g_{4}^{[0]}(\tau )-\int dy^{3}\frac{\partial _{3}[(\ _{2}%
\widetilde{\Psi }(\tau ))^{2}]}{4(\ ~_{2}^{\shortmid }\widetilde{\Im }%
^{\star }(\tau ))}\}},...,$ as above.

The nonholonomic s-frames in (\ref{sol1nhrfsmfh}) are computed as%
\begin{eqnarray}
\widetilde{\mathbf{e}}^{3}(\tau ) &=&dy^{3}+\widetilde{w}_{k_{1}}(\hbar
,\kappa ,\tau ,x^{i_{1}},y^{3})dx^{k_{1}}=dy^{3}+\frac{\partial _{k_{1}}(\
_{2}\widetilde{\Psi }(\tau ))}{\partial _{3}(\ _{2}\Psi (\tau ))}dx^{k_{1}},
\notag \\
\widetilde{\mathbf{e}}^{4}(\tau ) &=&dt+\widetilde{n}_{k_{1}}(\hbar ,\kappa
,\tau ,x^{i_{1}},y^{3})dx^{k_{1}}  \label{sol1nrffh} \\
&=&dy^{4}+(\ _{1}\widetilde{n}_{k_{1}}(\tau )+\ _{2}\widetilde{n}%
_{k_{1}}(\tau )\int dy^{3}\frac{\partial _{3}[(\ _{2}\widetilde{\Psi }(\tau
))^{2}]}{4(\ ~_{2}^{\shortmid }\widetilde{\Im }^{\star }(\tau ))^{2}|%
\widetilde{g}_{4}^{[0]}(\tau )-\int dy^{3}\frac{\partial _{3}[(\ _{2}%
\widetilde{\Psi }(\tau ))^{2}]}{4(~_{2}^{\shortmid }\widetilde{\Im }^{\star
}(\tau ))}|^{5/2}})dx^{k_{1}},  \notag
\end{eqnarray}%
\begin{eqnarray*}
\ ^{\shortmid }\widetilde{\mathbf{e}}_{5}(\tau ) &=&dp_{5}+\ ^{\shortmid }%
\widetilde{w}_{k_{2}}(\hbar ,\kappa ,\tau ,x^{i_{2}},p_{5})dx^{k_{2}}=dp_{5}+%
\frac{\partial _{k_{2}}(\ _{3}^{\shortmid }\widetilde{\Psi }(\tau ))}{\
^{\shortmid }\partial ^{5}(\ _{3}^{\shortmid }\widetilde{\Psi }(\tau ))}%
dx^{k_{2}}, \\
\ ^{\shortmid }\widetilde{\mathbf{e}}_{6}(\tau ) &=&dp_{6}+\ ^{\shortmid }%
\widetilde{n}_{k_{2}}(\hbar ,\kappa ,\tau ,x^{i_{2}},p_{5})dx^{k_{2}} \\
&=&dp_{6}+(\ _{1}^{\shortmid }\widetilde{n}_{k_{2}}(\tau )+\ _{2}^{\shortmid
}\widetilde{n}_{k_{2}}(\tau )\int dp_{5}\frac{\ ^{\shortmid }\partial
^{5}[(\ _{3}^{\shortmid }\widetilde{\Psi }(\tau ))^{2}]}{4(\ _{3}^{\shortmid
}\widetilde{\Im }^{\star }(\tau ))^{2}|\ ^{\shortmid }\widetilde{g}%
_{[0]}^{6}(\tau )-\int dp_{5}\frac{\ ^{\shortmid }\partial ^{5}[(\
_{3}^{\shortmid }\widetilde{\Psi }(\tau ))^{2}]}{4(\ _{3}^{\shortmid }%
\widetilde{\Im }^{\star }(\tau ))}|^{5/2}})dx^{k_{2}},
\end{eqnarray*}%
\begin{eqnarray*}
\ ^{\shortmid }\widetilde{\mathbf{e}}_{7}(\tau ) &=&dp_{7}+\ ^{\shortmid }%
\widetilde{w}_{k_{3}}(\hbar ,\kappa ,\tau ,x^{i_{2}},p_{5},p_{7})d\
^{\shortmid }x^{k_{3}}=dp_{7}+\frac{\ ^{\shortmid }\partial _{k_{3}}(\
_{4}^{\shortmid }\widetilde{\Psi }(\tau ))}{\ ^{\shortmid }\partial ^{7}(\
_{4}^{\shortmid }\widetilde{\Psi }(\tau ))}d\ ^{\shortmid }x^{k_{3}}, \\
\ ^{\shortmid }\widetilde{\mathbf{e}}_{8}(\tau ) &=&dp_{8}+\ ^{\shortmid }%
\widetilde{n}_{k_{3}}(\hbar ,\kappa ,\tau ,x^{i_{2}},p_{5},p_{7})d\
^{\shortmid }x^{k_{3}} \\
&=&dp_{8}+(\ _{1}^{\shortmid }\widetilde{n}_{k_{3}}(\tau )+\ _{2}^{\shortmid
}\widetilde{n}_{k_{3}}(\tau )\int dp_{7}\frac{\ ^{\shortmid }\partial
^{7}[(\ _{4}^{\shortmid }\widetilde{\Psi }(\tau ))^{2}]}{4(\ _{4}^{\shortmid
}\widetilde{\Im }^{\star }(\tau ))^{2}|\ ^{\shortmid }\widetilde{g}%
_{[0]}^{8}(\tau )-\int dp_{7}\frac{\ ^{\shortmid }\partial ^{7}[(\
_{4}^{\shortmid }\widetilde{\Psi }(\tau ))^{2}]}{4(\ _{4}^{\shortmid }%
\widetilde{\Im }^{\star }(\tau ))}|^{5/2}})\ d\ ^{\shortmid }x^{k_{3}}.
\end{eqnarray*}
Such values are similar to those studied in section 4 of \cite{partner05}
but for different generating sources when the nonholonomic frame transforms
are constrained to relate certain tilde geometric data with 4+4 splitting to
canonical data with s-shells. In (\ref{sol1nhrfsmfh}) and (\ref{sol1nrffh}),
there are considered: 
\begin{eqnarray}
\mbox{generating functions: } &&\widetilde{\psi }(\tau )\simeq \widetilde{%
\psi }(\hbar ,\kappa ;\tau ,x^{k_{1}});\ _{2}\widetilde{\Psi }(\tau )\simeq
\ _{2}\widetilde{\Psi }(\hbar ,\kappa ;\tau ,x^{k_{1}},y^{3});
\label{integrfunctrffh} \\
&&\ _{3}^{\shortmid }\widetilde{\Psi }(\tau )\simeq \ _{3}^{\shortmid }%
\widetilde{\Psi }(\hbar ,\kappa ;\tau ,x^{k_{2}},p_{5});\ _{4}^{\shortmid }%
\widetilde{\Psi }(\tau )\simeq \ _{4}^{\shortmid }\widetilde{\Psi }(\hbar
,\kappa ;\tau ,\ ^{\shortmid }x^{k_{3}},p_{7});  \notag \\
\mbox{generating sources: \quad} &&\ _{1}^{\shortmid }\widetilde{\Im }%
^{\star }\mathcal{(\tau )}\simeq \ ~_{1}^{\shortmid }\widetilde{\Im }^{\star
}(\hbar ,\kappa ;\tau ,x^{k_{1}});\ ~_{2}^{\shortmid }\widetilde{\Im }%
^{\star }\mathcal{(\tau )}\simeq \ ~_{2}^{\shortmid }\widetilde{\Im }^{\star
}(\hbar ,\kappa ;\tau ,x^{k_{1}},y^{3});  \notag \\
&&\ _{3}^{\shortmid }\widetilde{\Im }^{\star }\mathcal{(\tau )}\simeq \
~_{3}^{\shortmid }\widetilde{\Im }^{\star }(\hbar ,\kappa ;\tau
,x^{k_{2}},p_{5});~_{4}^{\shortmid }\widetilde{\Im }^{\star }\mathcal{(\tau )%
}\simeq \ ~~_{4}^{\shortmid }\widetilde{\Im }^{\star }(\hbar ,\kappa ;\tau
,\ ^{\shortmid }x^{k_{3}},p_{7});  \notag \\
\mbox{integrating functions: } &&  \notag \\
\widetilde{g}_{4}^{[0]}(\tau ) &\simeq &\widetilde{g}_{4}^{[0]}(\hbar
,\kappa ;\tau ,x^{k_{1}}),\ _{1}\widetilde{n}_{k_{1}}\mathcal{(\tau )}\simeq
\ _{1}\widetilde{n}_{k_{1}}(\hbar ,\kappa ;\tau ,x^{j_{1}}),\ _{2}\widetilde{%
n}_{k_{1}}\mathcal{(\tau )}\simeq \ _{2}\widetilde{n}_{k_{1}}(\hbar ,\kappa
;\tau ,x^{j_{1}});  \notag \\
\ ^{\shortmid }\widetilde{g}_{[0]}^{6}\tau ) &\simeq &\ ^{\shortmid }%
\widetilde{g}_{[0]}^{6}(\hbar ,\kappa ;\tau ,x^{k_{2}}),\ _{1}\widetilde{n}%
_{k_{2}}\mathcal{(\tau )}\simeq \ _{1}\widetilde{n}_{k_{2}}(\hbar ,\kappa
;\tau ,x^{j_{2}}),\ _{2}\widetilde{n}_{k_{2}}\mathcal{(\tau )}\simeq \ _{2}%
\widetilde{n}_{k_{2}}(\hbar ,\kappa ;\tau ,x^{j_{2}});  \notag \\
\ ^{\shortmid }\widetilde{g}_{[0]}^{8}(\tau ) &\simeq &\ ^{\shortmid }%
\widetilde{g}_{[0]}^{8}(\hbar ,\kappa ;\tau ,\ ^{\shortmid }x^{j_{3}}),\ _{1}%
\widetilde{n}_{k_{3}}\mathcal{(\tau )}\simeq \ _{1}^{\shortmid }\widetilde{n}%
_{k_{3}}(\hbar ,\kappa ;\tau ,\ ^{\shortmid }x^{j_{3}}),\ _{2}\widetilde{n}%
_{k_{3}}\mathcal{(\tau )}\simeq \ _{2}^{\shortmid }\widetilde{n}%
_{k_{3}}(\hbar ,\kappa ;\tau ,\ ^{\shortmid }x^{j_{3}}).  \notag
\end{eqnarray}%
The functions $\widetilde{\psi }(\tau )$ are solutions of a respective
family of 2-d Poisson equations, 
\begin{equation*}
\partial _{11}^{2}\widetilde{\psi }(\hbar ,\kappa ;\tau ,x^{k_{1}})+\partial
_{22}^{2}\widetilde{\psi }(\hbar ,\kappa ;\tau ,x^{k_{1}})=2\ _{1}\widetilde{%
\Im }^{\star }(\hbar ,\kappa ;\tau ,x^{k_{1}}).
\end{equation*}

Nonassociative geometric parametric evolution of quasi-stationary
Finsler-Hamilton configurations defined above are characterized by four
types of additional geometric and thermodynamic flow variables:

\begin{enumerate}
\item The nonassociative geometric evolution of nonsymmetric metrics $\
_{\star }^{\shortmid }\widetilde{\mathfrak{a}}_{\alpha _{s}\beta _{s}}(\tau
)=\ _{\star }^{\shortmid }\widetilde{\mathfrak{a}}_{\alpha _{s}\beta
_{s}}(\hbar ,\kappa ;\tau ,\ ^{\shortmid }u^{\gamma _{s}})$ induced by
Finsler--Hamilton configurations is computed in explicit form by introducing
in (\ref{aux40aa}) using the coefficients of the s-metric and the
N-connection, respectively, (\ref{sol1nhrfsmfh}) and (\ref{sol1nrffh}).

\item Above classes of such solutions are with nontrivial geometric flows of
nonholonomic torsion which is not zero for tilde variables. We can define
certain classes of nonholonomic frame transforms and distortions to
canonical s-variables when the nonassociative geometric evoution is
described by families of LC-connections $\ _{s}^{\shortmid }\widetilde{%
\nabla }^{\star }(\tau )$ and$\ _{s}^{\shortmid }\widetilde{\nabla }(\tau ).$

\item We can compute necessary thermodynamic variables (\ref{nagthermodfh})
associated to nonassociative quasi-stationary solutions, or their time dual
ones defined as nonassociative locally anisotropic cosmological solutions
with additional cosmological flow. In next subsections, we shall provide
such examples for nonassociative BH and WH configurations.

\item The solutions for nonassociative Finsler-Lagrange-Ricci soliton
equations (\ref{naricsolfh}) consist self-similar configurations of (\ref%
{sol1nhrfsmfh}) and (\ref{sol1nrffh}) with. $\tau =\tau _{0}.$ We can
construct such quasi-stationary solutions directely or after a class of
generic off-diagonal solutions has be constructed for nonassociative
geometric evolution flows. Such Ricci soliton configurations can be
generated equivalently by solutions constructed using the AFCDM as it is
outlined in Appendix B to \cite{partner05}.

\item Finally, we note that $\tau $-families of nonassociative
quasi-stationary can be generated using Tables 13 and 14 (see respectively
ansatz (\ref{qcosm8d7}) and (\ref{lc8cstp7})) when the \ s- and
N-coefficients are considered with additional $\tau $-dependence and the
generating sources are correspondingly redefined for nonassociative
Finsler-Hamilton distortions, $\ _{s}^{\shortmid }\widehat{\Upsilon }%
\rightarrow \ _{s}^{\shortmid }\widetilde{\Im }^{\star }.$
\end{enumerate}

\subsubsection{Nonlinear symmetries and space and time duality encoding
nonassociative generalized Finsler data}

\label{ssnonlsym} Nonassociative quasi-stationary solutions encoding
Finsler-Hamilton configurations very important nonlinear symmetries. We
study nonholonomic geometric flow deformations of some families of \textbf{%
prime} s-metrics $\ _{s}^{\shortmid }\mathbf{\mathring{g}}(\tau ).$ They can
be arbitrary ones, i.e. not solutions of some (modified) Einstein equations,
but for understanding nonlinear off-diagonal interaction and evolution
models we can subject them to be certain trivial phase space extensions of
some physically important solutions in GR. A corresponding family of \textbf{%
target} s-metrics $\ _{s}^{\shortmid }\mathbf{g}(\tau )$ defining a
nonassociative Finsler-Hamilton flow evolution scenarios of quasi-stationary
metrics on $\ _{s}^{\star }\widetilde{\mathcal{M}}$ can be modelled by
R-flux deformations 
\begin{equation}
\ _{s}^{\shortmid }\mathbf{\mathring{g}}(\tau )\rightarrow \ _{s}^{\shortmid
}\widetilde{\mathbf{g}}(\tau )=[\ ^{\shortmid }\widetilde{g}_{\alpha
_{s}}(\tau )=\ ^{\shortmid }\widetilde{\eta }_{\alpha _{s}}(\tau )\
^{\shortmid }\mathring{g}_{\alpha _{s}}(\tau ),\ ^{\shortmid }\widetilde{N}%
_{i_{s-1}}^{a_{s}}(\tau )=\ ^{\shortmid }\widetilde{\eta }%
_{i_{s-1}}^{a_{s}}(\tau )\ ^{\shortmid }\mathring{N}_{i_{s-1}}^{a_{s}}(\tau
)],  \label{offdiagdefrfh}
\end{equation}%
with phase space gravitational $\eta $-polarizations generalized for $\tau $%
-dependencies of respective s-and N-coefficients in (\ref{sol1nhrfsmfh}) and
(\ref{sol1nrffh}). Such constructions consist 8-d phase space
generalizations of formulas related to (\ref{offdiagpm}). Correspondingly,
the generating functions and effective sources can be related to certain
effective shell $\tau $-running cosmological constants, 
\begin{eqnarray}
&&(\ _{s}\widetilde{\Psi }(\tau ),\ _{s}^{\shortmid }\widetilde{\Im }^{\star
}(\tau ))\leftrightarrow (\ _{s}^{\shortmid }\widetilde{\mathbf{g}}(\tau ),\
_{s}^{\shortmid }\widetilde{\Im }^{\star }(\tau ))\leftrightarrow (\
_{s}^{\shortmid }\widetilde{\eta }(\tau )\ ^{\shortmid }\mathring{g}_{\alpha
_{s}}(\tau )\sim (\ ^{\shortmid }\widetilde{\zeta }_{\alpha _{s}}(\tau
)(1+\kappa \ ^{\shortmid }\widetilde{\chi }_{\alpha _{s}}(\tau ))\
^{\shortmid }\mathring{g}_{\alpha _{s}}(\tau ),\ _{s}^{\shortmid }\widetilde{%
\Im }^{\star }(\tau ))\leftrightarrow  \notag \\
&&(\ _{s}\widetilde{\Phi }(\tau ),\ _{s}^{\shortmid }\widetilde{\Lambda }%
(\tau ))\leftrightarrow (\ _{s}^{\shortmid }\widetilde{\mathbf{g}},\
_{s}^{\shortmid }\widetilde{\Lambda }(\tau ))\leftrightarrow (\
_{s}^{\shortmid }\widetilde{\eta }(\tau )\ ^{\shortmid }\mathring{g}_{\alpha
_{s}}(\tau )\sim (\ ^{\shortmid }\widetilde{\zeta }_{\alpha _{s}}(\tau
)(1+\kappa \ ^{\shortmid }\widetilde{\chi }_{\alpha _{s}}(\tau ))\
^{\shortmid }\mathring{g}_{\alpha _{s}}(\tau ),\ _{s}^{\shortmid }\widetilde{%
\Lambda }(\tau )),  \label{nonlinsymr}
\end{eqnarray}%
where $\ _{s}^{\shortmid }\widetilde{\Lambda }_{0}=\ _{s}^{\shortmid }%
\widetilde{\Lambda }(\tau _{0})=const$ for deriving nonassociative Ricci
soliton symmetries.

In explicit form, the nonlinear symmetries of quasi-stationary solutions (%
\ref{sol1nhrfsmfh}) and (\ref{sol1nrffh}) are defined by formulas 
\begin{eqnarray}
\partial _{3}[(\ _{2}\widetilde{\Psi }(\tau ))^{2}] &=&-\int
dy^{3}(~_{2}^{\shortmid }\widetilde{\Im }^{\star }(\tau ))\partial _{3}%
\widetilde{g}_{4}(\tau )\simeq -\int dy^{3}(~_{2}^{\shortmid }\widetilde{\Im 
}^{\star }(\tau ))\partial _{3}(\ ^{\shortmid }\widetilde{\eta }_{4}(\tau )\ 
\mathring{g}_{4}(\tau ))  \notag \\
&\simeq &-\int dy^{3}(~_{2}^{\shortmid }\widetilde{\Im }^{\star }(\tau
))\partial _{3}[\ ^{\shortmid }\widetilde{\zeta }_{4}(\tau )(1+\kappa \
^{\shortmid }\widetilde{\chi }_{4}(\tau ))\ \mathring{g}_{4}(\tau )],
\label{nonlinsymrexfh} \\
(\ _{2}\widetilde{\Phi }(\tau ))^{2} &=&-4\ _{2}\widetilde{\Lambda }(\tau )%
\widetilde{g}_{4}(\tau )\simeq -4\ _{2}\widetilde{\Lambda }(\tau )\
^{\shortmid }\widetilde{\eta }_{4}(\tau )\ \mathring{g}_{4}(\tau )  \notag \\
&\simeq &-4\ _{2}\widetilde{\Lambda }(\tau )\ ^{\shortmid }\widetilde{\zeta }%
_{4}(\tau )(1+\kappa \ ^{\shortmid }\widetilde{\chi }_{4}(\tau ))\ \mathring{%
g}_{4}(\tau );  \notag
\end{eqnarray}%
\begin{eqnarray*}
~\ ^{\shortmid }\partial ^{5}[(\ _{3}^{\shortmid }\widetilde{\Psi }(\tau
))^{2}] &=&-\int dp_{5}(~_{3}^{\shortmid }\widetilde{\Im }^{\star }(\tau ))\
^{\shortmid }\partial ^{5}\ ^{\shortmid }\widetilde{g}^{6}(\tau )\simeq
-\int dp_{5}(~_{3}^{\shortmid }\widetilde{\Im }^{\star }(\tau ))\
^{\shortmid }\partial ^{5}(\ ^{\shortmid }\widetilde{\eta }^{6}(\tau )\
^{\shortmid }\mathring{g}^{6}(\tau )) \\
&\simeq &-\int dp_{5}(~_{3}^{\shortmid }\widetilde{\Im }^{\star }(\tau ))\
^{\shortmid }\partial ^{5}[\ ^{\shortmid }\widetilde{\zeta }^{6}(\tau
)(1+\kappa \ ^{\shortmid }\widetilde{\chi }^{6}(\tau ))\ \mathring{g}%
^{6}(\tau )], \\
(\ _{3}^{\shortmid }\widetilde{\Phi }(\tau ))^{2} &=&-4\ _{3}^{\shortmid }%
\widetilde{\Lambda }(\tau )\ ^{\shortmid }\widetilde{g}^{6}(\tau )\simeq \
-4\ _{3}^{\shortmid }\widetilde{\Lambda }(\tau )\ ^{\shortmid }\widetilde{%
\eta }^{6}(\tau )\ ^{\shortmid }\mathring{g}^{6}(\tau ) \\
&\simeq &-4\ _{3}^{\shortmid }\widetilde{\Lambda }(\tau )\ ^{\shortmid }%
\widetilde{\zeta }^{6}(\tau )(1+\kappa \ ^{\shortmid }\widetilde{\chi }%
^{6}(\tau ))\ ^{\shortmid }\mathring{g}^{6}(\tau );
\end{eqnarray*}%
\begin{eqnarray*}
~\ ^{\shortmid }\partial ^{7}[(\ _{4}^{\shortmid }\widetilde{\Psi }(\tau
))^{2}] &=&-\int dp_{7}(~_{4}^{\shortmid }\widetilde{\Im }^{\star }(\tau ))\
^{\shortmid }\partial ^{7}\ ^{\shortmid }\widetilde{g}^{8}(\tau )\simeq
-\int dp_{7}(~_{4}^{\shortmid }\widetilde{\Im }^{\star }(\tau ))\
^{\shortmid }\partial ^{7}(\ ^{\shortmid }\widetilde{\eta }^{8}(\tau )\
^{\shortmid }\mathring{g}^{8}(\tau )) \\
&\simeq &-\int dp_{7}(~_{4}^{\shortmid }\widetilde{\Im }^{\star }(\tau ))\
^{\shortmid }\partial ^{7}[\ ^{\shortmid }\widetilde{\zeta }^{8}(\tau
)(1+\kappa \ ^{\shortmid }\widetilde{\chi }^{8}(\tau ))\ \mathring{g}%
^{8}(\tau )], \\
(\ _{4}^{\shortmid }\widetilde{\Phi }(\tau ))^{2} &=&-4\ _{4}^{\shortmid }%
\widetilde{\Lambda }(\tau )\ ^{\shortmid }\widetilde{g}^{8}(\tau )\simeq \
-4\ _{4}^{\shortmid }\widetilde{\Lambda }(\tau )\ ^{\shortmid }\widetilde{%
\eta }^{8}(\tau )\ ^{\shortmid }\mathring{g}^{8}(\tau ) \\
&\simeq &-4\ _{4}^{\shortmid }\widetilde{\Lambda }(\tau )\ ^{\shortmid }%
\widetilde{\zeta }^{8}(\tau )(1+\kappa \ ^{\shortmid }\widetilde{\chi }%
^{8}(\tau ))\ ^{\shortmid }\mathring{g}^{8}(\tau ).
\end{eqnarray*}

For instance, nonlinear transforms (\ref{nonlinsymrexfh}), when 
\begin{equation*}
\ _{s}^{\shortmid }\widetilde{\mathbf{g}}[\hbar ,\kappa ,\tau ,\psi (\tau
),\ _{s}\widetilde{\Psi }(\tau ),\ _{s}^{\shortmid }\widetilde{\Im }^{\star
}(\tau )]\rightarrow \ _{s}^{\shortmid }\widetilde{\mathbf{g}}[\hbar ,\kappa
,\tau ,\psi (\tau ),\ _{s}\widetilde{\Phi }(\tau ),\ _{s}^{\shortmid }%
\widetilde{\Lambda }(\tau )],
\end{equation*}%
transform (\ref{nonassocrffh}) into an equivalent system of nonlinear PDEs
with effective $\tau $-running cosmological constants,%
\begin{equation}
\ ^{\shortmid }\widetilde{\mathbf{R}}_{\ \ \gamma _{s}}^{\beta _{s}}(\tau ,\
_{s}\widetilde{\Phi }(\tau ),\ _{s}^{\shortmid }\widetilde{\Im }^{\star
}(\tau ))={\delta }_{\ \ \gamma _{s}}^{\beta _{s}}\ _{s}^{\shortmid }%
\widetilde{\Lambda }(\tau ).  \label{nonassocrffhc}
\end{equation}%
The solutions of such equations, in various $\eta $- and $\chi $-variables,
are presented in sections 4.3, 4.4, and 4.5 of \cite{partner05}. For the
Part II of this work, those proofs and formulas can be redefined in abstract
geometric form for tilde variables encoding Finsler-Hamilton structures.

On shells $s=1,2$, the quasi-stationary solutions (\ref{nonassocrffh}) and (%
\ref{nonassocrffhc}) can be transformed into local anisotropic cosmological
solutions as we explained in section \ref{ssstdual} for formulas (\ref%
{qeltorsc}) with underlined s- and N-coefficients to emphasize dependencies
on a time like variable $y^{4}=0.$ The co-fiber variables $p_{a_{s}}$ can be
added in different forms which generates different classes of off-diagonal
cosmological solutions with $\tau $-dependence. For a toy 2+2 model, the
main ideas and solutions were provided in section (\ref{qeltorsd}). In
another turn, possible classes of nonassociative geometric flow equations
(quasi-stationary or locally anisotropic ones) can be generated in abstract
geometric form using Tables 12-16 from appendix \ref{tab1216}.

\subsection{Nonassociative Finsler-Hamilton evolution of phase space black
holes}

In this subsection, we study explicit examples how using the AFCDM we can
construct nonassociative BH solutions and study their Finsler-Hamilton flow
evolution and respective geometric thermodynamics properties. We reformulate
in modified Finsler variables the constructions from section 5.3 of \cite%
{partner05} for nonassociative flows of phase space Reisner-Nordstr\"{o}%
m-anti-de\ Sitter (in brief, RN AdS) BHs. The priority of the
Finsler-Hamilton configurations is that we can extend naturally the approach
to quantum deformations (which may include, or not, nonassociative data). We
generate and study physical properties of nonassociative generalized Finsler
flow deformations of RN BHs, when the effective cosmological constants are
determined by negative cosmological constants and respective prime metric
configurations.

\subsubsection{Prime metrics for defining commutative phase spaces and
RN-AdS BHs}

We consider a set of prime metric coefficients $\ \breve{g}_{1}=\breve{f}(%
\breve{r})^{-1},\ ^{\shortmid }\breve{g}_{2}=\ ^{\shortmid }\breve{g}_{3}=\
^{\shortmid }\breve{g}^{5}=\breve{r}^{2},\ \breve{g}_{4}=-\breve{f}(\breve{r}%
),\ ^{\shortmid }\breve{g}^{6}=\ ^{\shortmid }\breve{g}^{7}= -\ ^{\shortmid }%
\breve{g}^{8}=1$ and $\ ^{\shortmid }\breve{g}_{i_{s-1}}^{a_{s}}(\breve{r},t,%
\hat{x}^{2},\hat{x}^{3},\hat{x}^{5},p_{6,}p_{7},E)=0$ for a quadratic line
element 
\begin{equation}
d\ \breve{s}_{[5+3]}^{2}=\ ^{\shortmid }\breve{g}_{\alpha _{s}}(\
^{\shortmid }u^{\gamma _{s}})(\mathbf{\breve{e}}^{\alpha _{s}})^{2}=\frac{d%
\breve{r}^{2}}{\breve{f}(\breve{r})}-\breve{f}(\breve{r})dt^{2}+\breve{r}%
^{2}[(d^{2}\hat{x}^{2})^{2}+(d\hat{x}%
^{3})^{2}+(dp_{5})^{5}]+(dp_{6})^{2}+(dp_{7})^{2}-dE^{2}.  \label{pm5d8d}
\end{equation}%
The coordinates are defined in natural units $\hat{x}^{1}=\breve{r}=\sqrt{%
(x^{1})^{2}+(x^{2})^{2}+(x^{3})^{2}+(p_{5})^{2}}$ and $\hat{x}^{2}=\hat{x}%
^{2}(x^{2},x^{3},p_{5}),$ $\hat{x}^{3}=\hat{x}^{3}(x^{2},x^{3},p_{5})$ and $%
\hat{x}^{5}=\hat{x}^{5}(x^{2},x^{3},p_{5})$ chosen as coordinates for a
diagonal metric on an effective 3-d Einstein phase space $V_{[3]}$ of
constant scalar curvature (let say, $6\hat{k},$ for $\hat{k}=1$). The metric
function $\ \breve{f}(\breve{r})=1-\frac{\hat{m}}{\breve{r}^{2}}+\frac{%
\breve{r}^{2}}{l_{[5]}^{2}}+\frac{\hat{q}^{2}}{\breve{r}^{4}}$ in (\ref%
{pm5d8d}) is defined by integration constant $\hat{m}$ determined by the
mass of a BH, $\hat{M}=3\omega _{\lbrack 3]}\hat{m}/16\pi G_{[5]},$ for $%
\omega _{\lbrack 3]}$ denoting the volume of $V_{[3]};$ and the parameter $%
\hat{q}$ is related to the physical charge $\hat{Q}$ of the RN-AdS BH via
formula $\hat{q}=4\pi G_{[5]}\hat{Q}/\sqrt{3}\omega _{\lbrack 3]},$ see \cite%
{chamb99,zhang15}. In such a model a negative constant $\Lambda _{\lbrack
5]}=-6/l_{[5]}^{2}$ is related to the AdS radius $l_{[5]}$ which can be
naturally viewed as an effective truncation of the IIB supergravity on a 5-d
sphere, $\mathbb{S}^{5}.$ The 5-d part of the 8-d metric (\ref{pm5d8d}) can
be uplifted to 10-d. In such a case, it can be viewed as a near horizon
geometry of $\check{N}$ rotating black D3-branes in type IIB supergravity
when $l_{[10]}^{4}=2\check{N}\ell _{p}^{4}/\pi ^{2}\equiv \alpha ^{2}\check{N%
}$ , where $\ell _{p}$ is the 10-d Planck length.

To apply the AFCDM is convenient to consider certain nontrivial nonlinear
coordinates when $\mathbf{\breve{e}}^{\alpha _{s}}$ are N-elongated \ by
certain nontrivial $\ \breve{N}_{i_{1}}^{3}(\ ^{\shortmid }u^{\gamma _{s}})$
and $\ ^{\shortmid }\breve{N}_{5i_{2}}(\ ^{\shortmid }u^{\gamma _{s}}),$ see
formulas below. This allow to construct off-diagonal nonassociative
Finsler-Hamilton deformations in explicit form and without coordinate
singularities.

\subsubsection{Nonassociative Finsler-Hamilton $\protect\kappa $-linear
evolution of phase space RN-AdS BHs}

We consider nonassociative generic off-diagonal generalizations of s-metric $%
\ ^{\shortmid }\mathbf{\breve{g}}_{\alpha _{s}}$ (\ref{pm5d8d}) to certain
quasi-stationary $\ ^{\shortmid }\widetilde{\mathbf{g}}_{\alpha _{s}}$ under 
$\kappa $-linear geometric flow evolution with fixed 8-d phases space
cosmological constant $_{s}\widetilde{\Lambda }(\tau _{0})=\widetilde{%
\Lambda }_{[5]}<0,$ for $s=1,2,3,4,$ when $\widetilde{\Lambda }_{[5]}$ can
be different from $\Lambda _{\lbrack 5]}$ as a result of \ $\tau $%
-evolution. The s-coefficients $\ ^{\shortmid }\breve{g}_{\alpha _{s}}$ are
taken for a prime s-metric (instead of $\ _{s}^{\shortmid }\mathbf{\mathring{%
g}}(\tau )$ in (\ref{offdiagdefrfh})) for respective effective sources $\
_{s}^{\shortmid }\widetilde{\Im }^{\star }(\tau )$ related via nonlinear
symmetries (\ref{nonlinsymrexfh}) to $\ \widetilde{\Lambda }_{[5]}$ extended
on all phase space. Applying the AFCDM we generate a $\tau $-family of
quasi-stationary solutions of nonassociative Finsler-Hamilton flow equations
(\ref{nonassocrffh}), $\ ^{\shortmid }\breve{g}_{\alpha _{s}}\rightarrow \
_{s}^{\shortmid }\mathbf{\mathring{g}}(\tau ),$ 
\begin{eqnarray}
d\widetilde{s}^{2}(\tau ) &=&e^{\widetilde{\psi }(\hbar ,\kappa ;\tau ,%
\breve{r},\hat{x}^{2},\widetilde{\Lambda }_{[5]})}[(d\breve{r})^{2}+(d\hat{x}%
^{2})^{2}]-  \label{narnbhsfh} \\
&&\frac{1}{\widetilde{g}_{4}^{[0]}(\tau )-\frac{(\ _{2}\widetilde{\Phi }%
(\tau ))^{2}}{4\ \widetilde{\Lambda }_{[5]}}}\frac{(\ _{2}\widetilde{\Phi }%
(\tau ))^{2}[\partial _{3}(\ _{2}\widetilde{\Phi }(\tau ))]^{2}}{|\ 
\widetilde{\Lambda }_{[5]}\int dy^{3}(~_{2}^{\shortmid }\widetilde{\Im }%
^{\star }(\tau ))[\partial _{3}(\ _{2}\widetilde{\Phi }(\tau ))^{2}]|}(%
\widetilde{\mathbf{e}}^{3}(\tau ))^{2}+\left( \widetilde{g}_{4}^{[0]}(\tau )-%
\frac{(\ _{2}\widetilde{\Phi }(\tau ))^{2}}{4\ \widetilde{\Lambda }_{[5]}}%
\right) (\widetilde{\mathbf{e}}^{4}(\tau ))^{2}  \notag
\end{eqnarray}%
\begin{equation*}
-\frac{1}{\ ^{\shortmid }\widetilde{g}_{[0]}^{6}(\tau )-\frac{(\
_{3}^{\shortmid }\widetilde{\Phi }(\tau ))^{2}}{4\ \widetilde{\Lambda }_{[5]}%
}}\frac{(\ _{3}^{\shortmid }\widetilde{\Phi }(\tau ))^{2}[\ ^{\shortmid
}\partial ^{5}(\ _{3}^{\shortmid }\widetilde{\Phi }(\tau ))]^{2}}{|\ 
\widetilde{\Lambda }_{[5]}\int dp_{5}(~_{3}^{\shortmid }\widetilde{\Im }%
^{\star }(\tau ))\ \ ^{\shortmid }\partial ^{5}[(\ _{3}^{\shortmid }\Phi
(\tau ))^{2}]|}(\ ^{\shortmid }\widetilde{\mathbf{e}}_{5}(\tau ))^{2}+\left(
\ ^{\shortmid }\widetilde{g}_{[0]}^{6}(\tau )-\frac{(\ _{3}^{\shortmid }%
\widetilde{\Phi }(\tau ))^{2}}{4\ \widetilde{\Lambda }_{[5]}}\right) (\
^{\shortmid }\widetilde{\mathbf{e}}_{6}(\tau ))^{2}
\end{equation*}%
\begin{equation*}
-\frac{1}{\ ^{\shortmid }\widetilde{g}_{[0]}^{8}(\tau )-\frac{(\
_{4}^{\shortmid }\widetilde{\Phi }(\tau ))^{2}}{4\ \widetilde{\Lambda }_{[5]}%
}}\frac{(\ _{4}^{\shortmid }\widetilde{\Phi }(\tau ))^{2}[\ ^{\shortmid
}\partial ^{7}(\ _{4}^{\shortmid }\widetilde{\Phi }(\tau ))]^{2}}{|%
\widetilde{\ \Lambda }_{[5]}\int dp_{7}(~_{4}^{\shortmid }\widetilde{\Im }%
^{\star })\ \ ^{\shortmid }\partial ^{7}[(\ _{4}^{\shortmid }\widetilde{\Phi 
}(\tau ))^{2}]|}(\ ^{\shortmid }\widetilde{\mathbf{e}}_{7}(\tau
))^{2}+\left( \ ^{\shortmid }\widetilde{g}_{[0]}^{8}(\tau )-\frac{(\
_{4}^{\shortmid }\widetilde{\Phi }(\tau ))^{2}}{4\ \widetilde{\Lambda }_{[5]}%
}\right) (\ ^{\shortmid }\widetilde{\mathbf{e}}_{8}(\tau ))^{2}.
\end{equation*}%
In (\ref{narnbhsfh}), there local coordinates are defined as $\ ^{\shortmid
}u^{\gamma _{s}}=(\breve{r},t,\hat{x}^{2},\hat{x}^{3},\hat{x}%
^{5},p_{6,}p_{7},E)$ and the s-adapted frames are computed: 
\begin{eqnarray*}
\widetilde{\mathbf{e}}^{3}(\tau ) &=&d\hat{x}^{3}+\frac{\partial _{k_{1}}\
\int d\hat{x}^{3}(~_{2}^{\shortmid }\widetilde{\Im }^{\star }(\tau ))\hat{%
\partial}_{3}[(\ _{2}\widetilde{\Phi }(\tau ))^{2}]}{(~_{2}^{\shortmid }%
\widetilde{\Im }^{\star }(\tau ))\ \hat{\partial}_{3}[(\ _{2}\widetilde{\Phi 
}(\tau ))^{2}]}dx^{k_{1}}, \\
\widetilde{\mathbf{e}}^{4}(\tau ) &=&dt+(\ _{1}\widetilde{n}_{k_{1}}(\tau
)+\ _{2}\widetilde{n}_{k_{1}}(\tau )\frac{\int d\hat{x}^{3}\frac{(\ _{2}%
\widetilde{\Phi }(\tau ))^{2}[\ \hat{\partial}_{3}(\ _{2}\widetilde{\Phi }%
(\tau ))]^{2}}{|\ \Lambda _{5}(\tau )\int d\hat{x}^{3}(~_{2}^{\shortmid }%
\widetilde{\Im }^{\star }(\tau ))[\ \hat{\partial}_{3}(\ _{2}\widetilde{\Phi 
}(\tau ))^{2}]|}}{\left\vert \widetilde{g}_{4}^{[0]}(\tau )-\frac{(\ _{2}%
\widetilde{\Phi }(\tau ))^{2}}{4\ \widetilde{\Lambda }_{5}(\tau )}%
\right\vert ^{5/2}})dx^{k_{1}},
\end{eqnarray*}%
\begin{eqnarray*}
\ ^{\shortmid }\widetilde{\mathbf{e}}_{5}(\tau ) &=&d\hat{x}^{5}+\frac{%
\partial _{k_{2}}\ \int d\hat{x}^{5}(~_{3}^{\shortmid }\widetilde{\Im }%
^{\star }(\tau ))\ \hat{\partial}[(\ _{3}^{\shortmid }\widetilde{\Phi }(\tau
))^{2}]}{(~_{3}^{\shortmid }\widetilde{\Im }^{\star }(\tau ))\ \ \ \hat{%
\partial}_{5}[(\ _{3}^{\shortmid }\widetilde{\Phi }(\tau ))^{2}]}dx^{k_{2}},
\\
\ ^{\shortmid }\widetilde{\mathbf{e}}_{6}(\tau ) &=&dp_{6}+(\
_{1}^{\shortmid }\widetilde{n}_{k_{2}}(\tau )+\ _{2}^{\shortmid }\widetilde{n%
}_{k_{2}}(\tau )\frac{\int dp_{5}\frac{(\ _{3}^{\shortmid }\widetilde{\Phi }%
(\tau ))^{2}[\ ^{\shortmid }\partial ^{5}(\ _{3}^{\shortmid }\widetilde{\Phi 
}(\tau ))]^{2}}{|\ \Lambda _{\lbrack 5]}\int dp_{5}(~_{3}^{\shortmid }%
\widetilde{\Im }^{\star }(\tau ))[\ ^{\shortmid }\partial ^{5}(\
_{3}^{\shortmid }\widetilde{\Phi }(\tau ))^{2}]|}}{\left\vert \ ^{\shortmid }%
\widetilde{g}_{[0]}^{6}(\tau )-\frac{(\ _{3}^{\shortmid }\widetilde{\Phi }%
(\tau ))^{2}}{4\widetilde{\Lambda }_{[5]}}\right\vert ^{5/2}})dx^{k_{2}},
\end{eqnarray*}%
\begin{eqnarray*}
\ ^{\shortmid }\widetilde{\mathbf{e}}_{7}(\tau ) &=&dp_{7}+\frac{\partial
_{k_{3}}\ \int dp_{7}(~_{4}^{\shortmid }\widetilde{\Im }^{\star }(\tau ))\ \
^{\shortmid }\partial ^{7}[(\ _{4}^{\shortmid }\widetilde{\Phi }(\tau ))^{2}]%
}{(~_{4}^{\shortmid }\widetilde{\Im }^{\star }(\tau ))\ \ \ ^{\shortmid
}\partial ^{7}[(\ _{4}^{\shortmid }\widetilde{\Phi }(\tau ))^{2}]}d\
^{\shortmid }x^{k_{3}}, \\
\ ^{\shortmid }\widetilde{\mathbf{e}}_{8}(\tau ) &=&dE+(\ _{1}^{\shortmid }%
\widetilde{n}_{k_{3}}(\tau )+\ _{2}^{\shortmid }\widetilde{n}_{k_{3}}(\tau )%
\frac{\int dp_{7}\frac{(\ _{4}^{\shortmid }\widetilde{\Phi }(\tau ))^{2}[\
^{\shortmid }\partial ^{7}(\ _{4}^{\shortmid }\widetilde{\Phi }(\tau ))]^{2}%
}{|\ \widetilde{\Lambda }_{[5]}\int dp_{7}(~_{4}^{\shortmid }\widetilde{\Im }%
^{\star }(\tau ))[\ ^{\shortmid }\partial ^{7}(\ _{4}^{\shortmid }\widetilde{%
\Phi }(\tau ))^{2}]|}}{\left\vert \ ^{\shortmid }\widetilde{g}%
_{[0]}^{8}(\tau )-\frac{(\ _{4}^{\shortmid }\widetilde{\Phi }(\tau ))^{2}}{%
4\ \widetilde{\Lambda }_{[5]}}\right\vert ^{5/2}})d\ ^{\shortmid }x^{k_{3}}.
\end{eqnarray*}%
The integration functions from the formulas above are certain parametric
functions 
\begin{eqnarray*}
&&\widetilde{g}_{4}^{[0]}(\hbar ,\kappa ,\tau ,\breve{r},\hat{x}^{2}),\ _{1}%
\widetilde{n}_{k_{1}}(\hbar ,\kappa ,\tau ,\breve{r},\hat{x}^{2}),\ _{2}%
\widetilde{n}_{k_{1}}(\hbar ,\kappa ,\tau ,\breve{r},\hat{x}^{2}); \\
&&\ ^{\shortmid }\widetilde{g}_{[0]}^{5}(\hbar ,\kappa ,\tau ,\breve{r},\hat{%
x}^{2},\hat{x}^{3},\hat{x}^{5}),\ _{1}\widetilde{n}_{k_{2}}(\hbar ,\kappa
,\tau ,\breve{r},\hat{x}^{2},\hat{x}^{3},\hat{x}^{5}),\ _{2}\widetilde{n}%
_{k_{2}}(\hbar ,\kappa ,\tau ,\breve{r},\hat{x}^{2},\hat{x}^{3},\hat{x}^{5});
\\
&&\ ^{\shortmid }\widetilde{g}_{[0]}^{7}(\hbar ,\kappa ,\tau ,\breve{r},\hat{%
x}^{2},\hat{x}^{3},\hat{x}^{5},p_{7}),\ _{1}^{\shortmid }\widetilde{n}%
_{k_{3}}(\hbar ,\kappa ,\tau ,\breve{r},\hat{x}^{2},\hat{x}^{3},\hat{x}%
^{5},p_{7}),\ _{2}^{\shortmid }\widetilde{n}_{k_{3}}(\hbar ,\kappa ,\tau ,%
\breve{r},\hat{x}^{2},\hat{x}^{3},\hat{x}^{5},p_{7}).
\end{eqnarray*}

Using nonlinear symmetries (\ref{nonlinsymrexfh}), the quasi-stationary
solutions (\ref{narnbhsfh}) can be written in terms of of different other
type generating functions and $\eta $- /$\chi $-polarization functions, when 
\begin{eqnarray}
\ _{2}\widetilde{\Phi }(\tau ) &=&2\sqrt{|\ \widetilde{\Lambda }_{[5]}\ 
\widetilde{g}_{4}(\tau )|}=\ 2\sqrt{|\ \widetilde{\Lambda }_{[5]}\ 
\widetilde{\eta }_{4}(\tau )\breve{g}_{4})|}\simeq 2\sqrt{|\ \widetilde{%
\Lambda }_{[5]}\ \widetilde{\zeta }_{4}(\tau )\breve{g}_{4}|}[1-\frac{\kappa 
}{2}\widetilde{\chi }_{4}(\tau )],  \label{genf2} \\
\ \ _{3}^{\shortmid }\widetilde{\Phi }(\tau ) &=&2\sqrt{|\ \widetilde{%
\Lambda }_{[5]}~^{\shortmid }\widetilde{g}^{6}(\tau )|}=\ 2\sqrt{|\ 
\widetilde{\Lambda }_{[5]}~^{\shortmid }\widetilde{\eta }^{6}(\tau
)~^{\shortmid }\breve{g}^{6}|}\simeq 2\sqrt{|\ \widetilde{\Lambda }%
_{[5]}~^{\shortmid }\widetilde{\zeta }^{6}(\tau )~^{\shortmid }\breve{g}^{6}|%
}[1-\frac{\kappa }{2}~^{\shortmid }\widetilde{\chi }^{6}(\tau )],  \notag \\
\ \ _{4}^{\shortmid }\widetilde{\Phi }(\tau ) &=&2\sqrt{|\ \widetilde{%
\Lambda }_{[5]}~^{\shortmid }\widetilde{g}^{8}(\tau )|}=\ 2\sqrt{|\widetilde{%
\ \Lambda }_{[5]}~^{\shortmid }\widetilde{\eta }^{8}(\tau )~^{\shortmid }%
\breve{g}^{8}|}\simeq 2\sqrt{|\ \widetilde{\Lambda }_{[5]}~^{\shortmid }%
\widetilde{\zeta }^{8}(\tau )~^{\shortmid }\breve{g}^{8}|}[1-\frac{\kappa }{2%
}~^{\shortmid }\widetilde{\chi }^{8}(\tau )].  \notag
\end{eqnarray}%
For a subclass of solutions, the generating and integration functions are
written in $\kappa $--linearized form as in (\ref{offdiagdefrfh}), 
\begin{eqnarray*}
\widetilde{\psi }(\tau ) &\simeq &\widetilde{\psi }(\hbar ,\kappa ;\tau ,%
\breve{r},\hat{x}^{2})\simeq \widetilde{\psi }_{0}(\hbar ,\tau ,\breve{r},%
\hat{x}^{2})(1+\kappa \ _{\psi }\widetilde{\chi }(\hbar ,\tau ,\breve{r},%
\hat{x}^{2})),\mbox{ for } \\
\ \widetilde{\eta }_{2}(\tau ) &\simeq &\widetilde{\eta }_{2}(\hbar ,\kappa
;\tau ,\breve{r},\hat{x}^{2})\simeq \widetilde{\zeta }_{2}(\hbar ,\tau ,%
\breve{r},\hat{x}^{2})(1+\kappa \widetilde{\chi }_{2}(\hbar ,\tau ,\breve{r},%
\hat{x}^{2})),\mbox{ we can consider }\ \widetilde{\eta }_{2}(\tau )=%
\widetilde{\eta }_{1}(\tau ); \\
\widetilde{\eta }_{4}(\tau ) &\simeq &\ \widetilde{\eta }_{4}(\hbar ,\kappa
;\tau ,\breve{r},\hat{x}^{2},\hat{x}^{3})\simeq \widetilde{\zeta }_{4}(\hbar
,\tau ,\breve{r},\hat{x}^{2},\hat{x}^{3})(1+\kappa \ \widetilde{\chi }%
_{4}(\hbar ,\tau ,\breve{r},\hat{x}^{2},\hat{x}^{3})), \\
\ ^{\shortmid }\widetilde{\eta }^{6}(\tau ) &\simeq &\ ^{\shortmid }%
\widetilde{\eta }^{6}(\hbar ,\kappa ;\tau ,\breve{r},\hat{x}^{2},\hat{x}^{3},%
\hat{x}^{5})\simeq \ ^{\shortmid }\widetilde{\zeta }^{6}(\hbar ,\kappa ;\tau
,\breve{r},\hat{x}^{2},\hat{x}^{3},\hat{x}^{5})(1+\kappa \ ^{\shortmid }%
\widetilde{\chi }^{6}(\hbar ,\kappa ;\tau ,\breve{r},\hat{x}^{2},\hat{x}^{3},%
\hat{x}^{5})), \\
\ ^{\shortmid }\widetilde{\eta }^{8}(\tau ) &\simeq &\ ^{\shortmid }%
\widetilde{\eta }^{8}(\hbar ,\kappa ;\tau ,\breve{r},\hat{x}^{2},\hat{x}^{3},%
\hat{x}^{5},p_{7})\simeq \ ^{\shortmid }\widetilde{\zeta }^{8}(\hbar ,\kappa
;\tau ,\breve{r},\hat{x}^{2},\hat{x}^{3},\hat{x}^{5},p_{7})(1+\kappa \
^{\shortmid }\widetilde{\chi }^{8}(\hbar ,\kappa ;\tau ,\breve{r},\hat{x}%
^{2},\hat{x}^{3},\hat{x}^{5},p_{7})).
\end{eqnarray*}

Using formulas (\ref{genf2}), we can define solutions with different type
configurations. For instance, we can extract solutions with rotoid spacetime
configurations determined by nonassociative star product R-flux deformation
(considering $\chi $-polarizations), or to compute volume forms (\ref%
{volformfh}) for $\eta $-polarizations defined by tilde variables.

\subsubsection{The Bekenstein-Hawking entropy of $\protect\tau $-running
Finsler Hamilton phase space RN-AdS BEs configurations}

The nonholonomic configurations, the s-metrics (\ref{narnbhsfh}) define
higher dimension BH and/or BE configurations with conventional horizons
which can be used for formulating models of generalized Bekenstein-Hawking
thermodynamics \cite{bek1,bek2,haw1,haw2}. For simplicity, we generate a
family of solutions for 6-d $\tau $-running quasi-stationary configurations
evolving in a 8-d phase space when $\ _{1}^{\shortmid }\widetilde{n}%
_{k_{s}}=0$ and $\ _{2}^{\shortmid }\widetilde{n}_{k_{s}}=0.$ Such s-metrics
are parameterized in the form: 
\begin{eqnarray}
d\ _{\shortmid }^{\chi }\widetilde{s}_{[6\subset 8d]}^{2}(\tau ) &=&e^{%
\widetilde{\psi }_{0}}(1+\kappa \ ^{\widetilde{\psi }(\tau )}\ ^{\shortmid }%
\widetilde{\chi }(\tau ))[\ \breve{g}_{1}(\breve{r})d\breve{r}^{2}+\breve{g}%
_{2}(\breve{r})(d\hat{x}^{2})]  \label{sol4of} \\
&&-\{\frac{4[\hat{\partial}_{3}(|\widetilde{\zeta }_{4}(\tau )\breve{g}_{4}(%
\breve{r})|^{1/2})]^{2}}{\ \breve{g}_{4}(\breve{r})|\int d\hat{x}^{3}\{\
_{2}^{\shortmid }\widetilde{\Im }^{\star }(\tau )\hat{\partial}_{3}(%
\widetilde{\zeta }_{4}(\tau )\ \breve{g}_{4}(\breve{r}))\}|}-\kappa \lbrack 
\frac{\hat{\partial}_{3}(\widetilde{\chi }_{4}(\tau )|\widetilde{\zeta }%
_{4}(\tau )\ \breve{g}_{4}(\breve{r})|^{1/2})}{4\hat{\partial}_{3}(|%
\widetilde{\zeta }_{4}(\tau )\ \breve{g}_{4}(\breve{r})|^{1/2})}  \notag \\
&&-\frac{\int d\hat{x}^{3}\{\ _{2}^{\shortmid }\widetilde{\Im }^{\star
}(\tau )\hat{\partial}_{3}[(\widetilde{\zeta }_{4}(\tau )\ \breve{g}_{4}(%
\breve{r}))\widetilde{\chi }_{4}(\tau )]\}}{\int d\hat{x}^{3}\{\
_{2}^{\shortmid }\widetilde{\Im }^{\star }(\tau )\hat{\partial}_{3}(%
\widetilde{\zeta }_{4}(\tau )\ \breve{g}_{4}(\breve{r}))\}}]\}\ \breve{g}%
_{3}(\mathbf{e}^{3}(\tau ))^{2}+\ \widetilde{\zeta }_{4}(\tau )(1+\kappa \ 
\widetilde{\chi }_{4}(\tau ))\breve{g}_{4}(\breve{r})dt^{2}  \notag
\end{eqnarray}%
\begin{eqnarray*}
&&-\{\frac{4[\hat{\partial}_{5}(|\ ^{\shortmid }\widetilde{\zeta }^{6}(\tau
)\ \ \breve{g}^{6}|^{1/2})]^{2}}{~\breve{g}_{5}(\breve{r})|\int d\hat{x}%
^{5}\{\ _{3}^{\shortmid }\widetilde{\Im }^{\star }(\tau )\ ^{\shortmid
}\partial ^{7}(\ ^{\shortmid }\widetilde{\zeta }^{6}(\tau )~\breve{g}^{6})\}|%
}-\kappa \lbrack \frac{\ \hat{\partial}_{5}(\ ^{\shortmid }\widetilde{\chi }%
^{6}(\tau )|\ ^{\shortmid }\widetilde{\zeta }^{6}(\tau )\ \breve{g}%
^{6}|^{1/2})}{4\hat{\partial}_{5}(|\ ^{\shortmid }\widetilde{\zeta }%
^{6}(\tau )\ \breve{g}^{6}|^{1/2})} \\
- &&\frac{\int d\hat{x}^{5}\{\ _{3}^{\shortmid }\widetilde{\Im }^{\star
}(\tau )\ \hat{\partial}_{5}[(\ ^{\shortmid }\widetilde{\zeta }^{6}(\tau )%
\breve{g}^{6})\ ^{\shortmid }\widetilde{\chi }^{8}(\tau )]\}}{\int d\hat{x}%
^{5}\{\ _{3}^{\shortmid }\widetilde{\Im }^{\star }(\tau )\ \hat{\partial}%
_{5}[(\ ^{\shortmid }\widetilde{\zeta }^{6}(\tau )\breve{g}^{6})]\}}]\}\ 
\breve{g}_{5}(\breve{r})(\ ^{\shortmid }\widetilde{\mathbf{e}}^{5}(\tau
))^{2}+\ ^{\shortmid }\widetilde{\zeta }^{6}(\tau )\ (1+\kappa \ ^{\shortmid
}\widetilde{\chi }^{6}(\tau ))(dp_{6})^{2}+(dp_{7})^{2}-dE^{2},
\end{eqnarray*}%
where%
\begin{equation*}
\mathbf{e}^{3}(\tau )=d\hat{x}^{3}+[\frac{\hat{\partial}_{i_{1}}\int d\hat{x}%
^{3}\ \ _{2}^{\shortmid }\widetilde{\Im }^{\star }(\tau )\ \hat{\partial}_{3}%
\widetilde{\zeta }_{4}(\tau )}{\breve{N}_{i_{1}}^{3}\ \ _{2}^{\shortmid }%
\widetilde{\Im }^{\star }(\tau )\hat{\partial}_{3}\widetilde{\zeta }%
_{4}(\tau )}+\kappa (\frac{\hat{\partial}_{i_{1}}[\int d\hat{x}^{3}\ \
_{2}^{\shortmid }\widetilde{\Im }^{\star }(\tau )\hat{\partial}_{3}(%
\widetilde{\zeta }_{4}(\tau )\widetilde{\chi }_{4}(\tau ))]}{\hat{\partial}%
_{i_{1}}\ [\int d\hat{x}^{3}\ \ _{2}^{\shortmid }\widetilde{\Im }^{\star
}(\tau )\hat{\partial}_{3}\widetilde{\zeta }_{4}(\tau )]}-\frac{\hat{\partial%
}_{3}(\widetilde{\zeta }_{4}(\tau )\widetilde{\chi }_{4}(\tau ))}{\hat{%
\partial}_{3}\widetilde{\zeta }_{4}(\tau )})]\ \breve{N}%
_{i_{1}}^{3}dx^{i_{1}},
\end{equation*}%
\begin{eqnarray*}
\ \ ^{\shortmid }\mathbf{e}^{5}(\tau ) &=&d\hat{x}^{5}+[\frac{\hat{\partial}%
_{i_{2}}\ \int d\hat{x}^{5}\ _{3}^{\shortmid }\widetilde{\Im }^{\star }(\tau
)\ \ \hat{\partial}_{5}(\ ^{\shortmid }\widetilde{\zeta }^{6}(\tau ))}{\
^{\shortmid }\breve{N}_{i_{2}}^{5}\ _{3}^{\shortmid }\widetilde{\Im }^{\star
}(\tau )\ \ \hat{\partial}_{5}(\ ^{\shortmid }\widetilde{\zeta }^{6}(\tau ))}%
+ \\
&&\kappa (\frac{\hat{\partial}_{i_{2}}[\int d\hat{x}^{5}\ _{3}^{\shortmid }%
\widetilde{\Im }^{\star }(\tau )\ \ \hat{\partial}_{5}(\ ^{\shortmid }%
\widetilde{\zeta }^{6}(\tau )\ \ \breve{g}^{6})]}{\hat{\partial}_{i_{2}}\
[\int d\hat{x}^{5}\ _{3}^{\shortmid }\widetilde{\Im }^{\star }(\tau )\ \ 
\hat{\partial}_{5}(\ ^{\shortmid }\widetilde{\zeta }^{6}(\tau ))]}-\frac{\ 
\hat{\partial}_{5}(\ ^{\shortmid }\widetilde{\zeta }^{6}(\tau )\ \ \breve{g}%
^{6})}{\ \hat{\partial}_{5}(\ ^{\shortmid }\widetilde{\zeta }^{6}(\tau ))}%
)]\ \ ^{\shortmid }\breve{N}_{5i_{2}}d\ ^{\shortmid }x^{i_{2}}.
\end{eqnarray*}

The Finsler-Hamilton s-metrics (\ref{sol4of}) generate $\tau $-families of
rotoid configurations in coordinates $(\breve{r},\hat{x}^{2},\hat{x}^{3})$
(as nonholonomic deformations of the phase BH solution (\ref{pm5d8d})) if we
chose such generating functions: 
\begin{equation}
\ \ \widetilde{\chi }_{4}(\tau )=\ \widetilde{\chi }_{4}(\tau ,\breve{r},%
\hat{x}^{2},\hat{x}^{3})=2\underline{\chi }(\tau ,\breve{r},\hat{x}^{2})\sin
(\omega _{0}\hat{x}^{3}+\hat{x}_{0}^{3}),  \label{5dbe}
\end{equation}%
where $\underline{\chi }(\tau ,\breve{r},\hat{x}^{2})$ are smooth functions
(which can be approximated to some constants) and $(\omega _{0},\hat{x}%
_{0}^{3})$ is a couple of constants. In a conventional 5-d phase space on
the shells $s=1,2,3,$ trivially imbedded into a 8-d phase space, such a
solution posses a distinct ellipsoidal type horizon with respective
eccentricity $\kappa $ stated by the equations 
\begin{equation*}
\widetilde{\zeta }_{4}(\tau )(1+\kappa \ \widetilde{\chi }_{4}(\tau ))\breve{%
g}_{4}(\breve{r})=0\mbox{ i.e. }(1+\kappa \ \widetilde{\chi }_{4})\breve{f}(%
\breve{r})=(1-\frac{\hat{m}}{\breve{r}^{2}}-\frac{\widetilde{\Lambda }_{[5]}%
}{6}\breve{r}^{2}+\frac{\hat{q}^{2}}{\breve{r}^{4}}+\kappa \ \widetilde{\chi 
}_{4})=0,
\end{equation*}%
for $\widetilde{\zeta }_{4}\neq 0.$ For small parametric deformations and
configurations with $-\frac{\Lambda _{\lbrack 5]}}{6}\breve{r}^{2}+\frac{%
\hat{q}^{2}}{\breve{r}^{4}}\approx 0,$ we can approximate for a fixed $\tau
_{0},$ $\breve{r}\simeq \hat{m}^{1/2}/(1-\frac{\kappa \ }{2}\ \widetilde{%
\chi }_{4}).$ Such parametric formulas define for a rotoid horizon encoding
data for small gravitational R-flux polarizations and Finsler-Hamilton
configurations. In the limits of zero eccentricity, such e BE configurations
transform into a 5-d BH embedded into nonassociative 8-d phase space.

Extending the concept of Bekenstein-Hawking entropy for phase spaces
determined by quadratic linear elements (\ref{pm5d8d}), we can define such
thermodynamic values (computations are similar to those for formulas
(8)-(15) \ in \cite{zhang15} but with different constants and using
notations for Finsler-Hamilton spaces): 
\begin{eqnarray}
\ ^{0}\breve{S} &=&\frac{\ ^{0}\breve{A}}{4G_{[5]}}=\frac{\omega _{\lbrack
3]}\breve{r}_{h}}{4G_{[5]}}\mbox{ and }\ ^{0}\breve{T}=\frac{1}{2\pi \breve{r%
}_{h}}(\epsilon +2\frac{\breve{r}_{h}^{2}}{l_{[5]}^{2}})-\frac{2G_{[10]}^{2}%
\hat{Q}^{2}}{3\pi ^{9}l_{[5]}^{8}\breve{r}_{h}^{5}},\mbox{ for }  \notag \\
\hat{M} &=&\frac{3\omega _{\lbrack 3]}\hat{m}}{16\pi G_{[5]}}(\epsilon 
\breve{r}_{h}^{2}+\frac{\breve{r}_{h}^{4}}{l_{[5]}^{2}}+\frac{4G_{[5]}\hat{Q}%
^{2}l_{[5]}^{2}}{3\pi ^{2}\breve{r}_{h}^{2}}),  \label{bhth58}
\end{eqnarray}%
where $\breve{r}_{h}$ and $^{0}\breve{A}$ are, respectively the horizon and
area of horizon of 5-d BH, $G_{[5]}=G_{[10]}/(\pi ^{3}l_{[5]}^{5})$ and $%
G_{[10]}=\ell _{p}^{8}.$ Such formulas can be generalized for rotoid
deformations $\breve{r}_{h}\rightarrow \hat{m}^{1/2}/(1-\frac{\kappa \ }{2}%
\widetilde{\chi }_{4})$ and $^{0}\breve{A}$ $\rightarrow \ ^{rot}\breve{A},$
with a tilde $\widetilde{\chi }_{4}(\tau )$ (\ref{5dbe}), when we compute
for respective BE configurations:%
\begin{equation}
\breve{S}(\tau )=\ ^{0}\breve{S}(1+\frac{\kappa \ }{2}\widetilde{\chi }%
_{4}(\tau ))\mbox{ and }\breve{T}(\tau )=\ ^{0}\breve{T}+\kappa \left( -%
\frac{\epsilon }{4\pi \breve{r}_{h}}+\frac{\breve{r}_{h}}{2\pi l_{[5]}^{2}}-%
\frac{5G_{[10]}^{2}\hat{Q}^{2}}{3\pi ^{9}l_{[5]}^{8}\breve{r}_{h}^{5}}%
\right) \widetilde{\chi }_{4}(\tau ).  \label{rotbhbhthermv}
\end{equation}%
As in section 5.3.3 of \cite{partner05} (see formulas (97) and (98) in that
work), the modified Hawking temperatures $\breve{T}(\tau )$ and$\ ^{0}\breve{%
T}$ are stated by requiring the absence of the potential conical singularity
of the Euclidean BH at the horizon in the phase space. Such conditions can
be imposed on Finsler-Hamilton configurations possessing certain phase space
horizons.

\subsubsection{G. Perelman thermodynamics of nonassociative flows of
Finsler-Hamilton phase RN-AdS BHs}

The Bekenstein-Hawking thermodynamic paradigm does not allow to characterize
and study physical properties of general classes of quasi-stationary
solutions (\ref{narnbhsfh}) or (\ref{sol4of}) if there are not imposed
special conditions for nonassociative deformations, for instance, when there
are generated BE configurations of type (\ref{5dbe}). In our partner works 
\cite{partner04,partner05,partner06} we generalized in nonassociative form
the G. Perelman approach for the geometric flows. This allows also to define
and compute statistical thermodynamic variables (\ref{nagthermodfh}) for
Finsler-Hamilton configurations.

We explain how to compute such values for any data $\ ^{\shortmid }\mathbf{%
\breve{g}}_{\alpha _{s}}$ (\ref{pm5d8d}), $\ $and $\ _{s}^{\shortmid }%
\widetilde{\Im }^{\star }(\tau )$ which are via nonlinear symmetries (\ref%
{nonlinsymrexfh}) to $\ \widetilde{\Lambda }_{[5]},$ and 
\begin{equation*}
|\ \widetilde{\Lambda }_{[5]}\ \widetilde{\eta }_{4}(\tau )\breve{g}_{4})|=|%
\widetilde{\ \Lambda }_{[5]}\ \widetilde{\zeta }_{4}(\tau )\breve{g}%
_{4}|(1-\kappa \widetilde{\chi }_{4}(\tau )),|\ \widetilde{\Lambda }%
_{[5]}~^{\shortmid }\widetilde{\eta }^{6}(\tau )~^{\shortmid }\breve{g}%
^{6}|=|\ \widetilde{\Lambda }_{[5]}~^{\shortmid }\widetilde{\zeta }^{6}(\tau
)~^{\shortmid }\breve{g}^{6}|(1-\kappa ~^{\shortmid }\widetilde{\chi }%
^{6}(\tau ))
\end{equation*}%
defining a a subclass of s-metrics (\ref{sol4of}). We obtain such
thermodynamic functionals: 
\begin{eqnarray*}
\ _{s}^{\shortmid }\widetilde{\mathcal{W}}_{\kappa }^{\star }(\tau )
&=&\int\nolimits_{\tau ^{\prime }}^{\tau }\frac{d\tau }{32(\pi \tau )^{4}}%
\frac{2\tau \widetilde{\Lambda }_{[5]}^{2}-1}{\widetilde{\Lambda }_{[5]}^{2}}%
\ _{\eta }^{\shortmid }\mathcal{\breve{V}}(\tau ),\ \ _{s}^{\shortmid }%
\widetilde{\mathcal{Z}}_{\kappa }^{\star }(\tau )=\exp \left[
\int\nolimits_{\tau ^{\prime }}^{\tau }\frac{d\tau }{(2\pi \tau )^{4}}\frac{1%
}{\widetilde{\Lambda }_{[5]}^{2}}\ _{\eta }^{\shortmid }\mathcal{\breve{V}}%
(\tau )\right] , \\
\ _{s}^{\shortmid }\widetilde{\mathcal{E}}_{\kappa }^{\star }(\tau )
&=&-\int\nolimits_{\tau ^{\prime }}^{\tau }\frac{d\tau }{64\pi ^{4}\tau ^{3}}%
\frac{\tau \widetilde{\Lambda }_{[5]}-1}{\widetilde{\Lambda }_{[5]}^{2}}\
_{\eta }^{\shortmid }\mathcal{\breve{V}}(\tau ),\ _{s}^{\shortmid }%
\widetilde{\mathcal{S}}_{\kappa }^{\star }(\tau )=-\int\nolimits_{\tau
^{\prime }}^{\tau }\frac{d\tau }{64(\pi \tau )^{4}}\frac{\tau \widetilde{%
\Lambda }_{[5]}-2}{\widetilde{\Lambda }_{[5]}^{2}}\ _{\eta }^{\shortmid }%
\mathcal{\breve{V}}(\tau ),
\end{eqnarray*}%
for a running Finsler-Hamilton phase space volume functional 
\begin{equation*}
\ _{\eta }^{\shortmid }\mathcal{\breve{V}}(\tau )=\int_{\ _{s}^{\shortmid }%
\widehat{\Xi }}\ ^{\shortmid }\delta \ _{\eta }^{\shortmid }\widetilde{%
\mathcal{V}}(\ \ _{s}^{\shortmid }\widetilde{\Im }^{\star }(\tau
),~^{\shortmid }\breve{g}_{\alpha _{s}}),\mbox{
for }s=1,2,3.
\end{equation*}%
The thermodynamic properties of such nonassociative geometric flow systems
are studied in details in section 5.3.4 of \cite{partner05}. In this
subsection, we use the tilde variables for Finsler-Hamilton configurations.
They characterize different classes of nonassociative BH solutions and their 
$\tau $-evolution.

\section{Conclusions and perspectives}

\label{sec8} This is a status report on the anholonomic frames and
connection deformation method, AFCDM, and some important results for
constructing generic off-diagonal solutions in 4-d gravity theories and
higher dimension generalizations. The approach includes new methods and
original solutions for nonassociative star product deformed gravity and
related 8-d Finsler-Lagrange-Hamilton phase space structures when an
abstract geometric and s-adapted formalism for MGTs and nonassociative and
nonholonomic flows is summarized in Tables 1-16 from Appendix. We reviewed
and presented new classes of exact/parametric solutions constructed by using
the AFCDM for constructing exact and parametric solutions in general
relativity, GR, and modified gravity theories, MGTs \cite%
{sv00,sv00a,vp,vt,sv07,sv11,vvy13,bubuianu17,vacaru18,bubuianu18a,bubuianu20,partner02,partner03,partner04,partner05,partner06}%
.

\vskip5pt The main results solving the objectives of this work are as follow:

\vskip5pt

The first objective, \textbf{Obj1}, was to outline the geometry of
nonholonomic Lorentz manifolds with conventional (2+2)-splitting and
distortion of connections. In such an approach, the fundamental geometric
distinguished objects, d-objects, (for instance, the curvature and Ricci
d-tensors) are derived as distorted from the Levi-Civita, LC, connection to
a canonical distinguished connection, d-connection, structure and when the
geometric constructions are adapted to a prescribed nonlinear connection,
N-connection, structure. The difference between our nonholonomic dyadic
formulation and other geometric and analytic methods involving dyadic
variables (in most general forms, there are considered complex dyads, for
instance, the Newman--Penrose formalism, outlined in \cite%
{misner,hawking73,wald82}) is that we define and work with N-adapted
distortions and canonical d-connections. Such canonical nonholonomic
variables are important for proving general decoupling and integration
properties of (modified) Einstein equations (which was the \textbf{Obj2}).
Such proof is impossible for other dyadic/ tetradic formalisms involving
only the LC-connection structure if canonical d-connections are not
considered.

\vskip5pt

In section \ref{sec4}, we provided and studied explicit examples of new
classes of generic off-diagonal solutions in GR and MGTs constructed
following the AFCDM (stated by \textbf{Obj3}). We analyzed, in brief the
physical, properties of new Kerr de Sitter solutions and their deformations
to spheroidal configurations. Then, the geometric formalism was developed
for nonholonomic off-diagonal deformations of cylindrical systems and
applied for generating solutions describing locally anisotropic wormholes,
black torus and black ellipsoid systems. We proved that corresponding types
of nonlinear symmetries and time-space duality properties allow the use of
AFCDM and formulas for quasi-stationary off-diagonal configurations to
construct and analyze new classes of locally anisotropic cosmological
solitonic and spheroidal deformations, study vacuum gravitational 2-d
vertices and solitonic vacua for voids. Here we note that all examples of 16
classes of generic off-diagonal solutions constructed in explicit geometric
and analytic forms in this article are different from similar ones (derived
by more special parameterizations and AFCDM) in previous works \cite%
{sv00a,vs01a,vs01b,v01t,v01q,vp,vmon3,v13,v14}.

\vskip5pt

The nonassociative geometric flow theory on 8-d phase spaces can be
formulated in an equivalent form in nonholonomic canonical (hat) variables
and in Finsler-Hamilton variables as we defined in \ref{sec5} for \textbf{%
Obj4}. The first type of formulation allows to decouple and integrate in
general off-diagonal form such systems of nonlinear PDEs when a generalized
Finsler approach involve metric and affine structures derived from Lagrange
and Hamilton generating functions. This provides a possibility to connect in
future research the methods of generating off-diagonal solutions in MGTs to
general quantum deformations (of gravitational and matter field interactions
and geometric evolution scenarios, in general including nonassociative and
noncommutative data) determined by conventional Lagrangians or Hamiltonians.

\vskip5pt

The \textbf{Obj5} was achieved by considering canonical distortion relations
between hat and tilde connections and defining F- and W-functionals for
nonassociative Finsler-Hamilton variables. This allowed to define in
abstract geometric form the nonassociative Finsler-Hamilton geometric flow
equations which consisted a solution of \textbf{Obj6} (for parametric
solutions, it is possible also a N-adapted variational proofs using F- and
W-functionals). Then, formulating the generalized G. Perelman thermodynamics
for nonassociative Finsler-Hamilton geometric flows, provide a solutions of
the \textbf{Obj7}. Such a statistical and geometric thermodynamics is very
important because it allows to characterize very general classes of
off-diagonal solutions in MGTs and nonassociative geometric flows which
can't be considered in the framework of the Bekenstein-Hawking thermodynamic
paradigm (which can be applied only for some subclasses of solutions
involving hypersurface horizons, duality and holographic conditions).

\vskip5pt

The \textbf{Obj8}, for providing explicit examples of constructing
quasi-stationary solutions encoding 8-d Finsler-Hamilton data as
nonassociative geometric flows of higher dimension generalized solutions
from Part 1, was solved by applying abstract geometric methods and
re-defining the effective sources (to encode nonassociative data) from
Tables 12-14 from Appendix. Correspondingly, that allowed us to define and
use corresponding linear symmetries to construct an example of
temperature-like evolution of nonassociative black holes and black
ellipsoids and respective off-diagonal deformations, i.e. to solve the 
\textbf{Obj9}. We characterize such off-diagonal solutions in different
forms by computing respective Bekenstein-Hawking and G. Perelman
thermodynamic variables.

\vskip5pt

Recent progress in elaborating (non) associative/ commutative geometric and
quantum flow theories \cite{sv19,bubuianu20,ibvv20,bvv21,ibv22} with
applications in modern MGTs, accelerating cosmology and dark matter and dark
energy physics \cite{vmon3,stavr14,sv14,sv15,elizalde15,sv16,rajpoot17,sv18}
is reviewed in a series of works \cite%
{vacaru18,bubuianu18a,bubuianu17,vi17,stavrinos21}. Such constructions are
performed on nonholonomic Lorentz manifolds of 4-10 dimensions and (co)
tangent Lorentz bundles (i.e. phase spaces with respective velocity and/or
momentum-like coordinates). Even detailed proofs and solutions are provided
in the main part of this article only for 4-d nonholonomic dyadic
geometries, our methods, main formulas and exact/ parametric solutions can
be extended in straightforward forms using a corresponding abstract
geometric formalism to higher dimension spacetimes and phase spaces with
conventional (2+2)+(2+2)+2+... splitting. For pedagogical purposes (to
propose to interested researchers a complete classification and summary of
most important, general, and typical solutions), all such constructions for
higher dimensions are summarized in Tables 1-16 provided in Appendix (a
solution of \textbf{Obj10}). Those tables and related typical ansatz for
generic off-diagonal quasi-stationary/ locally anisotropic/ velocity /
momentum phase space solutions provide a general solution for the \textbf{%
Obj4} of this work. Nevertheless, we emphasize here that for constructing
explicit examples of exact and parametric solutions of nonlinear systems of
PDEs describing geometric and information flows and with applications of the
AFCDM in modern cosmology and astrophysics one should elaborate other
original papers and status reports.

\vskip5pt This review covers the Obj1-Obj4 as they were stated for SV's
Fulbright scholar program (2022-2023, USA) and the Obj5-Obj10 for his CAS
LMU "scholar at risk" fellowship in Munich, Germany (2024). A part of those
programs consisted in elaborating on a series of lectures and supplementary
material (as this status report) for a summary of results elaborated by his
research groups during the last 25 years in Eastern Europe and collaboration
with researchers from Western Countries, Romania and Turkey. To outline such
original and new methods and results is necessary to cite and discuss
certain tenths of works related to the AFCDM, further developments, and
applications. Papers by other authors are cited and discussed only if they
contain relevant former and important results.

\vskip5pt

As we proved in section \ref{sec4}, the AFCDM can be applied for
constructing and investigating nonholonomic off-diagonal deformations of
black hole, BH, solutions into other types of quasi-stationary and/or
locally anisotropic cosmological solutions in GR and MGTs. It is well known
that there are fundamental and rigorous proofs of mathematical theorems on
the uniqueness of smooth BH theorems in vacuum (for instance, for the Kerr
solution), rigidity of stationary BHs, on the stability of the Minkowski
space and various BH solutions, on the formation of trapped surfaces, cosmic
censorship theorems etc. Such results are outlined in many monographs
containing hundreds of pages with tedious mathematical proofs (see \cite%
{klainerman22,klainerman17,ionescu15} and references therein). It is not
clear if and how those rigorous mathematical theorems have connections to
MGTs, quantum gravity models, and modern accelerating cosmology. This is a
task for a new generation of mathematicians and theoretical physics and
quantum information theory researchers.

\vskip5pt

The AFCDM can be considered not only as a general geometric method for
decoupling nonlinear systems of PDEs in mathematical relativity, MGTs, and
certain geometric flow evolution equations related to gravity and geometric
thermodynamics; when such a general nonlinear decoupling allows formal
explicit integrations and generating/ finding off-diagonal solutions for
PDEs not reducing them to ODEs by some special diagonal ansatz. In
principle, the AFCDM provides a new methodology for constructing different
types of new classes of exact/ parametric solutions using symbolic and
formal geometric techniques, when the rigorous mathematical and physical
properties of generated solutions have to be stated by further assumptions.
We can model nonsingular and physically viable (at least for certain small
parameters) off-diagonal solutions for necessary smooth classes of
generating functions and generating sources, choosing specific types of
integration functions/ constants. They can define quasi-stationary
gravitational and (effective) matter field configurations, or describe
certain locally anisotropic cosmological scenarios. For certain types of
(cosmological) evolution problems, we can put and solve respective Cauchy
problems etc., analyze stability properties etc. (for instance, in \cite%
{sv03a,sv03b,sv05} we studied if black ellipsoids, BE, can be stable in GR
and (non) commutative MGTs).

\vskip5pt

In general, the new classes of constructed and studied in our works
solutions, involving generic off-diagonal metrics and modified
LC-connections, are not subjected to the conditions of theorems proven in
mathematical relativity for spacetimes with higher symmetries and metrics
with coefficients to be of some required smooth class of functions. It took
more than 50 years till physicists and mathematicians understood the
fundamental properties of the Schwarzschild and Kerr solutions in 4-d
gravity. The approach with transforms of systems of nonlinear PDEs into
nonlinear ODEs under various special assumptions on symmetries of
interactions, smooth classes of solutions, asymptotic/ boundary conditions
etc. had in the past a strong motivation being used less exact observations
in cosmology and astrophysics. The Universe was considered as an almost
spherical one being isotropic and homogeneous; and BH solutions were found
also for spherical configurations, with certain observed stability and
asymptotic conditions. The "high symmetry" paradigm, including certain
"fluctuations" with quantum anisotropies and structure formation, was
changed in modern accelerating cosmology and related dark matter and dark
energy physics. The Einstein gravity seems to be modified and new classes of
solutions may be generic off-diagonal, with coefficients of metrics
depending on all spacetime and possible phase space coordinates. Such
solutions provide more rich opportunities to elaborate on global and local
quasi-periodic structure formation scenarios, local anisotropic
configurations, and provide new types of geometric evolution of nonholonomic
gravitational systems and characterized by respective geometric
thermodynamic models studied in \cite%
{sv19,bubuianu20,ibvv20,bvv21,ibv22,vmon3,stavr14,sv14,sv15,elizalde15,sv16,rajpoot17,sv18}%
. Another important property is that such solutions can be generated in such
forms when they define realistic, viable and important physical models even
if they can be unstable, with singularities, nontrivial nonlinear
anisotropic cosmology, encoding quantum and extra dimension contributions
etc. In explicit form, this is possible by selecting respective classes of
generating and integration functions and making certain assumptions on the
type of effective sources, variation of physical constants and their
polarization etc.

\vskip5pt

Finally, we note that the AFCDM and solutions reviewed and/or constructed in
this work provide a commutative and associative background for developing a
research program on nonassociative geometric methods and exact/parametric
solutions applied in mathematical particle physics, string and M-theory,
quantum information and gravity \cite%
{partner01,partner02,partner03,partner04,partner05,partner06}.

\vskip6pt \textbf{Acknowledgments:}\ This status report article consists of
a bridge between SV's Fulbright in the USA and CAS LMU in Germany
fellowships and research programs. The reviewed geometric methods in physics
with new results and applications reflect also 40 years of research and
collaborations between scholars and young researchers from Eastern Europe,
Turkey, and Western Countries (more than 10 long-term NATO fellowships;
visiting programs at CERN, Geneva; Perimeter and Fields Institutes, Canada,
etc.). During the reviewing process of this work, a series of recent partner
papers developing the approaches on nonassociative/ noncommutative and
nonmetric modifications to Einstein-Yang-Mills-Higgs-Dirac theories, with
applications in quantum gravity and modern cosmology and astrophysics were
published \cite{bgrg24,bapny24,vapny24,vcqg25,bepjc24a,bepjc24b}.

\newpage

\appendix
\setcounter{equation}{0} \renewcommand{\theequation}
{A.\arabic{equation}} \setcounter{subsection}{0} 
\renewcommand{\thesubsection}
{A.\arabic{subsection}}

\label{appendixa}

\section{Tables 1-16 for the AFCDM and 4-10-d spacetime or 8-d phase space
solutions}

In this Appendix, we summarize the results on the AFCDM for constructing
exact/parametric solutions for various 4-d and extra dimensions, and/or with
additional velocity/momentum type coordinates in respective GR and MGTs. In
respective subsections, we provide some basic formulas and main references
on rigorous geometric proofs and existing applications published in modern
literature on mathematics and physics.

The main steps on certain general decoupling and integrating of generalized
Einstein equations with generic off-diagonal quasi-stationary and locally
anisotropic cosmological metrics in 4-d gravity theories outlined below in
Tables 1-3. The geometric proofs on finding solutions and various examples
for Lorentz manifolds $\mathbf{V},$ $dim\mathbf{V}=4$ with nonholonomic 2+2
splitting and canonical deformation of the LC-connection were sketched in
previous sections. We use geometric data $(\mathbf{g},\mathbf{N},\widehat{%
\mathbf{D}}[\mathbf{g},\mathbf{N}]= \nabla \lbrack \mathbf{g}]+\widehat{%
\mathbf{Z}}[\mathbf{g},\mathbf{N}])$ stated for a N-connection structure $%
\mathbf{N}$, when the canonical d-connection $\widehat{\mathbf{D}}$ is
defined in such a way that corresponding modified Einstein equations can be
decoupled and integrated in general off-diagonal form for a metrics $\mathbf{%
g}[u],$ with coefficients depending on all spacetime coordinates $u^{\alpha
}=(x^{i},y^{a})$. The new classes of solutions are determined by respective
generating and integration functions and generating (effective) sources $%
\widehat{\mathbf{Y}},$ all related via nonlinear symmetries to some
effective cosmological constants $\Lambda =(\ _{1}\Lambda ,\ _{2}\Lambda ).$
We can extract LC-configurations by imposing additional nonholonomic
constraints when $\widehat{\mathbf{D}}_{\mid \widehat{\mathcal{T}}%
\rightarrow 0}=\nabla $ and the related canonical both distortion, $\widehat{%
\mathbf{Z}},$ and torsion, $\widehat{\mathcal{T}},$ d-tensors vanish. The
generating source $\widehat{\mathbf{Y}}=(\ _{1}\Upsilon ,\ _{2}\Upsilon )$
encodes respectively the information on energy-momentum tensors for matter
fields and various possible contributions from other classical and/or
quantum gravity and matter field interactions. Geometric constructions and
detailed proofs are provided in \cite{sv07,sv11,vvy13,bubuianu17} and review 
\cite{vacaru18}.

In abstract geometric and N-adapted coordinate forms, the AFCDM can be
generalized for constructing solutions in various higher dimension gravity
theories. Such extensions and modifications can be performed in a formal
abstract geometric form, for instance, for $\dim \mathbf{V}%
=5,6,...,10,11,... $ (which is applicable in (super) string/gravity
theories), with a nonholonomic diadic splitting determined by a conventional
2+2+2+... splitting of dimensions for some 2-d shells $s=1,2,3,4$ etc. The
local signature of such metrics $^{s}\mathbf{g}$ is chosen to be $%
(+++-+++...+)$ on higher dimension Lorentz manifolds (this in order to
simplify the geometric formulas). For a corresponding set of geometric and
physical data $(\ ^{s}\mathbf{g},\ ^{s}\mathbf{N},\ ^{s}\widehat{\mathbf{D}}%
[\ ^{s}\mathbf{g}, \ ^{s}\mathbf{N}]=\nabla \lbrack \ ^{s}\mathbf{g}]+\ ^{s}%
\widehat{\mathbf{Z}}[\ ^{s}\mathbf{g},\ ^{s}\mathbf{N}],\ _{s}\Upsilon ,\
_{s}\Lambda ),$ the gravitational field equations can be decoupled and
integrated in very general forms for generic off-diagonal metrics depending
on all higher dimension spacetime coordinates $u^{\alpha
_{s}}=(x^{i_{1}},y^{a_{2}},y^{a_{3}},y^{a_{4}},...),$ when $%
i_{1}=1,2;a_{2}=3,4;a_{3}=5,6;a_{4}=7,8;..$. A respective nonholonomic
dyadic shell decomposition structure can be defined for all necessary
geometric and physical objects and computed in general off-diagonal form for
coordinate frames. The solutions may involve nontrivial torsion and/or
nonmetricity fields structures, and various contributions of
extra-dimensions, for instance, from string and M-theory theories, gauge
gravity models, noncommutative and nonassociative models with data encoded
into certain nontrivial effective sources $\ _{s}\Upsilon .$ For detailed
proofs and various examples and applications, we cite \cite%
{vp,vt,v01t,v01q,vs01a,vs01b,sv05,sv08,v13,v14,sv15,biv17,vi17,vmon3,partner02,partner03,partner04}%
, see a recent review of results in Appendix B and references to \cite%
{vacaru18}.

Various relativistic phase space theories can be elaborated on tangent
bundle, $T\mathbf{V,}$ and contangent bundle, $T^{\ast }\mathbf{V,}$ where $%
\mathbf{V}$ is a Lorentz manifold \cite%
{vmon3,vacaru18,bubuianu18a,bubuianu20,bubuianu19}. Here we study models
with $\dim \mathbf{V}=4,$ when $\dim T\mathbf{V}=8$ and $\dim T^{\ast }%
\mathbf{V}=8$. Theories on $T\mathbf{V}$ are relativistic generalizations of
the so-called Finsler--Lagrange geometry when the total space metrics (and
the coefficients of fundamental geometric objects, like the (non) linear
connections, curvature/torsion/ Ricci etc. tensors) depend both on spacetime
coordinates $x^{i}$ and on velocity type coordinates, $u^{a}=v^{a},$ for $%
u=(x,v)=\{u^{\alpha }=(x^{i},v^{a})\},$ where $i=1,2,3,4$ on the base
manifold $\mathbf{V}$ and $a=5,6,7,8$ for a typical fiber in the phase
space. The nonholonomic geometric constructions can be performed in shell
dyadic form with conventional (2+2)+(2+2) splitting of dimensions and local
coordinates $u^{\alpha _{s}}=(x^{i_{1}},y^{a_{2}},v^{a_{3}},v^{a_{4}}),$
when the canonical geometric data are defined by distortions 
\begin{equation*}
(\ _{s}\mathbf{g}(x,v),_{s}\mathbf{N}(x,v),_{s}\widehat{\mathbf{D}}[\ _{s}%
\mathbf{g},\ _{s}\mathbf{N}]=\nabla \lbrack \ _{s}\mathbf{g}]+\ _{s}\widehat{%
\mathbf{Z}}[\ _{s}\mathbf{g},\ _{s}\mathbf{N}],\ _{s}\Upsilon ,\ _{s}\Lambda
).
\end{equation*}

Standard phase space models ale elaborated on cotangent bundles $\mathcal{M}%
=T^{\ast }\mathbf{V}$, when the geometric/physical objects depend on
spacetime and momentum like variables $\
^{\shortmid}u=(x,p)=\{x^{i},p_{a}\}, $ where $\ ^{\shortmid
}p=p=(p_{3},p_{4}=E)$ are cofiber coordinates. Such dual coordinates can be
related to velocity type ones $v^{a}$ using, for instance, Legendre
transforms, and subjected to additional conditions to define almost
symplectic systems. On $\mathcal{M},$ we can construct various types of
phase space kinetic, geometric thermodynamic, or (nonholonomic)
gravitational models. The nonholonomic phase space constructions can be
performed in shell dyadic form with conventional (2+2)+(2+2) splitting of
dimensions and local coordinates $u^{\alpha
_{s}}=(x^{i_{1}},y^{a_{2}},p_{a_{3}},p_{a4}).$ The canonical geometric data
are defined as 
\begin{equation*}
(\ _{s}^{\shortmid }\mathbf{g}(x,p),_{s}^{\shortmid }\mathbf{N}(x,v),\
_{s}^{\shortmid }\widehat{\mathbf{D}}[\ _{s}^{\shortmid }\mathbf{g},\
_{s}^{\shortmid }\mathbf{N}]=\ ^{\shortmid }\nabla \lbrack \ _{s}^{\shortmid
}\mathbf{g}]+\ _{s}^{\shortmid }\widehat{\mathbf{Z}}[\ _{s}\mathbf{g},\
_{s}^{\shortmid }\mathbf{N}],\ _{s}^{\shortmid }\Upsilon ,\ _{s}^{\shortmid
}\Lambda ),
\end{equation*}%
which allows us to decouple and integrate in certain general forms
respective phase space modified gravitational equations. Generating sources $%
\ _{s}^{\shortmid }\Upsilon $ may encode, for instance, contributions from
nonassociative/ noncommutative terms in string theory, various
quasi-classical and quantum deformations etc. when certain momentum like
variables are introduced for respective geometric/ physical models. It is
possible to define certain nonholonomic variables when such phase space
geometries are models as Hamilton type relativistic spaces (which are dual
to respective Lagrange-Finsler geometries). Such geometric models can be
generalized in supersymmetric forms, for nonassociative and noncommutative
geometries; metric-affine theories with nonmetricity fields and torsion;
nonsymmetric metrics and generalized connections; subjected to deformation
quantization and generalized, for instance, to geometric and quantum
information flow theories \cite%
{sv00,vmon3,vacaru18,bubuianu18a,bubuianu20,bubuianu19,partner02,partner03,partner04}%
.

All above mentioned MGTs can be formulated in abstract form and in
nonholonomic canonical variables which allow to apply the AFCDM in order to
prove general decoupling and integration properties and construct exact and
parametric solutions defined by generic off-diagonal metrics and generalized
connections. If necessary, LC-configurations can be extracted by imposing
additional nonholonomic constraints. Proofs of such properties and explicit
generating of necessary types of quasi-stationary and/or locally anisotropic
cosmological solutions are formal geometric symbolic generalizations and
with higher dimension extensions of the constructions provided in the
sections of the main part of the paper, for 4-d Lorentz manifolds.

\subsection{4-d off-diagonal quasi-stationary and cosmological solutions,
tables 1-3}

The first three tables summarize the main steps on how to use 2+2
nonholonomic variables and corresponding ansatz for metrics which allow us
to construct quasi-stationary and, for respective $t$-dual symmetries,
locally anisotropic cosmological solutions.

\subsubsection{Metric ansatz and systems of nonlinear ODEs and PDEs}

Parameterizations of frames/coordinates for Lorentz manifolds with
N-connection h- and v-splitting of geometric objects and generating of
(effective) sources are provided in Table 1. There are stated two types of
generic off-diagonal metric ansatz. The first one is for generating
quasi-stationary metrics with dependence only on space coordinates and the
second one, for so-called locally anisotropic cosmological solutions, is
with dependence on the time-like coordinate and possible dependencies on two
other space like coordinates.

\vskip5pt General decoupling properties can be proven in explicit form for
generic off-diagonal ansatz with Killing symmetry on $\partial _{4},$ for
quasi-stationary configurations, or on $\partial _{3},$ for locally
anisotropic cosmological models, see respectively (\ref{dmq}) and (\ref{dmc}%
). All formulas derived for (\ref{dmc}) are certain $t$-dual to those for
quasi-stationary configurations, but with a change of local signature. To
emphasize this, we underline respective symbols of geometric objects.



{\scriptsize 
\begin{eqnarray*}
&&%
\begin{tabular}{l}
\hline\hline
\begin{tabular}{lll}
& {\ \textsf{Table 1:\ Diagonal and off-diagonal ansatz resulting in systems
of nonlinear ODEs and PDEs} } &  \\ 
& applying the Anholonomic Frame and Connection Deformation Method, \textbf{%
AFCDM}, &  \\ 
& \textit{for constructing generic off-diagonal exact and parametric
solutions} & 
\end{tabular}%
\end{tabular}
\\
&&{%
\begin{tabular}{lll}
\hline
diagonal ansatz: PDEs $\rightarrow $ \textbf{ODE}s &  & AFCDM: \textbf{PDE}s 
\textbf{with decoupling; \ generating functions} \\ 
radial coordinates $u^{\alpha }=(r,\theta ,\varphi ,t)$ & $u=(x,y):$ & 
\mbox{ nonholonomic 2+2
splitting, } $u^{\alpha }=(x^{1},x^{2},y^{3},y^{4}=t)$ \\ 
LC-connection $\mathring{\nabla}$ & [connections] & $%
\begin{array}{c}
\mathbf{N}:T\mathbf{V}=hT\mathbf{V}\oplus vT\mathbf{V,}\mbox{ locally }%
\mathbf{N}=\{N_{i}^{a}(x,y)\} \\ 
\mbox{ canonical connection distortion }\widehat{\mathbf{D}}=\nabla +%
\widehat{\mathbf{Z}};\widehat{\mathbf{D}}\mathbf{g=0,} \\ 
\widehat{\mathcal{T}}[\mathbf{g,N,}\widehat{\mathbf{D}}]%
\mbox{ canonical
d-torsion}%
\end{array}%
$ \\ 
$%
\begin{array}{c}
\mbox{ diagonal ansatz  }g_{\alpha \beta }(u) \\ 
=\left( 
\begin{array}{cccc}
\mathring{g}_{1} &  &  &  \\ 
& \mathring{g}_{2} &  &  \\ 
&  & \mathring{g}_{3} &  \\ 
&  &  & \mathring{g}_{4}%
\end{array}%
\right)%
\end{array}%
$ & $\mathbf{g}\Leftrightarrow $ & $%
\begin{array}{c}
g_{\alpha \beta }=%
\begin{array}{c}
g_{\alpha \beta }(x^{i},y^{a})\mbox{ general frames / coordinates} \\ 
\left[ 
\begin{array}{cc}
g_{ij}+N_{i}^{a}N_{j}^{b}h_{ab} & N_{i}^{b}h_{cb} \\ 
N_{j}^{a}h_{ab} & h_{ac}%
\end{array}%
\right] ,\mbox{ 2 x 2 blocks }%
\end{array}
\\ 
\mathbf{g}_{\alpha \beta }=[g_{ij},h_{ab}],\mathbf{g}=\mathbf{g}%
_{i}(x^{k})dx^{i}\otimes dx^{i}+\mathbf{g}_{a}(x^{k},y^{b})\mathbf{e}%
^{a}\otimes \mathbf{e}^{b}%
\end{array}%
$ \\ 
$\mathring{g}_{\alpha \beta }=\left\{ 
\begin{array}{cc}
\mathring{g}_{\alpha }(r) & \mbox{ for BHs} \\ 
\mathring{g}_{\alpha }(t) & \mbox{ for FLRW }%
\end{array}%
\right. $ & [coord.frames] & $g_{\alpha \beta }=\left\{ 
\begin{array}{cc}
g_{\alpha \beta }(x^{i},y^{3}) & \mbox{ quasi-stationary configurations} \\ 
\underline{g}_{\alpha \beta }(x^{i},y^{4}=t) & 
\mbox{locally anisotropic
cosmology}%
\end{array}%
\right. $ \\ 
&  &  \\ 
$%
\begin{array}{c}
\mbox{coord. transf. }e_{\alpha }=e_{\ \alpha }^{\alpha ^{\prime }}\partial
_{\alpha ^{\prime }}, \\ 
e^{\beta }=e_{\beta ^{\prime }}^{\ \beta }du^{\beta ^{\prime }},\mathring{g}%
_{\alpha \beta }=\mathring{g}_{\alpha ^{\prime }\beta ^{\prime }}e_{\ \alpha
}^{\alpha ^{\prime }}e_{\ \beta }^{\beta ^{\prime }} \\ 
\begin{array}{c}
\mathbf{\mathring{g}}_{\alpha }(x^{k},y^{a})\rightarrow \mathring{g}_{\alpha
}(r),\mbox{ or }\mathring{g}_{\alpha }(t), \\ 
\mathring{N}_{i}^{a}(x^{k},y^{a})\rightarrow 0.%
\end{array}%
\end{array}%
$ & [N-adapt. fr.] & $\left\{ 
\begin{array}{cc}
\begin{array}{c}
\mathbf{g}_{i}(x^{k}),\mathbf{g}_{a}(x^{k},y^{3}), \\ 
\mbox{ or }\mathbf{g}_{i}(x^{k}),\underline{\mathbf{g}}_{a}(x^{k},t),%
\end{array}
& \mbox{ d-metrics } \\ 
\begin{array}{c}
N_{i}^{3}=w_{i}(x^{k},y^{3}),N_{i}^{4}=n_{i}(x^{k},y^{3}), \\ 
\mbox{ or }\underline{N}_{i}^{3}=\underline{n}_{i}(x^{k},t),\underline{N}%
_{i}^{4}=\underline{w}_{i}(x^{k},t),%
\end{array}
& 
\end{array}%
\right. $ \\ 
$\mathring{\nabla},$ $Ric=\{\mathring{R}_{\ \beta \gamma }\}$ & Ricci tensors
& $\widehat{\mathbf{D}},\ \widehat{\mathcal{R}}ic=\{\widehat{\mathbf{R}}_{\
\beta \gamma }\}$ \\ 
$~^{m}\mathcal{L[\mathbf{\phi }]\rightarrow }\ ^{m}\mathbf{T}_{\alpha \beta }%
\mathcal{[\mathbf{\phi }]}$ & 
\begin{tabular}{l}
generating \\ 
sources%
\end{tabular}
& $%
\begin{array}{cc}
\widehat{\mathbf{\Upsilon }}_{\ \nu }^{\mu }=\mathbf{e}_{\ \mu ^{\prime
}}^{\mu }\mathbf{e}_{\nu }^{\ \nu ^{\prime }}\mathbf{\Upsilon }_{\ \nu
^{\prime }}^{\mu ^{\prime }}[\ ^{m}\mathcal{L}(\mathbf{\varphi ),}T_{\mu \nu
},\Lambda ] &  \\ 
=diag[\ _{1}\Upsilon (x^{i})\delta _{j}^{i},\ _{2}\Upsilon
(x^{i},y^{3})\delta _{b}^{a}], & \mbox{ quasi-stationary configurations} \\ 
=diag[\ _{1}\Upsilon (x^{i})\delta _{j}^{i},\ _{2}\underline{\Upsilon }%
(x^{i},t)\delta _{b}^{a}], & \mbox{ locally anisotropic cosmology}%
\end{array}%
$ \\ 
trivial equations for $\mathring{\nabla}$-torsion & LC-conditions & $%
\widehat{\mathbf{D}}_{\mid \widehat{\mathcal{T}}\rightarrow 0}=\mathbf{%
\nabla }\mbox{ extracting new classes of solutions in GR}$ \\ \hline\hline
\end{tabular}%
}
\end{eqnarray*}%
}


\subsubsection{Decoupling and integration of (modified) Einstein eqs \&
quasi-stationary configurations}

The key steps for applying the AFCDM for generating stationary off-diagonal
exact solutions of (modified) Einstein equations are outlined in Table 2.
Such solutions are, in general, with nontrivial nonholonomically induced
torsion (\ref{dtors}). They can be re-defined equivalently in terms of
generating functions $\Psi (x^{k},y^{3})$ or $\Phi (x^{k},y^{3})$ using
nonlinear symmetries (\ref{ntransf1}) and (\ref{ntransf2}), see also (\ref%
{nonlinsymrex}). Considering $\eta $--polarization functions, respective
d-metrics and N-connections can be parameterized to describe nonholonomic
deformations of a primary (for instance, BH) d-metric $\mathbf{\mathring{g}}$
into target generic off diagonal stationary solutions $\widehat{\mathbf{g}}$
(\ref{dmq}) \ (sew also (\ref{qeltors})) as $\mathbf{\mathring{g}}%
\rightarrow \widehat{\mathbf{g}}(x^{k},y^{3})=[g_{\alpha }(x^{k},y^{3})=
\eta _{\alpha }(x^{k},y^{3})\mathring{g}_{\alpha },\ \eta
_{i}^{a}(x^{k},y^{3})\mathring{N}_{i}^{a}]$. Zero torsion LC-configurations
in GR can be extracted for additional nonholonomic constraints which are
satisfied for a more special class of "integrable" generating functions $(%
\check{h}_{4}(x^{k},y^{3}),$ or $\check{\Psi}(x^{k},y^{3})$ and/or $\check{%
\Phi}(x^{k},y^{3}))$ for respective sources $\ _{2}\check{\Upsilon}%
(x^{k},y^{3})$ and $\ _{2}\Lambda $ (\ref{zerot1}).

\vskip5pt The main assumption on (effective) generating sources is that in
N-adapted form they can be parameterized in the form $\widehat{\mathbf{%
\Upsilon }}_{\ \beta }^{\alpha }=[\ ^{h}\Upsilon \delta _{\ \ j}^{i},\
^{v}\Upsilon \delta _{\ \ b}^{a}]$ (\ref{esourc}), when certain relations to
an energy-momentum tensor for matter (\ref{emdt}) can be established in
algebraic form (choosing respectively the coefficients of frame transforms).
This imposes some nonholonomic constraints on the h- and v-dynamics of
matter fields, determined by respective distributions of matter fields, and
related effective cosmological constants. For such assumptions, we can prove
general decoupling and integration properties of the nonholonomic canonical
deformed Einstein equations (\ref{cdeq1}). For quasi-stationary
configurations, we need additional assumption on two generating sources, for
instance, that $\ ^{h}\Upsilon = \ _{1}\widehat{\Upsilon }(x^{i})$ and $\
^{v}\Upsilon = \ _{2}\widehat{\Upsilon }(x^{i},y^{3}).$ If such conditions
are not satisfied, we can not integrate in explicit form the gravitational
field equations for a respectively chosen ansatz. To construct generic
off-diagonal solutions in explicit form, we prescribe such generating
sources and then show how to decouple and find exact/ parametric solutions
for an ansatz (\ref{dmq}).

\newpage 
{\scriptsize 
\begin{eqnarray*}
&&%
\begin{tabular}{l}
\hline\hline
\begin{tabular}{lll}
& {\large \textsf{Table 2:\ Off-diagonal quasi-stationary configurations}} & 
\\ 
& Exact solutions of $\widehat{\mathbf{R}}_{\mu \nu }=\mathbf{\Upsilon }%
_{\mu \nu }$ (\ref{cdeq1}) transformed into a system of nonlinear PDEs (\ref%
{eq1})-(\ref{e2c}) & 
\end{tabular}
\\ 
\end{tabular}
\\
&&%
\begin{tabular}{lll}
\hline\hline
&  &  \\ 
$%
\begin{array}{c}
\\ 
\end{array}%
\begin{array}{c}
\mbox{d-metric ansatz with} \\ 
\mbox{Killing symmetry }\partial _{4}=\partial _{t} \\ 
\mbox{general or spherical coordinates}%
\end{array}%
$ &  & $%
\begin{array}{c}
ds^{2}=g_{i}(x^{k})(dx^{i})^{2}+g_{a}(x^{k},y^{3})(dy^{a}+N_{i}^{a}(x^{k},y^{3})dx^{i})^{2},%
\mbox{ for } \\ 
g_{i}=e^{\psi {(x}^{k}{)}%
},g_{a}=h_{a}(x^{k},y^{3}),N_{i}^{3}=w_{i}(x^{k},y^{3}),N_{i}^{4}=n_{i}(x^{k},y^{3});
\\ 
g_{i}=e^{\psi {(r,\theta )}},\,\,\,\,g_{a}=h_{a}({r,\theta },\varphi ),\
N_{i}^{3}=w_{i}({r,\theta },\varphi ),\,\,\,\,N_{i}^{4}=n_{i}({r,\theta }%
,\varphi ),%
\end{array}%
$ \\ 
&  &  \\ 
Effective matter sources &  & $\mathbf{\Upsilon }_{\ \nu }^{\mu }=[\ _{1}%
\widehat{\Upsilon }({r,\theta })\delta _{j}^{i},\ _{2}\widehat{\Upsilon }({%
r,\theta },\varphi )\delta _{b}^{a}],\mbox{ if }x^{1}=r,x^{2}=\theta
,y^{3}=\varphi ,y^{4}=t$ \\ \hline
Nonlinear PDEs (\ref{eq1})-(\ref{e2c}) &  & $%
\begin{array}{c}
\psi ^{\bullet \bullet }+\psi ^{\prime \prime }=2\ \ _{1}\widehat{\Upsilon };
\\ 
\varpi ^{\ast }\ h_{4}^{\ast }=2h_{3}h_{4}\ _{2}\widehat{\Upsilon }; \\ 
\beta w_{i}-\alpha _{i}=0; \\ 
n_{k}^{\ast \ast }+\gamma n_{k}^{\ast }=0;%
\end{array}%
$ for $%
\begin{array}{c}
\varpi {=\ln |\partial _{3}h_{4}/\sqrt{|h_{3}h_{4}|}|,} \\ 
\alpha _{i}=(\partial _{3}h_{4})\ (\partial _{i}\varpi ),\beta =(\partial
_{3}h_{4})\ (\partial _{3}\varpi ),\  \\ 
\ \gamma =\partial _{3}\left( \ln |h_{4}|^{3/2}/|h_{3}|\right) , \\ 
\partial _{1}q=q^{\bullet },\partial _{2}q=q^{\prime },\partial
_{3}q=\partial q/\partial \varphi =q^{\ast }%
\end{array}%
$ \\ \hline
$%
\begin{array}{c}
\mbox{ Generating functions:}\ h_{3}(x^{k},y^{3}), \\ 
\Psi (x^{k},y^{3})=e^{\varpi },\Phi ((x^{k},y^{3})); \\ 
\mbox{integration functions:}\ h_{4}^{[0]}(x^{k}),\  \\ 
_{1}n_{k}(x^{i}),\ _{2}n_{k}(x^{i}); \\ 
\mbox{\& nonlinear symmetries}%
\end{array}%
$ &  & $%
\begin{array}{c}
\ (\Psi ^{2})^{\ast }=-\int dy^{3}\ _{2}\widehat{\Upsilon }h_{4}^{\ \ast },
\\ 
\Phi ^{2}=-4\ _{2}\Lambda h_{4},\mbox{ see }(\ref{nonlinsymrex}); \\ 
h_{4}=h_{4}^{[0]}-\Phi ^{2}/4\ _{2}\Lambda ,h_{4}^{\ast }\neq 0,\
_{2}\Lambda \neq 0=const%
\end{array}%
$ \\ \hline
Off-diag. solutions, $%
\begin{array}{c}
\mbox{d--metric} \\ 
\mbox{N-connec.}%
\end{array}%
$ &  & $%
\begin{array}{c}
\ g_{i}=e^{\ \psi (x^{k})}\mbox{ as a solution of 2-d Poisson eqs. }\psi
^{\bullet \bullet }+\psi ^{\prime \prime }=2~\ _{1}\widehat{\Upsilon }; \\ 
h_{3}=-(\Psi ^{\ast })^{2}/4\ _{2}\widehat{\Upsilon }^{2}h_{4},\mbox{ see }(%
\ref{g3}),(\ref{g4}); \\ 
h_{4}=h_{4}^{[0]}-\int dy^{3}(\Psi ^{2})^{\ast }/4\ _{2}\widehat{\Upsilon }%
=h_{4}^{[0]}-\Phi ^{2}/4\ _{2}\Lambda ; \\ 
\\ 
w_{i}=\partial _{i}\ \Psi /\ \partial _{3}\Psi =\partial _{i}\ \Psi ^{2}/\
\partial _{3}\Psi ^{2}|; \\ 
n_{k}=\ _{1}n_{k}+\ _{2}n_{k}\int dy^{3}(\Psi ^{\ast })^{2}/\ _{2}\widehat{%
\Upsilon }^{2}|h_{4}^{[0]}-\int dy^{3}(\Psi ^{2})^{\ast }/4\ _{2}\widehat{%
\Upsilon }^{2}|^{5/2}. \\ 
\\ 
\end{array}%
$ \\ \hline
LC-configurations (\ref{zerot1}) &  & $%
\begin{array}{c}
\partial _{\varphi }w_{i}=(\partial _{i}-w_{i}\partial _{3})\ln \sqrt{|h_{3}|%
},(\partial _{i}-w_{i}\partial _{3})\ln \sqrt{|h_{4}|}=0, \\ 
\partial _{k}w_{i}=\partial _{i}w_{k},\partial _{3}n_{i}=0,\partial
_{i}n_{k}=\partial _{k}n_{i}; \\ 
\mbox{ see d-metric }(\ref{qellc})\mbox{ for } \\ 
\Psi =\check{\Psi}(x^{i},y^{3}),(\partial _{i}\check{\Psi})^{\ast }=\partial
_{i}(\check{\Psi}^{\ast })\mbox{ and } \\ 
\Upsilon (x^{i},\varphi )=\Upsilon \lbrack \check{\Psi}]=\check{\Upsilon},%
\mbox{ or }\Upsilon =const. \\ 
\end{array}%
$ \\ \hline
N-connections, zero torsion &  & $%
\begin{array}{c}
w_{i}=\partial _{i}\check{A}=\left\{ 
\begin{array}{c}
\partial _{i}(\int dy^{3}\ \check{\Upsilon}\ \check{h}_{4}{}^{\ast }])/%
\check{\Upsilon}\ \check{h}_{4}{}^{\ast }; \\ 
\partial _{i}\check{\Psi}/\check{\Psi}^{\ast }; \\ 
\partial _{i}(\int dy^{3}\ \check{\Upsilon}(\check{\Phi}^{2})^{\ast })/(%
\check{\Phi})^{\ast }\check{\Upsilon};%
\end{array}%
\right. \\ 
\mbox{ and }n_{k}=\check{n}_{k}=\partial _{k}n(x^{i}).%
\end{array}%
$ \\ \hline
$%
\begin{array}{c}
\mbox{polarization functions} \\ 
\mathbf{\mathring{g}}\rightarrow \widehat{\mathbf{g}}\mathbf{=}[g_{\alpha
}=\eta _{\alpha }\mathring{g}_{\alpha },\ \eta _{i}^{a}\mathring{N}_{i}^{a}]%
\end{array}%
$ &  & $%
\begin{array}{c}
\\ 
ds^{2}=\eta _{1}(r,\theta )\mathring{g}_{1}(r,\theta )[dx^{1}(r,\theta
)]^{2}+\eta _{2}(r,\theta )\mathring{g}_{2}(r,\theta )[dx^{2}(r,\theta
)]^{2}+ \\ 
\eta _{3}(r,\theta ,\varphi )\mathring{g}_{3}(r,\theta )[d\varphi +\eta
_{i}^{3}(r,\theta ,\varphi )\mathring{N}_{i}^{3}(r,\theta )dx^{i}(r,\theta
)]^{2}+ \\ 
\eta _{4}(r,\theta ,\varphi )\mathring{g}_{4}(r,\theta )[dt+\eta
_{i}^{4}(r,\theta ,\varphi )\mathring{N}_{i}^{4}(r,\theta )dx^{i}(r,\theta
)]^{2}, \\ 
\end{array}%
$ \\ \hline
Prime metric defines a BH &  & $%
\begin{array}{c}
\\ 
\lbrack \mathring{g}_{i}(r,\theta ),\mathring{g}_{a}=\mathring{h}%
_{a}(r,\theta );\mathring{N}_{k}^{3}=\mathring{w}_{k}(r,\theta ),\mathring{N}%
_{k}^{4}=\mathring{n}_{k}(r,\theta )] \\ 
\mbox{diagonalizable by frame/ coordinate transforms.} \\ 
\end{array}%
$ \\ 
Example of a prime metric &  & $%
\begin{array}{c}
\\ 
\mathring{g}_{1}=(1-r_{g}/r)^{-1},\mathring{g}_{2}=r^{2},\mathring{h}%
_{3}=r^{2}\sin ^{2}\theta ,\mathring{h}_{4}=(1-r_{g}/r),r_{g}=const \\ 
\mbox{the Schwarzschild solution, or any BH solution.} \\ 
\mbox{ for new KdS solutions }(\ref{offdiagpm1})\mbox{ with }\mathbf{%
\mathring{g}\simeq \breve{g}}(x^{i},y^{3})\mathbf{=}(\breve{g}_{\alpha };%
\breve{N}_{i}^{a}); \\ 
\end{array}%
$ \\ \hline
Solutions for polarization funct. &  & $%
\begin{array}{c}
\eta _{i}=e^{\ \psi (x^{k})}/\mathring{g}_{i};\eta _{3}\mathring{h}_{3}=-%
\frac{4[(|\eta _{4}\mathring{h}_{4}|^{1/2})^{\ast }]^{2}}{|\int dy^{3}\ _{2}%
\widehat{\Upsilon }[(\eta _{4}\mathring{h}_{4})]^{\ast }|\ }; \\ 
\eta _{4}=\eta _{4}(x^{k},y^{3})\mbox{ as a generating
function}; \\ 
\ \eta _{i}^{3}\ \mathring{N}_{i}^{3}=\frac{\partial _{i}\ \int dy^{3}%
\widehat{\Upsilon }(\eta _{4}\ \mathring{h}_{4})^{\ast }}{\widehat{\Upsilon }%
\ (\eta _{4}\ \mathring{h}_{4})^{\ast }}; \\ 
\eta _{k}^{4}\ \mathring{N}_{k}^{4}=\ _{1}n_{k}+16\ \ _{2}n_{k}\int dy^{3}%
\frac{\left( [(\eta _{4}\mathring{h}_{4})^{-1/4}]^{\ast }\right) ^{2}}{|\int
dy^{3}\widehat{\Upsilon }[(\eta _{4}\ \mathring{h}_{4})]^{\ast }|\ }%
\end{array}%
\mbox{ see }(\ref{nkernew})$ or $(\ref{offdnceleps1});$ \\ \hline
Polariz. funct. with zero torsion &  & $%
\begin{array}{c}
\eta _{i}=e^{\ \psi (x^{k})}/\mathring{g}_{i};\eta _{4}=\check{\eta}%
_{4}(x^{k},y^{3})\mbox{ as a generating function}; \\ 
\eta _{3}=-\frac{4[(|\eta _{4}\mathring{h}_{4}|^{1/2})^{\ast }]^{2}}{%
\mathring{g}_{3}|\int dy^{3}\widehat{\Upsilon }[(\check{\eta}_{4}\mathring{h}%
_{4})]^{\ast }|\ };\eta _{i}^{3}=\partial _{i}\check{A}/\mathring{w}%
_{k},\eta _{k}^{4}=\frac{\ \partial _{k}n}{\mathring{n}_{k}}. \\ 
\end{array}%
$ \\ \hline\hline
\end{tabular}%
\end{eqnarray*}%
%
} 

If a $\widehat{\mathbf{\Upsilon }}_{\ \ \beta }^{\alpha }$ involves certain
small parameters encoding nonholonomic deformations and distortions and
contributions, for instance, from some effective classical and/or quantum
interactions, extra dimension etc., we can formulate a a physical
interpretation which is similar to "not-deformed" models. In general, for
different types of parameterizations of generating sources, it is not clear
if a solutions may have importance for physical theories. Nevertheless,
using the AFCDM we are able to investigate off-diagonal nonlinear
gravitational and (effective) matter field interactions and construct
respective classes of solutions in explicit form. This is more general than
in the case when the (modified) Einstein equations are transformed in
systems of nonlinear ODEs.

\vskip5pt Different parameterizations of quasi-stationary metrics involving
respective generating functions, effective sources and cosmological
constants, and defining nonholonomic deformations of certain prime d-metrics
into target d-metrics are stated by respective formulas (\ref{qeltors}), (%
\ref{offdiagcosmcsh}), (\ref{offdsolgenfgcosmc}), (\ref{offdiagpolfr}) and (%
\ref{offdncelepsilon}). Examples of physically important quasi-stationary
solutions are given by explicit models of off-diagonal deformations of new
KdS metrics; for cylindrical systems and their deformations; locally
anisotropic black holes; BH and BT deformed systems etc., see section \ref%
{sec4}.

\subsubsection{Decoupling and integration of gravitational PDEs generating
cosmological metrics}

In Table 3, we summarize the main steps for generating off-diagonal locally
anisotropic solutions of (modified) Einstein equations using the AFCDM.

{\scriptsize 
\begin{eqnarray*}
&&%
\begin{tabular}{l}
\hline\hline
\begin{tabular}{lll}
& {\large \textsf{Table 3:\ Off-diagonal locally anisotropic cosmological
models}} &  \\ 
& Exact solutions of $\widehat{\mathbf{R}}_{\mu \nu }=\underline{\mathbf{%
\Upsilon }}_{\mu \nu }$ (\ref{cdeq1}) transformed into a system of nonlinear
PDEs (\ref{dualcosm}) & 
\end{tabular}
\\ 
\end{tabular}
\\
&&%
\begin{tabular}{lll}
\hline\hline
$%
\begin{array}{c}
\mbox{d-metric ansatz with} \\ 
\mbox{Killing symmetry }\partial _{3}=\partial _{\varphi }%
\end{array}%
$ &  & $%
\begin{array}{c}
d\underline{s}^{2}=g_{i}(x^{k})(dx^{i})^{2}+\underline{g}%
_{a}(x^{k},y^{4})(dy^{a}+\underline{N}_{i}^{a}(x^{k},y^{4})dx^{i})^{2},%
\mbox{ for } \\ 
g_{i}=e^{\psi {(x}^{k}{)}},\,\,\,\,\underline{g}_{a}=\underline{h}_{a}({x}%
^{k},t),\ \underline{N}_{i}^{3}=\underline{n}_{i}({x}^{k},t),\,\,\,%
\underline{\,N}_{i}^{4}=\underline{w}_{i}({x}^{k},t),%
\end{array}%
$ \\ 
&  &  \\ 
Effective matter sources &  & $\underline{\mathbf{\Upsilon }}_{\ \nu }^{\mu
}=[~\ _{h}\underline{\Upsilon }({x}^{k})\delta _{j}^{i},~\ _{v}\underline{%
\Upsilon }({x}^{k},t)\delta _{b}^{a}];x^{1},x^{2},y^{3},y^{4}=t$ \\ \hline
Nonlinear PDEs &  & $%
\begin{array}{c}
\psi ^{\bullet \bullet }+\psi ^{\prime \prime }=2\ _{1}\underline{\Upsilon };
\\ 
\underline{\varpi }^{\diamond }\ \underline{h}_{3}^{\diamond }=2\underline{h}%
_{3}\underline{h}_{4}\ _{2}\underline{\Upsilon }; \\ 
\underline{n}_{k}^{\diamond \diamond }+\underline{\gamma }\underline{n}%
_{k}^{\diamond }=0; \\ 
\underline{\beta }\underline{w}_{i}-\underline{\alpha }_{i}=0;%
\end{array}%
$ for $%
\begin{array}{c}
\underline{\varpi }{=\ln |\partial _{t}\underline{{h}}_{3}/\sqrt{|\underline{%
h}_{3}\underline{h}_{4}|}|,} \\ 
\underline{\alpha }_{i}=(\partial _{t}\underline{h}_{3})\ (\partial _{i}%
\underline{\varpi }),\ \underline{\beta }=(\partial _{t}\underline{h}_{3})\
(\partial _{t}\underline{\varpi }), \\ 
\ \underline{\gamma }=\partial _{t}\left( \ln |\underline{h}_{3}|^{3/2}/|%
\underline{h}_{4}|\right) , \\ 
\partial _{1}q=q^{\bullet },\partial _{2}q=q^{\prime },\partial
_{4}q=\partial q/\partial t=q^{\diamond }%
\end{array}%
$ \\ \hline
$%
\begin{array}{c}
\mbox{ Generating functions:}\ \underline{h}_{4}({x}^{k},t), \\ 
\underline{\Psi }(x^{k},t)=e^{\underline{\varpi }},\underline{\Phi }({x}%
^{k},t); \\ 
\mbox{integr. functions:}\ h_{4}^{[0]}(x^{k}),\ _{1}n_{k}(x^{i}),\  \\ 
_{2}n_{k}(x^{i});\mbox{\& nonlinear symmetries}%
\end{array}%
$ &  & $%
\begin{array}{c}
\ (\underline{\Psi }^{2})^{\diamond }=-\int dt\ _{2}\underline{\Upsilon }%
\underline{h}_{3}^{\diamond }, \\ 
\underline{\Phi }^{2}=-4\ _{2}\underline{\Lambda }\underline{h}_{3}; \\ 
\underline{h}_{3}=\underline{h}_{3}^{[0]}-\underline{\Phi }^{2}/4\ _{2}%
\underline{\Lambda },\underline{h}_{3}^{\diamond }\neq 0,\ _{2}\underline{%
\Lambda }\neq 0=const%
\end{array}%
$ \\ \hline
Off-diag. solutions, $%
\begin{array}{c}
\mbox{d--metric} \\ 
\mbox{N-connec.}%
\end{array}%
$ &  & $%
\begin{array}{c}
\ g_{i}=e^{\ \psi (x^{k})}\mbox{ as a solution of 2-d Poisson eqs. }\psi
^{\bullet \bullet }+\psi ^{\prime \prime }=2\ _{1}\underline{\Upsilon }; \\ 
\overline{h}_{4}=-(\overline{\Psi }^{2})^{\diamond }/4\ _{2}\underline{%
\Upsilon }^{2}\underline{h}_{3}; \\ 
\underline{h}_{3}=h_{3}^{[0]}-\int dt(\underline{\Psi }^{2})^{\diamond }/4\
_{2}\underline{\Upsilon }=h_{3}^{[0]}-\underline{\Phi }^{2}/4\ _{2}%
\underline{\Lambda }; \\ 
\underline{n}_{k}=\ _{1}n_{k}+\ _{2}n_{k}\int dt(\underline{\Psi }^{\diamond
})^{2}/\ _{2}\underline{\Upsilon }^{2}\ |h_{3}^{[0]}-\int dt(\underline{\Psi 
}^{2})^{\diamond }/4\ _{2}\underline{\Upsilon }|^{5/2}; \\ 
\underline{w}_{i}=\partial _{i}\ \underline{\Psi }/\ \partial _{t}\underline{%
\Psi }=\partial _{i}\underline{\Psi }^{2}/\ \partial _{t}\underline{\Psi }%
^{2}. \\ 
\end{array}%
$ \\ \hline
LC-configurations &  & $%
\begin{array}{c}
\partial _{t}\underline{w}_{i}=(\partial _{i}-\underline{w}_{i}\partial
_{t})\ln \sqrt{|\underline{h}_{4}|},(\partial _{i}-\underline{w}_{i}\partial
_{4})\ln \sqrt{|\underline{h}_{3}|}=0, \\ 
\partial _{k}\underline{w}_{i}=\partial _{i}\underline{w}_{k},\partial _{t}%
\underline{n}_{i}=0,\partial _{i}\underline{n}_{k}=\partial _{k}\underline{n}%
_{i}; \\ 
\underline{\Psi }=\underline{\check{\Psi}}(x^{i},t),(\partial _{i}\underline{%
\check{\Psi}})^{\diamond }=\partial _{i}(\underline{\check{\Psi}}^{\diamond
})\mbox{ and } \\ 
\ _{2}\underline{\Upsilon }(x^{i},t)=\underline{\Upsilon }[\underline{\check{%
\Psi}}]=\underline{\check{\Upsilon}},\mbox{ or }\underline{\Upsilon }=const.
\\ 
\end{array}%
$ \\ \hline
N-connections, zero torsion &  & $%
\begin{array}{c}
\underline{n}_{k}=\underline{\check{n}}_{k}=\partial _{k}\underline{n}(x^{i})
\\ 
\mbox{ and }\underline{w}_{i}=\partial _{i}\underline{\check{A}}=\left\{ 
\begin{array}{c}
\partial _{i}(\int dt\ \underline{\check{\Upsilon}}\ \underline{\check{h}}%
_{3}^{\diamond }])/\underline{\check{\Upsilon}}\ \underline{\check{h}}%
_{3}^{\diamond }{}; \\ 
\partial _{i}\underline{\check{\Psi}}/\underline{\check{\Psi}}^{\diamond };
\\ 
\partial _{i}(\int dt\ \underline{\check{\Upsilon}}(\underline{\check{\Phi}}%
^{2})^{\diamond })/\underline{\check{\Phi}}^{\diamond }\underline{\check{%
\Upsilon}};%
\end{array}%
\right. .%
\end{array}%
$ \\ \hline
$%
\begin{array}{c}
\mbox{polarization functions} \\ 
\mathbf{\mathring{g}}\rightarrow \underline{\widehat{\mathbf{g}}}\mathbf{=}[%
\underline{g}_{\alpha }=\underline{\eta }_{\alpha }\underline{\mathring{g}}%
_{\alpha },\underline{\eta }_{i}^{a}\underline{\mathring{N}}_{i}^{a}]%
\end{array}%
$ &  & $%
\begin{array}{c}
ds^{2}=\underline{\eta }_{i}(x^{k},t)\underline{\mathring{g}}%
_{i}(x^{k},t)[dx^{i}]^{2}+\underline{\eta }_{3}(x^{k},t)\underline{\mathring{%
h}}_{3}(x^{k},t)[dy^{3}+\underline{\eta }_{i}^{3}(x^{k},t)\underline{%
\mathring{N}}_{i}^{3}(x^{k},t)dx^{i}]^{2} \\ 
+\underline{\eta }_{4}(x^{k},t)\underline{\mathring{h}}_{4}(x^{k},t)[dt+%
\underline{\eta }_{i}^{4}(x^{k},t)\underline{\mathring{N}}%
_{i}^{4}(x^{k},t)dx^{i}]^{2}, \\ 
\end{array}%
$ \\ \hline
$%
\begin{array}{c}
\mbox{ Prime metric defines } \\ 
\mbox{ a cosmological solution}%
\end{array}%
$ &  & $%
\begin{array}{c}
\lbrack \underline{\mathring{g}}_{i}(x^{k},t),\underline{\mathring{g}}_{a}=%
\underline{\mathring{h}}_{a}(x^{k},t);\underline{\mathring{N}}_{k}^{3}=%
\underline{\mathring{w}}_{k}(x^{k},t),\underline{\mathring{N}}_{k}^{4}=%
\underline{\mathring{n}}_{k}(x^{k},t)] \\ 
\mbox{diagonalizable by frame/ coordinate transforms.} \\ 
\end{array}%
$ \\ 
$%
\begin{array}{c}
\mbox{Example of a prime } \\ 
\mbox{ cosmological metric }%
\end{array}%
$ &  & $%
\begin{array}{c}
\mathring{g}_{1}=a^{2}(t)/(1-kr^{2}),\mathring{g}_{2}=a^{2}(t)r^{2}, \\ 
\underline{\mathring{h}}_{3}=a^{2}(t)r^{2}\sin ^{2}\theta ,\underline{%
\mathring{h}}_{4}=c^{2}=const,k=\pm 1,0; \\ 
\mbox{ any frame transform of a FLRW or a Bianchi metrics} \\ 
\end{array}%
$ \\ \hline
Solutions for polarization funct. &  & $%
\begin{array}{c}
\eta _{i}=e^{\ \psi (x^{k})}/\mathring{g}_{i};\underline{\eta }_{4}%
\underline{\mathring{h}}_{4}=-\frac{4[(|\underline{\eta }_{3}\underline{%
\mathring{h}}_{3}|^{1/2})^{\diamond }]^{2}}{|\int dt\ _{2}\underline{%
\Upsilon }[(\underline{\eta }_{3}\underline{\mathring{h}}_{3})]^{\diamond
}|\ };\mbox{ gener. funct. }\underline{\eta }_{3}=\underline{\eta }%
_{3}(x^{i},t); \\ 
\underline{\eta }_{k}^{3}\ \underline{\mathring{N}}_{k}^{3}=\ _{1}n_{k}+16\
\ _{2}n_{k}\int dt\frac{\left( [(\underline{\eta }_{3}\underline{\mathring{h}%
}_{3})^{-1/4}]^{\diamond }\right) ^{2}}{|\int dt\ _{2}\underline{\Upsilon }[(%
\underline{\eta }_{3}\underline{\mathring{h}}_{3})]^{\diamond }|\ };\ 
\underline{\eta }_{i}^{4}\ \underline{\mathring{N}}_{i}^{4}=\frac{\partial
_{i}\ \int dt\ _{2}\underline{\Upsilon }(\underline{\eta }_{3}\underline{%
\mathring{h}}_{3})^{\diamond }}{\ _{2}\underline{\Upsilon }(\underline{\eta }%
_{3}\underline{\mathring{h}}_{3})^{\diamond }},%
\end{array}%
$ \\ \hline
Polariz. funct. with zero torsion &  & $%
\begin{array}{c}
\eta _{i}=e^{\ \psi }/\mathring{g}_{i};\underline{\eta }_{4}=-\frac{4[(|%
\underline{\eta }_{3}\underline{\mathring{h}}_{3}|^{1/2})^{\diamond }]^{2}}{%
\underline{\mathring{g}}_{4}|\int dt\ _{2}\underline{\Upsilon }[(\underline{%
\eta }_{3}\underline{\mathring{h}}_{3})]^{\diamond }|\ };%
\mbox{ gener.
funct. }\underline{\eta }_{3}=\underline{\check{\eta}}_{3}({x}^{i},t); \\ 
\underline{\eta }_{k}^{4}=\partial _{k}\underline{\check{A}}/\mathring{w}%
_{k};\underline{\eta }_{k}^{3}=(\partial _{k}\underline{n})/\mathring{n}_{k}.
\\ 
\end{array}%
$ \\ \hline\hline
\end{tabular}%
\end{eqnarray*}%
}

{\scriptsize 
} 

Applying the nonholonomic deformation procedure (for simplicity, we consider
metrics determined by a generating function $\underline{h}_{4}({x}^{k},t),$
we construct a class of generic off--diagonal cosmological solutions with
Killing symmetry on $\partial _{3}$ determined by effective sources, $\ _{1}%
\underline{\Upsilon }$ and $\ _{2}\underline{\Upsilon },$ and a nontrivial
cosmological constant, $\underline{\Lambda },$ 
\begin{eqnarray}
ds^{2} &=&e^{\ \psi (x^{k},_{1}\underline{\Upsilon }%
)}[(dx^{1})^{2}+(dx^{2})^{2}]+\underline{h}_{3}[dy^{3}+(\ _{1}n_{k}+4\
_{2}n_{k}\int dt\frac{(\underline{h}_{3}{}^{\diamond })^{2}}{|\int dt\ _{2}%
\underline{\Upsilon }\underline{h}_{3}{}^{\diamond }|(\underline{h}%
_{3})^{5/2}})dx^{k}]  \notag \\
&&-\frac{(\underline{h}_{3}{}^{\diamond })^{2}}{|\int dt\ _{2}\underline{%
\Upsilon }\underline{h}_{3}{}^{\diamond }|\ \overline{h}_{3}}[dt+\frac{%
\partial _{i}(\int dt\ _{2}\underline{\Upsilon }\ \underline{h}%
_{3}{}^{\diamond }])}{\ \ _{2}\underline{\Upsilon }\ \underline{h}%
_{3}{}^{\diamond }}dx^{i}],  \label{cosm1d}
\end{eqnarray}%
Such a d-metric is equivalent to (\ref{qeltorsc}) like (\ref%
{offdsolgenfgcosmc}) is equivalent to (\ref{qeltors}). The d-metric (\ref%
{cosm1d}) can be written in terms of gravitational $\eta $-polarization
and/or $\chi $-polarization functions.

\subsection{Off-diagonal higher dimension quasi-stationary and cosmological
solutions, tables 4-6}

The procedure of generating off-diagonal solutions and basic formulas from
Tables 1-3 can be extended in abstract geometric form for higher dimension
Lorentz manifolds. Such examples and applications in modern cosmology and
astrophysics are provided in \cite{sv15a,biv17,vi17}. In explicit form, we
can consider 10-d spacetime models with 9 space coordinates, wich can be
derived in the framework of string gravity theories. The AFCDM is extend and
applied in a geometric abstract form using nonholonomic shell dyadic
decompositions of dimensions. In this subsection, we consider five shells
labelled as $s=1,2,3,4,5,$ where the first two ones are stated for 4-d
Lorentz manifolds as we considered in the main part of the paper and in the
previous subsection of the appendix. There are used nonholonomic $2+2+2+2+2$
splitting of dimensions when the N-adapted coordinates and indices of
geometric objects are labeled to describe increasing on order "anion"
shells. We shall not provide in further subsections of the appendix the
equations and formulas defining higher dimension LC-configurations using
10-d variants of zero torsion conditions (\ref{lcconstr}). The constructions
are incremental when the higher dimension generating and integrations
functions are nonholonomically constrained for higher dimensions. Such
solutions can be found by a respective extra dimension generalization of the
system of equations (\ref{zerot1}) and corresponding d-metrics (\ref{qellc}).

\subsubsection{Diagonal and off-diagonal ansatz for higher dimensions}

In this subsection, we parameterize the higher dimension coordinates in any
form for a space like 9-d hypersurface when the time like coordinate is
stated as $u^{4}=y^{4}=t.$ The signature of the metrics is of type $%
(+++-++...+).$ In principle, the geometric constructions and the procedure
of generating solutions do not depend on signature. Adding, or substituting
certain dyadic shells, the main formulas and solutions can be re-defined for
higher/lower dimensions, for instance, for 12-d, 8-d, or 6-d nonholonomic
Lorentz manifolds with the same parameterizations for the first two shells
(used for describing the 4-d spacetimes).

\vskip5pt

In \cite{biv17,vi17}, we elaborated on string gravity models with almost
symplectic structures in higher dimensions and provided examples of explicit
classes of generic off-diagonal solutions. Such configurations can be with
non-compactified extra dimensions even decompositions on a string constant
parameter are necessary for deriving modified Einstein equations. The
physical interpretation of solutions in such higher dimension gravity
theories is different from those defined on (co) tangent bundles if the
metrics are with different signatures and the extra dimension coordinates
are not considered of velocity/ momentum type.

\newpage

{\scriptsize 
\begin{eqnarray*}
&&%
\begin{tabular}{lll}
& {\ \textsf{Table 4:\ Diagonal and off-diagonal ansatz for 10-d Lorentz
manifolds} } &  \\ 
& and the Anholonomic Frame and Connection Deformation Method, \textbf{AFCDM}%
, &  \\ 
& \textit{for constructing generic off-diagonal exact and parametric
solutions} & 
\end{tabular}
\\
&&{%
\begin{tabular}{lll}
\hline
diagonal ansatz: PDEs $\rightarrow $ \textbf{ODE}s &  & AFCDM: \textbf{PDE}s 
\textbf{with decoupling; } \\ 
\begin{tabular}{l}
coordinates \\ 
$u^{\alpha _{s}}=(u^{1},..,u^{4}=t,..u^{10})$%
\end{tabular}
& $\ ^{s}u=(\ ^{s-1}x,\ ^{s}y)$ & $%
\begin{tabular}{l}
nonholonomic 2+2+2+2+2 splitting; shels $s=1,2,3,4,5$ \\ 
$u^{\alpha _{s}}=(x^{1},x^{2},y^{3},y^{4}=t,y^{5},y^{6},...,y^{9},y^{10});$
\\ 
$u^{\alpha
_{s}}=(x^{i_{1}},y^{a_{2}},y^{a_{3}},y^{a_{4}},y^{a_{5}});i_{1}=1,2;a_{2}=3,4;...,a_{5}=9,10; 
$ \\ 
$u^{\alpha _{s}}=(x^{i_{s-1}},y^{a_{s}});\ ^{s}u=(\ ^{s-1}x,\ ^{s}y)=(x,y),$ 
$s=1,2,3,4,5;$%
\end{tabular}%
$ \\ 
LC-connection $\mathring{\nabla}$ & 
\begin{tabular}{l}
N-connection; \\ 
canonical \\ 
d-connection%
\end{tabular}
& $%
\begin{array}{c}
\ ^{s}\mathbf{N}:T\ ^{s}\mathbf{V}=hT\mathbf{V}\oplus \ ^{2}vT\mathbf{V}%
\oplus \ ^{3}v\mathbf{T\mathbf{V}\oplus }\ ^{4}v\mathbf{T\mathbf{V}\oplus }\
^{5}v\mathbf{\mathbf{T\mathbf{V}},} \\ 
\mbox{ locally }\ ^{s}\mathbf{N}%
=\{N_{i_{s-1}}^{a_{s}}(x,y)=N_{i_{s-1}}^{a_{s}}(\ ^{s-1}x,\
^{s}y)=N_{i_{s-1}}^{a_{s}}(\ ^{s}u)\} \\ 
\ ^{s}\widehat{\mathbf{D}}=(\ ^{1}h\widehat{\mathbf{D}},\ ^{2}v\widehat{%
\mathbf{D}},\ ^{3}v\widehat{\mathbf{D}},\ ^{4}v\widehat{\mathbf{D}},\ ^{5}v%
\widehat{\mathbf{D}})=\{\Gamma _{\ \beta _{s}\gamma _{s}}^{\alpha _{s}}\};
\\ 
\mbox{ canonical connection distortion }\ ^{s}\widehat{\mathbf{D}}=\nabla +\
^{s}\widehat{\mathbf{Z}};\ ^{s}\widehat{\mathbf{D}}\ ^{s}\mathbf{g=0,} \\ 
\ ^{s}\widehat{\mathcal{T}}[\ ^{s}\mathbf{g,}\ ^{s}\mathbf{N,}\ ^{s}\widehat{%
\mathbf{D}}]\mbox{ canonical
d-torsion}%
\end{array}%
$ \\ 
$%
\begin{array}{c}
\mbox{ diagonal ansatz  } \\ 
\ ^{2}\mathring{g}=\mathring{g}_{\alpha _{2}\beta _{2}}(\ ^{s}u)= \\ 
\left( 
\begin{array}{cccc}
\mathring{g}_{1} &  &  &  \\ 
& \mathring{g}_{2} &  &  \\ 
&  & \mathring{g}_{3} &  \\ 
&  &  & \mathring{g}_{4}%
\end{array}%
\right) ; \\ 
\ ^{s}g=\mathring{g}_{\alpha _{s}\beta _{s}}(\ ^{s}u)= \\ 
\left( 
\begin{array}{cccc}
\ ^{2}\mathring{g} &  &  &  \\ 
& \mathring{g}_{5} &  &  \\ 
&  & \ddots &  \\ 
&  &  & \mathring{g}_{10}%
\end{array}%
\right)%
\end{array}%
$ & $\mathbf{g}\Leftrightarrow $ & $%
\begin{tabular}{l}
$g_{\alpha _{2}\beta _{2}}=%
\begin{array}{c}
g_{\alpha _{2}\beta _{2}}(x^{i_{1}},y^{a_{2}})%
\mbox{ general frames /
coordinates} \\ 
\left[ 
\begin{array}{cc}
g_{i_{1}j_{1}}+N_{i_{1}}^{a_{2}}N_{j_{1}}^{b_{2}}h_{a_{2}b_{2}} & 
N_{i_{1}}^{b_{2}}h_{c_{2}b_{2}} \\ 
N_{j_{1}}^{a_{2}}h_{a_{2}b_{2}} & h_{a_{2}c_{2}}%
\end{array}%
\right] ,\mbox{ 2 x 2 blocks }%
\end{array}%
$ \\ 
$\ ^{2}\mathbf{g=\{g}_{\alpha _{2}\beta
_{2}}=[g_{i_{1}j_{1}},h_{a_{2}b_{2}}]\},$ \\ 
$\ ^{2}\mathbf{g}=\mathbf{g}_{i_{1}}(x^{k_{1}})dx^{i_{1}}\otimes dx^{i_{1}}+%
\mathbf{g}_{a_{2}}(x^{k_{2}},y^{b_{2}})\mathbf{e}^{a_{2}}\otimes \mathbf{e}%
^{b_{2}}$ \\ 
$\vdots $ \\ 
$g_{\alpha _{s}\beta _{s}}=%
\begin{array}{c}
g_{\alpha _{s}\beta _{s}}(x^{i_{s-1}},y^{a_{s}})%
\mbox{ general frames /
coordinates} \\ 
\left[ 
\begin{array}{cc}
g_{i_{s}j_{s}}+N_{i_{s-1}}^{a_{s}}N_{j_{s-1}}^{b_{s}}h_{a_{s}b_{s}} & 
N_{i_{s-1}}^{b_{s}}h_{c_{s}b_{s}} \\ 
N_{j_{s-1}}^{a_{s}}h_{a_{s}b_{s}} & h_{a_{s}c_{s}}%
\end{array}%
\right] ,%
\end{array}%
$ \\ 
$\ ^{s}\mathbf{g=\{g}_{\alpha _{s}\beta
_{s}}=[g_{i_{s-1}j_{s-1}},h_{a_{s}b_{s}}]=[g_{i_{1}j_{1}},h_{a_{2}b_{2}},...,h_{a_{5}b_{5}}]\}, 
$ \\ 
$\ ^{s}\mathbf{g}=\mathbf{g}_{i_{s-1}}(x^{k_{s-1}})dx^{i_{s-1}}\otimes
dx^{i_{s-1}}+\mathbf{g}_{a_{s}}(x^{k_{s-1}},y^{b_{s}})\mathbf{e}%
^{a_{s}}\otimes \mathbf{e}^{b_{s}}$ \\ 
$=\mathbf{g}_{i_{1}}(x^{k_{1}})dx^{i_{1}}\otimes dx^{i_{1}}+\mathbf{g}%
_{a_{2}}(x^{k_{1}},y^{b_{2}})\mathbf{e}^{a_{2}}\otimes \mathbf{e}%
^{b_{2}}+... $ \\ 
$+\mathbf{g}_{a_{5}}(x^{k_{1}},y^{b_{2}},y^{b_{3}},y^{b_{4}},y^{b_{5}})%
\mathbf{e}^{a_{5}}\otimes \mathbf{e}^{b_{5}};$%
\end{tabular}%
$ \\ 
$\mathring{g}_{\alpha _{2}\beta _{2}}=\left\{ 
\begin{array}{cc}
\mathring{g}_{\alpha _{2}}(\ ^{2}r) & \mbox{ for BHs} \\ 
\mathring{g}_{\alpha _{2}}(t) & \mbox{ for FLRW }%
\end{array}%
\right. $ & [coord.frames] & $g_{\alpha _{2}\beta _{2}}=\left\{ 
\begin{array}{cc}
g_{\alpha _{2}\beta _{2}}(x^{i},y^{3}) & \mbox{ quasi-stationary config. }
\\ 
\underline{g}_{\alpha _{2}\beta _{2}}(x^{i},y^{4}=t) & 
\mbox{locally anisotropic
cosmology}%
\end{array}%
\right. $ \\ 
$\mathring{g}_{\alpha _{s}\beta _{s}}=\left\{ 
\begin{array}{cc}
\mathring{g}_{\alpha _{s}}(\ ^{s}r) & \mbox{ for BHs} \\ 
\mathring{g}_{\alpha _{s}}(t) & \mbox{ for FLRW }%
\end{array}%
\right. $ &  & $g_{\alpha _{5}\beta _{5}}=\left\{ 
\begin{array}{cc}
g_{\alpha _{5}\beta _{5}}(x^{i_{4}},y^{9}) &  \\ 
\underline{g}_{\alpha _{s}\beta _{s}}(x^{i_{4}},y^{10}) & 
\end{array}%
\right. $ \\ 
$%
\begin{array}{c}
\mbox{coord. transf. }e_{\alpha _{s}}=e_{\ \alpha _{s}}^{\alpha _{s}^{\prime
}}\partial _{\alpha _{s}^{\prime }}, \\ 
e^{\beta _{s}}=e_{\beta _{s}^{\prime }}^{\ \beta _{s}}du^{\beta _{s}^{\prime
}}, \\ 
\mathring{g}_{\alpha _{s}\beta _{s}}=\mathring{g}_{\alpha _{s}^{\prime
}\beta _{s}^{\prime }}e_{\ \alpha _{s}}^{\alpha _{s}^{\prime }}e_{\ \beta
_{s}}^{\beta _{s}^{\prime }} \\ 
\begin{array}{c}
\mathbf{\mathring{g}}_{\alpha _{s}}(x^{k_{s-1}},y^{a_{s}})\rightarrow 
\mathring{g}_{\alpha _{s}}(\ ^{s}r),\mbox{ or } \\ 
\mathring{g}_{\alpha _{s}}(t),\mathring{N}%
_{i_{s-1}}^{a_{s}}(x^{k_{s-1}},y^{a_{s}})\rightarrow 0.%
\end{array}%
\end{array}%
$ & [N-adapt. fr.] & 
\begin{tabular}{l}
$\left\{ 
\begin{array}{cc}
\begin{array}{c}
\mathbf{g}_{i_{1}}(x^{k_{1}}),\mathbf{g}_{a_{2}}(x^{k_{1}},y^{3}), \\ 
\mbox{ or }\mathbf{g}_{i_{1}}(x^{k_{1}}),\underline{\mathbf{g}}%
_{a_{2}}(x^{k_{1}},t),%
\end{array}
& \mbox{ d-metrics } \\ 
\begin{array}{c}
N_{i_{1}}^{3}=w_{i_{1}}(x^{k},y^{3}),N_{i_{1}}^{4}=n_{i_{1}}(x^{k},y^{3}),
\\ 
\mbox{ or }\underline{N}_{i_{1}}^{3}=\underline{n}_{i_{1}}(x^{k_{1}},t),%
\underline{N}_{i_{1}}^{4}=\underline{w}_{i_{1}}(x^{k_{1}},t),%
\end{array}
& 
\end{array}%
\right. $ \\ 
$\vdots $ \\ 
$\left\{ 
\begin{array}{cc}
\begin{array}{c}
\mathbf{g}_{i_{4}}(x^{k_{4}}),\mathbf{g}_{a_{5}}(x^{k4},y^{9}), \\ 
\mbox{ or }\mathbf{g}_{i_{4}}(x^{k_{1}}),\underline{\mathbf{g}}%
_{a_{5}}(x^{k_{4}},y^{10}),%
\end{array}
&  \\ 
\begin{array}{c}
N_{i_{4}}^{9}=w_{i_{4}}(x^{k_{4}},y^{9}),N_{i_{1}}^{10}=n_{i_{4}}(x^{k_{4}},y^{9}),
\\ 
\mbox{ or }\underline{N}_{i_{4}}^{9}=\underline{n}_{i_{4}}(x^{k_{4}},y^{10}),%
\underline{N}_{i_{4}}^{10}=\underline{w}_{i_{4}}(x^{k_{4}},y^{10}),%
\end{array}
& 
\end{array}%
\right. $%
\end{tabular}
\\ 
$\ ^{s}\mathring{\nabla},$ $\ ^{s}Ric=\{\mathring{R}_{\ \beta _{s}\gamma
_{s}}\}$ & Ricci tensors & $\ ^{s}\widehat{\mathbf{D}},\ \ ^{s}\widehat{%
\mathcal{R}}ic=\{\widehat{\mathbf{R}}_{\ \beta _{s}\gamma _{s}}\}$ \\ 
$~^{m}\mathcal{L[\mathbf{\phi }]\rightarrow }\ ^{m}\mathbf{T}_{\alpha
_{s}\beta _{s}}\mathcal{[\mathbf{\phi }]}$ & 
\begin{tabular}{l}
generating \\ 
sources%
\end{tabular}
& $%
\begin{array}{cc}
\widehat{\mathbf{\Upsilon }}_{\ \nu _{s}}^{\mu _{s}}=\mathbf{e}_{\ \mu
_{s}^{\prime }}^{\mu _{s}}\mathbf{e}_{\nu _{s}}^{\ \nu _{s}^{\prime }}%
\mathbf{\Upsilon }_{\ \nu _{s}^{\prime }}^{\mu _{s}^{\prime }}[\ ^{m}%
\mathcal{L}(\mathbf{\varphi ),}T_{\mu _{s}\nu _{s}},\ ^{s}\Lambda ] &  \\ 
\begin{array}{c}
=diag[\ _{1}\Upsilon (x^{i_{1}})\delta _{j_{1}}^{i_{1}},\ _{2}\Upsilon
(x^{i_{1}},y^{3})\delta _{b_{2}}^{a_{2}}, \\ 
\ _{3}\Upsilon (x^{i_{2}},y^{5})\delta _{b_{3}}^{a_{3}},\ _{4}\Upsilon
(x^{i_{3}},y^{7})\delta _{b_{4}}^{a_{4}},\ _{5}\Upsilon
(x^{i_{4}},y^{9})\delta _{b_{5}}^{a_{5}}], \\ 
\mbox{ quasi-stationary configurations};%
\end{array}
&  \\ 
\begin{array}{c}
=diag[\ _{1}\Upsilon (x^{i_{1}})\delta _{j_{1}}^{i_{1}},\ _{2}\underline{%
\Upsilon }(x^{i_{1}},t)\delta _{b_{2}}^{a_{2}}, \\ 
\ _{3}\underline{\Upsilon }(x^{i_{2}},y^{6})\delta _{b_{3}}^{a_{3}},\ _{4}%
\underline{\Upsilon }(x^{i_{3}},y^{8})\delta _{b_{4}}^{a_{4}},\ _{5}%
\underline{\Upsilon }(x^{i_{4}},y^{10})\delta _{b_{5}}^{a_{5}}], \\ 
\mbox{ locally anisotropic cosmology};%
\end{array}
& 
\end{array}%
$ \\ 
trivial eqs for $\ ^{s}\mathring{\nabla}$-torsion & LC-conditions & $\ ^{s}%
\widehat{\mathbf{D}}_{\mid \ ^{s}\widehat{\mathcal{T}}\rightarrow 0}=\ ^{s}%
\mathbf{\nabla .}$ \\ \hline\hline
\end{tabular}%
}
\end{eqnarray*}%
}

The formulas from \ Table 4 can be generalized for polarizations $\eta $-
and $\chi $-polarization functions extending on higher dimensions formulas (%
\ref{offdiagdefr}) and (\ref{offdncelepsilon}).

\subsubsection{Quasi-stationary higher dimension solutions}

The formulas for 4-d off-diagonal quasi-stationary solutions defined by
nonlinear quadratic elements (\ref{qeltors}), (\ref{offdiagcosmcsh}), (\ref%
{offdsolgenfgcosmc}), (\ref{offdiagpolfr}) and (\ref{epsilongenfdecomp}) can
be generalized in symbolic geometric form to 10-d spacetime Lorentz
manifolds as we summarize in Table 5.

{\scriptsize 
\begin{eqnarray*}
&&%
\begin{tabular}{l}
\hline\hline
\begin{tabular}{lll}
& {\large \textsf{Table 5:\ Higher dimension off-diagonal quasi-stationary
configurations}} &  \\ 
& Exact solutions of $\widehat{\mathbf{R}}_{\mu \nu }=\mathbf{\Upsilon }%
_{\mu \nu }$ (\ref{cdeq1}) transformed into a shall system of nonlinear PDEs
(\ref{eq1})-(\ref{e2c}) & 
\end{tabular}
\\ 
\end{tabular}
\\
&&%
\begin{tabular}{lll}
\hline\hline
&  &  \\ 
$%
\begin{array}{c}
\\ 
\end{array}%
\begin{array}{c}
\mbox{d-metric ansatz with} \\ 
\mbox{Killing symmetry }\partial _{4}=\partial _{t}%
\end{array}%
$ &  & $%
\begin{array}{c}
ds^{2}=g_{i_{1}}(x^{k_{1}})(dx^{i_{1}})^{2}+g_{a_{2}}(x^{k_{1}},y^{3})(dy^{a_{2}}+N_{i_{1}}^{a_{2}}(x^{k_{1}},y^{3})dx^{i_{1}})^{2}
\\ 
+g_{a_{3}}(x^{k_{2}},y^{5})(dy^{a_{3}}+N_{i_{2}}^{a_{3}}(x^{k_{2}},y^{5})dx^{i_{2}})^{2}
\\ 
+g_{a_{4}}(x^{k_{3}},y^{7})(dy^{a_{4}}+N_{i_{3}}^{a_{4}}(x^{k_{3}},y^{7})dx^{i_{3}})^{2}
\\ 
+g_{a_{5}}(x^{k_{4}},y^{9})(dy^{a_{5}}+N_{i_{4}}^{a_{5}}(x^{k_{4}},y^{9})dx^{i_{4}})^{2},%
\mbox{ for }g_{i_{1}}=e^{\psi {(x}^{k_{1}}{)}}, \\ 
g_{a_{2}}=h_{a_{2}}(x^{k_{1}},y^{3}),N_{i_{1}}^{3}=\
^{2}w_{i_{1}}=w_{i_{1}}(x^{k_{1}},y^{3}),N_{i_{1}}^{4}=\
^{2}n_{i_{1}}=n_{i_{1}}(x^{k_{1}},y^{3}), \\ 
g_{a_{3}}=h_{a_{3}}(x^{k_{2}},y^{5}),N_{i_{2}}^{5}=\
^{3}w_{i_{2}}=w_{i_{2}}(x^{k_{2}},y^{5}),N_{i_{2}}^{6}=\
^{3}n_{i_{2}}=n_{i_{2}}(x^{k_{2}},y^{5}), \\ 
g_{a_{4}}=h_{a_{4}}(x^{k_{3}},y^{7}),N_{i_{3}}^{7}=\
^{4}w_{i_{3}}=w_{i_{3}}(x^{k_{3}},y^{7}),N_{i_{3}}^{8}=\
^{4}n_{i_{3}}=n_{i_{3}}(x^{k_{3}},y^{7}), \\ 
g_{a_{5}}=h_{a_{5}}(x^{k_{4}},y^{9}),N_{i_{4}}^{9}=\
^{5}w_{i_{4}}=w_{i_{4}}(x^{k_{4}},y^{9}),N_{i_{4}}^{10}=\
^{5}n_{i_{4}}=n_{i_{4}}(x^{k_{4}},y^{9}),%
\end{array}%
$ \\ 
Effective matter sources &  & $%
\begin{tabular}{l}
$\mathbf{\Upsilon }_{\ \nu _{s}}^{\mu _{s}}=[\ _{1}\widehat{\Upsilon }({x}%
^{k_{1}})\delta _{j_{1}}^{i_{1}},\ _{2}\widehat{\Upsilon }({x}%
^{k_{1}},y^{3})\delta _{b_{2}}^{a_{2}},\ _{3}\widehat{\Upsilon }({x}%
^{k_{2}},y^{5})\delta _{b_{3}}^{a_{3}},$ \\ 
$\ _{4}\widehat{\Upsilon }({x}^{k_{3}},y^{7})\delta _{b_{4}}^{a_{4}},\ _{5}%
\widehat{\Upsilon }({x}^{k_{4}},y^{9})\delta _{b_{5}}^{a_{5}}],$%
\end{tabular}%
$ \\ \hline
Nonlinear PDEs (\ref{eq1})-(\ref{e2c}) &  & $%
\begin{tabular}{lll}
$%
\begin{array}{c}
\psi ^{\bullet \bullet }+\psi ^{\prime \prime }=2\ \ _{1}\widehat{\Upsilon };
\\ 
\ ^{2}\varpi ^{\ast }\ h_{4}^{\ast }=2h_{3}h_{4}\ _{2}\widehat{\Upsilon };
\\ 
\ ^{2}\beta \ ^{2}w_{i_{1}}-\ ^{2}\alpha _{i_{1}}=0; \\ 
\ ^{2}n_{k_{1}}^{\ast \ast }+\ ^{2}\gamma \ ^{2}n_{k_{1}}^{\ast }=0;%
\end{array}%
$ &  & $%
\begin{array}{c}
\ ^{2}\varpi {=\ln |\partial _{3}h_{4}/\sqrt{|h_{3}h_{4}|}|,} \\ 
\ ^{2}\alpha _{i_{1}}=(\partial _{3}h_{4})\ (\partial _{i_{1}}\ ^{2}\varpi ),
\\ 
\ ^{2}\beta =(\partial _{3}h_{4})\ (\partial _{3}\ ^{2}\varpi ),\  \\ 
\ \ ^{2}\gamma =\partial _{3}\left( \ln |h_{4}|^{3/2}/|h_{3}|\right) , \\ 
\partial _{1}q=q^{\bullet },\partial _{2}q=q^{\prime },\partial
_{3}q=q^{\ast }%
\end{array}%
$ \\ 
$%
\begin{array}{c}
\partial _{5}(\ ^{3}\varpi )\ \partial _{5}h_{6}=2h_{5}h_{6}\ _{3}\widehat{%
\Upsilon }; \\ 
\ ^{3}\beta \ ^{3}w_{i_{2}}-\ ^{3}\alpha _{i_{2}}=0; \\ 
\partial _{5}(\partial _{5}\ ^{3}n_{k_{2}})+\ ^{3}\gamma \partial _{5}(\
^{3}n_{k_{2}})=0;%
\end{array}%
$ &  & $%
\begin{array}{c}
\\ 
\ ^{3}\varpi {=\ln |\partial _{5}h_{6}/\sqrt{|h_{5}h_{6}|}|,} \\ 
\ ^{3}\alpha _{i_{2}}=(\partial _{5}h_{6})\ (\partial _{i_{2}}\ ^{3}\varpi ),
\\ 
\ ^{3}\beta =(\partial _{5}h_{6})\ (\partial _{5}\ ^{3}\varpi ),\  \\ 
\ \ ^{3}\gamma =\partial _{5}\left( \ln |h_{6}|^{3/2}/|h_{5}|\right) ,%
\end{array}%
$ \\ 
$\vdots $ &  & $\vdots $ \\ 
$%
\begin{array}{c}
\partial _{9}(\ ^{5}\varpi )\ \partial _{9}h_{10}=2h_{9}h_{10}\ _{5}\widehat{%
\Upsilon }; \\ 
\ ^{5}\beta \ ^{5}w_{i_{4}}-\ ^{5}\alpha _{i_{4}}=0; \\ 
\partial _{9}(\partial _{9}\ ^{5}n_{k_{4}})+\ ^{5}\gamma \partial _{9}(\
^{3}n_{k_{3}})=0;%
\end{array}%
$ &  & $%
\begin{array}{c}
\\ 
\ ^{5}\varpi {=\ln |\partial _{9}h_{10}/\sqrt{|h_{9}h_{10}|}|,} \\ 
\ ^{5}\alpha _{i}=(\partial _{9}h_{10})\ (\partial _{i}\varpi ), \\ 
\ ^{5}\beta =(\partial _{9}h_{10})\ (\partial _{9}\varpi ),\  \\ 
\ \ ^{5}\gamma =\partial _{9}\left( \ln |h_{10}|^{3/2}/|h_{9}|\right) ,%
\end{array}%
$%
\end{tabular}%
$ \\ \hline
$%
\begin{array}{c}
\mbox{ Gener.  functs:}\ h_{3}(x^{k_{1}},y^{3}), \\ 
\ ^{2}\Psi (x^{k_{1}},y^{3})=e^{\ ^{2}\varpi },\ ^{2}\Phi (x^{k_{1}},y^{3}),
\\ 
\mbox{integr. functs:}\ h_{4}^{[0]}(x^{k_{1}}),\  \\ 
_{1}n_{k_{1}}(x^{i_{1}}),\ _{2}n_{k_{1}}(x^{i_{1}}); \\ 
\mbox{ Gener.  functs:}h_{5}(x^{k_{2}},y^{5}), \\ 
\ ^{3}\Psi (x^{k_{2}},y^{5})=e^{\ ^{3}\varpi },\ ^{3}\Phi (x^{k_{2}},y^{5}),
\\ 
\mbox{integr. functs:}\ h_{6}^{[0]}(x^{k_{2}}),\  \\ 
_{1}^{3}n_{k_{2}}(x^{i_{2}}),\ _{2}^{3}n_{k_{2}}(x^{i_{2}}); \\ 
... \\ 
\mbox{ Gener.  functs:}h_{9}(x^{k_{4}},y^{9}), \\ 
\ ^{5}\Psi (x^{k_{3}},y^{9})=e^{\ ^{5}\varpi },\ ^{5}\Phi (x^{k_{4}},y^{9}),
\\ 
\mbox{integr. functs:}\ h_{10}^{[0]}(x^{k_{4}}),\  \\ 
_{1}^{5}n_{k_{4}}(x^{i_{4}}),\ _{2}^{5}n_{k_{4}}(x^{i_{4}}); \\ 
\mbox{\& nonlinear symmetries}%
\end{array}%
$ &  & $%
\begin{array}{c}
\ ((\ ^{2}\Psi )^{2})^{\ast }=-\int dy^{3}\ _{2}\widehat{\Upsilon }h_{4}^{\
\ast }, \\ 
(\ ^{2}\Phi )^{2}=-4\ _{2}\Lambda h_{4},\mbox{ see }(\ref{nonlinsymrex}), \\ 
h_{4}=h_{4}^{[0]}-(\ ^{2}\Phi )^{2}/4\ _{2}\Lambda ,h_{4}^{\ast }\neq 0,\
_{2}\Lambda \neq 0=const; \\ 
\\ 
\partial _{5}((\ ^{3}\Psi )^{2})=-\int dy^{5}\ _{3}\widehat{\Upsilon }%
\partial _{5}h_{6}^{\ }, \\ 
(\ ^{3}\Phi )^{2}=-4\ _{3}\Lambda h_{6}, \\ 
h_{6}=h_{6}^{[0]}-(\ ^{3}\Phi )^{2}/4\ _{3}\Lambda ,\partial _{5}h_{6}\neq
0,\ _{3}\Lambda \neq 0=const; \\ 
... \\ 
\partial _{9}((\ ^{5}\Psi )^{2})=-\int dy^{9}\ _{5}\widehat{\Upsilon }%
\partial _{9}h_{10}^{\ }, \\ 
(\ ^{5}\Phi )^{2}=-4\ _{5}\Lambda h_{10}, \\ 
h_{10}=h_{10}^{[0]}-(\ ^{5}\Phi )^{2}/4\ _{5}\Lambda ,\partial
_{9}h_{10}\neq 0,\ _{5}\Lambda \neq 0=const;%
\end{array}%
$ \\ \hline
Off-diag. solutions, $%
\begin{array}{c}
\mbox{d--metric} \\ 
\mbox{N-connec.}%
\end{array}%
$ &  & $%
\begin{tabular}{l}
$%
\begin{array}{c}
\ g_{i}=e^{\ \psi (x^{k})}\mbox{ as a solution of 2-d Poisson eqs. }\psi
^{\bullet \bullet }+\psi ^{\prime \prime }=2~\ _{1}\widehat{\Upsilon }; \\ 
h_{3}=-(\Psi ^{\ast })^{2}/4\ _{2}\widehat{\Upsilon }^{2}h_{4},\mbox{ see }(%
\ref{g3}),(\ref{g4}); \\ 
h_{4}=h_{4}^{[0]}-\int dy^{3}(\Psi ^{2})^{\ast }/4\ _{2}\widehat{\Upsilon }%
=h_{4}^{[0]}-\Phi ^{2}/4\ _{2}\Lambda ; \\ 
w_{i}=\partial _{i}\ \Psi /\ \partial _{3}\Psi =\partial _{i}\ \Psi ^{2}/\
\partial _{3}\Psi ^{2}|; \\ 
n_{k}=\ _{1}n_{k}+\ _{2}n_{k}\int dy^{3}(\Psi ^{\ast })^{2}/\ _{2}\widehat{%
\Upsilon }^{2}|h_{4}^{[0]}-\int dy^{3}(\Psi ^{2})^{\ast }/4\ _{2}\widehat{%
\Upsilon }^{2}|^{5/2};%
\end{array}%
$ \\ 
$%
\begin{array}{c}
h_{5}=-(\partial _{5}\ ^{3}\Psi )^{2}/4\ _{3}\widehat{\Upsilon }^{2}h_{6};
\\ 
h_{6}=h_{6}^{[0]}-\int dy^{5}\partial _{5}((\ ^{3}\Psi )^{2})/4\ _{3}%
\widehat{\Upsilon }=h_{6}^{[0]}-(\ ^{3}\Phi )^{2}/4\ _{3}\Lambda ; \\ 
w_{i_{2}}=\partial _{i_{2}}(\ ^{3}\Psi )/\ \partial _{5}(\ ^{3}\Psi
)=\partial _{i_{2}}(\ ^{3}\Psi )^{2}/\ \partial _{5}(\ ^{3}\Psi )^{2}|; \\ 
n_{k_{2}}=\ _{1}n_{k_{2}}+\ _{2}n_{k_{2}}\int dy^{5}(\partial _{5}\ ^{3}\Psi
)^{2}/\ _{2}\widehat{\Upsilon }^{2}|h_{6}^{[0]}-\int dy^{5}\partial _{5}((\
^{3}\Psi )^{2})/4\ _{3}\widehat{\Upsilon }^{2}|^{5/2};%
\end{array}%
$ \\ 
$...$ \\ 
$%
\begin{array}{c}
h_{9}=-(\partial _{9}\ ^{5}\Psi )^{2}/4\ _{5}\widehat{\Upsilon }^{2}h_{10};
\\ 
h_{10}=h_{10}^{[0]}-\int dy^{9}\partial _{9}((\ ^{5}\Psi )^{2})/4\ _{5}%
\widehat{\Upsilon }=h_{10}^{[0]}-(\ ^{5}\Phi )^{2}/4\ _{5}\Lambda ; \\ 
w_{i_{4}}=\partial _{i_{4}}(\ ^{5}\Psi )/\ \partial _{9}(\ ^{5}\Psi
)=\partial _{i_{4}}(\ ^{5}\Psi )^{2}/\ \partial _{9}(\ ^{5}\Psi )^{2}|; \\ 
n_{k_{2}}=\ _{1}n_{k_{2}}+\ _{2}n_{k_{2}}\int dy^{9}(\partial _{9}\ ^{5}\Psi
)^{2}/\ _{5}\widehat{\Upsilon }^{2}|h_{10}^{[0]}-\int dy^{9}\partial _{9}((\
^{5}\Psi )^{2})/4\ _{5}\widehat{\Upsilon }^{2}|^{5/2}.%
\end{array}%
$%
\end{tabular}%
$ \\ \hline\hline
\end{tabular}%
\end{eqnarray*}%
%
} 

As an example of 10-d quasi-stationary quadratic element extending the 4-d
formulas (\ref{offdsolgenfgcosmc}), we provide 
\begin{eqnarray}
d\widehat{s}_{[10d]}^{2} &=&\widehat{g}_{\alpha _{s}\beta
_{s}}(x^{k},y^{3},y^{5},y^{7},y^{9};h_{4},h_{6},h_{8,}h_{10};\ _{s}\widehat{%
\Upsilon };\ _{s}\Lambda )du^{\alpha _{s}}du^{\beta _{s}}  \label{qst10d} \\
&=&e^{\psi (x^{k},\ _{s}\widehat{\Upsilon })}[(dx^{1})^{2}+(dx^{2})^{2}]-%
\frac{(h_{4}^{\ast })^{2}}{|\int dy^{3}[\ _{2}\widehat{\Upsilon }%
h_{4}]^{\ast }|\ h_{4}}\{dy^{3}+\frac{\partial _{i_{1}}[\int dy^{3}(\ _{2}%
\widehat{\Upsilon })\ h_{4}^{\ast }]}{\ _{2}\widehat{\Upsilon }\ h_{4}^{\ast
}}dx^{i_{1}}\}^{2}+  \notag \\
&&h_{4}\{dt+[\ _{1}n_{k_{1}}+\ _{2}n_{k_{1}}\int dy^{3}\frac{(h_{4}^{\ast
})^{2}}{|\int dy^{3}[\ _{2}\widehat{\Upsilon }h_{4}]^{\ast }|\ (h_{4})^{5/2}}%
]dx_{1}^{k}\}+  \notag \\
&&\frac{(\partial _{5}h_{6})^{2}}{|\int dy^{5}\partial _{5}[\ _{3}\widehat{%
\Upsilon }h_{6}]|\ h_{6}}\{dy^{5}+\frac{\partial _{i_{2}}[\int dy^{5}(\ _{3}%
\widehat{\Upsilon })\ \partial _{5}h_{6}]}{\ _{3}\widehat{\Upsilon }\
\partial _{5}h_{6}}dx^{i_{2}}\}^{2}+  \notag \\
&&h_{6}\{dy^{6}+[\ _{1}n_{k_{2}}+\ _{2}n_{k_{2}}\int dy^{5}\frac{(\partial
_{5}h_{6})^{2}}{|\int dy^{5}\partial _{5}[\ _{3}\widehat{\Upsilon }h_{6}]|\
(h_{6})^{5/2}}]dx^{k_{2}}\}+  \notag \\
&&\frac{(\partial _{7}h_{8})^{2}}{|\int dy^{7}\partial _{7}[\ _{4}\widehat{%
\Upsilon }h_{8}]|\ h_{8}}\{dy^{7}+\frac{\partial _{i_{3}}[\int dy^{7}(\ _{4}%
\widehat{\Upsilon })\ \partial _{7}h_{8}]}{\ _{4}\widehat{\Upsilon }\
\partial _{7}h_{8}}dx^{i_{3}}\}^{2}+  \notag \\
&&h_{8}\{dy^{8}+[\ _{1}n_{k_{3}}+\ _{2}n_{k_{3}}\int dy^{7}\frac{(\partial
_{7}h_{8})^{2}}{|\int dy^{7}\partial _{7}[\ _{4}\widehat{\Upsilon }h_{8}]|\
(h_{8})^{5/2}}]dx^{k_{3}}\}+  \notag \\
&&\frac{(\partial _{9}h_{10})^{2}}{|\int dy^{9}\partial _{9}[\ _{5}\widehat{%
\Upsilon }h_{10}]|\ h_{10}}\{dy^{9}+\frac{\partial _{i_{4}}[\int dy^{9}(\
_{5}\widehat{\Upsilon })\ \partial _{9}h_{10}]}{\ _{5}\widehat{\Upsilon }\
\partial _{9}h_{10}}dx^{i_{3}}\}^{2}+  \notag \\
&&h_{10}\{dy^{10}+[\ _{1}n_{k_{4}}+\ _{2}n_{k_{4}}\int dy^{10}\frac{%
(\partial _{9}h_{10})^{2}}{|\int dy^{9}\partial _{9}[\ _{5}\widehat{\Upsilon 
}h_{10}]|\ (h_{10})^{5/2}}]dx^{k_{4}}\}.  \notag
\end{eqnarray}

The nonlinear symmetries (\ref{ntransf1}) and (\ref{ntransf2}) allow to
perform similar computations and express shell by shell $(\ ^{s}\Phi
)^{2}=-4\ _{s}\Lambda h_{a_{s}}.$ In similar forms, we can generate
s-adapted solutions of type (\ref{qst10d}) when, for instance, Killing
symmetries on $\partial _{6}$ are changed into $\partial _{5}$ (we can
consider any permutations with Killing symmetries on $\partial _{8}$ changed
into $\partial _{7}$ and/or $\partial _{10}$ changed into $\partial _{9}$).
Using $\eta $- and/or $\chi $-polarizations, above classes of
quasi-stationary higher dimension solutions can be considered for
transforming certain prime 10-d s-metrics into respective nonholonomic
deformed to target s-metrics of the same or lower dimensions.

\subsubsection{Locally anisotropic cosmological solutions with extra
dimensions}

The formulas for 4-d locally anisotropic cosmological solutions from Table 3
can be extended in geometric abstract forms for 10-d nonholonomic Lorentz
manifolds, see Table 6. Such construction and applications in modern
cosmology were provide in a series of our previous works \cite%
{vmon3,bubuianu17} when off-diagonal cosmological metrics are derived as
dual ones (for a time like coordinate) to quasi-stationary metrics. In 4-d,
we explain the geometric principles with respect to generating the d-metric (%
\ref{qeltorsc}). A series of works \cite%
{sv14,sv15a,elizalde15,sv16,rajpoot17,sv18} is devoted to locally
anisotropic cosmological scenarios for MGTs with massive terms and/or
contributions from string theories; ekpyrotic scenarious with quasi-periodic
and patern structure formation; cosmological accelerating and infaltionary
space-time quasicrystal structure. A recent work \cite{bvv21} the AFCDM is
applied for generating solutions for the Kaluza-Kelin gravity and
cosmological models emerging from geometric and quantum information flow
generalizations of the Einstein equations. For simplicity, we consider only
canonical d-connections with Killing symmetry on $\partial _{3}$ and $%
\partial _{7}$ when respective restrictions to shells $s=3$ and/or $s=4,$
can be considered, in similar forms, for Killing symmetries on $\partial
_{5},\partial _{6},\partial _{7}$ and $\partial _{8}.$

\newpage {\scriptsize 
\begin{eqnarray*}
&&%
\begin{tabular}{l}
\hline\hline
\begin{tabular}{lll}
& {\large \textsf{Table 6:\ Higher dimension off-diagonal cosmological
solutions}} &  \\ 
& Exact solutions of $\widehat{\mathbf{R}}_{\mu \nu }=\mathbf{\Upsilon }%
_{\mu \nu }$ (\ref{cdeq1}) transformed into a shall system of nonlinear PDEs
(\ref{eq1})-(\ref{e2c}) & 
\end{tabular}
\\ 
\end{tabular}
\\
&&%
\begin{tabular}{lll}
\hline\hline
&  &  \\ 
$%
\begin{array}{c}
\\ 
\end{array}%
\begin{array}{c}
\mbox{d-metric ansatz with} \\ 
\mbox{Killing symmetry }\partial _{3},\partial _{9}%
\end{array}%
$ &  & $%
\begin{array}{c}
ds^{2}=g_{i_{1}}(x^{k_{1}})(dx^{i_{1}})^{2}+\underline{g}%
_{a_{2}}(x^{k_{1}},t)(dy^{a_{2}}+\underline{N}%
_{i_{1}}^{a_{2}}(x^{k_{1}},t)dx^{i_{1}})^{2} \\ 
+g_{a_{3}}(x^{k_{2}},y^{5})(dy^{a_{3}}+N_{i_{2}}^{a_{3}}(x^{k_{2}},y^{5})dx^{i_{2}})^{2}
\\ 
+g_{a_{4}}(x^{k_{3}},y^{7})(dy^{a_{4}}+N_{i_{3}}^{a_{4}}(x^{k_{3}},y^{7})dx^{i_{3}})^{2}
\\ 
+g_{a_{5}}(x^{k_{4}},y^{10})(dy^{a_{5}}+N_{i_{4}}^{a_{5}}(x^{k_{4}},y^{10})dx^{i_{4}})^{2},%
\mbox{ for }g_{i_{1}}=e^{\psi {(x}^{k_{1}}{)}}, \\ 
\underline{g}_{a_{2}}=\underline{h}_{a_{2}}(x^{k_{1}},t),\underline{N}%
_{i_{1}}^{3}=\ ^{2}\underline{n}_{i_{1}}=\underline{n}_{i_{1}}(x^{k_{1}},t),%
\underline{N}_{i_{1}}^{4}=\ ^{2}\underline{w}_{i_{1}}=\underline{w}%
_{i_{1}}(x^{k_{1}},t), \\ 
g_{a_{3}}=h_{a_{3}}(x^{k_{2}},y^{6}),N_{i_{2}}^{5}=\
^{3}w_{i_{2}}=w_{i_{2}}(x^{k_{2}},y^{5}),N_{i_{2}}^{6}=\
^{3}n_{i_{2}}=n_{i_{2}}(x^{k_{2}},y^{5}), \\ 
g_{a_{4}}=h_{a_{4}}(x^{k_{3}},y^{7}),N_{i_{3}}^{7}=\
^{4}w_{i_{3}}=w_{i_{3}}(x^{k_{3}},y^{7}),N_{i_{3}}^{8}=\
^{4}n_{i_{3}}=n_{i_{3}}(x^{k_{3}},y^{7}), \\ 
g_{a_{5}}=h_{a_{5}}(x^{k_{4}},y^{9}),N_{i_{4}}^{9}=\
^{5}w_{i_{4}}=w_{i_{4}}(x^{k_{4}},y^{9}),N_{i_{4}}^{10}=\
^{5}n_{i_{4}}=n_{i_{4}}(x^{k_{4}},y^{9}),%
\end{array}%
$ \\ 
Effective matter sources &  & $%
\begin{tabular}{l}
$\mathbf{\Upsilon }_{\ \nu _{s}}^{\mu _{s}}=[\ _{1}\widehat{\Upsilon }({x}%
^{k_{1}})\delta _{j_{1}}^{i_{1}},\ _{2}\underline{\widehat{\Upsilon }}({x}%
^{k_{1}},t)\delta _{b_{2}}^{a_{2}},\ _{3}\widehat{\Upsilon }({x}%
^{k_{2}},y^{5})\delta _{b_{3}}^{a_{3}},$ \\ 
$\ _{4}\widehat{\Upsilon }({x}^{k_{3}},y^{7})\delta _{b_{4}}^{a_{4}},\ _{5}%
\widehat{\Upsilon }({x}^{k_{4}},y^{9})\delta _{b_{5}}^{a_{5}}],$%
\end{tabular}%
$ \\ \hline
Nonlinear PDEs (\ref{eq1})-(\ref{e2c}) &  & $%
\begin{tabular}{lll}
$%
\begin{array}{c}
\psi ^{\bullet \bullet }+\psi ^{\prime \prime }=2\ \ _{1}\widehat{\Upsilon };
\\ 
\ ^{2}\underline{\varpi }^{\diamond }\ \underline{h}_{3}^{\diamond }=2%
\underline{h}_{3}\underline{h}_{4}\ _{2}\widehat{\Upsilon }; \\ 
\ ^{2}\underline{n}_{k_{1}}^{\diamond \diamond }+\ ^{2}\underline{\gamma }\
^{2}\underline{n}_{k_{1}}^{\diamond }=0; \\ 
\ ^{2}\underline{\beta }\ ^{2}\underline{w}_{i_{1}}-\ ^{2}\underline{\alpha }%
_{i_{1}}=0;%
\end{array}%
$ &  & $%
\begin{array}{c}
\ ^{2}\underline{\varpi }{=\ln |\partial _{4}\underline{{h}}_{4}/\sqrt{|%
\underline{h}_{3}\underline{h}_{4}|}|,} \\ 
\ ^{2}\underline{\alpha }_{i_{1}}=(\partial _{4}\underline{h}_{3})\
(\partial _{i_{1}}\ ^{2}\underline{\varpi }), \\ 
\ ^{2}\underline{\beta }=(\partial _{4}\underline{h}_{4})\ (\partial _{3}\
^{2}\underline{\varpi }),\  \\ 
\ \ ^{2}\underline{\gamma }=\partial _{4}\left( \ln |\underline{h}%
_{3}|^{3/2}/|\underline{h}_{4}|\right) , \\ 
\partial _{1}q=q^{\bullet },\partial _{2}q=q^{\prime },\partial
_{4}q=\partial _{t}q=q^{\diamond }%
\end{array}%
$ \\ 
$%
\begin{array}{c}
\partial _{5}(\ ^{3}\varpi )\ \partial _{5}h_{6}=2h_{5}h_{6}\ _{3}\widehat{%
\Upsilon }; \\ 
\ ^{3}\beta \ ^{3}w_{i_{2}}-\ ^{3}\alpha _{i_{2}}=0; \\ 
\partial _{5}(\partial _{5}\ ^{3}n_{k_{2}})+\ ^{3}\gamma \partial _{5}(\
^{3}n_{k_{2}})=0;%
\end{array}%
$ &  & $%
\begin{array}{c}
\\ 
\ ^{3}\varpi {=\ln |\partial _{5}h_{6}/\sqrt{|h_{5}h_{6}|}|,} \\ 
\ ^{3}\alpha _{i_{2}}=(\partial _{5}h_{6})\ (\partial _{i_{2}}\ ^{3}\varpi ),
\\ 
\ ^{3}\beta =(\partial _{5}h_{6})\ (\partial _{5}\ ^{3}\varpi ),\  \\ 
\ \ ^{3}\gamma =\partial _{5}\left( \ln |h_{6}|^{3/2}/|h_{5}|\right) ,%
\end{array}%
$ \\ 
$\vdots $ &  & $\vdots $ \\ 
$%
\begin{array}{c}
\partial _{9}(\ ^{5}\varpi )\ \partial _{9}h_{10}=2h_{9}h_{10}\ _{5}\widehat{%
\Upsilon }; \\ 
\ ^{5}\beta \ ^{5}w_{i_{4}}-\ ^{5}\alpha _{i_{4}}=0; \\ 
\partial _{9}(\partial _{9}\ ^{5}n_{k_{4}})+\ ^{5}\gamma \partial _{9}(\
^{3}n_{k_{3}})=0;%
\end{array}%
$ &  & $%
\begin{array}{c}
\\ 
\ ^{5}\varpi {=\ln |\partial _{9}h_{10}/\sqrt{|h_{9}h_{10}|}|,} \\ 
\ ^{5}\alpha _{i}=(\partial _{9}h_{10})\ (\partial _{i}\ ^{5}\varpi ), \\ 
\ ^{5}\beta =(\partial _{9}h_{10})\ (\partial _{9}\ ^{5}\varpi ),\  \\ 
\ \ ^{5}\gamma =\partial _{9}\left( \ln |h_{10}|^{3/2}/|h_{9}|\right) ,%
\end{array}%
$%
\end{tabular}%
$ \\ \hline
$%
\begin{array}{c}
\mbox{ Gener.  functs:}\ \underline{h}_{4}(x^{k_{1}},t), \\ 
\ ^{2}\underline{\Psi }(x^{k_{1}},t)=e^{\ ^{2}\underline{\varpi }},\ ^{2}%
\underline{\Phi }(x^{k_{1}},t), \\ 
\mbox{integr. functs:}\ \underline{h}_{3}^{[0]}(x^{k_{1}}),\  \\ 
_{1}\underline{n}_{k_{1}}(x^{i_{1}}),\ _{2}\underline{n}_{k_{1}}(x^{i_{1}});
\\ 
\mbox{ Gener.  functs:}h_{5}(x^{k_{2}},y^{5}), \\ 
\ ^{3}\Psi (x^{k_{2}},y^{5})=e^{\ ^{3}\varpi },\ ^{3}\Phi (x^{k_{2}},y^{5}),
\\ 
\mbox{integr. functs:}\ h_{6}^{[0]}(x^{k_{2}}),\  \\ 
_{1}^{3}n_{k_{2}}(x^{i_{2}}),\ _{2}^{3}n_{k_{2}}(x^{i_{2}}); \\ 
... \\ 
\mbox{ Gener.  functs:}h_{9}(x^{k_{4}},y^{9}), \\ 
\ ^{5}\Psi (x^{k_{3}},y^{9})=e^{\ ^{5}\varpi },\ ^{5}\Phi (x^{k_{4}},y^{9}),
\\ 
\mbox{integr. functs:}\ h_{10}^{[0]}(x^{k_{4}}),\  \\ 
_{1}^{5}n_{k_{4}}(x^{i_{4}}),\ _{2}^{5}n_{k_{4}}(x^{i_{4}}); \\ 
\mbox{\& nonlinear symmetries}%
\end{array}%
$ &  & $%
\begin{array}{c}
\ ((\ ^{2}\underline{\Psi })^{2})^{\diamond }=-\int dt\ _{2}\widehat{%
\underline{\Upsilon }}\underline{h}_{3}^{\ \diamond }, \\ 
(\ ^{2}\underline{\Phi })^{2}=-4\ _{2}\underline{\Lambda }\underline{h}_{3},
\\ 
h_{3}=h_{3}^{[0]}-(\ ^{2}\underline{\Phi })^{2}/4\ _{2}\underline{\Lambda },%
\underline{h}_{3}^{\diamond }\neq 0,\ _{2}\underline{\Lambda }\neq 0=const;
\\ 
\\ 
\partial _{5}((\ ^{3}\Psi )^{2})=-\int dy^{5}\ _{3}\widehat{\Upsilon }%
\partial _{5}h_{6}^{\ }, \\ 
(\ ^{3}\Phi )^{2}=-4\ _{3}\Lambda h_{6}, \\ 
h_{6}=h_{6}^{[0]}-(\ ^{3}\Phi )^{2}/4\ _{3}\Lambda ,\partial _{5}h_{6}\neq
0,\ _{3}\Lambda \neq 0=const; \\ 
... \\ 
\partial _{9}((\ ^{5}\Psi )^{2})=-\int dy^{9}\ _{5}\widehat{\Upsilon }%
\partial _{9}h_{10}^{\ }, \\ 
(\ ^{5}\Phi )^{2}=-4\ _{5}\Lambda h_{10}, \\ 
h_{10}=h_{10}^{[0]}-(\ ^{5}\Phi )^{2}/4\ _{5}\Lambda ,\partial
_{9}h_{10}\neq 0,\ _{5}\Lambda \neq 0=const;%
\end{array}%
$ \\ \hline
Off-diag. solutions, $%
\begin{array}{c}
\mbox{d--metric} \\ 
\mbox{N-connec.}%
\end{array}%
$ &  & $%
\begin{tabular}{l}
$%
\begin{array}{c}
\ g_{i}=e^{\ \psi (x^{k})}\mbox{ as a solution of 2-d Poisson eqs. }\psi
^{\bullet \bullet }+\psi ^{\prime \prime }=2~\ _{1}\widehat{\Upsilon }; \\ 
\underline{h}_{4}=-(\underline{\Psi }^{\diamond })^{2}/4\ _{2}\widehat{%
\underline{\Upsilon }}^{2}\underline{h}_{3}; \\ 
\underline{h}_{3}=\underline{h}_{3}^{[0]}-\int dt(\underline{\Psi }%
^{2})^{\diamond }/4\ _{2}\widehat{\underline{\Upsilon }}=\underline{h}%
_{3}^{[0]}-\underline{\Phi }^{2}/4\ _{2}\underline{\Lambda }; \\ 
\underline{w}_{i_{1}}=\partial _{i_{1}}\ \underline{\Psi }/\ \partial 
\underline{\Psi }^{\diamond }=\partial _{i_{1}}\ \underline{\Psi }^{2}/\
\partial _{t}\underline{\Psi }^{2}|; \\ 
\underline{n}_{k_{1}}=\ _{1}n_{k_{1}}+\ _{2}n_{k_{1}}\int dt(\underline{\Psi 
}^{\diamond })^{2}/\ _{2}\widehat{\underline{\Upsilon }}^{2}|\underline{h}%
_{3}^{[0]}-\int dt(\underline{\Psi }^{2})^{\diamond }/4\ _{2}\widehat{%
\underline{\Upsilon }}^{2}|^{5/2};%
\end{array}%
$ \\ 
$%
\begin{array}{c}
h_{5}=-(\partial _{5}\ ^{3}\Psi )^{2}/4\ _{3}\widehat{\Upsilon }^{2}h_{6};
\\ 
h_{6}=h_{6}^{[0]}-\int dy^{5}\partial _{5}((\ ^{3}\Psi )^{2})/4\ _{3}%
\widehat{\Upsilon }=h_{6}^{[0]}-(\ ^{3}\Phi )^{2}/4\ _{3}\Lambda ; \\ 
w_{i_{2}}=\partial _{i_{2}}(\ ^{3}\Psi )/\ \partial _{5}(\ ^{3}\Psi
)=\partial _{i_{2}}(\ ^{3}\Psi )^{2}/\ \partial _{5}(\ ^{3}\Psi )^{2}|; \\ 
n_{k_{2}}=\ _{1}n_{k_{2}}+\ _{2}n_{k_{2}}\int dy^{5}(\partial _{5}\ ^{3}\Psi
)^{2}/\ _{2}\widehat{\Upsilon }^{2}|h_{6}^{[0]}-\int dy^{5}\partial _{5}((\
^{3}\Psi )^{2})/4\ _{3}\widehat{\Upsilon }^{2}|^{5/2};%
\end{array}%
$ \\ 
$...$ \\ 
$%
\begin{array}{c}
h_{9}=-(\partial _{9}\ ^{5}\Psi )^{2}/4\ _{5}\widehat{\Upsilon }^{2}h_{10};
\\ 
h_{10}=h_{10}^{[0]}-\int dy^{9}\partial _{9}((\ ^{5}\Psi )^{2})/4\ _{5}%
\widehat{\Upsilon }=h_{10}^{[0]}-(\ ^{5}\Phi )^{2}/4\ _{5}\Lambda ; \\ 
w_{i_{4}}=\partial _{i_{4}}(\ ^{5}\Psi )/\ \partial _{9}(\ ^{5}\Psi
)=\partial _{i_{4}}(\ ^{5}\Psi )^{2}/\ \partial _{9}(\ ^{5}\Psi )^{2}|; \\ 
n_{k_{2}}=\ _{1}n_{k_{2}}+\ _{2}n_{k_{2}}\int dy^{9}(\partial _{9}\ ^{5}\Psi
)^{2}/\ _{5}\widehat{\Upsilon }^{2}|h_{10}^{[0]}-\int dy^{9}\partial _{9}((\
^{5}\Psi )^{2})/4\ _{5}\widehat{\Upsilon }^{2}|^{5/2}.%
\end{array}%
$%
\end{tabular}%
$ \\ \hline\hline
\end{tabular}%
\end{eqnarray*}%
}

Let us consider a 10-d generalization of 4-d locally anisortopic
cosmological solutions (\ref{cosm1d}), see also (\ref{qeltorsc}), 
\begin{eqnarray}
d\widehat{s}_{[10d]}^{2} &=&\widehat{g}_{\alpha _{s}\beta
_{s}}(x^{k},t,y^{5},y^{7},y^{9};\underline{h}_{3},h_{6},h_{8,}h_{10};\ _{s}%
\underline{\widehat{\Upsilon }};\ _{s}\Lambda )du^{\alpha _{s}}du^{\beta
_{s}}  \label{lcs10d} \\
&=&e^{\psi (x^{k},\ _{s}\widehat{\Upsilon })}[(dx^{1})^{2}+(dx^{2})^{2}]+%
\underline{h}_{3}[dy^{3}+(\ _{1}n_{k_{1}}+4\ _{2}n_{k_{1}}\int dt\frac{(%
\underline{h}_{3}{}^{\diamond })^{2}}{|\int dt\ _{2}\underline{\Upsilon }%
\underline{h}_{3}{}^{\diamond }|(\underline{h}_{3})^{5/2}})dx^{k_{1}}] 
\notag \\
&&-\frac{(\underline{h}_{3}{}^{\diamond })^{2}}{|\int dt\ _{2}\underline{%
\Upsilon }\underline{h}_{3}{}^{\diamond }|\ \overline{h}_{3}}[dt+\frac{%
\partial _{i}(\int dt\ _{2}\underline{\Upsilon }\ \underline{h}%
_{3}{}^{\diamond }])}{\ \ _{2}\underline{\Upsilon }\ \underline{h}%
_{3}{}^{\diamond }}dx^{i}]+  \notag \\
&&\frac{(\partial _{5}\underline{h}_{6})^{2}}{|\int dy^{5}\partial _{5}[\
_{3}\underline{\Upsilon }\underline{h}_{6}]|\underline{h}_{6}}\{dy^{5}+\frac{%
\partial _{i_{2}}[\int dy^{5}(\ _{3}\underline{\Upsilon })\ \partial _{5}%
\underline{h}_{6}]}{\ _{3}\underline{\Upsilon }\ \partial _{5}\underline{h}%
_{6}}dx^{i_{2}}\}^{2}+  \notag \\
&&\underline{h}_{6}\{dy^{6}+[\ _{1}n_{k_{2}}+\ _{2}n_{k_{2}}\int dy^{5}\frac{%
(\partial _{5}\underline{h}_{6})^{2}}{|\int dy^{5}\partial _{5}[\ _{3}%
\underline{\Upsilon }\underline{h}_{6}]|\ (\underline{h}_{6})^{5/2}}%
]dx^{k_{2}}\}+  \notag \\
&&\frac{(\partial _{7}\underline{h}_{8})^{2}}{|\int dy^{7}\partial _{7}[\
_{4}\underline{\Upsilon }\underline{h}_{8}]|\ \underline{h}_{8}}\{dy^{7}+%
\frac{\partial _{i_{3}}[\int dy^{7}(\ _{4}\underline{\Upsilon })\ \partial
_{7}\underline{h}_{8}]}{\ _{4}\underline{\Upsilon }\ \partial _{7}\underline{%
h}_{8}}dx^{i_{3}}\}^{2}+  \notag \\
&&h_{8}\{dy^{8}+[\ _{1}n_{k_{3}}+\ _{2}n_{k_{3}}\int dy^{7}\frac{(\partial
_{7}\underline{h}_{8})^{2}}{|\int dy^{7}\partial _{7}[\ _{4}\underline{%
\Upsilon }\underline{h}_{8}]|\ (\underline{h}_{8})^{5/2}}]dx^{k_{3}}\}+ 
\notag \\
&&\frac{(\partial _{9}\underline{h}_{10})^{2}}{|\int dy^{9}\partial _{9}[\
_{5}\underline{\Upsilon }\underline{h}_{10}]|\ h_{10}}\{dy^{9}+\frac{%
\partial _{i_{4}}[\int dy^{9}(\ _{5}\widehat{\Upsilon })\ \partial _{9}%
\underline{h}_{10}]}{\ _{5}\underline{\Upsilon }\ \partial _{9}\underline{h}%
_{10}}dx^{i_{3}}\}^{2}+  \notag \\
&&\underline{h}_{10}\{dy^{10}+[\ _{1}n_{k_{4}}+\ _{2}n_{k_{4}}\int dy^{10}%
\frac{(\partial _{9}\underline{h}_{10})^{2}}{|\int dy^{9}\partial _{9}[\ _{5}%
\underline{\Upsilon }\underline{h}_{10}]|\ (\underline{h}_{10})^{5/2}}%
]dx^{k_{4}}\}.  \notag
\end{eqnarray}%
The s-metric (\ref{lcs10d}) possess the same extra shell Killing symmetries
on higher dimension coordinates. \ Such generic off-diagonal extra dimension
cosmological solutions are characterized by nonlinear symmetries of type (%
\ref{ntransf1}) and (\ref{ntransf2}), when (shell by shell) $(\ ^{s}%
\underline{\Phi })^{2}=-4\ _{s}\underline{\Lambda }\underline{h}_{a_{s}}.$
We can generate s-adapted solutions of type (\ref{lcs10d}) when, for
instance, Killing symmetries on $\partial _{6}$ are changed into $\partial
_{5}$ and we can consider any permutations with Killing symmetries on $%
\partial _{8}$ changed into $\partial _{7}$ and/or $\partial _{10}$ changed
into $\partial _{9}$. Using $\eta $- and/or $\chi $-polarizations with
generic dependence on a time like coordinate $t$, above classes of locally
anisortopic higher dimension solutions can be considered for transforming
certain prime cosmological 10-d s-metrics into respective nonholonomic
deformed to target s-metrics of the same or lower dimensions.

\subsection{Off-diagonal velocity depending quasi-stationary/cosmological
solutions, tables 7-11}

Such geometries and MGTs are modelled on tangent bundle $T\mathbf{V}$ to a
nonholonomic Lorentz manifold $\mathbf{V}.$ They include as particular
examples various relativistic generalizations of Finsler-Lagrange geometry
and theories with modified dispersion relations, MDR, with respective dual
symmetries for (co) fiber coordinates. The typical signature of total
metrics is of type $(+++-;+++-)$ for a Lorentz base with signature $(+++-).$
So, the dimension of geometric constructions and signature on such velocity
depending phase spaces is $\dim TV=8$ is different from that considered
above in 10-d gravity. To apply the AFCDM we need four shells of dyads (when 
$s=1,2,3,4$) with a corresponding (2+2)+(2+2) nonholonomic splitting of the
total dimension. The formulas are quite similar to those provided in
previous subsection when $y^{a_{s}}=v^{a_{s}},$ for $s=3$ and 4.
Nevertheless, the physical interpretation of such velocity phase space
models and respective exact/ parametric is different from those considered
for the higher dimension gravity. This is because the signature of metrics
is different, when $v^{8}$ is a time like coordinate on the typical fiber,
but $y^{8}$ was a space like coordinate in the space of velocities. If the
phase space solutions are with Killing symmetry on $\partial _{8},$ we can
fix $v^{8}=v_{[0]}^{8},$ and elaborate on phase space models with space like
velocity hypersurfaces. Another class of solutions can be with variable $%
v^{8}$ but a fixed, for instance, velocity $v^{7}=v_{[0]}^{7},$ which
provide examples of "velocity-rainbow" metrics in phase gravity. Both types
of s-metrics with mentioned behaviour in the velocity typical fiber may have
a Killing symmetry on $\partial _{4}$ (for locally anisotropic cosmological
solutions), or, for instance, on $\partial _{3}$, for quasi-stationary
solutions. As results, we obtain 4 different types of velocity-phase
s-metrics with typical quadratic elements and applications of the AFCDM
stated in subsections below and respective Tables 8-11.

\subsubsection{Diagonal and off-diagonal ansatz for velocity phase spaces}

The parametrization of local coordinates, N-connection and canonical
d-connection structures and s-metrics for velocity-phase spaces are sated in
Table 7. 
{\scriptsize 
\begin{eqnarray*}
&&%
\begin{tabular}{lll}
& {\ \textsf{Table 7:\ Diagonal and off-diagonal ansatz for 8-d tangent
Lorentz bundles} } &  \\ 
& and the Anholonomic Frame and Connection Deformation Method, \textbf{AFCDM}%
, &  \\ 
& \textit{for constructing generic off-diagonal exact and parametric
solutions} & 
\end{tabular}
\\
&&{%
\begin{tabular}{lll}
\hline
diagonal ansatz: PDEs $\rightarrow $ \textbf{ODE}s &  & AFCDM: \textbf{PDE}s 
\textbf{with decoupling; } \\ 
\begin{tabular}{l}
coordinates \\ 
$u^{\alpha _{s}}=(x^{1},x^{2},y^{3},y^{4}=t,$ \\ 
$v^{5},v^{6},v^{7},v^{8})$%
\end{tabular}
& $%
\begin{array}{c}
\ ^{s}u=(\ ^{s-1}x,\ ^{s}y) \\ 
s=1,2,3,4;%
\end{array}%
$ & $%
\begin{tabular}{l}
nonholonomic 2+2+2+2+2 splitting; shels $s=1,2,3,4$ \\ 
$u^{\alpha _{s}}=(x^{1},x^{2},y^{3},y^{4}=t,y^{5},y^{6},y^{7},y^{8});$ \\ 
$u^{\alpha _{s}}=(x^{i_{1}},y^{a_{2}},y^{a_{3}},y^{a_{4}});u^{\alpha
_{s}}=(x^{i_{s-1}},y^{a_{s}});$ \\ 
$\ $ $i_{1}=1,2;a_{2}=3,4;a_{3}=5,6;a_{4}=7,8;$%
\end{tabular}%
$ \\ 
LC-connection $\mathring{\nabla}$ & 
\begin{tabular}{l}
N-connection; \\ 
canonical \\ 
d-connection%
\end{tabular}
& $%
\begin{array}{c}
\ ^{s}\mathbf{N}:T\ ^{s}\mathbf{V}=hT\mathbf{V}\oplus \ ^{2}hT\mathbf{V}%
\oplus \ ^{3}v\mathbf{T\mathbf{V}\oplus }\ ^{4}v\mathbf{T\mathbf{V},} \\ 
\mbox{ locally }\ ^{s}\mathbf{N}%
=\{N_{i_{s-1}}^{a_{s}}(x,v)=N_{i_{s-1}}^{a_{s}}(\ ^{s-1}x,\
^{s}y)=N_{i_{s-1}}^{a_{s}}(\ ^{s}u)\} \\ 
\ ^{s}\widehat{\mathbf{D}}=(\ ^{1}h\widehat{\mathbf{D}},\ ^{2}v\widehat{%
\mathbf{D}},\ ^{3}v\widehat{\mathbf{D}},\ ^{4}v\widehat{\mathbf{D}}%
)=\{\Gamma _{\ \beta _{s}\gamma _{s}}^{\alpha _{s}}\}; \\ 
\mbox{ canonical connection distortion }\ ^{s}\widehat{\mathbf{D}}=\nabla +\
^{s}\widehat{\mathbf{Z}};\ ^{s}\widehat{\mathbf{D}}\ ^{s}\mathbf{g=0,} \\ 
\ ^{s}\widehat{\mathcal{T}}[\ ^{s}\mathbf{g,}\ ^{s}\mathbf{N,}\ ^{s}\widehat{%
\mathbf{D}}]\mbox{ canonical
d-torsion}%
\end{array}%
$ \\ 
$%
\begin{array}{c}
\mbox{ diagonal ansatz  } \\ 
\ ^{2}\mathring{g}=\mathring{g}_{\alpha _{2}\beta _{2}}(\ ^{s}u)= \\ 
\left( 
\begin{array}{cccc}
\mathring{g}_{1} &  &  &  \\ 
& \mathring{g}_{2} &  &  \\ 
&  & \mathring{g}_{3} &  \\ 
&  &  & \mathring{g}_{4}%
\end{array}%
\right) ; \\ 
\ ^{s}g=\mathring{g}_{\alpha _{s}\beta _{s}}(\ ^{s}u)= \\ 
\left( 
\begin{array}{cccc}
\ ^{2}\mathring{g} &  &  &  \\ 
& \mathring{g}_{5} &  &  \\ 
&  & \ddots &  \\ 
&  &  & \mathring{g}_{8}%
\end{array}%
\right)%
\end{array}%
$ & $\mathbf{g}\Leftrightarrow $ & $%
\begin{tabular}{l}
$g_{\alpha _{2}\beta _{2}}=%
\begin{array}{c}
g_{\alpha _{2}\beta _{2}}(x^{i_{1}},y^{a_{2}})%
\mbox{ general frames /
coordinates} \\ 
\left[ 
\begin{array}{cc}
g_{i_{1}j_{1}}+N_{i_{1}}^{a_{2}}N_{j_{1}}^{b_{2}}h_{a_{2}b_{2}} & 
N_{i_{1}}^{b_{2}}h_{c_{2}b_{2}} \\ 
N_{j_{1}}^{a_{2}}h_{a_{2}b_{2}} & h_{a_{2}c_{2}}%
\end{array}%
\right] ,%
\end{array}%
$ \\ 
$\ ^{2}\mathbf{g=\{g}_{\alpha _{2}\beta
_{2}}=[g_{i_{1}j_{1}},h_{a_{2}b_{2}}]\},$ \\ 
$\ ^{2}\mathbf{g}=\mathbf{g}_{i_{1}}(x^{k_{1}})dx^{i_{1}}\otimes dx^{i_{1}}+%
\mathbf{g}_{a_{2}}(x^{k_{2}},y^{b_{2}})\mathbf{e}^{a_{2}}\otimes \mathbf{e}%
^{b_{2}}$ \\ 
$\vdots $ \\ 
$g_{\alpha _{s}\beta _{s}}=%
\begin{array}{c}
g_{\alpha _{s}\beta _{s}}(x^{i_{s-1}},y^{a_{s}})%
\mbox{ general frames /
coordinates} \\ 
\left[ 
\begin{array}{cc}
g_{i_{s}j_{s}}+N_{i_{s-1}}^{a_{s}}N_{j_{s-1}}^{b_{s}}h_{a_{s}b_{s}} & 
N_{i_{s-1}}^{b_{s}}h_{c_{s}b_{s}} \\ 
N_{j_{s-1}}^{a_{s}}h_{a_{s}b_{s}} & h_{a_{s}c_{s}}%
\end{array}%
\right] ,%
\end{array}%
$ \\ 
$\ ^{s}\mathbf{g=\{g}_{\alpha _{s}\beta
_{s}}=[g_{i_{s-1}j_{s-1}},h_{a_{s}b_{s}}]$ \\ 
$=[g_{i_{1}j_{1}},h_{a_{2}b_{2}},h_{a_{3}b_{3}},h_{a_{4}b_{4}}]\}$ \\ 
$\ ^{s}\mathbf{g}=\mathbf{g}_{i_{s-1}}(x^{k_{s-1}})dx^{i_{s-1}}\otimes
dx^{i_{s-1}}+$ \\ 
$\mathbf{g}_{a_{s}}(x^{k_{s-1}},y^{b_{s}})\mathbf{e}^{a_{s}}\otimes \mathbf{e%
}^{b_{s}}$ \\ 
$=\mathbf{g}_{i_{1}}(x^{k_{1}})dx^{i_{1}}\otimes dx^{i_{1}}+\mathbf{g}%
_{a_{2}}(x^{k_{1}},y^{b_{2}})\mathbf{e}^{a_{2}}\otimes \mathbf{e}^{b_{2}}+$
\\ 
$\mathbf{g}_{a_{3}}(x^{k_{1}},y^{b_{2}},v^{b_{3}})\mathbf{e}^{a_{3}}\otimes 
\mathbf{e}^{b_{3}}+\mathbf{g}%
_{a_{4}}(x^{k_{1}},y^{b_{2}},v^{b_{3}},v^{b_{4}})\mathbf{e}^{a_{4}}\otimes 
\mathbf{e}^{b_{4}};$%
\end{tabular}%
$ \\ 
$\mathring{g}_{\alpha _{2}\beta _{2}}=\left\{ 
\begin{array}{cc}
\mathring{g}_{\alpha _{2}}(\ ^{2}r) & \mbox{ for BHs} \\ 
\mathring{g}_{\alpha _{2}}(t) & \mbox{ for FLRW }%
\end{array}%
\right. $ & [coord.frames] & $g_{\alpha _{2}\beta _{2}}=\left\{ 
\begin{array}{cc}
g_{\alpha _{2}\beta _{2}}(x^{i},y^{3}) & \mbox{ quasi-stationary config. }
\\ 
\underline{g}_{\alpha _{2}\beta _{2}}(x^{i},y^{4}=t) & 
\mbox{locally anisotropic
cosmology}%
\end{array}%
\right. $ \\ 
$\mathring{g}_{\alpha _{s}\beta _{s}}=\left\{ 
\begin{array}{cc}
\mathring{g}_{\alpha _{s}}(\ ^{s}r) & \mbox{ for BHs} \\ 
\mathring{g}_{\alpha _{s}}(t) & \mbox{ for FLRW }%
\end{array}%
\right. $ &  & $g_{\alpha _{5}\beta _{5}}=\left\{ 
\begin{array}{cc}
g_{\alpha _{5}\beta _{5}}(x^{i_{3}},v^{7}) &  \\ 
\underline{g}_{\alpha _{s}\beta _{s}}(x^{i_{3}},y^{8}) & 
\end{array}%
\right. $ \\ 
$%
\begin{array}{c}
\mbox{coord. transf. }e_{\alpha _{s}}=e_{\ \alpha _{s}}^{\alpha _{s}^{\prime
}}\partial _{\alpha _{s}^{\prime }}, \\ 
e^{\beta _{s}}=e_{\beta _{s}^{\prime }}^{\ \beta _{s}}du^{\beta _{s}^{\prime
}}, \\ 
\mathring{g}_{\alpha _{s}\beta _{s}}=\mathring{g}_{\alpha _{s}^{\prime
}\beta _{s}^{\prime }}e_{\ \alpha _{s}}^{\alpha _{s}^{\prime }}e_{\ \beta
_{s}}^{\beta _{s}^{\prime }} \\ 
\begin{array}{c}
\mathbf{\mathring{g}}_{\alpha _{s}}(x^{k_{s-1}},y^{a_{s}})\rightarrow 
\mathring{g}_{\alpha _{s}}(\ ^{s}r),\mbox{ or } \\ 
\mathring{g}_{\alpha _{s}}(t),\mathring{N}%
_{i_{s-1}}^{a_{s}}(x^{k_{s-1}},y^{a_{s}})\rightarrow 0.%
\end{array}%
\end{array}%
$ & [N-adapt. fr.] & 
\begin{tabular}{l}
$\left\{ 
\begin{array}{cc}
\begin{array}{c}
\mathbf{g}_{i_{1}}(x^{k_{1}}),\mathbf{g}_{a_{2}}(x^{k_{1}},y^{3}), \\ 
\mbox{ or }\mathbf{g}_{i_{1}}(x^{k_{1}}),\underline{\mathbf{g}}%
_{a_{2}}(x^{k_{1}},t),%
\end{array}
& \mbox{ d-metrics } \\ 
\begin{array}{c}
N_{i_{1}}^{3}=w_{i_{1}}(x^{k},y^{3}),N_{i_{1}}^{4}=n_{i_{1}}(x^{k},y^{3}),
\\ 
\mbox{ or }\underline{N}_{i_{1}}^{3}=\underline{n}_{i_{1}}(x^{k_{1}},t),%
\underline{N}_{i_{1}}^{4}=\underline{w}_{i_{1}}(x^{k_{1}},t),%
\end{array}
& 
\end{array}%
\right. $ \\ 
$\vdots $ \\ 
$\left\{ 
\begin{array}{cc}
\begin{array}{c}
\mathbf{g}_{i_{3}}(x^{k_{3}}),\mathbf{g}_{a_{4}}(x^{k_{3}},v^{7}), \\ 
\mbox{ or }\mathbf{g}_{i_{3}}(x^{k_{1}}),\underline{\mathbf{g}}%
_{a_{4}}(x^{k_{3}},v^{8}),%
\end{array}
&  \\ 
\begin{array}{c}
N_{i_{3}}^{7}=w_{i_{3}}(x^{k_{3}},v^{7}),N_{i_{3}}^{8}=n_{i_{3}}(x^{k_{3}},v^{7}),
\\ 
\mbox{ or }\underline{N}_{i_{3}}^{8}=\underline{n}_{i_{3}}(x^{k_{3}},v^{8}),%
\underline{N}_{i_{3}}^{8}=\underline{w}_{i_{3}}(x^{k_{3}},v^{8}),%
\end{array}
& 
\end{array}%
\right. $%
\end{tabular}
\\ 
$\ ^{s}\mathring{\nabla},$ $\ ^{s}Ric=\{\mathring{R}_{\ \beta _{s}\gamma
_{s}}\}$ & Ricci tensors & $\ ^{s}\widehat{\mathbf{D}},\ \ ^{s}\widehat{%
\mathcal{R}}ic=\{\widehat{\mathbf{R}}_{\ \beta _{s}\gamma _{s}}\}$ \\ 
$~^{m}\mathcal{L[\mathbf{\phi }]\rightarrow }\ ^{m}\mathbf{T}_{\alpha
_{s}\beta _{s}}\mathcal{[\mathbf{\phi }]}$ & 
\begin{tabular}{l}
generating \\ 
sources%
\end{tabular}
& $%
\begin{array}{cc}
\widehat{\mathbf{\Upsilon }}_{\ \nu _{s}}^{\mu _{s}}=\mathbf{e}_{\ \mu
_{s}^{\prime }}^{\mu _{s}}\mathbf{e}_{\nu _{s}}^{\ \nu _{s}^{\prime }}%
\mathbf{\Upsilon }_{\ \nu _{s}^{\prime }}^{\mu _{s}^{\prime }}[\ ^{m}%
\mathcal{L}(\mathbf{\varphi ),}T_{\mu _{s}\nu _{s}},\ ^{s}\Lambda ] &  \\ 
\begin{array}{c}
=diag[\ _{1}\Upsilon (x^{i_{1}})\delta _{j_{1}}^{i_{1}},\ _{2}\Upsilon
(x^{i_{1}},y^{3})\delta _{b_{2}}^{a_{2}}, \\ 
\ _{3}\Upsilon (x^{i_{2}},v^{5})\delta _{b_{3}}^{a_{3}},\ _{4}\Upsilon
(x^{i_{3}},v^{7})\delta _{b_{4}}^{a_{4}}], \\ 
\mbox{ quasi-stationary configurations};%
\end{array}
&  \\ 
\begin{array}{c}
=diag[\ _{1}\Upsilon (x^{i_{1}})\delta _{j_{1}}^{i_{1}},\ _{2}\underline{%
\Upsilon }(x^{i_{1}},t)\delta _{b_{2}}^{a_{2}}, \\ 
\ _{3}\underline{\Upsilon }(x^{i_{2}},v^{6})\delta _{b_{3}}^{a_{3}},\ _{4}%
\underline{\Upsilon }(x^{i_{3}},v^{8})\delta _{b_{4}}^{a_{4}}], \\ 
\mbox{ locally anisotropic cosmology};%
\end{array}
& 
\end{array}%
$ \\ 
trivial eqs for $\ ^{s}\mathring{\nabla}$-torsion & LC-conditions & $\ ^{s}%
\widehat{\mathbf{D}}_{\mid \ ^{s}\widehat{\mathcal{T}}\rightarrow 0}=\ ^{s}%
\mathbf{\nabla .}$ \\ \hline\hline
\end{tabular}%
}
\end{eqnarray*}%
}

Such parameterizations, with respective polarization functions and
generating sources can be considered for generalized relativistic Finsler
spaces encoding data for nonassociative / noncommutative / supersymmetric
theories etc. The generating and integration functions can be restricted to
define LC-configurations.

\subsubsection{Qusi-stationary solutions with fixed light velocity parameter}

Such quasi-stationary solutions are nonholonomic generalizations and
extensions on tangent Lorentz bundles with $v^{8}=const,$ when the velocity
phase space involve space like hypersurfaces. 
{\scriptsize 
\begin{eqnarray*}
&&%
\begin{tabular}{l}
\hline\hline
\begin{tabular}{lll}
& {\large \textsf{Table 8:\ Off-diagonal quasi-stationary spacetime and
space velocity configurations}} &  \\ 
& Exact solutions of $\widehat{\mathbf{R}}_{\mu _{s}\nu _{s}}=\mathbf{%
\Upsilon }_{\mu _{s}\nu _{s}}$ (\ref{cdeq1}) on $TV$ transformed into a
shall system of nonlinear PDEs (\ref{eq1})-(\ref{e2c}) & 
\end{tabular}
\\ 
\end{tabular}
\\
&&%
\begin{tabular}{lll}
\hline\hline
&  &  \\ 
$%
\begin{array}{c}
\mbox{d-metric ansatz with} \\ 
\mbox{Killing symmetry }\partial _{4}=\partial _{t},\partial _{8}%
\end{array}%
$ &  & $%
\begin{array}{c}
ds^{2}=g_{i_{1}}(x^{k_{1}})(dx^{i_{1}})^{2}+g_{a_{2}}(x^{k_{1}},y^{3})(dy^{a_{2}}+N_{i_{1}}^{a_{2}}(x^{k_{1}},y^{3})dx^{i_{1}})^{2}
\\ 
+g_{a_{3}}(x^{k_{2}},v^{5})(dy^{a_{3}}+N_{i_{2}}^{a_{3}}(x^{k_{2}},v^{5})dx^{i_{2}})^{2}
\\ 
+g_{a_{4}}(x^{k_{3}},v^{7})(dy^{a_{4}}+N_{i_{3}}^{a_{4}}(x^{k_{3}},v^{7})dx^{i_{3}})^{2},%
\mbox{ for }g_{i_{1}}=e^{\psi {(x}^{k_{1}}{)}}, \\ 
g_{a_{2}}=h_{a_{2}}(x^{k_{1}},y^{3}),N_{i_{1}}^{3}=\
^{2}w_{i_{1}}=w_{i_{1}}(x^{k_{1}},y^{3}),N_{i_{1}}^{4}=\
^{2}n_{i_{1}}=n_{i_{1}}(x^{k_{1}},y^{3}), \\ 
g_{a_{3}}=h_{a_{3}}(x^{k_{2}},v^{5}),N_{i_{2}}^{5}=\
^{3}w_{i_{2}}=w_{i_{2}}(x^{k_{2}},v^{5}),N_{i_{2}}^{6}=\
^{3}n_{i_{2}}=n_{i_{2}}(x^{k_{2}},v^{5}), \\ 
g_{a_{4}}=h_{a_{4}}(x^{k_{3}},v^{7}),N_{i_{3}}^{7}=\
^{4}w_{i_{3}}=w_{i_{3}}(x^{k_{3}},v^{7}),N_{i_{3}}^{8}=\
^{4}n_{i_{3}}=n_{i_{3}}(x^{k_{3}},v^{7}),%
\end{array}%
$ \\ 
Effective matter sources &  & $\mathbf{\Upsilon }_{\ \nu _{s}}^{\mu _{s}}=[\
_{1}\widehat{\Upsilon }({x}^{k_{1}})\delta _{j_{1}}^{i_{1}},\ _{2}\widehat{%
\Upsilon }({x}^{k_{1}},y^{3})\delta _{b_{2}}^{a_{2}},\ _{3}\widehat{\Upsilon 
}({x}^{k_{2}},v^{5})\delta _{b_{3}}^{a_{3}},_{4}\widehat{\Upsilon }({x}%
^{k_{3}},v^{7})\delta _{b_{4}}^{a_{4}},],$ \\ \hline
Nonlinear PDEs (\ref{eq1})-(\ref{e2c}) &  & $%
\begin{tabular}{lll}
$%
\begin{array}{c}
\psi ^{\bullet \bullet }+\psi ^{\prime \prime }=2\ \ _{1}\widehat{\Upsilon };
\\ 
\ ^{2}\varpi ^{\ast }\ h_{4}^{\ast }=2h_{3}h_{4}\ _{2}\widehat{\Upsilon };
\\ 
\ ^{2}\beta \ ^{2}w_{i_{1}}-\ ^{2}\alpha _{i_{1}}=0; \\ 
\ ^{2}n_{k_{1}}^{\ast \ast }+\ ^{2}\gamma \ ^{2}n_{k_{1}}^{\ast }=0;%
\end{array}%
$ &  & $%
\begin{array}{c}
\ ^{2}\varpi {=\ln |\partial _{3}h_{4}/\sqrt{|h_{3}h_{4}|}|,} \\ 
\ ^{2}\alpha _{i_{1}}=(\partial _{3}h_{4})\ (\partial _{i_{1}}\ ^{2}\varpi ),
\\ 
\ ^{2}\beta =(\partial _{3}h_{4})\ (\partial _{3}\ ^{2}\varpi ),\  \\ 
\ \ ^{2}\gamma =\partial _{3}\left( \ln |h_{4}|^{3/2}/|h_{3}|\right) , \\ 
\partial _{1}q=q^{\bullet },\partial _{2}q=q^{\prime },\partial
_{3}q=q^{\ast }%
\end{array}%
$ \\ 
$%
\begin{array}{c}
\partial _{5}(\ ^{3}\varpi )\ \partial _{5}h_{6}=2h_{5}h_{6}\ _{3}\widehat{%
\Upsilon }; \\ 
\ ^{3}\beta \ ^{3}w_{i_{2}}-\ ^{3}\alpha _{i_{2}}=0; \\ 
\partial _{5}(\partial _{5}\ ^{3}n_{k_{2}})+\ ^{3}\gamma \partial _{5}(\
^{3}n_{k_{2}})=0;%
\end{array}%
$ &  & $%
\begin{array}{c}
\\ 
\ ^{3}\varpi {=\ln |\partial _{5}h_{6}/\sqrt{|h_{5}h_{6}|}|,} \\ 
\ ^{3}\alpha _{i_{2}}=(\partial _{5}h_{6})\ (\partial _{i_{2}}\ ^{3}\varpi ),
\\ 
\ ^{3}\beta =(\partial _{5}h_{6})\ (\partial _{5}\ ^{3}\varpi ),\  \\ 
\ \ ^{3}\gamma =\partial _{5}\left( \ln |h_{6}|^{3/2}/|h_{5}|\right) ,%
\end{array}%
$ \\ 
$%
\begin{array}{c}
\partial _{7}(\ ^{4}\varpi )\ \partial _{7}h_{8}=2h_{7}h_{8}\ _{4}\widehat{%
\Upsilon }; \\ 
\ ^{4}\beta \ ^{4}w_{i_{3}}-\ ^{4}\alpha _{i_{3}}=0; \\ 
\partial _{7}(\partial _{7}\ ^{4}n_{k_{3}})+\ ^{4}\gamma \partial _{7}(\
^{4}n_{k_{3}})=0;%
\end{array}%
$ &  & $%
\begin{array}{c}
\\ 
\ ^{4}\varpi {=\ln |\partial _{7}h_{8}/\sqrt{|h_{7}h_{8}|}|,} \\ 
\ ^{4}\alpha _{i}=(\partial _{7}h_{8})\ (\partial _{i}\ ^{4}\varpi ), \\ 
\ ^{4}\beta =(\partial _{7}h_{8})\ (\partial _{7}\ ^{4}\varpi ),\  \\ 
\ \ ^{4}\gamma =\partial _{7}\left( \ln |h_{8}|^{3/2}/|h_{7}|\right) ,%
\end{array}%
$%
\end{tabular}%
$ \\ \hline
$%
\begin{array}{c}
\mbox{ Gener.  functs:}\ h_{3}(x^{k_{1}},y^{3}), \\ 
\ ^{2}\Psi (x^{k_{1}},y^{3})=e^{\ ^{2}\varpi },\ ^{2}\Phi (x^{k_{1}},y^{3}),
\\ 
\mbox{integr. functs:}\ h_{4}^{[0]}(x^{k_{1}}),\  \\ 
_{1}n_{k_{1}}(x^{i_{1}}),\ _{2}n_{k_{1}}(x^{i_{1}}); \\ 
\mbox{ Gener.  functs:}h_{5}(x^{k_{2}},v^{5}), \\ 
\ ^{3}\Psi (x^{k_{2}},v^{5})=e^{\ ^{3}\varpi },\ ^{3}\Phi (x^{k_{2}},v^{5}),
\\ 
\mbox{integr. functs:}\ h_{6}^{[0]}(x^{k_{2}}),\  \\ 
_{1}^{3}n_{k_{2}}(x^{i_{2}}),\ _{2}^{3}n_{k_{2}}(x^{i_{2}}); \\ 
\mbox{ Gener.  functs:}h_{7}(x^{k_{3}},v^{7}), \\ 
\ ^{5}\Psi (x^{k_{2}},v^{7})=e^{\ ^{4}\varpi },\ ^{4}\Phi (x^{k_{3}},v^{7}),
\\ 
\mbox{integr. functs:}\ h_{8}^{[0]}(x^{k_{3}}),\  \\ 
_{1}^{4}n_{k_{3}}(x^{i_{3}}),\ _{2}^{4}n_{k_{3}}(x^{i_{4}}); \\ 
\mbox{\& nonlinear symmetries}%
\end{array}%
$ &  & $%
\begin{array}{c}
\ ((\ ^{2}\Psi )^{2})^{\ast }=-\int dy^{3}\ _{2}\widehat{\Upsilon }h_{4}^{\
\ast }, \\ 
(\ ^{2}\Phi )^{2}=-4\ _{2}\Lambda h_{4},\mbox{ see }(\ref{nonlinsymrex}), \\ 
h_{4}=h_{4}^{[0]}-(\ ^{2}\Phi )^{2}/4\ _{2}\Lambda ,h_{4}^{\ast }\neq 0,\
_{2}\Lambda \neq 0=const; \\ 
\\ 
\partial _{5}((\ ^{3}\Psi )^{2})=-\int dv^{5}\ _{3}\widehat{\Upsilon }%
\partial _{5}h_{6}^{\ }, \\ 
(\ ^{3}\Phi )^{2}=-4\ _{3}\Lambda h_{6}, \\ 
h_{6}=h_{6}^{[0]}-(\ ^{3}\Phi )^{2}/4\ _{3}\Lambda ,\partial _{5}h_{6}\neq
0,\ _{3}\Lambda \neq 0=const; \\ 
\\ 
\partial _{7}((\ ^{4}\Psi )^{2})=-\int dv^{7}\ _{4}\widehat{\Upsilon }%
\partial _{7}h_{8}^{\ }, \\ 
(\ ^{4}\Phi )^{2}=-4\ _{4}\Lambda h_{8}, \\ 
h_{8}=h_{8}^{[0]}-(\ ^{4}\Phi )^{2}/4\ _{4}\Lambda ,\partial _{7}h_{8}\neq
0,\ _{4}\Lambda \neq 0=const;%
\end{array}%
$ \\ \hline
Off-diag. solutions, $%
\begin{array}{c}
\mbox{d--metric} \\ 
\mbox{N-connec.}%
\end{array}%
$ &  & $%
\begin{tabular}{l}
$%
\begin{array}{c}
\ g_{i}=e^{\ \psi (x^{k})}\mbox{ as a solution of 2-d Poisson eqs. }\psi
^{\bullet \bullet }+\psi ^{\prime \prime }=2~\ _{1}\widehat{\Upsilon }; \\ 
h_{3}=-(\Psi ^{\ast })^{2}/4\ _{2}\widehat{\Upsilon }^{2}h_{4},\mbox{ see }(%
\ref{g3}),(\ref{g4}); \\ 
h_{4}=h_{4}^{[0]}-\int dy^{3}(\Psi ^{2})^{\ast }/4\ _{2}\widehat{\Upsilon }%
=h_{4}^{[0]}-\Phi ^{2}/4\ _{2}\Lambda ; \\ 
w_{i}=\partial _{i}\ \Psi /\ \partial _{3}\Psi =\partial _{i}\ \Psi ^{2}/\
\partial _{3}\Psi ^{2}|; \\ 
n_{k}=\ _{1}n_{k}+\ _{2}n_{k}\int dy^{3}(\Psi ^{\ast })^{2}/\ _{2}\widehat{%
\Upsilon }^{2}|h_{4}^{[0]}-\int dy^{3}(\Psi ^{2})^{\ast }/4\ _{2}\widehat{%
\Upsilon }^{2}|^{5/2};%
\end{array}%
$ \\ 
$%
\begin{array}{c}
h_{5}=-(\partial _{5}\ ^{3}\Psi )^{2}/4\ _{3}\widehat{\Upsilon }^{2}h_{6};
\\ 
h_{6}=h_{6}^{[0]}-\int dv^{5}\partial _{5}((\ ^{3}\Psi )^{2})/4\ _{3}%
\widehat{\Upsilon }=h_{6}^{[0]}-(\ ^{3}\Phi )^{2}/4\ _{3}\Lambda ; \\ 
w_{i_{2}}=\partial _{i_{2}}(\ ^{3}\Psi )/\ \partial _{5}(\ ^{3}\Psi
)=\partial _{i_{2}}(\ ^{3}\Psi )^{2}/\ \partial _{5}(\ ^{3}\Psi )^{2}|; \\ 
n_{k_{2}}=\ _{1}n_{k_{2}}+\ _{2}n_{k_{2}}\int dv^{5}(\partial _{5}\ ^{3}\Psi
)^{2}/\ _{2}\widehat{\Upsilon }^{2}|h_{6}^{[0]}-\int dv^{5}\partial _{5}((\
^{3}\Psi )^{2})/4\ _{3}\widehat{\Upsilon }^{2}|^{5/2};%
\end{array}%
$ \\ 
$%
\begin{array}{c}
h_{7}=-(\partial _{7}\ ^{3}\Psi )^{2}/4\ _{4}\widehat{\Upsilon }^{2}h_{8};
\\ 
h_{8}=h_{8}^{[0]}-\int dv^{7}\partial _{9}((\ ^{4}\Psi )^{2})/4\ _{4}%
\widehat{\Upsilon }=h_{8}^{[0]}-(\ ^{4}\Phi )^{2}/4\ _{4}\Lambda ; \\ 
w_{i_{3}}=\partial _{i_{3}}(\ ^{4}\Psi )/\ \partial _{7}(\ ^{4}\Psi
)=\partial _{i_{3}}(\ ^{4}\Psi )^{2}/\ \partial _{7}(\ ^{4}\Psi )^{2}|; \\ 
n_{k_{3}}=\ _{1}n_{k_{3}}+\ _{2}n_{k_{3}}\int dv^{7}(\partial _{7}\ ^{4}\Psi
)^{2}/\ _{4}\widehat{\Upsilon }^{2}|h_{8}^{[0]}-\int dv^{7}\partial _{7}((\
^{4}\Psi )^{2})/4\ _{4}\widehat{\Upsilon }^{2}|^{5/2}.%
\end{array}%
$%
\end{tabular}%
$ \\ \hline\hline
\end{tabular}%
\end{eqnarray*}%
}As an example of 8-d quasi-stationary quadratic element with $v^{8}=const$
on $T\mathbf{V},$ we provide 
\begin{eqnarray}
d\widehat{s}_{[8d]}^{2} &=&\widehat{g}_{\alpha _{s}\beta
_{s}}(x^{k},y^{3},y^{5},y^{7};h_{4},h_{6},h_{8,};\ _{s}\widehat{\Upsilon };\
_{s}\Lambda )du^{\alpha _{s}}du^{\beta _{s}}  \label{qst8d7} \\
&=&e^{\psi (x^{k},\ _{s}\widehat{\Upsilon })}[(dx^{1})^{2}+(dx^{2})^{2}]-%
\frac{(h_{4}^{\ast })^{2}}{|\int dy^{3}[\ _{2}\widehat{\Upsilon }%
h_{4}]^{\ast }|\ h_{4}}\{dy^{3}+\frac{\partial _{i_{1}}[\int dy^{3}(\ _{2}%
\widehat{\Upsilon })\ h_{4}^{\ast }]}{\ _{2}\widehat{\Upsilon }\ h_{4}^{\ast
}}dx^{i_{1}}\}^{2}+  \notag \\
&&h_{4}\{dt+[\ _{1}n_{k_{1}}+\ _{2}n_{k_{1}}\int dy^{3}\frac{(h_{4}^{\ast
})^{2}}{|\int dy^{3}[\ _{2}\widehat{\Upsilon }h_{4}]^{\ast }|\ (h_{4})^{5/2}}%
]dx_{1}^{k}\}+  \notag \\
&&\frac{(\partial _{5}h_{6})^{2}}{|\int dy^{5}\partial _{5}[\ _{3}\widehat{%
\Upsilon }h_{6}]|\ h_{6}}\{dv^{5}+\frac{\partial _{i_{2}}[\int dy^{5}(\ _{3}%
\widehat{\Upsilon })\ \partial _{5}h_{6}]}{\ _{3}\widehat{\Upsilon }\
\partial _{5}h_{6}}dx^{i_{2}}\}^{2}+  \notag \\
&&h_{6}\{dv^{5}+[\ _{1}n_{k_{2}}+\ _{2}n_{k_{2}}\int dv^{5}\frac{(\partial
_{5}h_{6})^{2}}{|\int dy^{5}\partial _{5}[\ _{3}\widehat{\Upsilon }h_{6}]|\
(h_{6})^{5/2}}]dx^{k_{2}}\}+  \notag \\
&&\frac{(\partial _{7}h_{8})^{2}}{|\int dv^{7}\partial _{7}[\ _{4}\widehat{%
\Upsilon }h_{8}]|\ h_{8}}\{dv^{7}+\frac{\partial _{i_{3}}[\int dv^{7}(\ _{4}%
\widehat{\Upsilon })\ \partial _{7}h_{8}]}{\ _{4}\widehat{\Upsilon }\
\partial _{7}h_{8}}dx^{i_{3}}\}^{2}+  \notag \\
&&h_{8}\{dv^{8}+[\ _{1}n_{k_{3}}+\ _{2}n_{k_{3}}\int dv^{7}\frac{(\partial
_{7}h_{8})^{2}}{|\int dv^{7}\partial _{7}[\ _{4}\widehat{\Upsilon }h_{8}]|\
(h_{8})^{5/2}}]dx^{k_{3}}\}.  \notag
\end{eqnarray}%
Such s-metrics possess nonlinear symmetries which allow to re-define the
generating functions and generating sources and related them to conventions
cosmological constants. Solutions with gravitational $\eta $- and $\chi $%
-polarizations can be defined for respective off-diagonal deformations of
prime s-metrics into target ones. All formulas can be proven using abstract
geometric methods and corresponding applications of the AFCDM.

\subsubsection{ Quasi-stationary solutions with variable light velocity
parameter}

Another class of quasi-stationary extensions of a Lorentz manifold, $\mathbf{%
V},$ metrics is for quadratic line elements with $v^{7}=const$ which provide
examples of velocity rainbow s-metrics on $T\mathbf{V}.$ Considering a $%
v^{8}\leftrightarrow v^{7}$ changing of velocity phase coordinates in (\ref%
{qst8d7}), we construct an example of 8-d quasi-stationary quadratic element
with $v^{7}=const$ on $T\mathbf{V}$ defining an example of velocity rainbow
s-metric, 
\begin{eqnarray}
d\widehat{s}_{[8d]}^{2} &=&\widehat{g}_{\alpha _{s}\beta
_{s}}(x^{k},y^{3},y^{5},y^{7};h_{4},h_{6},h_{8,};\ _{s}\widehat{\Upsilon };\
_{s}\Lambda )du^{\alpha _{s}}du^{\beta _{s}}  \label{qst8d8} \\
&=&e^{\psi (x^{k},\ _{s}\widehat{\Upsilon })}[(dx^{1})^{2}+(dx^{2})^{2}]-%
\frac{(h_{4}^{\ast })^{2}}{|\int dy^{3}[\ _{2}\widehat{\Upsilon }%
h_{4}]^{\ast }|\ h_{4}}\{dy^{3}+\frac{\partial _{i_{1}}[\int dy^{3}(\ _{2}%
\widehat{\Upsilon })\ h_{4}^{\ast }]}{\ _{2}\widehat{\Upsilon }\ h_{4}^{\ast
}}dx^{i_{1}}\}^{2}+  \notag \\
&&h_{4}\{dt+[\ _{1}n_{k_{1}}+\ _{2}n_{k_{1}}\int dy^{3}\frac{(h_{4}^{\ast
})^{2}}{|\int dy^{3}[\ _{2}\widehat{\Upsilon }h_{4}]^{\ast }|\ (h_{4})^{5/2}}%
]dx_{1}^{k}\}+  \notag \\
&&\frac{(\partial _{5}h_{6})^{2}}{|\int dy^{5}\partial _{5}[\ _{3}\widehat{%
\Upsilon }h_{6}]|\ h_{6}}\{dv^{5}+\frac{\partial _{i_{2}}[\int dy^{5}(\ _{3}%
\widehat{\Upsilon })\ \partial _{5}h_{6}]}{\ _{3}\widehat{\Upsilon }\
\partial _{5}h_{6}}dx^{i_{2}}\}^{2}+  \notag \\
&&h_{6}\{dy^{6}+[\ _{1}n_{k_{2}}+\ _{2}n_{k_{2}}\int dv^{5}\frac{(\partial
_{5}h_{6})^{2}}{|\int dv^{5}\partial _{5}[\ _{3}\widehat{\Upsilon }h_{6}]|\
(h_{6})^{5/2}}]dx^{k_{2}}\}+  \notag \\
&&\underline{h}_{7}\{dv^{7}+[\ _{1}n_{k_{3}}+\ _{2}n_{k_{3}}\int dv^{8}\frac{%
(\partial _{8}\underline{h}_{7})^{2}}{|\int dv^{8}\partial _{8}[\ _{4}%
\underline{\widehat{\Upsilon }}h_{7}]|\ (\underline{h}_{7})^{5/2}}%
]dx^{k_{3}}\}+  \notag \\
&&\frac{(\partial _{8}\underline{h}_{7})^{2}}{|\int dv^{8}\partial _{8}[\
_{4}\underline{\widehat{\Upsilon }}\underline{h}_{7}]|\ \underline{h}_{7}}%
\{dv^{8}+\frac{\partial _{i_{3}}[\int dv^{8}(\ _{4}\underline{\widehat{%
\Upsilon }})\ \partial _{8}\underline{h}_{7}]}{\ _{4}\underline{\widehat{%
\Upsilon }}\ \partial _{8}\underline{h}_{7}}dx^{i_{3}}\}^{2}.  \notag
\end{eqnarray}%
The principles of generating such quasi-stationary and rainbow solutions are
summarized in Table 9.

{\scriptsize 
\begin{eqnarray*}
&&%
\begin{tabular}{l}
\hline\hline
\begin{tabular}{lll}
& {\large \textsf{Table 9:\ Off-diagonal quasi-stationary spacetimes with
velocity rainbows}} &  \\ 
& Exact solutions of $\widehat{\mathbf{R}}_{\mu _{s}\nu _{s}}=\mathbf{%
\Upsilon }_{\mu _{s}\nu _{s}}$ (\ref{cdeq1}) on $TV$ transformed into a
shall system of nonlinear PDEs (\ref{eq1})-(\ref{e2c}) & 
\end{tabular}
\\ 
\end{tabular}
\\
&&%
\begin{tabular}{lll}
\hline\hline
&  &  \\ 
$%
\begin{array}{c}
\mbox{d-metric ansatz with} \\ 
\mbox{Killing symmetry }\partial _{4}=\partial _{t},\partial _{7}%
\end{array}%
$ &  & $%
\begin{array}{c}
ds^{2}=g_{i_{1}}(x^{k_{1}})(dx^{i_{1}})^{2}+g_{a_{2}}(x^{k_{1}},y^{3})(dy^{a_{2}}+N_{i_{1}}^{a_{2}}(x^{k_{1}},y^{3})dx^{i_{1}})^{2}
\\ 
+g_{a_{3}}(x^{k_{2}},v^{5})(dy^{a_{3}}+N_{i_{2}}^{a_{3}}(x^{k_{2}},v^{5})dx^{i_{2}})^{2}
\\ 
+\underline{g}_{a_{4}}(x^{k_{3}},v^{8})(dy^{a_{4}}+\underline{N}%
_{i_{3}}^{a_{4}}(x^{k_{3}},v^{8})dx^{i_{3}})^{2},\mbox{ for }%
g_{i_{1}}=e^{\psi {(x}^{k_{1}}{)}}, \\ 
g_{a_{2}}=h_{a_{2}}(x^{k_{1}},y^{3}),N_{i_{1}}^{3}=\
^{2}w_{i_{1}}=w_{i_{1}}(x^{k_{1}},y^{3}),N_{i_{1}}^{4}=\
^{2}n_{i_{1}}=n_{i_{1}}(x^{k_{1}},y^{3}), \\ 
g_{a_{3}}=h_{a_{3}}(x^{k_{2}},v^{5}),N_{i_{2}}^{5}=\
^{3}w_{i_{2}}=w_{i_{2}}(x^{k_{2}},v^{5}),N_{i_{2}}^{6}=\
^{3}n_{i_{2}}=n_{i_{2}}(x^{k_{2}},v^{5}), \\ 
\underline{g}_{a_{4}}=\underline{h}_{a_{4}}(x^{k_{3}},v^{8}),\underline{N}%
_{i_{3}}^{7}=\ ^{4}\underline{n}_{i_{3}}=\underline{n}%
_{i_{3}}(x^{k_{3}},v^{8}),\underline{N}_{i_{3}}^{8}=\ ^{4}\underline{w}%
_{i_{3}}=\underline{w}_{i_{3}}(x^{k_{3}},v^{8}),%
\end{array}%
$ \\ 
Effective matter sources &  & $\mathbf{\Upsilon }_{\ \nu _{s}}^{\mu _{s}}=[\
_{1}\widehat{\Upsilon }({x}^{k_{1}})\delta _{j_{1}}^{i_{1}},\ _{2}\widehat{%
\Upsilon }({x}^{k_{1}},y^{3})\delta _{b_{2}}^{a_{2}},\ _{3}\widehat{\Upsilon 
}({x}^{k_{2}},v^{5})\delta _{b_{3}}^{a_{3}},_{4}\widehat{\underline{\Upsilon 
}}({x}^{k_{3}},v^{8})\delta _{b_{4}}^{a_{4}},],$ \\ \hline
Nonlinear PDEs (\ref{eq1})-(\ref{e2c}) &  & $%
\begin{tabular}{lll}
$%
\begin{array}{c}
\psi ^{\bullet \bullet }+\psi ^{\prime \prime }=2\ \ _{1}\widehat{\Upsilon };
\\ 
\ ^{2}\varpi ^{\ast }\ h_{4}^{\ast }=2h_{3}h_{4}\ _{2}\widehat{\Upsilon };
\\ 
\ ^{2}\beta \ ^{2}w_{i_{1}}-\ ^{2}\alpha _{i_{1}}=0; \\ 
\ ^{2}n_{k_{1}}^{\ast \ast }+\ ^{2}\gamma \ ^{2}n_{k_{1}}^{\ast }=0;%
\end{array}%
$ &  & $%
\begin{array}{c}
\ ^{2}\varpi {=\ln |\partial _{3}h_{4}/\sqrt{|h_{3}h_{4}|}|,} \\ 
\ ^{2}\alpha _{i_{1}}=(\partial _{3}h_{4})\ (\partial _{i_{1}}\ ^{2}\varpi ),
\\ 
\ ^{2}\beta =(\partial _{3}h_{4})\ (\partial _{3}\ ^{2}\varpi ),\  \\ 
\ \ ^{2}\gamma =\partial _{3}\left( \ln |h_{4}|^{3/2}/|h_{3}|\right) , \\ 
\partial _{1}q=q^{\bullet },\partial _{2}q=q^{\prime },\partial
_{3}q=q^{\ast }%
\end{array}%
$ \\ 
$%
\begin{array}{c}
\partial _{5}(\ ^{3}\varpi )\ \partial _{5}h_{6}=2h_{5}h_{6}\ _{3}\widehat{%
\Upsilon }; \\ 
\ ^{3}\beta \ ^{3}w_{i_{2}}-\ ^{3}\alpha _{i_{2}}=0; \\ 
\partial _{5}(\partial _{5}\ ^{3}n_{k_{2}})+\ ^{3}\gamma \partial _{5}(\
^{3}n_{k_{2}})=0;%
\end{array}%
$ &  & $%
\begin{array}{c}
\\ 
\ ^{3}\varpi {=\ln |\partial _{5}h_{6}/\sqrt{|h_{5}h_{6}|}|,} \\ 
\ ^{3}\alpha _{i_{2}}=(\partial _{5}h_{6})\ (\partial _{i_{2}}\ ^{3}\varpi ),
\\ 
\ ^{3}\beta =(\partial _{5}h_{6})\ (\partial _{5}\ ^{3}\varpi ),\  \\ 
\ \ ^{3}\gamma =\partial _{5}\left( \ln |h_{6}|^{3/2}/|h_{5}|\right) ,%
\end{array}%
$ \\ 
$%
\begin{array}{c}
\partial _{8}(\ ^{4}\underline{\varpi })\ \partial _{8}\underline{h}_{7}=2%
\underline{h}_{7}\underline{h}_{8}\ _{4}\underline{\widehat{\Upsilon }}; \\ 
\partial _{8}(\partial _{8}\ ^{4}\underline{n}_{k_{3}})+\ ^{4}\underline{%
\gamma }\partial _{8}(\ ^{4}\underline{n}_{k_{3}})=0; \\ 
\ ^{4}\underline{\beta }\ ^{4}\underline{w}_{i_{3}}-\ ^{4}\underline{\alpha }%
_{i_{3}}=0;%
\end{array}%
$ &  & $%
\begin{array}{c}
\\ 
\ ^{4}\underline{\varpi }{=\ln |\partial _{8}\underline{h}_{7}/\sqrt{|%
\underline{h}_{7}\underline{h}_{8}|}|,} \\ 
\ ^{4}\underline{\alpha }_{i}=(\partial _{8}\underline{h}_{7})\ (\partial
_{i}\ ^{4}\underline{\varpi }), \\ 
\ ^{4}\underline{\beta }=(\partial _{8}\underline{h}_{7})\ (\partial _{8}\
^{4}\underline{\varpi }),\  \\ 
\ \ ^{4}\underline{\gamma }=\partial _{8}\left( \ln |\underline{h}%
_{7}|^{3/2}/|\underline{h}_{8}|\right) ,%
\end{array}%
$%
\end{tabular}%
$ \\ \hline
$%
\begin{array}{c}
\mbox{ Gener.  functs:}\ h_{3}(x^{k_{1}},y^{3}), \\ 
\ ^{2}\Psi (x^{k_{1}},y^{3})=e^{\ ^{2}\varpi },\ ^{2}\Phi (x^{k_{1}},y^{3}),
\\ 
\mbox{integr. functs:}\ h_{4}^{[0]}(x^{k_{1}}),\  \\ 
_{1}n_{k_{1}}(x^{i_{1}}),\ _{2}n_{k_{1}}(x^{i_{1}}); \\ 
\mbox{ Gener.  functs:}h_{5}(x^{k_{2}},v^{5}), \\ 
\ ^{3}\Psi (x^{k_{2}},v^{5})=e^{\ ^{3}\varpi },\ ^{3}\Phi (x^{k_{2}},v^{5}),
\\ 
\mbox{integr. functs:}\ h_{6}^{[0]}(x^{k_{2}}),\  \\ 
_{1}^{3}n_{k_{2}}(x^{i_{2}}),\ _{2}^{3}n_{k_{2}}(x^{i_{2}}); \\ 
\mbox{ Gener.  functs:}\underline{h}_{8}(x^{k_{3}},v^{8}), \\ 
\ ^{4}\underline{\Psi }(x^{k_{2}},v^{8})=e^{\ ^{4}\underline{\varpi }},\ ^{4}%
\underline{\Phi }(x^{k_{3}},v^{8}), \\ 
\mbox{integr. functs:}\ h_{8}^{[0]}(x^{k_{3}}),\  \\ 
_{1}^{4}n_{k_{3}}(x^{i_{3}}),\ _{2}^{4}n_{k_{3}}(x^{i_{4}}); \\ 
\mbox{\& nonlinear symmetries}%
\end{array}%
$ &  & $%
\begin{array}{c}
\ ((\ ^{2}\Psi )^{2})^{\ast }=-\int dy^{3}\ _{2}\widehat{\Upsilon }h_{4}^{\
\ast }, \\ 
(\ ^{2}\Phi )^{2}=-4\ _{2}\Lambda h_{4},\mbox{ see }(\ref{nonlinsymrex}), \\ 
h_{4}=h_{4}^{[0]}-(\ ^{2}\Phi )^{2}/4\ _{2}\Lambda ,h_{4}^{\ast }\neq 0,\
_{2}\Lambda \neq 0=const; \\ 
\\ 
\partial _{5}((\ ^{3}\Psi )^{2})=-\int dv^{5}\ _{3}\widehat{\Upsilon }%
\partial _{5}h_{6}^{\ }, \\ 
(\ ^{3}\Phi )^{2}=-4\ _{3}\Lambda h_{6}, \\ 
h_{6}=h_{6}^{[0]}-(\ ^{3}\Phi )^{2}/4\ _{3}\Lambda ,\partial _{5}h_{6}\neq
0,\ _{3}\Lambda \neq 0=const; \\ 
\\ 
\partial _{8}((\ ^{4}\underline{\Psi })^{2})=-\int dv^{8}\ _{4}\underline{%
\widehat{\Upsilon }}\partial _{8}\underline{h}_{7}^{\ }, \\ 
(\ ^{4}\underline{\Phi })^{2}=-4\ _{4}\underline{\Lambda }\underline{h}_{7},
\\ 
\underline{h}_{7}=h_{7}^{[0]}-(\ ^{4}\underline{\Phi })^{2}/4\ _{4}%
\underline{\Lambda },\partial _{8}\underline{h}_{7}\neq 0,\ _{4}\underline{%
\Lambda }\neq 0=const;%
\end{array}%
$ \\ \hline
Off-diag. solutions, $%
\begin{array}{c}
\mbox{d--metric} \\ 
\mbox{N-connec.}%
\end{array}%
$ &  & $%
\begin{tabular}{l}
$%
\begin{array}{c}
\ g_{i}=e^{\ \psi (x^{k})}\mbox{ as a solution of 2-d Poisson eqs. }\psi
^{\bullet \bullet }+\psi ^{\prime \prime }=2~\ _{1}\widehat{\Upsilon }; \\ 
h_{3}=-(\Psi ^{\ast })^{2}/4\ _{2}\widehat{\Upsilon }^{2}h_{4},\mbox{ see }(%
\ref{g3}),(\ref{g4}); \\ 
h_{4}=h_{4}^{[0]}-\int dy^{3}(\Psi ^{2})^{\ast }/4\ _{2}\widehat{\Upsilon }%
=h_{4}^{[0]}-\Phi ^{2}/4\ _{2}\Lambda ; \\ 
w_{i}=\partial _{i}\ \Psi /\ \partial _{3}\Psi =\partial _{i}\ \Psi ^{2}/\
\partial _{3}\Psi ^{2}|; \\ 
n_{k_{1}}=\ _{1}n_{k_{1}}+\ _{2}n_{k}\int dy^{3}(\Psi ^{\ast })^{2}/\ _{2}%
\widehat{\Upsilon }^{2}|h_{4}^{[0]}-\int dy^{3}(\Psi ^{2})^{\ast }/4\ _{2}%
\widehat{\Upsilon }^{2}|^{5/2};%
\end{array}%
$ \\ 
$%
\begin{array}{c}
h_{5}=-(\partial _{5}\ ^{3}\Psi )^{2}/4\ _{3}\widehat{\Upsilon }^{2}h_{6};
\\ 
h_{6}=h_{6}^{[0]}-\int dv^{5}\partial _{5}((\ ^{3}\Psi )^{2})/4\ _{3}%
\widehat{\Upsilon }=h_{6}^{[0]}-(\ ^{3}\Phi )^{2}/4\ _{3}\Lambda ; \\ 
w_{i_{2}}=\partial _{i_{2}}(\ ^{3}\Psi )/\ \partial _{5}(\ ^{3}\Psi
)=\partial _{i_{2}}(\ ^{3}\Psi )^{2}/\ \partial _{5}(\ ^{3}\Psi )^{2}|; \\ 
n_{k_{2}}=\ _{1}n_{k_{2}}+\ _{2}n_{k_{2}}\int dv^{5}(\partial _{5}\ ^{3}\Psi
)^{2}/\ _{2}\widehat{\Upsilon }^{2}|h_{6}^{[0]}-\int dv^{5}\partial _{5}((\
^{3}\Psi )^{2})/4\ _{3}\widehat{\Upsilon }^{2}|^{5/2};%
\end{array}%
$ \\ 
$%
\begin{array}{c}
\underline{h}_{7}=\underline{h}_{7}^{[0]}-\int dv^{8}\partial _{8}((\ ^{4}%
\underline{\Psi })^{2})/4\ _{4}\underline{\widehat{\Upsilon }}=\underline{h}%
_{7}^{[0]}-(\ ^{4}\underline{\Phi })^{2}/4\ _{4}\underline{\Lambda }; \\ 
\underline{h}_{8}=-(\partial _{8}\ ^{4}\underline{\Psi })^{2}/4\ _{4}%
\underline{\widehat{\Upsilon }}^{2}\underline{h}_{7}; \\ 
n_{k_{3}}=\ _{1}n_{k_{3}}+\ _{2}n_{k_{3}}\int dv^{8}(\partial _{8}\ ^{4}%
\underline{\Psi })^{2}/\ _{4}\underline{\widehat{\Upsilon }}^{2}|\underline{h%
}_{7}^{[0]}-\int dv^{8}\partial _{8}((\ ^{4}\underline{\Psi })^{2})/4\ _{4}%
\underline{\widehat{\Upsilon }}^{2}|^{5/2}; \\ 
w_{i_{3}}=\partial _{i_{3}}(\ ^{4}\underline{\Psi })/\ \partial _{8}(\ ^{4}%
\underline{\Psi })=\partial _{i_{3}}(\ ^{4}\underline{\Psi })^{2}/\ \partial
_{8}(\ ^{4}\underline{\Psi })^{2}|.%
\end{array}%
$%
\end{tabular}%
$ \\ \hline\hline
\end{tabular}%
\end{eqnarray*}%
}Other types of quasi-stationary and velocity rainbow solutions can be
constructed using nonlinear transforms of generating functions,
gravitational polarizations and constraints to LC-configurations. All
nonhlonomic geometric constructions involve respective abstract geometric
proofs and modifications/ generalizations of formulas.

\subsubsection{Locally anisotropic cosmological solutions with phase space
velocity configurations}

Such cosmological models are 8-d versions of (\ref{lcs10d}) derived
following the AFCDM as in Table 6 but redefined on velocity phase spaces.
Respective classes of generic off-diagonal s-metrics are constructed
following the steps outlined below in Table 10.

{\scriptsize 
\begin{eqnarray*}
&&%
\begin{tabular}{l}
\hline\hline
\begin{tabular}{lll}
& {\large \textsf{Table 10:\ Off-diagonal cosmological spacetimes with space
velocity configurations}} &  \\ 
& Exact solutions of $\widehat{\mathbf{R}}_{\mu _{s}\nu _{s}}=\mathbf{%
\Upsilon }_{\mu _{s}\nu _{s}}$ (\ref{cdeq1}) on $TV$ transformed into a
shall system of nonlinear PDEs (\ref{eq1})-(\ref{e2c}) & 
\end{tabular}
\\ 
\end{tabular}
\\
&&%
\begin{tabular}{lll}
\hline\hline
&  &  \\ 
$%
\begin{array}{c}
\mbox{d-metric ansatz with} \\ 
\mbox{Killing symmetry }\partial _{4}=\partial _{t},\partial _{8}%
\end{array}%
$ &  & $%
\begin{array}{c}
ds^{2}=g_{i_{1}}(x^{k_{1}})(dx^{i_{1}})^{2}+\underline{g}%
_{a_{2}}(x^{k_{1}},t)(dy^{a_{2}}+\underline{N}%
_{i_{1}}^{a_{2}}(x^{k_{1}},t)dx^{i_{1}})^{2} \\ 
+g_{a_{3}}(x^{k_{2}},v^{5})(dy^{a_{3}}+N_{i_{2}}^{a_{3}}(x^{k_{2}},v^{5})dx^{i_{2}})^{2}
\\ 
+g_{a_{4}}(x^{k_{3}},v^{7})(dy^{a_{4}}+N_{i_{3}}^{a_{4}}(x^{k_{3}},v^{7})dx^{i_{3}})^{2},%
\mbox{ for }g_{i_{1}}=e^{\psi {(x}^{k_{1}}{)}}, \\ 
\underline{g}_{a_{2}}=\underline{h}_{a_{2}}(x^{k_{1}},t),\underline{N}%
_{i_{1}}^{3}=\ ^{2}\underline{n}_{i_{1}}=\underline{n}_{i_{1}}(x^{k_{1}},t),%
\underline{N}_{i_{1}}^{4}=\ ^{2}\underline{w}_{i_{1}}=\underline{w}%
_{i_{1}}(x^{k_{1}},t), \\ 
g_{a_{3}}=h_{a_{3}}(x^{k_{2}},v^{5}),N_{i_{2}}^{5}=\
^{3}w_{i_{2}}=w_{i_{2}}(x^{k_{2}},v^{5}),N_{i_{2}}^{6}=\
^{3}n_{i_{2}}=n_{i_{2}}(x^{k_{2}},v^{5}), \\ 
g_{a_{4}}=h_{a_{4}}(x^{k_{3}},v^{7}),N_{i_{3}}^{7}=\
^{4}w_{i_{3}}=w_{i_{3}}(x^{k_{3}},v^{7}),N_{i_{3}}^{8}=\
^{4}n_{i_{3}}=n_{i_{3}}(x^{k_{3}},v^{7}),%
\end{array}%
$ \\ 
Effective matter sources &  & $\mathbf{\Upsilon }_{\ \nu _{s}}^{\mu _{s}}=[\
_{1}\widehat{\Upsilon }({x}^{k_{1}})\delta _{j_{1}}^{i_{1}},\ _{2}\underline{%
\widehat{\Upsilon }}({x}^{k_{1}},t)\delta _{b_{2}}^{a_{2}},\ _{3}\widehat{%
\Upsilon }({x}^{k_{2}},v^{5})\delta _{b_{3}}^{a_{3}},_{4}\widehat{\Upsilon }(%
{x}^{k_{3}},v^{7})\delta _{b_{4}}^{a_{4}},],$ \\ \hline
Nonlinear PDEs (\ref{eq1})-(\ref{e2c}) &  & $%
\begin{tabular}{lll}
$%
\begin{array}{c}
\psi ^{\bullet \bullet }+\psi ^{\prime \prime }=2\ \ _{1}\widehat{\Upsilon };
\\ 
\ ^{2}\underline{\varpi }^{\diamond }\ \underline{h}_{3}^{\diamond }=2%
\underline{h}_{3}\underline{h}_{4}\ _{2}\widehat{\Upsilon }; \\ 
\ ^{2}\underline{n}_{k_{1}}^{\diamond \diamond }+\ ^{2}\underline{\gamma }\
^{2}\underline{n}_{k_{1}}^{\diamond }=0; \\ 
\ ^{2}\underline{\beta }\ ^{2}\underline{w}_{i_{1}}-\ ^{2}\underline{\alpha }%
_{i_{1}}=0;%
\end{array}%
$ &  & $%
\begin{array}{c}
\ ^{2}\underline{\varpi }{=\ln |\partial _{4}\underline{{h}}_{4}/\sqrt{|%
\underline{h}_{3}\underline{h}_{4}|}|,} \\ 
\ ^{2}\underline{\alpha }_{i_{1}}=(\partial _{4}\underline{h}_{3})\
(\partial _{i_{1}}\ ^{2}\underline{\varpi }), \\ 
\ ^{2}\underline{\beta }=(\partial _{4}\underline{h}_{4})\ (\partial _{3}\
^{2}\underline{\varpi }),\  \\ 
\ \ ^{2}\underline{\gamma }=\partial _{4}\left( \ln |\underline{h}%
_{3}|^{3/2}/|\underline{h}_{4}|\right) , \\ 
\partial _{1}q=q^{\bullet },\partial _{2}q=q^{\prime },\partial
_{4}q=\partial _{t}q=q^{\diamond }%
\end{array}%
$ \\ 
$%
\begin{array}{c}
\partial _{5}(\ ^{3}\varpi )\ \partial _{5}h_{6}=2h_{5}h_{6}\ _{3}\widehat{%
\Upsilon }; \\ 
\ ^{3}\beta \ ^{3}w_{i_{2}}-\ ^{3}\alpha _{i_{2}}=0; \\ 
\partial _{5}(\partial _{5}\ ^{3}n_{k_{2}})+\ ^{3}\gamma \partial _{5}(\
^{3}n_{k_{2}})=0;%
\end{array}%
$ &  & $%
\begin{array}{c}
\\ 
\ ^{3}\varpi {=\ln |\partial _{5}h_{6}/\sqrt{|h_{5}h_{6}|}|,} \\ 
\ ^{3}\alpha _{i_{2}}=(\partial _{5}h_{6})\ (\partial _{i_{2}}\ ^{3}\varpi ),
\\ 
\ ^{3}\beta =(\partial _{5}h_{6})\ (\partial _{5}\ ^{3}\varpi ),\  \\ 
\ \ ^{3}\gamma =\partial _{5}\left( \ln |h_{6}|^{3/2}/|h_{5}|\right) ,%
\end{array}%
$ \\ 
$%
\begin{array}{c}
\partial _{7}(\ ^{4}\varpi )\ \partial _{7}h_{8}=2h_{7}h_{8}\ _{4}\widehat{%
\Upsilon }; \\ 
\ ^{4}\beta \ ^{4}w_{i_{3}}-\ ^{4}\alpha _{i_{3}}=0; \\ 
\partial _{7}(\partial _{7}\ ^{4}n_{k_{3}})+\ ^{4}\gamma \partial _{7}(\
^{4}n_{k_{3}})=0;%
\end{array}%
$ &  & $%
\begin{array}{c}
\\ 
\ ^{4}\varpi {=\ln |\partial _{7}h_{8}/\sqrt{|h_{7}h_{8}|}|,} \\ 
\ ^{4}\alpha _{i}=(\partial _{7}h_{8})\ (\partial _{i}\ ^{4}\varpi ), \\ 
\ ^{4}\beta =(\partial _{7}h_{8})\ (\partial _{7}\ ^{4}\varpi ),\  \\ 
\ \ ^{4}\gamma =\partial _{7}\left( \ln |h_{8}|^{3/2}/|h_{7}|\right) ,%
\end{array}%
$%
\end{tabular}%
$ \\ \hline
$%
\begin{array}{c}
\mbox{ Gener.  functs:}\ \underline{h}_{4}(x^{k_{1}},t), \\ 
\ ^{2}\underline{\Psi }(x^{k_{1}},t)=e^{\ ^{2}\underline{\varpi }},\ ^{2}%
\underline{\Phi }(x^{k_{1}},t), \\ 
\mbox{integr. functs:}\ \underline{h}_{3}^{[0]}(x^{k_{1}}),\  \\ 
_{1}\underline{n}_{k_{1}}(x^{i_{1}}),\ _{2}\underline{n}_{k_{1}}(x^{i_{1}});
\\ 
\mbox{ Gener.  functs:}h_{5}(x^{k_{2}},v^{5}), \\ 
\ ^{3}\Psi (x^{k_{2}},v^{5})=e^{\ ^{3}\varpi },\ ^{3}\Phi (x^{k_{2}},v^{5}),
\\ 
\mbox{integr. functs:}\ h_{6}^{[0]}(x^{k_{2}}),\  \\ 
_{1}^{3}n_{k_{2}}(x^{i_{2}}),\ _{2}^{3}n_{k_{2}}(x^{i_{2}}); \\ 
\mbox{ Gener.  functs:}h_{7}(x^{k_{3}},v^{7}), \\ 
\ ^{4}\Psi (x^{k_{2}},v^{7})=e^{\ ^{4}\varpi },\ ^{4}\Phi (x^{k_{3}},v^{7}),
\\ 
\mbox{integr. functs:}\ h_{8}^{[0]}(x^{k_{3}}),\  \\ 
_{1}^{4}n_{k_{3}}(x^{i_{3}}),\ _{2}^{4}n_{k_{3}}(x^{i_{4}}); \\ 
\mbox{\& nonlinear symmetries}%
\end{array}%
$ &  & $%
\begin{array}{c}
\ ((\ ^{2}\underline{\Psi })^{2})^{\diamond }=-\int dt\ _{2}\widehat{%
\underline{\Upsilon }}\underline{h}_{3}^{\ \diamond }, \\ 
(\ ^{2}\underline{\Phi })^{2}=-4\ _{2}\underline{\Lambda }\underline{h}_{3},
\\ 
h_{3}=h_{3}^{[0]}-(\ ^{2}\underline{\Phi })^{2}/4\ _{2}\underline{\Lambda },%
\underline{h}_{3}^{\diamond }\neq 0,\ _{2}\underline{\Lambda }\neq 0=const;
\\ 
\\ 
\partial _{5}((\ ^{3}\Psi )^{2})=-\int dv^{5}\ _{3}\widehat{\Upsilon }%
\partial _{5}h_{6}^{\ }, \\ 
(\ ^{3}\Phi )^{2}=-4\ _{3}\Lambda h_{6}, \\ 
h_{6}=h_{6}^{[0]}-(\ ^{3}\Phi )^{2}/4\ _{3}\Lambda ,\partial _{5}h_{6}\neq
0,\ _{3}\Lambda \neq 0=const; \\ 
\\ 
\partial _{7}((\ ^{4}\Psi )^{2})=-\int dv^{7}\ _{4}\widehat{\Upsilon }%
\partial _{7}h_{8}^{\ }, \\ 
(\ ^{4}\Phi )^{2}=-4\ _{4}\Lambda h_{8}, \\ 
h_{8}=h_{8}^{[0]}-(\ ^{4}\Phi )^{2}/4\ _{4}\Lambda ,\partial _{7}h_{8}\neq
0,\ _{4}\Lambda \neq 0=const;%
\end{array}%
$ \\ \hline
Off-diag. solutions, $%
\begin{array}{c}
\mbox{d--metric} \\ 
\mbox{N-connec.}%
\end{array}%
$ &  & $%
\begin{tabular}{l}
$%
\begin{array}{c}
\ g_{i}=e^{\ \psi (x^{k})}\mbox{ as a solution of 2-d Poisson eqs. }\psi
^{\bullet \bullet }+\psi ^{\prime \prime }=2~\ _{1}\widehat{\Upsilon }; \\ 
\underline{h}_{4}=-(\underline{\Psi }^{\diamond })^{2}/4\ _{2}\widehat{%
\underline{\Upsilon }}^{2}\underline{h}_{3}; \\ 
\underline{h}_{3}=\underline{h}_{3}^{[0]}-\int dt(\underline{\Psi }%
^{2})^{\diamond }/4\ _{2}\widehat{\underline{\Upsilon }}=\underline{h}%
_{3}^{[0]}-\underline{\Phi }^{2}/4\ _{2}\underline{\Lambda }; \\ 
\underline{w}_{i_{1}}=\partial _{i_{1}}\ \underline{\Psi }/\ \partial 
\underline{\Psi }^{\diamond }=\partial _{i_{1}}\ \underline{\Psi }^{2}/\
\partial _{t}\underline{\Psi }^{2}|; \\ 
\underline{n}_{k_{1}}=\ _{1}n_{k_{1}}+\ _{2}n_{k_{1}}\int dt(\underline{\Psi 
}^{\diamond })^{2}/\ _{2}\widehat{\underline{\Upsilon }}^{2}|\underline{h}%
_{3}^{[0]}-\int dt(\underline{\Psi }^{2})^{\diamond }/4\ _{2}\widehat{%
\underline{\Upsilon }}^{2}|^{5/2};%
\end{array}%
$ \\ 
$%
\begin{array}{c}
h_{5}=-(\partial _{5}\ ^{3}\Psi )^{2}/4\ _{3}\widehat{\Upsilon }^{2}h_{6};
\\ 
h_{6}=h_{6}^{[0]}-\int dv^{5}\partial _{5}((\ ^{3}\Psi )^{2})/4\ _{3}%
\widehat{\Upsilon }=h_{6}^{[0]}-(\ ^{3}\Phi )^{2}/4\ _{3}\Lambda ; \\ 
w_{i_{2}}=\partial _{i_{2}}(\ ^{3}\Psi )/\ \partial _{5}(\ ^{3}\Psi
)=\partial _{i_{2}}(\ ^{3}\Psi )^{2}/\ \partial _{5}(\ ^{3}\Psi )^{2}|; \\ 
n_{k_{2}}=\ _{1}n_{k_{2}}+\ _{2}n_{k_{2}}\int dv^{5}(\partial _{5}\ ^{3}\Psi
)^{2}/\ _{2}\widehat{\Upsilon }^{2}|h_{6}^{[0]}-\int dv^{5}\partial _{5}((\
^{3}\Psi )^{2})/4\ _{3}\widehat{\Upsilon }^{2}|^{5/2};%
\end{array}%
$ \\ 
$%
\begin{array}{c}
h_{7}=-(\partial _{7}\ ^{3}\Psi )^{2}/4\ _{4}\widehat{\Upsilon }^{2}h_{8};
\\ 
h_{8}=h_{8}^{[0]}-\int dv^{7}\partial _{9}((\ ^{4}\Psi )^{2})/4\ _{4}%
\widehat{\Upsilon }=h_{8}^{[0]}-(\ ^{4}\Phi )^{2}/4\ _{4}\Lambda ; \\ 
w_{i_{3}}=\partial _{i_{3}}(\ ^{4}\Psi )/\ \partial _{7}(\ ^{4}\Psi
)=\partial _{i_{3}}(\ ^{4}\Psi )^{2}/\ \partial _{7}(\ ^{4}\Psi )^{2}|; \\ 
n_{k_{3}}=\ _{1}n_{k_{3}}+\ _{2}n_{k_{3}}\int dv^{7}(\partial _{7}\ ^{4}\Psi
)^{2}/\ _{4}\widehat{\Upsilon }^{2}|h_{8}^{[0]}-\int dv^{7}\partial _{7}((\
^{4}\Psi )^{2})/4\ _{4}\widehat{\Upsilon }^{2}|^{5/2}.%
\end{array}%
$%
\end{tabular}%
$ \\ \hline\hline
\end{tabular}%
\end{eqnarray*}%
}As an example of 8-d quasi-stationary quadratic element with $v^{8}=const$
on $T\mathbf{V},$ we provide 
\begin{eqnarray}
d\widehat{s}_{[8d]}^{2} &=&\widehat{g}_{\alpha _{s}\beta
_{s}}(x^{k},t,y^{5},y^{7};\underline{h}_{3},h_{6},h_{8,};\ _{1}\widehat{%
\Upsilon },\ _{2}\underline{\widehat{\Upsilon }},\ _{3}\widehat{\Upsilon },\
_{4}\widehat{\Upsilon };\ _{1}\Lambda ,\ _{2}\underline{\Lambda },\
_{3}\Lambda ,\ _{4}\Lambda )du^{\alpha _{s}}du^{\beta _{s}}  \label{lc8d7} \\
&=&e^{\psi (x^{k},\ _{s}\widehat{\Upsilon })}[(dx^{1})^{2}+(dx^{2})^{2}]+%
\underline{h}_{3}[dy^{3}+(\ _{1}n_{k_{1}}+4\ _{2}n_{k_{1}}\int dt\frac{(%
\underline{h}_{3}{}^{\diamond })^{2}}{|\int dt\ _{2}\underline{\Upsilon }%
\underline{h}_{3}{}^{\diamond }|(\underline{h}_{3})^{5/2}})dx^{k_{1}}] 
\notag \\
&&-\frac{(\underline{h}_{3}{}^{\diamond })^{2}}{|\int dt\ _{2}\underline{%
\Upsilon }\underline{h}_{3}{}^{\diamond }|\ \overline{h}_{3}}[dt+\frac{%
\partial _{i}(\int dt\ _{2}\underline{\Upsilon }\ \underline{h}%
_{3}{}^{\diamond }])}{\ \ _{2}\underline{\Upsilon }\ \underline{h}%
_{3}{}^{\diamond }}dx^{i}]+  \notag \\
&&\frac{(\partial _{5}h_{6})^{2}}{|\int dy^{5}\partial _{5}[\ _{3}\widehat{%
\Upsilon }h_{6}]|\ h_{6}}\{dv^{5}+\frac{\partial _{i_{2}}[\int dv^{5}(\ _{3}%
\widehat{\Upsilon })\ \partial _{5}h_{6}]}{\ _{3}\widehat{\Upsilon }\
\partial _{5}h_{6}}dx^{i_{2}}\}^{2}+  \notag \\
&&h_{6}\{dv^{5}+[\ _{1}n_{k_{2}}+\ _{2}n_{k_{2}}\int dv^{5}\frac{(\partial
_{5}h_{6})^{2}}{|\int dy^{5}\partial _{5}[\ _{3}\widehat{\Upsilon }h_{6}]|\
(h_{6})^{5/2}}]dx^{k_{2}}\}+  \notag \\
&&\frac{(\partial _{7}h_{8})^{2}}{|\int dv^{7}\partial _{7}[\ _{4}\widehat{%
\Upsilon }h_{8}]|\ h_{8}}\{dv^{7}+\frac{\partial _{i_{3}}[\int dv^{7}(\ _{4}%
\widehat{\Upsilon })\ \partial _{7}h_{8}]}{\ _{4}\widehat{\Upsilon }\
\partial _{7}h_{8}}dx^{i_{3}}\}^{2}+  \notag \\
&&h_{8}\{dv^{8}+[\ _{1}n_{k_{3}}+\ _{2}n_{k_{3}}\int dv^{7}\frac{(\partial
_{7}h_{8})^{2}}{|\int dv^{7}\partial _{7}[\ _{4}\widehat{\Upsilon }h_{8}]|\
(h_{8})^{5/2}}]dx^{k_{3}}\}.  \notag
\end{eqnarray}

Similar classes of locally cosmological phase velocity space solutions can
be derived for the same Killing symmetries on $\partial _{3}$ and $\partial
_{8}$ using respective nonlinear symmetries and generating and integration
functions.

\subsubsection{Locally anisotropic cosmological solutions with phase space
rainbow symmetries}

The locally anisotropic cosmological models from previous Table 10 can be
re-defined by phase space rainbow symmetries with the shells $s=3,4$ part as
in Table 9. The shells $s=1,2$ parts are as in Table 6 when the AFCDM is
redefined on $T\mathbf{V.}$ The procedure of constructing such classes of
solutions with Killing symmetries on $\partial _{3}$ and $\partial _{7}$ is
summarized below in Table 11. As an example of 8-d locally anisotropic
cosmological quadratic element with $v^{7}=const$ on $T\mathbf{V},$ and
defining rainbow configurations as for $s=3,4$ in (\ref{qst8d8}), we provide 
\begin{eqnarray}
d\widehat{s}_{[8d]}^{2} &=&\widehat{g}_{\alpha _{s}\beta
_{s}}(x^{k},t,v^{5},v^{8};\underline{h}_{3},h_{6},\underline{h}_{7,};\ _{1}%
\widehat{\Upsilon },\ _{2}\underline{\widehat{\Upsilon }},\ _{3}\widehat{%
\Upsilon },\ _{4}\underline{\widehat{\Upsilon }};\ _{1}\Lambda ,\ _{2}%
\underline{\Lambda },\ _{3}\Lambda ,\ _{4}\underline{\Lambda })du^{\alpha
_{s}}du^{\beta _{s}}  \label{lc8d8} \\
&=&e^{\psi (x^{k},\ _{s}\widehat{\Upsilon })}[(dx^{1})^{2}+(dx^{2})^{2}]+%
\underline{h}_{3}[dy^{3}+(\ _{1}n_{k_{1}}+4\ _{2}n_{k_{1}}\int dt\frac{(%
\underline{h}_{3}{}^{\diamond })^{2}}{|\int dt\ _{2}\underline{\Upsilon }%
\underline{h}_{3}{}^{\diamond }|(\underline{h}_{3})^{5/2}})dx^{k_{1}}] 
\notag \\
&&-\frac{(\underline{h}_{3}{}^{\diamond })^{2}}{|\int dt\ _{2}\underline{%
\Upsilon }\underline{h}_{3}{}^{\diamond }|\ \overline{h}_{3}}[dt+\frac{%
\partial _{i}(\int dt\ _{2}\underline{\Upsilon }\ \underline{h}%
_{3}{}^{\diamond }])}{\ \ _{2}\underline{\Upsilon }\ \underline{h}%
_{3}{}^{\diamond }}dx^{i}]+  \notag \\
&&\frac{(\partial _{5}h_{6})^{2}}{|\int dv^{5}\partial _{5}[\ _{3}\widehat{%
\Upsilon }h_{6}]|\ h_{6}}\{dv^{5}+\frac{\partial _{i_{2}}[\int dv^{5}(\ _{3}%
\widehat{\Upsilon })\ \partial _{5}h_{6}]}{\ _{3}\widehat{\Upsilon }\
\partial _{5}h_{6}}dx^{i_{2}}\}^{2}+  \notag \\
&&h_{6}\{dv^{5}+[\ _{1}n_{k_{2}}+\ _{2}n_{k_{2}}\int dv^{5}\frac{(\partial
_{5}h_{6})^{2}}{|\int dv^{5}\partial _{5}[\ _{3}\widehat{\Upsilon }h_{6}]|\
(h_{6})^{5/2}}]dx^{k_{2}}\}+  \notag \\
&&\underline{h}_{7}\{dv^{7}+[\ _{1}n_{k_{3}}+\ _{2}n_{k_{3}}\int dv^{8}\frac{%
(\partial _{8}\underline{h}_{7})^{2}}{|\int dv^{8}\partial _{8}[\ _{4}%
\underline{\widehat{\Upsilon }}h_{7}]|\ (\underline{h}_{7})^{5/2}}%
]dx^{k_{3}}\}+  \notag \\
&&\frac{(\partial _{8}\underline{h}_{7})^{2}}{|\int dv^{8}\partial _{8}[\
_{4}\underline{\widehat{\Upsilon }}\underline{h}_{7}]|\ \underline{h}_{7}}%
\{dv^{8}+\frac{\partial _{i_{3}}[\int dv^{8}(\ _{4}\underline{\widehat{%
\Upsilon }})\ \partial _{8}\underline{h}_{7}]}{\ _{4}\underline{\widehat{%
\Upsilon }}\ \partial _{8}\underline{h}_{7}}dx^{i_{3}}\}^{2}.  \notag
\end{eqnarray}%
The AFCDM modifications for generating such solutions are described as
follow:

{\scriptsize 
\begin{eqnarray*}
&&%
\begin{tabular}{l}
\hline\hline
\begin{tabular}{lll}
& {\large \textsf{Table 11:\ Off-diagonal cosmological spacetimes with
velocity rainbow symmetries}} &  \\ 
& Exact solutions of $\widehat{\mathbf{R}}_{\mu _{s}\nu _{s}}=\mathbf{%
\Upsilon }_{\mu _{s}\nu _{s}}$ (\ref{cdeq1}) on $TV$ transformed into a
shall system of nonlinear PDEs (\ref{eq1})-(\ref{e2c}) & 
\end{tabular}
\\ 
\end{tabular}
\\
&&%
\begin{tabular}{lll}
\hline\hline
&  &  \\ 
$%
\begin{array}{c}
\mbox{d-metric ansatz with} \\ 
\mbox{Killing symmetry }\partial _{3}=\partial _{t},\partial _{8}%
\end{array}%
$ &  & $%
\begin{array}{c}
ds^{2}=g_{i_{1}}(x^{k_{1}})(dx^{i_{1}})^{2}+\underline{g}%
_{a_{2}}(x^{k_{1}},t)(dy^{a_{2}}+\underline{N}%
_{i_{1}}^{a_{2}}(x^{k_{1}},t)dx^{i_{1}})^{2} \\ 
+g_{a_{3}}(x^{k_{2}},v^{5})(dy^{a_{3}}+N_{i_{2}}^{a_{3}}(x^{k_{2}},v^{5})dx^{i_{2}})^{2}
\\ 
+\underline{g}_{a_{4}}(x^{k_{3}},v^{8})(dy^{a_{4}}+\underline{N}%
_{i_{3}}^{a_{4}}(x^{k_{3}},v^{8})dx^{i_{3}})^{2},\mbox{ for }%
g_{i_{1}}=e^{\psi {(x}^{k_{1}}{)}}, \\ 
\underline{g}_{a_{2}}=\underline{h}_{a_{2}}(x^{k_{1}},t),\underline{N}%
_{i_{1}}^{3}=\ ^{2}\underline{n}_{i_{1}}=\underline{n}_{i_{1}}(x^{k_{1}},t),%
\underline{N}_{i_{1}}^{4}=\ ^{2}\underline{w}_{i_{1}}=\underline{w}%
_{i_{1}}(x^{k_{1}},t), \\ 
g_{a_{3}}=h_{a_{3}}(x^{k_{2}},v^{5}),N_{i_{2}}^{5}=\
^{3}w_{i_{2}}=w_{i_{2}}(x^{k_{2}},v^{5}),N_{i_{2}}^{6}=\
^{3}n_{i_{2}}=n_{i_{2}}(x^{k_{2}},v^{5}), \\ 
\underline{g}_{a_{4}}=\underline{h}_{a_{4}}(x^{k_{3}},v^{8}),\underline{N}%
_{i_{3}}^{7}=\ ^{4}\underline{n}_{i_{3}}=\underline{n}%
_{i_{3}}(x^{k_{3}},v^{8}),\underline{N}_{i_{3}}^{8}=\ ^{4}\underline{w}%
_{i_{3}}=\underline{w}_{i_{3}}(x^{k_{3}},v^{8}),%
\end{array}%
$ \\ 
Effective matter sources &  & $\mathbf{\Upsilon }_{\ \nu _{s}}^{\mu _{s}}=[\
_{1}\widehat{\Upsilon }({x}^{k_{1}})\delta _{j_{1}}^{i_{1}},\ _{2}\underline{%
\widehat{\Upsilon }}({x}^{k_{1}},t)\delta _{b_{2}}^{a_{2}},\ _{3}\widehat{%
\Upsilon }({x}^{k_{2}},v^{5})\delta _{b_{3}}^{a_{3}},\ _{4}\underline{%
\widehat{\Upsilon }}({x}^{k_{3}},v^{8})\delta _{b_{4}}^{a_{4}},],$ \\ \hline
Nonlinear PDEs (\ref{eq1})-(\ref{e2c}) &  & $%
\begin{tabular}{lll}
$%
\begin{array}{c}
\psi ^{\bullet \bullet }+\psi ^{\prime \prime }=2\ \ _{1}\widehat{\Upsilon };
\\ 
\ ^{2}\underline{\varpi }^{\diamond }\ \underline{h}_{3}^{\diamond }=2%
\underline{h}_{3}\underline{h}_{4}\ _{2}\widehat{\Upsilon }; \\ 
\ ^{2}\underline{n}_{k_{1}}^{\diamond \diamond }+\ ^{2}\underline{\gamma }\
^{2}\underline{n}_{k_{1}}^{\diamond }=0; \\ 
\ ^{2}\underline{\beta }\ ^{2}\underline{w}_{i_{1}}-\ ^{2}\underline{\alpha }%
_{i_{1}}=0;%
\end{array}%
$ &  & $%
\begin{array}{c}
\ ^{2}\underline{\varpi }{=\ln |\partial _{4}\underline{{h}}_{4}/\sqrt{|%
\underline{h}_{3}\underline{h}_{4}|}|,} \\ 
\ ^{2}\underline{\alpha }_{i_{1}}=(\partial _{4}\underline{h}_{3})\
(\partial _{i_{1}}\ ^{2}\underline{\varpi }), \\ 
\ ^{2}\underline{\beta }=(\partial _{4}\underline{h}_{4})\ (\partial _{3}\
^{2}\underline{\varpi }),\  \\ 
\ \ ^{2}\underline{\gamma }=\partial _{4}\left( \ln |\underline{h}%
_{3}|^{3/2}/|\underline{h}_{4}|\right) , \\ 
\partial _{1}q=q^{\bullet },\partial _{2}q=q^{\prime },\partial
_{4}q=\partial _{t}q=q^{\diamond }%
\end{array}%
$ \\ 
$%
\begin{array}{c}
\partial _{5}(\ ^{3}\varpi )\ \partial _{5}h_{6}=2h_{5}h_{6}\ _{3}\widehat{%
\Upsilon }; \\ 
\ ^{3}\beta \ ^{3}w_{i_{2}}-\ ^{3}\alpha _{i_{2}}=0; \\ 
\partial _{5}(\partial _{5}\ ^{3}n_{k_{2}})+\ ^{3}\gamma \partial _{5}(\
^{3}n_{k_{2}})=0;%
\end{array}%
$ &  & $%
\begin{array}{c}
\\ 
\ ^{3}\varpi {=\ln |\partial _{5}h_{6}/\sqrt{|h_{5}h_{6}|}|,} \\ 
\ ^{3}\alpha _{i_{2}}=(\partial _{5}h_{6})\ (\partial _{i_{2}}\ ^{3}\varpi ),
\\ 
\ ^{3}\beta =(\partial _{5}h_{6})\ (\partial _{5}\ ^{3}\varpi ),\  \\ 
\ \ ^{3}\gamma =\partial _{5}\left( \ln |h_{6}|^{3/2}/|h_{5}|\right) ,%
\end{array}%
$ \\ 
$%
\begin{array}{c}
\partial _{8}(\ ^{4}\underline{\varpi })\ \partial _{8}\underline{h}_{7}=2%
\underline{h}_{7}\underline{h}_{8}\ _{4}\underline{\widehat{\Upsilon }}; \\ 
\partial _{8}(\partial _{8}\ ^{4}\underline{n}_{k_{3}})+\ ^{4}\underline{%
\gamma }\partial _{8}(\ ^{4}\underline{n}_{k_{3}})=0; \\ 
\ ^{4}\underline{\beta }\ ^{4}\underline{w}_{i_{3}}-\ ^{4}\underline{\alpha }%
_{i_{3}}=0;%
\end{array}%
$ &  & $%
\begin{array}{c}
\\ 
\ ^{4}\underline{\varpi }{=\ln |\partial _{8}\underline{h}_{7}/\sqrt{|%
\underline{h}_{7}\underline{h}_{8}|}|,} \\ 
\ ^{4}\underline{\alpha }_{i}=(\partial _{8}\underline{h}_{7})\ (\partial
_{i}\ ^{4}\underline{\varpi }), \\ 
\ ^{4}\underline{\beta }=(\partial _{8}\underline{h}_{7})\ (\partial _{8}\
^{4}\underline{\varpi }),\  \\ 
\ \ ^{4}\underline{\gamma }=\partial _{8}\left( \ln |\underline{h}%
_{7}|^{3/2}/|\underline{h}_{8}|\right) ,%
\end{array}%
$%
\end{tabular}%
$ \\ \hline
$%
\begin{array}{c}
\mbox{ Gener.  functs:}\ \underline{h}_{4}(x^{k_{1}},t), \\ 
\ ^{2}\underline{\Psi }(x^{k_{1}},t)=e^{\ ^{2}\underline{\varpi }},\ ^{2}%
\underline{\Phi }(x^{k_{1}},t), \\ 
\mbox{integr. functs:}\ \underline{h}_{3}^{[0]}(x^{k_{1}}),\  \\ 
_{1}\underline{n}_{k_{1}}(x^{i_{1}}),\ _{2}\underline{n}_{k_{1}}(x^{i_{1}});
\\ 
\mbox{ Gener.  functs:}h_{5}(x^{k_{2}},v^{5}), \\ 
\ ^{3}\Psi (x^{k_{2}},v^{5})=e^{\ ^{3}\varpi },\ ^{3}\Phi (x^{k_{2}},v^{5}),
\\ 
\mbox{integr. functs:}\ h_{6}^{[0]}(x^{k_{2}}),\  \\ 
_{1}^{3}n_{k_{2}}(x^{i_{2}}),\ _{2}^{3}n_{k_{2}}(x^{i_{2}}); \\ 
\mbox{ Gener.  functs:}h_{7}(x^{k_{3}},v^{7}), \\ 
\ ^{4}\underline{\Psi }(x^{k_{2}},v^{8})=e^{\ ^{4}\underline{\varpi }},\ ^{4}%
\underline{\Phi }(x^{k_{3}},v^{8}), \\ 
\mbox{integr. functs:}\ h_{8}^{[0]}(x^{k_{3}}),\  \\ 
_{1}^{4}n_{k_{3}}(x^{i_{3}}),\ _{2}^{4}n_{k_{3}}(x^{i_{4}}); \\ 
\mbox{\& nonlinear symmetries}%
\end{array}%
$ &  & $%
\begin{array}{c}
\ ((\ ^{2}\underline{\Psi })^{2})^{\diamond }=-\int dt\ _{2}\widehat{%
\underline{\Upsilon }}\underline{h}_{3}^{\ \diamond }, \\ 
(\ ^{2}\underline{\Phi })^{2}=-4\ _{2}\underline{\Lambda }\underline{h}_{3},
\\ 
h_{3}=h_{3}^{[0]}-(\ ^{2}\underline{\Phi })^{2}/4\ _{2}\underline{\Lambda },%
\underline{h}_{3}^{\diamond }\neq 0,\ _{2}\underline{\Lambda }\neq 0=const;
\\ 
\\ 
\partial _{5}((\ ^{3}\Psi )^{2})=-\int dv^{5}\ _{3}\widehat{\Upsilon }%
\partial _{5}h_{6}^{\ }, \\ 
(\ ^{3}\Phi )^{2}=-4\ _{3}\Lambda h_{6}, \\ 
h_{6}=h_{6}^{[0]}-(\ ^{3}\Phi )^{2}/4\ _{3}\Lambda ,\partial _{5}h_{6}\neq
0,\ _{3}\Lambda \neq 0=const; \\ 
\\ 
\partial _{8}((\ ^{4}\underline{\Psi })^{2})=-\int dv^{8}\ _{4}\underline{%
\widehat{\Upsilon }}\partial _{8}\underline{h}_{7}^{\ }, \\ 
(\ ^{4}\underline{\Phi })^{2}=-4\ _{4}\underline{\Lambda }\underline{h}_{7},
\\ 
\underline{h}_{7}=h_{7}^{[0]}-(\ ^{4}\underline{\Phi })^{2}/4\ _{4}%
\underline{\Lambda },\partial _{8}\underline{h}_{7}\neq 0,\ _{4}\underline{%
\Lambda }\neq 0=const;%
\end{array}%
$ \\ \hline
Off-diag. solutions, $%
\begin{array}{c}
\mbox{d--metric} \\ 
\mbox{N-connec.}%
\end{array}%
$ &  & $%
\begin{tabular}{l}
$%
\begin{array}{c}
\ g_{i}=e^{\ \psi (x^{k})}\mbox{ as a solution of 2-d Poisson eqs. }\psi
^{\bullet \bullet }+\psi ^{\prime \prime }=2~\ _{1}\widehat{\Upsilon }; \\ 
\underline{h}_{4}=-(\underline{\Psi }^{\diamond })^{2}/4\ _{2}\widehat{%
\underline{\Upsilon }}^{2}\underline{h}_{3}; \\ 
\underline{h}_{3}=\underline{h}_{3}^{[0]}-\int dt(\underline{\Psi }%
^{2})^{\diamond }/4\ _{2}\widehat{\underline{\Upsilon }}=\underline{h}%
_{3}^{[0]}-\underline{\Phi }^{2}/4\ _{2}\underline{\Lambda }; \\ 
\underline{w}_{i_{1}}=\partial _{i_{1}}\ \underline{\Psi }/\ \partial 
\underline{\Psi }^{\diamond }=\partial _{i_{1}}\ \underline{\Psi }^{2}/\
\partial _{t}\underline{\Psi }^{2}|; \\ 
\underline{n}_{k_{1}}=\ _{1}n_{k_{1}}+\ _{2}n_{k_{1}}\int dt(\underline{\Psi 
}^{\diamond })^{2}/\ _{2}\widehat{\underline{\Upsilon }}^{2}|\underline{h}%
_{3}^{[0]}-\int dt(\underline{\Psi }^{2})^{\diamond }/4\ _{2}\widehat{%
\underline{\Upsilon }}^{2}|^{5/2};%
\end{array}%
$ \\ 
$%
\begin{array}{c}
h_{5}=-(\partial _{5}\ ^{3}\Psi )^{2}/4\ _{3}\widehat{\Upsilon }^{2}h_{6};
\\ 
h_{6}=h_{6}^{[0]}-\int dv^{5}\partial _{5}((\ ^{3}\Psi )^{2})/4\ _{3}%
\widehat{\Upsilon }=h_{6}^{[0]}-(\ ^{3}\Phi )^{2}/4\ _{3}\Lambda ; \\ 
w_{i_{2}}=\partial _{i_{2}}(\ ^{3}\Psi )/\ \partial _{5}(\ ^{3}\Psi
)=\partial _{i_{2}}(\ ^{3}\Psi )^{2}/\ \partial _{5}(\ ^{3}\Psi )^{2}|; \\ 
n_{k_{2}}=\ _{1}n_{k_{2}}+\ _{2}n_{k_{2}}\int dv^{5}(\partial _{5}\ ^{3}\Psi
)^{2}/\ _{2}\widehat{\Upsilon }^{2}|h_{6}^{[0]}-\int dv^{5}\partial _{5}((\
^{3}\Psi )^{2})/4\ _{3}\widehat{\Upsilon }^{2}|^{5/2};%
\end{array}%
$ \\ 
$%
\begin{array}{c}
\underline{h}_{7}=\underline{h}_{7}^{[0]}-\int dv^{8}\partial _{8}((\ ^{4}%
\underline{\Psi })^{2})/4\ _{4}\underline{\widehat{\Upsilon }}=\underline{h}%
_{7}^{[0]}-(\ ^{4}\underline{\Phi })^{2}/4\ _{4}\underline{\Lambda }; \\ 
\underline{h}_{8}=-(\partial _{8}\ ^{4}\underline{\Psi })^{2}/4\ _{4}%
\underline{\widehat{\Upsilon }}^{2}\underline{h}_{7}; \\ 
n_{k_{3}}=\ _{1}n_{k_{3}}+\ _{2}n_{k_{3}}\int dv^{8}(\partial _{8}\ ^{4}%
\underline{\Psi })^{2}/\ _{4}\underline{\widehat{\Upsilon }}^{2}|\underline{h%
}_{7}^{[0]}-\int dv^{8}\partial _{8}((\ ^{4}\underline{\Psi })^{2})/4\ _{4}%
\underline{\widehat{\Upsilon }}^{2}|^{5/2}; \\ 
w_{i_{3}}=\partial _{i_{3}}(\ ^{4}\underline{\Psi })/\ \partial _{8}(\ ^{4}%
\underline{\Psi })=\partial _{i_{3}}(\ ^{4}\underline{\Psi })^{2}/\ \partial
_{8}(\ ^{4}\underline{\Psi })^{2}|.%
\end{array}%
$%
\end{tabular}%
$ \\ \hline\hline
\end{tabular}%
\end{eqnarray*}%
}Velocity rainbow s-metrics (\ref{lc8d8}) can be considered also for
Finsler-Lagrange spaces for respective generating functions. We can impose
homogeneity and other type conditions in order to define more special
classes of relativistic generalized Finsler geometries. Such models can be
redefined for momentum variables on cotangent Lorentz bundles as in next
subsection.

\subsection{Phase space momentum depending quasi-stationary and cosmological
solutions, tables 12-16}

\label{tab1216}A series of recent works on nonassociative geometric and
quantum information flows, nonassociative and noncommutative gravity and
Hamilton-Cartan geometry and gravity are elaborated on nonholonomic phases
spaces $\ $modeled on a cotangent Lorentz bundle, $\mathcal{M}=T^{\ast }%
\mathbf{V},$ see details and review of results in \cite%
{vmon3,vacaru18,bubuianu18a,bubuianu20,bubuianu19,partner01,partner02,partner03,partner04}%
. In section (\ref{sstoy}), we studied a 2+2 toy model with conventional 2-d
cofiber coordinates. We generalize those constructions on 8-d phase spaces
with conventional dyadic splitting (2+2)+(2+2), when the locall coordinates
on shells $s=3$ and $s=4$ are momentum type $p_{a}$ and the local
coordinates on the total space are labeled $\ ^{\shortmid }u=(x,p)=\{\
^{\shortmid }u^{\alpha }=(x^{i},p_{a})\}=\{\ ^{\shortmid }u^{\alpha
_{s}}=(x^{i_{1}},y^{a_{2}},p_{a_{3}},p_{a_{4}})\}$ for $\ ^{\shortmid
}p=p=(p_{a_{3}},p_{7},p_{8}=E),$ where $E$ is an energy type variable. For
mechanical like models on cotangent bundles, the momentum like variables $%
(p_{a_{3}},p_{a_{4}})$ can be related to velocity type variables $%
(v^{b_{3}},v^{b_{4}})$, considered in previous subsection, via certain
Legendre transforms. Here it should be noted that theories with momentum
like variables admit a respective almost symplectic formulation (in this
work, we omit such considerations which are important, for instance, in
deformation quantization).

The N-connection structure defining a nonholonomic dyadic splitting on $%
\mathcal{M}$ is defined as dual nonholonomic distribution (extending the
definition (\ref{cncon}) from 4-d to 8-d) 
\begin{equation}
\ _{s}^{\shortmid }\mathbf{N}:\ T_{s}^{\ast }\mathbf{V}=\ ^{1}hV\oplus \
^{2}vV\oplus \ ^{3}cV\oplus \ ^{4}cV.  \label{cncon8}
\end{equation}%
In local dual 8-d coordinate form, a N-connection (\ref{cncon8}) can be
written as $\ _{s}^{\shortmid }\mathbf{N}=\ ^{\shortmid }N_{i_{s-1}a_{s}}(\
_{s}^{\shortmid }u)dx^{i_{s-1}}\otimes \partial /\partial p_{a_{s}}.$ Such
N-coefficients define N--elongated (equivalently, N-adapted) local bases
(partial derivatives), $\ ^{\shortmid }\mathbf{e}_{\nu _{s}},$ and co-bases
(differentials), $\ ^{\shortmid }\mathbf{e}^{\mu _{s}},$ when 
\begin{eqnarray}
\ ^{\shortmid }\mathbf{e}_{\nu _{s}} &=&(\ ^{\shortmid }\mathbf{e}%
_{i_{s-1}},\ ^{\shortmid }e^{a_{s}})=(\ ^{\shortmid }\mathbf{e}%
_{i_{s-1}}=\partial /\partial x^{i_{s-1}}-\ \ ^{\shortmid
}N_{i_{s-1}b_{s}}(\ _{s}^{\shortmid }u)\partial /\partial p_{b_{s}},\ \
^{\shortmid }e^{a_{s}}=\ ^{\shortmid }\partial ^{a_{s}}=\partial /\partial
p_{a_{s}}),\mbox{ and  }  \label{naderd8} \\
\ ^{\shortmid }\mathbf{e}^{\mu _{s}} &=&(e^{i_{s-1}},\ ^{\shortmid }\mathbf{e%
}^{a_{s}})=(e^{i_{s-1}}=dx^{i_{s-1}},\ \ ^{\shortmid }\mathbf{e}%
_{a_{s}}=dp_{a_{s}}+\ ^{\shortmid }N_{i_{s-1}a_{s}}(\ _{s}^{\shortmid
}u)dx^{i_{s-1}}),  \label{nadifd8}
\end{eqnarray}

Any phase space metric$\ ^{\shortmid }\mathbf{g}$ on $\ ^{\shortmid }V$ can
be represented equivalently as a s--metric (s, from shell) $\
_{s}^{\shortmid }\mathbf{g}=(\ ^{1}h\ g,\ ^{2}v\ g,\ ^{3}c\ ^{\shortmid }g,\
^{4}c\ ^{\shortmid }g),$ when 
\begin{equation}
\ \ ^{\shortmid }\mathbf{g}=\ \ ^{\shortmid }g_{i_{s-1}j_{s-1}}(x,p)\
e^{i_{s-1}}\otimes e^{j_{s-1}}+\ ^{\shortmid }g^{a_{s}b_{s}}(x,p)\ \
^{\shortmid }\mathbf{e}_{a_{s}}\otimes \ ^{\shortmid }\mathbf{e}_{b_{s}},=\
^{\shortmid }\underline{g}_{\alpha _{s}\beta _{s}}(\ _{s}^{\shortmid }u)d\
^{\shortmid }u^{\alpha _{s}}\otimes d\ ^{\shortmid }u^{\beta _{s}},
\label{dmd8}
\end{equation}%
where $\ ^{1}h\ ^{\shortmid }g=\{\ ^{\shortmid }g_{i_{1}j_{1}}\},\ ^{2}v\
^{\shortmid }g=\{\ ^{\shortmid }g_{a_{2}b_{2}}\},\ ^{3}c\ ^{\shortmid }g=\{\
^{\shortmid }g^{a_{3}b_{3}}\}$ and $\ \ ^{4}c\ ^{\shortmid }g=\{\
^{\shortmid }g^{a_{4}b_{4}}\}.$

We can define on $\mathcal{M}$ a\textbf{\ d--connection} structure $\
_{s}^{\shortmid }\mathbf{D}=(\ ^{s-1}h\ ^{\shortmid }D,\ ^{s}c\ ^{\shortmid
}D)$ is a linear connection preserving under parallelism the N--connection
splitting (\ref{cncon8}), 
\begin{eqnarray}
\ _{s}^{\shortmid }\mathbf{D} &=&\{\ ^{\shortmid }\mathbf{\Gamma }_{\ \alpha
_{s}\beta _{s}}^{\gamma _{s}}=(\ ^{\shortmid }L_{j_{s-1}k_{s-1}}^{i_{s-1}},\
^{\shortmid }\acute{L}_{a_{s}\ k_{s-1}}^{\ b_{s}};\ ^{\shortmid }\acute{C}%
_{j_{s-1}}^{i_{s-1}\ c_{s}},\ ^{\shortmid }C_{a_{s}}^{\ b_{s}c_{s}})\},%
\mbox{
where }  \label{dcond8} \\
\ ^{s-1}h\ ^{\shortmid }D &=&(\ ^{\shortmid }L_{j_{s-1}k_{s-1}}^{i_{s-1}},\
^{\shortmid }\acute{L}_{a_{s}\ k_{s-1}}^{\ b_{s}})\mbox{ and }\ ^{s}c\
^{\shortmid }D=(\ ^{\shortmid }\acute{C}_{j_{s-1}}^{i_{s-1}\ c_{s}},\
^{\shortmid }C_{a_{s}}^{\ b_{s}c_{s}}).  \notag
\end{eqnarray}%
Tthe c-indices in such N-adapted formulas are inverse to v-indices in
N-adapted formulas for $T\mathbf{V}$ considered for previous subsection.
Using d-operator $\ _{s}^{\shortmid }\mathbf{D,}$ we can define respective
fundamental geometric s-objects as in higher dimension Lorentz manifolds, or
on their tangent bundles but with abstract definitions on $\ T_{s}^{\ast }%
\mathbf{V}:$ 
\begin{eqnarray*}
\ _{s}^{\shortmid }\mathcal{T}(\ _{s}^{\shortmid }\mathbf{X,}\
_{s}^{\shortmid }\mathbf{Y}):= &&\ _{s}^{\shortmid }\mathbf{D}_{\
_{s}^{\shortmid }\mathbf{X}}\ _{s}^{\shortmid }\mathbf{Y}-\ _{s}^{\shortmid }%
\mathbf{D}_{\ _{s}^{\shortmid }\mathbf{Y}}\ _{s}^{\shortmid }\mathbf{X}-[\
_{s}^{\shortmid }\mathbf{X,\ _{s}^{\shortmid }Y}],%
\mbox{ torsion s-tensor,
s-torsion}; \\
\ _{s}^{\shortmid }\mathcal{R}(\ _{s}^{\shortmid }\mathbf{X,}\
_{s}^{\shortmid }\mathbf{Y}):= &&\ _{s}^{\shortmid }\mathbf{D}_{\
_{s}^{\shortmid }\mathbf{X}}\ ^{\shortmid }\mathbf{D}_{\ _{s}^{\shortmid }%
\mathbf{Y}}-\ _{s}^{\shortmid }\mathbf{D}_{\ _{s}^{\shortmid }\mathbf{Y}}\
_{s}^{\shortmid }\mathbf{D}_{\ _{s}^{\shortmid }\mathbf{X}}-\
_{s}^{\shortmid }\mathbf{D}_{\mathbf{[}\ _{s}^{\shortmid }\mathbf{X,}\
_{s}^{\shortmid }\mathbf{Y]}},\mbox{ curvature s-tensor, s-curvature}; \\
\ _{s}^{\shortmid }\mathcal{Q}(\ ^{\shortmid }\mathbf{X}):= &&\
_{s}^{\shortmid }\mathbf{D}_{\ _{s}^{\shortmid }\mathbf{X}}\ _{s}^{\shortmid
}\mathbf{g,}\mbox{ nonmetricity s-fields, s-nonmetricity},
\end{eqnarray*}%
where d-vectors $\ _{s}^{\shortmid }\mathbf{X}$ and$\ _{s}^{\shortmid }%
\mathbf{Y,}$ and their duals as 1-forms, can be decomposed respectively to
N-linear frames (\ref{naderd8}) and (\ref{nadifd8}).

Considering geometric s-objects and formulas (\ref{cncon8})-(\ref{dcond8}),
we can re-define all geometric constructions and formulas for nonholonomic
manifolds $\mathbf{V}$ and tangent bundles $T\mathbf{V}$ on cotangent
bundles $T_{s}^{\ast }\mathbf{V,}$ with shell dyadic structure. For details
and applications in MGTs and geometric and quantum information flow
theories, we cite \cite%
{vacaru18,bubuianu18a,bubuianu20,bubuianu19,partner03,partner04}.

The nonholonomic canonical Einstein equations on 8-d phase spaces with
momentum variables can be defined and proven using sympolic re-definitions
of variables and geomeric d-objects in (\ref{cdeq1}) and (\ref{lccond1}), 
\begin{eqnarray}
\ ^{\shortmid }\widehat{\mathbf{R}}_{\ \ \beta _{s}}^{\alpha _{s}} &=&\
^{\shortmid }\widehat{\mathbf{\Upsilon }}_{\ \ \beta _{s}}^{\alpha _{s}},
\label{cdeq1c8} \\
\ ^{\shortmid }\widehat{\mathbf{T}}_{\ \alpha _{s}\beta _{s}}^{\gamma _{s}}
&=&0,\mbox{ for effective geneating sources }\ ^{\shortmid }\widehat{\mathbf{%
\Upsilon }}_{\ \ \beta _{s}}^{\alpha _{s}}=[\ _{1}^{\shortmid }\Upsilon
\delta _{\ \ j_{1}}^{i_{1}},\ _{2}^{\shortmid }\Upsilon \delta _{\ \
b_{2}}^{a_{2}},\ _{3}^{\shortmid }\Upsilon \delta _{\ \ b_{3}}^{a_{3}},\
_{4}^{\shortmid }\Upsilon \delta _{\ \ b_{4}}^{a_{4}}].  \label{lccond1c8}
\end{eqnarray}

The equations (\ref{cdeq1c}) and (\ref{lccond1c}) can be solved by generic
off-diagonal ansatz with a Killing vector on respective shells. For
instance, a phase space analog of a quasi-stationary s-metric of type (\ref%
{qst8d7}) but with momenum space like hypersurface and fixed $p_{8}=E_{0}$
is parameterized 
\begin{eqnarray}
\ ^{\shortmid }\mathbf{\hat{g}} &=&g_{i_{1}}(x^{k_{1}})dx^{i_{1}}\otimes
dx^{i_{1}}+g_{a_{2}}(x^{k_{1}},y^{b_{2}})\mathbf{e}^{a_{2}}\otimes \mathbf{e}%
^{a_{2}}+\ ^{\shortmid }h^{a_{3}}(x^{k_{2}},p_{b_{3}})\ ^{\shortmid }\mathbf{%
e}_{a_{3}}\otimes \ ^{\shortmid }\mathbf{e}_{a_{3}}  \notag \\
&&+\ ^{\shortmid }h^{7}(\ ^{\shortmid }x^{k_{3}},p_{7})\ ^{\shortmid }%
\mathbf{e}_{7}\otimes \ ^{\shortmid }\mathbf{e}_{7}+\ ^{\shortmid }h^{8}(\
^{\shortmid }x^{k_{3}},p_{7})\ ^{\shortmid }\mathbf{e}_{8}\otimes \
^{\shortmid }\mathbf{e}_{8},  \notag \\
\mathbf{e}^{a_{2}}
&=&dy^{a_{2}}+N_{i_{1}}^{a_{2}}(x^{k_{1}},y^{b_{2}})dx^{i_{1}},\ ^{\shortmid
}\mathbf{e}_{a_{3}}=dp_{a_{3}}+\ ^{\shortmid }N_{i_{2}}(\ ^{\shortmid
}x^{k_{2}},p_{b_{2}})d\ ^{\shortmid }x^{i_{2}},  \notag \\
\ ^{\shortmid }\mathbf{e}_{7} &=&dp_{7}+\ ^{\shortmid }w_{i_{3}}(\
^{\shortmid }x^{k_{3}},p_{7})dx^{i_{3}},\ \ ^{\shortmid }\mathbf{e}_{8}=dE+\
^{\shortmid }n_{i_{3}}(\ ^{\shortmid }x^{k_{3}},p_{7})d\ ^{\shortmid
}x^{i_{3}},  \label{dmqc8}
\end{eqnarray}%
with Killing symmetry on coordinate vector $\ ^{\shortmid }\partial
^{8}=\partial ^{E}$.

The phase space analog of locally anisotropic cosmological s-metrics \ (\ref%
{qst8d8}) \ is stated by formulas, 
\begin{eqnarray}
\ ^{\shortmid }\mathbf{\hat{g}} &=&g_{i_{1}}(x^{k_{1}})dx^{i_{1}}\otimes
dx^{i_{1}}+g_{a_{2}}(x^{k_{1}},y^{b_{2}})\mathbf{e}^{a_{2}}\otimes \mathbf{e}%
^{a_{2}}+\ ^{\shortmid }h^{a_{3}}(x^{k_{2}},p_{b_{3}})\ ^{\shortmid }\mathbf{%
e}_{a_{3}}\otimes \ ^{\shortmid }\mathbf{e}_{a_{3}}  \notag \\
&&+\ ^{\shortmid }h^{7}(\ ^{\shortmid }x^{k_{3}},p_{7})\ ^{\shortmid }%
\mathbf{e}_{7}\otimes \ ^{\shortmid }\mathbf{e}_{7}+\ ^{\shortmid }h^{8}(\
^{\shortmid }x^{k_{3}},p_{7})\ ^{\shortmid }\mathbf{e}_{8}\otimes \
^{\shortmid }\mathbf{e}_{8},  \notag \\
\mathbf{e}^{a_{2}}
&=&dy^{a_{2}}+N_{i_{1}}^{a_{2}}(x^{k_{1}},y^{b_{2}})dx^{i_{1}},\ ^{\shortmid
}\mathbf{e}_{a_{3}}=dp_{a_{3}}+\ ^{\shortmid }N_{i_{2}}(\ ^{\shortmid
}x^{k_{2}},p_{b_{2}})d\ ^{\shortmid }x^{i_{2}},  \notag \\
\ \ ^{\shortmid }\mathbf{e}_{7} &=&dp_{7}+\ ^{\shortmid }n_{i_{3}}(\
^{\shortmid }x^{k_{3}},E)d\ ^{\shortmid }x^{i_{3}},\ ^{\shortmid }\mathbf{e}%
_{8}=dE+\ ^{\shortmid }w_{i_{3}}(\ ^{\shortmid }x^{k_{3}},E)dx^{i_{3}},
\label{dmcc8}
\end{eqnarray}%
with Killing symmetry on coordinate vector $\ ^{\shortmid }\partial ^{7}$.

\subsubsection{Diagonal and off-diagonal ansatz for momentum phase spaces}

The parametrization of local coordinates, N-connection and canonical
d-connection structures and s-metrics for velocity-phase spaces are sated in
Table 12.

\newpage

{\scriptsize 
\begin{eqnarray*}
&&%
\begin{tabular}{lll}
& {\ \textsf{Table 12:\ Diagonal and off-diagonal ansatz for 8-d cotangent
Lorentz bundles} } &  \\ 
& and the Anholonomic Frame and Connection Deformation Method, \textbf{AFCDM}%
, &  \\ 
& \textit{for constructing generic off-diagonal exact and parametric
solutions} & 
\end{tabular}
\\
&&{%
\begin{tabular}{lll}
\hline
diagonal ansatz: PDEs $\rightarrow $ \textbf{ODE}s &  & AFCDM: \textbf{PDE}s 
\textbf{with decoupling; } \\ 
\begin{tabular}{l}
coordinates \\ 
$\ ^{\shortmid }u^{\alpha _{s}}=(x^{1},x^{2},y^{3},y^{4}=t,$ \\ 
$p_{5},p_{6},p_{7},p_{8}=E)$%
\end{tabular}
& $%
\begin{array}{c}
\ _{s}^{\shortmid }u=(\ _{s-1}x,\ _{s}^{\shortmid }y) \\ 
s=1,2,3,4;%
\end{array}%
$ & $%
\begin{tabular}{l}
nonholonomic 2+2+2+2+2 splitting; shels $s=1,2,3,4$ \\ 
$\ ^{\shortmid }u^{\alpha
_{s}}=(x^{1},x^{2},y^{3},y^{4}=t,p_{5},p_{6},p_{7},p_{8}=E);$ \\ 
$\ ^{\shortmid }u^{\alpha _{s}}=(x^{i_{1}},y^{a_{2}},p_{a_{3}},p_{a_{4}});\
^{\shortmid }u^{\alpha _{s}}=(x^{i_{s-1}},\ ^{\shortmid }y^{a_{s}});$ \\ 
$\ $ $i_{1}=1,2;a_{2}=3,4;a_{3}=5,6;a_{4}=7,8;$%
\end{tabular}%
$ \\ 
LC-connection $\ ^{\shortmid }\mathring{\nabla}$ & 
\begin{tabular}{l}
N-connection; \\ 
canonical \\ 
d-connection%
\end{tabular}
& $%
\begin{array}{c}
\ \ _{s}^{\shortmid }\mathbf{N}:T\ _{s}^{\ast }\mathbf{V}=hT^{\ast }\mathbf{V%
}\oplus \ ^{2}vT^{\ast }\mathbf{V}\oplus \ ^{3}c\mathbf{T}^{\ast }\mathbf{%
\mathbf{V}\oplus }\ ^{4}c\mathbf{T}^{\ast }\mathbf{\mathbf{V},} \\ 
\mbox{ locally }\ \ _{s}^{\shortmid }\mathbf{N}=\{\ ^{\shortmid
}N_{i_{s-1}}^{a_{s}}(x,p)= \\ 
\ ^{\shortmid }N_{i_{s-1}}^{a_{s}}(\ _{s-1}x,\ _{s}^{\shortmid }y)=\
^{\shortmid }N_{i_{s-1}}^{a_{s}}(\ _{s}^{\shortmid }u)\} \\ 
\ \ _{s}^{\shortmid }\widehat{\mathbf{D}}=(\ ^{1}h\ ^{\shortmid }\widehat{%
\mathbf{D}},\ ^{2}v\ ^{\shortmid }\widehat{\mathbf{D}},\ ^{3}c\ ^{\shortmid }%
\widehat{\mathbf{D}},\ ^{4}c\ ^{\shortmid }\widehat{\mathbf{D}})=\{\
^{\shortmid }\Gamma _{\ \beta _{s}\gamma _{s}}^{\alpha _{s}}\}; \\ 
\mbox{ canonical s-connection distortion }\  \\ 
\ \ _{s}^{\shortmid }\widehat{\mathbf{D}}=\ ^{\shortmid }\nabla +\
_{s}^{\shortmid }\ \widehat{\mathbf{Z}};\ \ \ _{s}^{\shortmid }\widehat{%
\mathbf{D}}\ \ _{s}^{\shortmid }\mathbf{g=0,} \\ 
\ \ \ _{s}^{\shortmid }\widehat{\mathcal{T}}[\ _{s}^{\shortmid }\mathbf{g,}\
\ _{s}^{\shortmid }\mathbf{N,}\ \ _{s}^{\shortmid }\widehat{\mathbf{D}}]%
\mbox{ canonical
d-torsion}%
\end{array}%
$ \\ 
$%
\begin{array}{c}
\mbox{ diagonal ansatz  } \\ 
\ ^{2}\mathring{g}=\mathring{g}_{\alpha _{2}\beta _{2}}(\ ^{s}u)= \\ 
\left( 
\begin{array}{cccc}
\mathring{g}_{1} &  &  &  \\ 
& \mathring{g}_{2} &  &  \\ 
&  & \mathring{g}_{3} &  \\ 
&  &  & \mathring{g}_{4}%
\end{array}%
\right) ; \\ 
\ ^{s}g=\mathring{g}_{\alpha _{s}\beta _{s}}(\ ^{s}u)= \\ 
\left( 
\begin{array}{cccc}
\ ^{2}\mathring{g} &  &  &  \\ 
& \ ^{\shortmid }\mathring{g}^{5} &  &  \\ 
&  & \ddots &  \\ 
&  &  & \ ^{\shortmid }\mathring{g}^{8}%
\end{array}%
\right)%
\end{array}%
$ & $\ ^{\shortmid }\mathbf{g}\Leftrightarrow $ & $%
\begin{tabular}{l}
$g_{\alpha _{2}\beta _{2}}=%
\begin{array}{c}
g_{\alpha _{2}\beta _{2}}(x^{i_{1}},y^{a_{2}})%
\mbox{ general frames /
coordinates} \\ 
\left[ 
\begin{array}{cc}
g_{i_{1}j_{1}}+N_{i_{1}}^{a_{2}}N_{j_{1}}^{b_{2}}h_{a_{2}b_{2}} & 
N_{i_{1}}^{b_{2}}h_{c_{2}b_{2}} \\ 
N_{j_{1}}^{a_{2}}h_{a_{2}b_{2}} & h_{a_{2}c_{2}}%
\end{array}%
\right]%
\end{array}%
$ \\ 
$\ ^{2}\mathbf{g=\{g}_{\alpha _{2}\beta
_{2}}=[g_{i_{1}j_{1}},h_{a_{2}b_{2}}]\},$ \\ 
$\ ^{2}\mathbf{g}=\mathbf{g}_{i_{1}}(x^{k_{1}})dx^{i_{1}}\otimes dx^{i_{1}}+%
\mathbf{g}_{a_{2}}(x^{k_{2}},y^{b_{2}})\mathbf{e}^{a_{2}}\otimes \mathbf{e}%
^{b_{2}}$ \\ 
$\vdots $ \\ 
$\ ^{\shortmid }g_{\alpha _{s}\beta _{s}}=$ \\ 
$%
\begin{array}{c}
\ ^{\shortmid }g_{\alpha _{s}\beta _{s}}(x^{i_{s-1}},\ ^{\shortmid
}y^{a_{s}})\mbox{ general frames / coordinates} \\ 
\left[ 
\begin{array}{cc}
\ ^{\shortmid }g_{i_{s}j_{s}}+\ ^{\shortmid }N_{i_{s-1}}^{a_{s}}\
^{\shortmid }N_{j_{s-1}}^{b_{s}}\ ^{\shortmid }h_{a_{s}b_{s}} & \
^{\shortmid }N_{i_{s-1}}^{b_{s}}\ ^{\shortmid }h_{c_{s}b_{s}} \\ 
\ ^{\shortmid }N_{j_{s-1}}^{a_{s}}\ ^{\shortmid }h_{a_{s}b_{s}} & \
^{\shortmid }h_{a_{s}c_{s}}%
\end{array}%
\right]%
\end{array}%
$ \\ 
$\ _{s}^{\shortmid }\mathbf{g=\{\ ^{\shortmid }g}_{\alpha _{s}\beta _{s}}=[\
^{\shortmid }g_{i_{s-1}j_{s-1}},\ ^{\shortmid }h_{a_{s}b_{s}}]$ \\ 
$=[g_{i_{1}j_{1}},h_{a_{2}b_{2}},\ ^{\shortmid }h^{a_{3}b_{3}},\ ^{\shortmid
}h^{a_{4}b_{4}}]\}$ \\ 
$\ \ _{s}^{\shortmid }\mathbf{g}=\ _{s}^{\shortmid }\mathbf{g}%
_{i_{s-1}}(x^{k_{s-1}})dx^{i_{s-1}}\otimes dx^{i_{s-1}}+$ \\ 
$\ _{s}^{\shortmid }\mathbf{g}_{a_{s}}(x^{k_{s-1}},y^{b_{s}})\mathbf{e}%
^{a_{s}}\otimes \mathbf{e}^{b_{s}}$ \\ 
$=\ \mathbf{g}_{i_{1}}(x^{k_{1}})dx^{i_{1}}\otimes dx^{i_{1}}+\mathbf{g}%
_{a_{2}}(x^{k_{1}},y^{b_{2}})\mathbf{e}^{a_{2}}\otimes \mathbf{e}^{a_{2}}+$
\\ 
$\mathbf{\ ^{\shortmid }g}^{a_{3}}(x^{k_{1}},y^{b_{2}},p_{b_{3}})\mathbf{\
^{\shortmid }e}_{a_{3}}\otimes \mathbf{\ ^{\shortmid }e}_{a_{3}}$ \\ 
$+\mathbf{\ ^{\shortmid }g}_{a_{4}}(x^{k_{1}},y^{b_{2}},p_{b_{3}},p_{b_{4}})%
\mathbf{\ ^{\shortmid }e}_{a_{4}}\otimes \mathbf{\ ^{\shortmid }e}_{a_{4}};$%
\end{tabular}%
$ \\ 
$\mathring{g}_{\alpha _{2}\beta _{2}}=\left\{ 
\begin{array}{cc}
\mathring{g}_{\alpha _{2}}(\ ^{2}r) & \mbox{ for BHs} \\ 
\mathring{g}_{\alpha _{2}}(t) & \mbox{ for FLRW }%
\end{array}%
\right. $ & [coord.frames] & $g_{\alpha _{2}\beta _{2}}=\left\{ 
\begin{array}{cc}
g_{\alpha _{2}\beta _{2}}(x^{i},y^{3}) &  \\ 
\underline{g}_{\alpha _{2}\beta _{2}}(x^{i},y^{4}=t) & 
\end{array}%
\right. $ \\ 
$\mathbf{\ ^{\shortmid }}\mathring{g}_{\alpha _{s}\beta _{s}}=\left\{ 
\begin{array}{cc}
\mathbf{\ ^{\shortmid }}\mathring{g}_{\alpha _{s}}(\ _{s}^{\shortmid }r) & %
\mbox{ for BHs} \\ 
\mathbf{\ ^{\shortmid }}\mathring{g}_{\alpha _{s}}(t) & \mbox{ for FLRW }%
\end{array}%
\right. $ &  & $\mathbf{\ ^{\shortmid }}g_{\alpha _{s}\beta _{s}}=\left\{ 
\begin{array}{cc}
\mathbf{\ ^{\shortmid }}g_{\alpha _{s}\beta _{s}}(x^{i_{3}},p_{7}) &  \\ 
\mathbf{\ ^{\shortmid }}\underline{g}_{\alpha _{s}\beta _{s}}(x^{i_{3}},E) & 
\end{array}%
\right. $ \\ 
$%
\begin{array}{c}
\mbox{coord. transf. }\mathbf{\ ^{\shortmid }}e_{\alpha _{s}}=\mathbf{\
^{\shortmid }}e_{\ \alpha _{s}}^{\alpha _{s}^{\prime }}\mathbf{\ ^{\shortmid
}}\partial _{\alpha _{s}^{\prime }}, \\ 
\mathbf{\ ^{\shortmid }}e^{\beta _{s}}=\mathbf{\ ^{\shortmid }}e_{\beta
_{s}^{\prime }}^{\ \beta _{s}}d\mathbf{\ ^{\shortmid }}u^{\beta _{s}^{\prime
}}, \\ 
\mathbf{\ ^{\shortmid }}\mathring{g}_{\alpha _{s}\beta _{s}}=\mathbf{\
^{\shortmid }}\mathring{g}_{\alpha _{s}^{\prime }\beta _{s}^{\prime }}%
\mathbf{\ ^{\shortmid }}e_{\ \alpha _{s}}^{\alpha _{s}^{\prime }}\mathbf{\
^{\shortmid }}e_{\ \beta _{s}}^{\beta _{s}^{\prime }} \\ 
\begin{array}{c}
\mathbf{\ ^{\shortmid }\mathring{g}}_{\alpha _{s}}(\mathbf{\ ^{\shortmid }}%
x^{k_{s-1}},\mathbf{\ ^{\shortmid }}y^{a_{s}})\rightarrow \mathbf{\
^{\shortmid }}\mathring{g}_{\alpha _{s}}(\ _{s}^{\shortmid }r), \\ 
\mathbf{\ ^{\shortmid }}\mathring{g}_{\alpha _{s}}(t),\mathbf{\ ^{\shortmid }%
}\mathring{N}_{i_{s-1}}^{a_{s}}(x^{k_{s-1}},\mathbf{\ ^{\shortmid }}%
y^{a_{s}})\rightarrow 0.%
\end{array}%
\end{array}%
$ & [N-adapt. fr.] & 
\begin{tabular}{l}
$\left\{ 
\begin{array}{cc}
\begin{array}{c}
\mathbf{g}_{i_{1}}(x^{k_{1}}),\mathbf{g}_{a_{2}}(x^{k_{1}},y^{3}), \\ 
\mbox{ or }\mathbf{g}_{i_{1}}(x^{k_{1}}),\underline{\mathbf{g}}%
_{a_{2}}(x^{k_{1}},t),%
\end{array}
&  \\ 
\begin{array}{c}
N_{i_{1}}^{3}=w_{i_{1}}(x^{k},y^{3}),N_{i_{1}}^{4}=n_{i_{1}}(x^{k},y^{3}),
\\ 
\mbox{ or }\underline{N}_{i_{1}}^{3}=\underline{n}_{i_{1}}(x^{k_{1}},t),%
\underline{N}_{i_{1}}^{4}=\underline{w}_{i_{1}}(x^{k_{1}},t),%
\end{array}
& 
\end{array}%
\right. $ \\ 
$\vdots $ \\ 
$\left\{ 
\begin{array}{cc}
\begin{array}{c}
\mathbf{\ ^{\shortmid }g}_{i_{3}}(\mathbf{\ ^{\shortmid }}x^{k_{3}}),\mathbf{%
\ ^{\shortmid }g}_{a_{4}}(\mathbf{\ ^{\shortmid }}x^{k_{3}},p_{7}), \\ 
\mbox{ or }\mathbf{\ ^{\shortmid }g}_{i_{3}}(\mathbf{\ ^{\shortmid }}%
x^{k_{1}}),\mathbf{\ ^{\shortmid }}\underline{\mathbf{g}}_{a_{4}}(\mathbf{\
^{\shortmid }}x^{k_{3}},E),%
\end{array}
&  \\ 
\begin{array}{c}
\ ^{\shortmid }N_{i_{3}7}=\ ^{\shortmid }w_{i_{3}}(\ ^{\shortmid
}x^{k_{3}},p_{7}),\ ^{\shortmid }N_{i_{3}8}=\ ^{\shortmid }n_{i_{3}}\
^{\shortmid }x^{k_{3}},p_{7}) \\ 
\ ^{\shortmid }\underline{N}_{i_{3}8}=\ ^{\shortmid }\underline{n}_{i_{3}}\
^{\shortmid }x^{k_{3}},E),\ ^{\shortmid }\underline{N}_{i_{3}8}=\
^{\shortmid }\underline{w}_{i_{3}}\ ^{\shortmid }x^{k_{3}},E)%
\end{array}
& 
\end{array}%
\right. $%
\end{tabular}
\\ 
$\ _{s}^{\shortmid }\mathring{\nabla},$ $\ \ _{s}^{\shortmid }Ric=\{\mathbf{%
\ ^{\shortmid }}\mathring{R}_{\ \beta _{s}\gamma _{s}}\}$ & Ricci tensors & $%
\ \ _{s}^{\shortmid }\widehat{\mathbf{D}},\ \ \ _{s}^{\shortmid }\widehat{%
\mathcal{R}}ic=\{\mathbf{\ ^{\shortmid }}\widehat{\mathbf{R}}_{\ \beta
_{s}\gamma _{s}}\}$ \\ 
$\ _{s}^{\shortmid m}\mathcal{L[\mathbf{\phi }]\rightarrow }\ \
_{s}^{\shortmid m}\mathbf{T}_{\alpha _{s}\beta _{s}}\mathcal{[\mathbf{\phi }]%
}$ & 
\begin{tabular}{l}
generating \\ 
sources%
\end{tabular}
& $%
\begin{array}{cc}
\mathbf{\ ^{\shortmid }}\widehat{\mathbf{\Upsilon }}_{\ \nu _{s}}^{\mu _{s}}=%
\mathbf{\ ^{\shortmid }e}_{\ \mu _{s}^{\prime }}^{\mu _{s}}\mathbf{\
^{\shortmid }e}_{\nu _{s}}^{\ \nu _{s}^{\prime }}\mathbf{\ ^{\shortmid
}\Upsilon }_{\ \nu _{s}^{\prime }}^{\mu _{s}^{\prime }}[\ ^{m}\mathcal{L}(%
\mathbf{\varphi ),\ ^{\shortmid }}T_{\mu _{s}\nu _{s}},\ \ _{s}^{\shortmid
}\Lambda ] &  \\ 
\begin{array}{c}
=diag[\ _{1}\Upsilon (x^{i_{1}})\delta _{j_{1}}^{i_{1}},\ _{2}\Upsilon
(x^{i_{1}},y^{3})\delta _{b_{2}}^{a_{2}}, \\ 
\ _{3}^{\shortmid }\Upsilon (x^{i_{2}},p_{5})\delta _{b_{3}}^{a_{3}},\
_{4}^{\shortmid }\Upsilon (x^{i_{3}},p_{7})\delta _{b_{4}}^{a_{4}}], \\ 
\mbox{ quasi-stationary configurations};%
\end{array}
&  \\ 
\begin{array}{c}
=diag[\ _{1}\Upsilon (x^{i_{1}})\delta _{j_{1}}^{i_{1}},\ _{2}\underline{%
\Upsilon }(x^{i_{1}},t)\delta _{b_{2}}^{a_{2}}, \\ 
\ _{3}^{\shortmid }\underline{\Upsilon }(x^{i_{2}},p_{6})\delta
_{b_{3}}^{a_{3}},\ _{4}^{\shortmid }\underline{\Upsilon }(x^{i_{3}},E)\delta
_{b_{4}}^{a_{4}}], \\ 
\mbox{ locally anisotropic cosmology};%
\end{array}
& 
\end{array}%
$ \\ 
trivial eqs for $\ \ _{s}^{\shortmid }\mathring{\nabla}$-torsion & 
LC-conditions & $\ \ _{s}^{\shortmid }\widehat{\mathbf{D}}_{\mid \ \
_{s}^{\shortmid }\widehat{\mathcal{T}}\rightarrow 0}=\ \ _{s}^{\shortmid }%
\mathbf{\nabla .}$ \\ \hline\hline
\end{tabular}%
}
\end{eqnarray*}%
}Parameterizations of geometric s-objects on shells $s=2,3$ depend on the
type of shell Killing symmetries we prescribe for such nonholonomic phase
spaces with momentum like variables.

\subsubsection{Quasi-stationary solutions with fixed energy parameter}

Such quasi-stationary solutions are nonholonomic momentum type phase
configurations modeled on cotangent Lorentz bundles with $p_{8}=E=const,$
when the momentum phase space involves space like hypersurfaces. 
{\scriptsize 
\begin{eqnarray*}
&&%
\begin{tabular}{l}
\hline\hline
\begin{tabular}{lll}
& {\large \textsf{Table 13:\ Off-diagonal quasi-stationary and pase space
configurations with fixed energy}} &  \\ 
& Exact solutions of $\mathbf{\ ^{\shortmid }}\widehat{\mathbf{R}}_{\mu
_{s}\nu _{s}}=\mathbf{\ ^{\shortmid }\Upsilon }_{\mu _{s}\nu _{s}}$ (\ref%
{cdeq1c8}) on $T_{s}^{\ast }V$ transformed into a momentum version of
nonlinear PDEs (\ref{eq1})-(\ref{e2c}) & 
\end{tabular}
\\ 
\end{tabular}
\\
&&%
\begin{tabular}{lll}
\hline\hline
&  &  \\ 
$%
\begin{array}{c}
\mbox{d-metric ansatz with} \\ 
\mbox{Killing symmetry }\partial _{4}=\partial _{t},\mathbf{\ ^{\shortmid }}%
\partial ^{8}%
\end{array}%
$ &  & $%
\begin{array}{c}
ds^{2}=g_{i_{1}}(x^{k_{1}})(dx^{i_{1}})^{2}+g_{a_{2}}(x^{k_{1}},y^{3})(dy^{a_{2}}+N_{i_{1}}^{a_{2}}(x^{k_{1}},y^{3})dx^{i_{1}})^{2}
\\ 
+\mathbf{\ ^{\shortmid }}g^{a_{3}}(x^{k_{2}},p_{5})(dp_{a_{3}}+\mathbf{\
^{\shortmid }}N_{i_{2}a_{3}}(x^{k_{2}},p_{5})dx^{i_{2}})^{2} \\ 
+\mathbf{\ ^{\shortmid }}g^{a_{4}}(\mathbf{\ ^{\shortmid }}%
x^{k_{3}},p_{7})(dp_{a_{4}}+\mathbf{\ ^{\shortmid }}N_{i_{3}a_{4}}(\mathbf{\
^{\shortmid }}x^{k_{3}},p_{7})d\mathbf{\ ^{\shortmid }}x^{i_{3}})^{2},%
\mbox{
for }g_{i_{1}}=e^{\psi {(x}^{k_{1}}{)}}, \\ 
g_{a_{2}}=h_{a_{2}}(x^{k_{1}},y^{3}),N_{i_{1}}^{3}=\
^{2}w_{i_{1}}=w_{i_{1}}(x^{k_{1}},y^{3}),N_{i_{1}}^{4}=\
^{2}n_{i_{1}}=n_{i_{1}}(x^{k_{1}},y^{3}), \\ 
\mathbf{\ ^{\shortmid }}g^{a_{3}}=\mathbf{\ ^{\shortmid }}%
h^{a_{3}}(x^{k_{2}},p_{5}),\mathbf{\ ^{\shortmid }}N_{i_{2}5}=\ _{\shortmid
}^{3}w_{i_{2}}=\mathbf{\ ^{\shortmid }}w_{i_{2}}(x^{k_{2}},p_{5}), \\ 
\mathbf{\ ^{\shortmid }}N_{i_{2}6}=\ _{\shortmid }^{3}n_{i_{2}}=\mathbf{\
^{\shortmid }}n_{i_{2}}(x^{k_{2}},p_{5}), \\ 
\mathbf{\ ^{\shortmid }}g^{a_{4}}=\mathbf{\ ^{\shortmid }}h^{a_{4}}(\mathbf{%
\ ^{\shortmid }}x^{k_{3}},p_{7}),\mathbf{\ ^{\shortmid }}N_{i_{3}7}=\ \
_{\shortmid }^{4}w_{i_{3}}=\mathbf{\ ^{\shortmid }}%
w_{i_{3}}(x^{k_{3}},p_{7}), \\ 
\mathbf{\ ^{\shortmid }}N_{i_{3}8}=\ _{\shortmid }^{4}n_{i_{3}}=\mathbf{\
^{\shortmid }}n_{i_{3}}(x^{k_{3}},p_{7}),%
\end{array}%
$ \\ 
Effective matter sources &  & $\mathbf{\ ^{\shortmid }\Upsilon }_{\ \nu
_{s}}^{\mu _{s}}=[\ _{1}\widehat{\Upsilon }({x}^{k_{1}})\delta
_{j_{1}}^{i_{1}},\ _{2}\widehat{\Upsilon }({x}^{k_{1}},y^{3})\delta
_{b_{2}}^{a_{2}},\ _{3}^{\shortmid }\widehat{\Upsilon }({x}%
^{k_{2}},p_{5})\delta _{b_{3}}^{a_{3}},\ _{4}^{\shortmid }\widehat{\Upsilon }%
({x}^{k_{3}},p_{7})\delta _{b_{4}}^{a_{4}},],$ \\ \hline
Nonlinear PDEs (\ref{eq1})-(\ref{e2c}) &  & $%
\begin{tabular}{lll}
$%
\begin{array}{c}
\psi ^{\bullet \bullet }+\psi ^{\prime \prime }=2\ \ _{1}\widehat{\Upsilon };
\\ 
\ ^{2}\varpi ^{\ast }\ h_{4}^{\ast }=2h_{3}h_{4}\ _{2}\widehat{\Upsilon };
\\ 
\ ^{2}\beta \ ^{2}w_{i_{1}}-\ ^{2}\alpha _{i_{1}}=0; \\ 
\ ^{2}n_{k_{1}}^{\ast \ast }+\ ^{2}\gamma \ ^{2}n_{k_{1}}^{\ast }=0;%
\end{array}%
$ &  & $%
\begin{array}{c}
\ ^{2}\varpi {=\ln |\partial _{3}h_{4}/\sqrt{|h_{3}h_{4}|}|,} \\ 
\ ^{2}\alpha _{i_{1}}=(\partial _{3}h_{4})\ (\partial _{i_{1}}\ ^{2}\varpi ),
\\ 
\ ^{2}\beta =(\partial _{3}h_{4})\ (\partial _{3}\ ^{2}\varpi ),\  \\ 
\ \ ^{2}\gamma =\partial _{3}\left( \ln |h_{4}|^{3/2}/|h_{3}|\right) , \\ 
\partial _{1}q=q^{\bullet },\partial _{2}q=q^{\prime },\partial
_{3}q=q^{\ast }%
\end{array}%
$ \\ 
$%
\begin{array}{c}
\mathbf{\ ^{\shortmid }}\partial ^{5}(\ _{\shortmid }^{3}\varpi )\ \mathbf{\
^{\shortmid }}\partial ^{5}\mathbf{\ ^{\shortmid }}h^{6}=2\mathbf{\
^{\shortmid }}h^{5}\mathbf{\ ^{\shortmid }}h^{6}\ \ _{3}^{\shortmid }%
\widehat{\Upsilon }; \\ 
\ _{\shortmid }^{3}\beta \ _{\shortmid }^{3}w_{i_{2}}-\ _{\shortmid
}^{3}\alpha _{i_{2}}=0; \\ 
\mathbf{\ ^{\shortmid }}\partial ^{5}(\mathbf{\ ^{\shortmid }}\partial ^{5}\
_{\shortmid }^{3}n_{k_{2}})+\ _{\shortmid }^{3}\gamma \mathbf{\ ^{\shortmid }%
}\partial ^{5}(\ _{\shortmid }^{3}n_{k_{2}})=0;%
\end{array}%
$ &  & $%
\begin{array}{c}
\\ 
\ _{\shortmid }^{3}\varpi {=\ln |\mathbf{\ ^{\shortmid }}\partial ^{5}%
\mathbf{\ ^{\shortmid }}h^{6}/\sqrt{|\mathbf{\ ^{\shortmid }}h^{5}\mathbf{\
^{\shortmid }}h^{6}|}|,} \\ 
\ _{\shortmid }^{3}\alpha _{i_{2}}=(\mathbf{\ ^{\shortmid }}\partial ^{5}%
\mathbf{\ ^{\shortmid }}h^{6})\ (\partial _{i_{2}}\ _{\shortmid }^{3}\varpi
), \\ 
\ _{\shortmid }^{3}\beta =(\mathbf{\ ^{\shortmid }}\partial ^{5}\mathbf{\
^{\shortmid }}h^{6})\ (\mathbf{\ ^{\shortmid }}\partial ^{5}\ _{\shortmid
}^{3}\varpi ),\  \\ 
\ \ _{\shortmid }^{3}\gamma =\mathbf{\ ^{\shortmid }}\partial ^{5}\left( \ln
|\mathbf{\ ^{\shortmid }}h^{6}|^{3/2}/|\mathbf{\ ^{\shortmid }}h^{5}|\right)
,%
\end{array}%
$ \\ 
$%
\begin{array}{c}
\mathbf{\ ^{\shortmid }}\partial ^{7}(\ _{\shortmid }^{4}\varpi )\ \mathbf{\
^{\shortmid }}\partial ^{7}\ \mathbf{\ ^{\shortmid }}h^{8}=2\mathbf{\
^{\shortmid }}h^{7}\mathbf{\ ^{\shortmid }}h^{8}\ \ \ _{4}^{\shortmid }%
\widehat{\Upsilon }; \\ 
\ _{\shortmid }^{4}\beta \ _{\shortmid }^{4}w_{i_{3}}-\ _{\shortmid
}^{4}\alpha _{i_{3}}=0; \\ 
\mathbf{\ ^{\shortmid }}\partial ^{7}(\mathbf{\ ^{\shortmid }}\partial ^{7}\
_{\shortmid }^{4}n_{k_{3}})+\ _{\shortmid }^{4}\gamma \mathbf{\ ^{\shortmid }%
}\partial ^{7}(\ _{\shortmid }^{4}n_{k_{3}})=0;%
\end{array}%
$ &  & $%
\begin{array}{c}
\\ 
\ _{\shortmid }^{4}\varpi {=\ln |\mathbf{\ ^{\shortmid }}\partial ^{7}%
\mathbf{\ ^{\shortmid }}h^{8}/\sqrt{|\mathbf{\ ^{\shortmid }}h^{7}\mathbf{\
^{\shortmid }}h^{8}|}|,} \\ 
\ _{\shortmid }^{4}\alpha _{i_{3}}=(\mathbf{\ ^{\shortmid }}\partial ^{7}%
\mathbf{\ ^{\shortmid }}h^{8})\ (\mathbf{\ ^{\shortmid }}\partial _{i_{3}}\
_{\shortmid }^{4}\varpi ), \\ 
\ _{\shortmid }^{4}\beta =(\mathbf{\ ^{\shortmid }}\partial ^{7}\mathbf{\
^{\shortmid }}h^{8})\ (\mathbf{\ ^{\shortmid }}\partial ^{7}\ _{\shortmid
}^{4}\varpi ),\  \\ 
\ \ _{\shortmid }^{4}\gamma =\mathbf{\ ^{\shortmid }}\partial ^{7}\left( \ln
|\mathbf{\ ^{\shortmid }}h^{8}|^{3/2}/|\mathbf{\ ^{\shortmid }}h^{7}|\right)
,%
\end{array}%
$%
\end{tabular}%
$ \\ \hline
$%
\begin{array}{c}
\mbox{ Gener.  functs:}\ h_{3}(x^{k_{1}},y^{3}), \\ 
\ ^{2}\Psi (x^{k_{1}},y^{3})=e^{\ ^{2}\varpi },\ ^{2}\Phi (x^{k_{1}},y^{3}),
\\ 
\mbox{integr. functs:}\ h_{4}^{[0]}(x^{k_{1}}),\  \\ 
_{1}n_{k_{1}}(x^{i_{1}}),\ _{2}n_{k_{1}}(x^{i_{1}}); \\ 
\mbox{ Gener.  functs:}\mathbf{\ ^{\shortmid }}h^{5}(x^{k_{2}},p_{5}), \\ 
\ \ _{\shortmid }^{3}\Psi (x^{k_{2}},p_{5})=e^{\ \ _{\shortmid }^{3}\varpi
},\ \ \ _{\shortmid }^{3}\Phi (x^{k_{2}},p_{5}), \\ 
\mbox{integr. functs:}\ h_{6}^{[0]}(x^{k_{2}}),\  \\ 
_{1}^{3}n_{k_{2}}(x^{i_{2}}),\ _{2}^{3}n_{k_{2}}(x^{i_{2}}); \\ 
\mbox{ Gener.  functs:}\mathbf{\ ^{\shortmid }}h^{7}(\mathbf{\ ^{\shortmid }}%
x^{k_{3}},p_{7}), \\ 
\ \ _{\shortmid }^{4}\Psi (x^{k_{2}},p_{7})=e^{\ \ \ _{\shortmid }^{4}\varpi
},\ \ \ _{\shortmid }^{4}\Phi (\mathbf{\ ^{\shortmid }}x^{k_{3}},p_{7}), \\ 
\mbox{integr. functs:}\ h_{8}^{[0]}(\mathbf{\ ^{\shortmid }}x^{k_{3}}),\  \\ 
_{1}^{4}n_{k_{3}}(\mathbf{\ ^{\shortmid }}x^{i_{3}}),\ _{2}^{4}n_{k_{3}}(%
\mathbf{\ ^{\shortmid }}x^{i_{3}}); \\ 
\mbox{\& nonlinear symmetries}%
\end{array}%
$ &  & $%
\begin{array}{c}
\ ((\ ^{2}\Psi )^{2})^{\ast }=-\int dy^{3}\ _{2}\widehat{\Upsilon }h_{4}^{\
\ast }, \\ 
(\ ^{2}\Phi )^{2}=-4\ _{2}\Lambda h_{4},\mbox{ see }(\ref{nonlinsymrex}), \\ 
h_{4}=h_{4}^{[0]}-(\ ^{2}\Phi )^{2}/4\ _{2}\Lambda ,h_{4}^{\ast }\neq 0,\
_{2}\Lambda \neq 0=const; \\ 
\\ 
\mathbf{\ ^{\shortmid }}\partial ^{5}((\ \ _{\shortmid }^{3}\Psi
)^{2})=-\int dp_{5}\ _{3}^{\shortmid }\widehat{\Upsilon }\mathbf{\
^{\shortmid }}\partial ^{5}\mathbf{\ ^{\shortmid }}h^{6}, \\ 
(\ \ _{\shortmid }^{3}\Phi )^{2}=-4\ _{3}^{\shortmid }\Lambda \mathbf{\
^{\shortmid }}h^{6}, \\ 
\mathbf{\ ^{\shortmid }}h^{6}=\mathbf{\ ^{\shortmid }}h_{[0]}^{6}-(\ \
_{\shortmid }^{3}\Phi )^{2}/4\ _{3}\Lambda ,\mathbf{\ ^{\shortmid }}\partial
^{5}\mathbf{\ ^{\shortmid }}h^{6}\neq 0,\ _{3}^{\shortmid }\Lambda \neq
0=const; \\ 
\\ 
\mathbf{\ ^{\shortmid }}\partial ^{7}((\ _{\shortmid }^{4}\Psi )^{2})=-\int
dp_{7}\ _{4}^{\shortmid }\widehat{\Upsilon }\mathbf{\ ^{\shortmid }}\partial
^{7}\mathbf{\ ^{\shortmid }}h^{8}, \\ 
(\ _{\shortmid }^{4}\Phi )^{2}=-4\ _{4}^{\shortmid }\Lambda \mathbf{\
^{\shortmid }}h^{8}, \\ 
\mathbf{\ ^{\shortmid }}h^{8}=\mathbf{\ ^{\shortmid }}h_{[0]}^{8}-(\
_{\shortmid }^{4}\Phi )^{2}/4\ _{4}^{\shortmid }\Lambda ,\mathbf{\
^{\shortmid }}\partial ^{7}\mathbf{\ ^{\shortmid }}h^{8}\neq 0,\
_{4}^{\shortmid }\Lambda \neq 0=const;%
\end{array}%
$ \\ \hline
Off-diag. solutions, $%
\begin{array}{c}
\mbox{d--metric} \\ 
\mbox{N-connec.}%
\end{array}%
$ &  & $%
\begin{tabular}{l}
$%
\begin{array}{c}
\ g_{i}=e^{\ \psi (x^{k})}\mbox{ as a solution of 2-d Poisson eqs. }\psi
^{\bullet \bullet }+\psi ^{\prime \prime }=2~\ _{1}\widehat{\Upsilon }; \\ 
h_{3}=-(\Psi ^{\ast })^{2}/4\ _{2}\widehat{\Upsilon }^{2}h_{4},\mbox{ see }(%
\ref{g3}),(\ref{g4}); \\ 
h_{4}=h_{4}^{[0]}-\int dy^{3}(\Psi ^{2})^{\ast }/4\ _{2}\widehat{\Upsilon }%
=h_{4}^{[0]}-\Phi ^{2}/4\ _{2}\Lambda ; \\ 
w_{i}=\partial _{i}\ \Psi /\ \partial _{3}\Psi =\partial _{i}\ \Psi ^{2}/\
\partial _{3}\Psi ^{2}|; \\ 
n_{k}=\ _{1}n_{k}+\ _{2}n_{k}\int dy^{3}(\Psi ^{\ast })^{2}/\ _{2}\widehat{%
\Upsilon }^{2}|h_{4}^{[0]}-\int dy^{3}(\Psi ^{2})^{\ast }/4\ _{2}\widehat{%
\Upsilon }^{2}|^{5/2};%
\end{array}%
$ \\ 
$%
\begin{array}{c}
\mathbf{\ ^{\shortmid }}h^{5}=-(\mathbf{\ ^{\shortmid }}\partial ^{5}\
_{\shortmid }^{3}\Psi )^{2}/4\ _{3}^{\shortmid }\widehat{\Upsilon }^{2}%
\mathbf{\ ^{\shortmid }}h^{6}; \\ 
\mathbf{\ ^{\shortmid }}h^{6}=\mathbf{\ ^{\shortmid }}h_{[0]}^{6}-\int dp_{5}%
\mathbf{\ ^{\shortmid }}\partial ^{5}((\ \ _{\shortmid }^{3}\Psi )^{2})/4\
_{3}^{\shortmid }\widehat{\Upsilon }=\mathbf{\ ^{\shortmid }}h_{[0]}^{6}-(\
\ _{\shortmid }^{3}\Phi )^{2}/4\ _{3}^{\shortmid }\Lambda ; \\ 
w_{i_{2}}=\partial _{i_{2}}(\ _{\shortmid }^{3}\Psi )/\mathbf{\ ^{\shortmid }%
}\partial ^{5}(\ _{\shortmid }^{3}\Psi )=\partial _{i_{2}}(\ _{\shortmid
}^{3}\Psi )^{2}/\ \mathbf{\ ^{\shortmid }}\partial ^{5}(\ _{\shortmid
}^{3}\Psi )^{2}|; \\ 
n_{k_{2}}=\ _{1}n_{k_{2}}+\ _{2}n_{k_{2}}\int dp_{5}(\mathbf{\ ^{\shortmid }}%
\partial ^{5}\ _{\shortmid }^{3}\Psi )^{2}/\ _{3}^{\shortmid }\widehat{%
\Upsilon }^{2}|\mathbf{\ ^{\shortmid }}h_{[0]}^{6}- \\ 
\int dp_{5}\mathbf{\ ^{\shortmid }}\partial ^{5}((\ _{\shortmid }^{3}\Psi
)^{2})/4\ _{3}^{\shortmid }\widehat{\Upsilon }^{2}|^{5/2}%
\end{array}%
$ \\ 
$%
\begin{array}{c}
\mathbf{\ ^{\shortmid }}h^{7}=-(\mathbf{\ ^{\shortmid }}\partial ^{7}\
_{\shortmid }^{4}\Psi )^{2}/4\ _{\shortmid }^{4}\widehat{\Upsilon }^{2}%
\mathbf{\ ^{\shortmid }}h^{8}; \\ 
\mathbf{\ ^{\shortmid }}h^{8}=\mathbf{\ ^{\shortmid }}h_{[0]}^{8}-\int dp_{7}%
\mathbf{\ ^{\shortmid }}\partial ^{7}((\ _{\shortmid }^{4}\Psi )^{2})/4\
_{4}^{\shortmid }\widehat{\Upsilon }=h_{8}^{[0]}-(\ \ _{\shortmid }^{4}\Phi
)^{2}/4\ \ _{4}^{\shortmid }\Lambda ; \\ 
\mathbf{\ ^{\shortmid }}w_{i_{3}}=\mathbf{\ ^{\shortmid }}\partial
_{i_{3}}(\ \ _{\shortmid }^{4}\Psi )/\ \mathbf{\ ^{\shortmid }}\partial
^{7}(\ _{\shortmid }^{4}\Psi )=\mathbf{\ ^{\shortmid }}\partial _{i_{3}}(\ \
_{\shortmid }^{4}\Psi )^{2}/\ \mathbf{\ ^{\shortmid }}\partial ^{7}(\
_{\shortmid }^{4}\Psi )^{2}|; \\ 
\mathbf{\ ^{\shortmid }}n_{k_{3}}=\ _{1}^{\shortmid }n_{k_{3}}+\
_{2}^{\shortmid }n_{k_{3}}\int dp_{7}(\ _{\shortmid }^{4}\Psi )^{2}/\
_{4}^{\shortmid }\widehat{\Upsilon }^{2}|h_{8}^{[0]}-\int dp_{7}\mathbf{\
^{\shortmid }}\partial ^{7}((\ \ _{\shortmid }^{4}\Psi )^{2})/4\
_{4}^{\shortmid }\widehat{\Upsilon }^{2}|^{5/2}%
\end{array}%
$%
\end{tabular}%
$ \\ \hline\hline
\end{tabular}%
\end{eqnarray*}%
}As a $T^{\ast }\mathbf{V}$ analog of the nonlinear quadratic element (\ref%
{qst8d7}), with $v^{8}=const,$ and data from Table 8 we provide example of
8-d quasi-stationary quadratic element (\ref{dmqc8}) with $p_{8}=E=const%
\mathbf{V},$ 
\begin{eqnarray}
d\widehat{s}_{[8d]}^{2} &=&\widehat{g}_{\alpha _{s}\beta
_{s}}(x^{k},y^{3},p_{5},p_{7};h_{4},\mathbf{\ ^{\shortmid }}h^{6},\mathbf{\
^{\shortmid }}h^{8};\ _{s}^{\shortmid }\widehat{\Upsilon };\ _{s}^{\shortmid
}\Lambda )d\mathbf{\ ^{\shortmid }}u^{\alpha _{s}}d\mathbf{\ ^{\shortmid }}%
u^{\beta _{s}}  \label{qcosm8d7} \\
&=&e^{\psi (x^{k},\ _{s}\widehat{\Upsilon })}[(dx^{1})^{2}+(dx^{2})^{2}]-%
\frac{(h_{4}^{\ast })^{2}}{|\int dy^{3}[\ _{2}\widehat{\Upsilon }%
h_{4}]^{\ast }|\ h_{4}}\{dy^{3}+\frac{\partial _{i_{1}}[\int dy^{3}(\ _{2}%
\widehat{\Upsilon })\ h_{4}^{\ast }]}{\ _{2}\widehat{\Upsilon }\ h_{4}^{\ast
}}dx^{i_{1}}\}^{2}+  \notag \\
&&h_{4}\{dt+[\ _{1}n_{k_{1}}+\ _{2}n_{k_{1}}\int dy^{3}\frac{(h_{4}^{\ast
})^{2}}{|\int dy^{3}[\ _{2}\widehat{\Upsilon }h_{4}]^{\ast }|\ (h_{4})^{5/2}}%
]dx_{1}^{k}\}+  \notag \\
&&\frac{(\mathbf{\ ^{\shortmid }}\partial ^{5}\mathbf{\ ^{\shortmid }}%
h^{6})^{2}}{|\int dp_{5}\mathbf{\ ^{\shortmid }}\partial ^{5}[\
_{3}^{\shortmid }\widehat{\Upsilon }\mathbf{\ ^{\shortmid }}h^{6}]|\ \mathbf{%
\ ^{\shortmid }}h^{6}}\{dp_{5}+\frac{\partial _{i_{2}}[\int dp_{5}(\ \
_{3}^{\shortmid }\widehat{\Upsilon })\mathbf{\ ^{\shortmid }}\partial ^{5}%
\mathbf{\ ^{\shortmid }}h^{6}]}{\ \ _{3}^{\shortmid }\widehat{\Upsilon }%
\mathbf{\ ^{\shortmid }}\partial ^{5}\ \mathbf{^{\shortmid }}h^{6}}%
dx^{i_{2}}\}^{2}+  \notag \\
&&\mathbf{\ ^{\shortmid }}h^{6}\{dp_{5}+[\ _{1}n_{k_{2}}+\ _{2}n_{k_{2}}\int
dp_{5}\frac{(\mathbf{\ ^{\shortmid }}\partial ^{5}\mathbf{\ ^{\shortmid }}%
h^{6})^{2}}{|\int dp_{5}\mathbf{\ ^{\shortmid }}\partial ^{5}[\ \
_{3}^{\shortmid }\widehat{\Upsilon }\mathbf{\ ^{\shortmid }}h^{6}]|\ (%
\mathbf{\ ^{\shortmid }}h^{6})^{5/2}}]dx^{k_{2}}\}+  \notag \\
&&\frac{(\mathbf{\ ^{\shortmid }}\partial ^{7}\mathbf{\ ^{\shortmid }}%
h^{8})^{2}}{|\int dp_{7}\mathbf{\ ^{\shortmid }}\partial ^{7}[\ _{4}\widehat{%
\Upsilon }\mathbf{\ ^{\shortmid }}h^{8}]|\ \mathbf{\ ^{\shortmid }}h^{8}}%
\{dp_{7}+\frac{\partial _{i_{3}}[\int dp_{7}(\ _{4}\widehat{\Upsilon })\ 
\mathbf{\ ^{\shortmid }}\partial ^{7}\mathbf{\ ^{\shortmid }}h^{8}]}{\ \
_{4}^{\shortmid }\widehat{\Upsilon }\ \mathbf{\ ^{\shortmid }}\partial ^{7}%
\mathbf{\ ^{\shortmid }}h^{8}}d\mathbf{\ ^{\shortmid }}x^{i_{3}}\}^{2}+ 
\notag \\
&&\mathbf{\ ^{\shortmid }}h^{8}\{dE+[\ _{1}^{\shortmid }n_{k_{3}}+\
_{2}^{\shortmid }n_{k_{3}}\int dp_{7}\frac{(\mathbf{\ ^{\shortmid }}\partial
^{7}\mathbf{\ ^{\shortmid }}h^{8})^{2}}{|\int dp_{7}\mathbf{\ ^{\shortmid }}%
\partial ^{7}[\ \ _{4}^{\shortmid }\widehat{\Upsilon }\mathbf{\ ^{\shortmid }%
}h^{8}]|\ (\mathbf{\ ^{\shortmid }}h^{8})^{5/2}}]d\mathbf{\ ^{\shortmid }}%
x^{k_{3}}\}.  \notag
\end{eqnarray}%
Such s-metrics possess nonlinear symmetries in phase spaces which allow to
re-define the generating functions and generating sources and related them
to conventions cosmological constants $\ _{s}^{\shortmid }\Lambda .$

\subsubsection{Quasi-stationary and rainbow phase space solutions}

Another class of quasi-stationary momentum phase space solutions of type (%
\ref{dmqc8}) can be generated if in the \ s-metric (\ref{qcosm8d7}) we
change the Killing symmetry on $\mathbf{\ ^{\shortmid }}\partial ^{8}$ into $%
\ ^{\shortmid }\partial ^{7}$ and introduce in the shell $s=4$ dependencies
on $E$-variable (in literature called rainbow metrics). Corresponding
coefficients of the geometric s-objects will be underlined. Respectively,
for such phase space configurations, the constructions stated by Table 9 and
s-metric (\ref{qst8d8}) transform via duality transforms $%
v^{a_{s}}\rightarrow p_{a_{s}}$ into those for Table 14.

Chronologically, we note that rainbow s-metrics in generalized
Finsler-Lagrange and dual Cartan-Hamilton forms were constructed following
different nonholonomic parameterizations in \cite{vmon3,sv07}, see further
developments and review of results in \cite{stavr14,vacaru18,stavrinos21}.
The cosmological scenarios elaborated in \cite%
{sv14,sv15a,elizalde15,sv16,rajpoot17,sv18} can be re-defined on $%
T_{s}^{\ast }V.$ They can be exploited as some alternative models of dark
matter and dark energy theories when the structure formation and phase space
dynamics depend on certain $E$ type variables/ coordinates.

The rainbow type solutions (for toy models) (\ref{dmcc}), (\ref{qeltorsd})
and (\ref{4ds}) can be re-defined into quasi-stationary, or $t$-depending,
and/or $E$-depending s-metrics with effective shell cosmological constants,
and related generating functions, or $\eta $- and $\chi $-polarization
functions.

\newpage

{\scriptsize 
\begin{eqnarray*}
&&%
\begin{tabular}{l}
\hline\hline
\begin{tabular}{lll}
& {\large \textsf{Table 14:\ Off-diagonal quasi-stationary and pase space
configurations with variable energy}} &  \\ 
& Exact solutions of $\mathbf{\ ^{\shortmid }}\widehat{\mathbf{R}}_{\mu
_{s}\nu _{s}}=\mathbf{\ ^{\shortmid }\Upsilon }_{\mu _{s}\nu _{s}}$ (\ref%
{cdeq1c8}) on $T_{s}^{\ast }V$ transformed into a momentum version of
nonlinear PDEs (\ref{eq1})-(\ref{e2c}) & 
\end{tabular}
\\ 
\end{tabular}
\\
&&%
\begin{tabular}{lll}
\hline\hline
&  &  \\ 
$%
\begin{array}{c}
\mbox{d-metric ansatz with} \\ 
\mbox{Killing symmetry }\partial _{4}=\partial _{t},\mathbf{\ ^{\shortmid }}%
\partial ^{7}%
\end{array}%
$ &  & $%
\begin{array}{c}
ds^{2}=g_{i_{1}}(x^{k_{1}})(dx^{i_{1}})^{2}+g_{a_{2}}(x^{k_{1}},y^{3})(dy^{a_{2}}+N_{i_{1}}^{a_{2}}(x^{k_{1}},y^{3})dx^{i_{1}})^{2}
\\ 
+\mathbf{\ ^{\shortmid }}g^{a_{3}}(x^{k_{2}},p_{5})(dp_{a_{3}}+\mathbf{\
^{\shortmid }}N_{i_{2}a_{3}}(x^{k_{2}},p_{5})dx^{i_{2}})^{2} \\ 
+\mathbf{\ ^{\shortmid }}g^{a_{4}}(\mathbf{\ ^{\shortmid }}%
x^{k_{3}},p_{7})(dp_{a_{4}}+\mathbf{\ ^{\shortmid }}N_{i_{3}a_{4}}(\mathbf{\
^{\shortmid }}x^{k_{3}},p_{7})d\mathbf{\ ^{\shortmid }}x^{i_{3}})^{2},%
\mbox{
for }g_{i_{1}}=e^{\psi {(x}^{k_{1}}{)}}, \\ 
g_{a_{2}}=h_{a_{2}}(x^{k_{1}},y^{3}),N_{i_{1}}^{3}=\
^{2}w_{i_{1}}=w_{i_{1}}(x^{k_{1}},y^{3}), \\ 
N_{i_{1}}^{4}=\ ^{2}n_{i_{1}}=n_{i_{1}}(x^{k_{1}},y^{3}), \\ 
\mathbf{\ ^{\shortmid }}g^{a_{3}}=\mathbf{\ ^{\shortmid }}%
h^{a_{3}}(x^{k_{2}},p_{5}),\mathbf{\ ^{\shortmid }}N_{i_{2}5}=\ _{\shortmid
}^{3}w_{i_{2}}=\mathbf{\ ^{\shortmid }}w_{i_{2}}(x^{k_{2}},p_{5}), \\ 
\mathbf{\ ^{\shortmid }}N_{i_{2}6}=\ _{\shortmid }^{3}n_{i_{2}}=\mathbf{\
^{\shortmid }}n_{i_{2}}(x^{k_{2}},p_{5}), \\ 
\mathbf{\ ^{\shortmid }}\underline{g}^{a_{4}}=\mathbf{\ ^{\shortmid }}%
\underline{h}^{a_{4}}(\mathbf{\ ^{\shortmid }}x^{k_{3}},E),\mathbf{\
^{\shortmid }}\underline{N}_{i_{3}7}=\ \ _{\shortmid }^{4}\underline{n}%
_{i_{3}}=\mathbf{\ ^{\shortmid }}\underline{n}_{i_{3}}(x^{k_{3}},E), \\ 
\mathbf{\ ^{\shortmid }}\underline{N}_{i_{3}8}=\ _{\shortmid }^{4}\underline{%
w}_{i_{3}}=\mathbf{\ ^{\shortmid }}\underline{w}_{i_{3}}(x^{k_{3}},E),%
\end{array}%
$ \\ 
Effective matter sources &  & $\mathbf{\ ^{\shortmid }\Upsilon }_{\ \nu
_{s}}^{\mu _{s}}=[\ _{1}\widehat{\Upsilon }({x}^{k_{1}})\delta
_{j_{1}}^{i_{1}},\ _{2}\widehat{\Upsilon }({x}^{k_{1}},y^{3})\delta
_{b_{2}}^{a_{2}},\ _{3}^{\shortmid }\widehat{\Upsilon }({x}%
^{k_{2}},p_{5})\delta _{b_{3}}^{a_{3}},\ _{4}^{\shortmid }\underline{%
\widehat{\Upsilon }}({x}^{k_{3}},E)\delta _{b_{4}}^{a_{4}},],$ \\ \hline
Nonlinear PDEs (\ref{eq1})-(\ref{e2c}) &  & $%
\begin{tabular}{lll}
$%
\begin{array}{c}
\psi ^{\bullet \bullet }+\psi ^{\prime \prime }=2\ \ _{1}\widehat{\Upsilon };
\\ 
\ ^{2}\varpi ^{\ast }\ h_{4}^{\ast }=2h_{3}h_{4}\ _{2}\widehat{\Upsilon };
\\ 
\ ^{2}\beta \ ^{2}w_{i_{1}}-\ ^{2}\alpha _{i_{1}}=0; \\ 
\ ^{2}n_{k_{1}}^{\ast \ast }+\ ^{2}\gamma \ ^{2}n_{k_{1}}^{\ast }=0;%
\end{array}%
$ &  & $%
\begin{array}{c}
\ ^{2}\varpi {=\ln |\partial _{3}h_{4}/\sqrt{|h_{3}h_{4}|}|,} \\ 
\ ^{2}\alpha _{i_{1}}=(\partial _{3}h_{4})\ (\partial _{i_{1}}\ ^{2}\varpi ),
\\ 
\ ^{2}\beta =(\partial _{3}h_{4})\ (\partial _{3}\ ^{2}\varpi ),\  \\ 
\ \ ^{2}\gamma =\partial _{3}\left( \ln |h_{4}|^{3/2}/|h_{3}|\right) , \\ 
\partial _{1}q=q^{\bullet },\partial _{2}q=q^{\prime },\partial
_{3}q=q^{\ast }%
\end{array}%
$ \\ 
$%
\begin{array}{c}
\mathbf{\ ^{\shortmid }}\partial ^{5}(\ _{\shortmid }^{3}\varpi )\ \mathbf{\
^{\shortmid }}\partial ^{5}\mathbf{\ ^{\shortmid }}h^{6}=2\mathbf{\
^{\shortmid }}h^{5}\mathbf{\ ^{\shortmid }}h^{6}\ \ _{3}^{\shortmid }%
\widehat{\Upsilon }; \\ 
\ _{\shortmid }^{3}\beta \ _{\shortmid }^{3}w_{i_{2}}-\ _{\shortmid
}^{3}\alpha _{i_{2}}=0; \\ 
\mathbf{\ ^{\shortmid }}\partial ^{5}(\mathbf{\ ^{\shortmid }}\partial ^{5}\
_{\shortmid }^{3}n_{k_{2}})+\ _{\shortmid }^{3}\gamma \mathbf{\ ^{\shortmid }%
}\partial ^{5}(\ _{\shortmid }^{3}n_{k_{2}})=0;%
\end{array}%
$ &  & $%
\begin{array}{c}
\\ 
\ _{\shortmid }^{3}\varpi {=\ln |\mathbf{\ ^{\shortmid }}\partial ^{5}%
\mathbf{\ ^{\shortmid }}h^{6}/\sqrt{|\mathbf{\ ^{\shortmid }}h^{5}\mathbf{\
^{\shortmid }}h^{6}|}|,} \\ 
\ _{\shortmid }^{3}\alpha _{i_{2}}=(\mathbf{\ ^{\shortmid }}\partial ^{5}%
\mathbf{\ ^{\shortmid }}h^{6})\ (\partial _{i_{2}}\ _{\shortmid }^{3}\varpi
), \\ 
\ _{\shortmid }^{3}\beta =(\mathbf{\ ^{\shortmid }}\partial ^{5}\mathbf{\
^{\shortmid }}h^{6})\ (\mathbf{\ ^{\shortmid }}\partial ^{5}\ _{\shortmid
}^{3}\varpi ),\  \\ 
\ \ _{\shortmid }^{3}\gamma =\mathbf{\ ^{\shortmid }}\partial ^{5}\left( \ln
|\mathbf{\ ^{\shortmid }}h^{6}|^{3/2}/|\mathbf{\ ^{\shortmid }}h^{5}|\right)
,%
\end{array}%
$ \\ 
$%
\begin{array}{c}
\mathbf{\ ^{\shortmid }}\underline{\partial }^{8}(\ _{\shortmid }^{4}%
\underline{\varpi })\ \mathbf{\ ^{\shortmid }}\underline{\partial }^{8}%
\mathbf{\ ^{\shortmid }}\underline{h}^{7}=2\ \mathbf{^{\shortmid }}%
\underline{h}^{7}\mathbf{\ ^{\shortmid }}\underline{h}^{8}\ _{4}^{\shortmid }%
\underline{\widehat{\Upsilon }}; \\ 
\mathbf{\ ^{\shortmid }}\underline{\partial }^{8}(\mathbf{\ ^{\shortmid }}%
\underline{\partial }^{8}\ _{\shortmid }^{4}\underline{n}_{k_{3}})+\
_{\shortmid }^{4}\underline{\gamma }\mathbf{\ ^{\shortmid }}\underline{%
\partial }^{8}(\ _{\shortmid }^{4}\underline{n}_{k_{3}})=0; \\ 
\ _{\shortmid }^{4}\underline{\beta }\ _{\shortmid }^{4}\underline{w}%
_{i_{3}}-\ _{\shortmid }^{4}\underline{\alpha }_{i_{3}}=0;%
\end{array}%
$ &  & $%
\begin{array}{c}
\\ 
\ _{\shortmid }^{4}\underline{\varpi }{=\ln |\mathbf{\ ^{\shortmid }}%
\underline{{\partial }}^{8}\mathbf{\ ^{\shortmid }}\underline{{h}}^{7}/\sqrt{%
|\mathbf{\ ^{\shortmid }}\underline{h}^{7}\mathbf{\ ^{\shortmid }}\underline{%
h}^{8}|}|,} \\ 
\ _{\shortmid }^{4}\underline{\alpha }_{i_{3}}=(\mathbf{\ ^{\shortmid }}%
\underline{\partial }^{8}\mathbf{\ ^{\shortmid }}\underline{h}^{7})\ (%
\mathbf{\ ^{\shortmid }}\partial _{i_{3}}\ _{\shortmid }^{4}\underline{%
\varpi }), \\ 
\ _{\shortmid }^{4}\underline{\beta }=(\mathbf{\ ^{\shortmid }}\underline{%
\partial }^{8}\mathbf{\ ^{\shortmid }}\underline{h}^{7})\ (\mathbf{\
^{\shortmid }}\underline{\partial }^{8}\ _{\shortmid }^{4}\underline{\varpi }%
),\  \\ 
\ \ _{\shortmid }^{4}\underline{\gamma }=\mathbf{\ ^{\shortmid }}\underline{%
\partial }^{8}\left( \ln |\mathbf{\ ^{\shortmid }}\underline{h}^{7}|^{3/2}/|%
\mathbf{\ ^{\shortmid }}\underline{h}^{8}|\right) ,%
\end{array}%
$%
\end{tabular}%
$ \\ \hline
$%
\begin{array}{c}
\mbox{ Gener.  functs:}\ h_{3}(x^{k_{1}},y^{3}), \\ 
\ ^{2}\Psi (x^{k_{1}},y^{3})=e^{\ ^{2}\varpi },\ ^{2}\Phi (x^{k_{1}},y^{3}),
\\ 
\mbox{integr. functs:}\ h_{4}^{[0]}(x^{k_{1}}),\  \\ 
_{1}n_{k_{1}}(x^{i_{1}}),\ _{2}n_{k_{1}}(x^{i_{1}}); \\ 
\mbox{ Gener.  functs:}\mathbf{\ ^{\shortmid }}h^{5}(x^{k_{2}},p_{5}), \\ 
\ \ _{\shortmid }^{3}\Psi (x^{k_{2}},p_{5})=e^{\ \ _{\shortmid }^{3}\varpi
},\ \ \ _{\shortmid }^{3}\Phi (x^{k_{2}},p_{5}), \\ 
\mbox{integr. functs:}\ h_{6}^{[0]}(x^{k_{2}}),\  \\ 
_{1}^{3}n_{k_{2}}(x^{i_{2}}),\ _{2}^{3}n_{k_{2}}(x^{i_{2}}); \\ 
\mbox{ Gener.  functs:}\mathbf{\ ^{\shortmid }}h^{7}(\mathbf{\ ^{\shortmid }}%
x^{k_{3}},p_{7}), \\ 
\ \ _{\shortmid }^{4}\underline{\Psi }(x^{k_{2}},E)=e^{\ \ \ _{\shortmid
}^{4}\underline{\varpi }},\ \ \ _{\shortmid }^{4}\underline{\Phi }(\mathbf{\
^{\shortmid }}x^{k_{3}},E), \\ 
\mbox{integr. functs:}\ \underline{h}_{7}^{[0]}(\mathbf{\ ^{\shortmid }}%
x^{k_{3}}),\  \\ 
_{1}^{4}\underline{n}_{k_{3}}(\mathbf{\ ^{\shortmid }}x^{i_{3}}),\ _{2}^{4}%
\underline{n}_{k_{3}}(\mathbf{\ ^{\shortmid }}x^{i_{3}}); \\ 
\mbox{\& nonlinear symmetries}%
\end{array}%
$ &  & $%
\begin{array}{c}
\ ((\ ^{2}\Psi )^{2})^{\ast }=-\int dy^{3}\ _{2}\widehat{\Upsilon }h_{4}^{\
\ast }, \\ 
(\ ^{2}\Phi )^{2}=-4\ _{2}\Lambda h_{4},\mbox{ see }(\ref{nonlinsymrex}), \\ 
h_{4}=h_{4}^{[0]}-(\ ^{2}\Phi )^{2}/4\ _{2}\Lambda ,h_{4}^{\ast }\neq 0,\
_{2}\Lambda \neq 0=const; \\ 
\\ 
\mathbf{\ ^{\shortmid }}\partial ^{5}((\ \ _{\shortmid }^{3}\Psi
)^{2})=-\int dp_{5}\ _{3}^{\shortmid }\widehat{\Upsilon }\mathbf{\
^{\shortmid }}\partial ^{5}\mathbf{\ ^{\shortmid }}h^{6}, \\ 
(\ \ _{\shortmid }^{3}\Phi )^{2}=-4\ _{3}^{\shortmid }\Lambda \mathbf{\
^{\shortmid }}h^{6}, \\ 
\mathbf{\ ^{\shortmid }}h^{6}=\mathbf{\ ^{\shortmid }}h_{[0]}^{6}-(\ \
_{\shortmid }^{3}\Phi )^{2}/4\ _{3}\Lambda ,\mathbf{\ ^{\shortmid }}\partial
^{5}\mathbf{\ ^{\shortmid }}h^{6}\neq 0,\ _{3}^{\shortmid }\Lambda \neq
0=const; \\ 
\\ 
\mathbf{\ ^{\shortmid }}\underline{\partial }^{8}((\ _{\shortmid }^{4}%
\underline{\Psi })^{2})=-\int dE\ _{4}^{\shortmid }\underline{\widehat{%
\Upsilon }}\mathbf{\ ^{\shortmid }}\underline{\partial }^{8}\mathbf{\
^{\shortmid }}\underline{h}^{7}, \\ 
(\ _{\shortmid }^{4}\underline{\Phi })^{2}=-4\ _{4}^{\shortmid }\underline{%
\Lambda }\mathbf{\ ^{\shortmid }}\underline{h}^{7}, \\ 
\mathbf{\ ^{\shortmid }}\underline{h}^{7}=\mathbf{\ ^{\shortmid }}\underline{%
h}_{[0]}^{7}-(\ _{\shortmid }^{4}\underline{\Phi })^{2}/4\ _{4}^{\shortmid }%
\underline{\Lambda },\mathbf{\ ^{\shortmid }}\underline{\partial }^{8}%
\mathbf{\ ^{\shortmid }}\underline{h}^{7}\neq 0,\ _{4}^{\shortmid }%
\underline{\Lambda }\neq 0=const;%
\end{array}%
$ \\ \hline
Off-diag. solutions, $%
\begin{array}{c}
\mbox{d--metric} \\ 
\mbox{N-connec.}%
\end{array}%
$ &  & $%
\begin{tabular}{l}
$%
\begin{array}{c}
\ g_{i}=e^{\ \psi (x^{k})}\mbox{ as a solution of 2-d Poisson eqs. }\psi
^{\bullet \bullet }+\psi ^{\prime \prime }=2~\ _{1}\widehat{\Upsilon }; \\ 
h_{3}=-(\Psi ^{\ast })^{2}/4\ _{2}\widehat{\Upsilon }^{2}h_{4},\mbox{ see }(%
\ref{g3}),(\ref{g4}); \\ 
h_{4}=h_{4}^{[0]}-\int dy^{3}(\Psi ^{2})^{\ast }/4\ _{2}\widehat{\Upsilon }%
=h_{4}^{[0]}-\Phi ^{2}/4\ _{2}\Lambda ; \\ 
w_{i}=\partial _{i}\ \Psi /\ \partial _{3}\Psi =\partial _{i}\ \Psi ^{2}/\
\partial _{3}\Psi ^{2}|; \\ 
n_{k}=\ _{1}n_{k}+\ _{2}n_{k}\int dy^{3}(\Psi ^{\ast })^{2}/\ _{2}\widehat{%
\Upsilon }^{2}|h_{4}^{[0]}-\int dy^{3}(\Psi ^{2})^{\ast }/4\ _{2}\widehat{%
\Upsilon }^{2}|^{5/2};%
\end{array}%
$ \\ 
$%
\begin{array}{c}
\mathbf{\ ^{\shortmid }}h^{5}=-(\mathbf{\ ^{\shortmid }}\partial ^{5}\
_{\shortmid }^{3}\Psi )^{2}/4\ _{3}^{\shortmid }\widehat{\Upsilon }^{2}%
\mathbf{\ ^{\shortmid }}h^{6}; \\ 
\mathbf{\ ^{\shortmid }}h^{6}=\mathbf{\ ^{\shortmid }}h_{[0]}^{6}-\int dp_{5}%
\mathbf{\ ^{\shortmid }}\partial ^{5}((\ \ _{\shortmid }^{3}\Psi )^{2})/4\
_{3}^{\shortmid }\widehat{\Upsilon }=\mathbf{\ ^{\shortmid }}h_{[0]}^{6}-(\
\ _{\shortmid }^{3}\Phi )^{2}/4\ _{3}^{\shortmid }\Lambda ; \\ 
w_{i_{2}}=\partial _{i_{2}}(\ _{\shortmid }^{3}\Psi )/\mathbf{\ ^{\shortmid }%
}\partial ^{5}(\ _{\shortmid }^{3}\Psi )=\partial _{i_{2}}(\ _{\shortmid
}^{3}\Psi )^{2}/\ \mathbf{\ ^{\shortmid }}\partial ^{5}(\ _{\shortmid
}^{3}\Psi )^{2}|; \\ 
n_{k_{2}}=\ _{1}n_{k_{2}}+\ _{2}n_{k_{2}}\int dp_{5}(\mathbf{\ ^{\shortmid }}%
\partial ^{5}\ _{\shortmid }^{3}\Psi )^{2}/\ _{3}^{\shortmid }\widehat{%
\Upsilon }^{2}|\mathbf{\ ^{\shortmid }}h_{[0]}^{6}- \\ 
\int dp_{5}\mathbf{\ ^{\shortmid }}\partial ^{5}((\ _{\shortmid }^{3}\Psi
)^{2})/4\ _{3}^{\shortmid }\widehat{\Upsilon }^{2}|^{5/2}%
\end{array}%
$ \\ 
$%
\begin{array}{c}
\mathbf{\ ^{\shortmid }}\underline{h}^{8}=-(\mathbf{\ ^{\shortmid }}%
\underline{\partial }^{8}\ _{\shortmid }^{4}\underline{\Psi })^{2}/4\
_{\shortmid }^{4}\underline{\widehat{\Upsilon }}^{2}\mathbf{\ ^{\shortmid }}%
\underline{h}^{7}; \\ 
\mathbf{\ ^{\shortmid }}\underline{h}^{7}=\mathbf{\ ^{\shortmid }}\underline{%
h}_{[0]}^{7}-\int dE\mathbf{\ ^{\shortmid }}\underline{\partial }^{8}((\
_{\shortmid }^{4}\underline{\Psi })^{2})/4\ _{4}^{\shortmid }\underline{%
\widehat{\Upsilon }}=\underline{h}_{[0]}^{7}-(\ _{\shortmid }^{4}\underline{%
\Phi })^{2}/4\ \ _{4}^{\shortmid }\underline{\Lambda }; \\ 
\mathbf{\ ^{\shortmid }}\underline{n}_{k_{3}}=\ _{1}^{\shortmid }\underline{n%
}_{k_{3}}+\ _{2}^{\shortmid }\underline{n}_{k_{3}}\int dE(\ _{\shortmid }^{4}%
\underline{\Psi })^{2}/\ _{4}^{\shortmid }\underline{\widehat{\Upsilon }}%
^{2}|\underline{h}_{[0]}^{7}- \\ 
\int dE\mathbf{\ ^{\shortmid }}\underline{\partial }^{8}((\ _{\shortmid }^{4}%
\underline{\Psi })^{2})/4\ _{4}^{\shortmid }\underline{\widehat{\Upsilon }}%
^{2}|^{5/2}; \\ 
\mathbf{\ ^{\shortmid }}\underline{w}_{i_{3}}=\mathbf{\ ^{\shortmid }}%
\partial _{i_{3}}(\ \ _{\shortmid }^{4}\underline{\Psi })/\ \mathbf{\
^{\shortmid }}\partial ^{8}(\ _{\shortmid }^{4}\underline{\Psi })=\mathbf{\
^{\shortmid }}\partial _{i_{3}}(\ _{\shortmid }^{4}\underline{\Psi })^{2}/%
\mathbf{\ ^{\shortmid }}\partial ^{8}(\ _{\shortmid }^{4}\underline{\Psi }%
)^{2}|.%
\end{array}%
$%
\end{tabular}%
$ \\ \hline\hline
\end{tabular}%
\end{eqnarray*}%
}A typical quasi-stationary rainbow metric on $T^{\ast }\mathbf{V}$
constructed for changing indices $7\longleftrightarrow 8$ and respective
dependencies on coordinates and Killing symmetry on $s=4$ in (\ref{qcosm8d7}%
) is defined by such a s-metric with explicit dependence on $E$-variable: 
\begin{eqnarray}
d\widehat{s}_{[8d]}^{2} &=&\widehat{g}_{\alpha _{s}\beta
_{s}}(x^{k},y^{3},p_{5},E;h_{4},\mathbf{\ ^{\shortmid }}h^{6},\mathbf{\
^{\shortmid }}\underline{h}^{7};\ _{1}^{\shortmid }\widehat{\Upsilon },\
_{2}^{\shortmid }\widehat{\Upsilon },\ _{3}^{\shortmid }\widehat{\Upsilon }%
,\ _{4}^{\shortmid }\underline{\widehat{\Upsilon }};\ _{1}^{\shortmid
}\Lambda ,\ _{2}^{\shortmid }\Lambda ,\ _{3}^{\shortmid }\Lambda ,\
_{4}^{\shortmid }\underline{\Lambda })d\mathbf{\ ^{\shortmid }}u^{\alpha
_{s}}d\mathbf{\ ^{\shortmid }}u^{\beta _{s}}  \label{qcosmrain8d8} \\
&=&e^{\psi (x^{k},\ _{s}\widehat{\Upsilon })}[(dx^{1})^{2}+(dx^{2})^{2}]-%
\frac{(h_{4}^{\ast })^{2}}{|\int dy^{3}[\ _{2}\widehat{\Upsilon }%
h_{4}]^{\ast }|\ h_{4}}\{dy^{3}+\frac{\partial _{i_{1}}[\int dy^{3}(\ _{2}%
\widehat{\Upsilon })\ h_{4}^{\ast }]}{\ _{2}\widehat{\Upsilon }\ h_{4}^{\ast
}}dx^{i_{1}}\}^{2}+  \notag \\
&&h_{4}\{dt+[\ _{1}n_{k_{1}}+\ _{2}n_{k_{1}}\int dy^{3}\frac{(h_{4}^{\ast
})^{2}}{|\int dy^{3}[\ _{2}\widehat{\Upsilon }h_{4}]^{\ast }|\ (h_{4})^{5/2}}%
]dx_{1}^{k}\}+  \notag \\
&&\frac{(\mathbf{\ ^{\shortmid }}\partial ^{5}\mathbf{\ ^{\shortmid }}%
h^{6})^{2}}{|\int dp_{5}\mathbf{\ ^{\shortmid }}\partial ^{5}[\
_{3}^{\shortmid }\widehat{\Upsilon }\mathbf{\ ^{\shortmid }}h^{6}]|\ \mathbf{%
\ ^{\shortmid }}h^{6}}\{dp_{5}+\frac{\partial _{i_{2}}[\int dp_{5}(\ \
_{3}^{\shortmid }\widehat{\Upsilon })\mathbf{\ ^{\shortmid }}\partial ^{5}%
\mathbf{\ ^{\shortmid }}h^{6}]}{\ \ _{3}^{\shortmid }\widehat{\Upsilon }%
\mathbf{\ ^{\shortmid }}\partial ^{5}\ \mathbf{^{\shortmid }}h^{6}}%
dx^{i_{2}}\}^{2}+  \notag \\
&&\mathbf{\ ^{\shortmid }}h^{6}\{dp_{5}+[\ _{1}n_{k_{2}}+\ _{2}n_{k_{2}}\int
dp_{5}\frac{(\mathbf{\ ^{\shortmid }}\partial ^{5}\mathbf{\ ^{\shortmid }}%
h^{6})^{2}}{|\int dp_{5}\mathbf{\ ^{\shortmid }}\partial ^{5}[\ \
_{3}^{\shortmid }\widehat{\Upsilon }\mathbf{\ ^{\shortmid }}h^{6}]|\ (%
\mathbf{\ ^{\shortmid }}h^{6})^{5/2}}]dx^{k_{2}}\}+  \notag \\
&&\mathbf{\ ^{\shortmid }}\underline{h}^{7}\{dp_{7}+[\ _{1}^{\shortmid }%
\underline{n}_{k_{3}}+\ _{2}^{\shortmid }\underline{n}_{k_{3}}\int dp_{7}%
\frac{(\mathbf{\ ^{\shortmid }}\underline{\partial }^{8}\mathbf{\
^{\shortmid }}\underline{h}^{7})^{2}}{|\int dE\mathbf{\ ^{\shortmid }}%
\underline{\partial }^{8}[\ \ _{4}^{\shortmid }\underline{\widehat{\Upsilon }%
}\mathbf{\ ^{\shortmid }}\underline{h}^{7}]|\ (\mathbf{\ ^{\shortmid }}%
\underline{h}^{7})^{5/2}}]d\mathbf{\ ^{\shortmid }}x^{k_{3}}\}+  \notag \\
&&\frac{(\mathbf{\ ^{\shortmid }}\underline{\partial }^{8}\mathbf{\
^{\shortmid }}\underline{h}^{7})^{2}}{|\int dE\mathbf{\ ^{\shortmid }}%
\underline{\partial }^{8}[\ _{4}\underline{\widehat{\Upsilon }}\mathbf{\
^{\shortmid }}\underline{h}^{7}]|\ \mathbf{\ ^{\shortmid }}\underline{h}^{7}}%
\{dE+\frac{\partial _{i_{3}}[\int dE(\ _{4}\underline{\widehat{\Upsilon }})\ 
\mathbf{\ ^{\shortmid }}\underline{\partial }^{8}\mathbf{\ ^{\shortmid }}%
\underline{h}^{7}]}{\ \ _{4}^{\shortmid }\underline{\widehat{\Upsilon }}\ 
\mathbf{\ ^{\shortmid }}\underline{\partial }^{8}\mathbf{\ ^{\shortmid }}%
\underline{h}^{7}}d\mathbf{\ ^{\shortmid }}x^{i_{3}}\}^{2}.  \notag
\end{eqnarray}%
Such rainbow s-metrics can be re-parameterized for another types of
generating functions and/or with gravitational polarization functions using
respective nonlinear symmetries.

\subsubsection{Locally anisotropic cosmological solutions with fixed energy
parameter}

For dual fiber to cofiber transforms, the procedure for constructing locally
anisotropic cosmological phase space solutions described in Table 10
transforms into a method of generating such solutions with off-diagonal
dependence on momentum like variables. Such generalizations and applications
of the AFCDM are summarized in Table 15. As a $T^{\ast }\mathbf{V}$ analog
of the nonlinear quadratic element (\ref{qst8d7}), with $v^{8}=const,$ and
data from Table 8 we provide example of 8-d quasi-stationary quadratic
element (\ref{dmqc8}) with $p_{8}=E=const\mathbf{V},$

\begin{eqnarray}
d\widehat{s}_{[8d]}^{2} &=&\widehat{g}_{\alpha _{s}\beta
_{s}}(x^{k},t,p_{5},p_{7};\underline{h}_{3},\mathbf{\ ^{\shortmid }}h^{6},%
\mathbf{\ ^{\shortmid }}h^{8};\ _{s}^{\shortmid }\widehat{\Upsilon };\
_{1}\Lambda ,\ _{2}\underline{\Lambda },\ _{3}^{\shortmid }\Lambda ,\
_{4}^{\shortmid }\Lambda )d\mathbf{\ ^{\shortmid }}u^{\alpha _{s}}d\mathbf{\
^{\shortmid }}u^{\beta _{s}}  \label{lc8cstp7} \\
&=&e^{\psi (x^{k},\ _{s}\widehat{\Upsilon })}[(dx^{1})^{2}+(dx^{2})^{2}]+%
\underline{h}_{3}[dy^{3}+(\ _{1}n_{k_{1}}+4\ _{2}n_{k_{1}}\int dt\frac{(%
\underline{h}_{3}{}^{\diamond })^{2}}{|\int dt\ _{2}\underline{\Upsilon }%
\underline{h}_{3}{}^{\diamond }|(\underline{h}_{3})^{5/2}})dx^{k_{1}}] 
\notag \\
&&-\frac{(\underline{h}_{3}{}^{\diamond })^{2}}{|\int dt\ _{2}\underline{%
\Upsilon }\underline{h}_{3}{}^{\diamond }|\ \overline{h}_{3}}[dt+\frac{%
\partial _{i}(\int dt\ _{2}\underline{\Upsilon }\ \underline{h}%
_{3}{}^{\diamond }])}{\ \ _{2}\underline{\Upsilon }\ \underline{h}%
_{3}{}^{\diamond }}dx^{i}]+  \notag \\
&&\frac{(\mathbf{\ ^{\shortmid }}\partial ^{5}\mathbf{\ ^{\shortmid }}%
h^{6})^{2}}{|\int dp_{5}\mathbf{\ ^{\shortmid }}\partial ^{5}[\
_{3}^{\shortmid }\widehat{\Upsilon }\mathbf{\ ^{\shortmid }}h^{6}]|\ \mathbf{%
\ ^{\shortmid }}h^{6}}\{dp_{5}+\frac{\partial _{i_{2}}[\int dp_{5}(\ \
_{3}^{\shortmid }\widehat{\Upsilon })\mathbf{\ ^{\shortmid }}\partial ^{5}%
\mathbf{\ ^{\shortmid }}h^{6}]}{\ \ _{3}^{\shortmid }\widehat{\Upsilon }%
\mathbf{\ ^{\shortmid }}\partial ^{5}\ \mathbf{^{\shortmid }}h^{6}}%
dx^{i_{2}}\}^{2}+  \notag \\
&&\mathbf{\ ^{\shortmid }}h^{6}\{dp_{5}+[\ _{1}n_{k_{2}}+\ _{2}n_{k_{2}}\int
dp_{5}\frac{(\mathbf{\ ^{\shortmid }}\partial ^{5}\mathbf{\ ^{\shortmid }}%
h^{6})^{2}}{|\int dp_{5}\mathbf{\ ^{\shortmid }}\partial ^{5}[\ \
_{3}^{\shortmid }\widehat{\Upsilon }\mathbf{\ ^{\shortmid }}h^{6}]|\ (%
\mathbf{\ ^{\shortmid }}h^{6})^{5/2}}]dx^{k_{2}}\}+  \notag \\
&&\frac{(\mathbf{\ ^{\shortmid }}\partial ^{7}\mathbf{\ ^{\shortmid }}%
h^{8})^{2}}{|\int dp_{7}\mathbf{\ ^{\shortmid }}\partial ^{7}[\ _{4}\widehat{%
\Upsilon }\mathbf{\ ^{\shortmid }}h^{8}]|\ \mathbf{\ ^{\shortmid }}h^{8}}%
\{dp_{7}+\frac{\partial _{i_{3}}[\int dp_{7}(\ _{4}\widehat{\Upsilon })\ 
\mathbf{\ ^{\shortmid }}\partial ^{7}\mathbf{\ ^{\shortmid }}h^{8}]}{\ \
_{4}^{\shortmid }\widehat{\Upsilon }\ \mathbf{\ ^{\shortmid }}\partial ^{7}%
\mathbf{\ ^{\shortmid }}h^{8}}d\mathbf{\ ^{\shortmid }}x^{i_{3}}\}^{2}+ 
\notag \\
&&\mathbf{\ ^{\shortmid }}h^{8}\{dE+[\ _{1}^{\shortmid }n_{k_{3}}+\
_{2}^{\shortmid }n_{k_{3}}\int dp_{7}\frac{(\mathbf{\ ^{\shortmid }}\partial
^{7}\mathbf{\ ^{\shortmid }}h^{8})^{2}}{|\int dp_{7}\mathbf{\ ^{\shortmid }}%
\partial ^{7}[\ \ _{4}^{\shortmid }\widehat{\Upsilon }\mathbf{\ ^{\shortmid }%
}h^{8}]|\ (\mathbf{\ ^{\shortmid }}h^{8})^{5/2}}]d\mathbf{\ ^{\shortmid }}%
x^{k_{3}}\}.  \notag
\end{eqnarray}%
The procedure of generating such s-metrics is described as follow:

{\scriptsize 
\begin{eqnarray*}
&&%
\begin{tabular}{l}
\hline\hline
\begin{tabular}{lll}
& {\large \textsf{Table 15:\ Off-diagonal cosmological and pase space
configurations with fixed energy}} &  \\ 
& Exact solutions of $\mathbf{\ ^{\shortmid }}\widehat{\mathbf{R}}_{\mu
_{s}\nu _{s}}=\mathbf{\ ^{\shortmid }\Upsilon }_{\mu _{s}\nu _{s}}$ (\ref%
{cdeq1c8}) on $T_{s}^{\ast }V$ transformed into a momentum version of
nonlinear PDEs (\ref{eq1})-(\ref{e2c}) & 
\end{tabular}
\\ 
\end{tabular}
\\
&&%
\begin{tabular}{lll}
\hline\hline
&  &  \\ 
$%
\begin{array}{c}
\mbox{d-metric ansatz with} \\ 
\mbox{Killing symmetry }\partial _{3}=\partial _{t},\mathbf{\ ^{\shortmid }}%
\partial ^{8}%
\end{array}%
$ &  & $%
\begin{array}{c}
ds^{2}=g_{i_{1}}(x^{k_{1}})(dx^{i_{1}})^{2}+\underline{g}%
_{a_{2}}(x^{k_{1}},t)(dy^{a_{2}}+\underline{N}%
_{i_{1}}^{a_{2}}(x^{k_{1}},t)dx^{i_{1}})^{2} \\ 
+\mathbf{\ ^{\shortmid }}g^{a_{3}}(x^{k_{2}},p_{5})(dp_{a_{3}}+\mathbf{\
^{\shortmid }}N_{i_{2}a_{3}}(x^{k_{2}},p_{5})dx^{i_{2}})^{2} \\ 
+\mathbf{\ ^{\shortmid }}g^{a_{4}}(\mathbf{\ ^{\shortmid }}%
x^{k_{3}},p_{7})(dp_{a_{4}}+\mathbf{\ ^{\shortmid }}N_{i_{3}a_{4}}(\mathbf{\
^{\shortmid }}x^{k_{3}},p_{7})d\mathbf{\ ^{\shortmid }}x^{i_{3}})^{2},%
\mbox{
for }g_{i_{1}}=e^{\psi {(x}^{k_{1}}{)}}, \\ 
\underline{g}_{a_{2}}=\underline{h}_{a_{2}}(x^{k_{1}},t),\underline{N}%
_{i_{1}}^{3}=\ ^{2}\underline{n}_{i_{1}}=\underline{n}%
_{i_{1}}(x^{k_{1}},t),N_{i_{1}}^{4}=\ ^{2}\underline{w}_{i_{1}}=\underline{w}%
_{i_{1}}(x^{k_{1}},t), \\ 
\mathbf{\ ^{\shortmid }}g^{a_{3}}=\mathbf{\ ^{\shortmid }}%
h^{a_{3}}(x^{k_{2}},p_{5}),\mathbf{\ ^{\shortmid }}N_{i_{2}5}=\ _{\shortmid
}^{3}w_{i_{2}}=\mathbf{\ ^{\shortmid }}w_{i_{2}}(x^{k_{2}},p_{5}), \\ 
\mathbf{\ ^{\shortmid }}N_{i_{2}6}=\ _{\shortmid }^{3}n_{i_{2}}=\mathbf{\
^{\shortmid }}n_{i_{2}}(x^{k_{2}},p_{5}), \\ 
\mathbf{\ ^{\shortmid }}g^{a_{4}}=\mathbf{\ ^{\shortmid }}h^{a_{4}}(\mathbf{%
\ ^{\shortmid }}x^{k_{3}},p_{7}),\mathbf{\ ^{\shortmid }}N_{i_{3}7}=\ \
_{\shortmid }^{4}w_{i_{3}}=\mathbf{\ ^{\shortmid }}%
w_{i_{3}}(x^{k_{3}},p_{7}), \\ 
N_{i_{3}8}=\ _{\shortmid }^{4}n_{i_{3}}=\mathbf{\ ^{\shortmid }}%
n_{i_{3}}(x^{k_{3}},p_{7}),%
\end{array}%
$ \\ 
Effective matter sources &  & $\mathbf{\ ^{\shortmid }\Upsilon }_{\ \nu
_{s}}^{\mu _{s}}=[\ _{1}\widehat{\Upsilon }({x}^{k_{1}})\delta
_{j_{1}}^{i_{1}},\ _{2}\underline{\widehat{\Upsilon }}({x}^{k_{1}},t)\delta
_{b_{2}}^{a_{2}},\ _{3}^{\shortmid }\widehat{\Upsilon }({x}%
^{k_{2}},p_{5})\delta _{b_{3}}^{a_{3}},\ _{4}^{\shortmid }\widehat{\Upsilon }%
({x}^{k_{3}},p_{7})\delta _{b_{4}}^{a_{4}},],$ \\ \hline
Nonlinear PDEs (\ref{eq1})-(\ref{e2c}) &  & $%
\begin{tabular}{lll}
$%
\begin{array}{c}
\psi ^{\bullet \bullet }+\psi ^{\prime \prime }=2\ \ _{1}\widehat{\Upsilon };
\\ 
\ ^{2}\underline{\varpi }^{\diamond }\ \underline{h}_{3}^{\diamond }=2%
\underline{h}_{3}\underline{h}_{4}\ _{2}\widehat{\Upsilon }; \\ 
\ ^{2}\underline{n}_{k_{1}}^{\diamond \diamond }+\ ^{2}\underline{\gamma }\
^{2}\underline{n}_{k_{1}}^{\diamond }=0; \\ 
\ ^{2}\underline{\beta }\ ^{2}\underline{w}_{i_{1}}-\ ^{2}\underline{\alpha }%
_{i_{1}}=0;%
\end{array}%
$ &  & $%
\begin{array}{c}
\ ^{2}\underline{\varpi }{=\ln |\partial _{4}\underline{{h}}_{4}/\sqrt{|%
\underline{h}_{3}\underline{h}_{4}|}|,} \\ 
\ ^{2}\underline{\alpha }_{i_{1}}=(\partial _{4}\underline{h}_{3})\
(\partial _{i_{1}}\ ^{2}\underline{\varpi }), \\ 
\ ^{2}\underline{\beta }=(\partial _{4}\underline{h}_{4})\ (\partial _{3}\
^{2}\underline{\varpi }),\  \\ 
\ \ ^{2}\underline{\gamma }=\partial _{4}\left( \ln |\underline{h}%
_{3}|^{3/2}/|\underline{h}_{4}|\right) , \\ 
\partial _{1}q=q^{\bullet },\partial _{2}q=q^{\prime }, \\ 
\partial _{4}q=\partial _{t}q=q^{\diamond }%
\end{array}%
$ \\ 
$%
\begin{array}{c}
\mathbf{\ ^{\shortmid }}\partial ^{5}(\ _{\shortmid }^{3}\varpi )\ \mathbf{\
^{\shortmid }}\partial ^{5}\mathbf{\ ^{\shortmid }}h^{6}=2\mathbf{\
^{\shortmid }}h^{5}\mathbf{\ ^{\shortmid }}h^{6}\ \ _{3}^{\shortmid }%
\widehat{\Upsilon }; \\ 
\ _{\shortmid }^{3}\beta \ _{\shortmid }^{3}w_{i_{2}}-\ _{\shortmid
}^{3}\alpha _{i_{2}}=0; \\ 
\mathbf{\ ^{\shortmid }}\partial ^{5}(\mathbf{\ ^{\shortmid }}\partial ^{5}\
_{\shortmid }^{3}n_{k_{2}})+\ _{\shortmid }^{3}\gamma \mathbf{\ ^{\shortmid }%
}\partial ^{5}(\ _{\shortmid }^{3}n_{k_{2}})=0;%
\end{array}%
$ &  & $%
\begin{array}{c}
\\ 
\ _{\shortmid }^{3}\varpi {=\ln |\mathbf{\ ^{\shortmid }}\partial ^{5}%
\mathbf{\ ^{\shortmid }}h^{6}/\sqrt{|\mathbf{\ ^{\shortmid }}h^{5}\mathbf{\
^{\shortmid }}h^{6}|}|,} \\ 
\ _{\shortmid }^{3}\alpha _{i_{2}}=(\mathbf{\ ^{\shortmid }}\partial ^{5}%
\mathbf{\ ^{\shortmid }}h^{6})\ (\partial _{i_{2}}\ _{\shortmid }^{3}\varpi
), \\ 
\ _{\shortmid }^{3}\beta =(\mathbf{\ ^{\shortmid }}\partial ^{5}\mathbf{\
^{\shortmid }}h^{6})\ (\mathbf{\ ^{\shortmid }}\partial ^{5}\ _{\shortmid
}^{3}\varpi ),\  \\ 
\ \ _{\shortmid }^{3}\gamma =\mathbf{\ ^{\shortmid }}\partial ^{5}\left( \ln
|\mathbf{\ ^{\shortmid }}h^{6}|^{3/2}/|\mathbf{\ ^{\shortmid }}h^{5}|\right)
,%
\end{array}%
$ \\ 
$%
\begin{array}{c}
\mathbf{\ ^{\shortmid }}\partial ^{7}(\ _{\shortmid }^{4}\varpi )\ \mathbf{\
^{\shortmid }}\partial ^{7}\ \mathbf{\ ^{\shortmid }}h^{8}=2\mathbf{\
^{\shortmid }}h^{7}\mathbf{\ ^{\shortmid }}h^{8}\ \ \ _{4}^{\shortmid }%
\widehat{\Upsilon }; \\ 
\ _{\shortmid }^{4}\beta \ _{\shortmid }^{4}w_{i_{3}}-\ _{\shortmid
}^{4}\alpha _{i_{3}}=0; \\ 
\mathbf{\ ^{\shortmid }}\partial ^{7}(\mathbf{\ ^{\shortmid }}\partial ^{7}\
_{\shortmid }^{4}n_{k_{3}})+\ _{\shortmid }^{4}\gamma \mathbf{\ ^{\shortmid }%
}\partial ^{7}(\ _{\shortmid }^{4}n_{k_{3}})=0;%
\end{array}%
$ &  & $%
\begin{array}{c}
\\ 
\ _{\shortmid }^{4}\varpi {=\ln |\mathbf{\ ^{\shortmid }}\partial ^{7}%
\mathbf{\ ^{\shortmid }}h^{8}/\sqrt{|\mathbf{\ ^{\shortmid }}h^{7}\mathbf{\
^{\shortmid }}h^{8}|}|,} \\ 
\ _{\shortmid }^{4}\alpha _{i_{3}}=(\mathbf{\ ^{\shortmid }}\partial ^{7}%
\mathbf{\ ^{\shortmid }}h^{8})\ (\mathbf{\ ^{\shortmid }}\partial _{i_{3}}\
_{\shortmid }^{4}\varpi ), \\ 
\ _{\shortmid }^{4}\beta =(\mathbf{\ ^{\shortmid }}\partial ^{7}\mathbf{\
^{\shortmid }}h^{8})\ (\mathbf{\ ^{\shortmid }}\partial ^{7}\ _{\shortmid
}^{4}\varpi ),\  \\ 
\ \ _{\shortmid }^{4}\gamma =\mathbf{\ ^{\shortmid }}\partial ^{7}\left( \ln
|\mathbf{\ ^{\shortmid }}h^{8}|^{3/2}/|\mathbf{\ ^{\shortmid }}h^{7}|\right)
,%
\end{array}%
$%
\end{tabular}%
$ \\ \hline
$%
\begin{array}{c}
\mbox{ Gener.  functs:}\ \underline{h}_{4}(x^{k_{1}},t), \\ 
\ ^{2}\underline{\Psi }(x^{k_{1}},t)=e^{\ ^{2}\underline{\varpi }},\ ^{2}%
\underline{\Phi }(x^{k_{1}},t) \\ 
\mbox{integr. functs:}\ h_{4}^{[0]}(x^{k_{1}}),\  \\ 
_{1}n_{k_{1}}(x^{i_{1}}),\ _{2}n_{k_{1}}(x^{i_{1}}); \\ 
\mbox{ Gener.  functs:}\mathbf{\ ^{\shortmid }}h^{5}(x^{k_{2}},p_{5}), \\ 
\ \ _{\shortmid }^{3}\Psi (x^{k_{2}},p_{5})=e^{\ \ _{\shortmid }^{3}\varpi
},\ _{\shortmid }^{3}\Phi (x^{k_{2}},p_{5}) \\ 
\mbox{integr. functs:}\ h_{6}^{[0]}(x^{k_{2}}),\  \\ 
_{1}^{3}n_{k_{2}}(x^{i_{2}}),\ _{2}^{3}n_{k_{2}}(x^{i_{2}}); \\ 
\mbox{ Gener.  functs:}\mathbf{\ ^{\shortmid }}h^{7}(\mathbf{\ ^{\shortmid }}%
x^{k_{3}},p_{7}), \\ 
\ \ _{\shortmid }^{4}\Psi (x^{k_{2}},p_{7})=e^{\ \ \ _{\shortmid }^{4}\varpi
},\ \ \ _{\shortmid }^{4}\Phi (\mathbf{\ ^{\shortmid }}x^{k_{3}},p_{7}), \\ 
\mbox{integr. functs:}\ h_{8}^{[0]}(\mathbf{\ ^{\shortmid }}x^{k_{3}}),\  \\ 
_{1}^{4}n_{k_{3}}(\mathbf{\ ^{\shortmid }}x^{i_{3}}),\ _{2}^{4}n_{k_{3}}(%
\mathbf{\ ^{\shortmid }}x^{i_{3}}); \\ 
\mbox{\& nonlinear symmetries}%
\end{array}%
$ &  & $%
\begin{array}{c}
\ ((\ ^{2}\underline{\Psi })^{2})^{\diamond }=-\int dt\ _{2}\widehat{%
\underline{\Upsilon }}\underline{h}_{3}^{\ \diamond }, \\ 
(\ ^{2}\underline{\Phi })^{2}=-4\ _{2}\underline{\Lambda }\underline{h}_{3},
\\ 
h_{3}=h_{3}^{[0]}-(\ ^{2}\underline{\Phi })^{2}/4\ _{2}\underline{\Lambda },%
\underline{h}_{3}^{\diamond }\neq 0,\ _{2}\underline{\Lambda }\neq 0=const;
\\ 
\\ 
\mathbf{\ ^{\shortmid }}\partial ^{5}((\ \ _{\shortmid }^{3}\Psi
)^{2})=-\int dp_{5}\ _{3}^{\shortmid }\widehat{\Upsilon }\mathbf{\
^{\shortmid }}\partial ^{5}\mathbf{\ ^{\shortmid }}h^{6}, \\ 
(\ \ _{\shortmid }^{3}\Phi )^{2}=-4\ _{3}^{\shortmid }\Lambda \mathbf{\
^{\shortmid }}h^{6}, \\ 
\mathbf{\ ^{\shortmid }}h^{6}=\mathbf{\ ^{\shortmid }}h_{[0]}^{6}-(\ \
_{\shortmid }^{3}\Phi )^{2}/4\ _{3}\Lambda ,\mathbf{\ ^{\shortmid }}\partial
^{5}\mathbf{\ ^{\shortmid }}h^{6}\neq 0,\ _{3}^{\shortmid }\Lambda \neq
0=const; \\ 
\\ 
\mathbf{\ ^{\shortmid }}\partial ^{7}((\ _{\shortmid }^{4}\Psi )^{2})=-\int
dp_{7}\ _{4}^{\shortmid }\widehat{\Upsilon }\mathbf{\ ^{\shortmid }}\partial
^{7}\mathbf{\ ^{\shortmid }}h^{8}, \\ 
(\ _{\shortmid }^{4}\Phi )^{2}=-4\ _{4}^{\shortmid }\Lambda \mathbf{\
^{\shortmid }}h^{8}, \\ 
\mathbf{\ ^{\shortmid }}h^{8}=\mathbf{\ ^{\shortmid }}h_{[0]}^{8}-(\
_{\shortmid }^{4}\Phi )^{2}/4\ _{4}^{\shortmid }\Lambda ,\mathbf{\
^{\shortmid }}\partial ^{7}\mathbf{\ ^{\shortmid }}h^{8}\neq 0,\
_{4}^{\shortmid }\Lambda \neq 0=const;%
\end{array}%
$ \\ \hline
Off-diag. solutions, $%
\begin{array}{c}
\mbox{d--metric} \\ 
\mbox{N-connec.}%
\end{array}%
$ &  & $%
\begin{tabular}{l}
$%
\begin{array}{c}
\ g_{i}=e^{\ \psi (x^{k})}\mbox{ as a solution of 2-d Poisson eqs. }\psi
^{\bullet \bullet }+\psi ^{\prime \prime }=2~\ _{1}\widehat{\Upsilon }; \\ 
\underline{h}_{4}=-(\underline{\Psi }^{\diamond })^{2}/4\ _{2}\widehat{%
\underline{\Upsilon }}^{2}\underline{h}_{3}; \\ 
\underline{h}_{3}=\underline{h}_{3}^{[0]}-\int dt(\underline{\Psi }%
^{2})^{\diamond }/4\ _{2}\widehat{\underline{\Upsilon }}=\underline{h}%
_{3}^{[0]}-\underline{\Phi }^{2}/4\ _{2}\underline{\Lambda }; \\ 
\underline{w}_{i_{1}}=\partial _{i_{1}}\ \underline{\Psi }/\ \partial 
\underline{\Psi }^{\diamond }=\partial _{i_{1}}\ \underline{\Psi }^{2}/\
\partial _{t}\underline{\Psi }^{2}|; \\ 
\underline{n}_{k_{1}}=\ _{1}n_{k_{1}}+\ _{2}n_{k_{1}}\int dt(\underline{\Psi 
}^{\diamond })^{2}/\ _{2}\widehat{\underline{\Upsilon }}^{2}|\underline{h}%
_{3}^{[0]}-\int dt(\underline{\Psi }^{2})^{\diamond }/4\ _{2}\widehat{%
\underline{\Upsilon }}^{2}|^{5/2};%
\end{array}%
$ \\ 
$%
\begin{array}{c}
\mathbf{\ ^{\shortmid }}h^{5}=-(\mathbf{\ ^{\shortmid }}\partial ^{5}\
_{\shortmid }^{3}\Psi )^{2}/4\ _{3}^{\shortmid }\widehat{\Upsilon }^{2}%
\mathbf{\ ^{\shortmid }}h^{6}; \\ 
\mathbf{\ ^{\shortmid }}h^{6}=\mathbf{\ ^{\shortmid }}h_{[0]}^{6}-\int dp_{5}%
\mathbf{\ ^{\shortmid }}\partial ^{5}((\ \ _{\shortmid }^{3}\Psi )^{2})/4\
_{3}^{\shortmid }\widehat{\Upsilon }=\mathbf{\ ^{\shortmid }}h_{[0]}^{6}-(\
\ _{\shortmid }^{3}\Phi )^{2}/4\ _{3}^{\shortmid }\Lambda ; \\ 
w_{i_{2}}=\partial _{i_{2}}(\ _{\shortmid }^{3}\Psi )/\mathbf{\ ^{\shortmid }%
}\partial ^{5}(\ _{\shortmid }^{3}\Psi )=\partial _{i_{2}}(\ _{\shortmid
}^{3}\Psi )^{2}/\ \mathbf{\ ^{\shortmid }}\partial ^{5}(\ _{\shortmid
}^{3}\Psi )^{2}|; \\ 
n_{k_{2}}=\ _{1}n_{k_{2}}+\ _{2}n_{k_{2}}\int dp_{5}(\mathbf{\ ^{\shortmid }}%
\partial ^{5}\ _{\shortmid }^{3}\Psi )^{2}/\ _{3}^{\shortmid }\widehat{%
\Upsilon }^{2}|\mathbf{\ ^{\shortmid }}h_{[0]}^{6}- \\ 
\int dp_{5}\mathbf{\ ^{\shortmid }}\partial ^{5}((\ _{\shortmid }^{3}\Psi
)^{2})/4\ _{3}^{\shortmid }\widehat{\Upsilon }^{2}|^{5/2};%
\end{array}%
$ \\ 
$%
\begin{array}{c}
\mathbf{\ ^{\shortmid }}h^{7}=-(\mathbf{\ ^{\shortmid }}\partial ^{7}\
_{\shortmid }^{4}\Psi )^{2}/4\ _{\shortmid }^{4}\widehat{\Upsilon }^{2}%
\mathbf{\ ^{\shortmid }}h^{8}; \\ 
\mathbf{\ ^{\shortmid }}h^{8}=\mathbf{\ ^{\shortmid }}h_{[0]}^{8}-\int dp_{7}%
\mathbf{\ ^{\shortmid }}\partial ^{7}((\ _{\shortmid }^{4}\Psi )^{2})/4\
_{4}^{\shortmid }\widehat{\Upsilon }=h_{8}^{[0]}-(\ \ _{\shortmid }^{4}\Phi
)^{2}/4\ \ _{4}^{\shortmid }\Lambda ; \\ 
\mathbf{\ ^{\shortmid }}w_{i_{3}}=\mathbf{\ ^{\shortmid }}\partial
_{i_{3}}(\ \ _{\shortmid }^{4}\Psi )/\ \mathbf{\ ^{\shortmid }}\partial
^{7}(\ _{\shortmid }^{4}\Psi )=\mathbf{\ ^{\shortmid }}\partial _{i_{3}}(\ \
_{\shortmid }^{4}\Psi )^{2}/\ \mathbf{\ ^{\shortmid }}\partial ^{7}(\
_{\shortmid }^{4}\Psi )^{2}|; \\ 
\mathbf{\ ^{\shortmid }}n_{k_{3}}=\ _{1}^{\shortmid }n_{k_{3}}+\
_{2}^{\shortmid }n_{k_{3}}\int dp_{7}(\ _{\shortmid }^{4}\Psi )^{2}/\
_{4}^{\shortmid }\widehat{\Upsilon }^{2}|h_{8}^{[0]}- \\ 
\int dp_{7}\mathbf{\ ^{\shortmid }}\partial ^{7}((\ \ _{\shortmid }^{4}\Psi
)^{2})/4\ _{4}^{\shortmid }\widehat{\Upsilon }^{2}|^{5/2}.%
\end{array}%
$%
\end{tabular}%
$ \\ \hline\hline
\end{tabular}%
\end{eqnarray*}%
} The spacetime part in (\ref{lc8cstp7}) is equivalent to the spacetime part
of (\ref{lc8d7}) (in both cases, on shells $s=1,2$).

\subsubsection{Locally anisotropic cosmological solutions with variable
energy parameter}

The Table 16 is a momentum phase version of the Table 11. In this
subsection, it is summarized the AFCDM for constructing locally anisotropic
cosmological rainbow solutions. {\scriptsize 
\begin{eqnarray*}
&&%
\begin{tabular}{l}
\hline\hline
\begin{tabular}{lll}
& {\large \textsf{Table 16:\ Off-diagonal cosmological and pase space
configurations with variable energy}} &  \\ 
& Exact solutions of $\mathbf{\ ^{\shortmid }}\widehat{\mathbf{R}}_{\mu
_{s}\nu _{s}}=\mathbf{\ ^{\shortmid }\Upsilon }_{\mu _{s}\nu _{s}}$ (\ref%
{cdeq1c8}) on $T_{s}^{\ast }V$ transformed into a momentum version of
nonlinear PDEs (\ref{eq1})-(\ref{e2c}) & 
\end{tabular}
\\ 
\end{tabular}
\\
&&%
\begin{tabular}{lll}
\hline\hline
&  &  \\ 
$%
\begin{array}{c}
\mbox{d-metric ansatz with} \\ 
\mbox{Killing symmetry }\partial _{3}=\partial _{t},\mathbf{\ ^{\shortmid }}%
\partial ^{7}%
\end{array}%
$ &  & $%
\begin{array}{c}
ds^{2}=g_{i_{1}}(x^{k_{1}})(dx^{i_{1}})^{2}+\underline{g}%
_{a_{2}}(x^{k_{1}},t)(dy^{a_{2}}+\underline{N}%
_{i_{1}}^{a_{2}}(x^{k_{1}},t)dx^{i_{1}})^{2} \\ 
+\mathbf{\ ^{\shortmid }}g^{a_{3}}(x^{k_{2}},p_{5})(dp_{a_{3}}+\mathbf{\
^{\shortmid }}N_{i_{2}a_{3}}(x^{k_{2}},p_{5})dx^{i_{2}})^{2} \\ 
+\mathbf{\ ^{\shortmid }}g^{a_{4}}(\mathbf{\ ^{\shortmid }}%
x^{k_{3}},p_{7})(dp_{a_{4}}+\mathbf{\ ^{\shortmid }}N_{i_{3}a_{4}}(\mathbf{\
^{\shortmid }}x^{k_{3}},p_{7})d\mathbf{\ ^{\shortmid }}x^{i_{3}})^{2},%
\mbox{
for }g_{i_{1}}=e^{\psi {(x}^{k_{1}}{)}}, \\ 
\underline{g}_{a_{2}}=\underline{h}_{a_{2}}(x^{k_{1}},t),\underline{N}%
_{i_{1}}^{3}=\ ^{2}\underline{n}_{i_{1}}=\underline{n}_{i_{1}}(x^{k_{1}},t),%
\underline{N}_{i_{1}}^{4}=\ ^{2}\underline{w}_{i_{1}}=\underline{w}%
_{i_{1}}(x^{k_{1}},t), \\ 
\mathbf{\ ^{\shortmid }}g^{a_{3}}=\mathbf{\ ^{\shortmid }}%
h^{a_{3}}(x^{k_{2}},p_{5}),\mathbf{\ ^{\shortmid }}N_{i_{2}5}=\ _{\shortmid
}^{3}w_{i_{2}}=\mathbf{\ ^{\shortmid }}w_{i_{2}}(x^{k_{2}},p_{5}), \\ 
\mathbf{\ ^{\shortmid }}N_{i_{2}6}=\ _{\shortmid }^{3}n_{i_{2}}=\mathbf{\
^{\shortmid }}n_{i_{2}}(x^{k_{2}},p_{5}), \\ 
\mathbf{\ ^{\shortmid }}\underline{g}^{a_{4}}=\mathbf{\ ^{\shortmid }}%
\underline{h}^{a_{4}}(\mathbf{\ ^{\shortmid }}x^{k_{3}},E),\mathbf{\
^{\shortmid }}\underline{N}_{i_{3}7}=\ \ _{\shortmid }^{4}\underline{n}%
_{i_{3}}=\mathbf{\ ^{\shortmid }}\underline{n}_{i_{3}}(x^{k_{3}},E), \\ 
\mathbf{\ ^{\shortmid }}\underline{N}_{i_{3}8}=\ _{\shortmid }^{4}\underline{%
w}_{i_{3}}=\mathbf{\ ^{\shortmid }}\underline{w}_{i_{3}}(x^{k_{3}},E),%
\end{array}%
$ \\ 
Effective matter sources &  & $\mathbf{\ ^{\shortmid }\Upsilon }_{\ \nu
_{s}}^{\mu _{s}}=[\ _{1}\widehat{\Upsilon }({x}^{k_{1}})\delta
_{j_{1}}^{i_{1}},\ _{2}\widehat{\Upsilon }({x}^{k_{1}},y^{3})\delta
_{b_{2}}^{a_{2}},\ _{3}^{\shortmid }\widehat{\Upsilon }({x}%
^{k_{2}},p_{5})\delta _{b_{3}}^{a_{3}},\ _{4}^{\shortmid }\underline{%
\widehat{\Upsilon }}({x}^{k_{3}},E)\delta _{b_{4}}^{a_{4}},],$ \\ \hline
Nonlinear PDEs (\ref{eq1})-(\ref{e2c}) &  & $%
\begin{tabular}{lll}
$%
\begin{array}{c}
\psi ^{\bullet \bullet }+\psi ^{\prime \prime }=2\ \ _{1}\widehat{\Upsilon };
\\ 
\ ^{2}\underline{\varpi }^{\diamond }\ \underline{h}_{3}^{\diamond }=2%
\underline{h}_{3}\underline{h}_{4}\ _{2}\widehat{\Upsilon }; \\ 
\ ^{2}\underline{n}_{k_{1}}^{\diamond \diamond }+\ ^{2}\underline{\gamma }\
^{2}\underline{n}_{k_{1}}^{\diamond }=0; \\ 
\ ^{2}\underline{\beta }\ ^{2}\underline{w}_{i_{1}}-\ ^{2}\underline{\alpha }%
_{i_{1}}=0;%
\end{array}%
$ &  & $%
\begin{array}{c}
\ ^{2}\underline{\varpi }{=\ln |\partial _{4}\underline{{h}}_{4}/\sqrt{|%
\underline{h}_{3}\underline{h}_{4}|}|,} \\ 
\ ^{2}\underline{\alpha }_{i_{1}}=(\partial _{4}\underline{h}_{3})\
(\partial _{i_{1}}\ ^{2}\underline{\varpi }), \\ 
\ ^{2}\underline{\beta }=(\partial _{4}\underline{h}_{4})\ (\partial _{3}\
^{2}\underline{\varpi }),\  \\ 
\ \ ^{2}\underline{\gamma }=\partial _{4}\left( \ln |\underline{h}%
_{3}|^{3/2}/|\underline{h}_{4}|\right) , \\ 
\partial _{1}q=q^{\bullet },\partial _{2}q=q^{\prime }, \\ 
\partial _{4}q=\partial _{t}q=q^{\diamond }%
\end{array}%
$ \\ 
$%
\begin{array}{c}
\mathbf{\ ^{\shortmid }}\partial ^{5}(\ _{\shortmid }^{3}\varpi )\ \mathbf{\
^{\shortmid }}\partial ^{5}\mathbf{\ ^{\shortmid }}h^{6}=2\mathbf{\
^{\shortmid }}h^{5}\mathbf{\ ^{\shortmid }}h^{6}\ \ _{3}^{\shortmid }%
\widehat{\Upsilon }; \\ 
\ _{\shortmid }^{3}\beta \ _{\shortmid }^{3}w_{i_{2}}-\ _{\shortmid
}^{3}\alpha _{i_{2}}=0; \\ 
\mathbf{\ ^{\shortmid }}\partial ^{5}(\mathbf{\ ^{\shortmid }}\partial ^{5}\
_{\shortmid }^{3}n_{k_{2}})+\ _{\shortmid }^{3}\gamma \mathbf{\ ^{\shortmid }%
}\partial ^{5}(\ _{\shortmid }^{3}n_{k_{2}})=0;%
\end{array}%
$ &  & $%
\begin{array}{c}
\\ 
\ _{\shortmid }^{3}\varpi {=\ln |\mathbf{\ ^{\shortmid }}\partial ^{5}%
\mathbf{\ ^{\shortmid }}h^{6}/\sqrt{|\mathbf{\ ^{\shortmid }}h^{5}\mathbf{\
^{\shortmid }}h^{6}|}|,} \\ 
\ _{\shortmid }^{3}\alpha _{i_{2}}=(\mathbf{\ ^{\shortmid }}\partial ^{5}%
\mathbf{\ ^{\shortmid }}h^{6})\ (\partial _{i_{2}}\ _{\shortmid }^{3}\varpi
), \\ 
\ _{\shortmid }^{3}\beta =(\mathbf{\ ^{\shortmid }}\partial ^{5}\mathbf{\
^{\shortmid }}h^{6})\ (\mathbf{\ ^{\shortmid }}\partial ^{5}\ _{\shortmid
}^{3}\varpi ),\  \\ 
\ \ _{\shortmid }^{3}\gamma =\mathbf{\ ^{\shortmid }}\partial ^{5}\left( \ln
|\mathbf{\ ^{\shortmid }}h^{6}|^{3/2}/|\mathbf{\ ^{\shortmid }}h^{5}|\right)
,%
\end{array}%
$ \\ 
$%
\begin{array}{c}
\mathbf{\ ^{\shortmid }}\underline{\partial }^{8}(\ _{\shortmid }^{4}%
\underline{\varpi })\ \mathbf{\ ^{\shortmid }}\underline{\partial }^{8}%
\mathbf{\ ^{\shortmid }}\underline{h}^{7}=2\ \mathbf{^{\shortmid }}%
\underline{h}^{7}\mathbf{\ ^{\shortmid }}\underline{h}^{8}\ _{4}^{\shortmid }%
\underline{\widehat{\Upsilon }}; \\ 
\mathbf{\ ^{\shortmid }}\underline{\partial }^{8}(\mathbf{\ ^{\shortmid }}%
\underline{\partial }^{8}\ _{\shortmid }^{4}\underline{n}_{k_{3}})+\
_{\shortmid }^{4}\underline{\gamma }\mathbf{\ ^{\shortmid }}\underline{%
\partial }^{8}(\ _{\shortmid }^{4}\underline{n}_{k_{3}})=0; \\ 
\ _{\shortmid }^{4}\underline{\beta }\ _{\shortmid }^{4}\underline{w}%
_{i_{3}}-\ _{\shortmid }^{4}\underline{\alpha }_{i_{3}}=0;%
\end{array}%
$ &  & $%
\begin{array}{c}
\\ 
\ _{\shortmid }^{4}\underline{\varpi }{=\ln |\mathbf{\ ^{\shortmid }}%
\underline{{\partial }}^{8}\mathbf{\ ^{\shortmid }}\underline{{h}}^{7}/\sqrt{%
|\mathbf{\ ^{\shortmid }}\underline{h}^{7}\mathbf{\ ^{\shortmid }}\underline{%
h}^{8}|}|,} \\ 
\ _{\shortmid }^{4}\underline{\alpha }_{i_{3}}=(\mathbf{\ ^{\shortmid }}%
\underline{\partial }^{8}\mathbf{\ ^{\shortmid }}\underline{h}^{7})\ (%
\mathbf{\ ^{\shortmid }}\partial _{i_{3}}\ _{\shortmid }^{4}\underline{%
\varpi }), \\ 
\ _{\shortmid }^{4}\underline{\beta }=(\mathbf{\ ^{\shortmid }}\underline{%
\partial }^{8}\mathbf{\ ^{\shortmid }}\underline{h}^{7})\ (\mathbf{\
^{\shortmid }}\underline{\partial }^{8}\ _{\shortmid }^{4}\underline{\varpi }%
),\  \\ 
\ \ _{\shortmid }^{4}\underline{\gamma }=\mathbf{\ ^{\shortmid }}\underline{%
\partial }^{8}\left( \ln |\mathbf{\ ^{\shortmid }}\underline{h}^{7}|^{3/2}/|%
\mathbf{\ ^{\shortmid }}\underline{h}^{8}|\right) ,%
\end{array}%
$%
\end{tabular}%
$ \\ \hline
$%
\begin{array}{c}
\mbox{ Gener.  functs:}\ \underline{h}_{4}(x^{k_{1}},t), \\ 
\ ^{2}\underline{\Psi }(x^{k_{1}},t)=e^{\ ^{2}\underline{\varpi }},\ ^{2}%
\underline{\Phi }(x^{k_{1}},t), \\ 
\mbox{integr. functs:}\ \underline{h}_{3}^{[0]}(x^{k_{1}}),\  \\ 
_{1}\underline{n}_{k_{1}}(x^{i_{1}}),\ _{2}\underline{n}_{k_{1}}(x^{i_{1}});
\\ 
\mbox{ Gener.  functs:}\mathbf{\ ^{\shortmid }}h^{5}(x^{k_{2}},p_{5}), \\ 
\ \ _{\shortmid }^{3}\Psi (x^{k_{2}},p_{5})=e^{\ _{\shortmid }^{3}\varpi },\
_{\shortmid }^{3}\Phi (x^{k_{2}},p_{5}) \\ 
\mbox{integr. functs:}\ h_{6}^{[0]}(x^{k_{2}}),\  \\ 
_{1}^{3}n_{k_{2}}(x^{i_{2}}),\ _{2}^{3}n_{k_{2}}(x^{i_{2}}); \\ 
\mbox{ Gener.  functs:}\mathbf{\ ^{\shortmid }}h^{7}(\mathbf{\ ^{\shortmid }}%
x^{k_{3}},p_{7}), \\ 
\ \ _{\shortmid }^{4}\underline{\Psi }(x^{k_{2}},E)=e^{\ _{\shortmid }^{4}%
\underline{\varpi }},\ _{\shortmid }^{4}\underline{\Phi }(\mathbf{\
^{\shortmid }}x^{k_{3}},E) \\ 
\mbox{integr. functs:}\ \underline{h}_{7}^{[0]}(\mathbf{\ ^{\shortmid }}%
x^{k_{3}}),\  \\ 
_{1}^{4}\underline{n}_{k_{3}}(\mathbf{\ ^{\shortmid }}x^{i_{3}}),\ _{2}^{4}%
\underline{n}_{k_{3}}(\mathbf{\ ^{\shortmid }}x^{i_{3}}); \\ 
\mbox{\& nonlinear symmetries}%
\end{array}%
$ &  & $%
\begin{array}{c}
\ ((\ ^{2}\underline{\Psi })^{2})^{\diamond }=-\int dt\ _{2}\widehat{%
\underline{\Upsilon }}\underline{h}_{3}^{\ \diamond }, \\ 
(\ ^{2}\underline{\Phi })^{2}=-4\ _{2}\underline{\Lambda }\underline{h}_{3},
\\ 
h_{3}=h_{3}^{[0]}-(\ ^{2}\underline{\Phi })^{2}/4\ _{2}\underline{\Lambda },%
\underline{h}_{3}^{\diamond }\neq 0,\ _{2}\underline{\Lambda }\neq 0=const;
\\ 
\\ 
\mathbf{\ ^{\shortmid }}\partial ^{5}((\ \ _{\shortmid }^{3}\Psi
)^{2})=-\int dp_{5}\ _{3}^{\shortmid }\widehat{\Upsilon }\mathbf{\
^{\shortmid }}\partial ^{5}\mathbf{\ ^{\shortmid }}h^{6}, \\ 
(\ \ _{\shortmid }^{3}\Phi )^{2}=-4\ _{3}^{\shortmid }\Lambda \mathbf{\
^{\shortmid }}h^{6}, \\ 
\mathbf{\ ^{\shortmid }}h^{6}=\mathbf{\ ^{\shortmid }}h_{[0]}^{6}-(\ \
_{\shortmid }^{3}\Phi )^{2}/4\ _{3}\Lambda ,\mathbf{\ ^{\shortmid }}\partial
^{5}\mathbf{\ ^{\shortmid }}h^{6}\neq 0,\ _{3}^{\shortmid }\Lambda \neq
0=const; \\ 
\\ 
\mathbf{\ ^{\shortmid }}\underline{\partial }^{8}((\ _{\shortmid }^{4}%
\underline{\Psi })^{2})=-\int dE\ _{4}^{\shortmid }\underline{\widehat{%
\Upsilon }}\mathbf{\ ^{\shortmid }}\underline{\partial }^{8}\mathbf{\
^{\shortmid }}\underline{h}^{7}, \\ 
(\ _{\shortmid }^{4}\underline{\Phi })^{2}=-4\ _{4}^{\shortmid }\underline{%
\Lambda }\mathbf{\ ^{\shortmid }}\underline{h}^{7}, \\ 
\mathbf{\ ^{\shortmid }}\underline{h}^{7}=\mathbf{\ ^{\shortmid }}\underline{%
h}_{[0]}^{7}-(\ _{\shortmid }^{4}\underline{\Phi })^{2}/4\ _{4}^{\shortmid }%
\underline{\Lambda },\mathbf{\ ^{\shortmid }}\underline{\partial }^{8}%
\mathbf{\ ^{\shortmid }}\underline{h}^{7}\neq 0,\ _{4}^{\shortmid }%
\underline{\Lambda }\neq 0=const;%
\end{array}%
$ \\ \hline
Off-diag. solutions, $%
\begin{array}{c}
\mbox{d--metric} \\ 
\mbox{N-connec.}%
\end{array}%
$ &  & $%
\begin{tabular}{l}
$%
\begin{array}{c}
\ g_{i}=e^{\ \psi (x^{k})}\mbox{ as a solution of 2-d Poisson eqs. }\psi
^{\bullet \bullet }+\psi ^{\prime \prime }=2~\ _{1}\widehat{\Upsilon }; \\ 
\underline{h}_{4}=-(\underline{\Psi }^{\diamond })^{2}/4\ _{2}\widehat{%
\underline{\Upsilon }}^{2}\underline{h}_{3}; \\ 
\underline{h}_{3}=\underline{h}_{3}^{[0]}-\int dt(\underline{\Psi }%
^{2})^{\diamond }/4\ _{2}\widehat{\underline{\Upsilon }}=\underline{h}%
_{3}^{[0]}-\underline{\Phi }^{2}/4\ _{2}\underline{\Lambda }; \\ 
\underline{w}_{i_{1}}=\partial _{i_{1}}\ \underline{\Psi }/\ \partial 
\underline{\Psi }^{\diamond }=\partial _{i_{1}}\ \underline{\Psi }^{2}/\
\partial _{t}\underline{\Psi }^{2}|; \\ 
\underline{n}_{k_{1}}=\ _{1}n_{k_{1}}+\ _{2}n_{k_{1}}\int dt(\underline{\Psi 
}^{\diamond })^{2}/\ _{2}\widehat{\underline{\Upsilon }}^{2}|\underline{h}%
_{3}^{[0]}-\int dt(\underline{\Psi }^{2})^{\diamond }/4\ _{2}\widehat{%
\underline{\Upsilon }}^{2}|^{5/2};%
\end{array}%
$ \\ 
$%
\begin{array}{c}
\mathbf{\ ^{\shortmid }}h^{5}=-(\mathbf{\ ^{\shortmid }}\partial ^{5}\
_{\shortmid }^{3}\Psi )^{2}/4\ _{3}^{\shortmid }\widehat{\Upsilon }^{2}%
\mathbf{\ ^{\shortmid }}h^{6}; \\ 
\mathbf{\ ^{\shortmid }}h^{6}=\mathbf{\ ^{\shortmid }}h_{[0]}^{6}-\int dp_{5}%
\mathbf{\ ^{\shortmid }}\partial ^{5}((\ \ _{\shortmid }^{3}\Psi )^{2})/4\
_{3}^{\shortmid }\widehat{\Upsilon }=\mathbf{\ ^{\shortmid }}h_{[0]}^{6}-(\
\ _{\shortmid }^{3}\Phi )^{2}/4\ _{3}^{\shortmid }\Lambda ; \\ 
w_{i_{2}}=\partial _{i_{2}}(\ _{\shortmid }^{3}\Psi )/\mathbf{\ ^{\shortmid }%
}\partial ^{5}(\ _{\shortmid }^{3}\Psi )=\partial _{i_{2}}(\ _{\shortmid
}^{3}\Psi )^{2}/\ \mathbf{\ ^{\shortmid }}\partial ^{5}(\ _{\shortmid
}^{3}\Psi )^{2}|; \\ 
n_{k_{2}}=\ _{1}n_{k_{2}}+\ _{2}n_{k_{2}}\int dp_{5}(\mathbf{\ ^{\shortmid }}%
\partial ^{5}\ _{\shortmid }^{3}\Psi )^{2}/\ _{3}^{\shortmid }\widehat{%
\Upsilon }^{2}|\mathbf{\ ^{\shortmid }}h_{[0]}^{6}- \\ 
\int dp_{5}\mathbf{\ ^{\shortmid }}\partial ^{5}((\ _{\shortmid }^{3}\Psi
)^{2})/4\ _{3}^{\shortmid }\widehat{\Upsilon }^{2}|^{5/2};%
\end{array}%
$ \\ 
$%
\begin{array}{c}
\mathbf{\ ^{\shortmid }}\underline{h}^{8}=-(\mathbf{\ ^{\shortmid }}%
\underline{\partial }^{8}\ _{\shortmid }^{4}\underline{\Psi })^{2}/4\
_{\shortmid }^{4}\underline{\widehat{\Upsilon }}^{2}\mathbf{\ ^{\shortmid }}%
\underline{h}^{7}; \\ 
\mathbf{\ ^{\shortmid }}\underline{h}^{7}=\mathbf{\ ^{\shortmid }}\underline{%
h}_{[0]}^{7}-\int dE\mathbf{\ ^{\shortmid }}\underline{\partial }^{8}((\
_{\shortmid }^{4}\underline{\Psi })^{2})/4\ _{4}^{\shortmid }\underline{%
\widehat{\Upsilon }}=\underline{h}_{[0]}^{7}-(\ _{\shortmid }^{4}\underline{%
\Phi })^{2}/4\ \ _{4}^{\shortmid }\underline{\Lambda }; \\ 
\mathbf{\ ^{\shortmid }}\underline{n}_{k_{3}}=\ _{1}^{\shortmid }\underline{n%
}_{k_{3}}+\ _{2}^{\shortmid }\underline{n}_{k_{3}}\int dE(\ _{\shortmid }^{4}%
\underline{\Psi })^{2}/\ _{4}^{\shortmid }\underline{\widehat{\Upsilon }}%
^{2}|\underline{h}_{[0]}^{7}- \\ 
\int dE\mathbf{\ ^{\shortmid }}\underline{\partial }^{8}((\ _{\shortmid }^{4}%
\underline{\Psi })^{2})/4\ _{4}^{\shortmid }\underline{\widehat{\Upsilon }}%
^{2}|^{5/2}; \\ 
\mathbf{\ ^{\shortmid }}\underline{w}_{i_{3}}=\mathbf{\ ^{\shortmid }}%
\partial _{i_{3}}(\ \ _{\shortmid }^{4}\underline{\Psi })/\ \mathbf{\
^{\shortmid }}\partial ^{8}(\ _{\shortmid }^{4}\underline{\Psi })=\mathbf{\
^{\shortmid }}\partial _{i_{3}}(\ _{\shortmid }^{4}\underline{\Psi })^{2}/%
\mathbf{\ ^{\shortmid }}\partial ^{8}(\ _{\shortmid }^{4}\underline{\Psi }%
)^{2}|.%
\end{array}%
$%
\end{tabular}%
$ \\ \hline\hline
\end{tabular}%
\end{eqnarray*}%
}As an example of such s-metrics we provide below a cosmological rainbow
metric with the $s=1,2$ part being equivalent to 
\begin{eqnarray}
&&d\widehat{s}_{[8d]}^{2} =\widehat{g}_{\alpha _{s}\beta
_{s}}(x^{k},t,p_{5},E;\underline{h}_{3},\mathbf{\ ^{\shortmid }}h^{6},%
\mathbf{\ ^{\shortmid }}\underline{h}^{7};\ _{1}^{\shortmid }\widehat{%
\Upsilon },\ _{2}^{\shortmid }\underline{\widehat{\Upsilon }},\
_{3}^{\shortmid }\widehat{\Upsilon },\ _{4}^{\shortmid }\underline{\widehat{%
\Upsilon }};\ _{1}^{\shortmid }\Lambda ,\ _{2}^{\shortmid }\underline{%
\Lambda },\ _{3}^{\shortmid }\Lambda ,\ _{4}^{\shortmid }\underline{\Lambda }%
)d\mathbf{\ ^{\shortmid }}u^{\alpha _{s}}d\mathbf{\ ^{\shortmid }}u^{\beta
_{s}}  \label{lc8cstp8} \\
&&=e^{\psi (x^{k},\ _{s}\widehat{\Upsilon })}[(dx^{1})^{2}+(dx^{2})^{2}]+%
\underline{h}_{3}[dy^{3}+(\ _{1}n_{k_{1}}+4\ _{2}n_{k_{1}}\int dt\frac{(%
\underline{h}_{3}{}^{\diamond })^{2}}{|\int dt\ _{2}\underline{\widehat{%
\Upsilon }}\underline{h}_{3}{}^{\diamond }|(\underline{h}_{3})^{5/2}}%
)dx^{k_{1}}]+  \notag \\
&&\frac{(\underline{h}_{3}{}^{\diamond })^{2}}{|\int dt\ _{2}\underline{%
\widehat{\Upsilon }}\underline{h}_{3}{}^{\diamond }|\ \overline{h}_{3}}[dt+%
\frac{\partial _{i}(\int dt\ _{2}\underline{\Upsilon }\ \underline{h}%
_{3}{}^{\diamond }])}{\ \ _{2}\underline{\widehat{\Upsilon }}\ \underline{h}%
_{3}{}^{\diamond }}dx^{i}]+ \frac{(\mathbf{\ ^{\shortmid }}\partial ^{5}%
\mathbf{\ ^{\shortmid }}h^{6})^{2}}{|\int dp_{5}\mathbf{\ ^{\shortmid }}%
\partial ^{5}[\ _{3}^{\shortmid }\widehat{\Upsilon }\mathbf{\ ^{\shortmid }}%
h^{6}]|\ \mathbf{\ ^{\shortmid }}h^{6}}\{dp_{5}+\frac{\partial _{i_{2}}[\int
dp_{5}(\ \ _{3}^{\shortmid }\widehat{\Upsilon })\mathbf{\ ^{\shortmid }}%
\partial ^{5}\mathbf{\ ^{\shortmid }}h^{6}]}{\ \ _{3}^{\shortmid }\widehat{%
\Upsilon }\mathbf{\ ^{\shortmid }}\partial ^{5}\ \mathbf{^{\shortmid }}h^{6}}%
dx^{i_{2}}\}^{2}  \notag \\
&&+\mathbf{\ ^{\shortmid }}h^{6}\{dp_{5}+[\ _{1}n_{k_{2}}+\
_{2}n_{k_{2}}\int dp_{5}\frac{(\mathbf{\ ^{\shortmid }}\partial ^{5}\mathbf{%
\ ^{\shortmid }}h^{6})^{2}}{|\int dp_{5}\mathbf{\ ^{\shortmid }}\partial
^{5}[\ \ _{3}^{\shortmid }\widehat{\Upsilon }\mathbf{\ ^{\shortmid }}%
h^{6}]|\ (\mathbf{\ ^{\shortmid }}h^{6})^{5/2}}]dx^{k_{2}}\}+  \notag \\
&&\mathbf{\ ^{\shortmid }}\underline{h}^{7}\{dp_{7}+[\ _{1}^{\shortmid }%
\underline{n}_{k_{3}}+\ _{2}^{\shortmid }\underline{n}_{k_{3}}\int dp_{7}%
\frac{(\mathbf{\ ^{\shortmid }}\underline{\partial }^{8}\mathbf{\
^{\shortmid }}\underline{h}^{7})^{2}}{|\int dE\mathbf{\ ^{\shortmid }}%
\underline{\partial }^{8}[\ \ _{4}^{\shortmid }\underline{\widehat{\Upsilon }%
}\mathbf{\ ^{\shortmid }}\underline{h}^{7}]|\ (\mathbf{\ ^{\shortmid }}%
\underline{h}^{7})^{5/2}}]d\mathbf{\ ^{\shortmid }}x^{k_{3}}\}+  \notag \\
&&\frac{(\mathbf{\ ^{\shortmid }}\underline{\partial }^{8}\mathbf{\
^{\shortmid }}\underline{h}^{7})^{2}}{|\int dE\mathbf{\ ^{\shortmid }}%
\underline{\partial }^{8}[\ _{4}\underline{\widehat{\Upsilon }}\mathbf{\
^{\shortmid }}\underline{h}^{7}]|\ \mathbf{\ ^{\shortmid }}\underline{h}^{7}}%
\{dE+\frac{\partial _{i_{3}}[\int dE(\ _{4}\underline{\widehat{\Upsilon }})\ 
\mathbf{\ ^{\shortmid }}\underline{\partial }^{8}\mathbf{\ ^{\shortmid }}%
\underline{h}^{7}]}{\ \ _{4}^{\shortmid }\underline{\widehat{\Upsilon }}\ 
\mathbf{\ ^{\shortmid }}\underline{\partial }^{8}\mathbf{\ ^{\shortmid }}%
\underline{h}^{7}}d\mathbf{\ ^{\shortmid }}x^{i_{3}}\}^{2}.  \notag
\end{eqnarray}%
The locally anisotropic cosmological s-metric (\ref{lc8cstp8}) is an example
of phase space rainbow s-metric (\ref{dmcc8}) \ constructed on $T^{\ast }%
\mathbf{V.}$

\vskip5pt

\textbf{Data Availability Statement:} No data are necessary to be associated
in the manuscript.

\end{document}